\pdfoutput=1
\documentclass[12pt,a4paper]{article}
\usepackage{jheplikemod,tocloft,ifpdf,rotating,amsmath,mathrsfs,multirow,bbm,mathtools,xcolor,xparse,stackengine}
\usepackage{tikz}
\usetikzlibrary{calc,decorations.markings,shapes.geometric,ext.topaths.arcthrough,shapes,decorations.pathmorphing,patterns.meta}
\widowpenalty=500\clubpenalty=1000\unitlength=1mm\textheight=9.05in\hypersetup{pdftitle={},pdfcreator={},linkcolor=[rgb]{0.15,0.35,0.75},colorlinks=true,citecolor=[rgb]{0.675,0,0.2},urlcolor=[rgb]{0.15,0.35,0.65}}
\let\olditemize\itemize\renewcommand{\itemize}{\vspace{-2pt}\olditemize\setlength{\itemsep}{12pt}\setlength{\parskip}{-0pt}\setlength{\parsep}{-8pt}}
\let\oldenumerate\enumerate\renewcommand{\enumerate}{\vspace{-4pt}\oldenumerate\setlength{\itemsep}{1pt}\setlength{\parskip}{0pt}\setlength{\parsep}{0pt}}
\makeatletter\renewcommand\section{\addtocontents{toc}{\protect\addvspace{-2.25\p@}}\@startsection {section}{1}{\z@}{0.5ex \@plus .2ex \@minus 0.2ex}{0.3ex \@plus.1ex\@minus .5ex}{\normalfont\large\bfseries}}
\renewcommand\subsection{\addtocontents{toc}{\protect\addvspace{-2.5\p@}}\@startsection {subsection}{1}{\z@}{0.5ex \@plus .2ex \@minus 0.2ex}{0.3ex \@plus.1ex\@minus .5ex}{\normalfont\bfseries}}
\renewcommand\subsubsection{\addtocontents{toc}{\protect\addvspace{-2.5\p@}}\@startsection {subsubsection}{1}{\z@}{0.5ex \@plus .2ex \@minus 0.2ex}{0.3ex \@plus.1ex\@minus .5ex}{\normalfont\bfseries}}

\newcommand{\eq}[1]{\vspace{-0.5pt}\begin{equation}#1\vspace{-0.5pt}\end{equation}}

\newcommand{\fwbox}[2]{\text{\makebox[#1][c]{$\hspace{-150pt}\displaystyle#2\hspace{-150pt}$}}}
\newcommand{\fwboxL}[2]{\text{\makebox[#1][l]{$#2$}}}
\newcommand{\fwboxR}[2]{\text{\makebox[#1][r]{$#2$}}}
\newcommand{\equivR}{\fwbox{14.5pt}{\hspace{-0pt}\fwboxR{0pt}{\raisebox{0.47pt}{\hspace{1.25pt}:\hspace{-4pt}}}=\fwboxL{0pt}{}}}
\newcommand{\equivL}{\fwbox{14.5pt}{\fwboxR{0pt}{}=\fwboxL{0pt}{\raisebox{0.47pt}{\hspace{-4pt}:\hspace{1.25pt}}}}}
\newcommand{\fig}[3]{\raisebox{#1}{\ensuremath{\vcenter{\hbox{\includegraphics[scale=#2]{#3}}}}}}

\newcommand{\bigger}[1]{\raisebox{-0.95pt}{\scalebox{1.25}{$#1$}}}
\newcommand{\mi}{\raisebox{0.75pt}{\scalebox{0.75}{$\hspace{-0.5pt}\,{-}\,\hspace{-0.5pt}$}}}
\newcommand{\pl}{\raisebox{0.75pt}{\scalebox{0.75}{$\hspace{-0.5pt}\,+\,\hspace{-0.5pt}$}}}
\renewcommand{\phi}{\varphi}
\renewcommand{\bar}{\overline}
\renewcommand{\hat}{\widehat}
\renewcommand{\tilde}{\widetilde}
\renewcommand{\check}{\widehat}

\newcommand{\mone}{\fwboxR{2pt}{\text{-}}1\fwboxL{2pt}{~}}
\newcommand{\dzero}{{\color{dim}0}}
\newcommand{\tmi}{\fwboxR{0pt}{\text{-}}}
\stackMath

\renewcommand{\r}[1]{{\color{hred}#1}}
\newcommand{\g}[1]{{\color{hgreen}#1}}
\renewcommand{\b}[1]{{\color{hblue}#1}}
\renewcommand{\t}[1]{{\color{hteal}#1}}

\newcommand{\indices}[2]{{\hspace{-0.0pt}}^{\smash{#1}}_{\phantom{\smash{#1}}\smash{#2}}}

\definecolor{hblue}{rgb}{0.2353,0.2353,0.60}
\definecolor{hred}{rgb}{0.575,0.0,0.225}
\definecolor{hgreen}{rgb}{0.0,0.5,0.4}
\definecolor{hteal}{rgb}{0.0,0.545,0.7451}
\definecolor{hpurple}{rgb}{0.4431,0.2862,0.7960}

\definecolor{dualblue}{rgb}{0.169,0.22,0.565}
\definecolor{hblue}{rgb}{0,0,0.575}
\definecolor{hred}{rgb}{0.575,0.0,0.225}
\definecolor{dim}{rgb}{0.55,0.55,0.55}
\definecolor{deemph}{rgb}{0.25,0.25,0.25}
\definecolor{rindou1}{rgb}{0.4431,0.2862,0.7960}
\definecolor{rindou2}{rgb}{0.0078,0.1215,0.4392}
\definecolor{lapis}{rgb}{0.0.0470,0.2941,0.5568}
\definecolor{emerald}{rgb}{0.31, 0.78, 0.47}
\definecolor{pinegreen}{rgb}{0.0, 0.47, 0.44}
\definecolor{jade}{rgb}{0.0, 0.66, 0.42}
\definecolor{teal}{rgb}{0.0, 0.5, 0.5}

\NewDocumentCommand{\tikzBox}{O{0pt} m m}{\IfFileExists{figures/#2.pdf}{\raisebox{-#1}{\ensuremath{\vcenter{\hbox{\includegraphics[scale=1]{figures/#2}}}}}}{}}
\newcommand{\dynkLabelK}[2]{\stackunder[0pt]{\text{\rule[-5pt]{0pt}{10pt}{\normalsize$#1$}}}{\rule[-4pt]{0pt}{10pt}\text{{\footnotesize$[#2]$}}}}

\newcommand{\dynkLabel}[2]{\stackunder[0pt]{\text{\rule[-5pt]{0pt}{10pt}{\footnotesize$#1$}}}{\rule[-4pt]{0pt}{10pt}\text{{\footnotesize$[#2]$}}}}

\def\figScale{1}
\def\edgeLength{0.75*\figScale}
\def\lineThickness{(1.5pt)}

\def\dynkEdge{0.75}

\tikzset{phantomEdge/.style={line width=2.5*\lineThickness,rounded corners=3pt,line cap=round,decoration={},postaction={decorate}}}
\tikzset{map/.style={line width=0.65*\lineThickness,line cap=round,rounded corners=5.5pt}}
\tikzset{edge/.style={line width=\lineThickness,line cap=round,rounded corners=5.5pt}}

\tikzset{mapArrow/.style={decoration={markings,mark connection node=arrownode,mark=at position 0.5 with {\node[draw,transform shape, rounded corners=0pt,scale=0.1025*\figScale,shape=dart,aspect=0.5,fill,rounded corners=0pt] (arrownode) {};}},postaction={decorate}}}

\tikzset{midArrow/.style={decoration={markings,mark connection node=arrownode,mark=at position 0.5 with {\node[draw,transform shape, rounded corners=0pt,scale=0.1025*\figScale,shape=dart,aspect=0.5,fill,rounded corners=0pt] (arrownode) {};}},postaction={decorate}}}

\tikzset{midArrowPhantom/.style={decoration={markings,mark connection node=arrownode,mark=at position 0.5 with {\node[draw,transform shape, rounded corners=0pt,scale=0.1025*\figScale,shape=dart,aspect=0.5,fill,rounded corners=0pt] (arrownode) {};}},postaction={decorate}}}

\tikzset{endArrow/.style={decoration={markings,mark connection node=arrownode,mark=at position 0.75 with {\node[draw,transform shape, rounded corners=0pt,scale=0.1025*\figScale,shape=dart,aspect=0.5,fill,rounded corners=0pt] (arrownode) {};}},postaction={decorate}}}

\tikzset{startArrow/.style={decoration={markings,mark connection node=arrownode,mark=at position 0.25 with {\node[draw,transform shape, rounded corners=0pt,scale=0.1025*\figScale,shape=dart,aspect=0.5,fill,rounded corners=0pt] (arrownode) {};}},postaction={decorate}}}

\tikzdeclarepattern{name=shadelines,parameters={\pgfkeysvalueof{/pgf/pattern keys/size},\pgfkeysvalueof{/pgf/pattern keys/angle},\pgfkeysvalueof{/pgf/pattern keys/line width},},bounding box={(0,-0.5*\pgfkeysvalueof{/pgf/pattern keys/line width}) and(\pgfkeysvalueof{/pgf/pattern keys/size},0.5*\pgfkeysvalueof{/pgf/pattern keys/line width})},tile size={(\pgfkeysvalueof{/pgf/pattern keys/size},\pgfkeysvalueof{/pgf/pattern keys/size})},tile transformation={rotate=\pgfkeysvalueof{/pgf/pattern keys/angle}},defaults={size/.initial=5pt,angle/.initial=45,line width/.initial=.4pt,},code={\draw [line width=\pgfkeysvalueof{/pgf/pattern keys/line width}](0,0) -- (\pgfkeysvalueof{/pgf/pattern keys/size},0);},}

\tikzset{bosonProp/.style={decorate, decoration={snake,amplitude=.4mm,segment length=2.2mm,post length=0.05mm},line width=\lineThickness,line cap=round,rounded corners=0.0pt}}
\tikzset{vert/.style={circle,minimum size=2*\figScale,draw=black,line width=0.5*\lineThickness,fill=black,inner sep=1.75pt}}

\tikzset{dynkS/.style={circle,minimum size=10*\figScale,draw=black,line width=0.5*\lineThickness,fill=white,inner sep=1pt}}
\tikzset{dynkL/.style={circle,minimum size=10*\figScale,draw=black,line width=0.5*\lineThickness,fill=black!90,inner sep=1pt}}

\tikzset{squareClebsch/.style={regular polygon,regular polygon sides=4,minimum size=5*\figScale,draw=black,line width=0.5*\lineThickness,fill=black!10,inner sep=1pt}}
\tikzset{clebsch/.style={circle,minimum size=5*\figScale,draw=black,line width=0.5*\lineThickness,fill=black!10,inner sep=1pt}}
\tikzset{clebschR/.style={draw,circle,minimum size=4*\figScale,line width=0.5*\lineThickness,fill,inner sep=1pt}}
\tikzset{clebschD/.style={draw,circle,minimum size=6*\figScale,line width=0.75*\lineThickness,fill=white,inner sep=1pt}}
\tikzset{clebschM/.style={draw,circle,minimum size=4*\figScale,line width=0.5*\lineThickness,fill=black!10,inner sep=1pt}}

\tikzset{smallDot/.style={draw,circle,minimum size=0.25*\figScale,fill,inner sep=0.5pt}}

\newcommand{\floatingEdge}[2]{\draw[#1,white,line width=2.5*\lineThickness]#2;\draw[#1]#2;}
\newcommand{\tbox}[1]{\begin{tikzpicture}[baseline=-2.5pt]#1\end{tikzpicture}}

\newcommand{\adR}{\mathfrak{g}}

\newcommand{\egTensorsAa}{\fig{0pt}{1}{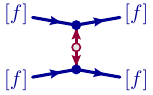}}
\newcommand{\egTensorsAb}{\fig{0pt}{1}{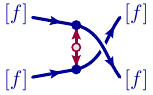}}
\newcommand{\egTensorsAc}{\fig{0pt}{1}{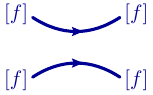}}
\newcommand{\egTensorsAd}{\fig{0pt}{1}{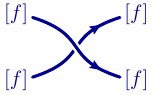}}
\newcommand{\egTensorsAe}{\fig{0pt}{1}{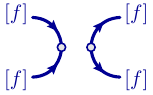}}
\newcommand{\fierzUnwindA}{\fig{0pt}{1}{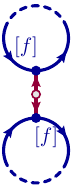}}
\newcommand{\fierzUnwindB}{\fig{0pt}{1}{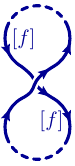}}
\newcommand{\fierzUnwindC}{\fig{0pt}{1}{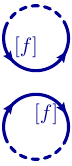}}

\NewDocumentCommand{\arrowTo}{O{black} m O{1}  O{0.5} m o O{0.2} O{-1}} {\tikzset{directedEdge/.style={line width=\lineThickness,rounded corners=3pt,line cap=round,decoration={markings,mark connection node=arrownode,mark=at position \IfNoValueTF{#4}{0.5}{#4} with {\node[draw,transform shape, scale=0.1025*\figScale,shape=dart,aspect=0.5,fill,rounded corners=0pt] (arrownode) {};}},postaction={decorate}}}
\coordinate(end)at($(#2)$);
\coordinate(in)at($(#2)+(#5:#3*\edgeLength)$);
\draw[white,line width=2.5*\lineThickness,rounded corners=3pt,line cap=round]($(in)!.25!(end)$)--($(in)!.75!(end)$);
\draw[#1,directedEdge](in)--(end);
\coordinate(arrowMark)at($(arrownode)+(#8*90+#5:#7)$);
\IfValueTF{#6}{\node[#1]at(arrowMark){{\footnotesize$#6$}};}{}}

\NewDocumentCommand{\arrowFrom}{O{black} m O{1}  O{0.5} m o O{0.2} O{-1} } {\tikzset{directedEdge/.style={line width=\lineThickness,rounded corners=3pt,line cap=round,decoration={markings,mark connection node=arrownode,mark=at position \IfNoValueTF{#4}{0.5}{#4} with {\node[draw,transform shape, scale=0.1025*\figScale,shape=dart,aspect=0.5,fill,rounded corners=0pt] (arrownode) {};}},postaction={decorate}}}
\coordinate(in)at($(#2)$);
\coordinate(end)at($(#2)+(#5:#3*\edgeLength)$);
\draw[white,line width=2.5*\lineThickness,rounded corners=3pt,line cap=round]($(in)!.25!(end)$)--($(in)!.75!(end)$);
\draw[#1,directedEdge](in)--(end);
\coordinate(arrowMark)at($(arrownode)+(#8*90+#5:#7)$);
\IfValueTF{#6}{\node[#1]at(arrowMark){{\footnotesize$#6$}};}{}}

\title{\texorpdfstring{~\\[00pt]{\LARGE \mbox{The Colour Dependence of Amplitudes}}\\[-02pt]}{The Colour Dependence of Amplitudes}}
\author{\vspace{-24pt}Jacob~L.~Bourjaily,}\emailAdd{bourjaily@psu.edu}
\author{\hspace{-2pt}Michael~Plesser,}\emailAdd{plesser@psu.edu}
\author{\hspace{-2pt}Cristian~Vergu}\emailAdd{cfv5175@psu.edu}
\affiliation{Institute for Gravitation and the Cosmos, Department of Physics,\\Pennsylvania State University, University Park, PA 16802, USA}
\abstract{%
~\\[-24pt]
We describe how to construct a spanning set of linearly-independent, automatically orthogonal colour tensors for scattering amplitudes involving coloured particles transforming under {arbitrary} representations of any gauge theory, sufficient to all orders of perturbation theory (or beyond). These tensors are constructed from any choice of a single, trivalent tree graph, with Clebsch-Gordan coefficients at vertices connecting the external particles' representations to internal, irreducible representations via tensor products. We describe how the colour dependence of any Feynman diagram can be systematically decomposed into these bases, and how amplitudes expressed in these bases compare with other choices of tensors such as multi-traces or `$f$-graphs'.
}
\preprint{}

\begin{document}
\maketitle\thispagestyle{empty}

\vspace{-50pt}
%================================================================================================================
%    1. Introduction 
%         
%================================================================================================================
%\newpage
\setcounter{page}{1}\vspace{-4pt}%
\pagenumbering{roman}\clearpage
\pagenumbering{arabic}
\vspace{-0pt}\section{Introduction and Overview}\label{sec:introduction}\vspace{-0pt}
%================================================================================================================
%

Scattering amplitudes in quantum field theory depend on all the observable data describing asymptotic states: their momenta, spin/helicity, and all the observable quantum numbers by which a state can be distinguished. These non-dynamical labels fall broadly into two categories: `colour' and `flavour', corresponding to local (`gauged') and global quantum numbers. In recent decades, a great deal has been learned about the kinematic dependence of scattering amplitudes, leading to revolutionary advances in our understanding of perturbation theory (see \emph{e.g.}~\cite{Bern:1987tw,Bern:1991aq,Bern1991ColorDO,Bern:1993qk,Bern:2004ky,Bern:2004bt,BCF,BCFW}), deep insights into the mathematical structure underlying scattering amplitudes (see \emph{e.g.}~\cite{Arkani-Hamed:2008owk,ArkaniHamed:2009dn,ArkaniHamed:2009sx,ArkaniHamed:2009dg,Arkani-Hamed:2013jha,Arkani-Hamed:2014via,Arkani-Hamed:2016byb}), and great advances in our ability to make perturbative predictions (see \emph{e.g.}~\cite{ArkaniHamed:2010gh,Bourjaily:2013mma,Bourjaily:2015jna,Bourjaily:2017wjl,Bourjaily:2010wh,Bourjaily:2023uln,Bourjaily:2011hi,Bourjaily:2015bpz,Bourjaily:2016evz,He:2024cej,Bourjaily:2025iad}).

The colour-dependence of scattering amplitudes, however, has largely been set aside or taken for granted---as structure to later adorn the more rich dynamics of the kinematic-dependent parts of amplitudes. One notable exception, of course, is the discovery of colour-kinematic duality \cite{BCJ,Bern:2019prr} (see also \emph{e.g.}~\mbox{\cite{Carrasco:2011mn,Carrasco:2012ca,Bjerrum-Bohr:2013iza,Monteiro:2013rya,He:2015wgf,Mogull:2015adi,Johansson:2017bfl,Bern:2017yxu,Bern:2017ucb,Edison:2022jln,Porkert:2022efy}}); but even in this case, the colour dependence of amplitudes is extremely generic---with complete agnosticism about the particulars of any gauge group, exploiting only universal relations such as the Jacobi identity. This is the attitude has been shared by a large fraction of the literature on colour-dressed amplitudes in recent years (see \emph{e.g.}~\mbox{\cite{Arkani-Hamed:2014bca,Bourjaily:2018omh,Bourjaily:2019gqu,Bourjaily:2021hcp,Bourjaily:2021iyq}}).

For amplitudes involving particular gauge groups or matter content, two cases have dominated our understanding: the case of  $\mathfrak{su}_{N_c}$ gauge theory in the large $N_c$ limit, and the phenomenologically important gauge groups of $\mathfrak{su}_{2}$ or $\mathfrak{su}_3$. Outside of these cases, relatively little is known in detail about the colour-dependence of scattering amplitudes today. 

Recently, we explored the sizes of colour tensor bases from the perspective of representation theory \cite{Bourjaily:2024jbt}; in particular, we described how, for any fixed gauge group, the number of independent colour tensors grows at most exponentially (not factorially) in the limit of large multiplicity. While this counting was illuminating, it was far from clear how one could go about directly constructing the tensors required to represent amplitudes involving arbitrary coloured particles---which we address here.\\

\subsection{\emph{Spiritus Movens}: The Colour-Dependence of Scattering Amplitudes}\label{subsec_motivation}

The various external states of any gauge theory must be organized into groups of particles distinguishable (and distinguished) by discrete labels called their `colours'. Consider for example the Feynman rule for the interaction between two fermions labelled by the colours $\{\b{c_1},\b{c_2}\}$ and a gluon with colour denoted by `$\r{a}$':
\eq{\tikzBox{feynman_gauge_matter_vertex}{\coordinate(in)at(-120:\edgeLength);\coordinate(out)at(120:\edgeLength);\coordinate(boson)at(0:\edgeLength);\arrowTo[hblue]{0,0}{-120}\arrowFrom[hblue]{0,0}{120}\draw[hred,bosonProp,line width=\lineThickness,line cap=round](1.0*\edgeLength,0)--(0,0);\node[vert]at(0,0){};\node[]at(-120:1.25*\edgeLength){{\footnotesize$|s_1;\b{c_1}\rangle$}};\node[anchor=west,inner sep=0pt]at(0:1.15*\edgeLength){{\footnotesize$|\mu;\r{a}\rangle$}};\node[]at(120:1.25*\edgeLength){{\footnotesize$|s_2;\b{c_2}\rangle$}};}\;\;\bigger{\Leftrightarrow}\;\;i\,g\,\gamma^{\mu}_{s_1\,s_2}{{T}}^{\smash{\b{c_1}\,\r{a}}}_{\phantom{\smash{c_1\,a}}\,\b{c_2}}\,.\label{rep_in_feynman_rule}}
Here, $g$ is the (gauge) coupling and $T^{\smash{\b{c_1}\,\r{a}}}_{\phantom{\smash{c_1\,a}}\,\b{c_2}}$ is some fixed (non-`running') constant depending on the particular set of colours $\{\b{c_1},\r{a},\b{c_2}\}$ of the particles involved; for each colour of gluon $\r{a}$, we may view the collection of these numbers spanned by the possible colours of fermions as a {matrix}
\eq{\mathbf{T}^{\r{a}}\equivR\mathbf{T}^{\smash{\b{[c]}}\,\r{a}}_{\phantom{\smash{[c]a}}\b{[c]}}\equivR\big\{T^{\smash{\b{c_1}\r{a}}}_{\phantom{\smash{c_1a}}\b{c_2}}\big\}_{\b{c_i}\in\b{[c]}}}
of size dictated by the range of possible colours $\b{c_i}$ for fermions which interact with the gluon of colour $\r{a}$; we denote the range of fermion colours `$\b{[c]}$'. It is a general theorem that for any local, unitary quantum field theory \cite{Benincasa:2007xk}, the collection of these matrices must furnish a \emph{representation} of some Lie algebra $\mathfrak{g}$. We call this representation `$\mathbf{T}$'; the individual matrices $\{\mathbf{T}^{\r{a}}\}_{\r{a}\in\r{[\adR]}}$ are called the \emph{generators} of the representation $\mathbf{T}$; their size is the representation's \emph{dimension}, $\b{[c]}\equivL\{1,\ldots,\mathrm{dim}(\mathbf{T})\}$; and the number of such generators is called the \emph{dimension of the Lie algebra}, denoted $\mathrm{dim}(\mathfrak{g})$; these are indexed by $\r{a}\!\in\!\r{[\adR]}\equivL\{1,\ldots,\mathrm{dim}(\mathfrak{g})\}$. These matrices are always constant, and they completely factor-out from any kinematic dependence of a Feynman diagram.

%\newpage 
Consider a process involving $k$ incoming and $m$ outgoing coloured particles of arbitrary spin and transforming under arbitrary representations of some gauge group's Lie algebra $\mathfrak{g}$:
\eq{|p_1,\epsilon_1,{\b{r}}\rangle\cdots|p_k,\epsilon_k,{\g{s}}\rangle\!\to|p_{\text{-}1},\epsilon_{\text{-}1},{\t{t}}\rangle\cdots|p_{\text{-}m},\epsilon_{\text{-}m},{\r{u}}\rangle\,.}
Any Feynman diagram $\Gamma$ contributing to this process can be expressed as some (gauge/scheme/regularization/renormalization and) kinematic-dependent function $f_{\Gamma}$\footnote{Whatever scale dependence(s) there may be in a running coupling should be viewed as part of $f_{\Gamma}$, as this running affects all interactions equally.} and some colour-\emph{dependent} but kinematic-independent constant $C_{\Gamma}$:\\[-10pt]
\eq{\Gamma\;\mapsto\;C_{\Gamma}{}^{\b{r}\cdots\g{s}}_{\phantom{r\cdots s}\t{t}\cdots\r{u}}\,\,f^\Gamma\big(\{p_1,\epsilon_1\},\ldots,\{p_k,\epsilon_k\};\{p_{\text{-}1},\epsilon_{\text{-}1}\},\ldots,\{p_{\text{-}m},\epsilon_{\text{-}m}\}\big)\,.\vspace{-4pt}}
Summing over all the Feynman diagrams contributing to the process would give the (perhaps regulated and/or renormalized) scattering \emph{amplitude} involving these particular coloured particles:\\[-10pt]
\eq{\fwbox{0pt}{\mathcal{A}\big(|p_1,\epsilon_1,{\b{r}}\rangle\cdots|p_k,\epsilon_k,{\g{s}}\rangle\!\to|p_{\text{-}1},\epsilon_{\text{-}1},{\t{t}}\rangle\cdots|p_{\text{-}m},\epsilon_{\text{-}m},{\r{u}}\rangle\big)=\sum_\Gamma C_\Gamma{}\,f^\Gamma\,.}\vspace{-5pt}\label{particular_coloured_amplitude}}
So far, the separation of each diagram's colour and kinematic dependence is fairly trivial. Modest insight can be found, however, when we realize that because the kinematic part is independent of the colours of the particles involved, the colour-dependent parts are naturally collectable into a rank-$(k,m)$ colour-\emph{tensor}:\\[-10pt]
\eq{\mathbf{C}_\Gamma(\b{\mathbf{R}}\cdots\g{\mathbf{S}}|\t{\mathbf{T}}\cdots\r{\mathbf{U}})\equivR\mathbf{C}_\Gamma{}^{\smash{\b{[r]}\cdots\g{[s]}}}_{\phantom{\smash{[r]\cdots[s]}}\smash{\t{[t]}\cdots\r{[u]}}}\equivR\big\{C_{\Gamma}{}^{\b{r}\cdots\g{s}}_{\phantom{r\cdots s}\t{t}\cdots\r{u}}\big\}_{\substack{
\fwboxL{38pt}{\hspace{0pt}\fwbox{4pt}{\b{r}}\!\in\!\hspace{-0.5pt}\fwboxL{8pt}{\smash{\b{[r]}}},\ldots,\hspace{-0.5pt}\fwbox{4pt}{\g{s}}\!\in\!\hspace{-0.5pt}\fwboxL{8pt}{{\g{[s]}}}}\\
\fwboxL{38pt}{\hspace{0pt}\fwbox{4pt}{\t{t}}\!\in\!\hspace{-0.5pt}\fwboxL{8pt}{{\t{[t]}}},\ldots,\hspace{-0.5pt}\fwbox{4pt}{\r{u}}\!\in\!\hspace{-0.5pt}\fwboxL{8pt}{{\r{[u]}}}}
}}\hspace{20pt}\,.}
Here, we have explicitly indicated the representations associated to each collection of external particles, with ranges of indices spanning each representation's dimension. 

This suggests that we consider the amplitude itself as a tensor, simultaneously encoding the particular amplitudes (\ref{particular_coloured_amplitude}) for each of the possible colours of the external states:
\vspace{-9pt}\eq{\mathcal{A}\big(\b{\mathbf{R}}\cdots\g{\mathbf{S}}|\t{\mathbf{T}}\cdots\r{\mathbf{U}}\big)\equivR\sum_\Gamma \mathbf{C}_\Gamma(\b{\mathbf{R}}\cdots\g{\mathbf{S}}|\t{\mathbf{T}}\cdots\r{\mathbf{U}}) f^\Gamma\,.\label{tensor_amp_0}\vspace{-10pt}}

Because gauge transformations would change both these tensors \emph{and} the corresponding kinematic functions, any such presentation is far from unique. Closely related (but not equivalent) to this non-uniqueness is the fact that the colour tensors $\{\mathbf{C}_\Gamma\}$ themselves satisfy non-trivial, linear relations; these follow both from general aspects of Lie algebra theory (such as the Jacobi identity) but also the {specific} identities that may exist for particular gauge theories and the particular particles' representations.

Suppose that one were to choose a \emph{basis} $\mathcal{B}$ of \emph{linearly-independent} colour tensors. In terms of these, we could express any colour tensor arising from a Feynman graph:\\[-10pt]
\eq{\mathbf{C}_\Gamma(\b{\mathbf{R}}\cdots\g{\mathbf{S}}|\t{\mathbf{T}}\cdots\r{\mathbf{U}})\equivL\sum_{\mathbf{B}\in\mathcal{B}}c_\Gamma^{\mathbf{B}}\,\mathbf{B}(\b{\mathbf{R}}\cdots\g{\mathbf{S}}|\t{\mathbf{T}}\cdots\r{\mathbf{U}})\,.\vspace{-4pt}}
When expanded into this basis, the amplitude (\ref{tensor_amp_0}) becomes expressed in terms of the what are called \emph{partial amplitudes} $A_{\mathbf{B}}$ defined \emph{relative to the basis} according to:
\vspace{-6pt}\eq{\begin{split}\mathcal{A}\big(\b{\mathbf{R}}\cdots\g{\mathbf{S}}|\t{\mathbf{T}}\cdots\r{\mathbf{U}}\big)&=\sum_\Gamma \Big(\sum_{\mathbf{B}\in\mathcal{B}}c_\Gamma^{\mathbf{B}}\,\mathbf{B}(\b{\mathbf{R}}\cdots\g{\mathbf{S}}|\t{\mathbf{T}}\cdots\r{\mathbf{U}})\Big)f^\Gamma\\
&=\sum_{\mathbf{B}\in\mathcal{B}}\mathbf{B}(\b{\mathbf{R}}\cdots\g{\mathbf{S}}|\t{\mathbf{T}}\cdots\r{\mathbf{U}})\Big(\sum_\Gamma c_\Gamma^{\mathbf{B}}\,f^\Gamma\Big)\\
&\equivL\sum_{\mathbf{B}\in\mathcal{B}}\mathbf{B}(\b{\mathbf{R}}\cdots\g{\mathbf{S}}|\t{\mathbf{T}}\cdots\r{\mathbf{U}})\,A_{\mathbf{B}}\,.\\[-14pt]
\end{split}\vspace{-6pt}}
Gauge invariance of the amplitude ensures that partial amplitudes $A_{\mathbf{B}}$ defined with respect to any choice of basis $\mathcal{B}$ will be gauge-invariant: if there were any gauge \emph{non}-invariance in the $A_{\mathbf{B}}$, this would either contradict the gauge-independence of the amplitude or imply linear dependence among the colour tensors $\mathbf{B}\!\in\!\mathcal{B}$. Moreover, it is easy to see that partial amplitudes defined \emph{with respect to \emph{some} basis} are always uniquely well-defined (at least with respect to \emph{that} basis).\\[-5pt]

For many amplitudes involving coloured particles, there exist familiar collections of colour tensors used by physicists for the representation of amplitudes. These include, for example, tensors constructed from traces over generators of some representation (often, the `fundamental') for processes involving gluons, the more refined tensors introduced in \cite{DelDuca:1999rs} for adjoint-scattering at tree-level, and its generalization to cases involving both adjoints and charged fermions in \cite{Johansson:2015oia} (see also \cite{Melia:2015ika,Ochirov:2019mtf}).

Many of these familiar sets of colour tensors, however, are limited in their scope: defined only for particular gauge groups or matter representations; valid to only certain orders of perturbation theory; or only useful perturbatively in the limit of large numbers of independent colours. But even when such tensors are available and useful, they rarely encode an actual \emph{basis}: they are often \emph{incomplete} (perturbatively or non-perturbatively) or fail to be linearly-independent (especially for large multiplicity or low-rank) \cite{Bourjaily:2024jbt}. As we will see, for example, there are far fewer independent colour tensors for almost all processes in $\mathfrak{su}_2$ gauge theory than for $\mathfrak{su}_{N_c\to\infty}$ gauge theory, allowing for substantial improvements in concision and indicating considerable non-uniqueness among the partial amplitudes computed in the general case.

A more pervasive concern about most familiar choices of colour tensors, however, is that they are far from \emph{orthogonal} in the space of particles' colours---meaning that, in the computation of colour-summed amplitudes or cross sections, partial amplitudes interfere with \emph{dense} overlap. This problem was addressed by Zeppenfeld in \cite{Zeppenfeld:1988bz} in the fairly limited case of all-adjoint scattering. Similarly, for amplitudes involving various matter and adjoints in $\mathfrak{a}_{\r{k}}$ gauge theory, there is the `multiplet basis' of \mbox{\cite{Keppeler2012OrthogonalMB,Sjodahl2015,Sjodahl2024}}, which is similar to those described here. 

In this work, we address both of these concerns in the general case, describing the direct construction of \emph{manifestly} independent, non-perturbatively \emph{complete} bases of automatically \emph{colour-orthogonal} colour tensors for processes involving \emph{arbitrary} representations of \emph{any} gauge theory.\\

\subsection{Summary of Main Result: Bases of Orthogonal Colour Tensors}\label{subsec:summary_of_result}

A complete set of linearly independent, mutually colour-orthogonal colour tensors for any process involving coloured particles can be constructed as follows. Let $\Gamma$ denote {any} \emph{fixed} trivalent tree with some \emph{fixed} (but arbitrary) assignment of the relevant particles' representations to the leaves of the graph and some arbitrary choice of orientation to all internal edges. 

For a process involving coloured particles transforming under (not necessarily irreducible) representations $(\mathbf{\b{R}}\,\mathbf{\g{S}}\cdots)\!\to\!(\cdots\mathbf{\t{T}}\,\mathbf{\r{U}})$, say, we may choose, for example, the fixed graph
\eq{\tikzBox{example_graph_unlabelled}{
\coordinate(left)at(-1.75*\edgeLength,0);\coordinate(left2)at(-0.75*\edgeLength,0);\coordinate(right2)at(0.75*\edgeLength,0);\coordinate(right)at(1.75*\edgeLength,0);
\arrowTo[hblue]{left}{-130};\node[anchor=20,inner sep=1.5pt] at(in){{\footnotesize$\b{[r]}$}};
\arrowTo[hgreen]{left}{130};\node[anchor=-20,inner sep=1.5pt] at(in){{\footnotesize$\g{[s]}$}};
\draw[edge,midArrow](left)--(left2);
\draw[edge](left2)--(-0.75*\edgeLength,\edgeLength);
\draw[draw=none,line width=\lineThickness,midArrowPhantom](-0.75*\edgeLength,\edgeLength)--(-0.75*\edgeLength,0.255*\edgeLength);
\draw[edge](right2)--(0.75*\edgeLength,\edgeLength);
\draw[draw=none,line width=\lineThickness,midArrowPhantom](0.75*\edgeLength,0.275*\edgeLength)--(0.75*\edgeLength,\edgeLength);
\draw[edge,dashed](-0.492*\edgeLength,0)--(0.55*\edgeLength,0);
\arrowTo[black]{right}{180}
\arrowFrom[hteal]{right}{50};\node[anchor=-160,inner sep=1.5pt] at(end){{\footnotesize$\t{[t]}$}};
\arrowFrom[hred]{right}{-50};\node[anchor=160,inner sep=1.5pt] at(end){{\footnotesize$\r{[u]}$}};
\node[clebsch]at(left){{\scriptsize$\phantom{}$}};
\node[clebsch]at(left2){{\scriptsize$\phantom{}$}};
\node[clebsch]at(right2){{\scriptsize$\phantom{}$}};
\node[clebsch]at(right){{\scriptsize$\phantom{}$}};
\node[]at(0,0.5*\edgeLength){{\footnotesize$\cdots$}};
}\,.}

This single graph encodes a collection of specific colour tensors according to how it is labelled in the following way. The internal edges of the graph should be labelled by irreducible representations of the Lie algebra of the gauge group, which arise via the decomposition of tensor-products into irreducible representations, as encoded by the vertices of the graph. That is, each vertex of the graph encodes a particular Clebsch-Gordan tensor for the decomposition of a tensor-product-representation into irreducible representations of the gauge group's Lie algebra. If the tensor-product representation `$\mathbf{\b{R}}\otimes\!\mathbf{\g{S}}$' decomposes into the irreducible representation $\mathbf{a}$ some multiplicity of $m\indices{\mathbf{\b{R}}\mathbf{\g{S}}}{\mathbf{a}}$ times---that is, if
\vspace{-4pt}\eq{\mathbf{\b{R}}\!\otimes\!\mathbf{\g{S}}\simeq\underbrace{\mathbf{a}\oplus\cdots\oplus\mathbf{a}}_{\fwboxL{0pt}{\hspace{-7.5pt}\equivL\!m\indices{\mathbf{\b{R}}\mathbf{\g{S}}}{\mathbf{a}}\text{ copies}}}\oplus\cdots\equivL\bigoplus_{\text{irreps}\,\mathbf{a}}\mathbf{a}^{\oplus m\indices{\mathbf{\b{R}}\mathbf{\g{S}}}{\mathbf{a}}}}
---then we must use a discrete label $\mu\!\in\![m\indices{\mathbf{\b{R}}\mathbf{\g{S}}}{\mathbf{a}}]$ to identify which particular Clebsch-Gordan tensor is encoded at the vertex. Because the tensor product representation of any pair of (finite-dimensional) representations will decompose into only a finite number of irreducible representations (each with finite multiplicity), there will be only a finite number of graphs representing non-vanishing tensors.

Thus, for each choice of graph, particular basis tensors will be identified by the assignment of an irreducible representation to each internal edge and a discrete multiplicity index chosen for each vertex. For each set of such labels, we construct a specific colour tensor
\vspace{-10pt}\eq{\mathcal{B}_{\mu\nu\cdots\rho\sigma}^{\hspace{4.5pt}\mathbf{a}\cdots\mathbf{c}}(\b{\mathbf{R}}\,\g{\mathbf{S}}\cdots\!|\!\cdots\t{\mathbf{T}}\,\r{\mathbf{U}})\indices{\b{[r]}\,\g{[s]}\cdots}{\cdots\t{[t]}\,\r{[u]}}\;\;\bigger{\Leftrightarrow}\;\;\tikzBox{graphical_diagram_for_basis}{\coordinate(left)at(-1.75*\edgeLength,0);\coordinate(left2)at(-0.75*\edgeLength,0);\coordinate(right2)at(0.75*\edgeLength,0);\coordinate(right)at(1.75*\edgeLength,0);
\arrowTo[hblue]{left}{-130};\node[anchor=20,inner sep=1.5pt] at(in){{\footnotesize$\b{[r]}$}};
\arrowTo[hgreen]{left}{130};\node[anchor=-20,inner sep=1.5pt] at(in){{\footnotesize$\g{[s]}$}};
\draw[edge,midArrow](left)--(left2);\node[anchor=90,inner sep=2pt]at(arrownode){{\footnotesize$[a]$}};
%\arrowTo[black]{left2}{90};
\draw[edge](left2)--(-0.75*\edgeLength,\edgeLength);
\draw[draw=none,line width=\lineThickness,midArrowPhantom](-0.75*\edgeLength,\edgeLength)--(-0.75*\edgeLength,0.255*\edgeLength);
\draw[edge](right2)--(0.75*\edgeLength,\edgeLength);
\draw[draw=none,line width=\lineThickness,midArrowPhantom](0.75*\edgeLength,0.275*\edgeLength)--(0.75*\edgeLength,\edgeLength);
\draw[edge,dashed](-0.492*\edgeLength,0)--(0.55*\edgeLength,0);
\arrowTo[black]{right}{180}
\node[anchor=90,inner sep=2pt]at(arrownode){{\footnotesize$[c]$}};
\arrowFrom[hteal]{right}{50};\node[anchor=-150,inner sep=1.5pt] at(end){{\footnotesize$\t{[t]}$}};
\arrowFrom[hred]{right}{-50};\node[anchor=150,inner sep=1.5pt] at(end){{\footnotesize$\r{[u]}$}};
\node[clebsch]at(left){{\scriptsize$\phantom{\nu}$}};\node at(left){{\scriptsize$\mu$}};
\node[clebsch]at(left2){{\scriptsize$\phantom{\nu}$}};\node at(left2){{\scriptsize$\nu$}};
\node[clebsch]at(right2){{\scriptsize$\phantom{\nu}$}};\node at(right2){{\scriptsize$\rho$}};
\node[clebsch]at(right){{\scriptsize$\phantom{\nu}$}};\node at(right){{\scriptsize$\sigma$}};
\node[]at(0,0.5*\edgeLength){{\footnotesize$\cdots$}};
}\,.\label{graphical_diagram_for_basis_tensors}\vspace{-10pt}}
The set of all such tensors
\vspace{-4pt}\eq{\mathcal{B}\equivR\!\left\{\mathcal{B}_{\mu\nu\cdots\rho\sigma}^{\hspace{4.5pt}\mathbf{a}\cdots\mathbf{c}}(\b{\mathbf{R}}\,\g{\mathbf{S}}\cdots\!|\!\cdots\t{\mathbf{T}}\,\r{\mathbf{U}})\right\}_{\substack{\fwboxL{0pt}{\mathbf{a},\ldots,\mathbf{c}\!\in\!\text{irreps}}\\\fwboxL{0pt}{\mu\!\in\![m\indices{\mathbf{\b{R}}\mathbf{\g{S}}}{\mathbf{a}}],\ldots}}}\vspace{-5pt}}
form a complete {basis} of linearly-independent, mutually orthogonal colour tensors. 

We will clarify the precise construction of these tensors below, after which their completeness and orthogonality will be straightforward to show. In particular, their orthogonality makes it easy to decompose other choices of colour tensors (such as those arising in the Feynman expansion) into these bases, and the various transformations between different choices of basis will prove to be mathematically rich.

\subsection{Organization and Outline}

This work is organized as follows. In \mbox{section~\ref{sec:notation_and_conventions}} we introduce graphical notation that formalizes the tensors associated with graphs such as (\ref{graphical_diagram_for_basis_tensors}), and make precise the connection between such graphs and similar colour-graphs that have played a key role in the literature such as trace diagrams, `$\r{f}$-graphs', and, more generally, the `birdtrack' diagrams of Cvitanovic \cite{Cvitanovi1976GroupTF}. Although mostly review, \mbox{section~\ref{sec:notation_and_conventions}} will formalize and clarify the particular choices and conventions made by us and used throughout this work. 

We have been careful to avoid making any \emph{unnecessary} choices regarding the representations of Lie algebras such as their unitarity or orthonormality. While it is always \emph{possible} to diagonalize generators of any representation so that, for example,
\vspace{-4pt}\eq{\fwbox{0pt}{\fwboxL{435pt}{\text{[sic]}}}\fwbox{0pt}{\mathrm{tr}_{\mathbf{\b{R}}}(\r{a}\,\r{b})\,\equivR\sum_{\b{r_i}\in\b{[r]}}\mathbf{\b{R}}^{\smash{\b{r_1}\,\r{a}}}_{\phantom{\smash{r_1\,a}}\smash{\b{r_2}}}\mathbf{\b{R}}^{\smash{\b{r_2}\,\r{b}}}_{\phantom{\smash{r_2\,b}}\smash{\b{r_1}}}\propto\delta^{\r{a\,b}}\,,}\nonumber\vspace{-6pt}}
we do \emph{not assume that the generators have been chosen to make this true}. (The reader is free to choose her generators however she likes.) We discuss our reasons for not making such common assumptions in \mbox{appendix~\ref{appendix:concrete_representations}}. But the effect of not making this choice is a necessary degree of care and precision in notation not often visible in discussions of representation theory in the physics literature. As such, while much of the material in \mbox{section~\ref{sec:notation_and_conventions}} may seem pedantic, it allows us to sharpen and clarify many subtleties which we expect may be clarifying to some readers. 

We review basic aspects of (matrix) representations of Lie algebras in \mbox{section~\ref{subsec:lie_algebra_representations}}, the adjoint representation and structure constants in \mbox{section~\ref{subsubsec:adjoint_representation_and_structure_constants}}, and discuss conjugate representations and similarity transformations in \mbox{section~\ref{subsubsec:conjugation_and_similarity}}. In \mbox{section~\ref{subsec:products_and_projections}}, we detail the construction of tensor-product representations and Clebsch-Gordan coefficients, and how these can be sewn together according to graphs. Importantly, it turns out that virtually all colour tensors used by physicists can be understood as constructible in this way, allowing us to connect the graphical notation used here to other conventions for encoding colour tensors that have appeared in the literature; this is described in \mbox{section~\ref{subsec:everything_is_clebsch}}. Upon formalizing the notions of conjugation and  contraction between conjugate tensors in \mbox{section~\ref{subsec:vacuum_graphs_and_contractions}}, we will have all the tools required for our main results described in \mbox{section~\ref{sec:clebsch_Gordan_colour_tensors}}.

In \mbox{section~\ref{subsec:general_basis_construction}}, we provide a more detailed description of how the colour tensor bases outlined in \ref{subsec:summary_of_result} above are constructed, and provide explicit proofs of their non-perturbative completeness and colour-orthogonality. The new colour tensor bases we describe are constructed from some choice of some fixed underlying trivalent tree, with some particular choice of assigning the particles' representations to external edges. Different choices of graphs or different choices of assignments of representations to its leaves result in generally distinct bases; the relation between bases constructed from different graphs is described in \mbox{section~\ref{subsec:duality_and_crossing}} together with a broader, `categorical' view of how these objects can be viewed. In \mbox{section~\ref{subsec:decomposition_and_reduction}} we describe how arbitrary colour tensors appearing in the Feynman expansion, say, can be systematically reduced into these new colour bases. 

In \mbox{section~\ref{sec:other_colour_bases}} we describe a number of concrete examples of these novel colour tensors. To compare these against more familiar choices of colour tensors used in the representation of scattering amplitudes,  in \mbox{section \ref{subsec:review_of_familiar_colour_tensors}} we define and outline a number of particular examples---namely, `$f$-graphs' for adjoint scattering and the tree-level tensors chosen by DDM in \cite{DelDuca:1999rs}, multi-trace tensors (and the limit of large-rank gauge groups), and the generalization of DDM to the case scattering adjoints and variously-charged matter representations of \cite{Johansson:2015oia} (see also \cite{Melia:2015ika}). With this formalism and notation in place, the rest of \mbox{section~\ref{sec:other_colour_bases}} describes in considerable detail the construction of our new bases relevant for particular cases, comparing against more familiar tensors. We conclude with a discussion of open problems and future applications in \mbox{section~\ref{sec:conclusions}}. 

For the sake of clarity, completeness, and concreteness, we have included two short appendices. In the first of these, \mbox{appendix~\ref{appendix:concrete_representations}}, we outline the construction of concrete tensors we have constructed in \textsc{Mathematica} to verify the details of our examples. For many of the concrete examples discussed in \mbox{section~\ref{sec:other_colour_bases}}, we make reference to specific irreducible representations of simple Lie algebras, the identification of which is done using Dynkin labels which are defined only with respect to a specific ordering of roots in a particular weight system. We clarify the choices made by us for the identification of these irreducible representations in \mbox{appendix~\ref{appendix:weight_system_conventions}}.

\newpage
\section[Building Colour Tensors: Notation and Graphical Conventions]{Colour Tensors: Notation \& Graphical Conventions}\label{sec:notation_and_conventions}

The purpose of this section is to review some basic aspects of Lie algebra representations, to clarify specific conventions used by us here, and to introduce some useful graphical notation. Such graphical notation is common in the discussion of Lie algebra representations (see e.g.~\cite{Cvitanovi1976GroupTF}) and we have been careful to ensure that our conventions cohere with many of those commonly encountered in physics literature. Readers interested in more background on the theory of Lie algebra representations should consult, \emph{e.g.} \cite{Slansky:1981yr,Cornwell:1997ke,DiFrancesco:1997nk,fulton2013}.

We are careful to make clear a number of subtleties that are rarely emphasized or otherwise avoided through convenient and common choices of convention. For example, while it is always possible to choose a linear combination of generators for any representation (of any semi-simple Lie algebra) so that the Killing metric is diagonal (rendering the adjoint representation's generators proportional to those of its conjugate), we do not assume this choice has been made. Allowing for this more general situation requires a deal of care and precision in our discussion, and we hope that some readers find this clarifying. 

Although the material discussed in this section may not be required for many readers, we hope that provides clarity and precision to the examples discussed in the subsequent sections.\\

\subsection{\emph{Representing} the Representations of Lie Algebras}\label{subsec:lie_algebra_representations}

We consider any (matrix) \emph{representation} `$\mathbf{\b{R}}$' of a Lie algebra $\mathfrak{g}$ as being associated with a collection of particular matrices or `{generators}' $\mathbf{\b{R}}^{\smash{\r{[\adR]}}}\!\equivL\{\mathbf{\b{R}}^{\r{a}}\}_{\r{a}\in\r{[\adR]}}$ which encode how this representation acts on the basis elements of some particular vector space. These matrices are indexed by $\r{a}\!\in\!\r{[\adR]}\equivR\!\{1,\ldots,\mathrm{dim}(\mathfrak{g})\}$ and satisfy the property that the ordinary matrix commutator of any pair of generators lies within the {linear span} of the representation's generators. That is, for every pair of generators labelled by $\r{a},\r{b}\!\in\!\r{[\adR]}$, we have 
\eq{\big[\mathbf{\b{R}}^{\r{a}},\mathbf{\b{R}}^{\r{b}}\big]\equivR\mathbf{\b{R}}^{\r{a}}\!.\mathbf{\b{R}}^{\r{b}}-\mathbf{\b{R}}^{\r{b}}\!.\mathbf{\b{R}}^{\r{a}}\equivL\, \sum_{\t{c}\,\in\r{[\adR]}}\r{{f}}^{\r{a\,b}}_{\phantom{a\,b\,}\t{c}}\,\mathbf{\b{R}}^{\t{c}}\,,\label{defining_commutation_relations}}%\equivR\sum_{\t{c}\in\r{[\adR]}}\,,}
for some set of coefficients---called `\emph{structure constants}'---denoted `$\r{f}^{\r{a\,b}}_{\phantom{a\,b\,}\r{c}}$' above. Notice that we have used `$.$' to denote ordinary matrix multiplication. From the definition of the structure constants, it is obvious that $\r{f}^{\r{a\,b}}_{\phantom{a\,b\,}\t{c}}={-}\r{f}^{\r{b\,a}}_{\phantom{a\,b\,}\t{c}}$ for all $\r{a},\r{b}\!\in\!\r{[\adR]}$, but no further symmetry is required or assumed. Letting `$\b{[r]}$' collectively denote the range of indices $\b{[r]}\equivL\{1,\ldots,\mathrm{dim}(\mathbf{\b{R}})\}$ for each generator (viewed as a matrix of coefficients relative to some choice of {basis}), we may consider the representation $\mathbf{\b{R}}$ as given by a rank-three (or rank-`$(2,1)$') tensor
\eq{\mathbf{\b{R}}\;\bigger{\Leftrightarrow}\;\mathbf{\b{R}}^{\smash{\b{[r]}\r{[\adR]}}}_{\phantom{\smash{[r][\adR]}}\smash{\b{[r]}}}\equivR\!\left\{\mathbf{\b{R}}^{\b{r_1}\,\r{a}}_{\phantom{r_1\,a}\b{r_2}}\right\}_{\substack{\b{r_i}\in\b{[r]}\\\r{{}_{\phantom{i}}a}\in\r{[\adR]}}}\,;}
in terms of this tensor, matrix multiplication appearing in (\ref{defining_commutation_relations}) would be defined by 
\eq{\mathbf{\b{R}}^{\r{a}}\!.\mathbf{\b{R}}^{\r{b}}\equivR\!\sum_{\b{r}\in\b{[r]}}\b{\mathbf{R}}^{\,\b{\smash{[r]}}\r{a}}_{\phantom{\,\smash{[r]}a\,}\b{r}}\,\b{\mathbf{R}}^{\,\b{r}\,\r{b}}_{\phantom{\,r\,b\,}\b{\smash{[r]}}}\,.\label{dot_prod_defined}}

It is rather useful to introduce a diagrammatic notation for tensors such as $\mathbf{\b{R}}$ and $\r{f}$; we choose to represent these tensors by the graphs
\eq{\b{\mathbf{R}}^{\smash{\b{[r]}\r{[\adR]}}}_{\phantom{\smash{[r][\adR]\,}}\smash{\b{[r]}}}\;\bigger{\Leftrightarrow}\;\tikzBox{rep_vertex}{\arrowTo[hblue]{0,0}{200}\node[anchor=20,inner sep=0pt] at(in){{\footnotesize$\b{[r]}$}};\arrowTo[hred]{0,0}{90}\node[anchor=-90,inner sep=2pt] at(in){{\footnotesize$\r{[\adR]}$}};\arrowFrom[hblue]{0,0}{-20}\node[anchor=160,inner sep=0pt] at(end){{\footnotesize$\b{[r]}$}};\node[hblue,clebschR]at(0,0){};
}\,\quad\text{and}\quad\r{{f}}^{\smash{\r{[\adR]}}\r{[\adR]}}_{\phantom{\smash{[\adR][\adR]\,}}\smash{\r{[\adR]}}}\;\bigger{\Leftrightarrow}\;\tikzBox{f_vertex}{\arrowTo[hred]{0,0}{200}\node[anchor=20,inner sep=0pt] at(in){{\footnotesize$\r{[\adR]}$}};\arrowTo[hred]{0,0}{90}\node[anchor=-90,inner sep=2pt] at(in){{\footnotesize$\r{[\adR]}$}};\arrowFrom[hred]{0,0}{-20}\node[anchor=160,inner sep=0pt] at(end){{\footnotesize$\r{[\adR]}$}};\node[hred,clebschR]at(0,0){};
}\,.\label{repn_vertex_diagram}}
Here, we distinguish raised/lowered tensor arguments by incoming/outgoing arrows, respectively, and use colour (or other labeling) to identify the ranges of arguments allowed for each `slot' of the tensor. The order of arguments of the tensor can be read-off from the figure, starting at the lower-left, reading clockwise around the diagram. (This convention is occasionally cumbersome, as we will see.)

Diagrammatically, the matrix product of any \emph{particular} pair of generators (\ref{dot_prod_defined}) would then be encoded as\\[-14pt]
\eq{\mathbf{\b{R}}^{\r{a}}\!.\mathbf{\b{R}}^{\r{b}}\;\bigger{\Leftrightarrow}\;
\tikzBox[-10pt]{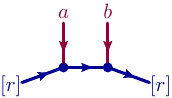}{\arrowTo[hblue]{0,0}{200}\node[anchor=20,inner sep=0pt] at(in){{\footnotesize$\b{[r]}$}};\arrowTo[hred]{0,0}{90}\node[anchor=-90,inner sep=2pt] at(in){{\footnotesize$\r{a}$}};\arrowFrom[hblue]{0,0}{0}%\node[anchor=160,inner sep=0pt] at(end){{\footnotesize$\b{[r]}$}};
\arrowFrom[hblue]{end}{-20}\node[anchor=160,inner sep=0pt] at(end){{\footnotesize$\b{[r]}$}};
\arrowTo[hred]{in}{90}\node[anchor=-90,inner sep=2pt] at(in){{\footnotesize$\r{b}$}};
\node[hblue,clebschR]at(0,0){};\node[hblue,clebschR]at(\edgeLength,0){};
}\,.\label{matrix_product_diagram}}
Notice that we have used an internal edge to denote the sum over indices $\b{r}\!\in\!\b{[r]}$. Such notation is naturally generalized to represent the entire rank-four tensor\\[-14pt]
\eq{\mathbf{\b{R}}.\mathbf{\b{R}}\;\bigger{\Leftrightarrow}\;\;(\mathbf{\b{R}}.\mathbf{\b{R}})^{\smash{\b{[r]}\r{[\adR][\adR]}}}_{\phantom{\smash{[r][\adR][\adR]}}\,\smash{\b{[r]}}}\;\bigger{\Leftrightarrow}\;
\tikzBox[-10pt]{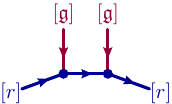}{\arrowTo[hblue]{0,0}{200}\node[anchor=20,inner sep=0pt] at(in){{\footnotesize$\b{[r]}$}};\arrowTo[hred]{0,0}{90}\node[anchor=-90,inner sep=2pt] at(in){{\footnotesize$\r{[\adR]}$}};\arrowFrom[hblue]{0,0}{0}%\node[anchor=160,inner sep=0pt] at(end){{\footnotesize$\b{[r]}$}};
\arrowFrom[hblue]{end}{-20}\node[anchor=160,inner sep=0pt] at(end){{\footnotesize$\b{[r]}$}};
\arrowTo[hred]{in}{90}\node[anchor=-90,inner sep=2pt] at(in){{\footnotesize$\r{[\adR]}$}};
\node[hblue,clebschR]at(0,0){};\node[hblue,clebschR]at(\edgeLength,0){};
}\,;\label{matrix_product_tensor_diagram}}
We run into a problem, however, when trying to upgrade this notation to the commutator, as our convention requires that the ordering of indices of the tensor be encoded clockwise around the diagram; that is, our conventions require that we encode the commutator\\[-14pt]
\eq{\big[\mathbf{\b{R}},\mathbf{\b{R}}\big]\;\bigger{\Leftrightarrow}\;\big[\mathbf{\b{R}},\mathbf{\b{R}}\big]^{\smash{\b{[r]}\r{[\adR][\adR]}}}_{\phantom{\smash{[r][\adR][\adR]}}\,\smash{\b{[r]}}}\;\bigger{\Leftrightarrow}\;
\tikzBox[-5pt]{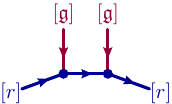}{\coordinate(v1)at(0,0);\coordinate(v2)at(\edgeLength,0);
\arrowTo[hblue]{v1}{200}\node[anchor=20,inner sep=0pt] at(in){{\footnotesize$\b{[r]}$}};\arrowTo[hred]{v1}{90}\node[anchor=-90,inner sep=2pt] at(in){{\footnotesize$\r{[\adR]}$}};\arrowFrom[hblue]{v1}{0}\arrowFrom[hblue]{end}{-20}
\node[anchor=160,inner sep=0pt] at(end){{\footnotesize$\b{[r]}$}};\arrowTo[hred]{v2}{90}\node[anchor=-90,inner sep=2pt] at(in){{\footnotesize$\r{[\adR]}$}};
\node[hblue,clebschR](c1)at(v1){};
\node[hblue,clebschR](c2)at(v2){};
}-\tikzBox[-5pt]{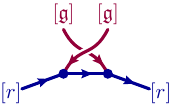}{\coordinate(v1)at(0,0);\coordinate(v2)at(\edgeLength,0);\coordinate(a)at(0,\edgeLength);\coordinate(b)at(\edgeLength,\edgeLength);
\arrowTo[hblue]{v1}{200}\node[anchor=20,inner sep=0pt] at(in){{\footnotesize$\b{[r]}$}};\floatingEdge{hred,edge,endArrow}{(a).. controls ($(v1)+(0.25,0.25)$) and ($(b)-(0.25,0.25)$) .. (v2);}
\floatingEdge{hred,edge,endArrow}{(b).. controls ($(v2)+(-.25,.25)$) and ($(a)+(.25,-.25)$) .. (v1);}\node[anchor=-90,inner sep=2pt] at(b){{\footnotesize$\r{[\adR]}$}};\arrowFrom[hblue]{v1}{0}\arrowFrom[hblue]{end}{-20}
\node[anchor=160,inner sep=0pt] at(end){{\footnotesize$\b{[r]}$}};
\node[anchor=-90,inner sep=2pt] at(a){{\footnotesize$\r{[\adR]}$}};
\node[hblue,clebschR](c1)at(v1){};
\node[hblue,clebschR](c2)at(v2){};
}\,.
\label{commutator_tensor_diagram}}

With these building blocks, the defining property of any (matrix) \emph{representation} $\mathbf{\b{R}}$ given in (\ref{defining_commutation_relations}) would translate to the statement that there exists a rank-three tensor $\r{f}$ such that\\[-14pt]
\eq{\tikzBox{r_dot_r_tensor_2}{\coordinate(v1)at(0,0);\coordinate(v2)at(\edgeLength,0);
\arrowTo[hblue]{v1}{200}\node[anchor=20,inner sep=0pt] at(in){{\footnotesize$\b{[r]}$}};\arrowTo[hred]{v1}{90}\node[anchor=-90,inner sep=2pt] at(in){{\footnotesize$\r{[\adR]}$}};\arrowFrom[hblue]{v1}{0}\arrowFrom[hblue]{end}{-20}
\node[anchor=160,inner sep=0pt] at(end){{\footnotesize$\b{[r]}$}};\arrowTo[hred]{v2}{90}\node[anchor=-90,inner sep=2pt] at(in){{\footnotesize$\r{[\adR]}$}};
\node[hblue,clebschR](c1)at(v1){};
\node[hblue,clebschR](c2)at(v2){};
}-\tikzBox{r_dot_r_tensor_twist}{\coordinate(v1)at(0,0);\coordinate(v2)at(\edgeLength,0);\coordinate(a)at(0,\edgeLength);\coordinate(b)at(\edgeLength,\edgeLength);
\arrowTo[hblue]{v1}{200}\node[anchor=20,inner sep=0pt] at(in){{\footnotesize$\b{[r]}$}};\floatingEdge{hred,edge,endArrow}{(a).. controls ($(v1)+(0.25,0.25)$) and ($(b)-(0.25,0.25)$) .. (v2);}
\floatingEdge{hred,edge,endArrow}{(b).. controls ($(v2)+(-.25,.25)$) and ($(a)+(.25,-.25)$) .. (v1);}\node[anchor=-90,inner sep=2pt] at(b){{\footnotesize$\r{[\adR]}$}};\arrowFrom[hblue]{v1}{0}\arrowFrom[hblue]{end}{-20}
\node[anchor=160,inner sep=0pt] at(end){{\footnotesize$\b{[r]}$}};
\node[anchor=-90,inner sep=2pt] at(a){{\footnotesize$\r{[\adR]}$}};
\node[hblue,clebschR](c1)at(v1){};
\node[hblue,clebschR](c2)at(v2){};
}%
=
\tikzBox[0.75pt]{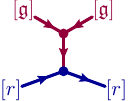}{\coordinate(v1)at(0,0);\coordinate(v2)at(\edgeLength,0);\coordinate(v3)at(0,0.85*\edgeLength);\draw[hred,edge,midArrow](v3)--(v1);
\arrowTo[hblue]{v1}{200}\node[anchor=20,inner sep=0pt] at(in){{\footnotesize$\b{[r]}$}};
\arrowTo[hred]{v3}[0.75]{30}\node[anchor=210,inner sep=0pt] at(in){{\footnotesize$\r{[\adR]}$}};
\arrowTo[hred]{v3}[0.75]{150}\node[anchor=-30,inner sep=0pt] at(in){{\footnotesize$\r{[\adR]}$}};
\arrowFrom[hblue]{v1}{-20}\node[anchor=160,inner sep=0pt] at(end){{\footnotesize$\b{[r]}$}};
\node[hblue,clebschR](c1)at(v1){};
\node[hred,clebschR](c3)at(v3){};
}.
\label{diagrammatic_defn_of_rep}}

To be clear, \emph{any} rank-three tensor satisfying the property that $[\mathbf{\b{R}},\mathbf{\b{R}}]\in\mathrm{span}(\mathbf{\b{R}})$ defines a representation of some Lie algebra; the set of coefficients $\r{f}$ which encode this linear relationship are uniquely determined. 

\paragraph{\emph{Concrete} Representations}~\\[-14pt]

To be clear, we consider a \emph{concrete} representation $\mathbf{\b{R}}$ to be some \emph{explicit} rank-three tensor---that is, not merely some $\mathrm{dim}(\mathbf{\b{R}})$-dimensional vector space \emph{upon which} the algebra acts; but the action of a particular, ordered list of generators on a specific choice of basis for some  $\mathrm{dim}(\mathbf{\b{R}})$-dimensional vector space. That is, we have in mind a literal array of $\mathrm{dim}(\mathbf{\b{R}})\!\times\!\mathrm{dim}(\r{\mathfrak{g}})\!\times\!\mathrm{dim}(\mathbf{\b{R}})$ \emph{numbers} (usually, integers).

Given any concrete representation (such that no linear combination of generators is identically zero), it is straightforward to determine the structure constants $\r{f}\indices{\r{a\,b}}{\t{c}}$ by simple linear algebra. Importantly, these are \emph{uniquely} determined by $\mathbf{\b{R}}$ provided it satisfies the defining property of a representation---namely, that $[\mathbf{\b{R}},\mathbf{\b{R}}]\!\in\!\mathrm{span}(\mathbf{\b{R}})$. While this statement would be left invariant if different linear combinations of generators were chosen via
\eq{\mathbf{\b{\tilde{R}}}\indices{\r{[\tilde{\smash{\adR}}]}}{}\equivR\sum_{\r{a}\in\r{[\adR]}}\mathbf{Q}\indices{\r{[\tilde{\smash{\adR}}]}}{\r{a}}\mathbf{\b{R}}^{\r{a}}\vspace{-6pt}}
for any $\mathbf{Q}\!\in\!\mathrm{GL}(\mathrm{dim}(\r{\adR}))$, the resulting structure constants would generally be different. And so, when discussing \emph{other} possible representations of $\mathfrak{g}$, we require that these satisfy (\ref{diagrammatic_defn_of_rep}) with \emph{identical structure constants}; and thus, the particular ordering and scaling of generators of different representations must cohere with one another. In contrast, representations related to one another by similarity transformations on their target-space indices automatically satisfy commutation relations with identical structure constants (as we review momentarily).\\

%\newpage
\subsubsection{Structure Constants and the {Adjoint} Representation}\label{subsubsec:adjoint_representation_and_structure_constants}

Given any particular (or concrete) representation $\mathbf{\b{R}}$, there will always be an infinity of distinct representations which satisfy the same commutation relations with identical structure constants. One of the most important of these is the {adjoint} representation, denoted `$\mathbf{\r{ad}}$' identified directly with the structure constants $\r{f}$ themselves:
\eq{\mathbf{\r{ad}}\indices{\r{[\adR][\adR]}}{\r{[\adR]}}\,\equivR\,\r{{f}}^{\smash{\r{[\adR]}}\r{[\adR]}}_{\phantom{\smash{[\adR][\adR]\,}}\smash{\r{[\adR]}}}\;\bigger{\Leftrightarrow}\;\tikzBox{f_tensor_diagram}{\arrowTo[hred]{0,0}{200}\node[anchor=20,inner sep=0pt] at(in){{\footnotesize$\r{[\adR]}$}};\arrowTo[hred]{0,0}{90}\node[anchor=-90,inner sep=2pt] at(in){{\footnotesize$\r{[\adR]}$}};\arrowFrom[hred]{0,0}{-20}\node[anchor=160,inner sep=0pt] at(end){{\footnotesize$\r{[\adR]}$}};\node[hred,clebschR]at(0,0){};
}\,.}
To see that this rank-three tensor \emph{also} satisfies commutation relations with \emph{identical} coefficients, we merely note that the identity
\eq{\big[[\mathbf{A},\mathbf{B}],\mathbf{C}\big]-\big[[\mathbf{A},\mathbf{C}],\mathbf{B}\big]=\big[\mathbf{A},[\mathbf{B},\mathbf{C}]\big]\,}
holds for \emph{arbitrary} matrices $\mathbf{A},\mathbf{B},\mathbf{C}$---and follows telescopically from the the definition of the commutator. Applying this identity to the generators of some particular representation $\mathbf{\b{R}}$, and using the defining property (\ref{defining_commutation_relations}) shows that, for all triples of generators labelled by $\{\r{a},\r{b},\r{c}\}$,
\eq{\begin{split}
\big[[\mathbf{\b{R}}^{\r{a}},\mathbf{\b{R}}^{\r{b}}],\mathbf{\b{R}}^{\r{c}}\big]-\big[[\mathbf{\b{R}}^{\r{a}},\mathbf{\b{R}}^\r{c}],\mathbf{\b{R}}^{\r{b}}\big]&=\big[\mathbf{\b{R}}^{\r{a}},[\mathbf{\b{R}}^\r{b},\mathbf{\b{R}}^{\r{c}}]\big]\\
=\sum_{e\in\r{[\adR]}}\Big(\r{f}\indices{\r{a}\,\r{b}}{e}\big[\mathbf{\b{R}}^{e},\mathbf{\b{R}}^{\r{c}}\big]-\r{f}\indices{\r{a}\,\r{c}}{e}\big[\mathbf{\b{R}}^{e},\mathbf{\b{R}}^{\r{b}}\big]\Big)&=\sum_{e\in\r{[\adR]}}\big[\mathbf{\b{R}}^{\r{a}},\mathbf{\b{R}}^{{e}}\big]\r{f}\indices{\r{b}\,\r{c}}{e}\\
=\sum_{\substack{e\in\r{[\adR]}\\\t{d}\in\r{[\adR]}}}\Big(\r{f}\indices{\r{a}\,\r{b}}{e}\r{f}\indices{{e}\,\r{c}}{\t{d}}\mathbf{\b{R}}^\t{d}-\r{f}\indices{\r{a}\,\r{c}}{e}\r{f}\indices{{e}\,\r{b}}{\t{d}}\mathbf{\b{R}}^\t{d})&=\sum_{\substack{e\in\r{[\adR]}\\\t{d}\in\r{[\adR]}}}\r{f}\indices{\r{b}\,\r{c}}{e}\r{f}\indices{\r{a}\,{e}}{\t{d}}\mathbf{\b{R}}^\t{d}\,.
\end{split}}
Provided no $\mathbf{\b{R}}^{\t{d}}$ is identically zero, this implies the \emph{Jacobi identity} 
\eq{\sum_{\substack{e\in\r{[\adR]}}}\Big(\r{f}\indices{\r{a}\,\r{b}}{e}\r{f}\indices{{e}\,\r{c}}{\t{d}}-\r{f}\indices{\r{a}\,\r{c}}{e}\r{f}\indices{{e}\,\r{b}}{\t{d}}-\r{f}\indices{\r{b}\,\r{c}}{e}\r{f}\indices{\r{a}\,{e}}{\t{d}}\big)=0\,.}
This is equivalent to the diagrammatic statement that the structure constants $\r{f}$ themselves satisfy
\eq{\tikzBox{adjoint_jacobi_1}{\coordinate(v1)at(0,0);\coordinate(v2)at(\edgeLength,0);
\arrowTo[hred]{v1}{200}\node[anchor=20,inner sep=2pt] at(in){{\footnotesize$\r{a}$}};\arrowTo[hred]{v1}{90}\node[anchor=-90,inner sep=2pt] at(in){{\footnotesize$\r{b}$}};\arrowFrom[hred]{v1}{0}\arrowFrom[hred]{end}{-20}
\node[anchor=160,inner sep=2pt] at(end){{\footnotesize$\t{d}$}};\arrowTo[hred]{v2}{90}\node[anchor=-90,inner sep=2pt] at(in){{\footnotesize$\r{c}$}};
\node[hred,clebschR](c1)at(v1){};
\node[hred,clebschR](c2)at(v2){};
}-\tikzBox{adjoint_jacobi_2}{\coordinate(v1)at(0,0);\coordinate(v2)at(\edgeLength,0);\coordinate(a)at(0,\edgeLength);\coordinate(b)at(\edgeLength,\edgeLength);
\arrowTo[hred]{v1}{200}\node[anchor=20,inner sep=2pt] at(in){{\footnotesize$\r{a}$}};\floatingEdge{hred,edge,endArrow}{(a).. controls ($(v1)+(0.25,0.25)$) and ($(b)-(0.25,0.25)$) .. (v2);}
\floatingEdge{hred,edge,endArrow}{(b).. controls ($(v2)+(-.25,.25)$) and ($(a)+(.25,-.25)$) .. (v1);}\node[anchor=-90,inner sep=2pt] at(b){{\footnotesize$\r{c}$}};\arrowFrom[hred]{v1}{0}\arrowFrom[hred]{end}{-20}
\node[anchor=160,inner sep=2pt] at(end){{\footnotesize$\t{d}$}};
\node[anchor=-90,inner sep=2pt] at(a){{\footnotesize$\r{b}$}};
\node[hred,clebschR](c1)at(v1){};
\node[hred,clebschR](c2)at(v2){};
}%
=
\tikzBox{adjoint_jacobi_3}{\coordinate(v3)at(0,0.75*\edgeLength);\coordinate(v1)at(0,0);\coordinate(v2)at(\edgeLength,0);\draw[hred,edge,midArrow](v3)--(v1);
\arrowTo[hred]{v1}{200}\node[anchor=20,inner sep=2pt] at(in){{\footnotesize$\r{a}$}};
\arrowTo[hred]{v3}[0.75]{30}\node[anchor=210,inner sep=2pt] at(in){{\footnotesize$\r{c}$}};
\arrowTo[hred]{v3}[0.75]{150}\node[anchor=-30,inner sep=2pt] at(in){{\footnotesize$\r{b}$}};
\arrowFrom[hred]{v1}{-20}\node[anchor=160,inner sep=2pt] at(end){{\footnotesize$\t{d}$}};
\node[hred,clebschR](c1)at(v1){};
\node[hred,clebschR](c3)at(v3){};
}
\label{diagrammatic_check_of_ad_is_rep}}
and therefore furnish another (possibly distinct) representation, called `$\mathbf{\r{ad}}$'.\\

\subsubsection{Similarity Transformations and Conjugate Representations}\label{subsubsec:conjugation_and_similarity}

Although we always have some concrete representation $\mathbf{\b{R}}$ in mind to define our structure constants and the adjoint representation, any change of basis---or \emph{similarity transformation}---of the $\mathrm{dim}(\mathbf{\b{R}})$-dimensional space on which the generators act would also constitute a representation: namely, different matrices that would satisfy the same commutation relations with identical structure constants. Because of this, it is common to consider a representation as a `module' acting abstractly on some vector space \emph{without} a basis chosen; for our purposes, however, {the} various colours of particles make reference to some specific choice of basis, and the scattering amplitude between certain coloured particles is not the same thing as one involving some linear combination of these coloured particles. As such, some care is required to discuss how representations reflect this choice of basis. 

A change of basis from coordinates indexed by $\b{[r]}$ to one indexed by $\g{[r']}$ would be accomplished by some invertible matrix 
\eq{\mathbf{M}\indices{\b{[r]}}{\g{[r']}}\;\bigger{\Leftrightarrow}\;\tikzBox{change_of_basis_M_vertex}{\node(v0)at(0,0){$\fwbox{10pt}{~}$};\arrowTo[hblue]{v0.180}[1]{180}\node[anchor=0,inner sep=0pt] at(in){{\footnotesize$\b{[r]}$}};
\arrowFrom[hgreen]{v0.0}[1]{0}\node[anchor=180,inner sep=0pt] at(end){{\footnotesize$\g{[r']}$}};\node[clebschM](v0)at(0,0){$\fwbox{13pt}{\mathbf{M}}$};
}}
with an inverse denoted $\mathbf{\bar{M}}$ and defined so that 
\eq{\begin{split}&\mathbf{M}.\mathbf{\bar{M}}=\fwbox{80pt}{\sum_{\g{r'}\in\g{[r']}}\mathbf{{M}}\indices{\b{[r]}}{\g{r'}}\mathbf{\bar{M}}\indices{\g{r'}}{\b{[r]}}}=\fwbox{30pt}{\delta\indices{\b{[r]}}{\b{[r]}}}\;\bigger{\Leftrightarrow}\;\fwbox{125pt}{\tikzBox{M_dot_Mbar}{\node(v0)at(0,0){$\fwbox{10pt}{~}$};\arrowTo[hblue]{v0.180}{180}\node[anchor=0,inner sep=0pt] at(in){{\footnotesize$\b{[r]}$}};\arrowFrom[hgreen]{v0.0}{0}\node(v1)at($(end)+(10pt,0)$){$\fwbox{10pt}{~}$};\arrowFrom[hblue]{v1.0}{0}\node[clebschM]at(0,0){$\fwbox{13pt}{\mathbf{M}}$};\node[clebschM]at($(in)-(10pt,0)$){$\fwbox{13pt}{\mathbf{\bar{M}}}$};\node[anchor=180,inner sep=0pt] at(end){{\footnotesize$\b{[r]}$}};}}=\fwbox{45pt}{\tikzBox{identity_on_r_indices}{\arrowTo[hblue]{0,0}{180}\node[anchor=0,inner sep=0pt] at(in){{\footnotesize$\b{[r]}$}};\node[anchor=180,inner sep=0pt] at(end){{\footnotesize$\b{[r]}$}};}};\\
&\mathbf{\bar{M}}.\mathbf{{M}}=\fwbox{80pt}{\sum_{\b{r}\in\b{[r]}}\mathbf{\bar{M}}\indices{\g{[r']}}{\b{r}}\mathbf{{M}}\indices{\b{r}}{\g{[r']}}}=\fwbox{30pt}{\delta\indices{\g{[r']}}{\g{[r']}}}\;\bigger{\Leftrightarrow}\;\fwbox{125pt}{\tikzBox{Mbar_dot_M}{\node(v0)at(0,0){$\fwbox{10pt}{~}$};\arrowTo[hgreen]{v0.180}{180}\node[anchor=0,inner sep=0pt] at(in){{\footnotesize$\g{[r']}$}};\arrowFrom[hblue]{v0.0}{0}\node(v1)at($(end)+(10pt,0)$){$\fwbox{10pt}{~}$};\arrowFrom[hgreen]{v1.0}{0}\node[clebschM]at(0,0){$\fwbox{13pt}{\mathbf{\bar{M}}}$};\node[clebschM]at($(in)-(10pt,0)$){$\fwbox{13pt}{\mathbf{{M}}}$};\node[anchor=180,inner sep=0pt] at(end){{\footnotesize$\g{[r']}$}};}}=\fwbox{45pt}{\tikzBox{identity_on_rpime_indices}{\arrowTo[hgreen]{0,0}{180}\node[anchor=0,inner sep=0pt] at(in){{\footnotesize$\g{[r']}$}};\node[anchor=180,inner sep=0pt] at(end){{\footnotesize$\g{[r']}$}};}}\,.
\end{split}\label{similarities_leave_the_prop_invariant}}
This change of basis transforms each generator of the representation $\mathbf{\b{R}}$ according to
\eq{\begin{split}\mathbf{\b{R}}\indices{\b{[r]}\r{a}}{\b{[r]}}\mapsto\mathbf{\g{R'}}\indices{\g{[r']}\r{a}}{\g{[r']}}\,\equivR\sum_{\b{r_i}\in\b{[r]}}\mathbf{\bar{M}}\indices{\g{[r']}}{\b{r_1}}\mathbf{\b{R}}\indices{\b{r_1}\r{a}}{\b{r_2}}\mathbf{M}\indices{\b{r_2}}{\g{[r']}};\end{split}}
or, at the level of rank-three tensors, 
\eq{\begin{split}\mathbf{\g{R'}}\,\equivR \mathbf{\bar{M}}.\mathbf{\b{R}}.\mathbf{M}\;\bigger{\Leftrightarrow}\;\tikzBox[-10pt]{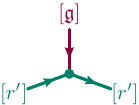}{\arrowTo[hgreen]{0,0}{200}\node[anchor=20,inner sep=0pt] at(in){{\footnotesize$\g{[r']}$}};\arrowTo[hred]{0,0}{90}\node[anchor=-90,inner sep=2pt] at(in){{\footnotesize$\r{[\adR]}$}};\arrowFrom[hgreen]{0,0}{-20}\node[anchor=160,inner sep=0pt] at(end){{\footnotesize$\g{[r']}$}};\node[hgreen,clebschR]at(0,0){};
}\equivR%
\tikzBox[-10pt]{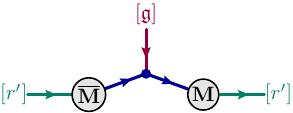}{\arrowTo[hblue]{0,0}{200}\node(m1)at($(in)+(200:8pt)$){$\fwbox{8pt}{~}$};\arrowTo[hgreen]{m1.180}{180}\node[anchor=0,inner sep=0pt] at(in){{\footnotesize$\g{[r']}$}};\arrowTo[hred]{0,0}{90}\node[anchor=-90,inner sep=2pt] at(in){{\footnotesize$\r{[\adR]}$}};\arrowFrom[hblue]{0,0}{-20}\node(m2)at($(end)+(-20:8pt)$){$\fwbox{8pt}{~}$};\arrowFrom[hgreen]{m2.0}{0}\node[anchor=180,inner sep=0pt] at(end){{\footnotesize${\g{[r']}}$}};\node[hblue,clebschR]at(0,0){};\node[clebschM]at(m1){\footnotesize{$\fwbox{10pt}{\mathbf{\bar{M}}}$}};\node[clebschM]at(m2){\footnotesize{$\fwbox{10pt}{\mathbf{{M}}}$}};
}.
\end{split}}
It is not hard to see that the commutation relations defining a representation are unchanged under such a similarity transformation. That is, if $\mathbf{\b{R}}$ is a representation, then so is $\mathbf{\g{R'}}$ for any choice of invertible matrix $\mathbf{M}$.

Representations such as $\mathbf{\g{R'}}$ and $\mathbf{\b{R}}$, related by a change of basis for their target space will be called \emph{similar}, but they are not \emph{identical} as tensors. We write $\mathbf{\g{R'}}\!\simeq\!\mathbf{\b{R}}$ whenever two representations are related by a similarity transformation.

\paragraph{Conjugate and Self-Conjugate Representations}~\\[-14pt]

Given any particular representation $\mathbf{\b{R}}$, one can always define a unique \emph{conjugate} representation $\mathbf{\b{\bar{R}}}$ via transposition of each generator: 
\eq{\mathbf{\b{\bar{R}}}\equivR\big\{\mathbf{\b{\bar{R}}}\indices{\r{a}}{}\big\}_{\r{a}\in\r{[\adR]}}\quad\text{with}\quad\mathbf{\b{\bar{R}}}\indices{\r{a}}{}\equivR({-}\mathbf{\b{R}}\indices{\r{a}}{})^{T}\,.}
Or, more succinctly, we may write
\eq{\mathbf{\b{\bar{R}}}\;\bigger{\Leftrightarrow}\;\mathbf{\b{\bar{R}}}^{\smash{\b{[\bar{r}]}\r{[\adR]}}}_{\phantom{\smash{[\bar{r}][\adR]}}\smash{\b{[\bar{r}]}}}\,\,\equivR{-}\mathbf{\b{R}}^{\phantom{\smash{[r]}}\smash{\r{[\adR]}\b{[r]}}}_{\smash{\b{[\b{r}]}\phantom{[\adR]}}}\;\bigger{\Leftrightarrow}\;\tikzBox{Rbar_vertex}{\arrowTo[hblue]{0,0}{200}\node[anchor=20,inner sep=0pt] at(in){{\footnotesize$\b{[\bar{r}]}$}};\arrowTo[hred]{0,0}{90}\node[anchor=-90,inner sep=2pt] at(in){{\footnotesize$\r{[\adR]}$}};\arrowFrom[hblue]{0,0}{-20}\node[anchor=160,inner sep=0pt] at(end){{\footnotesize$\b{[\bar{r}]}$}};\node[hblue,clebschR]at(0,0){};
}\,\,\equivR-\tikzBox{Rbar_as_reversed_R}{\arrowFrom[hblue]{0,0}{200}\node[anchor=20,inner sep=0pt] at(end){{\footnotesize$\b{[r]}$}};\arrowTo[hred]{0,0}{90}\node[anchor=-90,inner sep=2pt] at(in){{\footnotesize$\r{[\adR]}$}};\arrowTo[hblue]{0,0}{-20}\node[anchor=160,inner sep=0pt] at(in){{\footnotesize$\b{[r]}$}};\node[hblue,clebschR]at(0,0){};
}\,.\label{conjugate_reps}}
Here, the minus sign is required by the antisymmetry of the commutator (defining the Lie algebra representation $\mathbf{\b{R}}$ according to (\ref{defining_commutation_relations})) under transposition. 

In colour tensors related to scattering amplitudes, time-reversal transforms an incoming particle of representation $\mathbf{\b{R}}$ to an outgoing particle of the conjugate representation $\mathbf{\b{\bar{R}}}$. That is, 
\eq{C(\cdots\mathbf{\b{R}}|\cdots)\indices{\cdots\b{[r]}}{\cdots}\;\bigger{\Leftrightarrow}\; C(\cdots|\mathbf{\b{\bar{R}}}\cdots)\indices{\cdots}{\b{[\bar{r}]}\cdots}\,.}
We'd like to formalize this relationship by \emph{defining}
\eq{\tikzBox{RRbar_in_in_vertex}{ 
\arrowTo[hblue]{0,0}{180};\node[anchor=0,inner sep=0pt] at(in){{\footnotesize$\b{[{r}]}$}};
\arrowTo[hblue]{0,0}{0};\node[anchor=180,inner sep=0pt] at(in){{\footnotesize$\b{[\bar{r}]}$}};
\node[hblue,clebschC]at(0,0){$$}; 
}\,\equivR\delta^{\smash{\b{[r][\bar{r}]}}}\quad\text{and}\quad
\tikzBox{RbarR_out_out_vertex}{
\arrowFrom[hblue]{0,0}{180};\node[anchor=0,inner sep=0pt] at(end){{\footnotesize$\b{[\bar{{r}}]}$}};
\arrowFrom[hblue]{0,0}{0};\node[anchor=180,inner sep=0pt] at(end){{\footnotesize$\b{[{r}]}$}};
\node[hblue,clebschC]at(0,0){}; 
}\,\equivR\delta_{\smash{\b{[\bar{r}][{r}]}}}\,.\label{one_line_reversal}}
More generally, we would like to treat the reversal of a line as equivalent to conjugation of its representation:
\vspace{-4pt}\eq{\tikzBox[4.175pt]{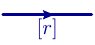}{\useasboundingbox(-0.8,-0.5)rectangle(0.8,0.25);\draw[hblue,edge,midArrow](-0.75,0)--(.75,0);\node[anchor=90,inner sep=2pt]at(arrownode){{\footnotesize$\b{[{r}]}$}};}\;\bigger{\Leftrightarrow}\;
\tikzBox[4.175pt]{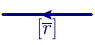}{\useasboundingbox(-0.8,-0.5)rectangle(0.8,0.25);\draw[hblue,edge,midArrow](0.75,0)--(-.75,0);\node[anchor=90,inner sep=2pt]at(arrownode){{\footnotesize$\b{[\bar{r}]}$}};}\,.\label{reversal_conjugation_rule}}
(The careful reader will notice a conflict between this identification and our definition of the conjugate representation in (\ref{conjugate_reps}) due to the minus sign on the right hand side of (\ref{conjugate_reps}). This is okay because we will soon re-interpret vertices as representing Clebsch-Gordan tensors, which are themselves only defined up to scaling.)

Obviously, $\mathbf{\b{\bar{R}}}$ and $\mathbf{\b{R}}$ have the same dimension. If they are \emph{similar}, $\mathbf{\b{\bar{R}}}\!\simeq\!\mathbf{\b{R}}$, then we call the representation \emph{self-conjugate} or `real'. In this case, there must exist some change of basis matrix $\mathbf{M}\indices{\b{[{r}]}}{\b{[\bar{r}]}}$ such that 
\vspace{-6pt}\eq{\mathbf{\b{\bar{R}}}=\bar{\mathbf{M}}.\mathbf{\b{{R}}}.\mathbf{M}\label{similarity_relation_for_real}\vspace{-4pt}}
To be clear, the matrix $\mathbf{M}$ is only defined---at best\footnote{If the change of basis is unique up to an overall scale, then the representation is irreducible.}---up to some overall scale, as (\ref{similarity_relation_for_real}) is invariant under $(\mathbf{M},\mathbf{\bar{M}})\!\mapsto\!(c\mathbf{M},c^{{-}1}\mathbf{\bar{M}})$ for any $c\!\neq\!0$. Let us suppose that this scaling has been fixed. Then we may define a `metric tensor' on $\mathbf{\b{R}}$ via
\eq{\begin{split}g_{\mathbf{\b{R}}}^{\smash{\b{[r][r]}}}\equivR\sum_{\b{\bar{r}}\in\b{[\bar{r}]}}\mathbf{M}\indices{\b{[r]}}{\b{\bar{r}}}\delta\indices{\b{\bar{r}[r]}}{}\;\bigger{\Leftrightarrow}\;
\tikzBox{metric_on_real_rep}{\arrowTo[hblue]{0,0}{180}\node[anchor=0,inner sep=0pt] at(in){{\footnotesize$\b{[{r}]}$}};\arrowTo[hblue]{0,0}{0}\node[anchor=180,inner sep=0pt] at(in){{\footnotesize$\b{[{r}]}$}};
\node[hblue,clebschM]at(0,0){};}%
\,\,\equivR%
\tikzBox[3.175pt]{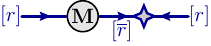}{\arrowTo[hblue]{0,0}{180}\node(m)at($(end)+(0:8pt)$){$\fwbox{8pt}{~}$};\node[anchor=0,inner sep=0pt] at(in){{\footnotesize$\b{[{r}]}$}};\arrowFrom[hblue]{m.0}{0}\node[anchor=90,inner sep=2pt]at(arrownode){{\footnotesize$\b{[\bar{r}]}$}};\arrowTo[hblue]{end}{0}\node[anchor=180,inner sep=0pt] at(in){{\footnotesize$\b{[{r}]}$}};
\node[clebschM]at(m){\footnotesize{$\fwbox{10pt}{\mathbf{M}}$}};\node[hblue,clebschC]at(end){};}\phantom{\,.}\\
g^{\mathbf{\b{R}}}_{\smash{\b{[r][r]}}}\equivR\sum_{\b{\bar{r}}\in\b{[\bar{r}]}}\delta\indices{}{\b{[r]\bar{r}}}\mathbf{\bar{M}}\indices{\,\b{\bar{r}}\!}{\b{[r]}}\;\bigger{\Leftrightarrow}\;
\tikzBox{inverse_metric_on_real_rep}{\arrowFrom[hblue]{0,0}{180}\node[anchor=0,inner sep=0pt] at(end){{\footnotesize$\b{[{r}]}$}};\arrowFrom[hblue]{0,0}{0}\node[anchor=180,inner sep=0pt] at(end){{\footnotesize$\b{[{r}]}$}};
\node[hblue,clebschM]at(0,0){};}%
\,\,\equivR%
\tikzBox[3.175pt]{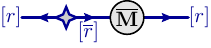}{\arrowFrom[hblue]{0,0}{180}\node[anchor=0,inner sep=0pt] at(end){{\footnotesize$\b{[{r}]}$}};\arrowFrom[hblue]{0,0}{0}\node[anchor=90,inner sep=2pt]at(arrownode){{\footnotesize$\b{[\bar{r}]}$}};
\node(m)at($(end)+(0:8pt)$){$\fwbox{8pt}{~}$};\arrowFrom[hblue]{m.0}{0}\node[anchor=180,inner sep=0pt] at(end){{\footnotesize$\b{[{r}]}$}};
\node[clebschM]at(m){\footnotesize{$\fwbox{10pt}{\mathbf{\bar{M}}}$}};\node[hblue,clebschC]at(0,0){};}\,;
\\[-8pt]\end{split}}
and similarly for $\mathbf{\b{\bar{R}}}$:
\eq{\begin{split}
g^{\mathbf{\b{\bar{R}}}}_{\smash{\b{[\bar{r}][\bar{r}]}}}\equivR\sum_{\b{{r}}\in\b{[{r}]}}\delta\indices{}{\b{[\bar{r}]r}}\mathbf{M}\indices{\b{r}}{\b{[\bar{r}]}}\;\bigger{\Leftrightarrow}\;
\tikzBox{inverse_metric_on_conjugate_of_real}{\arrowFrom[hblue]{0,0}{180}\node[anchor=0,inner sep=0pt] at(end){{\footnotesize$\b{[\bar{r}]}$}};\arrowFrom[hblue]{0,0}{0}\node[anchor=180,inner sep=0pt] at(end){{\footnotesize$\b{[\bar{r}]}$}};
\node[hblue,clebschM]at(0,0){};}%
\,\,\equivR%
\tikzBox[3.175pt]{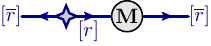}{\arrowFrom[hblue]{0,0}{180}\node[anchor=0,inner sep=0pt] at(end){{\footnotesize$\b{[\bar{r}]}$}};\arrowFrom[hblue]{0,0}{0}\node[anchor=90,inner sep=2pt]at(arrownode){{\footnotesize$\b{[{r}]}$}};
\node(m)at($(end)+(0:8pt)$){$\fwbox{8pt}{~}$};\arrowFrom[hblue]{m.0}{0}\node[anchor=180,inner sep=0pt] at(end){{\footnotesize$\b{[\bar{r}]}$}};
\node[clebschM]at(m){\footnotesize{$\fwbox{10pt}{\mathbf{{M}}}$}};\node[hblue,clebschC]at(0,0){};}\phantom{\,.}\\
g_{\mathbf{\b{\bar{R}}}}^{\smash{\b{[\bar{r}][\bar{r}]}}}\equivR\sum_{\b{{r}}\in\b{[{r}]}}\mathbf{\bar{M}}\indices{\,\b{[\bar{r}]}\!}{\b{r}}\delta\indices{\b{r[\bar{r}]}}{}\;\bigger{\Leftrightarrow}\;
\tikzBox{metric_on_conjugate_of_real_rep}{\arrowTo[hblue]{0,0}{180}\node[anchor=0,inner sep=0pt] at(in){{\footnotesize$\b{[\bar{r}]}$}};\arrowTo[hblue]{0,0}{0}\node[anchor=180,inner sep=0pt] at(in){{\footnotesize$\b{[\bar{r}]}$}};
\node[hblue,clebschM]at(0,0){};}%
\,\,\equivR%
\tikzBox[3.175pt]{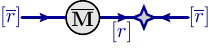}{\arrowTo[hblue]{0,0}{180}\node(m)at($(end)+(0:8pt)$){$\fwbox{8pt}{~}$};\node[anchor=0,inner sep=0pt] at(in){{\footnotesize$\b{[\bar{r}]}$}};\arrowFrom[hblue]{m.0}{0}\node[anchor=90,inner sep=2pt]at(arrownode){{\footnotesize$\b{[{r}]}$}};\arrowTo[hblue]{end}{0}\node[anchor=180,inner sep=0pt] at(in){{\footnotesize$\b{[\bar{r}]}$}};
\node[clebschM]at(m){\footnotesize{$\fwbox{10pt}{\mathbf{\bar{M}}}$}};\node[hblue,clebschC]at(end){};}\,.
\\[-8pt]\end{split}}
To be clear, there is no reason for these metrics to be diagonal. For real representations (such as the adjoint $\mathbf{\r{ad}}$) with $\mathbf{\b{R}}\!\simeq\!\mathbf{\b{\bar{R}}}$, colour tensors will be related via, for example,
\eq{C(\cdots|\mathbf{\b{R}}\cdots)\indices{\cdots}{\b{[r]}\cdots}=\sum_{\b{\bar{r}}\in\b{[\bar{r}]}}\delta\indices{}{\b{[r]\bar{r}}}C(\cdots\mathbf{\b{\bar{R}}}|\cdots)\indices{\cdots\,\b{\bar{r}}}{\cdots}=\sum_{\b{r}\in\b{[r]}}g^{\mathbf{\b{R}}}_{\b{[r]r}}C(\cdots\mathbf{\b{{R}}}|\cdots)\indices{\cdots\,\b{{r}}}{\cdots}\,.}

\paragraph{The Killing Metric, Dynkin Indices, and Casimirs}~\\[-16pt]

One representation which is always real for all Lie algebras is the adjoint: $\mathbf{\r{ad}}\!\simeq\!\r{\mathbf{\bar{ad}}}$. In this case the metric tensor is called the \emph{Killing metric} and denoted  $g_{\mathbf{\r{ad}}}$, whose overall scale must be defined or fixed by convention. It is a general theorem that any tensor constructed from a graph involving two incoming adjoint indices must be proportional to the {Killing metric}: 
\eq{\tikzBox{generic_adjoint_in_in_graph}{\arrowTo[hred]{-15pt,0}{180}\node[anchor=0,inner sep=0pt] at(in){{\footnotesize$\r{[\adR]}$}};\arrowTo[hred]{15pt,0}{0}\node[anchor=180,inner sep=0pt] at(in){{\footnotesize$\r{[\adR]}$}};
\fill[left color=white, right color=blue!0,postaction={pattern={shadelines[size=1.0pt,line width=0.35pt,angle=20]},pattern color=black!10}] (0,0) circle (15pt);
\node[circle,minimum size=30pt,draw=black,line width=0.5*\lineThickness,fill=none,inner sep=1pt]at(0,0){$$};
}\propto g_{\r{\mathbf{ad}}}^{\smash{\r{[\adR][\adR]}}}\label{two_point_proportionality}}% 
---where the constant of proportionality depends on the graph in question. One familiar example of such a graph, constructed from any particular representation gives the so-called \emph{Dynkin index} $T(\mathbf{\b{R}})$ of the representation:
\eq{\mathrm{tr}_{\mathbf{\b{R}}}(\r{1\,2})\equivL\,T(\mathbf{\b{R}})\,g_{\r{\mathbf{ad}}}^{\smash{\r{[\adR][\adR]}}}\;\bigger{\Leftrightarrow}\;\tikzBox{dynkin_index_graph}{\draw[hred,edge,midArrow](-1.15,0)--(-0.35,0);\node[anchor=0,inner sep=0pt] at(-1.15,0){{\footnotesize$\r{[\adR]}$}};\node[anchor=180,inner sep=0pt] at(1.15,0){{\footnotesize$\r{[\adR]}$}};\draw[hblue,edge,midArrow](-0.35,0)to[arc through={clockwise,(90:0.35)}](0.35,0);\node[anchor=-90,inner sep=2pt] at(arrownode){{\footnotesize$\b{[{r}]}$}};\draw[hblue,edge,midArrow](0.35,0)to[arc through={clockwise,(-90:0.35)}](-0.35,0);\node[anchor=90,inner sep=2pt] at(arrownode){{\footnotesize$\b{[{r}]}$}};\draw[hred,edge,midArrow](1.15,0)--(0.35,0);\node[hblue,clebschR]at(-0.35,0){};
\node[hblue,clebschR]at(0.35,0){};}\equivL\,T(\mathbf{\b{R}})\tikzBox{killing_metric_vertex}{
\arrowTo[hred]{0,0}{180};\node[anchor=0,inner sep=2pt] at(in){{\footnotesize$\r{[{\adR}]}$}};
\arrowTo[hred]{0,0}{0};\node[anchor=180,inner sep=2pt] at(in){{\footnotesize$\r{[{\adR}]}$}};
\node[hred,clebschM]at(0,0){};
}\,.\label{dynkin_index_defined}}
This relation provides a simple way to determine---or, more accurately, to \emph{define} as a concrete set of numbers---the Killing metric. The conventional scaling of the metric can be fixed by effectively declaring the index of any particular representation to have some particular value, which in turn will determine all other representations' indices. In the physics literature, it is common to set the index of some so-called \emph{fundamental} (or \emph{defining}) representation $\mathbf{\b{F}}$ to be 1 or $\frac{1}{2}$, while in the mathematics literature it is more common to choose $T(\mathbf{\r{ad}})$ to be 1.

In this work, for whatever concrete examples are required or discussed, we have followed the physicists' convention and taken $T(\mathbf{\b{F}})\!\to\!1$ for the `fundamental' representations defined in \mbox{appendix~\ref{appendix:weight_system_conventions}} for each of the simple Lie algebras. That is, we have chosen some irreducible representation $\mathbf{\b{F}}$ to have the property that 
\eq{g_{\mathbf{\r{ad}}}^{\smash{\r{[\adR][\adR]}}}\,\equivR \mathrm{tr}_{\mathbf{\b{F}}}(\r{1\,2})\,.} 
With this determined, one can construct a fully antisymmetric tensor (also (confusingly) called `structure constants'):
\eq{\begin{split}\r{f}\indices{\r{[\adR][\adR][\adR]}}{}\,\equivR\sum_{\r{a}\in\r{[\adR]}}{\r{f}}\indices{\r{[\adR][\adR]}}{\r{a}}g_{\mathbf{\r{ad}}}^{\smash{\r{a[\adR]}}}&=\frac{1}{T(\b{\mathbf{R}})}\big(\mathrm{tr}_{\mathbf{\b{R}}}(\r{1\,2\,3})-\mathrm{tr}_{\mathbf{\b{R}}}(\r{2\,1\,3})\big)\\[-5pt]
&\equivR\mathrm{tr}_{\mathbf{\b{F}}}(\r{1\,2\,3})-\mathrm{tr}_{\mathbf{\b{F}}}(\r{2\,1\,3}).\label{definition_of_antisymmetric_structure_constants}\end{split}}
The fact that the left-hand side is independent of the representation $\mathbf{\b{R}}$ reflects the statement that once any defining representation $\mathbf{\b{R}}$ is chosen the overall scale of all other representations is uniquely determined. Put another way, we are free to determine the overall scale of $\r{f}\indices{\r{[\adR][\adR][\adR]}}{}$ arbitrarily/conventionally; but this necessarily fixes the Dynkin index $T(\mathbf{\b{R}})$ of all representations. We should emphasize that this is in contrast to the case of $\r{f}\indices{\r{[\adR][\adR]}}{\r{[\adR]}}$, whose overall scale is uniquely determined by the commutation relations of any set of generators (\ref{defining_commutation_relations}).\\

\subsubsection{Irreducible Representations, Casimirs, and Projectors}

Recall the universal proportionality property of the adjoint described in (\ref{two_point_proportionality}). Is there an analogous statement to be made for \emph{any} representation $\mathbf{\b{R}}$? That is, will it be true that for any graph $\Gamma$ involving two copies of some representation that
\eq{\Gamma\indices{\b{[r]}}{\b{[r]}}\propto\delta\indices{\b{[r]}}{\b{[r]}}\;\bigger{\Leftrightarrow}\;\tikzBox{generic_rep_in_in_vertex}{\arrowTo[hblue]{-15pt,0}{180}\node[anchor=0,inner sep=0pt] at(in){{\footnotesize$\b{[r]}$}};\arrowFrom[hblue]{15pt,0}{0}\node[anchor=180,inner sep=0pt] at(end){{\footnotesize$\b{[{r}]}$}};
\fill[left color=white, right color=blue!0,postaction={pattern={shadelines[size=1.0pt,line width=0.35pt,angle=20]},pattern color=black!10}] (0,0) circle (15pt);
\node[circle,minimum size=30pt,draw=black,line width=0.5*\lineThickness,fill=none,inner sep=1pt]at(0,0){$\Gamma$};
}\propto \tikzBox{generic_rep_id}{\arrowTo[hblue]{0,0}{180}\node[anchor=180,inner sep=0pt] at(end){{\footnotesize$\b{[r]}$}};\node[anchor=0,inner sep=0pt] at(in){{\footnotesize$\b{[r]}$}};}\,?\label{irrep_two_point_proportionality}}
The answer turns out to be no, \emph{unless restricted to those representations that are} so-called `\textbf{irreducible}'. In fact, Schur's lemma allows us to \emph{define} a representation's irreducibility by the requirement that (\ref{irrep_two_point_proportionality}) holds for all graphs $\Gamma$. We use lower-case Roman letters such as `$\mathbf{\b{r}}$' to denote {necessarily} {irreducible} representations (and use capital letters for representations more generally). 

Closely related to (\ref{irrep_two_point_proportionality}), and a similar consequence of Schur's lemma, is that 
\eq{\Gamma\indices{\b{[r]}}{\r{[s]}}\propto\delta\indices{\mathbf{\b{r}}}{\mathbf{\r{s}}}\;\bigger{\Leftrightarrow}\;\tikzBox{r_to_s_irreps}{\arrowTo[hblue]{-15pt,0}{180}\node[anchor=0,inner sep=0pt] at(in){{\footnotesize$\b{[r]}$}};
\arrowFrom[hred]{15pt,0}{0}\node[anchor=180,inner sep=0pt] at(end){{\footnotesize$\r{[{s}]}$}};
\fill[left color=white, right color=blue!0,postaction={pattern={shadelines[size=1.0pt,line width=0.35pt,angle=20]},pattern color=black!10}] (0,0) circle (15pt);
\node[circle,minimum size=30pt,draw=black,line width=0.5*\lineThickness,fill=none,inner sep=1pt]at(0,0){$\Gamma$};
}\;.\label{irrep_two_point_orthogonality}}
That is, any graph connecting \emph{distinct} irreducible representations must vanish. This `orthogonality' among irreducible representations will prove useful to us later on. 

For irreducible representations, the constants of proportionality appearing in (\ref{irrep_two_point_proportionality}) are called \emph{Casimirs} of that representation. They are `invariants' in the sense that they necessarily commute with all generators. For any particular irreducible representation, there will be many possible Casimirs defined by various choices of graphs $\Gamma$ drawn in (\ref{irrep_two_point_proportionality}). The simplest and most familiar of these must be the \emph{quadratic Casimir} defined via
\eq{\sum_{\substack{\b{r}\in\b{[r]}\\\r{a},\r{b}\in\r{[\adR]}}}\mathbf{\b{r}}\indices{\b{[r]}\r{a}}{\b{r}}\mathbf{\b{r}}\indices{\b{r}\,\r{b}}{\b{[r]}}g^{\mathbf{\r{ad}}}_{\r{a\,b}}\equivL\,C_2(\mathbf{\b{r}})\delta\indices{\b{[r]}}{\b{[r]}}\;\bigger{\Leftrightarrow}\;\tikzBox{quadratic_casimir_diagram}{\coordinate(v1)at(0,0);\coordinate(v2)at(\edgeLength,0);\coordinate(m)at($(v1)+(0.5*\edgeLength,0.65*\edgeLength)$);
\arrowTo[hblue]{v1}{200}\node[anchor=20,inner sep=0pt] at(in){{\footnotesize$\b{[r]}$}};\arrowFrom[hblue]{v1}{0}\arrowFrom[hblue]{end}{-20}
\node[anchor=160,inner sep=0pt] at(end){{\footnotesize$\b{[r]}$}};
\draw[hred,edge,midArrow](m) to [bend right=45](v1);\draw[hred,edge,midArrow](m) to [bend left=45](v2);
\node[hblue,clebschR](c1)at(v1){};
\node[hblue,clebschR](c2)at(v2){};
\node[hred,clebschM]at(m){};
}\equivL\,C_2(\mathbf{\b{r}})\tikzBox{generic_rep_id}{\arrowTo[hblue]{0,0}{180}\node[anchor=180,inner sep=0pt] at(end){{\footnotesize$\b{[r]}$}};\node[anchor=0,inner sep=0pt] at(in){{\footnotesize$\b{[r]}$}};}\,.\label{casimir_c2_defined}}

For a Lie algebra $\mathfrak{g}$ with \emph{rank} $k$, there are in general $k$ independent Casimirs for irreducible representations. That is, any particular irreducible representation of a Lie algebra may be uniquely identified by a $k$-tuple of Casimirs. The detailed construction of Casimirs will not be important to us here, but it is important to note that their existence allows us to uniquely label all possible irreducible representations of any Lie algebra. With a suitable choice of these Casimirs and their ordering, these become \emph{Dynkin labels} $\vec{w}(\mathbf{\b{r}})\!\in\!\mathbb{Z}^k_{\geq0}$, which identifies each irreducible representation with its \emph{highest weight} relative to the root system defining the algebra. This fact is best understood from the theory of weight systems and roots, but it would go beyond the scope of this work to review here. One point is worth emphasizing, however: there is no universally accepted ordering to the root systems of simple Lie algebras, and so Dynkin labels for irreducible representations may differ by permutations of their indices. When we are required to identify a representation using such labels, we make use of the conventions clarified in \mbox{appendix~\ref{appendix:weight_system_conventions}}. 

What is important to us, however, is that some set of Casimirs will have independent eigenvalues on the subspaces of each irreducible representation of the Lie algebra. As such, they can be used to generate concrete {projection} tensors which connect arbitrary representations (\emph{e.g.}~tensor-product representations) into direct (outer) sums of particular subspaces of irreducible representations (possibly with multiplicity):
\eq{\mathbf{\b{R}}\simeq\!\!\bigoplus_{\text{irreps }\mathbf{\r{s}}}\!\!\mathbf{\r{s}}^{\oplus m\indices{\mathbf{\b{R}}}{\mathbf{\r{s}}}}\,\equivL\,\bigoplus_{\text{irreps }\mathbf{\r{s}}}\!\!m\indices{\mathbf{\b{R}}}{\mathbf{\r{s}}}\,\mathbf{\r{s}}\,.\label{decomposition_into_irreps}}
That is, for each irreducible representation $\mathbf{\r{s}}$ appearing in the decomposition of the arbitrary representation $\mathbf{\b{R}}$, there must exist tensors $\check{\mathbf{C}}^\mu(\mathbf{\r{s}}|\mathbf{\b{R}})\indices{\r{[s]}}{\b{[r]}}$ and $\mathbf{C}_\mu(\mathbf{\b{R}}|\mathbf{\r{s}})\indices{\b{[r]}}{\r{[s]}}$ such that 
\eq{\check{\mathbf{C}}^\mu(\mathbf{\r{s}}|\mathbf{\b{R}}).\mathbf{\b{R}}.\mathbf{C}_\mu(\mathbf{\b{R}}|\mathbf{\r{s}})=\mathbf{\r{s}}\label{single_decomp_into_irrep_clebsch}}
for each $\mu\!\in\!\{1,\ldots,m\indices{\mathbf{\b{R}}}{\mathbf{\r{s}}}\}$.

Actually, the criterion in (\ref{single_decomp_into_irrep_clebsch}) is stronger than we'd like---as it requires that any rescaling of $\mathbf{C}_\mu\!\mapsto\!c\,\mathbf{C}_{\mu}$ be compensated by a rescaling of $\hat{\mathbf{C}}^{\mu}\!\mapsto\!c^{{-}1}\hat{\mathbf{C}}^\mu$. We can allow for these tensors to be scaled independently by merely requiring that they satisfy, separately, the homogeneous conditions
\eq{\begin{split}\mathbf{\b{R}}.\mathbf{C}_{\mu}(\mathbf{\b{R}}|\mathbf{\r{s}})&=\mathbf{C}_{\mu}(\mathbf{\b{R}}|\mathbf{\r{s}}).\mathbf{\r{s}}\\
\fwboxR{0pt}{\text{and}\hspace{10pt}}\hat{\mathbf{C}}^\mu(\mathbf{\r{s}}|\mathbf{\b{R}}).\mathbf{\b{R}}&=\mathbf{\r{s}}.\hat{\mathbf{C}}^{\mu}(\mathbf{\r{s}}|\mathbf{\b{R}})\,.\end{split}}

Given some particular, concrete representations $\mathbf{\b{R}}$ and $\mathbf{\r{s}}$, how exactly can we construct the tensor $\mathbf{C}^\mu$? Notice that, $\mathbf{C}^\mu(\mathbf{\b{R}}|\mathbf{\r{s}})\indices{\b{[r]}}{\r{[s]}}$ depends explicitly on the choice of bases made for these representations, and that similarity transformations on the representations $\mathbf{\b{R}}$ and $\mathbf{\r{s}}$ fundamentally change the tensor. Let us now describe how they may be constructed. We follow mostly the construction described in \cite{Cvitanovi1976GroupTF}.

Suppose that one computes some Lie-algebra invariant $\Gamma\indices{\b{[r]}}{\b{[r]}}$ such the one associated with $C_2(\mathbf{\b{R}})$, say, and discovers it \emph{not} to be proportional to the identity. In particular, suppose that it has a collection of distinct eigenvalues $\{\lambda_1,\ldots,\lambda_n\}$. Then we may define 
\eq{\mathbf{P}_{\!i}\,\equivR\prod_{j\neq i}\frac{1}{\lambda_i{-}\lambda_j}\big(\Gamma\indices{\b{[r]}}{\b{[r]}}{-}\lambda_i\,\delta\indices{\b{[r]}}{\b{[r]}}\big)\,.}
These projectors satisfy (by construction) a completeness relation
\eq{\sum_i\mathbf{P}_{\!i}=\delta\indices{\b{[r]}}{\b{[r]}}}
and---via the existence of an eigenspace decomposition and the spectral theorem---orthogonality
\eq{\mathbf{P}_{\!i}.\mathbf{P}_{\!j}=\delta_{i\,j}\mathbf{P}_{\!j}\fwboxL{0pt}{\hspace{10pt}\text{(no sum)}}\,.}
Usefully, the rank of each projector is given by its trace:
\eq{\mathrm{rank}(\mathbf{P}_{\!i})=\mathrm{tr}(\mathbf{P}_{\!i})<\mathrm{dim}(\mathbf{\b{R}})\,.}

The $\mathbf{P}_{\!i}$ are called projectors because they literally project onto subspaces of the $\mathrm{dim}(\mathbf{\b{R}})$ space with definite eigenvalues for the operator encoded by $\Gamma$. To be clear, they are still $\mathrm{dim}(\mathbf{\b{R}})\!\times\!\mathrm{dim}(\mathbf{\b{R}})$ matrices, but they are of lower rank.  

\newpage
Now, according to the defining property of any (not necessarily irreducible) representation (\ref{defining_commutation_relations}), because $\mathbf{P}_{\!i}.\mathbf{P}_{\!i}\!=\!\mathbf{P}_{\!i}$ for any $i$, it is easy to see that whatever commutation relations are satisfied by $\mathbf{\b{R}}$ will also be satisfied by
\eq{(\mathbf{P}_{\!i}.\mathbf{\b{R}}.\mathbf{P}_{\!i})\indices{\b{[r]}\,\r{[\adR]}}{\b{[r]}}\label{projected_generators}}
for each $i$. That is, from the reducible representation $\mathbf{\b{R}}$ we may construct the more reduced, lower-dimensional representations associated with the generators given in (\ref{projected_generators}).

As the rank of the matrix $\mathbf{P}_i$ is less than $\mathrm{dim}(\mathbf{\b{R}})$, we can always write it in terms of matrices of full rank via
\eq{\mathbf{P}_{\!i}\indices{\b{[r]}}{\b{[r]}}\equivL \sum_{\t{a}\in\t{[a]}}\mathbf{{C}}_i\indices{\b{[r]}}{\t{a}}\check{\mathbf{C}}_i\indices{\t{a}}{\b{[r]}}\,,}
(say, by taking the first $\mathrm{rank}(\mathbf{P}_i)$ independent rows or columns) where $\t{[a]}$ now only runs over the range $\{1,\ldots,\mathrm{rank}(\mathbf{P}_{\!i})\}$ of independent rows or columns. Moreover, the property that $\mathbf{P}_{\!i}.\mathbf{P}_{\!i}\!=\!\mathbf{P}_{\!i}$ readily implies that 
\eq{\check{\mathbf{{C}}}_i.\mathbf{C}_i\equivR\sum_{\b{r}\in\b{[r]}}\check{\mathbf{{C}}}_i\indices{\t{[a]}}{\b{r}}\mathbf{C}_i\indices{\b{r}}{\t{[a]}}=\delta\indices{\t{[a]}}{\t{[a]}}\,}
from which it is easy to see that the particular rank-three tensor
\eq{(\check{\mathbf{C}}_i.\mathbf{\b{R}}.\mathbf{C}_i)\indices{\t{[a]}\,\r{[\adR]}}{\t{[a]}}\label{projected_rep}}
will also satisfy the same commutation relations as $\mathbf{\b{R}}$, but now involving generators of size $\mathrm{rank}(\mathbf{P}_{\!i})\!\times\!\mathrm{rank}(\mathbf{P}_{\!i})$. 

To be clear, the partitioning of $\mathbf{P}_{\!i}\equivL\,\mathbf{C}_i.\check{\mathbf{{C}}}_i$ is only well-defined up to a similarity transform: this pairing is indistinguishable from
\eq{(\mathbf{C}_i,\mathbf{\bar{C}}_i)\mapsto(\mathbf{C}_i.\mathbf{M},\mathbf{\bar{M}}.\check{\mathbf{{C}}}_i)\,,}
for any invertible matrix $\mathbf{M}\indices{\t{[a]}}{\t{[a]}}$. This is of course identical to the fact that any rank-three tensor related to (\ref{projected_rep}) by similarity will also constitute a representation with identical structure constants.\\

By virtue of the existence of a set of Casimirs sufficient to identify all irreducible representations, there exist projectors (or sequences thereof) by which the specific projectors (\ref{single_decomp_into_irrep_clebsch}) can be constructed. Moreover, provided some `standard' ($\mathrm{GL}(\mathrm{dim}(\mathbf{\t{a}}))$-fixed) representation matrices have been chosen for each irreducible representation, the only ambiguity left in the determination of the projectors $(\check{\mathbf{C}},\mathbf{C})$ is the overall scaling by some constant: $(\check{\mathbf{C}},\mathbf{C})\mapsto(c\,\check{\mathbf{C}},c^{{-}1}\mathbf{C})$ for any constant $c$.\\

\newpage
\subsection{Tensor Products and Clebsch-Gordan Coefficients}\label{subsec:products_and_projections}

Given any two representations $\mathbf{\b{R}}$ and $\mathbf{\g{S}}$, one may define the \emph{tensor product} representation denoted `$(\mathbf{\b{R}}\!\otimes\!\mathbf{\g{S}})$' via
\eq{(\mathbf{\b{R}}\!\otimes\!\mathbf{\g{S}})\indices{\b{[r]}\g{[s]}\,\r{[\adR]}}{\b{[r]}\g{[s]}}\,\equivR\mathbf{\b{R}}\indices{\b{[r]}\r{[\adR]}}{\b{[r]}}\delta\indices{\g{[s]}}{\g{[s]}}+\delta\indices{\b{[r]}}{\b{[r]}}\mathbf{\g{S}}\indices{\g{[s]}\r{[\adR]}}{\g{[s]}}\,.\label{tensor_product_rep_defined}}
Via the canonical isomorphism $\mathbb{V}^{\mathrm{dim}(\mathbf{\b{R}})}\!\otimes\!\mathbb{V}^{\mathrm{dim}(\mathbf{\g{S}})}\mapsto\mathbb{V}^{\mathrm{dim}(\mathbf{\b{R}})\times\mathrm{dim}(\mathbf{\g{S}})}$, it is common to view the tensor product as defined over the $\mathrm{dim}(\mathbf{\b{R}})\!\times\!\mathrm{dim}(\mathbf{\g{S}})$-dimensional space of tuples of indices $\b{[r]}\g{[s]}\simeq[\b{r},\!\g{s}]$; although according to our definition of a \emph{matrix representation} in (\ref{defining_commutation_relations}) this would be required, we find that the partitioned sets of indices are more convenient for what follows.

The decomposition of the tensor product representation into irreducible representations of the Lie algebra $\mathfrak{g}$ is accomplished via what are known as \emph{Clebsch-Gordan coefficients}. These can be constructed in the same way as above for the decomposition of arbitrary (not necessarily irreducible) representations. That is, we may construct a pair of tensors $\mathbf{{C}}_\mu(\mathbf{\b{R}}\,\mathbf{\g{S}}|\mathbf{\r{t}})$ and $\check{\mathbf{{C}}}^\mu(\mathbf{\r{t}}|\mathbf{\b{R}}\,\mathbf{\g{S}})$ which decomposes the tensor product into the irreducible representation $\mathbf{\r{t}}$ according to
\eq{\check{\mathbf{C}}^\mu(\mathbf{\r{t}}|\mathbf{\b{R}}\,\mathbf{\g{S}}).(\mathbf{\b{R}}\!\otimes\!\mathbf{\g{S}}).\mathbf{{C}}_\mu(\mathbf{\b{R}}\,\mathbf{\g{S}}|\mathbf{\r{t}})=\mathbf{\r{t}}\quad\text{for each $\mu$,}\label{clebsches_job}}
where, as always $\mu$ runs over the multiplicity $\mu\!\in\!\{1,\ldots,m\indices{\mathbf{\b{R}}\mathbf{\g{S}}}{\mathbf{\r{t}}}\}$, using the shorthand $m\indices{\mathbf{\b{R}}\mathbf{\g{S}}}{\mathbf{\r{t}}}\equivR m\indices{\mathbf{\b{R}}\otimes\mathbf{\g{S}}}{\mathbf{\r{t}}}$.

As before, the relation (\ref{clebsches_job}) is invariant under $(\check{\mathbf{C}},\mathbf{C})\mapsto(c\,\check{\mathbf{C}},c^{{-}1}\mathbf{C})$ for any constant $c$; that is, while the overall scaling of $\mathbf{C}$ may be arbitrarily chosen, that of $\check{\mathbf{C}}$ cannot be independently adjusted. It turns out to be useful to allow for independent scaling choices, which is easily accomplished by considering that these tensors be instead {defined} \emph{merely} by the homogeneous relations:
\eq{\begin{split}
(\mathbf{\b{R}}\!\otimes\!\mathbf{\g{S}}).\mathbf{{C}}_\mu(\mathbf{\b{R}}\,\mathbf{\g{S}}|\mathbf{\r{t}})&\equivR\,\mathbf{{C}}_\mu(\mathbf{\b{R}}\,\mathbf{\g{S}}|\mathbf{\r{t}}).\mathbf{\r{t}}\,;\\
\text{and}\qquad\mathbf{\bar{C}}^\mu(\mathbf{\r{t}}|\mathbf{\b{R}}\,\mathbf{\g{S}}).(\mathbf{\b{R}}\!\otimes\!\mathbf{\g{S}})&\equivR\,\mathbf{\r{t}}.\mathbf{\bar{C}}^\mu(\mathbf{\r{t}}|\mathbf{\b{R}}\,\mathbf{\g{S}})\,.
\end{split}}
These relations allow us to choose any overall scale we'd like separately for $\mathbf{C}$ and $\mathbf{\bar{C}}$. Our convention here is that $\mathbf{\bar{C}}$ should in fact be \emph{defined} by swapping incoming and outgoing representations with their conjugates: 
\eq{\check{\mathbf{C}}^\mu\,\propto\,\mathbf{\bar{C}}^{\mu}(\mathbf{\r{t}}|\mathbf{\b{R}}\,\mathbf{\g{S}})\indices{\r{[t]}}{\b{[r]}\g{[s]}}\equivR\,\sum_{\substack{\b{\bar{r}}\in\b{[\bar{r}]},\g{\bar{s}}\in\g{[\bar{s}]}\\\r{\bar{t}}\in\r{[\bar{t}]}}}\mathbf{{C}}^{\mu}(\mathbf{\b{\bar{R}}}\,\mathbf{\g{\bar{S}}}|\mathbf{\r{\bar{t}}})\indices{\b{\bar{r}}\,\g{\bar{s}}}{\,\r{\bar{t}}}\,\,\delta\indices{}{\b{\bar{r}}\,\b{[r]}}\,\delta\indices{}{\g{\bar{s}}\,\g{[s]}}\,\delta\indices{\r{\bar{t}}\,\r{[t]}}{}\,.\label{conjugate_clebsch_defined}}
Of course, $\mathbf{\bar{C}}\propto\check{\mathbf{C}}$, but these need not be identical. 

\newpage
Importantly (and independently of the scales chosen for these tensors), we may always choose Clebsch-Gordan tensors to be orthogonal with respect to their multiplicities. Specifically, we may always arrange them so that
\eq{\big\langle\mathbf{\bar{C}}^\nu(\mathbf{\r{t}}|\mathbf{\b{R}}\,\mathbf{\g{S}})|\mathbf{{C}}_\mu(\mathbf{\b{R}}\,\mathbf{\g{S}}|\mathbf{\r{t}})\big\rangle\,\equivR\sum_{\substack{\b{r}\in\b{[r]},\g{s}\in\g{[s]}\\\r{t}\in\r{[t]}}}\mathbf{\bar{C}}^\nu(\mathbf{\r{t}}|\mathbf{\b{R}}\,\mathbf{\g{S}})\indices{\r{t}}{\b{r}\g{s}}\mathbf{{C}}_\mu(\mathbf{\b{R}}\,\mathbf{\g{S}}|\mathbf{\r{t}})\indices{\b{r}\g{s}}{\r{t}}\propto\delta\indices{\nu}{\mu}\,.\label{orthogonality_of_clebsches}\vspace{-5pt}}
(Here, the constant of proportionality can depend on the index $\mu$.)

In many cases of interest to us below, there are no multiplicities to consider; and the few instances where multiplicities do matter, independent orthogonal Clebsch-Gordan tensors can be constructed by decomposing tensor-powers into symmetric or antisymmetric parts. For example, for $\mathfrak{a}_{\r{k}>1}$ Lie algebras, the multiplicity $m\indices{\mathbf{\r{ad\,ad}}}{\mathbf{\r{ad}}}{=}2$, which can be cleanly separated into one copy arising each from the symmetric and antisymmetric tensor products. 

Because of the importance of $\mathfrak{a}_{\r{k}>1}$ for physics, it may be worth clarifying how this works in somewhat greater detail. Whenever one considers the tensor product representations of two identical representations, it is worthwhile to exploit the canonical isomorphism between $\mathbb{V}^{\mathrm{d}}\!\otimes\!\mathbb{V}^{\mathrm{d}}\mapsto\mathbb{V}^{\mathrm{d}\times\mathrm{d}}\mapsto\mathbb{V}^{\mathrm{d}}\!\wedge\!\mathbb{V}^{\mathrm{d}}\!\oplus\!\mathrm{Sym}\big(\mathbb{V}^{\mathrm{d}},\mathbb{V}^{\mathrm{d}}\big)$; in terms of this, we may view the Lie algebra tensor product representation as given by the outer sum of
\eq{\mathbf{\b{R}}\!\otimes\!\mathbf{\b{R}}\equivL\,(\mathbf{\b{R}}\!\wedge\!\mathbf{\b{R}})\oplus(\mathbf{\b{R}}\!\odot\!\mathbf{\b{R}})\,,}
where we have used `$\wedge$' and `$\odot$' to denote the image of the tensor product on the \emph{antisymmetric} and \emph{symmetric} product spaces, respectively. In the case of the adjoint representation $\mathbf{\r{ad}}$ of $\mathfrak{a}_{\r{k}>1}$, each of these tensor-products includes in its decomposition a single copy of $\mathbf{\r{ad}}$, allowing us to define the two Clebsches $\mathbf{C}_{[\mu]}$ according to 
\eq{\begin{split}
(\mathbf{\r{ad}}\!\wedge\!\mathbf{\r{ad}}).\mathbf{C}_1(\mathbf{\r{ad}}\,\mathbf{\r{ad}}|\mathbf{\r{ad}})&=\,\mathbf{C}_1(\mathbf{\r{ad}}\,\mathbf{\r{ad}}|\mathbf{\r{ad}}).\mathbf{\r{ad}}\,,\\
(\mathbf{\r{ad}}\!\odot\!\mathbf{\r{ad}}).\mathbf{C}_2(\mathbf{\r{ad}}\,\mathbf{\r{ad}}|\mathbf{\r{ad}})&=\,\mathbf{C}_2(\mathbf{\r{ad}}\,\mathbf{\r{ad}}|\mathbf{\r{ad}}).\mathbf{\r{ad}}\,,\end{split}}
As we will clarify in more detail below, it turns out that the first of these can always be taken to be identical to the adjoint representation itself
\vspace{-4pt}\eq{\mathbf{C}_1(\mathbf{\r{ad}}\,\mathbf{\r{ad}}|\mathbf{\r{ad}})\equivR\,\mathbf{\r{ad}}=\r{f}\indices{\r{[\adR][\adR]}}{\r{[\adR]}}\;\bigger{\Leftrightarrow}\;
\tikzBox{f_as_clebsch_1}{\arrowTo[hred]{0,0}{200}\node[anchor=20,inner sep=0pt] at(in){{\footnotesize$\r{[\adR]}$}};\arrowTo[hred]{0,0}{90}\node[anchor=-90,inner sep=2pt] at(in){{\footnotesize$\r{[\adR]}$}};\arrowFrom[hred]{0,0}{-20}\node[anchor=160,inner sep=0pt] at(end){{\footnotesize$\r{[\adR]}$}};\node[hred,clebschR]at(0,0){};
}\,,\vspace{-4pt}}
while the second can be chosen to be related to the so-called `$\r{d}$ tensor'
\vspace{-4pt}\eq{\mathbf{C}_2(\mathbf{\r{ad}}\,\mathbf{\r{ad}}|\mathbf{\r{ad}})\equivR \r{d}\indices{\r{[\adR][\adR]}}{\r{[\adR]}}\equivR\,\sum_{\r{a}\in\r{[\adR]}}\r{d}\indices{\r{[\adR][\adR]a}}{}g^{\mathbf{\r{ad}}}_{\smash{\r{a\,[\adR]}}}\;\bigger{\Leftrightarrow}\;
\tikzBox{d_tensor_diagram}{\arrowTo[hred]{0,0}{200}\node[anchor=20,inner sep=0pt] at(in){{\footnotesize$\r{[\adR]}$}};\arrowTo[hred]{0,0}{90}\node[anchor=-90,inner sep=2pt] at(in){{\footnotesize$\r{[\adR]}$}};\arrowFrom[hred]{0,0}{-20}\node[anchor=160,inner sep=0pt] at(end){{\footnotesize$\r{[\adR]}$}};\node[hred,clebschD]at(0,0){};
}\,,\vspace{-4pt}}
where $\r{d}\indices{\r{[\adR][\adR][\adR]}}{}$ is the fully symmetric tensor typically defined in terms of the fundamental representation `$\mathbf{\b{F}}$' analogously to the antisymmetric tensor $\r{f}\indices{\r{[\adR][\adR][\adR]}}{}$ in (\ref{definition_of_antisymmetric_structure_constants}):
\eq{\r{d}\indices{\r{[\adR][\adR][\adR]}}{}\equivR\mathrm{tr}_{\mathbf{\b{F}}}(\r{1\,2\,3})+\mathrm{tr}_{\mathbf{\b{F}}}(\r{2\,1\,3})
\,.\label{d_tensor_defined}}
It is easy to see that these tensors $\mathbf{C}_\mu$ are automatically orthogonal.

Unfortunately, this trick of symmetrization/anti-symmetrization does not always suffice; in particular, this doesn't work for cases where $\mathbf{\b{R}}$ and $\mathbf{\g{S}}$ are distinct. For example, in the case of the Lie algebra $\mathfrak{a}_{\r{3}}$ we find that the tensor product of the irreducible representations of dimension $\mathbf{{384}}$ and $\mathbf{{729}}$ includes \emph{seven} copies of the irreducible representation of dimension $\mathbf{{960}}$:
\eq{\dynkLabelK{\mathbf{384}}{131}\!\otimes\!\dynkLabelK{\mathbf{729}}{222}\supset\dynkLabelK{\mathbf{960}}{313}^{\oplus 7}\,.\label{high_multiplicity_example}}
(Here, we have indicated the Dynkin labels for each of the irreducible representations; however, in this case, such labelling is superfluous as these are the unique irreducible representations of $\mathfrak{a}_{\r{3}}$ of these dimensions.) 

In many cases like that of (\ref{high_multiplicity_example}), the construction of orthogonal Clebsch-Gordan tensors is highly non-trivial and computationally cumbersome. Conveniently, however, these cases rarely arise in situations that are most interesting and relevant to physics. When there are no multiplicities greater than one appearing in a tensor product, we often suppress the `$\mu$' index label, referring only to `the' Clebsch-Gordan tensor---as opposed to one among many. And for cases such as the $\r{f}$ versus $\r{d}$ tensors relevant (only) to $\mathfrak{a}_{\r{k}>1}$, we use the diagrammatic notation introduced above to uniquely identify each Clebsch.\\

\subsection{Universality of Clebsch-Gordan Colour Tensors}\label{subsec:everything_is_clebsch}

Although such Clebsch-Gordan coefficients may seem quite removed from the more obvious tensors such as $\mathbf{\b{R}}$, $\mathbf{\r{ad}}$, or even $g_{\mathbf{\r{ad}}}^{\smash{\r{[\adR][\adR]}}}$, it turns out that all of these more familiar tensors can be interpreted as Clebsch-Gordan coefficients. For example, for any representation $\mathbf{\b{R}}$, the tensor product representation $\mathbf{\b{R}}\!\otimes\!\mathbf{\r{ad}}$ \emph{always} includes at least one copy of $\mathbf{\b{R}}$; and it is always possible to choose the normalization of our Clebsches so that
\eq{\mathbf{C}_1(\mathbf{\b{R}}\,\mathbf{\r{ad}}|\mathbf{\b{R}})\indices{\b{[r]}\r{[\adR]}}{\b{[r]}}\,\equivR\mathbf{\b{R}}\indices{\b{[r]}\r{[\adR]}}{\b{[r]}}\,.\label{rep_is_clebsch}}
To be clear, if $m\indices{\mathbf{\b{R}}\,\mathbf{\r{ad}}}{\mathbf{\b{R}}}\!>\!1$, there will be other Clebsch-Gordan coefficients as well. This happens, for example, for $\mathbf{C}(\mathbf{\r{ad}}\,\mathbf{\r{ad}}|\mathbf{\r{ad}})$ in the case of $\mathfrak{a}_{\r{k}>1}$ gauge-theory. (Another example is the $\mathbf{350}$-dimensional representation of $\mathfrak{d}_{\r{4}}$, which has multiplicity with the adjoint of 3.) Regardless of multiplicity, however, there is always one instance that can be identified by (\ref{rep_is_clebsch}). In such cases, we choose to suppress its multiplicity index and need not further embellish our diagrammatic notation. 

All the examples involving representation tensors appearing in this work so far can therefore be immediately interpreted as those involving (particular, if multiple exist) Clebsch-Gordan coefficients. This allows us to directly re-interpret the many various colour tensors constructed out of generators as those constructed from Clebsch-Gordan coefficients, unifying the various diagrammatic notations used for colour tensors in the literature.

Another category of tensors we've encountered before are the rank-two tensors relating incoming representation indices to outgoing conjugate-representation indices and the `metric' tensors for real representations (such as the Killing metric for the adjoint representation). These too can be understood as examples of Clebsch-Gordan coefficients: those for the projection of a pair of representations into the trivial representation, denoted `$\mathbf{1}$'.

Recall that for every representation $\mathbf{\b{R}}$ of every simple Lie algebra, there is a unique conjugate representation $\mathbf{\b{\bar{R}}}$ defined (up to similarity) by the condition that $m\indices{\mathbf{\b{R\,\bar{R}}}}{\mathbf{1}}\!=\!1$. That is, there always exists a set of Clebsch-Gordan coefficients $\mathbf{C}(\mathbf{\b{R}}\,\mathbf{\b{\bar{R}}}|\mathbf{1})$ which we may conventionally take to be given by the identity:
\eq{\mathbf{C}(\mathbf{\b{R}}\,\mathbf{\b{\bar{R}}}|\mathbf{1})\indices{\b{[r]\,[\bar{r}]}}{1}\,\equivR\delta\indices{\b{[r]\,[\bar{r}]}}{1}=\delta\indices{\b{[r]\,[\bar{r}]}}\;\bigger{\Leftrightarrow}\;
\tikzBox[-9.125pt]{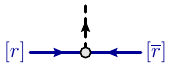}{\arrowTo[hblue]{0,0}[1.25]{180}\node[anchor=0,inner sep=2pt] at(in){{\footnotesize$\b{[r]}$}};\arrowTo[hblue]{0,0}[1.25]{0}\node[anchor=180,inner sep=2pt] at(in){{\footnotesize$\b{[\bar{r}]}$}};\arrowFrom[draw=none]{0,0}[1.125]{90}\draw[dashed,edge](0,0.05)--(0,1.15*\edgeLength);\node[clebsch]at(in){};}\;\equivR\;
\tikzBox{d_tensor_as_another_clebsch}{\arrowTo[hblue]{0,0}[1.]{180}\node[anchor=0,inner sep=2pt] at(in){{\footnotesize$\b{[r]}$}};\arrowTo[hblue]{0,0}[1.]{0}\node[anchor=180,inner sep=2pt] at(in){{\footnotesize$\b{[\bar{r}]}$}};\node[hblue,clebschC]at(end){};}\,.
\label{conjugation_as_clebsch}}
Notice this is consistent with our convention outlined in (\ref{reversal_conjugation_rule}). To be clear, for \emph{real} representations $\mathbf{\b{R}}\!\simeq\!\mathbf{\b{\bar{R}}}$, there is often a non-trivial similarity transformation between the coordinates indexed by $\b{[\bar{r}]}$---which are effectively \emph{defined} by (\ref{conjugation_as_clebsch})---and those indexed by $\b{[r]}$; in such cases, there will be some non-trivial metric tensor involved. For the adjoint representation, for example this would be the Killing metric 
\eq{\mathbf{C}(\mathbf{\r{ad}}\,\mathbf{\r{{ad}}}|\mathbf{1})\indices{\r{[\adR][\adR]}}{1}\,\equivR\, g_{\mathbf{\r{ad}}}^{\smash{\r{[\adR][\adR]}}}\;\bigger{\Leftrightarrow}\;
\tikzBox[-9.125pt]{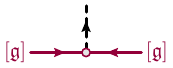}{\arrowTo[hred]{0,0}[1.25]{180}\node[anchor=0,inner sep=2pt] at(in){{\footnotesize$\r{[\adR]}$}};\arrowTo[hred]{0,0}[1.25]{0}\node[anchor=180,inner sep=2pt] at(in){{\footnotesize$\r{[\adR]}$}};\arrowFrom[draw=none]{0,0}[1.125]{90}\draw[dashed,edge](0,0.05)--(0,1.15*\edgeLength);\node[hred,clebschM]at(in){};}\;\equivR\;
\tikzBox{RRbar_vertex}{\arrowTo[hred]{0,0}[1.]{180}\node[anchor=0,inner sep=2pt] at(in){{\footnotesize$\r{[\adR]}$}};\arrowTo[hred]{0,0}[1.]{0}\node[anchor=180,inner sep=2pt] at(in){{\footnotesize$\r{[\adR]}$}};\node[hred,clebschM]at(end){};}\,.
\label{killing_metric_as_clebsch}}
To be clear, we still would \emph{define} 
\eq{\mathbf{C}(\mathbf{\r{ad}}\,\mathbf{\r{\bar{ad}}}|\mathbf{1})\indices{\r{[\adR][\bar{\adR}]}}{1}\,\equivR\, \delta^{\smash{\r{[\adR][\bar{\adR}]}}}\;\bigger{\Leftrightarrow}\;
\tikzBox[-9.125pt]{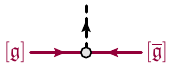}{\arrowTo[hred]{0,0}[1.25]{180}\node[anchor=0,inner sep=2pt] at(in){{\footnotesize$\r{[\adR]}$}};\arrowTo[hred]{0,0}[1.25]{0}\node[anchor=180,inner sep=2pt] at(in){{\footnotesize$\r{[\bar{\adR}]}$}};\arrowFrom[draw=none]{0,0}[1.125]{90}\draw[dashed,edge](0,0.05)--(0,1.15*\edgeLength);\node[clebsch]at(in){};}\;\equivR\,
\tikzBox{adj_adjbar_vertex}{\arrowTo[hred]{0,0}[1.]{180}\node[anchor=0,inner sep=2pt] at(in){{\footnotesize$\r{[\adR]}$}};\arrowTo[hred]{0,0}[1.]{0}\node[anchor=180,inner sep=2pt] at(in){{\footnotesize$\r{[\bar{\adR}]}$}};\node[hred,clebschC]at(end){};}\equivR
\tikzBox{adj_propagator}{\arrowTo[hred]{0,0}[1.5]{180}\node[anchor=0,inner sep=2pt] at(in){{\footnotesize$\r{[\adR]}$}};\node[anchor=180,inner sep=2pt] at(end){{\footnotesize$\r{[\adR]}$}};}\,.\nonumber}

The discussion above makes it clear that all of the tensors (and diagrammatics) we have encountered above can be universally understood as involving Clebsches---provided that their normalizations are chosen judiciously according to (\ref{rep_is_clebsch}) and (\ref{conjugation_as_clebsch}). This is good, because we would like to compare the more familiar colour tensors used by physicists in the representations of amplitudes to the more novel constructions discussed below.\\

\newpage
\subsection{Conjugation, Contraction, Vacuum Graphs, and Orthogonality}\label{subsec:vacuum_graphs_and_contractions}

We have seen many examples above where tensors are contracted by summing over internal indices. Generally, for any pair of colour tensors $C_1,C_2$ involving some representation $\mathbf{Q}$ outgoing and incoming, respectively, we may define their contraction via
\vspace{-22pt}\eq{\begin{split}~\\[0pt]\hspace{-16pt}\Bigg\{\begin{array}{@{}l@{}}C_1(\fwbox{10pt}{\mathbf{\b{R}}}\cdots|\fwbox{10pt}{\mathbf{Q}}\cdots\fwbox{10pt}{\mathbf{\r{U}}})\indices{\b{[r]}\,\cdots\,}{[q]\,\cdots\,\r{[u]}},\\
C_2(\fwbox{10pt}{\mathbf{Q}}\cdots\fwbox{10pt}{\mathbf{\g{S}}}|\fwbox{10pt}{\mathbf{\t{T}}}\cdots)\indices{[q]\,\cdots\,\g{[s]}}{\t{[t]}\,\cdots}\end{array}\Bigg\}&\hspace{-4pt}\mapsto\!D(\mathbf{\b{R}}\cdots\mathbf{\g{S}}|\mathbf{\t{T}}\cdots\mathbf{\r{U}})\indices{\b{[r]}\,\cdots\,\g{[s]}}{\t{[t]}\,\cdots\,\r{[u]}}\\[-5pt]
\hspace{20pt}&\hspace{10pt}\equivR\!\!\sum_{q\in[q]}C_1(\cdots|\mathbf{Q}\cdots)\indices{\b{[r]}\,\cdots\,}{q\,\cdots\,\r{[u]}}C_2(\mathbf{Q}\cdots|\cdots)\indices{q\,\cdots\,\g{[s]}}{\t{[t]}\,\cdots}\hspace{-40pt}\\[-8pt]\end{split}}
which we may represent graphically via
\vspace{-12pt}\eq{\fwboxR{0pt}{\bigger{\Leftrightarrow}\hspace{10pt}}\Bigg\{
\tikzBox[0pt]{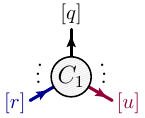}{
\arrowTo[hblue]{-150:10pt}[0.6]{-150}\node[anchor=10,inner sep=2pt] at(in){{\footnotesize$\b{[r]}$}};
\node[]at($(-155-15:16pt)$){${.}$};\node[]at($(-155-30:16pt)$){${.}$};\node[]at($(-155-45:16pt)$){${.}$};
\arrowFrom[black]{90:10pt}[0.6]{90}\node[anchor=-90,inner sep=2pt] at(end){{\footnotesize${[q]}$}};
\node[]at($(-25+15:16pt)$){${.}$};\node[]at($(-25+30:16pt)$){${.}$};\node[]at($(-25+45:16pt)$){${.}$};
\arrowFrom[hred]{-30:10pt}[0.6]{-30}\node[anchor=170,inner sep=2pt] at(end){{\footnotesize${\r{[u]}}$}};
\fill[left color=white, right color=blue!0,postaction={pattern={shadelines[size=1.0pt,line width=0.35pt,angle=20]},pattern color=black!10}] (0,0) circle (9.5pt);
\node[circle,minimum size=10pt,draw=black,line width=0.5*\lineThickness,fill=none,inner sep=1pt]at(0,0){$C_1$};
},
\tikzBox[0pt]{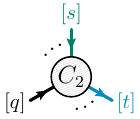}{
\arrowTo[black]{-150:10pt}[0.6]{-150}\node[anchor=10,inner sep=2pt] at(in){{\footnotesize${[q]}$}};
\node[]at($(95+15:16pt)$){${.}$};\node[]at($(95+30:16pt)$){${.}$};\node[]at($(95+45:16pt)$){${.}$};
\arrowTo[hgreen]{90:10pt}[0.6]{90}\node[anchor=-90,inner sep=2pt] at(in){{\footnotesize$\g{[s]}$}};
\node[]at($(-35-15:16pt)$){${.}$};\node[]at($(-35-30:16pt)$){${.}$};\node[]at($(-35-45:16pt)$){${.}$};
\arrowFrom[hteal]{-30:10pt}[0.6]{-30}\node[anchor=170,inner sep=2pt] at(end){{\footnotesize${\t{[t]}}$}};
\fill[left color=white, right color=blue!0,postaction={pattern={shadelines[size=1.0pt,line width=0.35pt,angle=20]},pattern color=black!10}] (0,0) circle (9.5pt);
\node[circle,minimum size=10pt,draw=black,line width=0.5*\lineThickness,fill=none,inner sep=1pt]at(0,0){$C_2$};
}\Bigg\}\!\mapsto\!
\tikzBox[0pt]{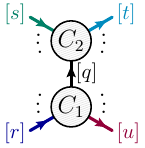}{\arrowTo[hblue]{-150:10pt}[0.6]{-150}\node[anchor=10,inner sep=2pt] at(in){{\footnotesize$\b{[r]}$}};
\node[]at($(-155-15:16pt)$){${.}$};\node[]at($(-155-30:16pt)$){${.}$};\node[]at($(-155-45:16pt)$){${.}$};
\arrowFrom[black]{90:10pt}[0.6]{90}\node[anchor=180,inner sep=2pt] at(arrownode){{\footnotesize${[q]}$}};
\node[]at($(-25+15:16pt)$){${.}$};\node[]at($(-25+30:16pt)$){${.}$};\node[]at($(-25+45:16pt)$){${.}$};
\arrowFrom[hred]{-30:10pt}[0.6]{-30}\node[anchor=170,inner sep=2pt] at(end){{\footnotesize${\r{[u]}}$}};
\coordinate(top)at($(0,0.6*\edgeLength)+(0,19pt)$);
\coordinate(topL)at($(0,0.6*\edgeLength)+(0,19pt)+(150:10pt)$);
\coordinate(topR)at($(0,0.6*\edgeLength)+(0,19pt)+(30:10pt)$);
\arrowTo[hgreen]{topL}[0.6]{150}\node[anchor=-10,inner sep=2pt] at(in){{\footnotesize${\g{[s]}}$}};
\node[]at($(top)+(155+15:16pt)$){${.}$};\node[]at($(top)+(155+30:16pt)$){${.}$};\node[]at($(top)+(155+45:16pt)$){${.}$};
\node[]at($(top)+(25-15:16pt)$){${.}$};\node[]at($(top)+(25-30:16pt)$){${.}$};\node[]at($(top)+(25-45:16pt)$){${.}$};
\arrowFrom[hteal]{topR}[0.6]{30}\node[anchor=190,inner sep=2pt] at(end){{\footnotesize${\t{[t]}}$}};
\fill[left color=white, right color=blue!0,postaction={pattern={shadelines[size=1.0pt,line width=0.35pt,angle=20]},pattern color=black!10}] (0,0) circle (9.5pt);
\node[circle,minimum size=10pt,draw=black,line width=0.5*\lineThickness,fill=none,inner sep=1pt]at(0,0){$C_1$};
\fill[left color=white, right color=blue!0,postaction={pattern={shadelines[size=1.0pt,line width=0.35pt,angle=20]},pattern color=black!10}] (top) circle (9.5pt);
\node[circle,minimum size=10pt,draw=black,line width=0.5*\lineThickness,fill=none,inner sep=1pt]at(top){$C_2$};
}\,.}
In such cases, it is easy to see that the \emph{particular} tensors involved in the representation $\mathbf{Q}$ are not important, as any similarity transform would leave the summand invariant (see equation (\ref{similarities_leave_the_prop_invariant})). In particular, \emph{if the representation $\mathbf{Q}$ were self-conjugate}, then as $\mathbf{Q}\!\simeq\!\bar{\mathbf{Q}}$, this tensor would be identical to that with the orientation of $\mathbf{Q}$ reversed. If $\mathbf{Q}\!\nsim\!\bar{\mathbf{Q}}$, however, this information must be preserved. For the sake of clarification, among the four possible choices,
\eq{\fwbox{0pt}{\tikzBox[0pt]{merged_graph_example}{\arrowTo[hblue]{-150:10pt}[0.6]{-150}\node[anchor=10,inner sep=2pt] at(in){{\footnotesize$\b{[r]}$}};
\node[]at($(-155-15:16pt)$){${.}$};\node[]at($(-155-30:16pt)$){${.}$};\node[]at($(-155-45:16pt)$){${.}$};
\arrowFrom[black]{90:10pt}[0.6]{90}\node[anchor=180,inner sep=2pt] at(arrownode){{\footnotesize${[q]}$}};
\node[]at($(-25+15:16pt)$){${.}$};\node[]at($(-25+30:16pt)$){${.}$};\node[]at($(-25+45:16pt)$){${.}$};
\arrowFrom[hred]{-30:10pt}[0.6]{-30}\node[anchor=170,inner sep=2pt] at(end){{\footnotesize${\r{[u]}}$}};
\coordinate(top)at($(0,0.6*\edgeLength)+(0,19pt)$);
\coordinate(topL)at($(0,0.6*\edgeLength)+(0,19pt)+(150:10pt)$);
\coordinate(topR)at($(0,0.6*\edgeLength)+(0,19pt)+(30:10pt)$);
\arrowTo[hgreen]{topL}[0.6]{150}\node[anchor=-10,inner sep=2pt] at(in){{\footnotesize${\g{[s]}}$}};
\node[]at($(top)+(155+15:16pt)$){${.}$};\node[]at($(top)+(155+30:16pt)$){${.}$};\node[]at($(top)+(155+45:16pt)$){${.}$};
\node[]at($(top)+(25-15:16pt)$){${.}$};\node[]at($(top)+(25-30:16pt)$){${.}$};\node[]at($(top)+(25-45:16pt)$){${.}$};
\arrowFrom[hteal]{topR}[0.6]{30}\node[anchor=190,inner sep=2pt] at(end){{\footnotesize${\t{[t]}}$}};
\fill[left color=white, right color=blue!0,postaction={pattern={shadelines[size=1.0pt,line width=0.35pt,angle=20]},pattern color=black!10}] (0,0) circle (9.5pt);
\node[circle,minimum size=10pt,draw=black,line width=0.5*\lineThickness,fill=none,inner sep=1pt]at(0,0){$C_1$};
\fill[left color=white, right color=blue!0,postaction={pattern={shadelines[size=1.0pt,line width=0.35pt,angle=20]},pattern color=black!10}] (top) circle (9.5pt);
\node[circle,minimum size=10pt,draw=black,line width=0.5*\lineThickness,fill=none,inner sep=1pt]at(top){$C_2$};
}\bigger{=}
\tikzBox[0pt]{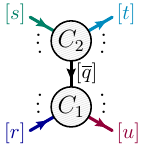}{\arrowTo[hblue]{-150:10pt}[0.6]{-150}\node[anchor=10,inner sep=2pt] at(in){{\footnotesize$\b{[r]}$}};
\node[]at($(-155-15:16pt)$){${.}$};\node[]at($(-155-30:16pt)$){${.}$};\node[]at($(-155-45:16pt)$){${.}$};
\arrowTo[black]{90:10pt}[0.6]{90}\node[anchor=180,inner sep=2pt] at(arrownode){{\footnotesize${[\bar{q}]}$}};
\node[]at($(-25+15:16pt)$){${.}$};\node[]at($(-25+30:16pt)$){${.}$};\node[]at($(-25+45:16pt)$){${.}$};
\arrowFrom[hred]{-30:10pt}[0.6]{-30}\node[anchor=170,inner sep=2pt] at(end){{\footnotesize${\r{[u]}}$}};
\coordinate(top)at($(0,0.6*\edgeLength)+(0,19pt)$);
\coordinate(topL)at($(0,0.6*\edgeLength)+(0,19pt)+(150:10pt)$);
\coordinate(topR)at($(0,0.6*\edgeLength)+(0,19pt)+(30:10pt)$);
\arrowTo[hgreen]{topL}[0.6]{150}\node[anchor=-10,inner sep=2pt] at(in){{\footnotesize${\g{[s]}}$}};
\node[]at($(top)+(155+15:16pt)$){${.}$};\node[]at($(top)+(155+30:16pt)$){${.}$};\node[]at($(top)+(155+45:16pt)$){${.}$};
\node[]at($(top)+(25-15:16pt)$){${.}$};\node[]at($(top)+(25-30:16pt)$){${.}$};\node[]at($(top)+(25-45:16pt)$){${.}$};
\arrowFrom[hteal]{topR}[0.6]{30}\node[anchor=190,inner sep=2pt] at(end){{\footnotesize${\t{[t]}}$}};
\fill[left color=white, right color=blue!0,postaction={pattern={shadelines[size=1.0pt,line width=0.35pt,angle=20]},pattern color=black!10}] (0,0) circle (9.5pt);
\node[circle,minimum size=10pt,draw=black,line width=0.5*\lineThickness,fill=none,inner sep=1pt]at(0,0){$C_1$};
\fill[left color=white, right color=blue!0,postaction={pattern={shadelines[size=1.0pt,line width=0.35pt,angle=20]},pattern color=black!10}] (top) circle (9.5pt);
\node[circle,minimum size=10pt,draw=black,line width=0.5*\lineThickness,fill=none,inner sep=1pt]at(top){$C_2$};
}\bigger{\neq}\tikzBox[0pt]{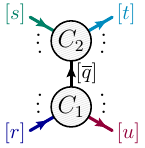}{\arrowTo[hblue]{-150:10pt}[0.6]{-150}\node[anchor=10,inner sep=2pt] at(in){{\footnotesize$\b{[r]}$}};
\node[]at($(-155-15:16pt)$){${.}$};\node[]at($(-155-30:16pt)$){${.}$};\node[]at($(-155-45:16pt)$){${.}$};
\arrowFrom[black]{90:10pt}[0.6]{90}\node[anchor=180,inner sep=2pt] at(arrownode){{\footnotesize${[\bar{q}]}$}};
\node[]at($(-25+15:16pt)$){${.}$};\node[]at($(-25+30:16pt)$){${.}$};\node[]at($(-25+45:16pt)$){${.}$};
\arrowFrom[hred]{-30:10pt}[0.6]{-30}\node[anchor=170,inner sep=2pt] at(end){{\footnotesize${\r{[u]}}$}};
\coordinate(top)at($(0,0.6*\edgeLength)+(0,19pt)$);
\coordinate(topL)at($(0,0.6*\edgeLength)+(0,19pt)+(150:10pt)$);
\coordinate(topR)at($(0,0.6*\edgeLength)+(0,19pt)+(30:10pt)$);
\arrowTo[hgreen]{topL}[0.6]{150}\node[anchor=-10,inner sep=2pt] at(in){{\footnotesize${\g{[s]}}$}};
\node[]at($(top)+(155+15:16pt)$){${.}$};\node[]at($(top)+(155+30:16pt)$){${.}$};\node[]at($(top)+(155+45:16pt)$){${.}$};
\node[]at($(top)+(25-15:16pt)$){${.}$};\node[]at($(top)+(25-30:16pt)$){${.}$};\node[]at($(top)+(25-45:16pt)$){${.}$};
\arrowFrom[hteal]{topR}[0.6]{30}\node[anchor=190,inner sep=2pt] at(end){{\footnotesize${\t{[t]}}$}};
\fill[left color=white, right color=blue!0,postaction={pattern={shadelines[size=1.0pt,line width=0.35pt,angle=20]},pattern color=black!10}] (0,0) circle (9.5pt);
\node[circle,minimum size=10pt,draw=black,line width=0.5*\lineThickness,fill=none,inner sep=1pt]at(0,0){$C_1$};
\fill[left color=white, right color=blue!0,postaction={pattern={shadelines[size=1.0pt,line width=0.35pt,angle=20]},pattern color=black!10}] (top) circle (9.5pt);
\node[circle,minimum size=10pt,draw=black,line width=0.5*\lineThickness,fill=none,inner sep=1pt]at(top){$C_2$};
}\bigger{=}
\tikzBox[0pt]{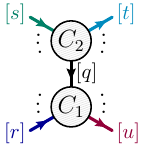}{\arrowTo[hblue]{-150:10pt}[0.6]{-150}\node[anchor=10,inner sep=2pt] at(in){{\footnotesize$\b{[r]}$}};
\node[]at($(-155-15:16pt)$){${.}$};\node[]at($(-155-30:16pt)$){${.}$};\node[]at($(-155-45:16pt)$){${.}$};
\arrowTo[black]{90:10pt}[0.6]{90}\node[anchor=180,inner sep=2pt] at(arrownode){{\footnotesize${[q]}$}};
\node[]at($(-25+15:16pt)$){${.}$};\node[]at($(-25+30:16pt)$){${.}$};\node[]at($(-25+45:16pt)$){${.}$};
\arrowFrom[hred]{-30:10pt}[0.6]{-30}\node[anchor=170,inner sep=2pt] at(end){{\footnotesize${\r{[u]}}$}};
\coordinate(top)at($(0,0.6*\edgeLength)+(0,19pt)$);
\coordinate(topL)at($(0,0.6*\edgeLength)+(0,19pt)+(150:10pt)$);
\coordinate(topR)at($(0,0.6*\edgeLength)+(0,19pt)+(30:10pt)$);
\arrowTo[hgreen]{topL}[0.6]{150}\node[anchor=-10,inner sep=2pt] at(in){{\footnotesize${\g{[s]}}$}};
\node[]at($(top)+(155+15:16pt)$){${.}$};\node[]at($(top)+(155+30:16pt)$){${.}$};\node[]at($(top)+(155+45:16pt)$){${.}$};
\node[]at($(top)+(25-15:16pt)$){${.}$};\node[]at($(top)+(25-30:16pt)$){${.}$};\node[]at($(top)+(25-45:16pt)$){${.}$};
\arrowFrom[hteal]{topR}[0.6]{30}\node[anchor=190,inner sep=2pt] at(end){{\footnotesize${\t{[t]}}$}};
\fill[left color=white, right color=blue!0,postaction={pattern={shadelines[size=1.0pt,line width=0.35pt,angle=20]},pattern color=black!10}] (0,0) circle (9.5pt);
\node[circle,minimum size=10pt,draw=black,line width=0.5*\lineThickness,fill=none,inner sep=1pt]at(0,0){$C_1$};
\fill[left color=white, right color=blue!0,postaction={pattern={shadelines[size=1.0pt,line width=0.35pt,angle=20]},pattern color=black!10}] (top) circle (9.5pt);
\node[circle,minimum size=10pt,draw=black,line width=0.5*\lineThickness,fill=none,inner sep=1pt]at(top){$C_2$};
}}}
only the first and last pair are equal to each other; each pair encodes a generally distinct tensor. Always remember that reversing an arrow is identical (by \emph{definition}) to exchanging a representation with its conjugate. 

Although the resulting tensor is independent of the particular basis choice for coordinates for the tensor $\mathbf{Q}$, it is clear that the new tensor $D$ still carries \emph{some} information about the internal representation $\mathbf{Q}$. This information may be preserved in the graph by labeling the representations associated with each edge, or by indicating the representation in some other way. 

This residual information on the internal representation is especially evident for a very important class of colour `tensors': \emph{vacuum graphs}. These are tensors of zero rank; they are simply representation-theoretic `constants' defined precisely by the representations involved in their definition, and the graphs in which they are sewn together. 

Perhaps the simplest vacuum graph can be obtained from the `propagator' of any representation
\eq{\delta^{\smash{\b{[r]}}}_{\phantom{\smash{[r]\,}}\smash{\b{[r]}}}\;\bigger{\Leftrightarrow}\;
\tikzBox{generic_propagator}{\arrowFrom[hblue]{0,0}[1.5]{0}\node[anchor=180,inner sep=0pt] at(end){{\footnotesize$\b{[r]}$}};\node[anchor=0,inner sep=0pt] at(in){{\footnotesize$\b{[r]}$}};
}\;\bigger{\Leftrightarrow}\;
\tikzBox[6.175pt]{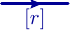}{\arrowFrom[hblue]{0,0}[1.5]{0}\node[anchor=90,inner sep=2pt] at(arrownode){{\footnotesize$\b{[r]}$}};
}\,}
via contracting its indices: 
\eq{\fwbox{0pt}{\hspace{-20pt}\mathrm{dim}(\mathbf{\b{R}})\equivR|\b{[r]}|=\sum_{\b{r}\in\b{[r]}}\delta^{\b{r}}_{\phantom{r\,}\b{r}}\;\bigger{\Leftrightarrow}\;
\tikzBox{rep_self_trace}{
\draw[hblue,edge,midArrow](0,0)arc(180:-180:0.5*\edgeLength);\node[anchor=180,inner sep=2pt] at(arrownode){{\footnotesize$\b{[r]}$}};
}=\tikzBox{rep_self_trace_bar}{
\draw[hblue,edge,midArrow](0,0)arc(-180:180:0.5*\edgeLength);\node[anchor=180,inner sep=2pt] at(arrownode){{\footnotesize$\b{[\bar{r}]}$}};
}\bigger{\Leftrightarrow}\sum_{\b{\bar{r}}\in\b{[\bar{r}]}}\delta_{\smash{\b{\bar{r}}}}^{\smash{\phantom{\b{\bar{r}}}\b{\bar{r}}}}=|\b{[\bar{r}]}|=\mathrm{dim}(\b{\mathbf{\bar{R}}}).}} 
Here, it is clear that this constant depends only upon the size of the representation, and not upon the $GL(\mathrm{dim}(\mathbf{R}))$ representative chosen for its particular components. 

Another familiar example of a vacuum graph involves only the vertex defining any particular representation:
\eq{\Theta(\mathbf{\b{r}}\,\r{\mathbf{ad}}|\mathbf{\b{r}})
\bigger{\;\Leftrightarrow\;}\tikzBox[-3pt]{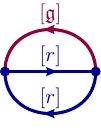}{\coordinate(v1)at(0,0);\coordinate(v2)at(2.05*\edgeLength,0);
\draw[hred,edge,midArrow](v2).. controls (2.05*\edgeLength,0.95) and (0,0.95) .. (v1);\node[anchor=-90,inner sep=2pt] at (arrownode){{\footnotesize$\r{[\adR]}$}};
\draw[hblue,edge,midArrow](v1) to [bend left=0] (v2);\node[anchor=-90,inner sep=2pt] at (arrownode){{\footnotesize$\b{[r]}$}};
\draw[hblue,edge,midArrow](v2).. controls (2.05*\edgeLength,-0.95) and (0,-0.95) .. (v1);\node[anchor=-90,inner sep=2pt] at (arrownode){{\footnotesize$\b{[r]}$}};
\node[clebschR,hblue]at(v1){};
\node[clebschR,hblue]at(v2){};
}=\hspace{-3pt}\tikzBox[0pt]{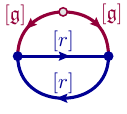}{\coordinate(v1)at(0,0);\coordinate(v2)at(2.05*\edgeLength,0);
\draw[hred,edge,midArrow](1.025*\edgeLength,0.75).. controls  (1.5375*\edgeLength,0.75)and(2.05*\edgeLength,0.55) .. (v2);
\node[anchor=210,inner sep=2pt] at (arrownode){{\footnotesize$\r{[\adR]}$}};
\draw[hred,edge,midArrow](1.025*\edgeLength,0.75).. controls (0.6125*\edgeLength,0.75) and (0,0.55) .. (v1);
\node[anchor=-30,inner sep=2pt] at (arrownode){{\footnotesize$\r{[\adR]}$}};
\draw[hblue,edge,midArrow](v1) to [bend left=0] (v2);\node[anchor=-90,inner sep=2pt] at (arrownode){{\footnotesize$\b{[r]}$}};
\draw[hblue,edge,midArrow](v2).. controls (2.05*\edgeLength,-0.95) and (0,-0.95) .. (v1);\node[anchor=-90,inner sep=2pt] at (arrownode){{\footnotesize$\b{[r]}$}};
\node[hred,clebschM]at(1.025*\edgeLength,0.75){};
\node[clebschR,hblue]at(v1){};
\node[clebschR,hblue]at(v2){};
}} 
We may compute this representation-dependent number by cutting open any edge. In particular, by cutting an $\mathbf{\b{r}}$ edge and using the definition of the quadratic Casimir (\ref{casimir_c2_defined}), or cutting the adjoint edge and using the definition of the Dynkin index defined in (\ref{dynkin_index_defined}) we find the well-known result that
\eq{\Theta(\mathbf{\b{r}}\,\r{\mathbf{ad}}|\mathbf{\b{r}})=C_2(\mathbf{\b{r}})\,\mathrm{dim}(\mathbf{\b{r}})=T(\mathbf{\b{r}})\,\mathrm{dim}(\mathfrak{\r{g}})\,.}

Alternatively, we could imagine inserting internal conjugations, considering $\Theta(\mathbf{\b{r}}\,\mathbf{\r{ad}}|\mathbf{\b{r}})$ as given by
\eq{\Theta(\mathbf{\b{r}}\,\r{\mathbf{ad}}|\mathbf{\b{r}})
\bigger{\;\Leftrightarrow\;}\tikzBox[-3pt]{theta_rrg_v1}{\coordinate(v1)at(0,0);\coordinate(v2)at(2.05*\edgeLength,0);
\draw[hred,edge,midArrow](v2).. controls (2.05*\edgeLength,0.95) and (0,0.95) .. (v1);\node[anchor=-90,inner sep=2pt] at (arrownode){{\footnotesize$\r{[\adR]}$}};
\draw[hblue,edge,midArrow](v1) to [bend left=0] (v2);\node[anchor=-90,inner sep=2pt] at (arrownode){{\footnotesize$\b{[r]}$}};
\draw[hblue,edge,midArrow](v2).. controls (2.05*\edgeLength,-0.95) and (0,-0.95) .. (v1);\node[anchor=-90,inner sep=2pt] at (arrownode){{\footnotesize$\b{[r]}$}};
\node[clebschR,hblue]at(v1){};
\node[clebschR,hblue]at(v2){};
}=\hspace{-3pt}\tikzBox[0pt]{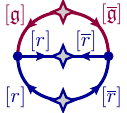}{\coordinate(v1)at(0,0);\coordinate(v2)at(2.05*\edgeLength,0);
\draw[hred,edge,midArrow](1.025*\edgeLength,0.75).. controls  (1.5375*\edgeLength,0.75)and(2.05*\edgeLength,0.55) .. (v2);
\node[anchor=210,inner sep=2pt] at (arrownode){{\footnotesize$\r{[\bar{\adR}]}$}};
\draw[hred,edge,midArrow](1.025*\edgeLength,0.75).. controls (0.6125*\edgeLength,0.75) and (0,0.55) .. (v1);
\node[anchor=-30,inner sep=2pt] at (arrownode){{\footnotesize$\r{[\adR]}$}};
\draw[hblue,edge,midArrow](v1) to [bend left=0] (1.025*\edgeLength,0);\node[anchor=-90,inner sep=2pt] at (arrownode){{\footnotesize$\b{[r]}$}};
\draw[hblue,edge,midArrow](v2) to [bend left=0] (1.025*\edgeLength,0);\node[anchor=-90,inner sep=2pt] at (arrownode){{\footnotesize$\b{[\bar{r}]}$}};
\draw[hblue,edge,midArrow](1.025*\edgeLength,-0.75).. controls (0.6125*\edgeLength,-0.75) and (0,-0.55) .. (v1);\node[anchor=30,inner sep=2pt] at (arrownode){{\footnotesize$\b{[r]}$}};
\draw[hblue,edge,midArrow](1.025*\edgeLength,-0.75).. controls  (1.5375*\edgeLength,-0.75)and(2.05*\edgeLength,-0.55) .. (v2);\node[anchor=150,inner sep=2pt] at (arrownode){{\footnotesize$\b{[\bar{r}]}$}};
\node[hred,clebschC]at(1.025*\edgeLength,0.75){};
\node[hblue,clebschC]at(1.025*\edgeLength,-0.75){};
\node[hblue,clebschC]at(1.025*\edgeLength,0){};
\node[clebschR,hblue]at(v1){};
\node[clebschR,hblue]at(v2){};
}} 
from which we can recognize the complete contraction of indices between the \emph{conjugate} tensors $\mathbf{C}(\mathbf{\b{r}}\,\mathbf{\r{ad}}|\mathbf{\b{r}})\indices{\b{[r]}\,\r{[\adR]}}{{\b{[r]}}}$ and 
\eq{\begin{split}\mathbf{C}(\b{\bar{\mathbf{{r}}}}\,\r{\bar{\mathbf{{ad}}}}|\b{\bar{\mathbf{{r}}}})\indices{\b{[\bar{r}]}\,\r{[\bar{\adR}]}}{\b{[\bar{r}]}}&\equivR\sum_{\substack{\b{r_i}\in\b{[r]}\\\r{a}\in\r{[\adR]}}}\delta\indices{\b{[\bar{r}]}\,\b{r_1}}{}\delta\indices{\r{[\bar{\adR}]}\,\r{a}}{}\delta\indices{}{\b{[\bar{r}]}\,\b{r_2}}\mathbf{C}(\mathbf{\b{r}}|\mathbf{\r{ad}}\,\mathbf{\b{r}})\indices{\b{r_2}}{\r{a}\,\b{r_1}}\\
&=\sum_{\substack{\b{r_i}\in\b{[r]}\\\r{{a}}\in\r{[{\adR}]}}}\delta\indices{\b{[\bar{r}]}\,\b{r_1}}{}\delta\indices{\r{[\bar{\adR}]}\,\r{a}}{}\delta\indices{}{\b{[\bar{r}]}\,\b{r_2}}\mathbf{C}(\mathbf{\b{r}}\,\mathbf{\r{\bar{ad}}}|\,\mathbf{\b{r}})\indices{\b{r_2}\,\r{\bar{a}}}{\b{r_1}}\\
&=\sum_{\substack{\b{r_i}\in\b{[r]}\\\r{{a_i}}\in\r{[{\adR}]}}}\delta\indices{\b{[\bar{r}]}\,\b{r_1}}{}\delta\indices{\r{[\bar{\adR}]}\,\r{a_1}}{}g^{\mathbf{\r{ad}}}_{\smash{\r{a_1\,a_2}}}\delta\indices{}{\b{[\bar{r}]}\,\b{r_2}}\mathbf{C}(\mathbf{\b{r}}\,\mathbf{\r{{ad}}}|\,\mathbf{\b{r}})\indices{\b{r_2}\,\r{{a_2}}}{\b{r_1}}\,.\end{split}} 

In general, it is useful to define the \emph{conjugate} tensor to any tensor via
\eq{\begin{split}\bar{C(\mathbf{\b{R}}\cdots\mathbf{\g{S}}|\mathbf{\t{T}}\cdots\mathbf{\r{U}})\indices{\b{[r]}\cdots\g{[s]}}{\t{[t]}\cdots\r{[u]}}}\,&\equivR C(\mathbf{\r{U}}\cdots\mathbf{\t{T}}|\mathbf{\g{S}}\cdots\mathbf{\b{R}})\indices{\r{[u]}\cdots\t{[t]}}{\g{[s]}\cdots\b{[r]}}\\
&\hspace{-150pt}=\hspace{-7pt}\sum_{\substack{\b{\bar{r}}\in\fwbox{7pt}{\b{[\bar{r}]}},\ldots,\g{\bar{s}}\in\fwbox{7pt}{\g{[\bar{s}]}}\hspace{-10pt}\\
\t{\bar{t}}\in\fwbox{7pt}{\t{[\bar{t}]}},\ldots,\r{\bar{u}}\in\fwbox{7pt}{\r{[\bar{u}]}}\hspace{-10pt}}}\hspace{-10pt}\big(\delta\indices{\r{[u]}\r{\bar{u}}}{}\!\cdots\delta\indices{\t{[t]}\t{\bar{t}}}{}\big)\big(\delta\indices{}{\g{[s]}\g{\bar{s}}}\!\cdots\delta\indices{}{\b{[r]}\b{\bar{r}}}\big)C(\mathbf{\b{\bar{R}}}\cdots\mathbf{\g{\bar{S}}}|\mathbf{\t{\bar{T}}}\cdots\mathbf{\r{\bar{U}}})\indices{\b{\bar{r}}\cdots\g{\bar{s}}}{\t{\bar{t}}\cdots\r{\bar{u}}}.\end{split}}
Graphically, this corresponds to flipping the diagram, and reversing the orientation of all edges (including those internal to the diagram). For any pair of tensors
\eq{C(\mathbf{\b{R}}\cdots\mathbf{\g{S}}|\mathbf{\t{T}}\cdots\mathbf{\r{U}})\quad\text{and}\quad C'(\mathbf{\r{U}}\cdots\mathbf{\t{T}}|\mathbf{\g{S}}\cdots\mathbf{\b{R}})}
we can define the \emph{complete contraction} between them using a bra/ket-like notation:
\eq{\langle C'|C\rangle\equivR\hspace{-7pt}\sum_{\substack{\fwbox{5pt}{\r{r}}\in\fwbox{7pt}{\r{[r]}},\ldots,\fwbox{5pt}{\g{s}}\in\fwbox{7pt}{\g{[s]}}\hspace{-10pt}\\\fwbox{5pt}{\t{t}}\in\fwbox{7pt}{\t{[t]}},\ldots,\fwbox{5pt}{\r{u}}\in\fwbox{7pt}{\r{[u]}}\hspace{-10pt}}}\hspace{-10pt}C'(\mathbf{\r{U}}\cdots\mathbf{\t{T}}|\mathbf{\g{S}}\cdots\mathbf{\b{R}})\indices{\r{u}\cdots\t{t}}{\g{s}\cdots\b{r}}C(\mathbf{\b{R}}\cdots\mathbf{\g{S}}|\mathbf{\t{T}}\cdots\mathbf{\r{U}})\indices{\b{r}\cdots\g{s}}{\t{t}\cdots\r{u}}\,.} 

The orthogonality condition of Clebsch-Gordan coefficients (\ref{orthogonality_of_clebsches}) is one of many examples where \emph{all} the representations' indices have been contracted, leaving only a residual dependence on the discrete data of the representations involved and the multiplicity indices $\mu$ required to define the particular Clebsch-Gordan tensor. If we let 
\eq{\hspace{-55pt}\Theta\indices{\nu}{\mu}(\mathbf{\b{R}}\,\mathbf{\g{S}}|\mathbf{\r{T}})\equivR\big\langle\mathbf{\bar{C}}^\nu(\mathbf{\r{T}}|\mathbf{\b{R}}\,\mathbf{\g{S}})|\mathbf{{C}}_\mu(\mathbf{\b{R}}\,\mathbf{\g{S}}|\mathbf{\r{T}})\big\rangle\,\equivR\hspace{-8pt}\sum_{\substack{\b{r}\in\b{[r]},\g{s}\in\g{[s]}\\\r{t}\in\r{[t]}}}\hspace{-8pt}\mathbf{\bar{C}}^\nu(\mathbf{\r{T}}|\mathbf{\b{R}}\,\mathbf{\g{S}})\indices{\r{t}}{\b{r}\g{s}}\mathbf{{C}}_\mu(\mathbf{\b{R}}\,\mathbf{\g{S}}|\mathbf{\r{T}})\indices{\b{r}\g{s}}{\r{t}}\hspace{-40pt}\label{orthogonality_of_clebsches_graphics_v2}} 
which we may represent graphically as
\eq{\Theta\indices{\nu}{\mu}(\mathbf{\b{R}}\,\mathbf{\g{S}}|\mathbf{\r{T}})\bigger{\;\Leftrightarrow\;}\tikzBox{theta_graph_general}{\coordinate(v1)at(0,0);\coordinate(v2)at(2.05*\edgeLength,0);
\draw[hgreen,edge,midArrow](v2).. controls (2.05*\edgeLength,0.95) and (0,0.95) .. (v1);\node[anchor=-90,inner sep=2pt] at (arrownode){{\footnotesize$\g{[s]}$}};
\draw[hred,edge,midArrow](v1) to [bend left=0] (v2);\node[anchor=-90,inner sep=2pt] at (arrownode){{\footnotesize$\r{[t]}$}};
\draw[hblue,edge,midArrow](v2).. controls (2.05*\edgeLength,-0.95) and (0,-0.95) .. (v1);\node[anchor=-90,inner sep=2pt] at (arrownode){{\footnotesize$\b{[r]}$}};
\node[clebsch]at(v1){{\footnotesize$\phantom{\nu}$}};\node at(v1){{\footnotesize$\mu$}};
\node[clebsch]at(v2){{\footnotesize$\nu$}};
}\label{theta_graph_defined}}
then orthogonality is the requirement that 
\eq{\Theta\indices{\nu}{\mu}(\mathbf{\b{R}}\,\mathbf{\g{S}}|\mathbf{\r{T}})\equivL\,\Theta_{\mu}(\mathbf{\b{R}}\,\mathbf{\g{S}}|\mathbf{\t{T}})\,\delta\indices{\nu}{\mu}\fwboxL{0pt}{\hspace{10pt}\text{(no sum)\,.}}}
(As always, the constant of proportionality can depend on the discrete data labeling the vacuum graph.)

\paragraph{Colour-Orthogonality Among Sets of Colour Graphs}~\\[-12pt]

Recall that the colour tensor bases described here are constructed from tensors of Clebsch-Gordan coefficients arranged according to some fixed tree; for example, 
\vspace{-10pt}\eq{\mathcal{B}_{\mu\nu\cdots\rho\sigma}^{\hspace{4.5pt}\mathbf{a}\cdots\mathbf{c}}(\b{\mathbf{R}}\,\g{\mathbf{S}}\cdots\!|\!\cdots\t{\mathbf{T}}\,\r{\mathbf{U}})\indices{\b{[r]}\,\g{[s]}\cdots}{\cdots\t{[t]}\,\r{[u]}}\;\;\bigger{\Leftrightarrow}\;\;\tikzBox{graphical_diagram_for_basis}{\coordinate(left)at(-1.75*\edgeLength,0);\coordinate(left2)at(-0.75*\edgeLength,0);\coordinate(right2)at(0.75*\edgeLength,0);\coordinate(right)at(1.75*\edgeLength,0);
\arrowTo[hblue]{left}{-130};\node[anchor=20,inner sep=1.5pt] at(in){{\footnotesize$\b{[r]}$}};
\arrowTo[hgreen]{left}{130};\node[anchor=-20,inner sep=1.5pt] at(in){{\footnotesize$\g{[s]}$}};
\draw[edge,midArrow](left)--(left2);\node[anchor=90,inner sep=2pt]at(arrownode){{\footnotesize$[a]$}};
%\arrowTo[black]{left2}{90};
\draw[edge](left2)--(-0.75*\edgeLength,\edgeLength);
\draw[draw=none,line width=\lineThickness,midArrowPhantom](-0.75*\edgeLength,\edgeLength)--(-0.75*\edgeLength,0.255*\edgeLength);
\draw[edge](right2)--(0.75*\edgeLength,\edgeLength);
\draw[draw=none,line width=\lineThickness,midArrowPhantom](0.75*\edgeLength,0.275*\edgeLength)--(0.75*\edgeLength,\edgeLength);
\draw[edge,dashed](-0.492*\edgeLength,0)--(0.55*\edgeLength,0);
\arrowTo[black]{right}{180}
\node[anchor=90,inner sep=2pt]at(arrownode){{\footnotesize$[c]$}};
\arrowFrom[hteal]{right}{50};\node[anchor=-150,inner sep=1.5pt] at(end){{\footnotesize$\t{[t]}$}};
\arrowFrom[hred]{right}{-50};\node[anchor=150,inner sep=1.5pt] at(end){{\footnotesize$\r{[u]}$}};
\node[clebsch]at(left){{\scriptsize$\phantom{\nu}$}};\node at(left){{\scriptsize$\mu$}};
\node[clebsch]at(left2){{\scriptsize$\phantom{\nu}$}};\node at(left2){{\scriptsize$\nu$}};
\node[clebsch]at(right2){{\scriptsize$\phantom{\nu}$}};\node at(right2){{\scriptsize$\rho$}};
\node[clebsch]at(right){{\scriptsize$\phantom{\nu}$}};\node at(right){{\scriptsize$\sigma$}};
\node[]at(0,0.5*\edgeLength){{\footnotesize$\cdots$}};
%\node[clebsch]at(right){{\footnotesize$\nu$}};
}\,.\label{graphical_diagram_for_basis_tensors_revisit}\vspace{-5pt}}
These basis tensors are labelled by the internal irreducible representations $\mathbf{a},\ldots,\mathbf{c}$ and the particular Clebsch-Gordan tensors chosen for each triple---indicated by the discrete indices $\mu,\nu,\ldots,\rho,\sigma$. These tensors are defined in the obvious way in terms of three-representation-dependent Clebsch-Gordan tensors, constructed recursively via
\eq{\mathcal{B}_{\mu\nu\cdots\sigma}^{\hspace{2.5pt}\mathbf{\r{a}}\mathbf{b}\cdots}(\b{\mathbf{R}}\,\g{\mathbf{S}}\,\t{\mathbf{T}}\cdots\!|\!\cdots)\equivR\sum_{\r{a}\in\r{[a]}}C_{\mu}(\b{\mathbf{R}}\,\g{\mathbf{S}}|\mathbf{\r{a}})^{\smash{\b{[r]}\g{[s]}}}_{\phantom{\smash{[r][s]}}\r{a}}\mathcal{B}^{\hspace{1.5pt}\smash{\mathbf{b}\cdots}}_{\nu\cdots\sigma}(\r{\mathbf{a}}\,\t{\mathbf{T}}\cdots|\!\cdots)^{\smash{\r{a}\t{[t]}\cdots}}_{\phantom{\smash{a[t]\cdots}}\cdots}\,.}

For each such colour tensor, there is a natural dual tensor involving conjugated representations incoming and outgoing, or---equivalently---involving incoming and outgoing representations reversed:
\eq{\begin{split}\bar{\mathcal{B}_{\mu\nu\cdots\rho\sigma}^{\hspace{4.5pt}\mathbf{a}\cdots\mathbf{c}}}\equivL\,\mathcal{B}^{\smash{\mu\nu\cdots\rho\sigma}}_{\hspace{4.5pt}\mathbf{a}\cdots\mathbf{c}}\;\;\bigger{\Leftrightarrow}\;\;\tikzBox{dual_basis_direct}{\coordinate(left)at(-1.75*\edgeLength,0);\coordinate(left2)at(-0.75*\edgeLength,0);\coordinate(right2)at(0.75*\edgeLength,0);\coordinate(right)at(1.75*\edgeLength,0);
\arrowTo[hblue]{left}{-120};\node[anchor=60,inner sep=2pt] at(in){{\footnotesize$\b{[\bar{r}]}$}};
\arrowTo[hgreen]{left}{120};\node[anchor=-60,inner sep=2pt] at(in){{\footnotesize$\g{[\bar{s}]}$}};
\draw[edge,midArrow](left)--(left2);\node[anchor=90,inner sep=2pt]at(arrownode){{\footnotesize$[\bar{a}]$}};
%\arrowTo[black]{left2}{90};
\draw[edge](left2)--(-0.75*\edgeLength,\edgeLength);
\draw[draw=none,line width=\lineThickness,midArrowPhantom](-0.75*\edgeLength,\edgeLength)--(-0.75*\edgeLength,0.255*\edgeLength);
\draw[edge](right2)--(0.75*\edgeLength,\edgeLength);
\draw[draw=none,line width=\lineThickness,midArrowPhantom](0.75*\edgeLength,0.275*\edgeLength)--(0.75*\edgeLength,\edgeLength);
\draw[edge,dashed](-0.492*\edgeLength,0)--(0.55*\edgeLength,0);
\arrowTo[black]{right}{180}
\node[anchor=90,inner sep=2pt]at(arrownode){{\footnotesize$[\bar{c}]$}};
\arrowFrom[hteal]{right}{60};\node[anchor=-120,inner sep=2pt] at(end){{\footnotesize$\t{[\bar{t}]}$}};
\arrowFrom[hred]{right}{-60};\node[anchor=120,inner sep=2pt] at(end){{\footnotesize$\r{[\bar{u}]}$}};
\node[clebsch]at(left){{\scriptsize$\phantom{\nu}$}};\node at(left){{\scriptsize$\mu$}};
\node[clebsch]at(left2){{\scriptsize$\phantom{\nu}$}};\node at(left2){{\scriptsize$\nu$}};
\node[clebsch]at(right2){{\scriptsize$\phantom{\nu}$}};\node at(right2){{\scriptsize$\rho$}};
\node[clebsch]at(right){{\scriptsize$\phantom{\nu}$}};\node at(right){{\scriptsize$\sigma$}};
\node[]at(0,0.5*\edgeLength){{\footnotesize$\cdots$}};
%\node[clebsch]at(right){{\footnotesize$\nu$}};
}\equivR
\tikzBox{dual_basis_flipped}{\coordinate(left)at(-1.75*\edgeLength,0);\coordinate(left2)at(-0.75*\edgeLength,0);\coordinate(right2)at(0.75*\edgeLength,0);\coordinate(right)at(1.75*\edgeLength,0);
\arrowTo[hred]{left}{-120};\node[anchor=60,inner sep=2pt] at(in){{\footnotesize$\r{[u]}$}};
\arrowTo[hteal]{left}{120};\node[anchor=-60,inner sep=2pt] at(in){{\footnotesize$\t{[t]}$}};
\draw[edge,midArrow](left)--(left2);\node[anchor=90,inner sep=2pt]at(arrownode){{\footnotesize$[c]$}};
%\arrowTo[black]{left2}{90};
\draw[edge](left2)--(-0.75*\edgeLength,\edgeLength);
\draw[draw=none,line width=\lineThickness,midArrowPhantom](0.75*\edgeLength,0.275*\edgeLength)--(0.75*\edgeLength,\edgeLength);
\draw[edge](right2)--(0.75*\edgeLength,\edgeLength);
\draw[draw=none,line width=\lineThickness,midArrowPhantom](-0.75*\edgeLength,\edgeLength)--(-0.75*\edgeLength,0.255*\edgeLength);
\draw[edge,dashed](-0.492*\edgeLength,0)--(0.55*\edgeLength,0);
\arrowTo[black]{right}{180}
\node[anchor=90,inner sep=2pt]at(arrownode){{\footnotesize$[a]$}};
\arrowFrom[hgreen]{right}{60};\node[anchor=-120,inner sep=2pt] at(end){{\footnotesize$\g{[s]}$}};
\arrowFrom[hblue]{right}{-60};\node[anchor=120,inner sep=2pt] at(end){{\footnotesize$\b{[r]}$}};
\node[clebsch]at(left){{\scriptsize$\phantom{\nu}$}};\node at(left){{\scriptsize$\sigma$}};
\node[clebsch]at(left2){{\scriptsize$\phantom{\nu}$}};\node at(left2){{\scriptsize$\rho$}};
\node[clebsch]at(right2){{\scriptsize$\phantom{\nu}$}};\node at(right2){{\scriptsize$\nu$}};
\node[clebsch]at(right){{\scriptsize$\phantom{\nu}$}};\node at(right){{\scriptsize$\mu$}};
\node[]at(0,0.5*\edgeLength){{\footnotesize$\cdots$}};
}\,.\end{split}}
We prove in section~\ref{subsec:general_basis_construction} that these tensors are \emph{automatically orthogonal} (provided the building-block, three-representation Clebsch-Gordan coefficients have been orthogonalized in their multiplicity indices $\mu$). That is, letting the complete contraction between colour tensors involving conjugated representations be denoted $\langle C|D\rangle$, we have 
\eq{\langle\mathcal{B}^{\smash{\mu'\!\cdots\sigma'}}_{\hspace{0.5pt}\mathbf{a'}\!\cdots\mathbf{c'}}|\mathcal{B}_{\mu\cdots\sigma}^{\hspace{0.5pt}\mathbf{a}\cdots\mathbf{c}}\rangle\propto\big(\delta^{\mu'}_{\mu}\cdots\delta^{\sigma'}_{\sigma}\big)\big(\delta^{\mathbf{a}}_{\mathbf{a'}}\cdots\delta^{\mathbf{c}}_{\mathbf{c'}}\big)\,.}
Moreover, the same orthogonality is ensured for any basis constructed using \emph{any} \textbf{fixed} trivalent tree of Clebsch-Gordan coefficients connecting the external particles' colour representations via the exchange of intermediate irreducible representations. The choice of how the external representations are arranged on the external legs of the tree is also arbitrary (but fixed)---and this choice can be viewed as part of the characteristic defining the tree graph itself and the corresponding basis of colour tensors. 

Due to this orthogonality, the decomposition of any particular colour tensor into this basis can be easily achieved via
\eq{C(\b{\mathbf{R}}\,\g{\mathbf{S}}\cdots\!|\!\cdots\t{\mathbf{T}}\,\r{\mathbf{U}})=\sum_{\substack{\mathbf{a}\cdots\mathbf{c}\\\mu\cdots\sigma}}\frac{\langle\mathcal{B}^{\smash{\mu\cdots\sigma}}_{\hspace{0.5pt}\mathbf{a}\cdots\mathbf{c}}|C\rangle}{\langle\mathcal{B}^{\smash{\mu\cdots\sigma}}_{\hspace{0.5pt}\mathbf{a}\cdots\mathbf{c}}|\mathcal{B}_{\mu\cdots\sigma}^{\hspace{0.5pt}\mathbf{a}\cdots\mathbf{c}}\rangle}\mathcal{B}_{\mu\cdots\sigma}^{\hspace{0.5pt}\mathbf{a}\cdots\mathbf{c}}(\b{\mathbf{R}}\,\g{\mathbf{S}}\cdots\!|\!\cdots\t{\mathbf{T}}\,\r{\mathbf{U}})\,.}
As we describe below, such a decomposition can also be implemented graphically and recursively for any colour tensor constructed from representation matrices, colour `traces', structure constants, or other graphs of Clebsch-Gordan coefficients.

%================================================================================================================
%    Discussion of the New Basis 
%================================================================================================================

\newpage
\section{Clebsch-Gordan Colour Tensor Bases}\label{sec:clebsch_Gordan_colour_tensors}

As discussed at length in ref.~\cite{Bourjaily:2024jbt}, the number of independent colour tensors involving arbitrary external particle representations is given by 
\eq{\mathrm{rank}\Big[\mathrm{span}\{C(\mathbf{\b{R}}\cdots\mathbf{\g{S}}|\mathbf{\t{T}}\cdots\mathbf{\r{U}})\}\Big]=m\indices{\b{\mathbf{R}}\otimes\cdots\otimes\g{\mathbf{S}}\otimes\t{\mathbf{\bar{T}}}\otimes\cdots\otimes\mathbf{\r{\bar{U}}}}{\mathbf{1}}\,.}
This number can be computed recursively, by decomposing arbitrary pairs of external representations into irreducible representations of the Lie algebra. This can be done in many inequivalent ways due to the associativity of the tensor product and the commutativity of this counting. Any sequence of pairwise decompositions can be viewed as encoding a trivalent tree with vertices connecting external representations to each other via internal, irreducible representations. By viewing the vertices of such a diagram as Clebsch-Gordan coefficients for the decomposition of arbitrary products into irreducible representations, we construct a spanning set of explicit colour tensors manifestly of full rank. Importantly, the resulting basis of colour tensors consists of tensors which \emph{share the same underlying tree} with a \emph{fixed} ordering of external particle representations. 

There are of course many trivalent tree graphs involving $n$ external edges; and for each of these, there are many distinct arrangements of possible external representations. If all the external edges represent distinct representations, then there will be $(2n{-}5)!!$ different choices of trees connecting the fixed (but arbitrarily arranged) external representations. 

One word of caution is in order about how the external particle representations can be ordered `distinctly' in such a graph. Although when all the external particle representations involved are distinct it is hard to overlook that each Clebsch-Gordan coefficient requires a \emph{precise} ordering of its arguments---as $C(\mathbf{\b{R}}\,\mathbf{\g{S}}|\mathbf{\t{T}})\!\neq\!C(\mathbf{\g{S}}\,\mathbf{\b{R}}|\mathbf{\t{T}})$ as tensors---one must be careful, when $\mathbf{\b{R}}\!\simeq\!\mathbf{\g{S}}$. For example, when encountering a Clebsch-Gordan coefficient such as $C_\mu(\mathbf{\b{R}}\,\mathbf{\b{R}}|\mathbf{\t{T}})$, one must be careful to remember that these tensors need not enjoy any symmetry under the exchange of the first two representations:
\vspace{-20pt}\eq{\Big\{C_{\mu}(\mathbf{\b{R}}\,\mathbf{\b{R}}|\mathbf{\t{T}})\indices{\b{r_1r_2}}{\t{t}}\Big\}_{\substack{\b{r_i}\in\b{[r]}\\\hspace{2.95pt}\t{t}\in\t{[t]}}}\neq\Big\{C_{\mu}(\mathbf{\b{R}}\,\mathbf{\b{R}}|\mathbf{\t{T}})\indices{\b{r_2r_1}}{\t{t}}\Big\}_{\substack{\b{r_i}\in\b{[r]}\\\hspace{2.95pt}\t{t}\in\t{[t]}}}\;\bigger{\Leftrightarrow}\;\tikzBox{rrt_clebsch_perm_1}{\coordinate(left)at(-0.5*\edgeLength,0);\coordinate(right)at(0.5*\edgeLength,0);
\arrowTo[hblue]{left}{-120};\node[anchor=20,inner sep=2pt] at(in){{\footnotesize$\b{[r]}$}};
\arrowTo[hblue]{left}{120};\node[anchor=-20,inner sep=2pt] at(in){{\footnotesize$\b{[r]}$}};
\draw[hteal,edge,midArrow](left)--(right);\node[anchor=180,inner sep=0pt]at(right){{\footnotesize$\t{[t]}$}};
\node[clebsch]at(left){{\footnotesize$\phantom{\nu}$}};\node at(left){{\footnotesize$\mu$}};
}\neq\hspace{-5pt}\tikzBox{rrt_clebsch_perm_2}{\coordinate(left)at(-0.5*\edgeLength,0);\coordinate(right)at(0.5*\edgeLength,0);
\floatingEdge{hblue,edge,endArrow}{($(left)+(120:\edgeLength)-(0.85*\edgeLength,-0.05)$).. controls ($(left)+(100:0.9*\edgeLength)-(0.55,0)$) and ($(left)+(-120:2.2*\edgeLength)-(0.025,0)$).. ($(left)$);}\node[anchor=-20,inner sep=2pt] at($(left)+(120:\edgeLength)-(0.85*\edgeLength,0)$){{\footnotesize$\b{[r]}$}};
\floatingEdge{hblue,edge,endArrow}{($(left)+(-120:\edgeLength)-(0.85*\edgeLength,0.05)$).. controls ($(left)+(-100:0.9*\edgeLength)-(0.55,0)$) and ($(left)+(120:2.2*\edgeLength)-(0.025,0)$).. ($(left)$);}\node[anchor=20,inner sep=2pt] at($(left)+(-120:\edgeLength)-(0.85*\edgeLength,0)$){{\footnotesize$\b{[r]}$}};
\draw[hteal,edge,midArrow](left)--(right);\node[anchor=180,inner sep=0pt]at(right){{\footnotesize$\t{[t]}$}};
\node[clebsch]at(left){{\footnotesize$\phantom{\nu}$}};\node at(left){{\footnotesize$\mu$}};
}\,.\nonumber\vspace{-15pt}}
Thus, one must view the ordering of (especially isomorphic) representation labels as being important to the meaning of the resulting tensor. This issue is effectively avoided through our convention that the sequence of arguments of the tensor should read from the bottom-left of the diagram, proceeding clockwise around the graph; but this convention makes some graphs---trace tensors defined with arbitrary orderings, for example---extremely cumbersome to draw. In such cases, we often prefer to eschew diagrammatic notation all-together, or otherwise view the sequence-ordering of arguments as additional information adorning the tensor diagram. 

In general, completeness ensures that  
\vspace{-20pt}\eq{\tikzBox{rrt_clebsch_perm_2}{\coordinate(left)at(-0.5*\edgeLength,0);\coordinate(right)at(0.5*\edgeLength,0);
\floatingEdge{hblue,edge,endArrow}{($(left)+(120:\edgeLength)-(0.85*\edgeLength,-0.05)$).. controls ($(left)+(100:0.9*\edgeLength)-(0.55,0)$) and ($(left)+(-120:2.2*\edgeLength)-(0.025,0)$).. ($(left)$);}\node[anchor=-20,inner sep=2pt] at($(left)+(120:\edgeLength)-(0.85*\edgeLength,0)$){{\footnotesize$\b{[r]}$}};
\floatingEdge{hblue,edge,endArrow}{($(left)+(-120:\edgeLength)-(0.85*\edgeLength,0.05)$).. controls ($(left)+(-100:0.9*\edgeLength)-(0.55,0)$) and ($(left)+(120:2.2*\edgeLength)-(0.025,0)$).. ($(left)$);}\node[anchor=20,inner sep=2pt] at($(left)+(-120:\edgeLength)-(0.85*\edgeLength,0)$){{\footnotesize$\b{[r]}$}};
\draw[hteal,edge,midArrow](left)--(right);\node[anchor=180,inner sep=0pt]at(right){{\footnotesize$\t{[t]}$}};
\node[clebsch]at(left){{\footnotesize$\phantom{\nu}$}};\node at(left){{\footnotesize$\mu$}};
}\in\mathrm{span}_{\nu}\left\{\tikzBox{rrt_clebsch_perm_1prime}{\coordinate(left)at(-0.5*\edgeLength,0);\coordinate(right)at(0.5*\edgeLength,0);
\arrowTo[hblue]{left}{-120};\node[anchor=20,inner sep=2pt] at(in){{\footnotesize$\b{[r]}$}};
\arrowTo[hblue]{left}{120};\node[anchor=-20,inner sep=2pt] at(in){{\footnotesize$\b{[r]}$}};
\draw[hteal,edge,midArrow](left)--(right);\node[anchor=180,inner sep=0pt]at(right){{\footnotesize$\t{[t]}$}};
\node[clebsch]at(left){{\footnotesize$\phantom{\nu}$}};\node at(left){{\footnotesize$\nu$}};
}\hspace{-0pt}\right\};\vspace{-15pt}}
which is to say that there must exist some coefficients $\mathscr{R}^{\smash{\phantom{\mu}\nu}}_{\smash{\mu}}$ such that
\vspace{-20pt}\eq{\tikzBox{rrt_clebsch_perm_2}{\coordinate(left)at(-0.5*\edgeLength,0);\coordinate(right)at(0.5*\edgeLength,0);
\floatingEdge{hblue,edge,endArrow}{($(left)+(120:\edgeLength)-(0.85*\edgeLength,-0.05)$).. controls ($(left)+(100:0.9*\edgeLength)-(0.55,0)$) and ($(left)+(-120:2.2*\edgeLength)-(0.025,0)$).. ($(left)$);}\node[anchor=-20,inner sep=2pt] at($(left)+(120:\edgeLength)-(0.85*\edgeLength,0)$){{\footnotesize$\b{[r]}$}};
\floatingEdge{hblue,edge,endArrow}{($(left)+(-120:\edgeLength)-(0.85*\edgeLength,0.05)$).. controls ($(left)+(-100:0.9*\edgeLength)-(0.55,0)$) and ($(left)+(120:2.2*\edgeLength)-(0.025,0)$).. ($(left)$);}\node[anchor=20,inner sep=2pt] at($(left)+(-120:\edgeLength)-(0.85*\edgeLength,0)$){{\footnotesize$\b{[r]}$}};
\draw[hteal,edge,midArrow](left)--(right);\node[anchor=180,inner sep=0pt]at(right){{\footnotesize$\t{[t]}$}};
\node[clebsch]at(left){{\footnotesize$\phantom{\nu}$}};\node at(left){{\footnotesize$\mu$}};
}\,\equivL\,\sum_{\nu\in[m\indices{\mathbf{\b{R}}\,\mathbf{\b{R}}}{\mathbf{\t{T}}}]}\hspace{-4pt}\mathscr{R}^{\smash{\phantom{\mu}\nu}}_{\smash{\mu}}\tikzBox{rrt_clebsch_perm_1prime}{\coordinate(left)at(-0.5*\edgeLength,0);\coordinate(right)at(0.5*\edgeLength,0);
\arrowTo[hblue]{left}{-120};\node[anchor=20,inner sep=2pt] at(in){{\footnotesize$\b{[r]}$}};
\arrowTo[hblue]{left}{120};\node[anchor=-20,inner sep=2pt] at(in){{\footnotesize$\b{[r]}$}};
\draw[hteal,edge,midArrow](left)--(right);\node[anchor=180,inner sep=0pt]at(right){{\footnotesize$\t{[t]}$}};
\node[clebsch]at(left){{\footnotesize$\phantom{\nu}$}};\node at(left){{\footnotesize$\nu$}};
}\,.\vspace{-15pt}\label{definition_of_r_move}}
To be clear, the above statement is independent of the representation $\mathbf{\t{T}}$ (which itself may or may not be isomorphic to $\mathbf{\b{R}}$). 

It will be useful later to extend the notion of transposing tensor arguments to the general case, letting
\eq{{C}_\mu(\mathbf{\b{R}}\,\mathbf{\g{S}}|\mathbf{\t{T}})\indices{\b{[r]}\g{[s]}}{\t{[t]}}\equivL \sum_{\nu\in[m\indices{\mathbf{\b{R}}\,\mathbf{\g{S}}}{\mathbf{\t{T}}}]}\hspace{-5pt}\mathscr{R}^{\phantom{\mu}\nu}_{\mu}(\mathbf{\b{R}}\,\mathbf{\g{S}})C_\nu(\mathbf{\g{S}}\,\mathbf{\b{R}}|\mathbf{\t{T}})\indices{\g{[s]}\b{[r]}}{\t{[t]}}\,.}
Of course, when $\mathbf{\b{R}}\!\nsim\!\mathbf{\g{S}}$, this operation is fairly trivial.

\subsection{Bases of Colour Tensors for Arbitrary Coloured Particles}\label{subsec:general_basis_construction}

The generalization of the discussion above to the case of an arbitrary collection of external representations should be clear: any particular, fixed, trivalent tree constructed from Clebsch-Gordan tensors connecting the external representations via internal, irreducible representations will form a complete basis of colour tensors. Although many of the key ideas have been outlined above, let us briefly sketch the essential arguments involved in proving the non-perturbative completeness and colour-orthogonality of the tensors constructed in this way.

Consider, for example, the case of 4 particles with distinct representations. There are three choices of graph and correspondingly three choices of colour tensor bases constructed from Clebsch-Gordan coefficients:
\eq{\begin{array}{c@{$\;\;\;\;\;\;$}c@{$\;\;\;\;\;\;$}c}\mathcal{B}^{\hspace{2.2pt}\mathbf{{a}}}_{\hspace{-1pt}\mu\nu}(\mathbf{\b{R}}\,\mathbf{\g{S}}|\mathbf{\t{T}}\,\mathbf{\r{U}})&\mathcal{C}^{\hspace{1.0pt}\mathbf{{b}}}_{\hspace{-1pt}\rho\sigma}(\mathbf{\b{R}}\,\mathbf{\g{S}}|\mathbf{\t{T}}\,\mathbf{\r{U}})&\mathcal{D}^{\hspace{2.2pt}\mathbf{{c}}}_{\hspace{-1pt}\tau\upsilon}(\mathbf{\b{R}}\,\mathbf{\g{S}}|\mathbf{\t{T}}\,\mathbf{\r{U}})\\
\tikzBox{four_point_generic_basis_s}{\arrowTo[hblue]{0,0}{-142};\node[anchor=38,inner sep=0pt] at(in){{\footnotesize$\b{[r]}$}};\arrowTo[hgreen]{0,0}{142}\node[anchor=-38,inner sep=0pt] at(in){{\footnotesize$\g{[s]}$}};\arrowFrom[black]{0,0}[1.3]{0}\node[clebsch]at(in){{\scriptsize$\phantom{\nu}$}};\node[]at(in){{\scriptsize$\mu$}};\node[anchor=90,inner sep=2pt] at(arrownode){{\footnotesize${[a]}$}};\arrowFrom[hteal]{end}{38}
\node[anchor=-142,inner sep=0pt] at(end){{\footnotesize$\t{[t]}$}};\arrowFrom[hred]{in}{-38}\node[anchor=142,inner sep=0pt] at(end){{\footnotesize$\r{[u]}$}};\node[clebsch]at(in){{\scriptsize$\phantom{\nu}$}};\node[]at(in){{\scriptsize$\nu$}};
}&\tikzBox{four_point_generic_basis_t}{
\arrowFrom[hred]{0,-0.6*\edgeLength}[1.]{-10}\node[anchor=170,inner sep=2pt] at(end){{\footnotesize$\r{[u]}$}};
\arrowFrom[hteal]{0,0.6*\edgeLength}[1.]{10}\node[anchor=-170,inner sep=2pt] at(end){{\footnotesize$\t{[t]}$}};
\arrowTo[hblue]{0,-0.6*\edgeLength}[1.]{-170}\node[anchor=10,inner sep=2pt] at(in){{\footnotesize$\b{[r]}$}};
\arrowTo[hgreen]{0,0.6*\edgeLength}[1.]{170}\node[anchor=-10,inner sep=2pt] at(in){{\footnotesize$\g{[s]}$}};
\arrowTo[black]{0,0.6*\edgeLength}[1.2][0.55]{-90}\node[clebsch]at(in){{\scriptsize$\phantom{\nu}$}};\node[]at(in){{\scriptsize$\rho$}};\node[clebsch]at(end){{\scriptsize$\phantom{\nu}$}};\node[]at(end){{\scriptsize$\sigma$}};
\node[anchor=0,inner sep=2pt]at(arrownode){{\footnotesize$[b]$}};
}&\tikzBox{four_point_gneric_basis_u}{
\arrowTo[hblue]{0,0}{-142};\node[anchor=38,inner sep=0pt] at(in){{\footnotesize$\b{[r]}$}};
\floatingEdge{hgreen,edge,startArrow}{($(142:\edgeLength)+(0,0.125)$).. controls ($(0.5*\edgeLength,0.65*\edgeLength)$) and ($(0.8*\edgeLength,0.5*\edgeLength)$) .. ($(1.3*\edgeLength,0)$);}\node[anchor=-38,inner sep=0pt] at(142:\edgeLength){{\footnotesize$\g{[s]}$}};
\floatingEdge{hteal,edge,endArrow}{(0,0).. controls ($(0.5*\edgeLength,0.5*\edgeLength)$) and ($(0.8*\edgeLength,0.65*\edgeLength)$) .. ($(0,0.125)+(38:\edgeLength)+(1.3*\edgeLength,0)$);}\node[anchor=-142,inner sep=0pt] at($(38:\edgeLength)+(1.3*\edgeLength,0)$){{\footnotesize$\t{[t]}$}};
\arrowFrom[black]{0,0}[1.3]{0}\node[clebsch]at(in){{\scriptsize$\phantom{\nu}$}};\node[]at(in){{\scriptsize$\tau$}};\node[anchor=90,inner sep=2pt] at(arrownode){{\footnotesize${[c]}$}};
\arrowFrom[hred]{end}{-38}\node[anchor=142,inner sep=0pt] at(end){{\footnotesize$\r{[u]}$}};\node[clebsch]at(in){{\scriptsize$\phantom{\nu}$}};\node[]at(in){{\scriptsize$\upsilon$}};}
\end{array}\label{different_bases_for_four_particles}} 
In each case, the individual tensors are labelled by a single irreducible representation and a pair of multiplicities; but \emph{which} particular irreducible representations appear, how many appear, and the multiplicity-ranges for each pair of Clebsches can be very different for each. Only their total number (and their total linear-spans as tensors) must agree. 

It is not hard to see that each of these bases are individually self-orthogonal. To see this, one may compute the overlaps:
\eq{\begin{array}{c@{$\!$}c@{$\!$}c}\langle\mathcal{B}_{\hspace{1.5pt}\mathbf{{b}}}^{\hspace{-0pt}\rho\sigma}|\mathcal{B}^{\hspace{2.2pt}\mathbf{{a}}}_{\hspace{-1pt}\mu\nu}\rangle&\langle\mathcal{C}_{\hspace{2.2pt}\mathbf{{a}}}^{\hspace{-1pt}\mu\nu}|\mathcal{C}^{\hspace{1.2pt}\mathbf{{b}}}_{\hspace{-1pt}\rho\sigma}\rangle&\langle\mathcal{D}_{\hspace{2.2pt}\mathbf{{a}}}^{\hspace{-0pt}\mu\nu}|\mathcal{D}^{\hspace{1.2pt}\mathbf{{c}}}_{\hspace{-1pt}\tau\upsilon}\rangle\\
\tikzBox{generic_s_basis_self_contraction}{
\floatingEdge{hgreen,edge,endArrow}{(3.25*\edgeLength,0).. controls ($(4.05*\edgeLength,1.65*\edgeLength)$) and ($(-1.35*\edgeLength,1.65*\edgeLength)$) .. ($(-0.5*\edgeLength,0)$);}\node[anchor=0,inner sep=10pt] at(arrownode){{\footnotesize$\g{[s]}$}};
\floatingEdge{hblue,edge,endArrow}{(3.25*\edgeLength,0).. controls ($(4.05*\edgeLength,-1.65*\edgeLength)$) and ($(-1.35*\edgeLength,-1.65*\edgeLength)$) .. ($(-0.5*\edgeLength,0)$);}\node[anchor=0,inner sep=10pt] at(arrownode){{\footnotesize$\b{[r]}$}};
\floatingEdge{hteal,edge,midArrow}{(0.75*\edgeLength,0).. controls ($(1.1*\edgeLength,0.55*\edgeLength)$) and ($(1.65*\edgeLength,0.55*\edgeLength)$) .. ($(2.*\edgeLength,0)$);}\node[anchor=-90,inner sep=2pt] at(arrownode){{\footnotesize$\t{[t]}$}};
\floatingEdge{hred,edge,midArrow}{(0.75*\edgeLength,0).. controls ($(1.1*\edgeLength,-0.55*\edgeLength)$) and ($(1.65*\edgeLength,-0.55*\edgeLength)$) .. ($(2.*\edgeLength,0)$);}\node[anchor=90,inner sep=2pt] at(arrownode){{\footnotesize$\r{[u]}$}};
\arrowFrom[black]{-0.5*\edgeLength,0}[1.25][0.5]{0}\node[anchor=-90,inner sep=2pt] at(arrownode){{\footnotesize${[a]}$}};
\arrowFrom[black]{2*\edgeLength,0}[1.25][0.5]{0}\node[anchor=-90,inner sep=2pt] at(arrownode){{\footnotesize${[b]}$}};
\node[clebsch]at($(-0.5*\edgeLength,0)$){{\scriptsize$\phantom{\nu}$}};\node[]at($(-0.5*\edgeLength,0)$){{\scriptsize$\mu$}};
\node[clebsch]at($(0.75*\edgeLength,0)$){{\scriptsize$\phantom{\nu}$}};\node[]at($(0.75*\edgeLength,0)$){{\scriptsize$\nu$}};
\node[clebsch]at($(2*\edgeLength,0)$){{\scriptsize$\phantom{\nu}$}};\node[]at($(2*\edgeLength,0)$){{\scriptsize$\sigma$}};
\node[clebsch]at($(3.25*\edgeLength,0)$){{\scriptsize$\phantom{\nu}$}};\node[]at($(3.25*\edgeLength,0)$){{\scriptsize$\rho$}};
}&\tikzBox{generic_t_basis_self_contraction}{
\floatingEdge{hblue,edge,endArrow}{($(3.25*\edgeLength,0)$).. controls($(4.05*\edgeLength,1.65*\edgeLength)$) and  ($(-1.35*\edgeLength,1.65*\edgeLength)$).. (-0.5*\edgeLength,0);}\node[anchor=0,inner sep=10pt] at(arrownode){{\footnotesize${\color{hblue}{[r]}}$}};
\floatingEdge{hred,edge,endArrow}{($(-0.5*\edgeLength,0)$).. controls  ($(-1.35*\edgeLength,-1.65*\edgeLength)$) and ($(4.05*\edgeLength,-1.65*\edgeLength)$) .. (3.25*\edgeLength,0);}\node[anchor=180,inner sep=10pt] at(arrownode){{\footnotesize$\r{[u]}$}};
\floatingEdge{hgreen,edge,midArrow}{($(2.*\edgeLength,0)$).. controls ($(1.65*\edgeLength,0.55*\edgeLength)$) and ($(1.1*\edgeLength,0.55*\edgeLength)$) .. (0.75*\edgeLength,0);}\node[anchor=-90,inner sep=2pt] at(arrownode){{\footnotesize$\g{[s]}$}};
\floatingEdge{hteal,edge,midArrow}{($(0.75*\edgeLength,0)$).. controls  ($(1.1*\edgeLength,-0.55*\edgeLength)$) and ($(1.65*\edgeLength,-0.55*\edgeLength)$).. (2*\edgeLength,0);}\node[anchor=90,inner sep=2pt] at(arrownode){{\footnotesize$\t{[t]}$}};
\arrowFrom[black]{-0.5*\edgeLength,0}[1.25][0.5]{0}\node[anchor=-90,inner sep=2pt] at(arrownode){{\footnotesize${[b]}$}};
\arrowFrom[black]{2*\edgeLength,0}[1.25][0.5]{0}\node[anchor=-90,inner sep=2pt] at(arrownode){{\footnotesize${[a]}$}};
\node[clebsch]at($(-0.5*\edgeLength,0)$){{\scriptsize$\phantom{\nu}$}};\node[]at($(-0.5*\edgeLength,0)$){{\scriptsize$\rho$}};
\node[clebsch]at($(0.75*\edgeLength,0)$){{\scriptsize$\phantom{\nu}$}};\node[]at($(0.75*\edgeLength,0)$){{\scriptsize$\sigma$}};
\node[clebsch]at($(2*\edgeLength,0)$){{\scriptsize$\phantom{\nu}$}};\node[]at($(2*\edgeLength,0)$){{\scriptsize$\nu$}};
\node[clebsch]at($(3.25*\edgeLength,0)$){{\scriptsize$\phantom{\nu}$}};\node[]at($(3.25*\edgeLength,0)$){{\scriptsize$\mu$}};
}&\tikzBox{generic_u_basis_self_contraction}{
\floatingEdge{hteal,edge,endArrow}{(-0.5*\edgeLength,0).. controls ($(-1.35*\edgeLength,1.65*\edgeLength)$) and ($(4.05*\edgeLength,1.65*\edgeLength)$).. ($(3.25*\edgeLength,0)$);}\node[anchor=180,inner sep=10pt] at(arrownode){{\footnotesize$\t{[t]}$}};
\floatingEdge{hblue,edge,endArrow}{(3.25*\edgeLength,0).. controls ($(4.05*\edgeLength,-1.65*\edgeLength)$) and ($(-1.35*\edgeLength,-1.65*\edgeLength)$) .. ($(-0.5*\edgeLength,0)$);}\node[anchor=0,inner sep=10pt] at(arrownode){{\footnotesize$\b{[r]}$}};
\floatingEdge{hgreen,edge,midArrow}{($(2.*\edgeLength,0)$).. controls ($(1.65*\edgeLength,0.55*\edgeLength)$) and ($(1.1*\edgeLength,0.55*\edgeLength)$).. (0.75*\edgeLength,0);}\node[anchor=-90,inner sep=2pt] at(arrownode){{\footnotesize$\g{[s]}$}};
\floatingEdge{hred,edge,midArrow}{($(0.75*\edgeLength,0)$).. controls  ($(1.1*\edgeLength,-0.55*\edgeLength)$) and ($(1.65*\edgeLength,-0.55*\edgeLength)$) .. (2*\edgeLength,0);}\node[anchor=90,inner sep=2pt] at(arrownode){{\footnotesize$\r{[u]}$}};
\arrowFrom[black]{-0.5*\edgeLength,0}[1.25][0.5]{0}\node[anchor=-90,inner sep=2pt] at(arrownode){{\footnotesize${[c]}$}};
\arrowFrom[black]{2*\edgeLength,0}[1.25][0.5]{0}\node[anchor=-90,inner sep=2pt] at(arrownode){{\footnotesize${[a]}$}};
\node[clebsch]at($(-0.5*\edgeLength,0)$){{\scriptsize$\phantom{\nu}$}};\node[]at($(-0.5*\edgeLength,0)$){{\scriptsize$\tau$}};
\node[clebsch]at($(0.75*\edgeLength,0)$){{\scriptsize$\phantom{\nu}$}};\node[]at($(0.75*\edgeLength,0)$){{\scriptsize$\upsilon$}};
\node[clebsch]at($(2*\edgeLength,0)$){{\scriptsize$\phantom{\nu}$}};\node[]at($(2*\edgeLength,0)$){{\scriptsize$\nu$}};
\node[clebsch]at($(3.25*\edgeLength,0)$){{\scriptsize$\phantom{\nu}$}};\node[]at($(3.25*\edgeLength,0)$){{\scriptsize$\mu$}};}
\end{array}\label{four_point_bases_self_overlap}} 
From the orthogonality of two-point graphs under bubble-deletion (\ref{irrep_two_point_orthogonality}), and the orthogonality of the individual Clebsch-Gordan coefficients, it is easy to conclude that 
\eq{\langle\mathcal{B}_{\hspace{1.5pt}\mathbf{{b}}}^{\hspace{-0pt}\rho\sigma}|\mathcal{B}^{\hspace{2.2pt}\mathbf{{a}}}_{\hspace{-1pt}\mu\nu}\rangle\propto\delta\indices{\mathbf{a}}{\mathbf{b}}\delta\indices{\mu}{\rho}\delta\indices{\nu}{\sigma}\,,}
and similarly for the bases $\mathcal{C},\mathcal{D}$. To be clear: the constants of proportionality above can (and generally do) depend on the irreducible representations and multiplicity indices. 

As each basis is self-orthogonal, one can determine the coefficients of expansion from one to another by computing the complete contractions such as
\eq{\hspace{-20pt}\langle\mathcal{C}_{\hspace{1.0pt}\mathbf{{b}}}^{\hspace{-1pt}\rho\sigma}|\mathcal{B}^{\hspace{2.2pt}\mathbf{{a}}}_{\hspace{-1pt}\mu\nu}\rangle\bigger{\;\Leftrightarrow}\;\tikzBox{ts_overlap_diagram}{
\floatingEdge{hblue,edge,midArrow}{(0,-1*\edgeLength).. controls ($(-0.85*\edgeLength,-1*\edgeLength)$) and ($(-1*\edgeLength,0.15*\edgeLength)$) .. ($(-1*\edgeLength,0)$);}\node[anchor=40,inner sep=1pt] at(arrownode){{\footnotesize$\b{[r]}$}};
\floatingEdge{hred,edge,midArrow}{(1*\edgeLength,0).. controls ($(1*\edgeLength,-0.85*\edgeLength)$) and ($(0.15*\edgeLength,-1*\edgeLength)$) .. ($(0,-1*\edgeLength)$);}\node[anchor=140,inner sep=1pt] at(arrownode){{\footnotesize$\r{[u]}$}};
\floatingEdge{black,edge,endArrow}{(0,1*\edgeLength).. controls ($(0.35,0.35*\edgeLength)$) and ($(0.35,-0.35*\edgeLength)$) .. ($(0,-1*\edgeLength)$);}\node[anchor=-10,inner sep=2pt] at(arrownode){{\footnotesize${[b]}$}};
\arrowFrom[black]{-1*\edgeLength,0}[2.0][0.4]{0}\node[anchor=-90,inner sep=2pt] at(arrownode){{\footnotesize${[a]}$}};
\floatingEdge{hteal,edge,midArrow}{(1*\edgeLength,0).. controls ($(1*\edgeLength,0.85*\edgeLength)$) and ($(0.15*\edgeLength,1*\edgeLength)$) .. ($(0,1*\edgeLength)$);}\node[anchor=-140,inner sep=2pt] at(arrownode){{\footnotesize$\t{[t]}$}};
\floatingEdge{hgreen,edge,midArrow}{(0,1*\edgeLength).. controls ($(-0.85*\edgeLength,1*\edgeLength)$) and ($(-1*\edgeLength,0.15*\edgeLength)$) .. ($(-1*\edgeLength,0)$);}\node[anchor=-40,inner sep=2pt] at(arrownode){{\footnotesize$\g{[s]}$}};
\node[clebsch]at($(1*\edgeLength,0)$){{\scriptsize$\phantom{\nu}$}};\node[]at($(1*\edgeLength,0)$){{\scriptsize$\nu$}};
\node[clebsch]at($(-1*\edgeLength,0)$){{\scriptsize$\phantom{\nu}$}};\node[]at($(-1*\edgeLength,0)$){{\scriptsize$\mu$}};
\node[clebsch]at($(0,-1*\edgeLength)$){{\scriptsize$\phantom{\nu}$}};\node[]at($(0,-1*\edgeLength)$){{\scriptsize$\rho$}};\node[clebsch]at($(0,1*\edgeLength)$){{\scriptsize$\phantom{\nu}$}};\node[]at($(0,1*\edgeLength)$){{\scriptsize$\sigma$}};
}
\quad\langle\mathcal{D}_{\hspace{2.2pt}\mathbf{{c}}}^{\hspace{-1pt}\tau\upsilon}|\mathcal{B}^{\hspace{2.2pt}\mathbf{{a}}}_{\hspace{-1pt}\mu\nu}\rangle\bigger{\;\Leftrightarrow}\;\tikzBox{us_overlap_diagram}{
\floatingEdge{hblue,edge,midArrow}{(0,-1*\edgeLength).. controls ($(-0.85*\edgeLength,-1*\edgeLength)$) and ($(-1*\edgeLength,0.15*\edgeLength)$) .. ($(-1*\edgeLength,0)$);}\node[anchor=40,inner sep=1pt] at(arrownode){{\footnotesize$\b{[r]}$}};
\floatingEdge{hteal,edge,midArrow}{(1*\edgeLength,0).. controls ($(1*\edgeLength,-0.85*\edgeLength)$) and ($(0.15*\edgeLength,-1*\edgeLength)$) .. ($(0,-1*\edgeLength)$);}\node[anchor=140,inner sep=1pt] at(arrownode){{\footnotesize$\t{[t]}$}};
\floatingEdge{black,edge,endArrow}{(0,1*\edgeLength).. controls ($(0.35,0.35*\edgeLength)$) and ($(0.35,-0.35*\edgeLength)$) .. ($(0,-1*\edgeLength)$);}\node[anchor=-10,inner sep=2pt] at(arrownode){{\footnotesize${[c]}$}};
\arrowFrom[black]{-1*\edgeLength,0}[2.0][0.4]{0}\node[anchor=-90,inner sep=2pt] at(arrownode){{\footnotesize${[a]}$}};
\floatingEdge{hred,edge,midArrow}{(1*\edgeLength,0).. controls ($(1*\edgeLength,0.85*\edgeLength)$) and ($(0.15*\edgeLength,1*\edgeLength)$) .. ($(0,1*\edgeLength)$);}\node[anchor=-140,inner sep=2pt] at(arrownode){{\footnotesize$\r{[u]}$}};
\floatingEdge{hgreen,edge,midArrow}{(0,1*\edgeLength).. controls ($(-0.85*\edgeLength,1*\edgeLength)$) and ($(-1*\edgeLength,0.15*\edgeLength)$) .. ($(-1*\edgeLength,0)$);}\node[anchor=-40,inner sep=2pt] at(arrownode){{\footnotesize$\g{[s]}$}};
\node[clebsch]at($(1*\edgeLength,0)$){{\scriptsize$\phantom{\nu}$}};\node[]at($(1*\edgeLength,0)$){{\scriptsize$\nu$}};
\node[clebsch]at($(-1*\edgeLength,0)$){{\scriptsize$\phantom{\nu}$}};\node[]at($(-1*\edgeLength,0)$){{\scriptsize$\mu$}};
\node[clebsch]at($(0,-1*\edgeLength)$){{\scriptsize$\phantom{\nu}$}};\node[]at($(0,-1*\edgeLength)$){{\scriptsize$\tau$}};\node[clebsch]at($(0,1*\edgeLength)$){{\scriptsize$\phantom{\nu}$}};\node[]at($(0,1*\edgeLength)$){{\scriptsize$\upsilon$}};
}
\quad\langle\mathcal{D}_{\hspace{2.2pt}\mathbf{{c}}}^{\hspace{-1pt}\tau\upsilon}|\mathcal{C}^{\hspace{0.5pt}\mathbf{{b}}}_{\hspace{-1pt}\rho\sigma}\rangle\bigger{\;\Leftrightarrow}\;\tikzBox{ut_overlap_diagram}{
\floatingEdge{hblue,edge,midArrow}{(0,-1*\edgeLength).. controls ($(-0.85*\edgeLength,-1*\edgeLength)$) and ($(-1*\edgeLength,0.15*\edgeLength)$) .. ($(-1*\edgeLength,0)$);}\node[anchor=40,inner sep=1pt] at(arrownode){{\footnotesize$\b{[r]}$}};
\floatingEdge{hteal,edge,midArrow}{(1*\edgeLength,0).. controls ($(1*\edgeLength,-0.85*\edgeLength)$) and ($(0.15*\edgeLength,-1*\edgeLength)$) .. ($(0,-1*\edgeLength)$);}\node[anchor=140,inner sep=1pt] at(arrownode){{\footnotesize$\t{[t]}$}};
\floatingEdge{black,edge,endArrow}{(0,1*\edgeLength).. controls ($(0.35,0.35*\edgeLength)$) and ($(0.35,-0.35*\edgeLength)$) .. ($(0,-1*\edgeLength)$);}\node[anchor=-10,inner sep=2pt] at(arrownode){{\footnotesize${[c]}$}};
\arrowFrom[black]{-1*\edgeLength,0}[2.0][0.4]{0}\node[anchor=-90,inner sep=2pt] at(arrownode){{\footnotesize${[b]}$}};
\floatingEdge{hgreen,edge,midArrow}{($(0,1*\edgeLength)$).. controls ($(0.15*\edgeLength,1*\edgeLength)$) and ($(1*\edgeLength,0.85*\edgeLength)$) .. (1*\edgeLength,0);}\node[anchor=-140,inner sep=2pt] at(arrownode){{\footnotesize$\g{[s]}$}};
\floatingEdge{hred,edge,midArrow}{($(-1*\edgeLength,0)$).. controls ($(-1*\edgeLength,0.15*\edgeLength)$) and ($(-0.85*\edgeLength,1*\edgeLength)$) .. (0,1*\edgeLength);}\node[anchor=-40,inner sep=2pt] at(arrownode){{\footnotesize$\r{[u]}$}};
\node[clebsch]at($(1*\edgeLength,0)$){{\scriptsize$\phantom{\nu}$}};\node[]at($(1*\edgeLength,0)$){{\scriptsize$\sigma$}};
\node[clebsch]at($(-1*\edgeLength,0)$){{\scriptsize$\phantom{\nu}$}};\node[]at($(-1*\edgeLength,0)$){{\scriptsize$\rho$}};
\node[clebsch]at($(0,-1*\edgeLength)$){{\scriptsize$\phantom{\nu}$}};\node[]at($(0,-1*\edgeLength)$){{\scriptsize$\tau$}};\node[clebsch]at($(0,1*\edgeLength)$){{\scriptsize$\phantom{\nu}$}};\node[]at($(0,1*\edgeLength)$){{\scriptsize$\upsilon$}};
}\hspace{-20pt}
\label{changes_between_general_four_point_bases}}
From these, any particular set of basis tensors can be expanded into any other; for example, 
\eq{\mathcal{B}^{\hspace{2.2pt}\mathbf{{a}}}_{\hspace{-1pt}\mu\nu}=\sum_{\substack{\mathbf{b},\rho,\sigma}}\frac{\langle\mathcal{C}_{\hspace{1.0pt}\mathbf{{b}}}^{\hspace{-1pt}\rho\sigma}|\mathcal{B}^{\hspace{2.2pt}\mathbf{{a}}}_{\hspace{-1pt}\mu\nu}\rangle}{\langle\mathcal{C}_{\hspace{2.2pt}\mathbf{{b}}}^{\hspace{-1pt}\rho\sigma}|\mathcal{C}^{\hspace{1.2pt}\mathbf{{b}}}_{\hspace{-1pt}\rho\sigma}\rangle}\mathcal{C}^{\hspace{1.2pt}\mathbf{{b}}}_{\hspace{-1pt}\rho\sigma}=\sum_{\substack{\mathbf{c},\tau,\upsilon}}\frac{\langle\mathcal{D}_{\hspace{2.2pt}\mathbf{{c}}}^{\hspace{-1pt}\tau\upsilon}|\mathcal{B}^{\hspace{2.2pt}\mathbf{{a}}}_{\hspace{-1pt}\mu\nu}\rangle}{\langle\mathcal{D}_{\hspace{2.2pt}\mathbf{{c}}}^{\hspace{-0pt}\tau\upsilon}|\mathcal{D}^{\hspace{1.2pt}\mathbf{{c}}}_{\hspace{-1pt}\tau\upsilon}\rangle}\mathcal{D}^{\hspace{1.2pt}\mathbf{{c}}}_{\hspace{-1pt}\tau\upsilon}\,.}

It is worth mentioning that, beyond the application of these graphs to the expansion coefficients of one basis into another, the `vacuum' graph
\eq{\mathscr{F}^{\rho\sigma}_{\,\mu\nu}\big(\mathbf{q}\,\b{\mathbf{R}}\,\g{\mathbf{S}}\,\t{\mathbf{T}}\,\r{\mathbf{U}}\,\mathbf{v}\big)\equivR\raisebox{8.5pt}{\text{`}}\left(\begin{array}{@{}c@{$\,$}c@{$\,$}c@{}}\mathbf{\b{R}}&\mathbf{\g{S}}&\mathbf{q}\\
\mathbf{\t{T}}&\mathbf{\r{U}}&\mathbf{v}\end{array}\right)\raisebox{-6pt}{${\hspace{-2pt}}^{\rho\sigma}_{\mu\nu}$}\raisebox{8.5pt}{\text{'}}\equivR\tikzBox{sixJ_plus_graph}{\floatingEdge{hblue,edge,midArrow}{(0,-1*\edgeLength).. controls ($(-0.85*\edgeLength,-1*\edgeLength)$) and ($(-1*\edgeLength,0.15*\edgeLength)$) .. ($(-1*\edgeLength,0)$);}\node[anchor=40,inner sep=1pt] at(arrownode){{\footnotesize$\b{[r]}$}};
\floatingEdge{hred,edge,midArrow}{(1*\edgeLength,0).. controls ($(1*\edgeLength,-0.85*\edgeLength)$) and ($(0.15*\edgeLength,-1*\edgeLength)$) .. ($(0,-1*\edgeLength)$);}\node[anchor=140,inner sep=1pt] at(arrownode){{\footnotesize$\r{[u]}$}};
\floatingEdge{black,edge,endArrow}{(0,1*\edgeLength).. controls ($(0.35,0.35*\edgeLength)$) and ($(0.35,-0.35*\edgeLength)$) .. ($(0,-1*\edgeLength)$);}\node[anchor=-10,inner sep=2pt] at(arrownode){{\footnotesize${[v]}$}};
\arrowFrom[black]{-1*\edgeLength,0}[2.0][0.4]{0}\node[anchor=-90,inner sep=2pt] at(arrownode){{\footnotesize${[q]}$}};
\floatingEdge{hteal,edge,midArrow}{(1*\edgeLength,0).. controls ($(1*\edgeLength,0.85*\edgeLength)$) and ($(0.15*\edgeLength,1*\edgeLength)$) .. ($(0,1*\edgeLength)$);}\node[anchor=-140,inner sep=2pt] at(arrownode){{\footnotesize$\t{[t]}$}};
\floatingEdge{hgreen,edge,midArrow}{(0,1*\edgeLength).. controls ($(-0.85*\edgeLength,1*\edgeLength)$) and ($(-1*\edgeLength,0.15*\edgeLength)$) .. ($(-1*\edgeLength,0)$);}\node[anchor=-40,inner sep=2pt] at(arrownode){{\footnotesize$\g{[s]}$}};
\node[clebsch]at($(1*\edgeLength,0)$){{\scriptsize$\phantom{\nu}$}};\node[]at($(1*\edgeLength,0)$){{\scriptsize$\nu$}};
\node[clebsch]at($(-1*\edgeLength,0)$){{\scriptsize$\phantom{\nu}$}};\node[]at($(-1*\edgeLength,0)$){{\scriptsize$\mu$}};
\node[clebsch]at($(0,-1*\edgeLength)$){{\scriptsize$\phantom{\nu}$}};\node[]at($(0,-1*\edgeLength)$){{\scriptsize$\rho$}};\node[clebsch]at($(0,1*\edgeLength)$){{\scriptsize$\phantom{\nu}$}};\node[]at($(0,1*\edgeLength)$){{\scriptsize$\sigma$}};
}\label{sixJ_plus_symbol}}
can be defined with any choice of six (irreducible) representations and for any possible values of the discrete multiplicity indices
\eq{\mu\!\in\!\big[m\indices{\mathbf{\b{R}}\,\mathbf{\g{S}}}{\mathbf{q}}\big],\quad\nu\!\in\!\big[m\indices{\t{\bar{\mathbf{T}}}\,\r{\bar{\mathbf{U}}}}{\bar{\mathbf{q}}}\big],\quad\rho\!\in\!\big[m\indices{\b{\bar{\mathbf{R}}}\,\r{\mathbf{U}}}{\bar{\mathbf{v}}}\big],\quad\sigma\!\in\!\big[m\indices{\g{\bar{\mathbf{S}}}\,\t{\mathbf{T}}}{\mathbf{v}}\big]\,.}
In the case of $\mathfrak{a}_{\r{1}}\!\simeq\!\mathfrak{su}_{\r{2}}$ gauge theory, for which all representations are real and $m\indices{\b{\mathbf{r}}\,\g{\mathbf{s}}}{\t{\mathbf{t}}}\!=\!1$ for all irreducible representations, which are uniquely labeled by a single Dynkin (highest-)weight $w\!\in\!\mathbb{Z}_{\geq0}$, these are related to what are called `six-$J$ coefficients'.\\

\subsubsection{Proof of (Non-Perturbative) Completeness}

From the discussion in \cite{Bourjaily:2024jbt}, the number of independent tensors for any process involving external representations $\mathbf{\b{R}},\mathbf{\g{S}},\ldots,\mathbf{\t{T}},\mathbf{\r{U}}$ is computed by the multiplicity of the irreducible representation $\mathbf{1}$ in the decomposition of the tensor product representation
\eq{(\mathbf{\b{R}}\!\otimes\!\mathbf{\g{S}}\!\otimes\!\cdots\!\otimes\!\mathbf{\t{T}}\!\otimes\!\mathbf{\r{U}})\simeq\mathbf{1}^{m\indices{\mathbf{\b{R}}\,\mathbf{\g{S}}\cdots\mathbf{\t{T}}\,\mathbf{\r{U}}}{\mathbf{1}}}\bigoplus\ldots\,.}
As the tensor-product representation $(\mathbf{\b{R}}\!\otimes\!\mathbf{\g{S}}\!\otimes\!\cdots\!\otimes\!\mathbf{\t{T}}\!\otimes\!\mathbf{\r{U}})$ may be constructed associatively by any sequence of pairings---e.g.
\eq{(\mathbf{\b{R}}\!\otimes\!\mathbf{\g{S}}\!\otimes\!\cdots\!\otimes\!\mathbf{\t{T}}\!\otimes\!\mathbf{\r{U}})\simeq(\mathbf{\b{R}}\!\otimes\!\mathbf{\g{S}})\!\otimes(\cdots\!\otimes\!\mathbf{\t{T}}\!\otimes\!\mathbf{\r{U}})}
and any pairing may be separately decomposed into irreducible representations---e.g., taking
\eq{(\mathbf{\b{R}}\!\otimes\!\mathbf{\g{S}})\simeq\bigoplus_{\text{irreps\,}\mathbf{q}}\mathbf{q}^{m\indices{\mathbf{\b{R}}\,\mathbf{\g{S}}}{\mathbf{q}}}}
we have that
\eq{(\mathbf{\b{R}}\!\otimes\!\mathbf{\g{S}}\!\otimes\!\cdots\!\otimes\!\mathbf{\t{T}}\!\otimes\!\mathbf{\r{U}})\simeq\bigoplus_{\text{irreps\,}\mathbf{q}}(m\indices{\mathbf{\b{R}}\,\mathbf{\g{S}}}{\mathbf{q}})\mathbf{q}\otimes(\cdots\!\otimes\!\mathbf{\t{T}}\!\otimes\!\mathbf{\r{U}})}
and we may write
\eq{m\indices{\mathbf{\b{R}}\,\mathbf{\g{S}}\cdots\mathbf{\t{T}}\,\mathbf{\r{U}}}{\mathbf{1}}=\sum_{\text{irreps}~\mathbf{q}}m\indices{\mathbf{\b{R}}\,\mathbf{\g{S}}}{\mathbf{q}}m\indices{\mathbf{q}\cdots\mathbf{\t{T}}\,\mathbf{\r{U}}}{\mathbf{1}}\,.}
Continuing in this manner by choosing any pair of representations from those remaining will result in a formula for the total number of tensors given by a sum over multiplicities involving only three representations. 

Any particular sequence of choices of representations to successively pair together can easily be seen as encoding a particular trivalent tree. For $n$ external representations, these graphs involve $(n{-}2)$ vertices and $(n{-}3)$ nested sums over `internal', irreducible representations. Interpreting these graphs as encoding particular \emph{tensors} labeled by the irreducible representations appearing in internal edges and the multiplicity indices for each set of representations, it is clear that these form a complete basis of colour tensors.\\

\subsubsection{Proof of Colour-Orthogonality}

We saw above that the three possible bases of colour tensors for four external representations constructed via Clebsch-Gordan tensors were individually self-orthogonal (\ref{four_point_bases_self_overlap}). We would like to show that this property holds for a basis of colour tensors constructed from any \emph{fixed} tree of Clebsch-Gordan tensors. The proof is fairly straight-forward, following inductively on the number of external legs from the orthogonality of the individual Clebsch-Gordan tensors involving three representations
\eq{\langle\bar{C}^{\nu}\!(\b{\bar{\mathbf{{R'}}}}\,\g{\bar{\mathbf{{S'}}}}|\t{\bar{\mathbf{{T'}}}})|C_{\mu}(\b{\mathbf{R}}\,\g{\mathbf{S}}|\t{\mathbf{T}})\rangle\propto\delta\indices{\nu}{\mu}\delta\indices{\b{\mathbf{R}'}}{\mathbf{\b{R}}}\delta\indices{\g{\mathbf{S}'}}{\mathbf{\g{S}}}\delta\indices{\t{\mathbf{T}'}}{\mathbf{\t{T}}}\,.}
(To be clear: orthogonality in $\mu,\nu$ indices is \emph{not} automatic, but we may assume that all three-particle Clebsch tensors have been \emph{chosen} or \emph{constructed} so that this orthogonality is satisfied. This is always possible; and although potentially computationally cumbersome in general, this is not a problem for cases of greatest significance to physics---where most three-particle Clebsch-Gordan tensors are unique or easily orthogonalized via symmetrization/anti-symmetrization of their arguments.)

For any fixed trivalent tree $\Gamma$ involving $n$ external legs, the particular colour tensors forming the Clebsch-Gordan basis are labelled by $(n{-}3)$ irreducible representations appearing on internal edges, and $(n{-}2)$ multiplicity indices (representing the span of tensors at each vertex). That is, considering the set of tensors
\vspace{-6pt}\eq{\mathcal{B}(\Gamma)\equivR\big\{\mathcal{B}_{\mu\cdots\sigma}^{\hspace{0.5pt}\mathbf{a}\cdots\mathbf{c}}\big\}_{\substack{\fwboxL{0pt}{\text{irreps}\,\mathbf{a},\ldots,\mathbf{c}}\\\fwboxL{0pt}{\text{multiplicities }\mu,\ldots,\sigma}}}\fwboxL{0pt}{\hspace{80pt},}\vspace{-6pt}}
we need to show that 
\eq{\langle\mathcal{B}^{\smash{\mu'\!\cdots\sigma'}}_{\hspace{0.5pt}\mathbf{a'}\!\cdots\mathbf{c'}}|\mathcal{B}_{\mu\cdots\sigma}^{\hspace{0.5pt}\mathbf{a}\cdots\mathbf{c}}\rangle\propto\big(\delta^{\mu'}_{\mu}\cdots\delta^{\sigma'}_{\sigma}\big)\big(\delta^{\mathbf{a}}_{\mathbf{a'}}\cdots\delta^{\mathbf{c}}_{\mathbf{c'}}\big)\,.}

Suppose for the sake of induction that colour-orthogonality is true for all fixed graphs involving $n\!\geq\!3$ external representations (This, we reiterate, is a choice one can freely make for the trivalent Clebsch-Gordan coefficients.) We must show that orthogonality is implied for any graph involving $(n{+}1)$ external representations. Let $\hat{\Gamma}$ represent any fixed trivalent tree whose legs are labelled by $(n{+}1)$ external representations. Any such graph must be of the form 
\eq{\tikzBox{orthogonality_argument_1}{\node[smallDot]at(180:25.5pt){};\node[smallDot]at(195:25.5pt){};\node[smallDot]at(165:25.5pt){};\arrowTo[black]{-150:15pt}[0.75]{-150}\arrowTo[black]{150:15pt}[0.75]{150}\arrowFrom[hblue]{30:15pt}[0.75]{30}\node[anchor=210,inner sep=1pt] at(end){{\footnotesize$\b{[r]}$}};
\arrowFrom[hgreen]{-30:15pt}[0.75]{-30}\node[anchor=-210,inner sep=1pt] at(end){{\footnotesize$\g{[s]}$}};\fill[left color=white, right color=blue!0,postaction={pattern={shadelines[size=1.0pt,line width=0.35pt,angle=20]},pattern color=black!10}] (0,0) circle (15pt);
\node[circle,minimum size=30pt,draw=black,line width=0.5*\lineThickness,fill=none,inner sep=0pt]at(0,0){\raisebox{3.5pt}{$\hat{{\Gamma}}$}};
}\equivL\hspace{5pt}\tikzBox{orthogonality_argument_2}{\node[smallDot]at(180:25.5pt){};\node[smallDot]at(195:25.5pt){};\node[smallDot]at(165:25.5pt){};\arrowTo[black]{-150:15pt}[0.75]{-150}\arrowTo[black]{150:15pt}[0.75]{150}\arrowFrom[hblue]{37pt,0}{60}\node[anchor=210,inner sep=1pt] at($(end)-(0,3pt)$){{\footnotesize$\b{[r]}$}};
\arrowFrom[hgreen]{37pt,0}{-60}\node[anchor=-210,inner sep=1pt] at($(end)+(0,3pt)$){{\footnotesize$\g{[s]}$}};
\arrowTo[hred]{35pt,0}{180}\node[anchor=90,inner sep=2pt] at(arrownode){{\footnotesize$\r{[q]}$}};
\fill[left color=white, right color=blue!0,postaction={pattern={shadelines[size=1.0pt,line width=0.35pt,angle=20]},pattern color=black!10}] (0,0) circle (15pt);
\node[circle,minimum size=30pt,draw=black,line width=0.5*\lineThickness,fill=none,inner sep=1pt]at(0,0){$\Gamma$};
\node[clebsch]at(37pt,0){{\scriptsize$\phantom{\nu}$}};\node at(37pt,0){{\scriptsize$\mu$}};
}}
for \emph{some} pair of external edges (which we have, without loss of generality, taken to encode `outgoing' representations `$\mathbf{\b{R}}$' and `$\mathbf{\g{S}}$'). That is, as a tensor, 
\eq{\mathcal{B}({\hat{\Gamma}})\ni\hat{C}^{\smash{\,\cdots \mathbf{\r{q}}}}_{\,\cdots\mu}\equivL\sum_{\r{q}\in\r{[q]}}C_\Gamma(\cdots|\cdots\,\mathbf{\r{q}})\indices{\cdots}{\cdots\,\r{q}}C_{\mu}(\mathbf{\r{q}}|\mathbf{\b{R}}\,\mathbf{\g{S}})\indices{\r{q}}{\b{[r]}\g{[s]}}\,.}
From this, we can easily see that the orthogonality of Clebsch-Gordan tensors involving $n\!\geq\!3$ external legs ensures complete orthogonality.\\
 
\subsection{Comparing Bases Built from Different Trees: Duality \& Crossing}\label{subsec:duality_and_crossing}

From the discussions above, it is clear that there are in fact \emph{many} possible colour-orthogonal bases to choose for any process involving arbitrary representations: not only must one choose some representative trivalent tree, but also the ordering of the representations among the graph's external edges. Each choice results in a distinct, self-orthogonal basis of colour tensors which span the space of all possible colour tensors---including those which may not arise via perturbation theory. 

As illustrated in our discussion of four-particle Clebsch bases around equation (\ref{different_bases_for_four_particles}) above, the expansion of any particular basis tensor into any other basis follows from the general decomposition of any colour tensor whatsoever. As discussed in \mbox{section~\ref{subsec:summary_of_result}}, orthogonality of any choice of basis ensures that colour tensors may be decomposed via
\eq{C(\b{\mathbf{R}}\,\g{\mathbf{S}}\cdots\!|\!\cdots\t{\mathbf{T}}\,\r{\mathbf{U}})=\sum_{\substack{\mathbf{a}\cdots\mathbf{c}\\\mu\cdots\sigma}}\frac{\langle\mathcal{B}^{\smash{\mu\cdots\sigma}}_{\hspace{0.5pt}\mathbf{a}\cdots\mathbf{c}}|C\rangle}{\langle\mathcal{B}^{\smash{\mu\cdots\sigma}}_{\hspace{0.5pt}\mathbf{a}\cdots\mathbf{c}}|\mathcal{B}_{\mu\cdots\sigma}^{\hspace{0.5pt}\mathbf{a}\cdots\mathbf{c}}\rangle}\mathcal{B}_{\mu\cdots\sigma}^{\hspace{0.5pt}\mathbf{a}\cdots\mathbf{c}}(\b{\mathbf{R}}\,\g{\mathbf{S}}\cdots\!|\!\cdots\t{\mathbf{T}}\,\r{\mathbf{U}})\,.}
Thus, the colour basis given above in (\ref{different_bases_for_four_particles}) represented by colour tensors $\{\mathcal{B}_{\mu\cdots\sigma}^{\hspace{0.5pt}\mathbf{a}\cdots\mathbf{c}}\}$ of the basis generated by any particular graph may be expressed in terms of the colour tensors of the basis represented by some other choice of graph $\{\mathcal{C}_{\hspace{2pt}\tau\cdots\upsilon}^{\hspace{0.5pt}\mathbf{d}\cdots\mathbf{f}}\}$ via
\eq{\mathcal{B}_{\mu\cdots\sigma}^{\hspace{0.5pt}\mathbf{a}\cdots\mathbf{c}}=\sum_{\substack{\\[-2pt]\smash{\mathbf{d},\ldots\mathbf{f}}\\\smash{\tau,\ldots,\upsilon}}}\mathbf{M}^{\smash{\mathbf{a}\cdots\mathbf{c},\hspace{1pt}\tau\cdots\upsilon}}_{\smash{\mu\cdots\rho,\hspace{2pt}\mathbf{d}\cdots\mathbf{f}}}\,\,\mathcal{C}_{\hspace{2pt}\tau\cdots\upsilon}^{\hspace{0.5pt}\mathbf{d}\cdots\mathbf{f}}}
where the coefficients of the expansion are given by 
\eq{\mathbf{M}^{\smash{\mathbf{a}\cdots\mathbf{c},\hspace{1pt}\tau\cdots\upsilon}}_{\smash{\mu\cdots\rho,\hspace{2pt}\mathbf{d}\cdots\mathbf{f}}}\,\equivR\frac{\langle\mathcal{C}^{\hspace{1pt}\tau\cdots\upsilon}_{\hspace{1.5pt}\mathbf{d}\cdots\mathbf{f}}|\mathcal{B}_{\mu\cdots\sigma}^{\hspace{0.5pt}\mathbf{a}\cdots\mathbf{c}}\rangle}{\langle\mathcal{C}^{\hspace{1pt}\tau\cdots\upsilon}_{\hspace{1.5pt}\mathbf{d}\cdots\mathbf{f}}|\mathcal{C}_{\hspace{2pt}\tau\cdots\upsilon}^{\hspace{0.5pt}\mathbf{d}\cdots\mathbf{f}}\rangle}\,.}
We call these numbers `\emph{duality} coefficients' as they relate one basis to another. As in other examples we have seen, these numbers are computed in terms of `vacuum graphs' of Clebsch tensors---those with no external representation indices.

Note that the duality coefficients $\mathbf{M}$ relate bases defined with identical cyclic orderings of the external legs. It turns out that all these changes of bases can be generated via sequences of those arising from four or fewer legs; in particular
\eq{\mathscr{F}^{\smash{\fwbox{9pt}{\mathbf{a}},\fwbox{9pt}{\rho\sigma}}}_{\smash{\fwbox{9pt}{\mu\nu},\fwbox{9pt}{\mathbf{b}}}}\!\!:\!\!\!\left(\fig{-0pt}{1}{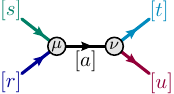}\right)\!\!\mapsto\!\!\left\{\fig{-0pt}{1}{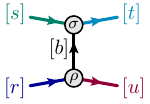}\right\}\,.}
Combined with the twist operation (\ref{definition_of_r_move}) $\mathscr{R}$, we can connect all possible graphs to one another. Moreover, this can be done in multiple distinct ways---for example: 
\eq{\tikzBox{su_duality_sequences}{
\node(s1) at(-1,0){\tbox{\useasboundingbox (-0.85,-0.75)rectangle(1.65,0.75);\arrowTo[hblue]{0,0}{-142};\node[anchor=38,inner sep=0pt] at(in){{\footnotesize$\b{[r]}$}};\arrowTo[hgreen]{0,0}{142}\node[anchor=-38,inner sep=0pt] at(in){{\footnotesize$\g{[s]}$}};\arrowFrom[black]{0,0}[1.]{0}\node[clebsch]at(in){{\scriptsize$\phantom{\nu}$}};\node[]at(in){{\scriptsize$\mu$}};\node[anchor=90,inner sep=2pt] at(arrownode){{\footnotesize${[a]}$}};\arrowFrom[hteal]{end}{38}
\node[anchor=-142,inner sep=0pt] at(end){{\footnotesize$\t{[t]}$}};\arrowFrom[hred]{in}{-38}\node[anchor=142,inner sep=0pt] at(end){{\footnotesize$\r{[u]}$}};\node[clebsch]at(in){{\scriptsize$\phantom{\nu}$}};\node[]at(in){{\scriptsize$\nu$}};
}};
\node(a1)at(3,1.25){\tbox{\useasboundingbox(-0.775-0.3,-0.75)rectangle(1.,0.75);
\arrowFrom[hred]{0,-0.6*\edgeLength}[1.]{-10}\node[anchor=170,inner sep=2pt] at(end){{\footnotesize$\r{[u]}$}};
\arrowFrom[hteal]{0,0.6*\edgeLength}[1.]{10}\node[anchor=-170,inner sep=2pt] at(end){{\footnotesize$\t{[t]}$}};
\arrowTo[hblue]{0,-0.6*\edgeLength}[1.]{-170}\node[anchor=10,inner sep=2pt] at(in){{\footnotesize$\b{[r]}$}};
\arrowTo[hgreen]{0,0.6*\edgeLength}[1.]{170}\node[anchor=-10,inner sep=2pt] at(in){{\footnotesize$\g{[s]}$}};
\arrowTo[black]{0,0.6*\edgeLength}[1.2][0.55]{-90}\node[clebsch]at(in){{\scriptsize$\phantom{\nu}$}};\node[]at(in){{\scriptsize$\rho$}};\node[clebsch]at(end){{\scriptsize$\phantom{\nu}$}};\node[]at(end){{\scriptsize$\sigma$}};
\node[anchor=0,inner sep=2pt]at(arrownode){{\footnotesize$[b]$}};}};
\node(b1)at(3,-1.25){\tbox{
\useasboundingbox(-0.775,-0.75)rectangle(1.5,0.75);\arrowTo[hblue]{0,0}{-142};\node[anchor=38,inner sep=0pt] at(in){{\footnotesize$\b{[r]}$}};\arrowTo[hgreen]{0,0}{142}\node[anchor=-38,inner sep=0pt] at(in){{\footnotesize$\g{[s]}$}};\arrowFrom[black]{0,0}[1.]{0}\node[clebsch]at(in){{\scriptsize$\phantom{\nu}$}};\node[]at(in){{\scriptsize$\mu$}};\node[anchor=90,inner sep=2pt] at(arrownode){{\footnotesize${[a]}$}};
\floatingEdge{hteal,edge,endArrow}{($(\edgeLength,0)$).. controls  ($(1.5*\edgeLength,-0.65)$) and ($(1.3*\edgeLength,0.55)$) .. ($(\edgeLength,0)+(38:\edgeLength)+(0,0.125)$);}
\floatingEdge{hred,edge,endArrow}{($(\edgeLength,0)$).. controls  ($(1.5*\edgeLength,0.65)$) and ($(1.3*\edgeLength,-0.55)$) .. ($(\edgeLength,0)+(-38:\edgeLength)+(0,-0.125)$);}
\node[anchor=-142,inner sep=0pt] at($(\edgeLength,0)+(38:\edgeLength)$){{\footnotesize$\t{[t]}$}};
\node[anchor=142,inner sep=0pt] at($(\edgeLength,0)+(-38:\edgeLength)$){{\footnotesize$\r{[u]}$}};
\node[clebsch]at(\edgeLength,0){{\scriptsize$\phantom{\nu}$}};\node[]at(\edgeLength,0){{\scriptsize$\nu$}};
}};
\node(a2)at(7,1.25){\tbox{\useasboundingbox(-0.85,-0.75)rectangle(0.85,0.75);
\arrowFrom[hred]{0,-0.6*\edgeLength}[1.]{-10}\node[anchor=170,inner sep=2pt] at(end){{\footnotesize$\r{[u]}$}};
\floatingEdge{hteal,edge,endArrow}{($(0,0.4*\edgeLength)$).. controls  ($(-0.75,0.8*\edgeLength)$) and ($(0.25,1.0*\edgeLength)$) .. ($(0,0.6*\edgeLength)+(10:\edgeLength)$);}\node[anchor=-170,inner sep=2pt] at ($(0,0.6*\edgeLength)+(10:\edgeLength)$){{\footnotesize$\t{[t]}$}};
\arrowTo[hblue]{0,-0.6*\edgeLength}[1.]{-170}\node[anchor=10,inner sep=2pt] at(in){{\footnotesize$\b{[r]}$}};
\floatingEdge{hgreen,edge,startArrow}{($(0,0.6*\edgeLength)+(170:\edgeLength)$).. controls  ($(-0.25,1.0*\edgeLength)$) and ($(0.75,0.8*\edgeLength)$) .. ($(0,0.4*\edgeLength)$);}\node[anchor=-10,inner sep=2pt] at ($(0,0.6*\edgeLength)+(170:\edgeLength)$){{\footnotesize$\g{[s]}$}};
\arrowTo[black]{0,0.4*\edgeLength}[1][0.55]{-90}\node[clebsch]at(in){{\scriptsize$\phantom{\nu}$}};\node[]at(in){{\scriptsize$\rho$}};\node[clebsch]at(end){{\scriptsize$\phantom{\nu}$}};\node[]at(end){{\scriptsize$\sigma$}};
\node[anchor=0,inner sep=2pt]at(arrownode){{\footnotesize$[b]$}};}
};
\node(b2)at(7,-1.25){\tbox{%\draw[red](-0.85,-0.75)rectangle(0.85,0.75);
\useasboundingbox(-0.85,-0.75)rectangle(0.85,0.75);
\floatingEdge{hteal,edge,endArrow}{(0,-0.6*\edgeLength).. controls ($(0.45,-0.35*\edgeLength)$) and ($(.6,0.25*\edgeLength)$) .. ($(0,0.6*\edgeLength)+(10:\edgeLength)$);}
\floatingEdge{hred,edge,endArrow}{(0,0.6*\edgeLength).. controls ($(0.45,0.35*\edgeLength)$) and ($(.6,-0.25*\edgeLength)$) .. ($(0,-0.6*\edgeLength)+(-10:\edgeLength)$);}
\arrowTo[hblue]{0,-0.6*\edgeLength}[1.]{-170}\node[anchor=10,inner sep=2pt] at(in){{\footnotesize$\b{[r]}$}};
\node[anchor=-170,inner sep=2pt] at($(0,0.6*\edgeLength)+(10:\edgeLength)$){{\footnotesize$\t{[t]}$}};
\node[anchor=170,inner sep=2pt] at($(0,-0.6*\edgeLength)+(-10:\edgeLength)$){{\footnotesize$\r{[u]}$}};
\arrowTo[hgreen]{0,0.6*\edgeLength}[1.]{170}\node[anchor=-10,inner sep=2pt] at(in){{\footnotesize$\g{[s]}$}};
\arrowTo[black]{0,0.6*\edgeLength}[1.2][0.55]{-90}\node[clebsch]at(in){{\scriptsize$\phantom{\nu}$}};\node[]at(in){{\scriptsize$\rho$}};\node[clebsch]at(end){{\scriptsize$\phantom{\nu}$}};\node[]at(end){{\scriptsize$\sigma$}};
\node[anchor=0,inner sep=2pt]at(arrownode){{\footnotesize$[b]$}};}};
\node(s2)at(11,0){\tbox{%\draw[red](-0.85,-0.75)rectangle(1.5,0.75);
\useasboundingbox (-0.85,-0.75)rectangle(1.5,0.75);
\arrowTo[hblue]{0,0}{-142};\node[anchor=38,inner sep=0pt] at(in){{\footnotesize$\b{[r]}$}};
\floatingEdge{hgreen,edge,startArrow}{($(142:\edgeLength)+(0,0.125)$).. controls ($(0.5*\edgeLength,0.65*\edgeLength)$) and ($(0.8*\edgeLength,0.5*\edgeLength)$) .. ($(1.0*\edgeLength,0)$);}\node[anchor=-38,inner sep=0pt] at(142:\edgeLength){{\footnotesize$\g{[s]}$}};
\floatingEdge{hteal,edge,endArrow}{(0,0).. controls ($(0.5*\edgeLength,0.5*\edgeLength)$) and ($(0.8*\edgeLength,0.65*\edgeLength)$) .. ($(0,0.125)+(38:\edgeLength)+(1.0*\edgeLength,0)$);}\node[anchor=-142,inner sep=0pt] at($(38:\edgeLength)+(1.0*\edgeLength,0)$){{\footnotesize$\t{[t]}$}};
\arrowFrom[black]{0,0}[1.0]{0}\node[clebsch]at(in){{\scriptsize$\phantom{\nu}$}};\node[]at(in){{\scriptsize$\tau$}};\node[anchor=90,inner sep=2pt] at(arrownode){{\footnotesize${[c]}$}};
\arrowFrom[hred]{end}{-38}\node[anchor=142,inner sep=0pt] at(end){{\footnotesize$\r{[u]}$}};\node[clebsch]at(in){{\scriptsize$\phantom{\nu}$}};\node[]at(in){{\scriptsize$\upsilon$}};
}};
\draw[-,map,black,midArrow](s1.25)to[bend left=20](a1.180);\node[anchor=-45,inner sep=2pt]at(arrownode){$\mathscr{F}$};
\draw[-,map,black,midArrow](s1.335)to[bend right=20](b1.180);\node[anchor=45,inner sep=2pt]at(arrownode){$\mathscr{R}$};
\draw[-,map,black,midArrow](a1.0)to[bend left=0](a2.180);\node[anchor=-90,inner sep=5pt]at(arrownode){$\mathscr{R}$};
\draw[-,map,black,midArrow](b1.0)to[bend left=0](b2.180);\node[anchor=90,inner sep=5pt]at(arrownode){$\mathscr{F}$};
\draw[-,map,black,midArrow](b2.0)to[bend right=20](s2.205);\node[anchor=135,inner sep=2pt]at(arrownode){$\mathscr{R}$};
\draw[-,map,black,midArrow](a2.0)to[bend left=20](s2.155);\node[anchor=225,inner sep=2pt]at(arrownode){$\mathscr{F}$};
}
\nonumber}
The above sequences of maps can be understood as the identity that
\eq{\mathscr{F}\!\circ\!\mathscr{R}\!\circ\!\mathscr{F}=\mathscr{R}\!\circ\!\mathscr{F}\!\circ\!\mathscr{R}\,.\label{four_point_duality_summary}}

For larger graphs, it is easy to see that sequences of four-leg $\mathscr{F}$ operations acting on single edges of the graph can connect any two plane graphs with consistent orderings of the external labels; for five particles, for example, we could connect all graphs via:
\eq{
\tikzBox{planar_five_duality}{
\node[anchor=center](v4)at(-2,-0.25){\tbox{%\draw[red](-0.75,-0.5)rectangle(0.75,1.35);
\useasboundingbox(-0.55,-0.5)rectangle(0.55,1.35);
\arrowTo[black]{0,-0.16}[0.75]{210};\node[anchor=20,inner sep=0.5pt] at(in){{\footnotesize${A}$}};
\arrowFrom[black!50]{0,-0.16}[0.75]{-30};
\arrowTo[hpurple]{0,-0.16}[0.75]{90};\node[clebsch]at(end){{\scriptsize$\phantom{}$}};
\arrowTo[black]{in}[0.75]{0};\node[anchor=180,inner sep=1pt] at(in){{\footnotesize${D}$}};
\arrowTo[hteal]{end}[0.75]{90};\node[clebsch]at(end){{\scriptsize$\phantom{}$}};
\arrowTo[black]{in}[0.75]{150}\node[anchor=-10,inner sep=1pt] at(in){{\footnotesize${B}$}};
\arrowTo[black]{end}[0.75]{30}\node[anchor=190,inner sep=1pt] at(in){{\footnotesize${C}$}};
\node[clebsch]at(end){{\scriptsize$\phantom{}$}};
}};
\node[anchor=center](v5)at(-5,1.75){\tbox{\useasboundingbox(-0.5,-0.5)rectangle(1,1);
\arrowTo[black]{0,0}[0.75]{230};\node[anchor=20,inner sep=0.5pt] at(in){{\footnotesize${A}$}};
\arrowTo[hgreen]{end}[0.75]{90}
\arrowTo[black]{in}[0.75]{150}\node[anchor=-10,inner sep=1pt] at(in){{\footnotesize${B}$}};
\arrowTo[black]{end}[0.75]{30}\node[anchor=190,inner sep=1pt] at(in){{\footnotesize${C}$}};
\node[clebsch]at(end){{\scriptsize$\phantom{}$}};
\arrowFrom[hpurple]{0,0}[0.75]{0}\node[clebsch]at(in){{\scriptsize$\phantom{}$}};
\arrowTo[black]{end}[0.75]{50}\node[anchor=200,inner sep=0.5pt] at(in){{\footnotesize${D}$}};
\arrowFrom[black!50]{end}[0.75]{-50}\node[clebsch]at(in){{\scriptsize$\phantom{}$}};
}};
\node(v0)at(90:3){\tbox{%\draw[red](-0.5,-0.5)rectangle(1.5,0.5);
\useasboundingbox(-0.5,-0.65)rectangle(1.5,0.65);
\arrowTo[black]{0,0}[0.75]{230};\node[anchor=20,inner sep=0.5pt] at(in){{\footnotesize${A}$}};
\arrowTo[black]{0,0}[0.75]{130};\node[anchor=-20,inner sep=0.5pt] at(in){{\footnotesize${B}$}};
\arrowFrom[hgreen]{end}[0.75]{0}\node[clebsch]at(in){{\scriptsize$\phantom{}$}};%\node at(in){{\scriptsize$\mu$}};
\arrowTo[black]{end}[0.55]{90}\node[anchor=-90,inner sep=2pt] at(in){{\footnotesize${C}$}};
\arrowFrom[hblue]{end}[0.75]{0}\node[clebsch]at(in){{\scriptsize$\phantom{}$}};%\node at(in){{\scriptsize$\nu$}};
\arrowTo[black]{end}[0.75]{50}\node[anchor=200,inner sep=0.5pt] at(in){{\footnotesize${D}$}};
\arrowFrom[black!50]{end}[0.75]{-50}
\node[clebsch]at(in){{\scriptsize$\phantom{}$}};%\node at(in){{\scriptsize$\rho$}};
%\node[anchor=160,inner sep=0.5pt] at(end){{\footnotesize${}$}};
}};
\node(v1)at(5,1.75){\tbox{%\draw[red](-0.5,-0.5)rectangle(1,1);
\useasboundingbox(-0.5,-0.5)rectangle(1,1);
\arrowTo[black]{0,0}[0.75]{230};\node[anchor=20,inner sep=0.5pt] at(in){{\footnotesize${A}$}};
\arrowTo[black]{0,0}[0.75]{130};\node[anchor=-20,inner sep=0.5pt] at(in){{\footnotesize${B}$}};
\arrowFrom[hred]{end}[0.75]{0}\node[clebsch]at(in){{\scriptsize$\phantom{}$}};%\node at(in){{\scriptsize$\mu$}};
\arrowFrom[black!50]{end}[0.75]{-50}
\arrowTo[hblue]{in}[0.75]{90}\node[clebsch]at(end){{\scriptsize$\phantom{}$}};%\node at(end){{\scriptsize$\sigma$}};
\arrowTo[black]{in}[0.75]{150}\node[anchor=-10,inner sep=1pt] at(in){{\footnotesize${C}$}};
\arrowTo[black]{end}[0.75]{30}\node[anchor=190,inner sep=1pt] at(in){{\footnotesize${D}$}};
\node[clebsch]at(end){{\scriptsize$\phantom{}$}};%\node at(end){{\scriptsize$\tau$}};
}};
\node[](v2)at(2.,-0.25){\tbox{%\draw[red](-0.85,-0.5)rectangle(0.85,1.35);
\useasboundingbox(-0.55,-0.5)rectangle(0.55,1.35);
\arrowTo[black]{0,-0.16}[0.75]{210};\node[anchor=20,inner sep=0.5pt] at(in){{\footnotesize${A}$}};
\arrowFrom[black!50]{0,-0.16}[0.75]{-30};
\arrowTo[hred]{0,-0.16}[0.75]{90};\node[clebsch]at(end){{\scriptsize$\phantom{}$}};
\arrowTo[black]{in}[0.75]{180};\node[anchor=0,inner sep=1pt] at(in){{\footnotesize${B}$}};
\arrowTo[hteal]{end}[0.75]{90};\node[clebsch]at(end){{\scriptsize$\phantom{}$}};
\arrowTo[black]{in}[0.75]{150}\node[anchor=-10,inner sep=1pt] at(in){{\footnotesize${C}$}};
\arrowTo[black]{end}[0.75]{30}\node[anchor=190,inner sep=1pt] at(in){{\footnotesize${D}$}};
\node[clebsch]at(end){{\scriptsize$\phantom{}$}};
}};
\node[anchor=90,inner sep=0pt]at(v0.-90){{\footnotesize$((A\!\otimes\!B)\!\otimes\!C)\!\otimes\!D$}};
\node[anchor=90,inner sep=0pt](l5)at(v5.-90){{\footnotesize$((A\!\otimes\!(B\!\otimes\!C))\!\otimes\!D$}};
\node[anchor=90,inner sep=0pt](l4)at(v4.-90){{\footnotesize$A\!\otimes\!((B\!\otimes\!C)\!\otimes\!D)$}};
\node[anchor=90,inner sep=0pt](l2)at(v2.-90){{\footnotesize$A\!\otimes\!(B\!\otimes\!(C\!\otimes\!D))$}};
\node[anchor=90,inner sep=0pt](l1)at(v1.-90){{\footnotesize$(A\!\otimes\!B)\!\otimes\!(C\!\otimes\!D)$}};
\draw[-,map,hblue,midArrow](v0.0)to[bend left=20](v1.155);\node[anchor=225,inner sep=2pt]at(arrownode){$\b{\mathscr{F}}$};
\draw[-,map,hred,midArrow](l1.195)to[bend left=20](v2.-10);\node[anchor=135,inner sep=2pt]at(arrownode){$\r{\mathscr{F}}$};
\draw[-,map,hgreen,midArrow](v0.180)to[bend right=20](v5.25);\node[anchor=-45,inner sep=2pt]at(arrownode){$\g{\mathscr{F}}$};
\draw[-,map,hpurple,midArrow](l5.-15)to[bend right=20](v4.190);\node[anchor=45,inner sep=2pt]at(arrownode){${\color{hpurple}\mathscr{F}}$};
\draw[-,map,hteal,midArrow]($(v4.-10)+(0.3,-0.25)$)to[bend right=20]($(v2.190)+(-0.3,-0.25)$);\node[anchor=90,inner sep=2pt]at(arrownode){${\color{hteal}\mathscr{F}}$};
}\label{graphical_5pt_duality_sequence}
}
This, when spelled out in terms of specific tensors, corresponds to the identity:
\eq{\mathscr{F}^{\smash{\hspace{2.5pt}\g{\mathbf{r}}\b{\mathbf{s}}\hspace{2.5pt},\upsilon\omega\tau}}_{\hspace{1.2pt}\smash{\mu\nu\hspace{-1pt}\rho,\hspace{2.5pt}\r{\mathbf{p}}\t{\mathbf{t}}}}=\sum_{\substack{\mathbf{\r{r}},\mathbf{\b{t}}\\\bar{\mu},\sigma,\tau}}\hspace{-2pt}\b{\mathscr{F}}^{\smash{\hspace{2.5pt}\g{\mathbf{r}}\b{\mathbf{s}}\hspace{2.5pt},\bar{\mu}\sigma\tau}}_{\smash{\mu\nu\hspace{-1pt}\rho,\hspace{2.5pt}\r{\mathbf{r}}\b{\mathbf{t}}}}
\,\r{\mathscr{F}}^{\smash{\hspace{3.5pt}\r{\mathbf{r}}\b{\mathbf{t}}\hspace{3.5pt},\upsilon\omega\tau}}_{\hspace{1.2pt}\smash{\bar{\mu}\sigma\hspace{-0pt}\tau,\hspace{2.5pt}\r{\mathbf{p}}\t{\mathbf{t}}}}=\hspace{-10pt}\sum_{\substack{\mathbf{\g{q}},\mathbf{{\color{hpurple}s}},\t{\mathbf{q}},{\color{hpurple}\mathbf{p}}\\\eta,\theta,\bar{\rho},\bar{\eta},\epsilon,\upsilon}}
\hspace{-10pt}\g{\mathscr{F}}^{\smash{\hspace{2.5pt}\g{\mathbf{r}}\b{\mathbf{s}}\hspace{2.5pt},\eta\theta\bar{\rho}}}_{\hspace{1.2pt}\smash{\mu\nu\hspace{-1pt}\rho,\hspace{2.5pt}\g{\mathbf{q}}{\color{hpurple}{\mathbf{s}}}}}
\,{\color{hpurple}\mathscr{F}}^{\smash{\hspace{2.5pt}\g{\mathbf{q}}{\color{hpurple}\mathbf{s}}\hspace{2.5pt},\bar{\eta}\epsilon\upsilon}}_{\hspace{1.2pt}\smash{\hspace{0.75pt}\eta\theta\hspace{-0.5pt}\bar{\rho},\hspace{2.5pt}\t{\mathbf{q}}{\color{hpurple}{\mathbf{p}}}}}
\,{\color{hteal}\mathscr{F}}^{\smash{\hspace{2.5pt}\t{\mathbf{q}}{\color{hpurple}\mathbf{p}}\hspace{1.05pt},\upsilon\omega\tau}}_{\hspace{1.2pt}\smash{\hspace{0.75pt}\bar{\eta}\epsilon\upsilon,\hspace{2.5pt}\r{\mathbf{p}}{\color{hteal}{\mathbf{t}}}}}
\,.\label{five_point_planar_relations_in_components}}

It is a well-known folk theorem that combining sequences of $\mathscr{F}$ operations with three-leg $\mathscr{R}$ operators acting on single vertices, any pair of plane graphs with arbitrary orderings of external edge labels (but the same number of edges and vertices) can be mapped to one another. 

Identities such as (\ref{four_point_duality_summary}) and (\ref{five_point_planar_relations_in_components}) are guaranteed to hold as a consequence of the completeness of the bases of tensors represented by each; but it is interesting to take a more `categorical' view of the situation.

We may instead view the $\mathscr{F}$ and $\mathscr{R}$ as general maps between abstract objects; and view the commutativity relations of (\ref{four_point_duality_summary}) or (\ref{five_point_planar_relations_in_components}) as \emph{constraints} which these maps must satisfy. The existence of Clebsch-Gordan colour tensor bases ensures that this category is non-trivially realizable. In this context, relations (\ref{four_point_duality_summary}) and (\ref{five_point_planar_relations_in_components}) are known as the hexagon and petagon relations, and are part of the \emph{coherrence conditions} of the category. The Biedenharn-Elliott relation among six-$J$ symbols and the famous 5-term Dilog identity are both instantiations of the categorical pentagon relation: the unification of seemingly unrelated relations is a hallmark of `categorification'. For higher-point trees one can construct similar webs of interrelations, but these turn out to generate no new constraints. For more on monoidal tensor categories we direct the reader to standard references such as \cite{MR3674995, beer2018categories, MR182649, Selinger2011}.

This attitude toward the duality maps is analogous to viewing Pl\"ucker relations for the Grassmannian as \emph{constraints} imposed upon abstract `coordinates' labelled by $k$-tuples: when these symbols are \emph{defined} in terms of actual determinants, then all Pl\"ucker relations are trivially satisfied as implications of Cramer's rule; but a more abstract view is occasionally powerful. It is also intriguing that similar structures arise in scattering amplitudes through connections to cluster algebras \cite{FockGoncharov2007, Golden:2013xva}. It would be interesting to pursue such connections in the future.

\subsection{Colour Tensor Decomposition and Reduction}\label{subsec:decomposition_and_reduction}

Completeness of any Clebsch-Gordan colour tensor basis ensures that \emph{any} tensor involving the same set of external representations can be represented with unique coefficients in that basis:
\eq{C(\mathbf{\b{R}}\cdots\mathbf{\g{S}}|\mathbf{\t{T}}\cdots\mathbf{\r{U}})\in\mathrm{span}\big\{\mathcal{B}_{\mu\cdots\sigma}^{\hspace{0.5pt}\mathbf{a}\cdots\mathbf{c}}\big\}_{\substack{\fwboxL{0pt}{\text{irreps }\mathbf{a},\ldots,\mathbf{c}}\\\fwboxL{0pt}{\text{multiplicies }\mu,\ldots,\sigma}}}\fwboxL{0pt}{\hspace{80pt},}\vspace{-6pt}}
where these basis graphs are chosen for any fixed choice of trivalent tree $\Gamma$. Moreover, orthogonality of the tensors within any such basis ensures that the coefficients of the expansion of $C$ into the basis $\{\mathcal{B}\}$ can be computed directly via
\eq{C(\b{\mathbf{R}}\cdots\g{\mathbf{S}}\cdots\!|\t{\mathbf{T}}\cdots\r{\mathbf{U}})=\sum_{\substack{\mathbf{a}\cdots\mathbf{c}\\\mu\cdots\sigma}}\frac{\langle\mathcal{B}^{\smash{\mu\cdots\sigma}}_{\hspace{0.5pt}\mathbf{a}\cdots\mathbf{c}}|C\rangle}{\langle\mathcal{B}^{\smash{\mu\cdots\sigma}}_{\hspace{0.5pt}\mathbf{a}\cdots\mathbf{c}}|\mathcal{B}_{\mu\cdots\sigma}^{\hspace{0.5pt}\mathbf{a}\cdots\mathbf{c}}\rangle}\mathcal{B}_{\mu\cdots\sigma}^{\hspace{0.5pt}\mathbf{a}\cdots\mathbf{c}}(\b{\mathbf{R}}\cdots\g{\mathbf{S}}|\t{\mathbf{T}}\cdots\r{\mathbf{U}})\,}%
where the non-trivial computation of 
\eq{\langle\mathcal{B}^{\smash{\mu\cdots\sigma}}_{\hspace{0.5pt}\mathbf{a}\cdots\mathbf{c}}|C\rangle}
can always be considered as a vacuum graph connecting the (conjugated) tree $\Gamma$ to a graph encoding the colour tensor $C$. 

While the statements above are quite general and powerful, it is useful to observe that these coefficients can also be determined by iterated sequences of smaller duality relations which \emph{decompose} a given graph into those of any chosen basis.\\

\subsubsection{Direct \emph{Decomposition} of Arbitrary Colour Tensors}

Suppose one is given some arbitrary colour tensor defined by some arbitrary graph. We would like to show how the operations $\mathscr{F}$ and $\mathscr{R}$ (viewed as local maps acting on single edges and vertices) can be used to systematically decompose any graph into those of any chosen basis. 

Actually, these two operations alone cannot connect \emph{arbitrary} graphs, as neither $\mathscr{F}$ nor $\mathscr{R}$ affects the Euler characteristic of a graph. We need one further operation: \emph{bubble-deletion} defined 
\eq{\sum_{\g{s}\in\g{[s]},\t{t}\in\t{[t]}}\mathbf{\bar{C}}^\nu(\mathbf{\t{T}}|\mathbf{\g{S}}\,\mathbf{\r{u}})\indices{\t{t}}{\g{s}\r{[u]}}\mathbf{C}_\mu(\mathbf{\b{r}}\,\mathbf{\g{S}}|\mathbf{\t{T}})\indices{\b{[r]}}{\g{s}\,\t{t}}\equivL\,\mathscr{B}_\mu(\mathbf{\b{R}}\,\mathbf{\g{S}}|\mathbf{\t{T}})\,\delta\indices{\b{[r]}}{\r{[u]}}\delta\indices{\nu}{\mu}\delta\indices{\mathbf{\b{r}}}{\mathbf{\r{u}}}}
which we can represent graphically as the statement that 
\vspace{-6pt}\eq{\tikzBox{bubble_deletion_general_1}{\draw[hblue,edge,midArrow](-1.15,0)--(-0.35,0);\node[anchor=0,inner sep=0pt] at(-1.15,0){{\footnotesize$\b{[r]}$}};\node[anchor=180,inner sep=0pt] at(1.35,0){{\footnotesize$\r{[u]}$}};\draw[hgreen,edge,midArrow](0.55,0)to[arc through={counterclockwise,(90:0.45)}](-0.35,0);\node[anchor=-90,inner sep=2pt] at(arrownode){{\footnotesize$\g{[{s}]}$}};\draw[hteal,edge,midArrow](-0.35,0)to[arc through={counterclockwise,(-90:0.45)}](0.55,0);\node[anchor=90,inner sep=2pt] at(arrownode){{\footnotesize$\t{[{t}]}$}};\draw[hred,edge,midArrow](0.55,0)--(1.35,0);\node[clebsch]at(-0.35,0){{\footnotesize$\phantom{\nu}$}};
\node[clebsch]at(0.55,0){{\footnotesize$\phantom{\nu}$}};\node[]at(0.55,0){{\footnotesize$\nu$}};\node[]at(-0.35,0){{\footnotesize$\mu$}};}
\hspace{5pt}\equivL\;\mathscr{B}_\mu(\mathbf{\b{r}}\,\mathbf{\g{S}}|\mathbf{\t{T}})\;
\tikzBox{identity_r_v2}{\arrowTo[hblue]{0,0}[1.25]{180}\node[anchor=180,inner sep=0pt] at(end){{\footnotesize$\b{[r]}$}};\node[anchor=0,inner sep=0pt] at(in){{\footnotesize$\b{[r]}$}};}}
which we can understand as an instance of  the more general implication of Schur's lemma, (\ref{irrep_two_point_orthogonality})---holding for any pair of irreducible representations $\mathbf{\b{r}}$ and $\mathbf{\r{u}}$. It is worth noting that these coefficients are related to the $\Theta$ coefficients defined in (\ref{theta_graph_defined}) many ways: 
\eq{\begin{split}\Theta_\mu(\mathbf{\b{r}}\,\mathbf{\g{s}}|\mathbf{\t{t}})&=\mathscr{B}_\mu(\mathbf{\b{r}}\,\mathbf{\g{s}}|\mathbf{\t{t}})\,\,\mathrm{dim}(\mathbf{\b{r}})\\&=\sum_{\nu\in[m\indices{\mathbf{\b{r}}\,\mathbf{\g{s}}}{\mathbf{\t{t}}}]}\mathscr{R}_{\mu}^{\phantom{\mu}\,\nu}(\mathbf{\b{r}}\,\mathbf{\g{s}})\mathscr{B}_\nu(\mathbf{\g{s}}\,\mathbf{\b{r}}|\mathbf{\t{t}})\,\,\mathrm{dim}(\mathbf{\g{s}})\\
&=\mathscr{B}_\mu(\mathbf{\t{t}}\,\mathbf{\g{\bar{s}}}|\mathbf{\b{r}})\,\,\mathrm{dim}(\mathbf{\g{t}})
\end{split}\vspace{-10pt}}
and so-on. 

Combined with the operations $\mathscr{F}$ and $\mathscr{R}$, any colour tensor graph can be reduced into any chosen basis. The proof is semi-trivial: we have already seen how sequences of $\mathscr{F}$ and $\mathscr{R}$ can be used to connect any pair of tree-level tensor diagrams, allowing one to `expand' any colour tensor encoded by a tree into a basis specified by some other fixed tree; therefore, we must only see that colour tensors encoded by graphs of higher Euler-characteristic can be reduced to trees. 

The argument is very simple. Consider an arbitrary tensor diagram of \emph{girth} $g$. Recall that the girth of a graph is the size of its smallest cycle. If a graph has girth 2, then use bubble-deletion to reduce it to trees; if its girth is greater than 2, then pick any edge along a minimal-length cycle, and apply the $\mathscr{F}$ operation (an `st-duality'); the resulting graph will be of strictly lower girth; repeat until there exists a 2-cycle, and then apply bubble deletion. To illustrate this point, consider the reduction of a girth-4 graph via:
\vspace{4pt}\eq{\hspace{-240pt}\tikzBox{box_to_tree_sequence_0}{
\arrowTo[hblue]{0,0}{-135}\node[anchor=20,inner sep=0pt] at(in){{\footnotesize$\b{[r]}$}};
\arrowFrom[black]{end}{90}\node[anchor=0,inner sep=2pt] at(arrownode){{\footnotesize${[a]}$}};
\arrowTo[hgreen]{end}{135}\node[anchor=-20,inner sep=0pt] at(in){{\footnotesize$\g{[s]}$}};
\arrowFrom[black]{end}{0}\node[anchor=-90,inner sep=2pt] at(arrownode){{\footnotesize${[b]}$}};
\arrowFrom[hteal]{end}{45}\node[anchor=-160,inner sep=0pt] at(end){{\footnotesize$\t{[t]}$}};
\arrowFrom[black]{in}{-90}\node[anchor=180,inner sep=2pt] at(arrownode){{\footnotesize${[c]}$}};
\arrowFrom[hred]{end}{-45}\node[anchor=160,inner sep=0pt] at(end){{\footnotesize$\r{[u]}$}};
\arrowFrom[black]{in}{180}\node[anchor=90,inner sep=2pt] at(arrownode){{\footnotesize${[d]}$}};
\node[clebsch]at(0,0){{\footnotesize$\phantom{\nu}$}};\node[]at(0,0){{\footnotesize$\mu$}};
\node[clebsch]at(0,\edgeLength){{\footnotesize$\phantom{\nu}$}};\node[]at(0,\edgeLength){{\footnotesize$\nu$}};
\node[clebsch]at(\edgeLength,\edgeLength){{\footnotesize$\phantom{\nu}$}};\node[]at(\edgeLength,\edgeLength){{\footnotesize$\rho$}};
\node[clebsch]at(\edgeLength,0){{\footnotesize$\phantom{\nu}$}};\node[]at(\edgeLength,0){{\footnotesize$\sigma$}};
}\hspace{-4.5pt}\overset{\mathscr{F}}{\bigger{\longrightarrow}}\hspace{-4.5pt}
\tikzBox{box_to_tree_sequence_1}{
\arrowTo[hblue]{0,0.35}{-120}\node[anchor=10,inner sep=0pt] at(in){{\footnotesize$\b{[r]}$}};
\arrowTo[hgreen]{end}{120}\node[anchor=-10,inner sep=0pt] at(in){{\footnotesize$\g{[s]}$}};
\arrowFrom[black]{end}{0}\node[anchor=-90,inner sep=2pt] at(arrownode){{\footnotesize${[e]}$}};
\node[clebsch]at(in){{\footnotesize$\phantom{\nu}$}};\node[]at(in){{\footnotesize$\tau$}};
\arrowFrom[black]{end}[1]{40}\node[anchor=-50,inner sep=2pt] at(arrownode){{\footnotesize${[b]}$}};
\arrowFrom[black]{in}[1]{-40}\node[anchor=50,inner sep=2pt] at(arrownode){{\footnotesize${[d]}$}};
\node[clebsch]at(in){{\footnotesize$\phantom{\nu}$}};\node[]at(in){{\footnotesize$\upsilon$}};
\arrowFrom[hred]{end}{-15}\node[anchor=167,inner sep=0pt] at(end){{\footnotesize$\r{[u]}$}};
\arrowFrom[black]{in}[1.28]{90}\node[anchor=180,inner sep=2pt] at(arrownode){{\footnotesize${[c]}$}};
\node[clebsch]at(in){{\footnotesize$\phantom{\nu}$}};\node[]at(in){{\footnotesize$\sigma$}};
\arrowFrom[hteal]{end}{15}\node[anchor=-167,inner sep=0pt] at(end){{\footnotesize$\t{[t]}$}};
\node[clebsch]at(in){{\footnotesize$\phantom{\nu}$}};\node[]at(in){{\footnotesize$\rho$}};
}\hspace{-4.5pt}\overset{\mathscr{F}}{\bigger{\longrightarrow}}\hspace{-4.5pt}
\tikzBox{box_to_tree_sequence_2}{
\arrowTo[hblue]{0,0.35}{-120}\node[anchor=10,inner sep=0pt] at(in){{\footnotesize$\b{[r]}$}};
\arrowTo[hgreen]{end}{120}\node[anchor=-10,inner sep=0pt] at(in){{\footnotesize$\g{[s]}$}};
\arrowFrom[black]{end}{0}\node[anchor=-90,inner sep=2pt] at(arrownode){{\footnotesize${[e]}$}};
\node[clebsch]at(in){{\footnotesize$\phantom{\nu}$}};\node[]at(in){{\footnotesize$\tau$}};
\draw[black,edge,midArrow](end)to[arc through={clockwise,($(end)+(45:0.65*\edgeLength)$)}]($(end)+(0:\edgeLength)$);
\node[anchor=-90,inner sep=2pt] at(arrownode){{\footnotesize${[b]}$}};
\draw[black,edge,midArrow]($(end)+(0:\edgeLength)$)to[arc through={clockwise,($(end)+(-45:0.65*\edgeLength)$)}]($(end)$);
\node[anchor=90,inner sep=2pt] at(arrownode){{\footnotesize${[d]}$}};
\node[clebsch]at(end){{\footnotesize$\phantom{\nu}$}};\node[]at(end){{\footnotesize$\upsilon$}};
\coordinate(right)at($(end)+(0:\edgeLength)$);
\arrowFrom[black]{right}{0};\node[anchor=-90,inner sep=2pt] at(arrownode){{\footnotesize${[f]}$}};
\node[clebsch]at(in){{\footnotesize$\phantom{\nu}$}};\node[]at(in){{\footnotesize$\phi$}};
\arrowFrom[hteal]{end}{60}\node[anchor=-170,inner sep=0pt] at(end){{\footnotesize$\t{[t]}$}};
\arrowFrom[hred]{in}{-60}\node[anchor=170,inner sep=0pt] at(end){{\footnotesize$\r{[u]}$}};
\node[clebsch]at(in){{\footnotesize$\phantom{\nu}$}};\node[]at(in){{\footnotesize$\omega$}};
}\hspace{-4.5pt}\overset{\mathscr{B}}{\bigger{\longrightarrow}}\hspace{-4.5pt}
\tikzBox{box_to_tree_sequence_3}{
\arrowTo[hblue]{0,0.35}{-120}\node[anchor=10,inner sep=0pt] at(in){{\footnotesize$\b{[r]}$}};
\arrowTo[hgreen]{end}{120}\node[anchor=-10,inner sep=0pt] at(in){{\footnotesize$\g{[s]}$}};
\arrowFrom[black]{end}{0}\node[anchor=-90,inner sep=2pt] at(arrownode){{\footnotesize${[e]}$}};
\node[clebsch]at(in){{\footnotesize$\phantom{\nu}$}};\node[]at(in){{\footnotesize$\tau$}};
\arrowFrom[hteal]{end}{60}\node[anchor=-170,inner sep=0pt] at(end){{\footnotesize$\t{[t]}$}};
\arrowFrom[hred]{in}{-60}\node[anchor=170,inner sep=0pt] at(end){{\footnotesize$\r{[u]}$}};
\node[clebsch]at(in){{\footnotesize$\phantom{\nu}$}};\node[]at(in){{\footnotesize$\omega$}};
}\fwboxL{0pt}{.}\hspace{-220pt}
\label{box_to_tree_reduction_sequence}}

It is clear that by repeating this process, any graph can be systematically decomposed into those with vanishing Euler characteristic---namely, trees. Then, it is a trivial fact that any tree can be connected to any other via sequences of $\mathscr{F}$ and $\mathscr{R}$ operations, thereby expanding the original graph into those of any chosen basis. 

It is also clear that there is no unique way to do this decomposition, and a different reduction sequence can land on the same tree at the end. The equivalence of these reductions implies relations between the coefficient six-$J$s, the most famous being the Biedenharn-Elliott identity \cite{Biedenharn1971}.

\newpage
\subsubsection{Computing Coefficients via Duality Sequences}

It is worthwhile to see how this works in practice. Suppose that we are given a colour tensor generated by some multi-loop Feynman diagram. The colour tensors arising via Feynman rules always involve tensors arising via graphs of representation matrices and structure constants---both of which can be understood as particular instances of Clebsch-Gordan tensors, sewn together. Thus, we may apply sequences of maps (in many distinct ways) to systematically decompose the tensor into a chosen basis.

The upshot of this approach is that the requisite maps, $\mathscr{F}$, $\mathscr{R}$, and $\mathscr{B}$ may be computed (and tabulated, say) once and for all for a wide array of possible arguments. Recall that these maps depend on various irreducible representations and multiplicity indices, which in principle span unbounded (although countable) ranges of possible arguments; but the ranges of arguments required for the decomposition of a given class of tensors---those involving external adjoint-charged particles, say---can be manageably limited in scope. 

More directly, we have seen that the coefficients of an arbitrary colour tensor $C$ into any colour-diagonal basis can be computed in terms of vacuum diagrams representing the complete contractions representing $\langle\mathcal{B}^{\smash{\mu\cdots\sigma}}_{\hspace{0.5pt}\mathbf{a}\cdots\mathbf{c}}|C\rangle$. These vacuum graphs are clearly reducible to `empty' trees via the same sequences of moves as described above. For example, the sequence of moves (\ref{box_to_tree_reduction_sequence}) can be applied to any \emph{subgraph}---it not being necessary that the edges corresponding to the representations $\mathbf{\b{r}},\ldots,\mathbf{\r{u}}$ in (\ref{box_to_tree_reduction_sequence}), for example, be `external'. This procedure reduces all vacuum graphs to sums of products of six-$J$ coefficients, which can be further reduced to a \emph{minimal} set of independent six-$J$ coefficients to be calculated once and for all, as has been discussed in \emph{e.g.}~\cite{birdtracks} and \cite{Searle1988}. Such symbolic approaches can be extremely powerful for dealing with tensors associated to large graphs or ones dressed with high-dimensional representations, settings where explicit tensor contractions may be prohibitively costly in comparison.

%================================================================================================================
%    Comparisons with Other Bases 
%================================================================================================================
\newpage
\section{Comparisons with Other Choices for Colour Tensor `Bases'}\label{sec:other_colour_bases}

Although Clebsch-Gordan coefficients are well known to physicists, the bases of colour tensors described here are several steps removed from those which appear directly in the Feynman rules of gauge theory or those frequently encountered in the study of scattering amplitudes among coloured particles. As described in \mbox{section~\ref{subsec_motivation}}, the Feynman rules of gauge theory (\ref{rep_in_feynman_rule}) \emph{directly} make reference to the generators of various representations of Lie algebras. As we have seen in \mbox{section~\ref{subsec:everything_is_clebsch}}, these representations, encoded by rank-three tensors appearing directly in the Lagrangian defining the gauge theory, can in fact be understood as particular Clebsch-Gordan tensors:
\eq{\mathbf{\b{R}}\indices{\b{[r]}\,\r{[\adR]}}{\b{[r]}}\equivL\, C_{1}(\mathbf{\b{R}}\,\mathbf{\r{ad}}|\mathbf{\b{R}})\indices{\b{[r]}\,\r{[\adR]}}{\b{[r]}}\,.}
But the \emph{particular} colour tensor involved in any one Feynman diagram is of course fairly meaningless: these tensors satisfy many linear relations arising from both general aspects of Lie algebras (such as the Jacobi identity) and many peculiar relations specific to particular representations of particular algebras. Such relations are necessitated by the gauge \emph{non-invariance} of individual Feynman diagrams diagrams.  

Due to this redundancy, a number of familiar collections of convenient colour tensors have been used to represent amplitudes with various degrees of completeness, redundancy, and generality, but almost universally suffering from colour non-orthogonality.\\

One salient feature of most of these frameworks is that they often involve tensors defined directly (or closely) in terms of the generators defining the representations of coloured particles. This endows them with a sense of naturalness that may appear absent from the Clebsch colour tensor bases described above in \mbox{section~\ref{sec:clebsch_Gordan_colour_tensors}}. These new tensors, by contrast, are defined in terms of irreducible representations often \emph{not} corresponding to those of the coloured particles involved in a process (or even present in the theory). 

Consider for example the case of $\mathfrak{so}_{10}$ gauge theory with a single particle charged in the fundamental $(\mathbf{\b{10}}$-dimensional) representation `$\mathbf{\b{F}}$'. For a two-to-two process involving this particle and an adjoint-charged gluon, we would need colour tensors of the form $C(\mathbf{\b{F}}\,\mathbf{\r{ad}}|\mathbf{\r{ad}}\,\mathbf{\b{F}})$. There are three linearly independent tensors for this process (to arbitrary orders of perturbation theory). As we will see below in \mbox{section~\ref{subsection:bases_for_fggf}}, one choice for a Clebsch colour basis of tensors would consist of
\eq{\fwbox{0pt}{\fwboxL{435pt}{(\mathfrak{so}_{{10}})}}\fwbox{0pt}{\left\{
\tikzBox{so10_clebsch_basis_1}{\arrowTo[hblue]{0,0}{-130};\node[anchor=10,inner sep=2pt] at(in){{\footnotesize$\b{[f]}$}};\arrowTo[hred]{0,0}{130}\node[anchor=-10,inner sep=2pt] at(in){{\footnotesize$\r{[\adR]}$}};\arrowFrom[hblue]{0,0}[1.5]{0}\node[hblue,clebschR]at(in){};\node[anchor=90,inner sep=2pt] at(arrownode){{\footnotesize$\b{[f]}$}};\arrowFrom[hred]{end}{50}
\node[anchor=-170,inner sep=2pt] at(end){{\footnotesize$\r{[\adR]}$}};\arrowFrom[hblue]{in}{-50}\node[anchor=170,inner sep=2pt] at(end){{\footnotesize$\b{[f]}$}};\node[hblue,clebschR]at(in){};
},
\tikzBox{so10_clebsch_basis_2}{\arrowTo[hblue]{0,0}{-130};\node[anchor=10,inner sep=2pt] at(in){{\footnotesize$\b{[f]}$}};\arrowTo[hred]{0,0}{130}\node[anchor=-10,inner sep=2pt] at(in){{\footnotesize$\r{[\adR]}$}};\arrowFrom[hgreen]{0,0}[1.5]{0}\node[clebsch]at(in){};\node[anchor=90,inner sep=2pt] at(arrownode){{\footnotesize$\g{[120]}$}};\arrowFrom[hred]{end}{50}
\node[anchor=-170,inner sep=2pt] at(end){{\footnotesize$\r{[\adR]}$}};\arrowFrom[hblue]{in}{-50}\node[anchor=170,inner sep=2pt] at(end){{\footnotesize$\b{[f]}$}};\node[clebsch]at(in){};
},
\tikzBox{so10_clebsch_basis_3}{\arrowTo[hblue]{0,0}{-130};\node[anchor=10,inner sep=2pt] at(in){{\footnotesize$\b{[f]}$}};\arrowTo[hred]{0,0}{130}\node[anchor=-10,inner sep=2pt] at(in){{\footnotesize$\r{[\adR]}$}};\arrowFrom[hteal]{0,0}[1.5]{0}\node[clebsch]at(in){};\node[anchor=90,inner sep=2pt] at(arrownode){{\footnotesize$\t{[320]}$}};\arrowFrom[hred]{end}{50}
\node[anchor=-170,inner sep=2pt] at(end){{\footnotesize$\r{[\adR]}$}};\arrowFrom[hblue]{in}{-50}\node[anchor=170,inner sep=2pt] at(end){{\footnotesize$\b{[f]}$}};\node[clebsch]at(in){};
}\right\}\,.
}}
While the first tensor above is obviously `natural', one may naturally wonder: what on earth do the $\mathbf{\g{120}}$-dimensional or $\mathbf{\t{320}}$-dimensional irreducible representations of $\mathfrak{so}_{10}$ have to do with this process?---and why should these tensors be useful?

In answer to the first question, it is worth noting that the representations $\mathbf{\g{120}}$ and $\mathbf{\t{320}}$ arise directly in the tensor product 
\eq{\fwbox{0pt}{\fwboxL{435pt}{(\mathfrak{so}_{{10}})}}\fwbox{0pt}{\dynkLabelK{\mathbf{\b{F}}}{10000}\!\otimes\!\dynkLabelK{\mathbf{\r{ad}}}{01000}\simeq\dynkLabelK{\mathbf{\b{F}}}{10000}\!\oplus\!\dynkLabelK{\mathbf{\g{120}}}{00100}\!\oplus\!\dynkLabelK{\mathbf{\t{320}}}{11000}\,.}}
As such, it is not so surprising that these irreducible representations should play a role---it is what one may expect from studying scattering in elementary quantum mechanics.

As for their usefulness, we remind the reader that they are manifestly colour-orthogonal, which would not be the case for the more arguably `natural' tensors, say:
\eq{\fwbox{0pt}{\fwboxL{435pt}{(\mathfrak{so}_{{10}})}}\fwbox{0pt}{\left\{\mathbf{S}^1,\mathbf{S}^2,\mathbf{S}^3\right\}\equivR\left\{
\tikzBox{so10_s_basis_1}{\arrowTo[hblue]{0,0}{-130};\node[anchor=10,inner sep=2pt] at(in){{\footnotesize$\b{[f]}$}};\arrowTo[hred]{0,0}{130}\node[anchor=-10,inner sep=2pt] at(in){{\footnotesize$\r{[\adR]}$}};\arrowFrom[hblue]{0,0}[1.5]{0}\node[hblue,clebschR]at(in){};\node[anchor=90,inner sep=2pt] at(arrownode){{\footnotesize$\b{[f]}$}};\arrowFrom[hred]{end}{50}
\node[anchor=-170,inner sep=2pt] at(end){{\footnotesize$\r{[\adR]}$}};\arrowFrom[hblue]{in}{-50}\node[anchor=170,inner sep=2pt] at(end){{\footnotesize$\b{[f]}$}};\node[hblue,clebschR]at(in){};
},
\tikzBox{so10_s_basis_2}{\coordinate(e1)at($(\edgeLength,0)+(30:\edgeLength)$);\coordinate(e2)at($(\edgeLength,0)+(-30:\edgeLength)$);\coordinate(i1)at($(0,0)+(-150:\edgeLength)$);\coordinate(i2)at($(0,0)+(150:\edgeLength)$);\arrowTo[hblue]{0,0}{-150}\node[anchor=10,inner sep=2pt] at(in){{\footnotesize$\b{[f]}$}};
%\arrowTo[hred]{0,0}{150}
\floatingEdge{hred,edge,endArrow}{(0,0).. controls ($(0.25,0.25)$) and ($(.5,.4)$) .. (e1);}
\floatingEdge{hred,edge,endArrow}{(i2).. controls ($(\edgeLength-0.5,0.4)$) and ($(\edgeLength-0.25,.25)$) .. (\edgeLength,0);}
\node[anchor=-10,inner sep=2pt] at(i2){{\footnotesize$\r{[\adR]}$}};
\arrowFrom[hblue]{0,0}{0}\node[hblue,clebschR]at(in){};
\arrowFrom[hblue]{end}{-30}\node[anchor=170,inner sep=2pt] at(end){{\footnotesize$\b{[f]}$}};
\node[anchor=-170,inner sep=2pt] at(e1){{\footnotesize$\r{[\adR]}$}};\node[hblue,clebschR]at(in){};
},\tikzBox{so10_s_basis_3}{\coordinate(e1)at($(0,0)+(30:\edgeLength)$);\coordinate(e2)at($(0,0)+(-30:\edgeLength)$);\arrowTo[hblue]{e2}[1.25]{180}\node[anchor=170,inner sep=2pt] at(end){{\footnotesize$\b{[f]}$}};\node[anchor=10,inner sep=2pt] at(in){{\footnotesize$\b{[f]}$}};
\arrowTo[hred]{e1}[1.25]{180}\node[anchor=-170,inner sep=2pt] at(end){{\footnotesize$\r{[\adR]}$}};\node[anchor=-10,inner sep=2pt] at(in){{\footnotesize$\r{[\adR]}$}};}
\right\}\,.}\label{so10_s_tensors_defined}\vspace{-4pt}}
These tensors are in fact densely overlapping in colour space:
\eq{\langle\mathbf{S}_i|\mathbf{S}^j\rangle=\left(\begin{array}{@{}ccc@{}}\frac{405}{2}&\frac{45}{2}&45\\
\frac{45}{2}&\frac{405}{2}&45\\
45&45&450\end{array}\right)\,.}

One could view the novel tensors in the Clebsch basis as being defined in terms of these more natural ones via
\begin{align}\hspace{-100pt}\tikzBox{so10_clebsch_basis_2}{\arrowTo[hblue]{0,0}{-130};\node[anchor=10,inner sep=2pt] at(in){{\footnotesize$\b{[f]}$}};\arrowTo[hred]{0,0}{130}\node[anchor=-10,inner sep=2pt] at(in){{\footnotesize$\r{[\adR]}$}};\arrowFrom[hgreen]{0,0}[1.5]{0}\node[clebsch]at(in){};\node[anchor=90,inner sep=2pt] at(arrownode){{\footnotesize$\g{[120]}$}};\arrowFrom[hred]{end}{50}
\node[anchor=-170,inner sep=2pt] at(end){{\footnotesize$\r{[\adR]}$}};\arrowFrom[hblue]{in}{-50}\node[anchor=170,inner sep=2pt] at(end){{\footnotesize$\b{[f]}$}};\node[clebsch]at(in){};
}&\hspace{-2pt}=\frac{1}{3}\left(\!2\!\tikzBox{so10_s_basis_2}{\coordinate(e1)at($(\edgeLength,0)+(30:\edgeLength)$);\coordinate(e2)at($(\edgeLength,0)+(-30:\edgeLength)$);\coordinate(i1)at($(0,0)+(-150:\edgeLength)$);\coordinate(i2)at($(0,0)+(150:\edgeLength)$);\arrowTo[hblue]{0,0}{-150}\node[anchor=10,inner sep=2pt] at(in){{\footnotesize$\b{[f]}$}};
%\arrowTo[hred]{0,0}{150}
\floatingEdge{hred,edge,endArrow}{(0,0).. controls ($(0.25,0.25)$) and ($(.5,.4)$) .. (e1);}
\floatingEdge{hred,edge,endArrow}{(i2).. controls ($(\edgeLength-0.5,0.4)$) and ($(\edgeLength-0.25,.25)$) .. (\edgeLength,0);}
\node[anchor=-10,inner sep=2pt] at(i2){{\footnotesize$\r{[\adR]}$}};
\arrowFrom[hblue]{0,0}{0}\node[hblue,clebschR]at(in){};
\arrowFrom[hblue]{end}{-30}\node[anchor=170,inner sep=2pt] at(end){{\footnotesize$\b{[f]}$}};
\node[anchor=-170,inner sep=2pt] at(e1){{\footnotesize$\r{[\adR]}$}};\node[hblue,clebschR]at(in){};
}-\tikzBox{so10_s_basis_3}{\coordinate(e1)at($(0,0)+(30:\edgeLength)$);\coordinate(e2)at($(0,0)+(-30:\edgeLength)$);\arrowTo[hblue]{e2}[1.25]{180}\node[anchor=170,inner sep=2pt] at(end){{\footnotesize$\b{[f]}$}};\node[anchor=10,inner sep=2pt] at(in){{\footnotesize$\b{[f]}$}};
\arrowTo[hred]{e1}[1.25]{180}\node[anchor=-170,inner sep=2pt] at(end){{\footnotesize$\r{[\adR]}$}};\node[anchor=-10,inner sep=2pt] at(in){{\footnotesize$\r{[\adR]}$}};}\right)\\
\hspace{-200pt}\tikzBox{so10_clebsch_basis_3}{\arrowTo[hblue]{0,0}{-130};\node[anchor=10,inner sep=2pt] at(in){{\footnotesize$\b{[f]}$}};\arrowTo[hred]{0,0}{130}\node[anchor=-10,inner sep=2pt] at(in){{\footnotesize$\r{[\adR]}$}};\arrowFrom[hteal]{0,0}[1.5]{0}\node[clebsch]at(in){};\node[anchor=90,inner sep=2pt] at(arrownode){{\footnotesize$\t{[320]}$}};\arrowFrom[hred]{end}{50}
\node[anchor=-170,inner sep=2pt] at(end){{\footnotesize$\r{[\adR]}$}};\arrowFrom[hblue]{in}{-50}\node[anchor=170,inner sep=2pt] at(end){{\footnotesize$\b{[f]}$}};\node[clebsch]at(in){};
}&\hspace{-2pt}=\frac{1}{3}\!\left(-\frac{1}{3}\!\tikzBox{so10_s_basis_1}{\arrowTo[hblue]{0,0}{-130};\node[anchor=10,inner sep=2pt] at(in){{\footnotesize$\b{[f]}$}};\arrowTo[hred]{0,0}{130}\node[anchor=-10,inner sep=2pt] at(in){{\footnotesize$\r{[\adR]}$}};\arrowFrom[hblue]{0,0}[1.5]{0}\node[hblue,clebschR]at(in){};\node[anchor=90,inner sep=2pt] at(arrownode){{\footnotesize$\b{[f]}$}};\arrowFrom[hred]{end}{50}
\node[anchor=-170,inner sep=2pt] at(end){{\footnotesize$\r{[\adR]}$}};\arrowFrom[hblue]{in}{-50}\node[anchor=170,inner sep=2pt] at(end){{\footnotesize$\b{[f]}$}};\node[hblue,clebschR]at(in){};
}\hspace{-2pt}+\tikzBox{so10_s_basis_2}{\coordinate(e1)at($(\edgeLength,0)+(30:\edgeLength)$);\coordinate(e2)at($(\edgeLength,0)+(-30:\edgeLength)$);\coordinate(i1)at($(0,0)+(-150:\edgeLength)$);\coordinate(i2)at($(0,0)+(150:\edgeLength)$);\arrowTo[hblue]{0,0}{-150}\node[anchor=10,inner sep=2pt] at(in){{\footnotesize$\b{[f]}$}};
%\arrowTo[hred]{0,0}{150}
\floatingEdge{hred,edge,endArrow}{(0,0).. controls ($(0.25,0.25)$) and ($(.5,.4)$) .. (e1);}
\floatingEdge{hred,edge,endArrow}{(i2).. controls ($(\edgeLength-0.5,0.4)$) and ($(\edgeLength-0.25,.25)$) .. (\edgeLength,0);}
\node[anchor=-10,inner sep=2pt] at(i2){{\footnotesize$\r{[\adR]}$}};
\arrowFrom[hblue]{0,0}{0}\node[hblue,clebschR]at(in){};
\arrowFrom[hblue]{end}{-30}\node[anchor=170,inner sep=2pt] at(end){{\footnotesize$\b{[f]}$}};
\node[anchor=-170,inner sep=2pt] at(e1){{\footnotesize$\r{[\adR]}$}};\node[hblue,clebschR]at(in){};
}\hspace{-2pt}+\tikzBox{so10_s_basis_3}{\coordinate(e1)at($(0,0)+(30:\edgeLength)$);\coordinate(e2)at($(0,0)+(-30:\edgeLength)$);\arrowTo[hblue]{e2}[1.25]{180}\node[anchor=170,inner sep=2pt] at(end){{\footnotesize$\b{[f]}$}};\node[anchor=10,inner sep=2pt] at(in){{\footnotesize$\b{[f]}$}};
\arrowTo[hred]{e1}[1.25]{180}\node[anchor=-170,inner sep=2pt] at(end){{\footnotesize$\r{[\adR]}$}};\node[anchor=-10,inner sep=2pt] at(in){{\footnotesize$\r{[\adR]}$}};}\right)\hspace{-200pt}\nonumber
\end{align}
and take these unusual tensors to be a simple change of basis that happens to realize orthogonality in colour space. While this is a valid viewpoint, it is by no means obvious how to generalize this case to more general scattering amplitudes involving more (or more diversely) coloured particles. For one thing, even when a set of `natural' colour tensors such as (\ref{so10_s_tensors_defined}) exist, their completeness or their linear independence can be far from obvious.\\

In this section, we'd like to discuss more examples analogous to the case above at greater length. In particular, we'd like to explore how these new tensors relate to the more familiar colour tensors that have often been used. But in order to do this, a brief review of those more familiar tensors may be useful.

\subsection{Familiar Sets of Colour Tensors for Representing Amplitudes}\label{subsec:review_of_familiar_colour_tensors}

\subsubsection{\texorpdfstring{Feynman Diagrams for Adjoint Scattering: `$f$-Graphs'}{Feynman Diagrams for Adjoint Scattering: `f-Graphs'}}

For amplitudes involving particles only charged in the adjoint representation of the gauge group's Lie algebra, the only tensor which appears directly in the Feynman rules will be $\mathbf{\r{ad}}$ itself. Recall that this representation is identical (as a tensor) to what are often called the \emph{structure} constants of the Lie algebra $\r{f}\indices{\r{[\adR]\,[\adR]}}{\r{[\adR]}}$. (The gauge-kinetic term may be chosen to involve a four-particle interaction vertex, but the colour dependence of this term is directly given as a sum of products of $\r{f}$'s.)

Thus, the colour tensors that arise directly from the Feynman rules are those representable as graphs of structure constants sewn together. These tensors are called `$\r{f}$-graphs'. At $L$ loops, these tensors will be encoded by graphs with Euler characteristic $L$. It has become fairly standard practice to express amplitudes involving adjoint-charged particles directly in terms of such tensors (see, for example, the works of \cite{Bern:2007ct,Carrasco:2011mn,Bourjaily:2018omh,Bourjaily:2019iqr,Bourjaily:2021iyq,Carrasco:2021otn,Bourjaily:2023uln}), viewed abstractly (or implicitly) as encoding tensors as graphs built from structure constants, with the understanding that the reader can use whatever concrete structure constants she may choose to realize these terms concretely. Expressing amplitudes in this way has the advantage that expressions remain equally valid for any particular gauge theory. Another advantage is that the tensors relevant to amplitudes arise \emph{perturbatively} in loop-order, with only a restricted subset of all possible tensors arising.

Of course, $\r{f}$-graphs are not always linearly independent: minimally, they satisfy the Jacobi relation. This implies that not all $\r{f}$-graphs are independent at any loop-order. Moreover, identities such as
\vspace{-6pt}\eq{\tikzBox{triangle_for_adjoint_id_0}{\draw[hred,edge,midArrow](120:0.5*\edgeLength)arc(120:0:0.5*\edgeLength);\draw[hred,edge,midArrow](0:0.5*\edgeLength)arc(0:-120:0.5*\edgeLength);\draw[hred,edge,midArrow](-120:0.5*\edgeLength)arc(-120:-240:0.5*\edgeLength);\arrowTo[hred]{0:0.5*\edgeLength}[0.7]{0}\node[anchor=180,inner sep=2pt] at(in){{\footnotesize$\r{[\adR]}$}};
\arrowTo[hred]{120:0.5*\edgeLength}[0.7]{120}\node[anchor=0,inner sep=2pt] at(in){{\footnotesize$\r{[\adR]}$}};\arrowTo[hred]{-120:0.5*\edgeLength}[0.7]{-120}\node[anchor=0,inner sep=2pt] at(in){{\footnotesize$\r{[\adR]}$}};\node[hred,clebschR]at(120:0.5*\edgeLength){};\node[hred,clebschR]at(0:0.5*\edgeLength){};\node[hred,clebschR]at(-120:0.5*\edgeLength){};
}=\frac{1}{2}C_{2}(\mathbf{\r{ad}})\tikzBox{triangle_for_adjoint_id_1}{\arrowTo[hred]{0,0}[1]{120}\node[anchor=-20,inner sep=2pt] at(in){{\footnotesize$\r{[\adR]}$}};\arrowTo[hred]{0,0}[1]{-120}\node[anchor=20,inner sep=2pt] at(in){{\footnotesize$\r{[\adR]}$}};\arrowTo[hred]{0,0}[1]{0}\node[anchor=180,inner sep=2pt] at(in){{\footnotesize$\r{[\adR]}$}};\node[hred,clebschR]at(0,0){};}\,\vspace{-6pt}}
and also those peculiar to particular Lie algebras allow $\r{f}$-graphs at distinct loop-orders to be linearly related to one another.

It is a general folk theorem (not entirely difficult to prove) that all tree-level $\r{f}$-graphs can be related to one another via simple applications of the Jacobi relation; in particular, a `basis' of independent $\r{f}$-graphs can be chosen of the form
\vspace{-6pt}\eq{\r{f}\indices{\r{a_1}\,\b{a_2\cdots a_{\text{-}2}}\,\r{a_{\text{-}1}}}{}\;\bigger{\Leftrightarrow}\;\tikzBox{ddm_basis_tensor_structure}{
\arrowTo[hred]{0,0}[0.65]{210}\node[anchor=10,inner sep=1.5pt] at(in){{\footnotesize$\r{a_1}$}};
\arrowTo[hred]{end}[0.65]{90}\node[anchor=-90,inner sep=1.5pt] at(in){{\footnotesize${\,\,\b{a_2}}$}};
\arrowFrom[hred]{end}[0.65]{0}\node[hred,clebschR]at(in){};
\arrowTo[hred]{end}[0.65]{90}\node[anchor=-90,inner sep=1.5pt] at(in){{\footnotesize${\,\,\b{a_3}}$}};
\arrowFrom[hred]{end}[0.65]{0}\node[hred,clebschR]at(in){};\draw[hred,edge,dashed](end)--($(end)+(0.65*\edgeLength,0)$);
\arrowFrom[hred]{$(end)+(0.65*\edgeLength,0)$}[0.65]{0}
\arrowTo[hred]{end}[0.65]{90};\node[anchor=-90,inner sep=1.5pt] at(in){{\footnotesize${\,\,\b{a_{\text{-}2}}}$}};
\node[anchor=-90,inner sep=1.5pt] at($(in)-(0.75,0)$){{\footnotesize${\,\,\b{\cdots}}$}};
\arrowTo[hred]{end}[0.65]{-30};
\node[hred,clebschR]at(end){};\node[anchor=170,inner sep=1.5pt] at(in){{\footnotesize$\r{a_{\text{-}1}}$}};
}\vspace{-6pt}\label{ddm_basis_graph}}
together with the $(n{-}2)!$ permutations in the slot-sequences of the blue indices (keeping the first and last slots fixed). Moreover, for any choice of initial and final legs $\{\r{a},\r{c}\}\!\subset\![n]$, the set of tensors obtained by permuting the remaining $(n{-}2)!$ other slot sequences $\b{\vec{b}}\!\in\!\mathfrak{S}([n]\backslash\{\r{a},\r{c}\})$ will form a `basis' of tensors independent under Jacobi relations. 

This basis of colour tensors is most familiar from the work of DDM \cite{DelDuca:1999rs}. The equality of amplitudes represented according to different choices of tree-level `bases' (distinguished by the choice of initial and final legs $\{\r{a},\r{c}\}\!\subset\![n]$ held fixed) requires that the partial amplitudes satisfy the Kleiss-Kuijf-relations \cite{KK} which generalize the famous $\mathfrak{u}_{1}$ decoupling identity \cite{Berends:1987me,Mangano:1987xk} (see also \cite{Arkani-Hamed:2014bca}).

We have used scare-quotes around `basis' above for two reasons. First, the number of such tensors increases factorially with multiplicity, whereas the number of independent colour tensors to \emph{arbitrary} loop-order is bounded to grow exponentially \cite{Bourjaily:2024jbt}; thus, for any simple Lie algebra, there will be a multiplicity $n$ for which not all \emph{tree-level} $\r{f}$-graphs will be linearly independent. The second reason for having caution against declaring these tensors a \emph{basis} is that in almost all cases, these tensors are \emph{incomplete}: new, linearly-independent $\r{f}$-graphs arise at higher loop-orders. 

It is worthwhile to mention a few cases where these issues are most transparent. For the case of $\mathfrak{a}_{1}$ gauge theory, the tree-level $\r{f}$-graph tensors are \emph{over}-complete for $n\!\geq\!6$ external legs, and \emph{incomplete} for any even number of particles. (Interestingly, for all even multiplicities, only one independent tensor exists beyond tree-level.) For $n{=}5$ particles and gauge theories with rank exceeding 1, all $6{=}(5{-}2)!$ tree-level $\r{f}$-graphs are independent; but new, linearly independent tensors arise at both one and two loops. (For most simple Lie algebras, no new tensors arise beyond two loops for 5 particles; but for $\mathfrak{a}_{\r{k}>1}$, these tensors are still \emph{incomplete}\footnote{If one considered both $\r{f}$-graphs and `$\r{d}$-graphs'---those built from the $\r{d}$ tensor defined in (\ref{d_tensor_defined})---then graphs with Euler characteristic $\leq\!2$ would be complete for 5 particles; interestingly, however, because $\r{\mathbf{ad}}$ is necessarily \emph{real}, it is not possible to define the $\r{d}$ tensor in terms of $\r{f}$ alone. This coheres with our understanding that $\r{d}$ tensor contributions to amplitudes signal \emph{anomalies} in the gauge theory, which can only arise from matter transforming in complex representations.}.)

More pervasively, these tensors are highly non-orthogonal in colour space. For example, for $\mathfrak{a}_{\r{1}}$ gauge theory, the $24{=}(6{-}2)!$ tree-level DDM tensors have overlap:
\vspace{-12pt}\eq{\fwboxR{0pt}{\frac{\langle \r{f}(\b{\vec{a}})|\r{f}(\b{\vec{b}})\rangle}{\langle\r{f}(\b{\vec{a}})|\r{f}(\b{\vec{a}})\rangle}=\frac{1}{8}\!}\left(\begin{array}{@{}c@{$\,\,$}c@{$\,\,$}c@{$\,\,$}c@{$\,\,$}c@{$\,\,$}c@{$\,\,$}c@{$\,\,$}c@{$\,\,$}c@{$\,\,$}c@{$\,\,$}c@{$\,\,$}c@{$\,\,$}c@{$\,\,$}c@{$\,\,$}c@{$\,\,$}c@{$\,\,$}c@{$\,\,$}c@{$\,\,$}c@{$\,\,$}c@{$\,\,$}c@{$\,\,$}c@{$\,\,$}c@{$\,\,$}c@{}}
8&4&4&2&2&\dzero&4&2&2&1&1&\dzero&2&1&\dzero&\dzero&4&3&1&\dzero&\dzero&3&3&3\\[-6pt]
4&8&2&\dzero&4&2&2&4&1&\dzero&2&1&1&\dzero&\dzero&3&3&3&2&1&\dzero&\dzero&4&3\\[-6pt]
4&2&8&4&\dzero&2&2&1&\dzero&\dzero&4&3&4&2&2&1&1&\dzero&\dzero&1&3&3&\dzero&3\\[-6pt]
2&\dzero&4&8&2&4&1&\dzero&\dzero&3&3&3&2&4&1&\dzero&2&1&1&2&4&3&\dzero&\dzero\\[-6pt]
2&4&\dzero&2&8&4&1&2&4&3&\dzero&\dzero&\dzero&1&3&3&\dzero&3&4&2&2&1&1&\dzero\\[-6pt]
\dzero&2&2&4&4&8&\dzero&1&3&3&\dzero&3&1&2&4&3&\dzero&\dzero&2&4&1&\dzero&2&1\\[-6pt]
4&2&2&1&1&\dzero&8&4&4&2&2&\dzero&\dzero&\dzero&2&1&3&4&\dzero&3&1&\dzero&3&3\\[-6pt]
2&4&1&\dzero&2&1&4&8&2&\dzero&4&2&\dzero&3&1&\dzero&3&3&\dzero&\dzero&2&1&3&4\\[-6pt]
2&1&\dzero&\dzero&4&3&4&2&8&4&\dzero&2&2&1&4&2&\dzero&1&3&3&\dzero&1&3&\dzero\\[-6pt]
1&\dzero&\dzero&3&3&3&2&\dzero&4&8&2&4&1&\dzero&2&4&1&2&4&3&1&2&\dzero&\dzero\\[-6pt]
1&2&4&3&\dzero&\dzero&2&4&\dzero&2&8&4&3&3&\dzero&1&3&\dzero&2&1&4&2&\dzero&1\\[-6pt]
\dzero&1&3&3&\dzero&3&\dzero&2&2&4&4&8&4&3&1&2&\dzero&\dzero&1&\dzero&2&4&1&2\\[-6pt]
2&1&4&2&\dzero&1&\dzero&\dzero&2&1&3&4&8&4&4&2&2&\dzero&3&\dzero&3&3&1&\dzero\\[-6pt]
1&\dzero&2&4&1&2&\dzero&3&1&\dzero&3&3&4&8&2&\dzero&4&2&\dzero&\dzero&3&4&2&1\\[-6pt]
\dzero&\dzero&2&1&3&4&2&1&4&2&\dzero&1&4&2&8&4&\dzero&2&3&3&3&\dzero&\dzero&1\\[-6pt]
\dzero&3&1&\dzero&3&3&1&\dzero&2&4&1&2&2&\dzero&4&8&2&4&3&4&\dzero&\dzero&1&2\\[-6pt]
4&3&1&2&\dzero&\dzero&3&3&\dzero&1&3&\dzero&2&4&\dzero&2&8&4&1&2&\dzero&1&4&2\\[-6pt]
3&3&\dzero&1&3&\dzero&4&3&1&2&\dzero&\dzero&\dzero&2&2&4&4&8&\dzero&1&1&2&2&4\\[-6pt]
1&2&\dzero&1&4&2&\dzero&\dzero&3&4&2&1&3&\dzero&3&3&1&\dzero&8&4&4&2&2&\dzero\\[-6pt]
\dzero&1&1&2&2&4&3&\dzero&3&3&1&\dzero&\dzero&\dzero&3&4&2&1&4&8&2&\dzero&4&2\\[-6pt]
\dzero&\dzero&3&4&2&1&1&2&\dzero&1&4&2&3&3&3&\dzero&\dzero&1&4&2&8&4&\dzero&2\\[-6pt]
3&\dzero&3&3&1&\dzero&\dzero&1&1&2&2&4&3&4&\dzero&\dzero&1&2&2&\dzero&4&8&2&4\\[-6pt]
3&4&\dzero&\dzero&1&2&3&3&3&\dzero&\dzero&1&1&2&\dzero&1&4&2&2&4&\dzero&2&8&4\\[-6pt]
3&3&3&\dzero&\dzero&1&3&4&\dzero&\dzero&1&2&\dzero&1&1&2&2&4&\dzero&2&2&4&4&8\end{array}\right)\vspace{-12pt}}
This overlap does not diminish in the large-rank limit. 

%Beyond tree-level, there are other various tensors which generalize $\r{f}$-graphs. 
%One set of tensors we will not review but wish to mention is the so-called $SU(N_c)$ Multiplet Basis. This is more or less morally equivalent to what we are describing at present, just specified to $SU(N_c)$ with fixed half-ladder tree topology. Since this fits nicely within our general framework it would be redundant to review, and so we merely point to the original works for reference \cite{Keppeler2012OrthogonalMB, Sjodahl2015, Sjodahl2024}. 

\subsubsection{Multi-Trace Colour Tensors and the Large-Rank Limit}

Even for amplitudes not involving any charged matter (and therefore without any reference to any particular representation other than $\mathbf{\r{ad}}$), it is often useful to use tensors constructed from \emph{some} representation to define tensors involving $\mathbf{\r{ad}}$. This follows from the simple fact that for \emph{any} representation $\mathbf{\b{R}}$, we have 
\eq{\r{f}\indices{\r{[\adR][\adR][\adR]}}{}\,=\frac{1}{T(\mathbf{\b{R}})}\big(\mathrm{tr}_{\mathbf{\b{R}}}(\r{1\,2\,3})-\mathrm{tr}_{\mathbf{\b{R}}}(\r{2\,1\,3})\big)\,;\vspace{-6pt}}
this is equivalent to the graphical statement that 
\vspace{-6pt}\eq{\tikzBox{triangle_for_adjoint_id_1}{\arrowTo[hred]{0,0}[1]{120}\node[anchor=-20,inner sep=2pt] at(in){{\footnotesize$\r{[\adR]}$}};\arrowTo[hred]{0,0}[1]{-120}\node[anchor=20,inner sep=2pt] at(in){{\footnotesize$\r{[\adR]}$}};\arrowTo[hred]{0,0}[1]{0}\node[anchor=180,inner sep=2pt] at(in){{\footnotesize$\r{[\adR]}$}};\node[hred,clebschR]at(0,0){};}=\frac{1}{T(\mathbf{\b{R}})}\!\left(\tikzBox{triangle_for_structure_1}{\draw[hblue,edge,midArrow](120:0.5*\edgeLength)arc(120:0:0.5*\edgeLength);\node[anchor=-120,inner sep=2pt] at(arrownode){{\footnotesize$\b{[r]}$}};\draw[hblue,edge,midArrow](0:0.5*\edgeLength)arc(0:-120:0.5*\edgeLength);\node[anchor=120,inner sep=2pt] at(arrownode){{\footnotesize$\b{[r]}$}};
\draw[hblue,edge,midArrow](-120:0.5*\edgeLength)arc(-120:-240:0.5*\edgeLength);\node[anchor=0,inner sep=2pt] at(arrownode){{\footnotesize$\b{[r]}$}};\arrowTo[hred]{0:0.5*\edgeLength}[0.7]{0}\node[anchor=180,inner sep=2pt] at(in){{\footnotesize$\r{[\adR]}$}};
\arrowTo[hred]{120:0.5*\edgeLength}[0.7]{120}\node[anchor=0,inner sep=2pt] at(in){{\footnotesize$\r{[\adR]}$}};\arrowTo[hred]{-120:0.5*\edgeLength}[0.7]{-120}\node[anchor=0,inner sep=2pt] at(in){{\footnotesize$\r{[\adR]}$}};\node[hblue,clebschR]at(120:0.5*\edgeLength){};\node[hblue,clebschR]at(0:0.5*\edgeLength){};\node[hblue,clebschR]at(-120:0.5*\edgeLength){};
}
-\hspace{-2pt}
\tikzBox{triangle_for_structure_2}{\draw[hblue,edge,midArrow](120:0.5*\edgeLength)arc(120:0:0.5*\edgeLength);\node[anchor=-120,inner sep=2pt] at(arrownode){{\footnotesize$\b{[r]}$}};\draw[hblue,edge,midArrow](0:0.5*\edgeLength)arc(0:-120:0.5*\edgeLength);\node[anchor=120,inner sep=2pt] at(arrownode){{\footnotesize$\b{[r]}$}};
\draw[hblue,edge,midArrow](-120:0.5*\edgeLength)arc(-120:-240:0.5*\edgeLength);\node[anchor=180,inner sep=2pt] at(arrownode){{\footnotesize$\b{[r]}$}};\arrowTo[hred]{0:0.5*\edgeLength}[0.7]{0}\node[anchor=180,inner sep=2pt] at(in){{\footnotesize$\r{[\adR]}$}};
\floatingEdge{hred,edge,startArrow}{($(-120:1.0*\edgeLength)-(0.6,0)$)..controls ($(-120:1.0*\edgeLength)-(.6,0)+(20:0.5*\edgeLength)$) and ($(120:1.75*\edgeLength)$)..(120:0.5*\edgeLength);}
\floatingEdge{hred,edge,startArrow}{($(120:1.0*\edgeLength)-(0.6,0)$)..controls ($(120:1.0*\edgeLength)-(.6,0)+(-20:0.5*\edgeLength)$) and ($(-120:1.75*\edgeLength)$)..(-120:0.5*\edgeLength);}
\node[anchor=0,inner sep=2pt] at($(120:1.0*\edgeLength)-(0.6,0)$){{\footnotesize$\r{[\adR]}$}};\node[anchor=0,inner sep=2pt] at($(-120:1.0*\edgeLength)-(0.6,0)$){{\footnotesize$\r{[\adR]}$}};
%\node[anchor=0,inner sep=2pt] at(in){{\footnotesize$\r{[\adR]}$}};\arrowTo[hred]{-120:0.5*\edgeLength}[0.7]{-120}\node[anchor=0,inner sep=2pt] at(in){{\footnotesize$\r{[\adR]}$}};\
\node[hblue,clebschR]at(120:0.5*\edgeLength){};\node[hblue,clebschR]at(0:0.5*\edgeLength){};\node[hblue,clebschR]at(-120:0.5*\edgeLength){};
}\hspace{-4pt}\right).\vspace{-6pt}\label{graphical_f_to_traces}}
From this, we can directly convert any tensor defined as an $\r{f}$-graph directly into those involving traces over the generators of any particular representation. 

Applying (\ref{graphical_f_to_traces}) directly to every vertex of an $\r{f}$-graph such as (\ref{ddm_basis_graph}) may at first appear to horribly complicate things. At tree-level, however, it is easy to show recursively (using the defining commutation relation (\ref{diagrammatic_defn_of_rep}) for any representation) that all tree-level $\r{f}$ graphs are expressible in terms of sums of \emph{single}-traces over generators of any representation:
\vspace{-5pt}\eq{\mathrm{tr}_{\mathbf{\b{R}}}(\r{a_1},\ldots,\r{a_n})\equivR\sum_{\b{r_i}\in\b{[r]}}\big(\b{\mathbf{R}}\indices{\b{r_1}\r{a_1}}{\b{r_2}}\!\!\cdots\b{\mathbf{R}}\indices{\b{r_n}\r{a_n}}{\b{r_1}}\big)\;\bigger{\Leftrightarrow}\;
\tikzBox{trace_over_generators_diagram}{
\draw[hblue,edge,midArrow](200:0.5*\edgeLength)arc(200:120:0.5*\edgeLength);\node[anchor=-30,inner sep=3pt] at(arrownode){{\footnotesize$\b{[r]}$}};
\draw[hblue,edge,midArrow](120:0.5*\edgeLength)arc(120:60:0.5*\edgeLength);\node[anchor=-90,inner sep=3pt] at(arrownode){{\footnotesize$\b{[r]}$}};
\draw[hblue,dotted,edge](60:0.5*\edgeLength)arc(60:-20:0.5*\edgeLength);%\node[anchor=-135,inner sep=2pt] at(arrownode){{\footnotesize$\b{[r]}$}};
\draw[hblue,edge,midArrow](0:0.5*\edgeLength)arc(0:-180:0.5*\edgeLength);\node[anchor=90,inner sep=3pt] at(arrownode){{\footnotesize$\b{[r]}$}};
\arrowTo[hred]{-20:0.5*\edgeLength}{-20}\node[anchor=mid west,inner sep=2pt] at(in){{\footnotesize$\r{a_n}$}};
\arrowTo[hred]{60:0.5*\edgeLength}{60}\node[anchor=240,inner sep=2pt] at(in){{\footnotesize$\r{a_3}$}};
\arrowTo[hred]{120:0.5*\edgeLength}{120}\node[anchor=-60,inner sep=2pt] at(in){{\footnotesize$\r{a_2}$}};
\arrowTo[hred]{200:0.5*\edgeLength}{200}\node[anchor=mid east,inner sep=2pt] at(in){{\footnotesize$\r{a_1}$}};\node[hred,smallDot]at(20:1.05*\edgeLength){};\node[hred,smallDot]at(10:1.05*\edgeLength){};\node[hred,smallDot]at(30:1.05*\edgeLength){};
\node[hblue,clebschR]at(200:0.5*\edgeLength){};\node[hblue,clebschR]at(-20:0.5*\edgeLength){};\node[hblue,clebschR]at(120:0.5*\edgeLength){};\node[hblue,clebschR]at(60:0.5*\edgeLength){};
}\label{trace_over_r_defined}\vspace{-5pt}}
which is naturally upgraded to the entire rank-$\r{n}$ tensor 
\vspace{-5pt}\eq{\mathrm{tr}_{\mathbf{\b{R}}}(\r{1}\,\r{2}\,\r{\cdots}\,\r{n})\equivR\mathrm{tr}_{\mathbf{\b{R}}}^{\smash{\r{[\adR][\adR][\adR]\cdots[\adR]}}}\equivR\big\{\mathrm{tr}_{\mathbf{\b{R}}}(\r{a_1},\ldots,\r{a_n})\big\}_{\r{a_i}\in\r{[\adR]}}\;\bigger{\Leftrightarrow}\;
\tikzBox{trace_over_generators_tensor}{
\draw[hblue,edge,midArrow](200:0.5*\edgeLength)arc(200:120:0.5*\edgeLength);\node[anchor=-30,inner sep=3pt] at(arrownode){{\footnotesize$\b{[r]}$}};
\draw[hblue,edge,midArrow](120:0.5*\edgeLength)arc(120:60:0.5*\edgeLength);\node[anchor=-90,inner sep=3pt] at(arrownode){{\footnotesize$\b{[r]}$}};
\draw[hblue,dotted,edge](60:0.5*\edgeLength)arc(60:-20:0.5*\edgeLength);%\node[anchor=-135,inner sep=2pt] at(arrownode){{\footnotesize$\b{[r]}$}};
\draw[hblue,edge,midArrow](0:0.5*\edgeLength)arc(0:-180:0.5*\edgeLength);\node[anchor=90,inner sep=3pt] at(arrownode){{\footnotesize$\b{[r]}$}};
\arrowTo[hred]{-20:0.5*\edgeLength}{-20}\node[anchor=mid west,inner sep=2pt] at(in){{\footnotesize$\r{[\adR]}$}};
\arrowTo[hred]{60:0.5*\edgeLength}{60}\node[anchor=240,inner sep=2pt] at(in){{\footnotesize$\r{[\adR]}$}};
\arrowTo[hred]{120:0.5*\edgeLength}{120}\node[anchor=-60,inner sep=2pt] at(in){{\footnotesize$\r{[\adR]}$}};
\arrowTo[hred]{200:0.5*\edgeLength}{200}\node[anchor=mid east,inner sep=2pt] at(in){{\footnotesize$\r{[\adR]}$}};\node[hred,smallDot]at(20:1.05*\edgeLength){};\node[hred,smallDot]at(10:1.05*\edgeLength){};\node[hred,smallDot]at(30:1.05*\edgeLength){};
\node[hblue,clebschR]at(200:0.5*\edgeLength){};\node[hblue,clebschR]at(-20:0.5*\edgeLength){};\node[hblue,clebschR]at(120:0.5*\edgeLength){};\node[hblue,clebschR]at(60:0.5*\edgeLength){};
}\,.%
\vspace{-5pt}}
These tensors manifestly enjoy a rotational symmetry among their indices. Reflections, however, are more subtle: reversing the arrows of the traced representation corresponds to tracing over the generators of the \emph{conjugate} representation $\b{\bar{\mathbf{R}}}$:
\vspace{-5pt}\eq{\mathrm{tr}_{\mathbf{\b{R}}}(\r{a_1},\ldots,\r{a_n})=(\text{-}1)^\r{n}\mathrm{tr}_{\mathbf{\b{\bar{R}}}}(\r{a_n},\ldots,\r{a_1})\,\;\bigger{\Leftrightarrow}\hspace{0pt}\tikzBox{dual_of_trace_diagram_1}{
\draw[hblue,edge,midArrow](200:0.5*\edgeLength)arc(200:120:0.5*\edgeLength);\node[anchor=-30,inner sep=3pt] at(arrownode){{\footnotesize$\b{[{r}]}$}};
\draw[hblue,edge,midArrow](120:0.5*\edgeLength)arc(120:60:0.5*\edgeLength);\node[anchor=-90,inner sep=3pt] at(arrownode){{\footnotesize$\b{[{r}]}$}};
\draw[hblue,dotted,edge](60:0.5*\edgeLength)arc(60:-20:0.5*\edgeLength);%\node[anchor=-135,inner sep=2pt] at(arrownode){{\footnotesize$\b{[r]}$}};
\draw[hblue,edge,midArrow](0:0.5*\edgeLength)arc(0:-180:0.5*\edgeLength);\node[anchor=90,inner sep=3pt] at(arrownode){{\footnotesize$\b{[{r}]}$}};
\arrowTo[hred]{-20:0.5*\edgeLength}{-20}\node[anchor=mid west,inner sep=2pt] at(in){{\footnotesize$\r{{a_n}}$}};
\arrowTo[hred]{60:0.5*\edgeLength}{60}\node[anchor=240,inner sep=2pt] at(in){{\footnotesize$\r{a_3}$}};
\arrowTo[hred]{120:0.5*\edgeLength}{120}\node[anchor=-60,inner sep=2pt] at(in){{\footnotesize$\r{{a_2}}$}};
\arrowTo[hred]{200:0.5*\edgeLength}{200}\node[anchor=mid east,inner sep=2pt] at(in){{\footnotesize$\r{{a_1}}$}};\node[hred,smallDot]at(20:1.05*\edgeLength){};\node[hred,smallDot]at(10:1.05*\edgeLength){};\node[hred,smallDot]at(30:1.05*\edgeLength){};
\node[hblue,clebschR]at(200:0.5*\edgeLength){};\node[hblue,clebschR]at(-20:0.5*\edgeLength){};\node[hblue,clebschR]at(120:0.5*\edgeLength){};\node[hblue,clebschR]at(60:0.5*\edgeLength){};
}\hspace{-10pt}=(\text{-}1)^{\r{n}}\hspace{-10pt}\tikzBox{dual_of_trace_diagram_2}{
\draw[hblue,dotted,edge](200:0.5*\edgeLength)arc(200:120:0.5*\edgeLength);%\node[anchor=-30,inner sep=3pt] at(arrownode){{\footnotesize$\b{[\bar{r}]}$}};
\draw[hblue,edge,midArrow](120:0.5*\edgeLength)arc(120:60:0.5*\edgeLength);\node[anchor=-90,inner sep=3pt] at(arrownode){{\footnotesize$\b{[\bar{r}]}$}};
\draw[hblue,edge,midArrow](60:0.5*\edgeLength)arc(60:-20:0.5*\edgeLength);\node[anchor=210,inner sep=2pt] at(arrownode){{\footnotesize$\b{[\bar{r}]}$}};
\draw[hblue,edge,midArrow](0:0.5*\edgeLength)arc(0:-180:0.5*\edgeLength);\node[anchor=90,inner sep=3pt] at(arrownode){{\footnotesize$\b{[\bar{r}]}$}};
\arrowTo[hred]{-20:0.5*\edgeLength}{-20}\node[anchor=mid west,inner sep=2pt] at(in){{\footnotesize$\r{{a_1}}$}};
\arrowTo[hred]{60:0.5*\edgeLength}{60}\node[anchor=240,inner sep=2pt] at(in){{\footnotesize$\r{a_2}$}};
\arrowTo[hred]{120:0.5*\edgeLength}{120}\node[anchor=-60,inner sep=2pt] at(in){{\footnotesize$\r{a_3}$}};
\arrowTo[hred]{200:0.5*\edgeLength}{200}\node[anchor=mid east,inner sep=2pt] at(in){{\footnotesize$\r{{a_n}}$}};
\node[hred,smallDot]at(170:1.05*\edgeLength){};\node[hred,smallDot]at(150:1.05*\edgeLength){};\node[hred,smallDot]at(160:1.05*\edgeLength){};
\node[hblue,clebschR]at(200:0.5*\edgeLength){};\node[hblue,clebschR]at(-20:0.5*\edgeLength){};\node[hblue,clebschR]at(120:0.5*\edgeLength){};\node[hblue,clebschR]at(60:0.5*\edgeLength){};
}.\label{conjugation_of_traces_rule}\vspace{-5pt}}
Upgrading this statement to the level of tensors, we find it useful to define 
\eq{\mathrm{tr}_{\mathbf{\b{R}}}(\r{n}\,\r{\cdots}\,\r{2}\,\r{1})\equivR\big\{\mathrm{tr}_{\mathbf{\b{R}}}(\r{a_n},\ldots,\r{a_2},\r{a_1})\big\}_{\r{a_i}\in\r{[\adR]}};}
that is, we consider `$\mathrm{tr}_{\mathbf{\b{R}}}(\r{a_1\cdots a_n})$' to denote the tensor with a \emph{permuted} sequence of `slots' dictated by the ordering of the sequence $\r{a_1\cdots a_n}$. 

If $\mathbf{\b{R}}\!\simeq\!\b{\bar{\mathbf{\b{{R}}}}}$, then single-trace tensors with reflected orderings of arguments will be identical up to an overall sign; as such, there are generally at most $(n{-}1)!/2$ possibly independent tensors. For complex representations, however, all $(n{-}1)!$ tensors will be distinct in general. To be clear, these tensors \emph{necessarily} fail to be independent beyond some multiplicity of external legs for any fixed Lie algebra as the number of independent tensors (given by $m\indices{\r{\mathbf{ad}}^{\otimes n}}{\mathbf{1}}$) can grow at most exponentially.

At tree-level, reality of the representation $\mathbf{\r{ad}}$ necessitates that only \emph{self-conjugate} combinations of tensors appear. In particular, 
\eq{\hspace{-230pt}\bar{\mathrm{tr}_{\mathbf{\b{R}}}(\r{a_1},\r{a_2},\ldots,\r{a_n})}\;\bigger{\Leftrightarrow}\hspace{-5pt}
\tikzBox{dual_of_trace_indices_diagram_1}{
\draw[hblue,edge,midArrow](200:0.5*\edgeLength)arc(200:120:0.5*\edgeLength);\node[anchor=-30,inner sep=3pt] at(arrownode){{\footnotesize$\b{[\bar{r}]}$}};
\draw[hblue,edge,midArrow](120:0.5*\edgeLength)arc(120:60:0.5*\edgeLength);\node[anchor=-90,inner sep=3pt] at(arrownode){{\footnotesize$\b{[\bar{r}]}$}};
\draw[hblue,dotted,edge](60:0.5*\edgeLength)arc(60:-20:0.5*\edgeLength);%\node[anchor=-135,inner sep=2pt] at(arrownode){{\footnotesize$\b{[r]}$}};
\draw[hblue,edge,midArrow](0:0.5*\edgeLength)arc(0:-180:0.5*\edgeLength);\node[anchor=90,inner sep=3pt] at(arrownode){{\footnotesize$\b{[\bar{r}]}$}};
\arrowTo[hred]{-20:0.5*\edgeLength}{-20}\node[anchor=mid west,inner sep=2pt] at(in){{\footnotesize$\r{\bar{a}_n}$}};
\arrowTo[hred]{60:0.5*\edgeLength}{60}\node[anchor=240,inner sep=2pt] at(in){{\footnotesize$\r{\bar{a}_3}$}};
\arrowTo[hred]{120:0.5*\edgeLength}{120}\node[anchor=-60,inner sep=2pt] at(in){{\footnotesize$\r{\bar{a}_2}$}};
\arrowTo[hred]{200:0.5*\edgeLength}{200}\node[anchor=mid east,inner sep=2pt] at(in){{\footnotesize$\r{\bar{a}_1}$}};\node[hred,smallDot]at(20:1.05*\edgeLength){};\node[hred,smallDot]at(10:1.05*\edgeLength){};\node[hred,smallDot]at(30:1.05*\edgeLength){};
\node[hblue,clebschR]at(200:0.5*\edgeLength){};\node[hblue,clebschR]at(-20:0.5*\edgeLength){};\node[hblue,clebschR]at(120:0.5*\edgeLength){};\node[hblue,clebschR]at(60:0.5*\edgeLength){};
}\hspace{-15pt}=(\text{-}1)^{\r{n}}\hspace{-15pt}\tikzBox{dual_of_trace_indices_diagram_2}{
\draw[hblue,edge,midArrow](120:0.5*\edgeLength)arc(120:200:0.5*\edgeLength);\node[anchor=-30,inner sep=3pt] at(arrownode){{\footnotesize$\b{[{r}]}$}};
\draw[hblue,edge,midArrow](60:0.5*\edgeLength)arc(60:120:0.5*\edgeLength);\node[anchor=-90,inner sep=3pt] at(arrownode){{\footnotesize$\b{[{r}]}$}};
\draw[hblue,dotted,edge](-20:0.5*\edgeLength)arc(-20:60:0.5*\edgeLength);%\node[anchor=-135,inner sep=2pt] at(arrownode){{\footnotesize$\b{[r]}$}};
\draw[hblue,edge,midArrow](-180:0.5*\edgeLength)arc(-180:0:0.5*\edgeLength);\node[anchor=90,inner sep=3pt] at(arrownode){{\footnotesize$\b{[{r}]}$}};
\arrowFrom[hred]{-20:0.5*\edgeLength}{-20}\node[anchor=mid west,inner sep=2pt] at(end){{\footnotesize$\r{{{a}_n}}$}};
\arrowFrom[hred]{60:0.5*\edgeLength}{60}\node[anchor=240,inner sep=2pt] at(end){{\footnotesize$\r{{a}_3}$}};
\arrowFrom[hred]{120:0.5*\edgeLength}{120}\node[anchor=-60,inner sep=2pt] at(end){{\footnotesize$\r{{{a}_2}}$}};
\arrowFrom[hred]{200:0.5*\edgeLength}{200}\node[anchor=mid east,inner sep=2pt] at(end){{\footnotesize$\r{{{a}_1}}$}};\node[hred,smallDot]at(20:1.05*\edgeLength){};\node[hred,smallDot]at(10:1.05*\edgeLength){};\node[hred,smallDot]at(30:1.05*\edgeLength){};
\node[hblue,clebschR]at(200:0.5*\edgeLength){};\node[hblue,clebschR]at(-20:0.5*\edgeLength){};\node[hblue,clebschR]at(120:0.5*\edgeLength){};\node[hblue,clebschR]at(60:0.5*\edgeLength){};
}\hspace{-15pt}=(\text{-}1)^{\r{n}}\hspace{-15pt}\tikzBox{dual_of_trace_indices_diagram_3}{
\draw[hblue,dotted,edge](200:0.5*\edgeLength)arc(200:120:0.5*\edgeLength);%\node[anchor=-30,inner sep=3pt] at(arrownode){{\footnotesize$\b{[\bar{r}]}$}};
\draw[hblue,edge,midArrow](120:0.5*\edgeLength)arc(120:60:0.5*\edgeLength);\node[anchor=-90,inner sep=3pt] at(arrownode){{\footnotesize$\b{[{r}]}$}};
\draw[hblue,edge,midArrow](60:0.5*\edgeLength)arc(60:-20:0.5*\edgeLength);\node[anchor=210,inner sep=2pt] at(arrownode){{\footnotesize$\b{[{r}]}$}};
\draw[hblue,edge,midArrow](0:0.5*\edgeLength)arc(0:-180:0.5*\edgeLength);\node[anchor=90,inner sep=3pt] at(arrownode){{\footnotesize$\b{[{r}]}$}};
\arrowFrom[hred]{-20:0.5*\edgeLength}{-20}\node[anchor=mid west,inner sep=2pt] at(end){{\footnotesize$\r{{a_1}}$}};
\arrowFrom[hred]{60:0.5*\edgeLength}{60}\node[anchor=240,inner sep=2pt] at(end){{\footnotesize$\r{a_2}$}};
\arrowFrom[hred]{120:0.5*\edgeLength}{120}\node[anchor=-60,inner sep=2pt] at(end){{\footnotesize$\r{a_3}$}};
\arrowFrom[hred]{200:0.5*\edgeLength}{200}\node[anchor=mid east,inner sep=2pt] at(end){{\footnotesize$\r{{a_n}}$}};
\node[hred,smallDot]at(170:1.05*\edgeLength){};\node[hred,smallDot]at(150:1.05*\edgeLength){};\node[hred,smallDot]at(160:1.05*\edgeLength){};
\node[hblue,clebschR]at(200:0.5*\edgeLength){};\node[hblue,clebschR]at(-20:0.5*\edgeLength){};\node[hblue,clebschR]at(120:0.5*\edgeLength){};\node[hblue,clebschR]at(60:0.5*\edgeLength){};
}\hspace{-1pt};\hspace{-15pt}\hspace{-180pt}%
}
which is to say, 
\eq{\bar{\mathrm{tr}_{\mathbf{\b{R}}}(\r{1\,2\cdots n})}=\mathrm{tr}_{\mathbf{\b{\bar{R}}}}(\r{\bar{1}\,\bar{2}\cdots\bar{n}})=(\text{-}1)^{\r{n}}\mathrm{tr}_{\mathbf{\b{R}}}(\r{\bar{n}\cdots\bar{2}\,\bar{1}})\,.}
Here, we may be overly pedantic to note that conjugated adjoint index means simply 
\eq{C\indices{\r{[\bar{\adR}]}}{}\equivR\sum_{\r{a}\in\r{[\adR]}}C\indices{}{\!\r{a}}\delta\indices{\r{a}\r{[\bar{\adR}]}}{}=\sum_{\r{a}\in\r{[\adR]}}\big(\sum_{\r{b}\in\r{[\adR]}}C\indices{\r{b}}{}g^{\mathbf{\r{ad}}}_{\smash{\r{b}\r{a}}}\big)\delta\indices{\r{a}\r{[\bar{\adR}]}}{}\,,}
and we generally do \emph{not} assume that these indices have been arranged so that $g_{\r{\mathbf{ad}}}^{\smash{\r{[\adR][\adR]}}}\!\propto\!\delta\indices{\r{[\adR][\adR]}}{}$.\\

In terms of these, we may define the \emph{complete contraction} of the tensor $\mathrm{tr}_{\mathbf{\b{R}}}(\r{1\cdots n})$ with its conjugate via
\eq{\begin{split}\langle\mathrm{tr}_{\mathbf{\b{R}}}(\r{1\,2\cdots n})|\mathrm{tr}_{\mathbf{\b{R}}}(\r{1\,2\cdots n})\rangle&\equivR\sum_{\r{a_i}\in\r{[\adR]}}\bar{\mathrm{tr}_{\mathbf{\b{R}}}(\r{1\,2\cdots n})}\indices{}{\r{a_1\cdots a_n}}\mathrm{tr}_{\mathbf{\b{R}}}(\r{1\,2\cdots n})\indices{\r{a_1\cdots a_n}}{}\\
&\bigger{\Leftrightarrow}\;\tikzBox{self_contraction_of_trace}{
\draw[hblue,edge,midArrow](200:0.5*\edgeLength)arc(200:120:0.5*\edgeLength);%\node[anchor=-30,inner sep=3pt] at(arrownode){{\footnotesize$\b{[r]}$}};
\draw[hblue,edge,midArrow](120:0.5*\edgeLength)arc(120:60:0.5*\edgeLength);%\node[anchor=-90,inner sep=3pt] at(arrownode){{\footnotesize$\b{[r]}$}};
\draw[hblue,dotted,edge](60:0.5*\edgeLength)arc(60:-20:0.5*\edgeLength);%\node[anchor=-135,inner sep=2pt] at(arrownode){{\footnotesize$\b{[r]}$}};
\draw[hblue,edge,midArrow](0:0.5*\edgeLength)arc(0:-180:0.5*\edgeLength);\node[anchor=90,inner sep=3pt] at(arrownode){{\footnotesize$\b{[r]}$}};
\arrowTo[hred]{-20:0.5*\edgeLength}{-20}%\node[anchor=mid west,inner sep=2pt] at(in){{\footnotesize$\r{a_n}$}};
\arrowTo[hred]{60:0.5*\edgeLength}{60}%\node[anchor=240,inner sep=2pt] at(in){{\footnotesize$\r{a_3}$}};
\arrowTo[hred]{120:0.5*\edgeLength}{120}\node[anchor=30,inner sep=2pt]at(arrownode){{\footnotesize$\r{[\adR]}$}};%\node[anchor=-60,inner sep=2pt] at(in){{\footnotesize$\r{a_2}$}};
\arrowTo[hred]{200:0.5*\edgeLength}{200}%\node[anchor=mid east,inner sep=2pt] at(in){{\footnotesize$\r{a_1}$}};
\node[hred,smallDot]at(20:1.05*\edgeLength){};\node[hred,smallDot]at(10:1.05*\edgeLength){};\node[hred,smallDot]at(30:1.05*\edgeLength){};
\node[hblue,clebschR]at(200:0.5*\edgeLength){};\node[hblue,clebschR]at(-20:0.5*\edgeLength){};\node[hblue,clebschR]at(120:0.5*\edgeLength){};\node[hblue,clebschR]at(60:0.5*\edgeLength){};
\draw[hblue,dotted,edge](60:1.5*\edgeLength)arc(60:-20:1.5*\edgeLength);
\draw[hblue,edge,midArrow](60:1.5*\edgeLength)arc(60:120:1.5*\edgeLength);\node[anchor=90,inner sep=3pt] at(arrownode){{\footnotesize$\b{[r]}$}};
\draw[hblue,edge,midArrow](120:1.5*\edgeLength)arc(120:200:1.5*\edgeLength);%\node[anchor=-30,inner sep=3pt] at(arrownode){{\footnotesize$\b{[r]}$}};
\draw[hblue,edge,midArrow](200:1.5*\edgeLength)arc(200:340:1.5*\edgeLength);%\node[anchor=90,inner sep=3pt] at(arrownode){{\footnotesize$\b{[r]}$}};
\node[hblue,clebschR]at(-20:1.5*\edgeLength){};\node[hblue,clebschR]at(60:1.5*\edgeLength){};\node[hblue,clebschR]at(200:1.5*\edgeLength){};\node[hblue,clebschR]at(120:1.5*\edgeLength){};
}\,.\end{split}}
According to the convention that the Killing metric be normalized so that $g_{\mathbf{\r{ad}}}^{\smash{\r{[\adR][\adR]}}}\equivR\mathrm{tr}_{\mathbf{\b{F}}}(\r{1\,2})$, we can easily compute these for $\b{n}$ copies of the \emph{fundamental} representation $\mathbf{\b{F}}$ of the classical Lie algebras:
\eq{\begin{split}\tikzBox{self_contraction_over_fundamental_trace}{
\draw[hblue,edge,midArrow](200:0.5*\edgeLength)arc(200:120:0.5*\edgeLength);%\node[anchor=-30,inner sep=3pt] at(arrownode){{\footnotesize$\b{[r]}$}};
\draw[hblue,edge,midArrow](120:0.5*\edgeLength)arc(120:60:0.5*\edgeLength);%\node[anchor=-90,inner sep=3pt] at(arrownode){{\footnotesize$\b{[r]}$}};
\draw[hblue,dotted,edge](60:0.5*\edgeLength)arc(60:-20:0.5*\edgeLength);%\node[anchor=-135,inner sep=2pt] at(arrownode){{\footnotesize$\b{[r]}$}};
\draw[hblue,edge,midArrow](0:0.5*\edgeLength)arc(0:-180:0.5*\edgeLength);\node[anchor=90,inner sep=3pt] at(arrownode){{\footnotesize$\b{[f]}$}};
\arrowTo[hred]{-20:0.5*\edgeLength}{-20}%\node[anchor=mid west,inner sep=2pt] at(in){{\footnotesize$\r{a_n}$}};
\arrowTo[hred]{60:0.5*\edgeLength}{60}%\node[anchor=240,inner sep=2pt] at(in){{\footnotesize$\r{a_3}$}};
\arrowTo[hred]{120:0.5*\edgeLength}{120}\node[anchor=30,inner sep=2pt]at(arrownode){{\footnotesize$\r{[\adR]}$}};%\node[anchor=-60,inner sep=2pt] at(in){{\footnotesize$\r{a_2}$}};
\arrowTo[hred]{200:0.5*\edgeLength}{200}%\node[anchor=mid east,inner sep=2pt] at(in){{\footnotesize$\r{a_1}$}};
\node[hred,smallDot]at(20:1.05*\edgeLength){};\node[hred,smallDot]at(10:1.05*\edgeLength){};\node[hred,smallDot]at(30:1.05*\edgeLength){};
\node[hblue,clebschR]at(200:0.5*\edgeLength){};\node[hblue,clebschR]at(-20:0.5*\edgeLength){};\node[hblue,clebschR]at(120:0.5*\edgeLength){};\node[hblue,clebschR]at(60:0.5*\edgeLength){};
\draw[hblue,dotted,edge](60:1.5*\edgeLength)arc(60:-20:1.5*\edgeLength);
\draw[hblue,edge,midArrow](60:1.5*\edgeLength)arc(60:120:1.5*\edgeLength);\node[anchor=90,inner sep=3pt] at(arrownode){{\footnotesize$\b{[f]}$}};
\draw[hblue,edge,midArrow](120:1.5*\edgeLength)arc(120:200:1.5*\edgeLength);
\draw[hblue,edge,midArrow](200:1.5*\edgeLength)arc(200:340:1.5*\edgeLength);
\node[hblue,clebschR]at(-20:1.5*\edgeLength){};\node[hblue,clebschR]at(60:1.5*\edgeLength){};\node[hblue,clebschR]at(200:1.5*\edgeLength){};\node[hblue,clebschR]at(120:1.5*\edgeLength){};
}&=\left\{\begin{array}{ll}
\left(\!\r{k}\frac{(\r{k}{+}2)}{(\r{k}{+}1)}\!\right)^{\!\b{n}}\!\!\big(1{-}\frac{1}{(-\r{k}(\r{k}{+}2))^{\b{n}-1}}\big)&\mathfrak{a}_{\r{k}}\\
\r{k}^{\b{n}}\Big(1{+}\frac{\r{k}(2\r{k}{+}1)}{(2\r{k})^{\b{n}}}{+}\frac{\r{k}(2\r{k}{+}3)}{({-}2\r{k})^{\b{n}}}\Big)&\mathfrak{b}_{\r{k}}\\
\frac{(2\r{k}{+}1)^{\b{n}}}{2^{\b{n}}}\big(1{-}\frac{1{-}\r{k}(1{+}({-}1)^{\b{n}})}{(2\r{k}{+}1)^{\b{n}{-}1}}\big)&\mathfrak{c}_{\r{k}}\\
\frac{(2\r{k}{-}1)^{\b{n}}}{2^{\b{n}}}\big(1{-}\frac{1{-}(\r{k}{+}1)(1{+}({-}1)^{\b{n}})}{(2\r{k}{-}1)^{\b{n}{-}1}}\big)&\mathfrak{d}_{\r{k}}\\
\end{array}\right.\,.
\end{split}}
Single trace tensors with permuted orderings have non-vanishing overlap in general, but are generally sub-leading in the limit of large rank. Thus, these tensors \emph{approach} colour-orthogonality in the large-rank limit \cite{Mangano:1987xk}.

Beyond tree-level, it is easy to see that single-trace tensors do not suffice. But we may define \emph{multi-trace} tensors in the natural way:
\eq{\mathrm{tr}_{\mathbf{R}}(\r{a_1\cdots}|\b{b_1\cdots}|\cdots|\g{c_1\cdots})\equivR\mathrm{tr}_{\mathbf{\b{R}}}(\r{a_1\cdots})\mathrm{tr}_{\mathbf{\b{R}}}(\b{b_1\cdots})\cdots\mathrm{tr}_{\mathbf{\b{R}}}(\g{c_1\cdots})\,.}
For each of the classical Lie algebras, Fierz identities (\ref{fierz_a}) and (\ref{fierz_bcd}) (derived in \mbox{section~\ref{section:clebsch_bases_for_ffff}} below) allow one to systematically decompose any $\r{f}$-graph into multi-trace tensors involving the fundamental representation of the algebra; thus---even for $\mathfrak{d}_{\r{k}}$ gauge theories, for which these tensors are \emph{incomplete}---these tensors necessarily span the subspace of tensors generated by the Feynman expansion for adjoint-coloured particles. 

In the case of $\mathfrak{a}_{\r{k}}$ gauge theory, the Fierz identity (\ref{fierz_in_terms_of_traces}) is easily seen to show that, in the expansion of $\r{f}$-graphs into single and multi-trace tensors, multi-traces are suppressed relative to single traces in the limit of large rank.\\

\subsubsection{Feynman Diagrams involving Arbitrarily-Charged Fermions}

Recall that the Feynman rules for charged matter must involve the generators of some representation of the Lie algebra associated with the gauge theory. Thus, for the scattering of charged matter with adjoints, Feynman diagrams are naturally endowed with colour tensors given by graphs of $\mathbf{\b{R}}$ and $\mathbf{\r{ad}}$ vertices. Provided there is at least one charged Fermion in the process, the defining relation of a representation (\ref{diagrammatic_defn_of_rep}) allows one to eliminate all 3-point $\r{\mathbf{ad}}$ vertices, resulting in tensors built entirely from $\mathbf{\b{R}}$ vertices. 

Partial amplitudes involving arbitrary combinations of adjoint-coloured gluons and charged matter (all with colours given by the same (but arbitrary) representation) were studied by Melia in \cite{Melia:2013bta,Melia:2013epa,Melia:2013xok}, and the corresponding colour tensors at tree-level were described in \cite{Johansson:2015oia} (see also \cite{Ochirov:2019mtf}). These colour tensors can be viewed as analogues of the $\r{f}$-graphs described by DDM in \cite{DelDuca:1999rs}; and as with the $\r{f}$-graphs, these tensors are far from orthogonal in colour-space. 

We will have more to say on this subject in a forthcoming work \cite{tools_for_multiflavour_amps}.\\

Finally, it is worthwhile to note that the issue of colour non-orthogonality has been addressed by several authors in recent years in the case of $\mathfrak{a}_{\r{k}}$ gauge theory. In particular, Clebsch-Gordan coefficients were used to define orthogonal colour tensors for processes in $\mathfrak{a}_{\r{k}}$ gauge theory in the works of \cite{Keppeler2012OrthogonalMB,Sjodahl2015,Sjodahl2024}. These tensors formed what was called the `multiplet basis', and can be viewed as a specialization of the general constructions we describe below.

\newpage
\subsection{Novel Colour Tensors for Scattering Adjoints and Fundamentals}

Let us consider scattering amplitudes involving a single particle charged under any irreducible representation. The only possible colour structure consists of the identity tensor
\eq{\mathcal{B}(\mathbf{\b{R}}|\mathbf{\b{R}})\equivR\left\{\delta\indices{\b{[r]}}{\b{[r]}}\right\}\,.}
This follows from Schur's lemma, and the irreducibility of the representation. 

For the scattering between any single particle charged under a representation $\mathbf{\b{R}}$ and a gluon (or other particle transforming under the adjoint representation), the basis of colour tensors will consist of 
\eq{\left\{\mathcal{B}_{\hspace{-1pt}\mu}(\mathbf{\b{R}}\,\mathbf{\r{ad}}|\mathbf{\b{R}})\right\}_{\mu\in[m\indices{\mathbf{\b{R}\,\r{ad}}}{\mathbf{\b{R}}}]}\,\equivL\,\Big\{\mathbf{\b{R}}\indices{\b{[r]}\,\r{[\adR]}}{\b{[r]}},\ldots\Big\}\;\bigger{\Leftrightarrow}\;\Bigg\{\tikzBox[-5pt]{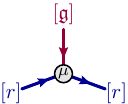}{\arrowTo[hblue]{0,0}{200}\node[anchor=20,inner sep=0pt] at(in){{\footnotesize$\b{[r]}$}};\arrowTo[hred]{0,0}{90}\node[anchor=-90,inner sep=2pt] at(in){{\footnotesize$\r{[\adR]}$}};\arrowFrom[hblue]{0,0}{-20}\node[anchor=160,inner sep=0pt] at(end){{\footnotesize$\b{[r]}$}};\node[clebsch]at(0,0){{\scriptsize$\phantom{\nu}$}};\node[]at(0,0){{\scriptsize${\mu}$}};
}\Bigg\}\,,
}
where the multiplicity takes values in $\mu\!\in\![m\indices{\mathbf{\b{R}\,\r{ad}}}{\mathbf{\b{R}}}]$. Within this space of colour tensors, one can \emph{always} be chosen to be the representation's generators. 

For the fundamental (or `defining') representations `$\mathbf{\b{F}}$' defined in \mbox{appendix~\ref{appendix:weight_system_conventions}}, the multiplicity $m\indices{\mathbf{\b{F}}\,\r{\mathbf{ad}}}{\mathbf{\b{F}}}$ is always 1, leaving us with a unique colour tensor associated with three-particle scattering:
\vspace{-5pt}\eq{\mathcal{B}(\mathbf{\b{F}}\,\mathbf{\r{ad}}|\mathbf{\b{F}})\,\equivR\,\Big\{\mathbf{\b{F}}\indices{\b{[f]}\,\r{[\adR]}}{\b{[f]}}\Big\}\;\bigger{\Leftrightarrow}\;\Bigg\{\hspace{0pt}\raisebox{-5pt}{\tikzBox[-9.125pt]{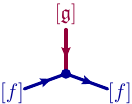}{\arrowTo[hblue]{0,0}{200}\node[anchor=20,inner sep=0pt] at(in){{\footnotesize$\b{[f]}$}};\arrowTo[hred]{0,0}{90}\node[anchor=-90,inner sep=2pt] at(in){{\footnotesize$\r{[\adR]}$}};\arrowFrom[hblue]{0,0}{-20}\node[anchor=160,inner sep=0pt] at(end){{\footnotesize$\b{[f]}$}};\node[hblue,clebschR]at(0,0){};
}}\,\hspace{0pt}\Bigg\}.
}
This uniqueness is shared for all representations of `fundamental' Dynkin weight for each of the simple Lie algebras, but is not true for more general charged matter. We saw earlier that for $\mathfrak{a}_{\r{k}>1}$ Lie algebras, $m\indices{\mathbf{\r{ad\,ad}}}{\mathbf{\r{ad}}}\!=\!2$, leading to a space of two linearly independent colour tensors:
\vspace{-5pt}\eq{\fwbox{0pt}{\fwboxL{435pt}{(\mathfrak{a}_{\r{k}>1})}}\fwbox{0pt}{\mathcal{B}_{\hspace{-1pt}\mu}(\mathbf{\r{ad}}\,\mathbf{\r{ad}}|\mathbf{\r{ad}})\,\equivR\,\Big\{\mathbf{\r{ad}}\indices{\r{[\adR]}\,\r{[\adR]}}{\r{[\adR]}},{\r{d}}\indices{\r{[\adR]}\,\r{[\adR]}}{\r{[\adR]}}\Big\}\;\bigger{\Leftrightarrow}\;\Bigg\{\raisebox{-5pt}{\hspace{0pt}\tikzBox[-9.125pt]{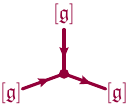}{\arrowTo[hred]{0,0}{200}\node[anchor=20,inner sep=0pt] at(in){{\footnotesize$\r{[\adR]}$}};\arrowTo[hred]{0,0}{90}\node[anchor=-90,inner sep=2pt] at(in){{\footnotesize$\r{[\adR]}$}};\arrowFrom[hred]{0,0}{-20}\node[anchor=160,inner sep=0pt] at(end){{\footnotesize$\r{[\adR]}$}};\node[hred,clebschR]at(0,0){};
}}\hspace{0pt},\hspace{0pt}\raisebox{-5pt}{\tikzBox[-9.125pt]{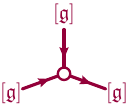}{\arrowTo[hred]{0,0}{200}\node[anchor=20,inner sep=0pt] at(in){{\footnotesize$\r{[\adR]}$}};\arrowTo[hred]{0,0}{90}\node[anchor=-90,inner sep=2pt] at(in){{\footnotesize$\r{[\adR]}$}};\arrowFrom[hred]{0,0}{-20}\node[anchor=160,inner sep=0pt] at(end){{\footnotesize$\r{[\adR]}$}};\node[hred,clebschD]at(0,0){};
}}\,\hspace{0pt}\Bigg\}.}
}
Although the coefficient of the second colour tensor could be viewed as arising from an anomaly, both tensors contribute to bases of colour tensors constructed even for non-anomalous scattering of adjoints. 

It is worth mentioning that for most all Lie algebras of rank $\r{k}\!>\!1$, there exist irreducible representations which appear in tensor products with \emph{arbitrarily large} multiplicities. For example, the $\mathbf{1024}$-dimensional representation of $\mathfrak{a}_{\r{4}}$ or the $\mathbf{4096}$-dimensional representation of $\mathfrak{d}_{\r{4}}$ (both with Dynkin labels $[1111]$) each require 4 independent colour tensors to describe its interactions with gluons. These novel colour tensors would not appear in the Feynman rules directly, but they could be generated perturbatively (or non-perturbatively).

\subsection{Novel Colour Tensors for Four-Particle Scattering Amplitudes}

In the completely general case of scattering four coloured particles, we have seen that a linearly-independent, orthogonal basis of colour tensors can be constructed as
\eq{\fwboxR{0pt}{\mathcal{B}^{\hspace{2.2pt}\mathbf{{a}}}_{\hspace{-1pt}\mu\nu}(\mathbf{\b{R}}\,\mathbf{\g{S}}|\mathbf{\t{T}}\,\mathbf{\r{U}})\;\bigger{\Leftrightarrow}\;}\tikzBox{default_basis_four_generic_reps}{\arrowTo[hblue]{0,0}{-130};\node[anchor=50,inner sep=2pt] at(in){{\footnotesize$\b{[r]}$}};\arrowTo[hgreen]{0,0}{130}\node[anchor=-50,inner sep=2pt] at(in){{\footnotesize$\g{[s]}$}};\arrowFrom[black]{0,0}[1.5]{0}\node[clebsch]at(in){{\scriptsize$\phantom{\nu}$}};\node[]at(in){{\scriptsize$\mu$}};\node[anchor=90,inner sep=2pt] at(arrownode){{\footnotesize${[a]}$}};\arrowFrom[hteal]{end}{50}
\node[anchor=-130,inner sep=2pt] at(end){{\footnotesize$\t{[t]}$}};\arrowFrom[hred]{in}{-50}\node[anchor=130,inner sep=2pt] at(end){{\footnotesize$\r{[u]}$}};\node[clebsch]at(in){{\scriptsize$\phantom{\nu}$}};\node[]at(in){{\scriptsize$\nu$}};
}\,.\label{four_particle_colour_basis_general}
}
The details of this basis depend strongly on the Lie algebra $\mathfrak{g}$ and the representations of particles involved. Although it is reasonably straight-forward to construct these tensors explicitly for any chosen representations, we would have no `standard' or familiar choice of tensors with which we could compare (whether or not they form a `basis'). For this reason, it is worthwhile to focus our attention to simpler cases---those involving combinations particles with some fixed representation and adjoints, say. For this, it is worthwhile to specialize further and consider matter charged only in the fundamental or `defining' representation $\mathbf{\b{F}}$ of the Lie algbera. 

Let us first consider the cases where an \emph{odd} number of $\mathbf{\b{F}}$-coloured particles are involved. For example, is there ever any support for colour tensors involving 3 fundamental particles and a single adjoint, say? As described in \cite{Bourjaily:2024jbt}, this is a simple question of representation theory, and turns out to only have support in the case of gauge theories involving one of six isolated simple Lie algebras: $\mathcal{B}(\mathbf{\b{F}}\,\mathbf{\b{F}}\,\mathbf{\b{F}}|\mathbf{\r{ad}})$ is non-empty for the simple Lie algebras $\{\mathfrak{a}_{\r{2}},\mathfrak{b}_{\r{2}},\mathfrak{e}_{\r{6}},\mathfrak{e}_{\r{8}},\mathfrak{f}_{\r{4}},\mathfrak{g}_{\r{2}}\}$, with ranks  $\{2,1,2,5,2,2\}$, respectively. Similarly, $\mathcal{B}(\mathbf{\b{F}}\,\mathbf{\r{ad}}|\mathbf{\r{ad}}\,\mathbf{\r{ad}})$ is non-empty only for the simple Lie algebras $\{\mathfrak{b}_{\r{2}},\mathfrak{b}_{\r{3}},\mathfrak{e}_{\r{8}},\mathfrak{g}_{\r{2}}\}$. Of course, for $\mathfrak{e}_{\r{8}}$, we take the fundamental to be the same as the adjoint, and so the case is fairly exceptional.

\subsubsection{\texorpdfstring{Clebsch Colour Bases for $C(\mathbf{\b{F}}\,\mathbf{\b{F}}|\mathbf{\b{F}}\,\mathbf{\b{F}})$}{Clebsch Colour Bases for C(FF|FF)}}\label{section:clebsch_bases_for_ffff}

Let us begin our analysis with the case of four particles all of which are coloured in the fundamental representation of some simple Lie algebra. In our language, we can span the complete space of colour tensors by sewing together Clebsch-Gordan factors according to:
\eq{\fwboxR{0pt}{\mathcal{B}^{\hspace{1pt}\mathbf{\g{q}}}(\mathbf{\b{F}}\,\mathbf{\b{F}}|\mathbf{\b{F}}\,\mathbf{\b{F}})\;\bigger{\Leftrightarrow}\;}\tikzBox{FFFF_basis_1}{\arrowTo[hblue]{0,0}{-130};\node[anchor=10,inner sep=2pt] at(in){{\footnotesize$\b{[f]}$}};\arrowTo[hblue]{0,0}{130}\node[anchor=-10,inner sep=2pt] at(in){{\footnotesize$\b{[f]}$}};\arrowFrom[hgreen]{0,0}[1.5]{0}\node[clebsch]at(in){};\node[anchor=90,inner sep=2pt] at(arrownode){{\footnotesize$\g{[q]}$}};\arrowFrom[hblue]{end}{50}
\node[anchor=-170,inner sep=2pt] at(end){{\footnotesize$\b{[f]}$}};\arrowFrom[hblue]{in}{-50}\node[anchor=170,inner sep=2pt] at(end){{\footnotesize$\b{[f]}$}};\node[clebsch]at(in){};
}\,.\label{ffff_colour_basis_defined}
}
As $m\indices{\mathbf{\b{F}}\,\mathbf{\b{F}}}{\mathbf{\g{q}}}\!=\!1$ for the fundamental representations of all simple Lie algebras as we have defined them (see \mbox{appendix~\ref{appendix:weight_system_conventions}}), each tensor is uniquely labelled by the irreducible representation $\mathbf{\g{q}}_{\g{i}}$ appearing in the tensor product representation's decomposition 
\eq{\mathbf{\b{F}}\!\otimes\!\mathbf{\b{F}}\equivL\bigoplus_{\g{i}}\mathbf{\g{q}}_{\g{i}}\,.\label{qReps_defined}}
It is easy to see that \emph{every} representation $\mathbf{\g{q}}_{\g{i}}$ appearing in (\ref{qReps_defined}) contributes, and so the number of independent colour tensors for $\mathbf{\b{F}}\mathbf{\b{F}}\!\to\!\mathbf{\b{F}}\mathbf{\b{F}}$ scattering is counted simply by the number of such irreducible representations.

\begin{table}[t]\vspace{-10pt}\caption{Irreducible representations appearing in the decomposition of the tensor product $\mathbf{\b{F}}\!\otimes\!\mathbf{\b{F}}\equivL\bigoplus_{\g{i}}\mathbf{\g{q}}_{\g{i}}$ for simple Lie algebras $\mathfrak{g}$ except $\mathfrak{e}_{\r{8}}$, which is discussed separately.}\vspace{-25pt}$$\begin{array}{|l@{$\,$}|ccc|}\multicolumn{1}{c}{~}&\fwbox{40pt}{\g{\mathbf{q}_{1}}}&\fwbox{70pt}{\g{\mathbf{q}_{2}}}&\multicolumn{1}{c}{\fwbox{60pt}{\g{\mathbf{q}_{3}}}}\\\hline\hline
\raisebox{-4pt}{$\mathfrak{a}_{\r{k}}$}&\dynkLabelK{\frac{1}{2}\r{k}(\hspace{-1pt}\r{k}\pl1\hspace{-1pt})}{01\cdots0}&\dynkLabelK{\frac{1}{2}(\hspace{-1pt}\r{k}\pl1\hspace{-1pt})(\hspace{-1pt}\r{k}\pl2\hspace{-1pt})}{20\cdots0}&~\\\hline
\raisebox{-4pt}{$\mathfrak{b}_{\r{k}}$}&\dynkLabelK{\mathbf{1}}{0\cdots0}&\dynkLabelK{\r{k}(\hspace{-1pt}2\r{k}\pl1\hspace{-1pt})}{010\cdots0}&\dynkLabelK{\r{k}(\hspace{-1pt}2\r{k}\pl3\hspace{-1pt})}{20\cdots0}\\\hline
\raisebox{-4pt}{$\mathfrak{c}_{\r{k}}$}&\dynkLabelK{\mathbf{1}}{0\cdots0}&\dynkLabelK{\r{k}(\hspace{-1pt}2\r{k}\pl1\hspace{-1pt})}{20\cdots0}&\dynkLabelK{(\hspace{-1pt}2\r{k}\pl1\hspace{-1pt})(\hspace{-1pt}\r{k}\mi1\hspace{-1pt})}{010\cdots0}\\\hline
\raisebox{-4pt}{$\mathfrak{d}_{\r{k}}$}&\dynkLabelK{\mathbf{1}}{0\cdots0}&\dynkLabelK{\r{k}(\hspace{-1pt}2\r{k}\mi1\hspace{-1pt})}{010\cdots0}&\dynkLabelK{(\hspace{-1pt}2\r{k}\mi1\hspace{-1pt})(\hspace{-1pt}\r{k}\pl1\hspace{-1pt})}{20\cdots0}\\\hline
\end{array}\;\;
\begin{array}{|l@{$\,$}|ccccc|}\multicolumn{1}{c}{~}&\fwbox{40pt}{\g{\mathbf{q}_{1}}}&\fwbox{20pt}{\g{\mathbf{q}_{2}}}&\multicolumn{1}{c}{\fwbox{30pt}{\g{\mathbf{q}_{3}}}}&\multicolumn{1}{c}{\fwbox{30pt}{\g{\mathbf{q}_{4}}}}&\multicolumn{1}{c}{\fwbox{30pt}{\g{\mathbf{q}_{5}}}}\\\hline\hline
\raisebox{-4pt}{$\mathfrak{e}_{\r{6}}$}&\dynkLabelK{\mathbf{\bar{27}}}{000010}&\dynkLabelK{\mathbf{351}}{010000}&\dynkLabelK{\mathbf{351'}}{200000}&~&~\\\hline
\raisebox{-4pt}{$\mathfrak{e}_{\r{7}}$}&\dynkLabelK{\mathbf{1}}{0000000}&\dynkLabelK{\mathbf{133}}{0000010}&\dynkLabelK{\mathbf{1539}}{0100000}&\dynkLabelK{\mathbf{1463}}{2000000}&~\\\hline
\raisebox{-4pt}{$\mathfrak{f}_{\r{4}}$}&\dynkLabelK{\mathbf{1}}{0000}&\dynkLabelK{\mathbf{52}}{0001}&\dynkLabelK{\mathbf{324}}{2000}&\dynkLabelK{\mathbf{273}}{0100}&\dynkLabelK{\mathbf{\b{26}}}{1000}~\\\hline
\raisebox{-4pt}{$\mathfrak{g}_{\r{2}}$}&\dynkLabelK{\mathbf{1}}{00}&\dynkLabelK{\mathbf{14}}{01}&\dynkLabelK{\mathbf{27}}{20}&\dynkLabelK{\mathbf{\b{7}}}{10}&~\\\hline
\end{array}$$\vspace{-25pt}\vspace{-0pt}\label{qReps_table}\end{table}

In \mbox{Table~\ref{qReps_table}}, we define the representations $\mathbf{\g{q}}_{\g{i}}$ which appear for each of the simple Lie algebras besides $\mathfrak{e}_{\r{8}}$---for which $\mathbf{\b{F}}\!\simeq\!\mathbf{\r{ad}}$ and 
\eq{\fwbox{0pt}{\fwboxL{435pt}{(\mathfrak{e}_{\r{8}})}}
\fwbox{0pt}{\begin{array}{c@{\oplus}c@{\oplus}c@{\oplus}c@{\oplus}ccccc}\fwboxR{0pt}{\mathbf{\b{F}}\!\otimes\!\mathbf{\b{F}}\equivL\,}\g{\mathbf{{~~\hspace{-4pt}}_{\phantom{1}}q}_{\g{1}}}&\g{\mathbf{\phantom{{~~\hspace{-4pt}}_{2}}q}_{\g{2}}}&\g{\mathbf{{~~\hspace{-3pt}}_{\phantom{3}}q}_{\g{3}}}&\g{\mathbf{{~~\hspace{-4pt}}_{\phantom{4}}q}_{\g{4}}}&\g{\mathbf{{~\hspace{-9pt}}_{\phantom{5}}q}_{5}}\\
\multicolumn{1}{c}{\dynkLabelK{\mathbf{1}}{00000000}}&\multicolumn{1}{c}{\dynkLabelK{\mathbf{248}}{10000000}}&\multicolumn{1}{c}{\dynkLabelK{\mathbf{3875}}{00000010}}&\multicolumn{1}{c}{\dynkLabelK{\mathbf{30380}}{01000000}}&\multicolumn{1}{c}{\dynkLabelK{\mathbf{27000}}{20000000}}\,.\end{array}}\label{e8_qReps_defined}}
For the classical Lie algebras, some care is required to specify these representations---as there are frequently distinct, irreducible representations that share the same dimension. For the sake of clarity, therefore we have listed both the dimension formulae and the (large-$\r{k}$) structure of these representations' Dynkin labels (with respect to the conventions outlined in \mbox{appendix~\ref{appendix:weight_system_conventions}}). (The precise representations' Dynkin labels may differ from the patterns suggested in \mbox{Table~\ref{qReps_table}} in cases of low rank.)

With what can we compare these tensors? A more familiar list of tensors---which in fact always span the space of possible colour tensors---would include the following:
\eq{\big\{\mathbf{Q}^1,\ldots,\mathbf{Q}^4\big\}\;\bigger{\Leftrightarrow}\;\Bigg\{\egTensorsAa,
\egTensorsAb,
\egTensorsAc,
\egTensorsAd\Bigg\}\,;\label{default_tensors_q_defined}}
and, if the fundamental representation is self-conjugate (if $\mathbf{\b{F}}\!\simeq\!\mathbf{\b{\bar{F}}}$) one may also include the tensor
\eq{\fwboxR{0pt}{\mathbf{Q}^5\;\bigger{\Leftrightarrow}\;}\egTensorsAe\,.}
%
%\newpage
%

By directly constructing each of the colour tensors in the basis (\ref{ffff_colour_basis_defined}), we have determined the coefficients appearing in the expansion
\vspace{-3pt}\eq{\mathbf{Q}^{\b{i}}\equivL\,\sum_{\g{j}}\mathbf{c}[\mathfrak{\r{g}}]^{\b{i}}_{\phantom{i}\g{j}}\mathcal{B}^{\hspace{1pt}\g{\mathbf{q}_j}}\,.\label{olde_to_new_ffff_coefficients_defined}\vspace{-6pt}}
for all of the simple Lie algebras. The results of this computation are documented in \mbox{Table~\ref{expansion_coefficients_ffff}}.

\begin{table}[t]\vspace{-10pt}\caption{Expansion coefficients ${\mathbf{c}}$ required to express $\mathbf{Q}^i$ tensors in terms of the basis $\mathcal{B}^{\hspace{1pt}\g{\mathbf{q}_i}}$.}\vspace{-30pt}$$\begin{array}{c}\begin{array}[t]{@{$\!$}c@{$\!$}c@{$\!$}c@{$\!$}c@{$\!$}c@{$\!$}}\begin{array}[t]{c}\mathfrak{a}_{\fwboxL{2pt}{\r{k}}}\\
\left(\begin{array}{c@{\;\;\;}c}\frac{\r{k}\hspace{-0pt}{+}\hspace{-0pt}2}{\r{k}\hspace{-0pt}{+}\hspace{-0pt}1}&\frac{\r{k}}{\r{k}\hspace{-0pt}{+}\hspace{-0pt}1}\\[-0.5pt]
\tmi\frac{\r{k}\hspace{-0pt}{+}\hspace{-0pt}2}{\r{k}\hspace{-0pt}{+}\hspace{-0pt}1}&\frac{\r{k}}{\r{k}\hspace{-0pt}{+}\hspace{-0pt}1}\\[-0.5pt]
\tmi1&1\\[-0.5pt]
1&1\end{array}\right)
\end{array}&\begin{array}[t]{c}\mathfrak{b}_{\fwboxL{2pt}{\r{k}}}\\
\left(\begin{array}{c@{\;\;\;}c@{\;\;\;}c}\tmi\frac{\r{k}}{2\r{k}\hspace{-0pt}{+}\hspace{-0pt}1}&\tmi\frac{1}{2}&\frac{1}{2}\\[-0.5pt]
\tmi\frac{\r{k}}{2\r{k}\hspace{-0pt}{+}\hspace{-0pt}1}&\frac{1}{2}&\frac{1}{2}\\[-0.5pt]
\frac{1}{2\r{k}\hspace{-0pt}{+}\hspace{-0pt}1}&1&1\\[-0.5pt]
\frac{1}{2\r{k}\hspace{-0pt}{+}\hspace{-0pt}1}&\tmi1&1\\[-0.5pt]
1&\dzero&\dzero\end{array}\right)
\end{array}&
\begin{array}[t]{c}\mathfrak{c}_{\fwboxL{2pt}{\r{k}}}\\
\left(\begin{array}{c@{\;\;\;}c@{\;\;\;}c}\tmi\frac{2\r{k}\hspace{-0pt}{+}\hspace{-0pt}1}{4\r{k}}&\frac{1}{2}&\frac{1}{2}\\[-0.5pt]
\frac{2\r{k}\hspace{-0pt}{+}\hspace{-0pt}1}{4\r{k}}&\frac{1}{2}&\tmi\frac{1}{2}\\[-0.5pt]
\frac{1}{2\r{k}}&1&\tmi1\\[-0.5pt]
\tmi\frac{1}{2\r{k}}&1&1\\[-0.5pt]
1&\dzero&\dzero\end{array}\right)
\end{array}&
\begin{array}[t]{c}\mathfrak{d}_{\fwboxL{2pt}{\r{k}}}\\
\left(\begin{array}{c@{\;\;\;}c@{\;\;\;}c}\tmi\frac{2\r{k}\hspace{-0pt}{-}\hspace{-0pt}1}{4\r{k}}&\tmi\frac{1}{2}&\frac{1}{2}\\[-0.5pt]
\tmi\frac{2\r{k}\hspace{-0pt}{-}\hspace{-0pt}1}{4\r{k}}&\frac{1}{2}&\frac{1}{2}\\[-0.5pt]
\frac{1}{2\r{k}}&1&1\\[-0.5pt]
\frac{1}{2\r{k}}&\tmi1&1\\[-0.5pt]
1&\dzero&\dzero\end{array}\right)
\end{array}&
\begin{array}[t]{c}\mathfrak{e}_{\fwboxL{2pt}{\r{6}}}\\
\left(\begin{array}{c@{\;\;\;}c@{\;\;\;}c}\tmi13&1&2\\[-0.5pt]
\tmi13&\tmi1&2\\[-0.5pt]
9&\tmi9&9\\[-0.5pt]
9&9&9\end{array}\right)
\end{array}
\end{array}\\
\begin{array}[t]{@{$\!$}c@{$\!$}c@{$\!$}c@{$\!$}c@{$\!$}}\begin{array}[t]{c}\mathfrak{e}_{\fwboxL{2pt}{\r{7}}}\\
\left(\begin{array}{c@{\;\;\;}c@{\;\;\;}c@{\;\;\;}c}\tmi\frac{19}{448}&\tmi7&1&1\\[-0.5pt]
\frac{19}{448}&\tmi7&\tmi1&1\\[-0.5pt]
\frac{1}{56}&8&\tmi24&8\\[-0.5pt]
\tmi\frac{1}{56}&8&24&8\\[-0.5pt]
1&\dzero&\dzero&\dzero\end{array}\right)
\end{array}&
\begin{array}[t]{c}\mathfrak{e}_{\fwboxL{2pt}{\r{8}}}\\
\left(\begin{array}{c@{\;\;\;}c@{\;\;\;}c@{\;\;\;}c@{\;\;\;}c}
\tmi\frac{1}{248}&\tmi\frac{1}{2}&\tmi1&\dzero&1\\[-0.5pt]
\tmi\frac{1}{248}&\frac{1}{2}&\tmi1&\dzero&1\\[-0.5pt]
\frac{1}{248}&1&5&\tmi1&30\\[-0.5pt]
\frac{1}{248}&\tmi1&5&1&30\\[-0.5pt]
1&\dzero&\dzero&\dzero&\dzero\end{array}\right)
\end{array}&
\begin{array}[t]{c}\mathfrak{f}_{\fwboxL{2pt}{\r{4}}}\\
\left(\begin{array}{c@{\;\;\;}c@{\;\;\;}c@{\;\;\;}c@{\;\;\;}c}
\tmi\frac{1}{13}&\tmi1&1&\dzero&\tmi1\\[-0.5pt]
\tmi\frac{1}{13}&1&1&\dzero&\tmi1\\[-0.5pt]
\frac{1}{26}&2&6&\tmi1&1\\[-0.5pt]
\frac{1}{26}&\tmi2&6&1&1\\[-0.5pt]
1&\dzero&\dzero&\dzero&\dzero\end{array}\right)
\end{array}&
\begin{array}[t]{c}\mathfrak{g}_{\fwboxL{2pt}{\r{2}}}\\
\left(\begin{array}{c@{\;\;\;}c@{\;\;\;}c@{\;\;\;}c}\tmi\frac{2}{7}&\dzero&1&1\\[-0.5pt]
\tmi\frac{2}{7}&\dzero&1&\tmi1\\[-0.5pt]
\frac{1}{7}&1&3&\tmi1\\[-0.5pt]
\frac{1}{7}&\tmi1&3&1\\[-0.5pt]
1&\dzero&\dzero&\dzero\end{array}\right)
\end{array}
\end{array}\end{array}$$\vspace{-25pt}\label{expansion_coefficients_ffff}\end{table}

%$$\left(\begin{array}{@{}c@{$\;\;$}c@{$\;\;$}c@{$\;\;$}c@{}}
%\dg&\frac{\dg}{\dF}&\dzero&\dg\\
%\frac{\dg}{\dF}&\dg&\dg&\dzero\\
%\dzero&\dg&\dF^2&\dF\\
%\dg&\dzero&\dF&\dF^2\end{array}\right)\;
%%
%\left(\begin{array}{@{}c@{$\;\;$}c@{$\;\;$}c@{$\;\;$}c@{$\;\;$}c@{}}
%\dg&\frac{\dg}{2}&\dzero&\dg&\tmi\dg\\
%\frac{\dg}{2}&\dg&\dg&\dzero&\tmi\dg\\
%\dzero&\dg&\dF^2&\dF&\dF\\
%\dg&\dzero&\dF&\dF^2&\dF\\
%\tmi\dg&\tmi\dg&\dF&\dF&\dF^2\end{array}\right)
%$$

Notice that not all the colour tensors $\mathbf{Q}^i$ described above are in fact linearly independent. For $\mathfrak{a}_{\r{k}}$, for example, we may construct the `null space' of the coefficients given in \mbox{Table~\ref{expansion_coefficients_ffff}} to discover the familiar \emph{Fierz identities}:
\vspace{-5pt}\eq{
\begin{split}\egTensorsAa
&=\phantom{\frac{1}{2}}\egTensorsAd-\fwbox{24pt}{\frac{1}{\r{k}{+}1}}\egTensorsAc\\[-10pt]
\fwbox{0pt}{\fwboxL{297pt}{(\mathfrak{a}_{\r{k}})}}\\[-10pt]
\egTensorsAb&=\phantom{\frac{1}{2}}\egTensorsAc-\fwbox{24pt}{\frac{1}{\r{k}{+}1}}
\egTensorsAd
\end{split}\label{fierz_a}}
These relations are especially useful in the context of using multi-trace (over the fundamental representation) tensors to encode the colour-dependence of all-adjoint scattering. Upon expanding the colour-dependence of a Feynman diagram into multi-traces, one generically encounters products of traces involving some number of internal `linkages' (including self-linkages). The Fierz identities allow us to expand any such object into multi-trace tensors without such internal linkages via
\vspace{-7pt}\eq{\fwbox{0pt}{\fwboxL{435pt}{(\mathfrak{a}_{\r{k}})}}\fwbox{0pt}{\fierzUnwindA=\fierzUnwindB-\frac{1}{\r{k}{+}1}\fierzUnwindC\,.}\label{fierz_in_terms_of_traces}\vspace{-10pt}}
The above identity shows that any $\r{f}$-graph (in $\mathfrak{a}$-type gauge theory), initially expressed in terms of (linked) products of traces involving via fundamental representation using (\ref{definition_of_antisymmetric_structure_constants}), can be systematically `unwound' into products of traces only involving external edges. Importantly, all `multi-trace' tensors that result from this will be sub-leading in $\r{k}$. Therefore, in the limit of large $\r{k}$, only single-trace colour tensors contribute to $\mathfrak{a}_{\r{k}}$ gauge theory.\\[-10pt]

Such identities also exist for $\mathfrak{b},\mathfrak{c},\mathfrak{d}$-type Lie algebras. In particular, we can detect the linear dependence among the five $\mathbf{Q}^i$ using the explicit decompositions given in \mbox{Table~\ref{expansion_coefficients_ffff}}:
\vspace{-15pt}\eq{
\begin{split}\egTensorsAa
&=\frac{1}{2}\egTensorsAd-\fwbox{24pt}{\frac{1}{2}}\egTensorsAe\\[-10pt]
\fwbox{0pt}{\fwboxL{297pt}{(\mathfrak{b}_{\r{k}},\mathfrak{d}_{\r{k}})}}\\[-10pt]
\egTensorsAb&=\frac{1}{2}\egTensorsAc-\fwbox{24pt}{\frac{1}{2}}
\egTensorsAe\\
\egTensorsAa
&=\frac{1}{2}\egTensorsAd-\fwbox{24pt}{\frac{1}{2}}\egTensorsAe\\[-10pt]
\fwbox{0pt}{\fwboxL{297pt}{(\mathfrak{c}_{\r{k}})}}\\[-10pt]
\egTensorsAb&=\frac{1}{2}\egTensorsAc+\fwbox{24pt}{\frac{1}{2}}
\egTensorsAe\\[-28pt]~
\end{split}\label{fierz_bcd}}
These identities can similarly be used to express any colour tensor arising via the Feynman expansion into the basis of multi-traces (over the fundamental representation) without any internal linkages. In contrast with $\mathfrak{a}_{\r{k}}$ gauge theory, however, there is no longer any hierarchy in orders of the rank $\r{k}$ of the gauge theory.

Interestingly, this does not appear possible in general for any of the exceptional algebras. Indeed, for $\mathfrak{e}_{\r{8}}$ or $\mathfrak{f}_{\r{4}}$ gauge theory, there is \emph{no} linear dependence among the tensors $\mathbf{Q}^i$; as such, there exists no Fierz-like relation. For the simple Lie algebras $\mathfrak{e}_{\r{6}}$, $\mathfrak{e}_{\r{7}}$, and $\mathfrak{g}_{\r{2}}$, there is linear dependence among the colour tensors $\mathbf{Q}^i$, but none of these allow one to systematically eliminate internal linkages among products of traces analogous to (\ref{fierz_in_terms_of_traces}):
\vspace{-00pt}\eq{\begin{split}~\\[-27pt]
\fwbox{0pt}{\fwboxL{438pt}{(\mathfrak{e}_{\r{6}})}}\fwbox{0pt}{\egTensorsAa=\egTensorsAb-\frac{1}{9}\egTensorsAc+\frac{1}{9}\egTensorsAd}\\
\fwbox{0pt}{\fwboxL{438pt}{(\mathfrak{e}_{\r{7}})}}\fwbox{0pt}{\egTensorsAa\hspace{-7pt}=\hspace{-2pt}\phantom{{-}}\hspace{-12pt}\egTensorsAb\hspace{-14pt}-\!\fwbox{14pt}{\frac{1}{24}}\hspace{-2pt}\egTensorsAc\hspace{-14pt}+\!\fwbox{14pt}{\frac{1}{24}}\hspace{-2pt}\egTensorsAd\hspace{-14pt}-\!\fwbox{14pt}{\frac{1}{12}}\hspace{-2pt}\egTensorsAe}\\
\fwbox{0pt}{\fwboxL{438pt}{(\mathfrak{g}_{\r{2}})}}\fwbox{0pt}{\egTensorsAa\hspace{-7pt}=\hspace{-2pt}{-}\hspace{-12pt}\egTensorsAb\hspace{-14pt}+\!\fwbox{14pt}{\frac{1}{3}}\hspace{-2pt}\egTensorsAc\hspace{-14pt}+\!\fwbox{14pt}{\frac{1}{3}}\hspace{-2pt}\egTensorsAd\hspace{-14pt}-\!\fwbox{14pt}{\frac{2}{3}}\hspace{-2pt}\egTensorsAe}\\[-28pt]~
\end{split}\label{fierz_e6e7g2}}
Importantly, none of these relations allow one to convert products of traces involving internal $\mathbf{\r{ad}}$ edges into those without such edges. As such, there is no analogous trace-expansion for these theories.%\footnote{There are low-multiplicity identities allowing one to reduce, for example, $\mathrm{tr}_{\mathbf{\b{F}}}(\g{a}\,\r{b}\,\r{c}\,\g{a},\r{d}\cdots)$ into traces not involving the internal linkages; this allows one to expand $\r{f}$-graph colour tensors for low multiplicity into multi-traces over fundamental representations without any internal linkages. But these systematically fail for larger traces required for higher multiplicity.}

%Interestingly, for the Lie algebras $\mathfrak{e}_{\r{8}}$ and $\mathfrak{f}_{\r{4}}$, no linear relations exist among the five colour tensors $\mathbf{Q}^i$; they may be taken as a basis, but it is not possible to construct any analogous Fierz relation to ensure that multi-trace tensors over fundamental representations span the space of $\r{f}$-graphs (which arise directly in the Feynman rules for gluon scattering, say). 

Of course, in choosing the basis for colour tensors, we are neither forced to choose any particular tree nor the ordering of the representations appearing on the leaves. Consider for example an alternate choice of basis dictated by the following graph of Clebsch-Gordan tensors:
\vspace{-5pt}\eq{\fwboxR{0pt}{\mathcal{C}^{\hspace{1pt}\mathbf{\t{r}}}(\mathbf{\b{F}}\,\mathbf{\b{F}}|\mathbf{\b{F}}\,\mathbf{\b{F}})\;\bigger{\Leftrightarrow}\;}\tikzBox{ffff_basis_alt_defined}{
\arrowTo[hblue]{0,0}{-150}\node[anchor=10,inner sep=2pt] at(in){{\footnotesize$\b{[f]}$}};\arrowFrom[hblue]{0,0}{-30};\node[anchor=170,inner sep=2pt] at(end){{\footnotesize$\b{[f]}$}};\arrowFrom[hteal]{0,0}[1]{90}\node[clebsch]at(in){};\node[anchor=0,inner sep=2pt] at(arrownode){{\footnotesize$\t{[r]}$}};\arrowTo[hblue]{end}{150}\node[anchor=-10,inner sep=2pt] at(in){{\footnotesize$\b{[f]}$}};\arrowFrom[hblue]{end}{30}\node[anchor=190,inner sep=2pt] at(end){{\footnotesize$\b{[f]}$}};\node[clebsch]at(in){};
}\,.
\label{ffff_basis2_defined}\vspace{-5pt}}
Like the basis tensors $\mathcal{B}^{\,\mathbf{\g{q}}_{\g{i}}}$ described in (\ref{ffff_colour_basis_defined}), these tensors are uniquely labelled by the internal representation $\mathbf{\t{r}}_{\t{i}}$ appearing in (\ref{ffff_basis2_defined}) and arising via
\vspace{-0pt}\eq{\mathbf{\b{F}}\!\otimes\!\mathbf{\b{\bar{F}}}\equivL\,\bigoplus_{\t{i}}\mathbf{\t{r}}_{\t{i}}\equivL\mathbf{1}\oplus\mathbf{\r{ad}}\oplus\mathbf{\t{r}}\bigoplus_{\t{i}=4}\t{\mathbf{\t{r}}_{i}}\,.\label{rReps_defined}\vspace{-0pt}}
For simple Lie algebras with self-conjugate fundamental representations---namely, all but $\mathfrak{a}_{\r{k}>1}$ and $\mathfrak{e}_{\r{6}}$---it is obvious that $\mathbf{\g{q}}_{\g{i}}=\mathbf{\t{r}}_{\t{i}}$. For $\mathfrak{a}_{\r{k}}$, no $\mathbf{\t{r}}$ appears in (\ref{rReps_defined}), as $\mathbf{\b{F}}\!\otimes\!\mathbf{\b{\bar{F}}}\!=\!\mathbf{1}\!\oplus\!\mathbf{\r{ad}}$. For $\mathfrak{e}_{\r{6}}$ we define 
\vspace{-2pt}\eq{\fwbox{0pt}{\fwboxL{435pt}{(\mathfrak{e}_{\r{6}})}}\fwbox{0pt}{\dynkLabelK{\mathbf{\b{F}}}{100000}\!\otimes\!\dynkLabelK{\mathbf{\b{\bar{F}}}}{0000010}=\dynkLabelK{\mathbf{1}}{000000}\oplus\dynkLabelK{\mathbf{\r{ad}}}{000001}\oplus\dynkLabelK{\mathbf{\t{r}}}{100010}}\vspace{-4pt}\label{e6_rRep_defined}}
where $\mathbf{\t{r}}$ is the (unique) $\mathbf{650}$-dimensional irreducible representation of $\mathfrak{e}_{\r{6}}$. 

Because $\mathbf{1}$ and $\mathbf{\r{ad}}$ always appear in the decomposition of $\mathbf{\b{F}}\!\otimes\!\mathbf{\b{\bar{F}}}$, the Clebsch-Gordan colour tensor basis always include $\{\mathbf{Q}^1,\mathbf{Q}^3\}$ among these `novel' colour basis tensors. Nevertheless, we can systematically decompose any colour tensor $C(\mathbf{\b{F}}\,\mathbf{\b{F}}|\mathbf{\b{F}}\,\mathbf{\b{F}})$ into this basis. For the `obvious' $\mathbf{Q}^i$ tensors, we can systematically decompose them according to
\vspace{-5pt}\eq{\mathbf{Q}^{\b{i}}\equivL\,\sum_{\t{j}}\tilde{\mathbf{c}}[\mathfrak{\r{g}}]^{\b{i}}_{\phantom{i}\t{j}}\mathcal{C}^{\hspace{1pt}\t{\mathbf{r}_j}}\,.\label{olde_to_new2_ffff_coefficients_defined}\vspace{-10pt}}
These coefficients are given for all simple Lie algebras in \mbox{Table~\ref{expansion_coefficients_ffff2}}.

\begin{table}[b]\vspace{-38pt}$$\begin{array}{c}\begin{array}[t]{@{$\!$}c@{$\!$}c@{$\!$}c@{$\!$}c@{$\!$}c@{$\!$}}\begin{array}[t]{c}\mathfrak{a}_{\fwboxL{2pt}{\r{k}}}\\
\left(\begin{array}{c@{\;\;\;}c}\dzero&1\\[-0.5pt]
\frac{\r{k}(\hspace{-1pt}\r{k}\hspace{-0pt}{+}\hspace{-0pt}2\hspace{-1pt})}{(\hspace{-1pt}\r{k}\hspace{-0pt}{+}\hspace{-0pt}1\hspace{-1pt})^2}&\tmi\frac{1}{\r{k}\hspace{-0pt}{+}\hspace{-0pt}1}\\[-0.5pt]
1&\dzero\\[-0.5pt]
\frac{1}{\r{k}\hspace{-0pt}{+}\hspace{-0pt}1}&1\end{array}\right)
\end{array}&
\begin{array}[t]{c}\mathfrak{b}_{\fwboxL{2pt}{\r{k}}}\\
\left(\begin{array}{c@{\;\;\;}c@{\;\;\;}c}\dzero&1&\dzero\\[-0.5pt]
\frac{\r{k}}{2\r{k}\hspace{-0pt}{+}\hspace{-0pt}1}&\frac{1}{2}&\tmi\frac{1}{2}\\[-0.5pt]
1&\dzero&\dzero\\[-0.5pt]
\frac{1}{2\r{k}\hspace{-0pt}{+}\hspace{-0pt}1}&1&1\\[-0.5pt]
\frac{1}{2\r{k}\hspace{-0pt}{+}\hspace{-0pt}1}&\tmi1&1\end{array}\right)
\end{array}&
\begin{array}[t]{c}\mathfrak{c}_{\fwboxL{2pt}{\r{k}}}\\
\left(\begin{array}{c@{\;\;\;}c@{\;\;\;}c}\dzero&1&\dzero\\[-0.5pt]
\frac{2\r{k}\hspace{-0pt}{+}\hspace{-0pt}1}{4\r{k}}&\tmi\frac{1}{2}&\frac{1}{2}\\[-0.5pt]
1&\dzero&\dzero\\[-0.5pt]
\frac{1}{2\r{k}}&1&1\\[-0.5pt]
\frac{1}{2\r{k}}&\tmi1&1\end{array}\right)
\end{array}&
\begin{array}[t]{c}\mathfrak{d}_{\fwboxL{2pt}{\r{k}}}\\
\left(\begin{array}{c@{\;\;\;}c@{\;\;\;}c}\dzero&1&\dzero\\[-0.5pt]
\frac{2\r{k}\hspace{-0pt}{-}\hspace{-0pt}1}{4\r{k}}&\frac{1}{2}&\tmi\frac{1}{2}\\[-0.5pt]
1&\dzero&\dzero\\[-0.5pt]
\frac{1}{2\r{k}}&1&1\\[-0.5pt]
\frac{1}{2\r{k}}&\tmi1&1\end{array}\right)
\end{array}&
\begin{array}[t]{c}\mathfrak{e}_{\fwboxL{2pt}{\r{6}}}\\
\left(\begin{array}{c@{\;\;\;}c@{\;\;\;}c}\dzero&1&\dzero\\[-0.5pt]
\frac{26}{243}&\frac{8}{9}&\tmi1\\[-0.5pt]
1&\dzero&\dzero\\[-0.5pt]
\frac{1}{27}&1&9\end{array}\right)
\end{array}
\end{array}\\
\begin{array}[t]{@{$\!$}c@{$\!$}c@{$\!$}c@{$\!$}c@{$\!$}}\begin{array}[t]{c}\mathfrak{e}_{\fwboxL{2pt}{\r{7}}}\\
\left(\begin{array}{c@{\;\;\;}c@{\;\;\;}c@{\;\;\;}c}\dzero&1&\dzero&\dzero\\[-0.5pt]
\frac{19}{448}&\frac{7}{8}&1&\tmi1\\[-0.5pt]
1&\dzero&\dzero&\dzero\\[-0.5pt]
\frac{1}{56}&1&24&8\\[-0.5pt]
\frac{1}{56}&\tmi1&24&\tmi8\end{array}\right)
\end{array}&
\begin{array}[t]{c}\mathfrak{e}_{\fwboxL{2pt}{\r{8}}}\\
\left(\begin{array}{c@{\;\;\;}c@{\;\;\;}c@{\;\;\;}c@{\;\;\;}c}\dzero&1&\dzero&\dzero&\dzero\\[-0.5pt]
\frac{1}{248}&\frac{1}{2}&1&\dzero&\tmi1\\[-0.5pt]
1&\dzero&\dzero&\dzero&\dzero\\[-0.5pt]
\frac{1}{248}&1&5&1&30\\[-0.5pt]
\frac{1}{248}&\tmi1&5&\tmi1&30\end{array}\right)
\end{array}&
\begin{array}[t]{c}\mathfrak{f}_{\fwboxL{2pt}{\r{4}}}\\
\left(\begin{array}{c@{\;\;\;}c@{\;\;\;}c@{\;\;\;}c@{\;\;\;}c}\dzero&1&\dzero&\dzero&\dzero\\[-0.5pt]
\frac{1}{13}&\frac{1}{2}&\tmi\frac{1}{2}&\dzero&1\\[-0.5pt]
1&\dzero&\dzero&\dzero&\dzero\\[-0.5pt]
\frac{1}{26}&1&3&1&1\\[-0.5pt]
\frac{1}{26}&\tmi1&3&\tmi1&1\end{array}\right)
\end{array}&
\begin{array}[t]{c}\mathfrak{g}_{\fwboxL{2pt}{\r{2}}}\\
\left(\begin{array}{c@{\;\;\;}c@{\;\;\;}c@{\;\;\;}c}\dzero&1&\dzero&\dzero\\[-0.5pt]
\frac{2}{7}&\dzero&\tmi1&1\\[-0.5pt]
1&\dzero&\dzero&\dzero\\[-0.5pt]
\frac{1}{7}&1&3&1\\[-0.5pt]
\frac{1}{7}&\tmi1&3&\tmi1\end{array}\right)
\end{array}
\end{array}\end{array}$$\vspace{-25pt}\caption{Expansion coefficients $\tilde{\mathbf{c}}$ required to express $\mathbf{Q}^i$ tensors in terms of the basis $\mathcal{C}^{\hspace{1pt}\t{\mathbf{r}_i}}$.}\label{expansion_coefficients_ffff2}\vspace{-12pt}\end{table}

\newpage
\subsubsection{\texorpdfstring{Clebsch Colour Bases for $C(\mathbf{\b{F}}\,\mathbf{\r{ad}}|\mathbf{\r{ad}}\,\mathbf{\b{F}})$}{Clebsch Colour Bases for C(Fg|gF)}}\label{subsection:bases_for_fggf}

Let us now consider the case of scattering two fundamental-charged particles with two adjoints. We'd like to construct a basis for colour tensors of the form $C(\mathbf{\b{F}}\,\mathbf{\r{ad}}|\mathbf{\r{ad}}\,\mathbf{\b{F}})$. Following our prescription, we can construct such a basis according to the tensors
\vspace{-5pt}\eq{\fwboxR{0pt}{\mathcal{B}^{\hspace{1pt}\mathbf{\g{s}}}(\mathbf{\b{F}}\,\mathbf{\r{ad}}|\mathbf{\r{ad}}\,\mathbf{\b{F}})\;\bigger{\Leftrightarrow}\;}\tikzBox{fggf_basis_diagram}{\arrowTo[hblue]{0,0}{-130};\node[anchor=10,inner sep=2pt] at(in){{\footnotesize$\b{[f]}$}};\arrowTo[hred]{0,0}{130}\node[anchor=-10,inner sep=2pt] at(in){{\footnotesize$\r{[\adR]}$}};\arrowFrom[hgreen]{0,0}[1.5]{0}\node[clebsch]at(in){};\node[anchor=90,inner sep=2pt] at(arrownode){{\footnotesize$\g{[s]}$}};\arrowFrom[hred]{end}{50}
\node[anchor=-170,inner sep=2pt] at(end){{\footnotesize$\r{[\adR]}$}};\arrowFrom[hblue]{in}{-50}\node[anchor=170,inner sep=2pt] at(end){{\footnotesize$\b{[f]}$}};\node[clebsch]at(in){};
}\,.\label{fggf_colour_basis_defined}\vspace{-5pt}
}
As before, $m\indices{\mathbf{\b{F}}\,\mathbf{\r{ad}}}{\mathbf{\g{s}}}\!=\!1$ for all irreducible representations $\mathbf{\g{s}}$ appearing in 
\vspace{-6pt}\eq{\mathbf{\b{F}}\!\otimes\mathbf{\r{{ad}}}\equivL\,\bigoplus_{\g{i}}\mathbf{\g{s}}_{\g{i}}\label{sReps_defined}\vspace{-10pt}}
and so every colour tensor is uniquely labelled simply by the irreducible representation $\mathbf{\g{s}}_{\g{i}}$ without any need to identify any multiplicity indices. We define the representations $\mathbf{\g{s}}_{\g{i}}$ that appear for each of the simple Lie algebras in \mbox{Table~\ref{sReps_table}}. For $\mathfrak{e}_{\r{8}}$, we have that $\mathbf{\g{s}}_{\g{i}}\!=\!\mathbf{\g{q}}_{\g{i}}$ given in (\ref{e8_qReps_defined}). As before, the Dynkin labels listed in the table may require modifications for low rank---\emph{e.g.}~for $\mathfrak{a}_{\r{1}}$, `$\mathbf{\g{s}}_{\g{3}}$' in \mbox{Table~\ref{fierz_in_terms_of_traces}} would have dimension 0; this should be understood to mean that, for this case, we only have
\vspace{-2pt}\eq{\fwbox{0pt}{\fwboxL{435pt}{(\mathfrak{a}_{\r{1}})}}\fwbox{0pt}{\dynkLabelK{\mathbf{\b{F}}}{1}\!\otimes\dynkLabelK{\mathbf{\r{ad}}}{2}\equivL\,\dynkLabelK{\mathbf{\g{s}}_{\g{1}}}{1}\oplus\dynkLabelK{\mathbf{\g{s}}_{\g{2}}}{3}\,.}\vspace{-4pt}}

\begin{table}[b]\vspace{-30pt}$$\begin{array}{|l@{$\,$}|ccc|}\multicolumn{1}{c}{~}&\fwbox{40pt}{\g{\mathbf{s}_{1}}}&\fwbox{80pt}{\g{\mathbf{s}_{2}}}&\multicolumn{1}{c}{\fwbox{80pt}{\g{\mathbf{s}_{3}}}}\\\hline\hline
\raisebox{-4pt}{$\mathfrak{a}_{\r{k}}$}&\dynkLabelK{\mathbf{\b{F}}}{10\cdots0}&\dynkLabelK{\frac{1}{2}\r{k}(\hspace{-1pt}\r{k}\pl1\hspace{-1pt})(\hspace{-1pt}\r{k}\pl3\hspace{-1pt})}{20\cdots01}&\dynkLabelK{\frac{1}{2}(\hspace{-1pt}\r{k}\pl1\hspace{-1pt})(\hspace{-1pt}\r{k}\mi1\hspace{-1pt})(\hspace{-1pt}\r{k}\pl2\hspace{-1pt})}{010\cdots01}\\\hline
\raisebox{-4pt}{$\mathfrak{b}_{\r{k}}$}&\dynkLabelK{\mathbf{\b{F}}}{10\cdots0}&\dynkLabelK{\frac{1}{3}\r{k}(\hspace{-1pt}2\r{k}\pl1\hspace{-1pt})(\hspace{-1pt}2\r{k}\mi1\hspace{-1pt})}{0010\cdots0}&\dynkLabelK{\frac{1}{3}(\hspace{-1pt}2\r{k}\pl1\hspace{-1pt})(\hspace{-1pt}2\r{k}\mi1\hspace{-1pt})(\hspace{-1pt}2\r{k}\pl3\hspace{-1pt})}{110\cdots0}\\\hline
\raisebox{-4pt}{$\mathfrak{c}_{\r{k}}$}&\dynkLabelK{\mathbf{\b{F}}}{10\cdots0}&\dynkLabelK{\frac{2}{3}\r{k}(\hspace{-1pt}\r{k}\pl1\hspace{-1pt})(\hspace{-1pt}2\r{k}\pl1\hspace{-1pt})}{30\cdots0}&\dynkLabelK{\frac{8}{3}\r{k}(\hspace{-1pt}\r{k}\pl1\hspace{-1pt})(\hspace{-1pt}\r{k}\mi1\hspace{-1pt})}{110\cdots0}\\\hline
\raisebox{-4pt}{$\mathfrak{d}_{\r{k}}$}&\dynkLabelK{\mathbf{\b{F}}}{10\cdots0}&\dynkLabelK{\frac{2}{3}\r{k}(\hspace{-1pt}\r{k}\mi1\hspace{-1pt})(\hspace{-1pt}2\r{k}\mi1\hspace{-1pt})}{0010\cdots0}&\dynkLabelK{\frac{8}{3}\r{k}(\hspace{-1pt}\r{k}\pl1\hspace{-1pt})(\hspace{-1pt}\r{k}\mi1\hspace{-1pt})}{110\cdots0}\\\hline
\end{array}\;\;
\begin{array}{|l@{$\,$}|ccc|}\multicolumn{1}{c}{~}&\fwbox{30pt}{\g{\mathbf{s}_{1}}}&\fwbox{30pt}{\g{\mathbf{s}_{2}}}&\multicolumn{1}{c}{\fwbox{30pt}{\g{\mathbf{s}_{3}}}}\\\hline\hline
\raisebox{-4pt}{$\mathfrak{e}_{\r{6}}$}&\dynkLabelK{\mathbf{\b{27}}}{100000}&\dynkLabelK{\mathbf{\bar{351}}}{000100}&\dynkLabelK{\mathbf{1728}}{100001}\\\hline
\raisebox{-4pt}{$\mathfrak{e}_{\r{7}}$}&\dynkLabelK{\mathbf{\b{56}}}{1000000}&\dynkLabelK{\mathbf{912}}{0000001}&\dynkLabelK{\mathbf{6480}}{1000010}\\\hline
\raisebox{-4pt}{$\mathfrak{f}_{\r{4}}$}&\dynkLabelK{\mathbf{\b{26}}}{1000}&\dynkLabelK{\mathbf{273}}{0100}&\dynkLabelK{\mathbf{1053}}{1001}\\\hline
\raisebox{-4pt}{$\mathfrak{g}_{\r{2}}$}&\dynkLabelK{\mathbf{\b{7}}}{10}&\dynkLabelK{\mathbf{27}}{20}&\dynkLabelK{\mathbf{64}}{11}\\\hline
\end{array}$$\vspace{-25pt}\caption{Irreducible representations appearing in the decomposition of the tensor product $\mathbf{\b{F}}\!\otimes\!\mathbf{\r{ad}}\equivL\bigoplus_{\g{i}}\mathbf{\g{s}}_{\g{i}}$ for simple Lie algebras $\mathfrak{g}$; the case of $\mathfrak{e}_{\r{8}}$ is discussed separately.}\vspace{-28pt}\label{sReps_table}\end{table}

There are relatively few familiar colour tensors with which we may compare. An obvious set of tensors (which are complete for all simple Lie algebras except $\mathfrak{e}_{\r{8}}$) would be the following:
\eq{\big\{\mathbf{S}^1,\mathbf{S}^2,\mathbf{S}^3\big\}\;\bigger{\Leftrightarrow}\;\Bigg\{
\tikzBox{fggf_reference_tensor_1}{\arrowTo[hblue]{0,0}{-150}\node[anchor=10,inner sep=2pt] at(in){{\footnotesize$\b{[f]}$}};
\arrowTo[hred]{0,0}{150}\node[anchor=-10,inner sep=2pt] at(in){{\footnotesize$\r{[\adR]}$}};
\arrowFrom[hblue]{0,0}{0}\node[hblue,clebschR]at(in){};
\arrowFrom[hblue]{end}{-30}\node[anchor=170,inner sep=2pt] at(end){{\footnotesize$\b{[f]}$}};
\arrowFrom[hred]{in}{30}\node[anchor=-170,inner sep=2pt] at(end){{\footnotesize$\r{[\adR]}$}};\node[hblue,clebschR]at(in){};
},
\tikzBox{fggf_reference_tensor_2}{\coordinate(e1)at($(\edgeLength,0)+(30:\edgeLength)$);\coordinate(e2)at($(\edgeLength,0)+(-30:\edgeLength)$);\coordinate(i1)at($(0,0)+(-150:\edgeLength)$);\coordinate(i2)at($(0,0)+(150:\edgeLength)$);\arrowTo[hblue]{0,0}{-150}\node[anchor=10,inner sep=2pt] at(in){{\footnotesize$\b{[f]}$}};
%\arrowTo[hred]{0,0}{150}
\floatingEdge{hred,edge,endArrow}{(0,0).. controls ($(0.25,0.25)$) and ($(.5,.4)$) .. (e1);}
\floatingEdge{hred,edge,endArrow}{(i2).. controls ($(\edgeLength-0.5,0.4)$) and ($(\edgeLength-0.25,.25)$) .. (\edgeLength,0);}
\node[anchor=-10,inner sep=2pt] at(i2){{\footnotesize$\r{[\adR]}$}};
\arrowFrom[hblue]{0,0}{0}\node[hblue,clebschR]at(in){};
\arrowFrom[hblue]{end}{-30}\node[anchor=170,inner sep=2pt] at(end){{\footnotesize$\b{[f]}$}};
\node[anchor=-170,inner sep=2pt] at(e1){{\footnotesize$\r{[\adR]}$}};\node[hblue,clebschR]at(in){};
},\tikzBox{fggf_reference_tensor_3}{\coordinate(e1)at($(0,0)+(30:\edgeLength)$);\coordinate(e2)at($(0,0)+(-30:\edgeLength)$);\arrowTo[hblue]{e2}[1.25]{180}\node[anchor=170,inner sep=2pt] at(end){{\footnotesize$\b{[f]}$}};\node[anchor=10,inner sep=2pt] at(in){{\footnotesize$\b{[f]}$}};
\arrowTo[hred]{e1}[1.25]{180}\node[anchor=-170,inner sep=2pt] at(end){{\footnotesize$\r{[\adR]}$}};\node[anchor=-10,inner sep=2pt] at(in){{\footnotesize$\r{[\adR]}$}};}
\Bigg\}\,.\label{default_tensors_s_defined}\vspace{-4pt}}
Notice that $\mathbf{S}^1$ is always included among the basis tensors $\mathcal{B}^{\hspace{1pt}\mathbf{\g{s}}_{\g{i}}}$ of (\ref{fggf_colour_basis_defined}) as $\mathbf{\b{F}}\equivL\,\mathbf{\g{s}}_{\g{1}}$ for each simple Lie algebra. For $\mathfrak{e}_{\r{8}}$, we take $\mathbf{\g{s}}_{\g{i}}\equivR\mathbf{\g{q}}_{\g{i}}$ defined in (\ref{e8_qReps_defined}).

As before, we can decompose the more familiar colour tensors $\mathbf{S}^{i}$ into this basis according to
\vspace{-2pt}\eq{\mathbf{S}^{\b{i}}\equivL\,\sum_{\g{j}}{\mathbf{c}}[\mathfrak{\r{g}}]^{\b{i}}_{\phantom{i}\g{j}}\mathcal{B}^{\hspace{1pt}\g{\mathbf{s}_j}}\,.\label{olde_to_new_fggf_coefficients_defined}\vspace{-4pt}}
These coefficients are enumerated in \mbox{Table~\ref{expansion_coefficients_fggf}}. In most cases the coefficients $\mathbf{c}[\r{\mathfrak{g}}]$ are full rank; as such, we may invert them to effectively \emph{define} the new colour tensors in terms of the more familiar ones.

\begin{table}[t]\vspace{-10pt}\caption{Expansion coefficients ${\mathbf{c}}$ required to express $\mathbf{S}^i$ tensors in terms of the basis $\mathcal{B}^{\hspace{1pt}\g{\mathbf{s}_i}}$.}
\vspace{-27.5pt}$$\begin{array}{c}\begin{array}[t]{@{$\!$}c@{$\!$}c@{$\!$}c@{$\!$}c@{$\!$}c@{$\!$}}
\begin{array}[t]{c}\mathfrak{a}_{\fwboxL{2pt}{\r{1}}}\\
\left(\begin{array}{c@{\;\;\;}c}1&\dzero\\[-0.5pt]
\tmi\frac{1}{3}&1\\[-0.5pt]
\frac{2}{3}&1\end{array}\right)
\end{array}&
\begin{array}[t]{c}\mathfrak{a}_{\fwboxL{2pt}{\r{k}\!>\!1}}\\
\left(\begin{array}{c@{\;\;\;}c@{\;\;\;}c}1&\dzero&\dzero\\[-0.5pt]
\tmi\frac{1}{\r{k}(\hspace{-1pt}\r{k}\hspace{-0pt}{+}\hspace{-0pt}2\hspace{-1pt})}&1&1\\[-0.5pt]
\frac{\r{k}\hspace{-0pt}{+}\hspace{-0pt}1}{\r{k}(\hspace{-1pt}\r{k}\hspace{-0pt}{+}\hspace{-0pt}2\hspace{-1pt})}&1&\tmi1\end{array}\right)
\end{array}&
\begin{array}[t]{c}\mathfrak{b}_{\fwboxL{2pt}{\r{k}}}\\
\left(\begin{array}{c@{\;\;\;}c@{\;\;\;}c}1&\dzero&\dzero\\[-0.5pt]
\frac{1}{2\r{k}}&1&1\\[-0.5pt]
\frac{1}{\r{k}}&\tmi1&2\end{array}\right)
\end{array}&
\begin{array}[t]{c}\mathfrak{c}_{\fwboxL{2pt}{\r{k}}}\\
\left(\begin{array}{c@{\;\;\;}c@{\;\;\;}c}1&\dzero&\dzero\\[-0.5pt]
\tmi\frac{1}{2\r{k}\hspace{-0pt}{+}\hspace{-0pt}1}&1&1\\[-0.5pt]
\frac{2}{2\r{k}\hspace{-0pt}{+}\hspace{-0pt}1}&1&\tmi2\end{array}\right)
\end{array}&
\begin{array}[t]{c}\mathfrak{d}_{\fwboxL{2pt}{\r{k}}}\\
\left(\begin{array}{c@{\;\;\;}c@{\;\;\;}c}1&\dzero&\dzero\\[-0.5pt]
\frac{1}{2\r{k}\hspace{-0pt}{-}\hspace{-0pt}1}&1&1\\[-0.5pt]
\frac{2}{2\r{k}\hspace{-0pt}{-}\hspace{-0pt}1}&\tmi1&2\end{array}\right)
\end{array}
\end{array}\\
\begin{array}[t]{@{$\!$}c@{$\!$}c@{$\!$}c@{$\!$}c@{$\!$}c@{$\!$}}
\begin{array}[t]{c}\mathfrak{e}_{\fwboxL{2pt}{\r{6}}}\\
\left(\begin{array}{c@{\;\;\;}c@{\;\;\;}c}1&\dzero&\dzero\\[-0.5pt]
\frac{4}{13}&1&1\\[-0.5pt]
\frac{9}{26}&\tmi\frac{3}{2}&6\end{array}\right)
\end{array}&
\begin{array}[t]{c}\mathfrak{e}_{\fwboxL{2pt}{\r{7}}}\\
\left(\begin{array}{c@{\;\;\;}c@{\;\;\;}c}1&\dzero&\dzero\\[-0.5pt]
\frac{7}{19}&1&1\\[-0.5pt]
\frac{8}{19}&\tmi2&12\end{array}\right)
\end{array}&
\begin{array}[t]{c}\mathfrak{e}_{\fwboxL{2pt}{\r{8}}}\\
\left(\begin{array}{c@{\;\;\;}c@{\;\;\;}c@{\;\;\;}c@{\;\;\;}c}
\dzero&1&\dzero&\dzero&\dzero\\[-0.5pt]
\tmi\frac{1}{248}&\frac{1}{2}&\tmi1&\dzero&1\\[-0.5pt]
\frac{1}{248}&1&5&1&30\end{array}\right)
\end{array}&
\begin{array}[t]{c}\mathfrak{f}_{\fwboxL{2pt}{\r{4}}}\\
\left(\begin{array}{c@{\;\;\;}c@{\;\;\;}c}1&\dzero&\dzero\\[-0.5pt]
\frac{1}{4}&1&1\\[-0.5pt]
\frac{1}{2}&\tmi2&6\end{array}\right)
\end{array}&
\begin{array}[t]{c}\mathfrak{g}_{\fwboxL{2pt}{\r{2}}}\\
\left(\begin{array}{c@{\;\;\;}c@{\;\;\;}c}1&\dzero&\dzero\\[-0.5pt]
\dzero&1&1\\[-0.5pt]
\frac{1}{2}&\tmi\frac{3}{2}&2\end{array}\right)
\end{array}
\end{array}\end{array}$$\vspace{-20pt}\label{expansion_coefficients_fggf}\vspace{-0pt}\end{table}

Because there are only two independent tensors for $\mathfrak{a}_\r{1}$, we may deduce the linear relation satisfied by the three colour tensors $\mathbf{S}^i$:
\eq{\fwbox{0pt}{\fwboxL{435pt}{(\mathfrak{a}_{\r{1}})}}\fwbox{0pt}{\tikzBox{fggf_tensors_a1_1}{\arrowTo[hblue]{0,0}{-150}\node[anchor=10,inner sep=2pt] at(in){{\footnotesize$\b{[f]}$}};
\arrowTo[hred]{0,0}{150}\node[anchor=-10,inner sep=2pt] at(in){{\footnotesize$\r{[\adR]}$}};
\arrowFrom[hblue]{0,0}{0}\node[hblue,clebschR]at(in){};
\arrowFrom[hblue]{end}{-30}\node[anchor=170,inner sep=2pt] at(end){{\footnotesize$\b{[f]}$}};
\arrowFrom[hred]{in}{30}\node[anchor=-170,inner sep=2pt] at(end){{\footnotesize$\r{[\adR]}$}};\node[hblue,clebschR]at(in){};
}+
\tikzBox{fggf_tensors_a1_2}{\coordinate(e1)at($(\edgeLength,0)+(30:\edgeLength)$);\coordinate(e2)at($(\edgeLength,0)+(-30:\edgeLength)$);\coordinate(i1)at($(0,0)+(-150:\edgeLength)$);\coordinate(i2)at($(0,0)+(150:\edgeLength)$);\arrowTo[hblue]{0,0}{-150}\node[anchor=10,inner sep=2pt] at(in){{\footnotesize$\b{[f]}$}};
\floatingEdge{hred,edge,endArrow}{(0,0).. controls ($(0.25,0.25)$) and ($(.5,.4)$) .. (e1);}
\floatingEdge{hred,edge,endArrow}{(i2).. controls ($(\edgeLength-0.5,0.4)$) and ($(\edgeLength-0.25,.25)$) .. (\edgeLength,0);}
\node[anchor=-10,inner sep=2pt] at(i2){{\footnotesize$\r{[\adR]}$}};
\arrowFrom[hblue]{0,0}{0}\node[hblue,clebschR]at(in){};
\arrowFrom[hblue]{end}{-30}\node[anchor=170,inner sep=2pt] at(end){{\footnotesize$\b{[f]}$}};
\node[anchor=-170,inner sep=2pt] at(e1){{\footnotesize$\r{[\adR]}$}};\node[hblue,clebschR]at(in){};
}=\tikzBox{fggf_tensors_a1_3}{\coordinate(e1)at($(0,0)+(30:\edgeLength)$);\coordinate(e2)at($(0,0)+(-30:\edgeLength)$);\arrowTo[hblue]{e2}[1.25]{180}\node[anchor=170,inner sep=2pt] at(end){{\footnotesize$\b{[f]}$}};\node[anchor=10,inner sep=2pt] at(in){{\footnotesize$\b{[f]}$}};
\arrowTo[hred]{e1}[1.25]{180}\node[anchor=-170,inner sep=2pt] at(end){{\footnotesize$\r{[\adR]}$}};\node[anchor=-10,inner sep=2pt] at(in){{\footnotesize$\r{[\adR]}$}};}}}
which reflects the familiar fact that the fundamental representation of $\mathfrak{a}_{\r{1}}$ \emph{also} may be used to define a Clifford algebra (satisfying \emph{anti}commutation relations).

As we did before, it is worth considering an alternative choice for the arrangement of the representations to define our basis tensors. In particular, we could have chosen 
\vspace{-5pt}\eq{\fwboxR{0pt}{\mathcal{C}_{\hspace{0pt}\mu}^{\hspace{1pt}\mathbf{\t{r}}}(\mathbf{\b{F}}\,\mathbf{\r{ad}}|\mathbf{\r{ad}}\,\mathbf{\b{F}})\;\bigger{\Leftrightarrow}\;}\tikzBox{fggf_alt_basis_diagram}{
\arrowTo[hblue]{0,0}{-150}\node[anchor=10,inner sep=2pt] at(in){{\footnotesize$\b{[f]}$}};\arrowFrom[hblue]{0,0}{-30};\node[anchor=170,inner sep=2pt] at(end){{\footnotesize$\b{[f]}$}};\arrowFrom[hteal]{0,0}[1]{90}\node[clebsch]at(in){};\node[anchor=0,inner sep=2pt] at(arrownode){{\footnotesize$\t{[r]}$}};\arrowTo[hred]{end}{150}\node[anchor=-10,inner sep=2pt] at(in){{\footnotesize$\r{[\adR]}$}};\arrowFrom[hred]{end}{30}\node[anchor=190,inner sep=2pt] at(end){{\footnotesize$\r{[\adR]}$}};\node[clebsch]at(in){{\scriptsize$\phantom{\nu}$}};\node[]at(in){{\scriptsize${\mu}$}};
}\,.
\label{fggf_basis2_defined}\vspace{-5pt}}
Here, the representations $\mathbf{\t{r}}_{\t{i}}$ are among the same ones defined above, but must \emph{also} appear in the decomposition of $\mathbf{\r{ad}}^{\otimes2}$ into irreducible representations
\vspace{-6pt}\eq{\mathbf{\r{ad}}\!\otimes\!\mathbf{\r{{ad}}}\equivL\,\bigoplus_{\b{i}}\mathbf{\b{t}}_{\b{i}}\,.\label{tReps_defined}\vspace{-7pt}}
We enumerate these representations in the following subsection (see \mbox{Table~\ref{tReps_table}}), but suffice it to say that only $\mathbf{1}$, $\mathbf{\r{ad}}$, and $\mathbf{\t{r}}$ (if this exists) defined in (\ref{rReps_defined}) appear also in (\ref{tReps_defined}) (with the exception of $\mathfrak{e}_{\r{8}}$, of course). 

For the $\mathfrak{a}_{\r{k}>1}$-type Lie algebras, for which only the representations $\mathbf{\g{r}}\!\in\!\{\mathbf{1},\mathbf{\r{ad}}\}$ appear in (\ref{fggf_basis2_defined}), it is crucial to note that the second vertex requires a multiplicity index, as $m\indices{\r{\mathbf{ad\,ad}}}{\r{\mathbf{ad}}}\!=\!2$---reflecting the fact that a non-vanishing $\r{d}$ tensor exists (see (\ref{d_tensor_defined})). Thus, the basis consists of the three tensors
\vspace{-4pt}\eq{\fwbox{0pt}{\fwboxL{435pt}{(\mathfrak{a}_{\r{k}>1})}}\fwbox{0pt}{\fwboxR{0pt}{\big\{\mathcal{C}_{\hspace{0pt}\mu}^{\hspace{1pt}\mathbf{\t{r}}}\big\}\;\bigger{\Leftrightarrow}\;}\left\{
\tikzBox{fggf_a_series_alt_basis_1}{
\coordinate(eph)at($(0,1.0*\edgeLength)$);\arrowTo[hblue]{0,0}{-150}\node[anchor=10,inner sep=2pt] at(in){{\footnotesize$\b{[f]}$}};\arrowFrom[hblue]{0,0}{-30};\node[anchor=170,inner sep=2pt] at(end){{\footnotesize$\b{[f]}$}};\draw[dashed,edge](in)--(eph);\node[clebsch]at(in){};\arrowTo[hred]{eph}{150}\node[anchor=-10,inner sep=2pt] at(in){{\footnotesize$\r{[\adR]}$}};\arrowFrom[hred]{end}{30}\node[anchor=190,inner sep=2pt] at(end){{\footnotesize$\r{[\adR]}$}};\node[clebsch]at(in){};
},
\tikzBox{fggf_a_series_alt_basis_2}{
\coordinate(eph)at($(0,1.0*\edgeLength)$);\arrowTo[hblue]{0,0}{-150}\node[anchor=10,inner sep=2pt] at(in){{\footnotesize$\b{[f]}$}};\arrowFrom[hblue]{0,0}{-30};\node[anchor=170,inner sep=2pt] at(end){{\footnotesize$\b{[f]}$}};\arrowFrom[hred]{in}{90}\node[anchor=0,inner sep=2pt] at(arrownode){{\footnotesize$\r{[\adR]}$}};\node[hblue,clebschR]at(in){};\arrowTo[hred]{end}{150}\node[anchor=-10,inner sep=2pt] at(in){{\footnotesize$\r{[\adR]}$}};\arrowFrom[hred]{end}{30}\node[anchor=190,inner sep=2pt] at(end){{\footnotesize$\r{[\adR]}$}};\node[hred,clebschR]at(in){};
},\tikzBox{fggf_a_series_alt_basis_3}{
\coordinate(eph)at($(0,1.0*\edgeLength)$);\arrowTo[hblue]{0,0}{-150}\node[anchor=10,inner sep=2pt] at(in){{\footnotesize$\b{[f]}$}};\arrowFrom[hblue]{0,0}{-30};\node[anchor=170,inner sep=2pt] at(end){{\footnotesize$\b{[f]}$}};\arrowFrom[hred]{in}{90}\node[anchor=0,inner sep=2pt] at(arrownode){{\footnotesize$\r{[\adR]}$}};\node[hblue,clebschR]at(in){};\arrowTo[hred]{end}{150}\node[anchor=-10,inner sep=2pt] at(in){{\footnotesize$\r{[\adR]}$}};\arrowFrom[hred]{end}{30}\node[anchor=190,inner sep=2pt] at(end){{\footnotesize$\r{[\adR]}$}};\node[hred,clebschD]at(in){};
}
\right\}\,.}\vspace{-4pt}}
For $\mathfrak{a}_{\r{1}}$, the third tensor above vanishes, leaving us with a basis spanned by the first two. For all other algebras (besides $\mathfrak{e}_{\r{8}}$,) we have 
\eq{\fwbox{0pt}{\fwboxL{435pt}{(\hspace{-1.25pt}\not{\hspace{-2pt}\mathfrak{a}_{\r{k}})}}}\fwbox{0pt}{\fwboxR{0pt}{\big\{\mathcal{C}_{\hspace{0pt}\mu}^{\hspace{1pt}\mathbf{\t{r}}}\big\}\;\bigger{\Leftrightarrow}\;}\left\{
\tikzBox{fggf_a_series_alt_basis_1}{
\coordinate(eph)at($(0,1.0*\edgeLength)$);\arrowTo[hblue]{0,0}{-150}\node[anchor=10,inner sep=2pt] at(in){{\footnotesize$\b{[f]}$}};\arrowFrom[hblue]{0,0}{-30};\node[anchor=170,inner sep=2pt] at(end){{\footnotesize$\b{[f]}$}};\draw[dashed,edge](in)--(eph);\node[clebsch]at(in){};\arrowTo[hred]{eph}{150}\node[anchor=-10,inner sep=2pt] at(in){{\footnotesize$\r{[\adR]}$}};\arrowFrom[hred]{end}{30}\node[anchor=190,inner sep=2pt] at(end){{\footnotesize$\r{[\adR]}$}};\node[clebsch]at(in){};
},
\tikzBox{fggf_a_series_alt_basis_2}{
\coordinate(eph)at($(0,1.0*\edgeLength)$);\arrowTo[hblue]{0,0}{-150}\node[anchor=10,inner sep=2pt] at(in){{\footnotesize$\b{[f]}$}};\arrowFrom[hblue]{0,0}{-30};\node[anchor=170,inner sep=2pt] at(end){{\footnotesize$\b{[f]}$}};\arrowFrom[hred]{in}{90}\node[anchor=0,inner sep=2pt] at(arrownode){{\footnotesize$\r{[\adR]}$}};\node[hblue,clebschR]at(in){};\arrowTo[hred]{end}{150}\node[anchor=-10,inner sep=2pt] at(in){{\footnotesize$\r{[\adR]}$}};\arrowFrom[hred]{end}{30}\node[anchor=190,inner sep=2pt] at(end){{\footnotesize$\r{[\adR]}$}};\node[hred,clebschR]at(in){};
},
\tikzBox{fggf_nota_alt_basis_3}{\coordinate(eph)at($(0,1.0*\edgeLength)$);\arrowTo[hblue]{0,0}{-150}\node[anchor=10,inner sep=2pt] at(in){{\footnotesize$\b{[f]}$}};\arrowFrom[hblue]{0,0}{-30};\node[anchor=170,inner sep=2pt] at(end){{\footnotesize$\b{[f]}$}};\arrowFrom[hteal]{in}{90}\node[anchor=0,inner sep=2pt] at(arrownode){{\footnotesize$\t{[r]}$}};\node[hblue,clebschR]at(in){};\arrowTo[hred]{end}{150}\node[anchor=-10,inner sep=2pt] at(in){{\footnotesize$\r{[\adR]}$}};\arrowFrom[hred]{end}{30}\node[anchor=190,inner sep=2pt] at(end){{\footnotesize$\r{[\adR]}$}};\node[clebsch]at(in){};
}
\right\}}}
where $\mathbf{\t{r}}$ is defined in (\ref{rReps_defined}) and is identified with $\mathbf{\g{q}}_{\g{3}}$ in \mbox{Table~\ref{qReps_table}} and in (\ref{e6_rRep_defined}) for $\mathfrak{e}_{\r{6}}$. 

As before, we can determine the coefficients of the tensors $\mathbf{S}^i$ in this basis:
\vspace{-3pt}\eq{\mathbf{S}^{\b{i}}\equivL\,\sum_{\t{j}}\tilde{\mathbf{c}}[\mathfrak{\r{g}}]^{\b{i}}_{\phantom{i}\t{j}}\mathcal{C}^{\hspace{1pt}\t{j}}\,.\label{olde_to_new2_fggf_coefficients_defined}\vspace{-5pt}}
We have computed these explicitly for all simple Lie algebras, and have enumerated the results in \mbox{Table~\ref{expansion_coefficients_fggf2}}.

\begin{table}[t]\vspace{-10pt}\caption{Expansion coefficients $\tilde{\mathbf{c}}$ required to express $\mathbf{S}^i$ tensors in terms of the basis $\mathcal{C}^{\hspace{1pt}\t{\mathbf{r}_i}}$.}
\vspace{-25pt}$$\begin{array}{c}\begin{array}[t]{@{$\!$}c@{$\!$}c@{$\!$}c@{$\!$}c@{$\!$}c@{$\!$}}
\begin{array}[t]{c}\mathfrak{a}_{\fwboxL{2pt}{\r{1}}}\\
\left(\begin{array}{c@{\;\;\;}c}\frac{1}{2}&\frac{1}{2}\\[-0.5pt]
\frac{1}{2}&\tmi\frac{1}{2}\\[-0.5pt]
1&\dzero\end{array}\right)
\end{array}&
\begin{array}[t]{c}\mathfrak{a}_{\fwboxL{2pt}{\r{k}\!>\!1}}\\
\left(\begin{array}{c@{\;\;\;}c@{\;\;\;}c}\frac{1}{\r{k}\hspace{-0pt}{+}\hspace{-0pt}1}&\frac{1}{2}&\frac{1}{2}\\[-0.5pt]
\frac{1}{\r{k}\hspace{-0pt}{+}\hspace{-0pt}1}&\tmi\frac{1}{2}&\frac{1}{2}\\[-0.5pt]
1&\dzero&\dzero\end{array}\right)
\end{array}&
\begin{array}[t]{c}\mathfrak{b}_{\fwboxL{2pt}{\r{k}}}\\
\left(\begin{array}{c@{\;\;\;}c@{\;\;\;}c}\frac{1}{2\r{k}\hspace{-0pt}{+}\hspace{-0pt}1}&\frac{1}{2}&\frac{1}{2}\\[-0.5pt]
\frac{1}{2\r{k}\hspace{-0pt}{+}\hspace{-0pt}1}&\tmi\frac{1}{2}&\frac{1}{2}\\[-0.5pt]
1&\dzero&\dzero\end{array}\right)
\end{array}&
\begin{array}[t]{c}\mathfrak{c}_{\fwboxL{2pt}{\r{k}}}\\
\left(\begin{array}{c@{\;\;\;}c@{\;\;\;}c}\frac{1}{2\r{k}}&\frac{1}{2}&\frac{1}{2}\\[-0.5pt]
\frac{1}{2\r{k}}&\tmi\frac{1}{2}&\frac{1}{2}\\[-0.5pt]
1&\dzero&\dzero\end{array}\right)
\end{array}&
\begin{array}[t]{c}\mathfrak{d}_{\fwboxL{2pt}{\r{k}}}\\
\left(\begin{array}{c@{\;\;\;}c@{\;\;\;}c}\frac{1}{2\r{k}}&\frac{1}{2}&\frac{1}{2}\\[-0.5pt]
\frac{1}{2\r{k}}&\tmi\frac{1}{2}&\frac{1}{2}\\[-0.5pt]
1&\dzero&\dzero\end{array}\right)
\end{array}
\end{array}\\
\begin{array}[t]{@{$\!$}c@{$\!$}c@{$\!$}c@{$\!$}c@{$\!$}c@{$\!$}}
\begin{array}[t]{c}\mathfrak{e}_{\fwboxL{2pt}{\r{6}}}\\
\left(\begin{array}{c@{\;\;\;}c@{\;\;\;}c}\frac{1}{27}&\frac{1}{2}&\frac{1}{2}\\[-0.5pt]
\frac{1}{27}&\tmi\frac{1}{2}&\frac{1}{2}\\[-0.5pt]
1&\dzero&\dzero\end{array}\right)
\end{array}&
\begin{array}[t]{c}\mathfrak{e}_{\fwboxL{2pt}{\r{7}}}\\
\left(\begin{array}{c@{\;\;\;}c@{\;\;\;}c}\frac{1}{56}&\frac{1}{2}&\frac{1}{2}\\[-0.5pt]
\frac{1}{56}&\tmi\frac{1}{2}&\frac{1}{2}\\[-0.5pt]
1&\dzero&\dzero\end{array}\right)
\end{array}&
\begin{array}[t]{c}\mathfrak{e}_{\fwboxL{2pt}{\r{8}}}\\
\left(\begin{array}{c@{\;\;\;}c@{\;\;\;}c@{\;\;\;}c@{\;\;\;}c}
\frac{1}{248}&\tmi\frac{1}{2}&1&\dzero&\tmi1\\[-0.5pt]
\frac{1}{248}&\frac{1}{2}&1&\dzero&\tmi1\\[-0.5pt]
1&\dzero&\dzero&\dzero&\dzero\end{array}\right)
\end{array}&
\begin{array}[t]{c}\mathfrak{f}_{\fwboxL{2pt}{\r{4}}}\\
\left(\begin{array}{c@{\;\;\;}c@{\;\;\;}c}\frac{1}{26}&\frac{1}{2}&\frac{1}{2}\\[-0.5pt]
\frac{1}{26}&\tmi\frac{1}{2}&\frac{1}{2}\\[-0.5pt]
1&\dzero&\dzero\end{array}\right)
\end{array}&
\begin{array}[t]{c}\mathfrak{g}_{\fwboxL{2pt}{\r{2}}}\\
\left(\begin{array}{c@{\;\;\;}c@{\;\;\;}c}\frac{1}{7}&\frac{1}{2}&\frac{1}{2}\\[-0.5pt]
\frac{1}{7}&\tmi\frac{1}{2}&\frac{1}{2}\\[-0.5pt]
1&\dzero&\dzero\end{array}\right)
\end{array}
\end{array}\end{array}$$\vspace{-20pt}\label{expansion_coefficients_fggf2}\vspace{-0pt}\end{table}

%\newpage
\subsubsection{\texorpdfstring{Clebsch Colour Bases for $C(\mathbf{\r{ad}}\,\mathbf{\r{ad}}|\mathbf{\r{ad}}\,\mathbf{\r{ad}})$}{Clebsch Colour Bases for C(gg|gg)}}

In the case of four adjoint-coloured particles, the basis can be expressed
\vspace{-5pt}\eq{\fwboxR{0pt}{\mathcal{B}_{\mu\nu}^{\hspace{4pt}\mathbf{\b{t}}}(\mathbf{\r{ad}}\,\mathbf{\r{ad}}|\mathbf{\r{ad}}\,\mathbf{\r{ad}})\;\bigger{\Leftrightarrow}\;}\tikzBox{gggg_basis_diagram}{\arrowTo[hred]{0,0}{-130};\node[anchor=10,inner sep=2pt] at(in){{\footnotesize$\r{[\adR]}$}};\arrowTo[hred]{0,0}{130}\node[anchor=-10,inner sep=2pt] at(in){{\footnotesize$\r{[\adR]}$}};\arrowFrom[hblue]{0,0}[1.5]{0}\node[clebsch]at(in){{\scriptsize$\phantom{\nu}$}};\node[]at(in){{\scriptsize$\mu$}};\node[anchor=90,inner sep=2pt] at(arrownode){{\footnotesize$\b{[t]}$}};\arrowFrom[hred]{end}{50}
\node[anchor=-170,inner sep=2pt] at(end){{\footnotesize$\r{[\adR]}$}};\arrowFrom[hred]{in}{-50}\node[anchor=170,inner sep=2pt] at(end){{\footnotesize$\r{[\adR]}$}};\node[clebsch]at(in){{\scriptsize$\phantom{\nu}$}};\node[]at(in){\scriptsize${\nu}$};
}\,.\label{gggg_colour_basis_defined}\vspace{-5pt}
}
These tensors are uniquely labelled by the irreducible representations $\mathbf{\b{t}}_{\b{i}}$ in 
\vspace{-4pt}\eq{\mathbf{\r{ad}}\!\otimes\!\mathbf{\r{{ad}}}\equivL\,\bigoplus_{\b{i}}\mathbf{\b{t}}_{\b{i}}\equivL\,\mathbf{1}\oplus\mathbf{\r{ad}}^{m}\bigoplus_{\b{j}=3}\mathbf{\b{t}}_{\b{j}}\,,\label{tReps_defined2}\vspace{-5pt}}
where $m{=}2$ for $\mathfrak{a}_{\r{k}>1}$ and $m{=}1$ for all other simple Lie algebras. We have enumerated the representations $\mathbf{\b{t}}_{\b{i}}$ appearing in (\ref{tReps_defined2}) in \mbox{Table~\ref{tReps_table}}. As in similar examples above, there is some ambiguity regarding these representations' Dynkin labels in the cases of low rank; in such cases, the dimension formulae should help to disambiguate each individual irreducible representation appearing in (\ref{tReps_defined2}).

\begin{table}[t]\vspace{-10pt}\caption{Irreducible representations $\mathbf{\b{t}}_{\b{i}}$ appearing in $\mathbf{\r{ad}}\!\otimes\!\mathbf{\r{ad}}\equivL\bigoplus_{\b{i}}\mathbf{\b{t}}_{\b{i}}$ for simple algebras.}\label{tReps_table}\vspace{-24pt}
\vspace{-04pt}$$\hspace{-10pt}\begin{array}{|c@{$\,$}|ccccc@{$\hspace{-10pt}$}c|}\multicolumn{1}{c}{~}&\fwbox{30pt}{{\mathbf{\b{t}}_{\b{1}}}}&\fwbox{30pt}{{\mathbf{\b{t}}_{\b{2}}}}&\fwbox{40pt}{{\mathbf{\b{t}}_{\b{3}}}}&\fwbox{70pt}{{\mathbf{\b{t}}_{\b{4}}}}&\multicolumn{1}{c}{\fwbox{60pt}{{\mathbf{\b{t}}_{\b{5}}}}}&\multicolumn{1}{c}{\fwbox{60pt}{{\mathbf{\b{t}}_{\b{6}}}}}\\\hline\hline
\raisebox{-4pt}{$\underset{\r{k}>2}{\mathfrak{a}_{\r{k}}}$}&
\dynkLabel{\mathbf{1}}{0\cdots0}&\dynkLabel{\mathbf{\r{ad}}}{10\cdots01}&
\dynkLabel{\frac{1}{4}\r{k}(\hspace{-1pt}\r{k}\pl1\hspace{-1pt})^2(\hspace{-1pt}\r{k}\pl4\hspace{-1pt})}{20\cdots02}&
\dynkLabel{\frac{1}{4}\r{k}(\hspace{-1pt}\r{k}\mi1\hspace{-1pt})(\hspace{-1pt}\r{k}\pl2\hspace{-1pt})(\hspace{-1pt}\r{k}\pl3\hspace{-1pt})}{20\cdots\cdots010}&
\dynkLabel{\frac{1}{4}\r{k}(\hspace{-1pt}\r{k}\mi1\hspace{-1pt})(\hspace{-1pt}\r{k}\pl2\hspace{-1pt})(\hspace{-1pt}\r{k}\pl3\hspace{-1pt})}{010\cdots02}&
\dynkLabel{\frac{1}{4}(\hspace{-1pt}\r{k}\pl1\hspace{-1pt})^2(\hspace{-1pt}\r{k}^2\!\mi4\hspace{-1pt})}{010\cdots010}
\\\hline
\raisebox{-4pt}{$\mathfrak{b}_{\r{k}}$}&\dynkLabel{\mathbf{1}}{0\cdots0}&\dynkLabel{\mathbf{\r{ad}}}{010\cdots0}&
\dynkLabel{\r{k}(\hspace{-1pt}2\r{k}\pl3\hspace{-1pt})}{20\cdots0}&
\dynkLabel{\frac{1}{2}\r{k}(\hspace{-1pt}\r{k}\mi1\hspace{-1pt})(\hspace{-1pt}2\r{k}\pl1\hspace{-1pt})(\hspace{-1pt}2\r{k}\pl3\hspace{-1pt})}{1010\cdots0}&
\dynkLabel{\frac{1}{3}(\hspace{-1pt}\r{k}^2\!\mi1\hspace{-1pt})(\hspace{-1pt}2\r{k}\pl1\hspace{-1pt})(\hspace{-1pt}2\r{k}\pl3\hspace{-1pt})}{020\cdots0}&
\dynkLabel{\frac{1}{6}\r{k}(\hspace{-1pt}\r{k}\mi1\hspace{-1pt})(\hspace{-1pt}4\r{k}^2\!\mi1\hspace{-1pt})}{00010\cdots0}
\\\hline
\raisebox{-4pt}{$\mathfrak{c}_{\r{k}}$}&\dynkLabel{\mathbf{1}}{0\cdots0}&\dynkLabel{\mathbf{\r{ad}}}{20\cdots0}&
\dynkLabel{(\hspace{-1pt}\r{k}\mi1\hspace{-1pt})(\hspace{-1pt}2\r{k}\pl1\hspace{-1pt})}{010\cdots0}&
\dynkLabel{\frac{1}{2}\r{k}(\hspace{-1pt}\r{k}\mi1\hspace{-1pt})(\hspace{-1pt}2\r{k}\pl1\hspace{-1pt})(\hspace{-1pt}2\r{k}\pl3\hspace{-1pt})}{210\cdots0}&
\dynkLabel{\frac{1}{3}\r{k}(\hspace{-1pt}4\r{k}^3\!\mi7\r{k}\pl3\hspace{-1pt})}{020\cdots0}&
\dynkLabel{\frac{1}{6}\r{k}(\hspace{-1pt}\r{k}\pl1\hspace{-1pt})(\hspace{-1pt}2\r{k}\pl1\hspace{-1pt})(\hspace{-1pt}2\r{k}\pl3\hspace{-1pt})}{40\cdots0}
\\\hline
\raisebox{-4pt}{$\underset{\r{k}>4}{\mathfrak{d}_{\r{k}}}$}&\dynkLabel{\mathbf{1}}{0\cdots0}&\dynkLabel{\mathbf{\r{ad}}}{010\cdots0}&\dynkLabel{(\hspace{-1pt}2\r{k}^2\!\pl\r{k}\mi1\hspace{-1pt})}{20\cdots0}&
\dynkLabel{\frac{1}{2}\r{k}(\hspace{-1pt}\r{k}\pl1\hspace{-1pt})(\hspace{-1pt}2\r{k}\mi1\hspace{-1pt})(\hspace{-1pt}2\r{k}\mi3\hspace{-1pt})}{1010\cdots0}&
\dynkLabel{\frac{1}{3}\r{k}(\hspace{-1pt}4\r{k}^3\!\mi7\r{k}\mi3\hspace{-1pt})}{020\cdots0}&
\dynkLabel{\frac{1}{6}\r{k}(\hspace{-1pt}\r{k}\mi1\hspace{-1pt})(\hspace{-1pt}2\r{k}\mi1\hspace{-1pt})(\hspace{-1pt}2\r{k}\mi3\hspace{-1pt})}{00010\cdots0}
\\\hline
\raisebox{-4pt}{$\mathfrak{e}_{\r{6}}$}&\dynkLabel{\mathbf{1}}{000000}&\dynkLabel{\mathbf{\r{ad}}}{000001}&
\dynkLabel{\mathbf{\b{650}}}{100010}&
\dynkLabel{\mathbf{\b{2925}}}{001000}&
\dynkLabel{\mathbf{\b{2430}}}{000002}&
\\\hline
\raisebox{-4pt}{$\mathfrak{e}_{\r{7}}$}&\dynkLabel{\mathbf{1}}{0\cdots0}&\dynkLabel{\mathbf{\r{ad}}}{0000010}&
\dynkLabel{\mathbf{\b{1539}}}{0100000}&
\dynkLabel{\mathbf{\b{8654}}}{0000100}&
\dynkLabel{\mathbf{\b{7371}}}{0000020}&
\\\hline
\raisebox{-4pt}{$\mathfrak{e}_{\r{8}}$}&\dynkLabel{\mathbf{1}}{0\cdots0}&\dynkLabel{\mathbf{\r{ad}}}{10000000}&
\dynkLabel{\mathbf{\b{3875}}}{00000010}&
\dynkLabel{\mathbf{\b{30380}}}{01000000}&
\dynkLabel{\mathbf{\b{27000}}}{20000000}&
\\\hline
\raisebox{-4pt}{$\mathfrak{f}_{\r{4}}$}&\dynkLabel{\mathbf{1}}{0000}&\dynkLabel{\mathbf{\r{ad}}}{0001}&
\dynkLabel{\mathbf{\b{324}}}{2000}&
\dynkLabel{\mathbf{\b{1274}}}{0010}&
\dynkLabel{\mathbf{\b{1053}}}{0002}&
\\\hline
\raisebox{-4pt}{$\mathfrak{g}_{\r{2}}$}&\dynkLabel{\mathbf{1}}{00}&\dynkLabel{\mathbf{\r{ad}}}{01}&
\dynkLabel{\mathbf{\b{27}}}{20}&
\dynkLabel{\mathbf{\b{77'}}}{30}&
\dynkLabel{\mathbf{\b{77}}}{02}&
\\\hline
\end{array}$$\vspace{-25pt}\end{table}

\newpage

There are three simple Lie algebras missing from \mbox{Table~\ref{tReps_table}}. These are the cases of $\mathfrak{a}_{\r{1}}$, $\mathfrak{a}_{\r{2}}$ and $\mathfrak{d}_{\r{4}}$, for which we may write
\vspace{-12pt}\eq{\begin{split}&\fwbox{0pt}{\fwboxL{435pt}{(\mathfrak{a}_{\r{1}})}\fwbox{0pt}{\fwboxL{300pt}{\hspace{-240pt}\mathbf{\r{ad}}\!\otimes\!\mathbf{\r{ad}}\equivL\,\begin{array}{c}\hspace{2pt}\mathbf{\b{t}}_{\b{1}}\\[-3pt]\dynkLabelK{\mathbf{1}}{0}\end{array}\oplus\begin{array}{c}\hspace{2pt}\mathbf{\b{t}}_{\b{2}}\\[-3pt]\dynkLabelK{\mathbf{\r{ad}}}{2}\end{array}\oplus\begin{array}{c}\mathbf{\hspace{2pt}\b{t}}_{\b{3}}\\[-3pt]\dynkLabelK{\mathbf{5}}{4}\end{array}}}}\\
&\fwbox{0pt}{\fwboxL{435pt}{(\mathfrak{a}_{\r{2}})}\fwbox{0pt}{\fwboxL{300pt}{\hspace{-240pt}\mathbf{\r{ad}}\!\otimes\!\mathbf{\r{ad}}\equivL\,\begin{array}{c}\hspace{2pt}\mathbf{\b{t}}_{\b{1}}\\[-5pt]\dynkLabelK{\mathbf{1}}{00}\end{array}\oplus\begin{array}{c}\hspace{2pt}\mathbf{\b{t}}_{\b{2}}\\[-3pt]\dynkLabelK{\mathbf{\r{ad}}}{11}\end{array}\oplus\begin{array}{c}\hspace{2pt}\mathbf{\b{t}}_{\b{3}}\\[-3pt]\dynkLabelK{\mathbf{27}}{22}\end{array}\oplus\begin{array}{c}\hspace{2pt}\mathbf{\b{t}}_{\b{4}}\\[-3pt]\dynkLabelK{\mathbf{10}}{30}\end{array}\oplus\begin{array}{c}\hspace{2pt}\mathbf{\b{t}}_{\b{5}}\\[-3pt]\dynkLabelK{\mathbf{\bar{10}}}{03}\end{array}
}}}\\
&\fwbox{0pt}{\fwboxL{435pt}{(\mathfrak{d}_{\r{4}})}\fwbox{0pt}{\fwboxL{300pt}{\hspace{-240pt}\mathbf{\r{ad}}\!\otimes\!\mathbf{\r{ad}}\equivL\,\begin{array}{c}\hspace{2pt}\mathbf{\b{t}}_{\b{1}}\\[-3pt]\dynkLabelK{\mathbf{1}}{0000}\end{array}\oplus\begin{array}{c}\hspace{2pt}\mathbf{\b{t}}_{\b{2}}\\[-3pt]\dynkLabelK{\mathbf{\r{ad}}}{0100}\end{array}\oplus\begin{array}{c}\hspace{2pt}\mathbf{\b{t}}_{\b{3}}\\[-3pt]\dynkLabelK{\mathbf{35}}{2000}\end{array}\oplus\hspace{2pt}\begin{array}{c}\mathbf{\b{t}}_{\b{4}}\\[-3pt]\dynkLabelK{\mathbf{350}}{1011}\end{array}\oplus\begin{array}{c}\hspace{2pt}\mathbf{\b{t}}_{\b{5}}\\[-3pt]\dynkLabelK{\mathbf{{300}}}{0200}\end{array}\oplus\begin{array}{c}\hspace{2pt}\mathbf{\b{t}}_{\b{6}}\\[-3pt]\dynkLabelK{\mathbf{{35}}}{0002}\end{array}\oplus\begin{array}{c}\hspace{2pt}\mathbf{\b{t}}_{\b{7}}\\[-3pt]\dynkLabelK{\mathbf{{35}}}{0020}\end{array}
}}}\\[-10pt]
\end{split}\label{remaining_tReps_defined}}
In the first two cases, these identifications are consistent with \mbox{Table~\ref{tReps_table}}, provided one drops the irreducible representations indicated with negative dimension; in the last case, the representation denoted `$\mathbf{\b{t}}_{\b{6}}$' for $\mathfrak{d}_{\r{k}}$ in  \mbox{Table~\ref{tReps_table}} would have dimension `$\mathbf{70}$', but should be represented instead by the sum of the two irreducible representations $\mathbf{\b{t}}_{\b{6}}\!\oplus\!\mathbf{\b{t}}_{\b{7}}$ given in (\ref{remaining_tReps_defined}) above. 

Thus, for the $\mathfrak{a}_{\r{k}>2}$ type gauge theories, there will be 9 tensors spanned by those in (\ref{gggg_colour_basis_defined}), which we order as follows:
\vspace{2pt}\eq{\fwbox{0pt}{\fwboxL{435pt}{(\mathfrak{a}_{\r{k}>1})}}\fwbox{0pt}{\left\{\mathcal{B}^{\b{1}},\ldots,\right\}\bigger{\Leftrightarrow}\left\{
\tikzBox{gggg_a_series_basis_1}{
\arrowTo[hred]{0,0}[0.75]{-120};%\node[anchor=0,inner sep=1pt] at(in){{\footnotesize$\r{[\adR]}$}};
\arrowTo[hred]{0,0}[0.75]{120}%\node[anchor=0,inner sep=0pt] at(in){{\footnotesize$\r{[\adR]}$}};
\coordinate(eph)at($(0,0)+(0.985*\edgeLength,0)$);
\draw[dashed,edge](end)--(eph);\node[clebsch]at(end){};\arrowFrom[hred]{eph}[0.75]{60}%\node[anchor=180,inner sep=0pt] at(end){{\footnotesize$\r{[\adR]}$}};
\arrowFrom[hred]{eph}[0.75]{-60}%\node[anchor=180,inner sep=0pt] at(end){{\footnotesize$\r{[\adR]}$}};
\node[clebsch]at(in){};
},
\tikzBox{gggg_a_series_basis_2}{
\arrowTo[hred]{0,0}[0.75]{-120}%\node[anchor=0,inner sep=1pt] at(in){{\footnotesize$\r{[\adR]}$}};
\arrowTo[hred]{0,0}[0.75]{120}%\node[anchor=0,inner sep=0pt] at(in){{\footnotesize$\r{[\adR]}$}};
\coordinate(eph)at($(0,0)+(0.985*\edgeLength,0)$);\arrowTo[hred]{eph}[0.985]{180}\node[anchor=90,inner sep=2pt]at(arrownode){{\footnotesize$\r{[\adR]}$}};
\node[hred,clebschR]at(in){};\arrowFrom[hred]{eph}[0.75]{60}%\node[anchor=180,inner sep=0pt] at(end){{\footnotesize$\r{[\adR]}$}};
\arrowFrom[hred]{eph}[0.75]{-60}%\node[anchor=180,inner sep=0pt] at(end){{\footnotesize$\r{[\adR]}$}};
\node[hred,clebschR]at(in){};
}
,\tikzBox{gggg_a_series_basis_3}{
\arrowTo[hred]{0,0}[0.75]{-120}%\node[anchor=0,inner sep=1pt] at(in){{\footnotesize$\r{[\adR]}$}};
\arrowTo[hred]{0,0}[0.75]{120}%\node[anchor=0,inner sep=0pt] at(in){{\footnotesize$\r{[\adR]}$}};
\coordinate(eph)at($(0,0)+(0.985*\edgeLength,0)$);\arrowTo[hred]{eph}[0.985]{180}\node[anchor=90,inner sep=2pt]at(arrownode){{\footnotesize$\r{[\adR]}$}};
\node[hred,clebschR]at(in){};\arrowFrom[hred]{eph}[0.75]{60}%\node[anchor=180,inner sep=0pt] at(end){{\footnotesize$\r{[\adR]}$}};
\arrowFrom[hred]{eph}[0.75]{-60}%\node[anchor=180,inner sep=0pt] at(end){{\footnotesize$\r{[\adR]}$}};
\node[hred,clebschD]at(in){};
},
\tikzBox{gggg_a_series_basis_4}{
\arrowTo[hred]{0,0}[0.75]{-120}%\node[anchor=0,inner sep=1pt] at(in){{\footnotesize$\r{[\adR]}$}};
\arrowTo[hred]{0,0}[0.75]{120}%\node[anchor=0,inner sep=0pt] at(in){{\footnotesize$\r{[\adR]}$}};
\coordinate(eph)at($(0,0)+(0.985*\edgeLength,0)$);\arrowTo[hred]{eph}[0.985]{180}\node[anchor=90,inner sep=2pt]at(arrownode){{\footnotesize$\r{[\adR]}$}};
\node[hred,clebschD]at(in){};\arrowFrom[hred]{eph}[0.75]{60}%\node[anchor=180,inner sep=0pt] at(end){{\footnotesize$\r{[\adR]}$}};
\arrowFrom[hred]{eph}[0.75]{-60}%\node[anchor=180,inner sep=0pt] at(end){{\footnotesize$\r{[\adR]}$}};
\node[hred,clebschR]at(in){};
},
\tikzBox{gggg_a_series_basis_5}{
\arrowTo[hred]{0,0}[0.75]{-120}%\node[anchor=0,inner sep=1pt] at(in){{\footnotesize$\r{[\adR]}$}};
\arrowTo[hred]{0,0}[0.75]{120}%\node[anchor=0,inner sep=0pt] at(in){{\footnotesize$\r{[\adR]}$}};
\coordinate(eph)at($(0,0)+(0.985*\edgeLength,0)$);\arrowTo[hred]{eph}[0.985]{180}\node[anchor=90,inner sep=2pt]at(arrownode){{\footnotesize$\r{[\adR]}$}};
\node[hred,clebschD]at(in){};\arrowFrom[hred]{eph}[0.75]{60}%\node[anchor=180,inner sep=0pt] at(end){{\footnotesize$\r{[\adR]}$}};
\arrowFrom[hred]{eph}[0.75]{-60}%\node[anchor=180,inner sep=0pt] at(end){{\footnotesize$\r{[\adR]}$}};
\node[hred,clebschD]at(in){};
},
\tikzBox{gggg_a_series_basis_6}{
\arrowTo[hred]{0,0}[0.75]{-120}%\node[anchor=0,inner sep=1pt] at(in){{\footnotesize$\r{[\adR]}$}};
\arrowTo[hred]{0,0}[0.75]{120}%\node[anchor=0,inner sep=0pt] at(in){{\footnotesize$\r{[\adR]}$}};
\coordinate(eph)at($(0,0)+(0.985*\edgeLength,0)$);\arrowTo[hblue]{eph}[0.985]{180}\node[anchor=90,inner sep=2pt]at(arrownode){{\footnotesize$\b{[t_3]}$}};
\node[clebsch]at(in){};\arrowFrom[hred]{eph}[0.75]{60}%\node[anchor=180,inner sep=0pt] at(end){{\footnotesize$\r{[\adR]}$}};
\arrowFrom[hred]{eph}[0.75]{-60}%\node[anchor=180,inner sep=0pt] at(end){{\footnotesize$\r{[\adR]}$}};
\node[clebsch]at(in){};
},\ldots
\right\}\!.}\vspace{-0pt}\label{gggg_basis_tensors_a_series}}
For $\mathfrak{a}_{\r{2}}$ gauge theory, there are only $5$ representations $\mathbf{\b{t}}_{\b{i}}$, leading to a basis of 8 tensors; and for $\mathfrak{a}_{\r{1}}$, only 3 basis elements are needed (as the $\r{d}$ tensor vanishes for $\mathfrak{a}_{\r{1}}$). 

For all other gauge theories, the $\r{d}$ tensor vanishes allowing us to uniquely identify each basis tensor in (\ref{gggg_colour_basis_defined}) simply by the representation $\mathbf{\b{t}}_{\b{i}}$ exchanged. This gives us a basis of 6 tensors for the classical algebras (7 tensors for $\mathfrak{d}_{\r{4}}$), and a basis of 5 tensors for the exceptional algebras. We choose to order these in the obvious way:
\vspace{-0pt}\eq{\fwbox{0pt}{\fwboxL{435pt}{(\hspace{-1.25pt}\not{\hspace{-2pt}\mathfrak{a}_{\r{k}})}}}\fwbox{0pt}{\fwboxR{0pt}{\left\{\mathcal{B}^{\b{1}},\ldots,\right\}\bigger{\Leftrightarrow}}\left\{
\tikzBox{gggg_basis_1}{
\arrowTo[hred]{0,0}[0.75]{-120};%\node[anchor=0,inner sep=1pt] at(in){{\footnotesize$\r{[\adR]}$}};
\arrowTo[hred]{0,0}[0.75]{120}%\node[anchor=0,inner sep=0pt] at(in){{\footnotesize$\r{[\adR]}$}};
\coordinate(eph)at($(0,0)+(0.985*\edgeLength,0)$);
\draw[dashed,edge](end)--(eph);\node[clebsch]at(end){};\arrowFrom[hred]{eph}[0.75]{60}%\node[anchor=180,inner sep=0pt] at(end){{\footnotesize$\r{[\adR]}$}};
\arrowFrom[hred]{eph}[0.75]{-60}%\node[anchor=180,inner sep=0pt] at(end){{\footnotesize$\r{[\adR]}$}};
\node[clebsch]at(in){};
},
\tikzBox{gggg_basis_2}{
\arrowTo[hred]{0,0}[0.75]{-120}%\node[anchor=0,inner sep=1pt] at(in){{\footnotesize$\r{[\adR]}$}};
\arrowTo[hred]{0,0}[0.75]{120}%\node[anchor=0,inner sep=0pt] at(in){{\footnotesize$\r{[\adR]}$}};
\coordinate(eph)at($(0,0)+(0.985*\edgeLength,0)$);\arrowTo[hred]{eph}[0.985]{180}\node[anchor=90,inner sep=2pt]at(arrownode){{\footnotesize$\r{[\adR]}$}};
\node[hred,clebschR]at(in){};\arrowFrom[hred]{eph}[0.75]{60}%\node[anchor=180,inner sep=0pt] at(end){{\footnotesize$\r{[\adR]}$}};
\arrowFrom[hred]{eph}[0.75]{-60}%\node[anchor=180,inner sep=0pt] at(end){{\footnotesize$\r{[\adR]}$}};
\node[hred,clebschR]at(in){};
}
,\tikzBox{gggg_basis_3}{
\arrowTo[hred]{0,0}[0.75]{-120}%\node[anchor=0,inner sep=1pt] at(in){{\footnotesize$\r{[\adR]}$}};
\arrowTo[hred]{0,0}[0.75]{120}%\node[anchor=0,inner sep=0pt] at(in){{\footnotesize$\r{[\adR]}$}};
\coordinate(eph)at($(0,0)+(0.985*\edgeLength,0)$);\arrowTo[hblue]{eph}[0.985]{180}\node[anchor=90,inner sep=2pt]at(arrownode){{\footnotesize$\b{[t_3]}$}};
\node[clebsch]at(in){};\arrowFrom[hred]{eph}[0.75]{60}%\node[anchor=180,inner sep=0pt] at(end){{\footnotesize$\r{[\adR]}$}};
\arrowFrom[hred]{eph}[0.75]{-60}%\node[anchor=180,inner sep=0pt] at(end){{\footnotesize$\r{[\adR]}$}};
\node[clebsch]at(in){};
},\ldots
%,\tikzBox{gggg_basis_6}{
%\arrowTo[hred]{0,0}[0.75]{-120}%\node[anchor=0,inner sep=1pt] at(in){{\footnotesize$\r{[\adR]}$}};
%\arrowTo[hred]{0,0}[0.75]{120}%\node[anchor=0,inner sep=0pt] at(in){{\footnotesize$\r{[\adR]}$}};
%\coordinate(eph)at($(0,0)+(0.985*\edgeLength,0)$);\arrowTo[hblue]{eph}[0.985]{180}\node[anchor=90,inner sep=2pt]at(arrownode){{\footnotesize$\b{[t_6]}$}};
%\node[clebsch]at(in){};\arrowFrom[hred]{eph}[0.75]{60}%\node[anchor=180,inner sep=0pt] at(end){{\footnotesize$\r{[\adR]}$}};
%\arrowFrom[hred]{eph}[0.75]{-60}%\node[anchor=180,inner sep=0pt] at(end){{\footnotesize$\r{[\adR]}$}};
%\node[clebsch]at(in){};
%}
\right\}\!.}\vspace{-0pt}\label{gggg_basis_tensors_general}} 

With what tensors should we compare those defined in (\ref{gggg_colour_basis_defined})? A natural set of familiar tensors would be the multi-traces involving the fundamental representation. Only for $\mathfrak{a}_{\r{k}>1}$ gauge theories are the four-particle multi-traces non-dihedrally invariant---because $\mathbf{\b{F}}\!\nsim\!\b{\bar{\mathbf{F}}}$.\footnote{Of course, the fundamental $\mathbf{\b{27}}\!\equivL\mathbf{\b{F}}$ of $\mathfrak{e}_{\r{6}}$ is also complex; however, traces involving four or fewer particles are in fact dihedrally symmetric.} Thus, we may consider the `reference' tensors given by 
\begin{align}
\fwbox{0pt}{\raisebox{-12pt}{$\fwboxL{37.5pt}{(\mathfrak{a}_{\r{k}})}$}}\big\{\mathbf{T}^{\r{1}},\ldots,\mathbf{T}^{\r{9}}\big\}\equivR\big\{&\mathrm{tr}_{\mathbf{\b{F}}}(\r{1\,2\,\bar{3}\,\bar{4}}),\mathrm{tr}_{\mathbf{\b{F}}}(\r{\bar{4}\,\bar{3}\,{2}\,{1}}),\mathrm{tr}_{\mathbf{\b{F}}}(\r{1\,2\,\bar{4}\,\bar{3}}),\mathrm{tr}_{\mathbf{\b{F}}}(\r{\bar{3}\,\bar{4}\,{2}\,{1}}),\\[-12pt]
&\mathrm{tr}_{\mathbf{\b{F}}}(\r{1\,\bar{3}\,{2}\,\bar{4}}),\mathrm{tr}_{\mathbf{\b{F}}}(\r{\bar{4}\,2\,{3}\,\bar{1}}),\mathrm{tr}_{\mathbf{\b{F}}}(\r{1\hspace{1.3pt}2}|\r{\bar{3}\hspace{1.3pt}\bar{4}}),\mathrm{tr}_{\mathbf{\b{F}}}(\r{1\hspace{1.3pt}\bar{3}}|\r{{2}\hspace{1.3pt}\bar{4}}),\mathrm{tr}_{\mathbf{\b{F}}}(\r{1\hspace{1.3pt}\bar{4}}|\r{{2}\hspace{1.3pt}\bar{3}})\big\};\,\nonumber\\[-24pt]~\nonumber
\end{align}
for $\mathfrak{a}_{\r{k}}$, and 
\begin{align}
\fwbox{0pt}{\raisebox{-12pt}{$\fwboxL{162.2pt}{(\hspace{-1.25pt}\not{\hspace{-2pt}\mathfrak{a}_{\r{k}})}}$}}\big\{\mathbf{T}^{\r{1}},\ldots,\mathbf{T}^{\r{6}}\big\}\equivR\big\{&\mathrm{tr}_{\mathbf{\b{F}}}(\r{1\,2\,\bar{3}\,\bar{4}}),\mathrm{tr}_{\mathbf{\b{F}}}(\r{1\,2\,\bar{4}\,\bar{3}}),\mathrm{tr}_{\mathbf{\b{F}}}(\r{1\,\bar{3}\,{2}\,\bar{4}}),\\[-12pt]
&\mathrm{tr}_{\mathbf{\b{F}}}(\r{1\hspace{1.3pt}2}|\r{\bar{3}\hspace{1.3pt}\bar{4}}),\mathrm{tr}_{\mathbf{\b{F}}}(\r{1\hspace{1.3pt}\bar{3}}|\r{{2}\hspace{1.3pt}\bar{4}}),\mathrm{tr}_{\mathbf{\b{F}}}(\r{1\hspace{1.3pt}\bar{4}}|\r{{2}\hspace{1.3pt}\bar{3}})\big\}\nonumber\\[-24pt]~\nonumber\end{align}
for all other simple Lie algebras, respectively. To be clear, the `$\r{\bar{a}}$' instructs us to convert the last two slots to \emph{outgoing} representations using the Killing metric $g^{\mathbf{\r{ad}}}_{\smash{\r{[\adR][\adR]}}}$  in order to compare with colour tensors of the form `$C(\mathbf{\r{ad}}\,\mathbf{\r{ad}}|\mathbf{\r{ad}}\,\mathbf{\r{ad}})$' as opposed to those of `$C(\mathbf{\r{ad}}\,\mathbf{\r{ad}}\,\mathbf{\r{ad}}\,\mathbf{\r{ad}}|\mathbf{1})$'. Thus, for example, we are considering the tensor
\eq{\mathrm{tr}_{\mathbf{\b{F}}}(\r{1\hspace{1.3pt}\bar{3}}|\r{2\hspace{1.3pt}\bar{4}})\indices{\r{[\adR][\adR]}}{\r{[\adR][\adR]}}\equivR\!\!\left\{\hspace{-1pt}\mathrm{tr}_{\mathbf{\b{F}}}(\r{1\hspace{1.3pt}\bar{3}}|\r{2\hspace{1.3pt}\bar{4}})\indices{\r{a_1a_2}}{\r{a_3a_4}}\!\right\}_{\!\r{a_i}\in\r{[\adR]}}\!\!=\!\!\left\{\hspace{-1pt}\delta\indices{\r{a_1}}{\r{a_3}}\delta\indices{\r{a_2}}{\r{a_4}}\hspace{-1pt}\right\}_{\!\r{a_i}\in\r{[\adR]}}\vspace{-6pt}}
as a reference as opposed to
\vspace{6pt}\eq{\fwbox{0pt}{\mathrm{tr}_{\mathbf{\b{F}}}(\r{1\hspace{1.3pt}{3}}|\r{2\hspace{1.3pt}{4}})\indices{\r{[\adR][\adR]}\r{[\adR][\adR]}}{}\equivR\!\!\left\{\hspace{-1pt}\mathrm{tr}_{\mathbf{\b{F}}}(\r{1\hspace{1.3pt}{3}}|\r{2\hspace{1.3pt}{4}})\indices{\r{a_1a_2}\r{a_3a_4}}{}\!\right\}_{\!\r{a_i}\in\r{[\adR]}}\!\equivR\!\!\left\{\hspace{-1pt}\mathrm{tr}_{\mathbf{\b{F}}}(\r{1\hspace{1.3pt}{2}}|\r{3\hspace{1.3pt}{4}})\indices{\r{a_1a_3}\r{a_2a_4}}{}\!\right\}_{\!\r{a_i}\in\r{[\adR]}}\!\!=\!\!\left\{\hspace{-1pt}g_{\mathbf{\r{ad}}}^{\r{a_1a_3}}g_{\mathbf{\r{ad}}}^{\r{a_2a_4}}\!\right\}_{\!\r{a_i}\in\r{[\adR]}}\!.}\nonumber\vspace{-0pt}}

It is instructive to compare the Clebsch-Gordan basis tensors to the multi-traces by determining the coefficients
\vspace{-4pt}\eq{\mathbf{T}^{\r{i}}\equivL\,\sum_{\b{j}}{\mathbf{c}}[\mathfrak{\r{g}}]^{\r{i}}_{\phantom{i}\b{j}}\mathcal{B}^{\hspace{1pt}\b{j}}\,.\label{traces_to_new_gggg_coefficients_defined}\vspace{-10pt}}
We have computed these coefficients (by direct construction) and enumerate the results in \mbox{Table~\ref{olde_to_new_gggg_coefficients}} for the simple Lie algebras with the exception of $\mathfrak{d}_{\r{4}}$, which we discuss momentarily. 

For most of the classical Lie algebras, we see that multi-traces (over the fundamental representation) suffice to span the space of all possible colour tensors; for these cases, the coefficients given in \mbox{Table~\ref{olde_to_new_gggg_coefficients}} are full rank, and can be inverted to effectively define the new colour tensors. For the exceptional algebras, the coefficients are less than full-rank, implying one linear relation satisfied by each set of trace tensors. 

\newpage
There is one simple Lie algebra, however, for which there are \emph{more} Clebsch-Gordan tensors than there are distinct multi-traces involving the fundamental representation: $\mathfrak{d}_{\r{4}}$. In this case, there are 7 basis tensors, but only 6 distinct multi-traces.

%For $\mathfrak{a}_{\r{1}}$:
%%
%\eq{\mathrm{tr}_{\mathbf{\b{F}}}(\r{1\,2\,\bar{3}\,\bar{4}}){+}\mathrm{tr}_{\mathbf{\b{F}}}(\r{1\,2\,\bar{4}\,\bar{3}})=\mathrm{tr}_{\mathbf{\b{F}}}(\r{1\,2|\bar{3}\,\bar{4}})}
%

\begin{table}[t]\vspace{-10pt}\caption{Expansion coefficients ${\mathbf{c}}$ required to express $\mathbf{T}^{\r{i}}$ tensors in terms of the basis $\mathcal{B}^{\hspace{0pt}\b{i}}$.}\label{olde_to_new_gggg_coefficients}
\vspace{-17.5pt}$$\begin{array}{@{}c@{}}
\fwbox{0pt}{\begin{array}{cccccc}\mathfrak{a}_{\r{1}}&\mathfrak{a}_{\r{2}}&\mathfrak{a}_{\r{k}>2}\\
\left(\begin{array}{@{\hspace{1pt}}c@{\;\;}c@{\;\;}c@{\hspace{0pt}}}\frac{1}{2}&\frac{1}{4}&\dzero\\
\frac{1}{2}&\frac{1}{4}&\dzero\\
\frac{1}{2}&\tmi\frac{1}{4}&\dzero\\
\frac{1}{2}&\tmi\frac{1}{4}&\dzero\\
\tmi\frac{1}{6}&\dzero&\frac{1}{4}\\
\tmi\frac{1}{6}&\dzero&\frac{1}{4}\\
1&\dzero&\dzero\\
\frac{1}{3}&\tmi\frac{1}{4}&\frac{1}{4}\\
\frac{1}{3}&\frac{1}{4}&\frac{1}{4}\end{array}\right) 
&
\left(\begin{array}{@{\hspace{0pt}}c@{\;\;}c@{\;\;}c@{\;\;}c@{\;\;}c@{\;\;}c@{\;\;}c@{\;\;}c@{\hspace{0pt}}}
\frac{1}{3}&\frac{1}{4}&\frac{1}{4}&\frac{1}{4}&\frac{1}{4}&\dzero&\dzero&\dzero\\
\frac{1}{3}&\frac{1}{4}&\tmi\frac{1}{4}&\tmi\frac{1}{4}&\frac{1}{4}&\dzero&\dzero&\dzero\\
\frac{1}{3}&\tmi\frac{1}{4}&\frac{1}{4}&\tmi\frac{1}{4}&\frac{1}{4}&\dzero&\dzero&\dzero\\
\frac{1}{3}&\tmi\frac{1}{4}&\tmi\frac{1}{4}&\frac{1}{4}&\frac{1}{4}&\dzero&\dzero&\dzero\\
\tmi\frac{1}{24}&\dzero&\dzero&\dzero&\tmi\frac{1}{5}&\frac{1}{4}&\frac{1}{4}&\frac{1}{4}\\
\tmi\frac{1}{24}&\dzero&\dzero&\dzero&\tmi\frac{1}{5}&\tmi\frac{1}{4}&\tmi\frac{1}{4}&\frac{1}{4}\\
1&\dzero&\dzero&\dzero&\dzero&\dzero&\dzero&\dzero\\
\frac{1}{8}&\tmi\frac{1}{6}&\dzero&\dzero&\frac{3}{10}&\frac{1}{4}&\tmi\frac{1}{4}&\frac{1}{4}\\
\frac{1}{8}&\frac{1}{6}&\dzero&\dzero&\frac{3}{10}&\tmi\frac{1}{4}&\frac{1}{4}&\frac{1}{4}\end{array}\right)
&
\left(\begin{array}{@{\hspace{-5pt}}c@{\hspace{-4pt}}c@{}c@{\;\;}c@{\,}c@{\,}c@{\;\;}c@{\;\;}c@{\;\;}c@{\hspace{0pt}}}\frac{1}{\r{k}\hspace{-0pt}{+}\hspace{-0pt}1}&\frac{1}{4}&\frac{1}{4}&\frac{1}{4}&\frac{1}{4}&\dzero&\dzero&\dzero&\dzero\\[-0.5pt]
\frac{1}{\r{k}\hspace{-0pt}{+}\hspace{-0pt}1}&\frac{1}{4}&\tmi\frac{1}{4}&\tmi\frac{1}{4}&\frac{1}{4}&\dzero&\dzero&\dzero&\dzero\\[-0.5pt]
\frac{1}{\r{k}\hspace{-0pt}{+}\hspace{-0pt}1}&\tmi\frac{1}{4}&\frac{1}{4}&\tmi\frac{1}{4}&\frac{1}{4}&\dzero&\dzero&\dzero&\dzero\\[-0.5pt]
\frac{1}{\r{k}\hspace{-0pt}{+}\hspace{-0pt}1}&\tmi\frac{1}{4}&\tmi\frac{1}{4}&\frac{1}{4}&\frac{1}{4}&\dzero&\dzero&\dzero&\dzero\\[-0.5pt]
\frac{\tmi1\phantom{\tmi}}{\r{k}(\hspace{-1pt}\r{k}\hspace{-0pt}{+}\hspace{-0pt}1\hspace{-1pt})(\hspace{-1pt}\r{k}\hspace{-0pt}{+}\hspace{-0pt}2\hspace{-1pt})}&\dzero&\dzero&\dzero&\frac{\tmi1\phantom{\tmi}}{(\hspace{-1pt}\r{k}\hspace{-0pt}{-}\hspace{-0pt}1\hspace{-1pt})(\hspace{-1pt}\r{k}\hspace{-0pt}{+}\hspace{-0pt}3\hspace{-1pt})}&\frac{1}{4}&\frac{1}{4}&\frac{1}{4}&\frac{1}{4}\\[-0.5pt]
\frac{\tmi1\phantom{\tmi}}{\r{k}(\hspace{-1pt}\r{k}\hspace{-0pt}{+}\hspace{-0pt}1\hspace{-1pt})(\hspace{-1pt}\r{k}\hspace{-0pt}{+}\hspace{-0pt}2\hspace{-1pt})}&\dzero&\dzero&\dzero&\frac{\tmi1\phantom{\tmi}}{(\hspace{-1pt}\r{k}\hspace{-0pt}{-}\hspace{-0pt}1\hspace{-1pt})(\hspace{-1pt}\r{k}\hspace{-0pt}{+}\hspace{-0pt}3\hspace{-1pt})}&\frac{1}{4}&\tmi\frac{1}{4}&\tmi\frac{1}{4}&\frac{1}{4}\\[-0.5pt]
1&\dzero&\dzero&\dzero&\dzero&\dzero&\dzero&\dzero&\dzero\\[-0.5pt]
\frac{1}{\r{k}(\hspace{-1pt}\r{k}\hspace{-0pt}{+}\hspace{-0pt}2\hspace{-1pt})}&\frac{\tmi1\phantom{\tmi}}{2(\hspace{-1pt}\r{k}\hspace{-0pt}{+}\hspace{-0pt}1\hspace{-1pt})}&\dzero&\dzero&\frac{\r{k}\hspace{-0pt}{+}\hspace{-0pt}1}{2(\hspace{-1pt}\r{k}\hspace{-0pt}{-}\hspace{-0pt}1\hspace{-1pt})(\hspace{-1pt}\r{k}\hspace{-0pt}{+}\hspace{-0pt}3\hspace{-1pt})}&\frac{1}{4}&\tmi\frac{1}{4}&\frac{1}{4}&\tmi\frac{1}{4}\\[-0.5pt]
\frac{1}{\r{k}(\hspace{-1pt}\r{k}\hspace{-0pt}{+}\hspace{-0pt}2\hspace{-1pt})}&\frac{1}{2(\hspace{-1pt}\r{k}\hspace{-0pt}{+}\hspace{-0pt}1\hspace{-1pt})}&\dzero&\dzero&\frac{\r{k}\hspace{-0pt}{+}\hspace{-0pt}1}{2(\hspace{-1pt}\r{k}\hspace{-0pt}{-}\hspace{-0pt}1\hspace{-1pt})(\hspace{-1pt}\r{k}\hspace{-0pt}{+}\hspace{-0pt}3\hspace{-1pt})}&\frac{1}{4}&\frac{1}{4}&\tmi\frac{1}{4}&\tmi\frac{1}{4}\end{array}\right)\end{array}}\\[-4pt]
\fwbox{0pt}{\begin{array}{@{}c@{}c@{}c@{}}\mathfrak{b}_{\r{k}}&\mathfrak{c}_{\r{k}}&\mathfrak{d}_{\r{k}>4}\\
\left(\begin{array}{@{\hspace{-3pt}}c@{}c@{\hspace{-2pt}}c@{\;}c@{\;\,}c@{\;\,}c@{}}\frac{1}{2\r{k}\hspace{-0pt}{+}\hspace{-0pt}1}&\frac{1}{4}&\frac{1}{4}&\dzero&\dzero&\dzero\\[-0.5pt]
\frac{1}{2\r{k}\hspace{-0pt}{+}\hspace{-0pt}1}&\tmi\frac{1}{4}&\frac{1}{4}&\dzero&\dzero&\dzero\\[-0.5pt]
\frac{1}{2\r{k}(\hspace{-1pt}2\r{k}\hspace{-0pt}{+}\hspace{-0pt}1\hspace{-1pt})}&\dzero&\frac{1}{2\r{k}(\hspace{-1pt}2\r{k}\hspace{-0pt}{-}\hspace{-0pt}1\hspace{-1pt})}&\dzero&\frac{1}{4}&\tmi\frac{1}{4}\\[-0.5pt]
1&\dzero&\dzero&\dzero&\dzero&\dzero\\[-0.5pt]
\frac{1}{\r{k}(\hspace{-1pt}2\r{k}\hspace{-0pt}{+}\hspace{-0pt}1\hspace{-1pt})}&\frac{\tmi1\phantom{\tmi}}{2\r{k}\hspace{-0pt}{-}\hspace{-0pt}1}&\frac{1}{2\r{k}\hspace{-0pt}{-}\hspace{-0pt}1}&\tmi\frac{1}{2}&\frac{1}{2}&\frac{1}{4}\\[-0.5pt]
\frac{1}{\r{k}(\hspace{-1pt}2\r{k}\hspace{-0pt}{+}\hspace{-0pt}1\hspace{-1pt})}&\frac{1}{2\r{k}\hspace{-0pt}{-}\hspace{-0pt}1}&\frac{1}{2\r{k}\hspace{-0pt}{-}\hspace{-0pt}1}&\frac{1}{2}&\frac{1}{2}&\frac{1}{4}\end{array}\right)
&
\left(\begin{array}{@{\hspace{-3pt}}c@{}c@{}c@{\;\,}c@{\;\,}c@{\;\,}c@{}}\frac{1}{2\r{k}}&\frac{1}{4}&\frac{1}{4}&\dzero&\dzero&\dzero\\[-0.5pt]
\frac{1}{2\r{k}}&\tmi\frac{1}{4}&\frac{1}{4}&\dzero&\dzero&\dzero\\[-0.5pt]
\frac{\tmi1\phantom{\tmi}}{2\r{k}(\hspace{-1pt}2\r{k}\hspace{-0pt}{+}\hspace{-0pt}1\hspace{-1pt})}&\dzero&\frac{\tmi1\phantom{\tmi}}{4(\hspace{-1pt}\r{k}\hspace{-0pt}{+}\hspace{-0pt}1\hspace{-1pt})}&\dzero&\frac{1}{4}&\tmi\frac{1}{4}\\[-0.5pt]
1&\dzero&\dzero&\dzero&\dzero&\dzero\\[-0.5pt]
\frac{1}{\r{k}(\hspace{-1pt}2\r{k}\hspace{-0pt}{+}\hspace{-0pt}1\hspace{-1pt})}&\frac{\tmi1\phantom{\tmi}}{2(\hspace{-1pt}\r{k}\hspace{-0pt}{+}\hspace{-0pt}1\hspace{-1pt})}&\frac{1}{2(\hspace{-1pt}\r{k}\hspace{-0pt}{+}\hspace{-0pt}1\hspace{-1pt})}&\tmi\frac{1}{2}&\tmi\frac{1}{2}&\tmi\frac{1}{4}\\[-0.5pt]
\frac{1}{\r{k}(\hspace{-1pt}2\r{k}\hspace{-0pt}{+}\hspace{-0pt}1\hspace{-1pt})}&\frac{1}{2(\hspace{-1pt}\r{k}\hspace{-0pt}{+}\hspace{-0pt}1\hspace{-1pt})}&\frac{1}{2(\hspace{-1pt}\r{k}\hspace{-0pt}{+}\hspace{-0pt}1\hspace{-1pt})}&\frac{1}{2}&\tmi\frac{1}{2}&\tmi\frac{1}{4}\end{array}\right)&
\left(\begin{array}{@{\hspace{-3pt}}c@{}c@{}c@{\;\,}c@{\;\,}c@{\;\,}c@{}}\frac{1}{2\r{k}}&\frac{1}{4}&\frac{1}{4}&\dzero&\dzero&\dzero\\[-0.5pt]
\frac{1}{2\r{k}}&\tmi\frac{1}{4}&\frac{1}{4}&\dzero&\dzero&\dzero\\[-0.5pt]
\frac{1}{2\r{k}(\hspace{-1pt}2\r{k}\hspace{-0pt}{-}\hspace{-0pt}1\hspace{-1pt})}&\dzero&\frac{1}{4(\hspace{-1pt}\r{k}\hspace{-0pt}{-}\hspace{-0pt}1\hspace{-1pt})}&\dzero&\frac{1}{4}&\tmi\frac{1}{4}\\[-0.5pt]
1&\dzero&\dzero&\dzero&\dzero&\dzero\\[-0.5pt]
\frac{1}{\r{k}(\hspace{-1pt}2\r{k}\hspace{-0pt}{-}\hspace{-0pt}1\hspace{-1pt})}&\frac{\tmi1\phantom{\tmi}}{2(\hspace{-1pt}\r{k}\hspace{-0pt}{-}\hspace{-0pt}1\hspace{-1pt})}&\frac{1}{2(\hspace{-1pt}\r{k}\hspace{-0pt}{-}\hspace{-0pt}1\hspace{-1pt})}&\tmi\frac{1}{2}&\frac{1}{2}&\frac{1}{4}\\[-0.5pt]
\frac{1}{\r{k}(\hspace{-1pt}2\r{k}\hspace{-0pt}{-}\hspace{-0pt}1\hspace{-1pt})}&\frac{1}{2(\hspace{-1pt}\r{k}\hspace{-0pt}{-}\hspace{-0pt}1\hspace{-1pt})}&\frac{1}{2(\hspace{-1pt}\r{k}\hspace{-0pt}{-}\hspace{-0pt}1\hspace{-1pt})}&\frac{1}{2}&\frac{1}{2}&\frac{1}{4}\end{array}\right)
\end{array}}\\[-4pt]
\fwbox{0pt}{\begin{array}{@{}c@{$\!\!$}c@{$\!\!$}c@{$\!\!$}c@{$\!\!$}c@{$\!\!$}c@{}}
\mathfrak{e}_{\r{6}}&\mathfrak{e}_{\r{7}}&\mathfrak{e}_{\r{8}}&\mathfrak{f}_{\r{4}}&\mathfrak{g}_{\r{2}}\\
\!\hspace{-2pt}\left(\begin{array}{@{\hspace{-2pt}}c@{\;\;}c@{\;\;}c@{\;\;}c@{\;\;}c@{}}\frac{1}{27}&\frac{1}{4}&\frac{1}{4}&\dzero&\dzero\\[-0.5pt]
\frac{1}{27}&\tmi\frac{1}{4}&\frac{1}{4}&\dzero&\dzero\\[-0.5pt]
\frac{4}{351}&\dzero&\tmi\frac{5}{16}&\dzero&\frac{1}{4}\\[-0.5pt]
1&\dzero&\dzero&\dzero&\dzero\\[-0.5pt]
\frac{1}{78}&\tmi\frac{1}{4}&\frac{9}{8}&\tmi\frac{1}{2}&\frac{3}{2}\\[-0.5pt]
\frac{1}{78}&\frac{1}{4}&\frac{9}{8}&\frac{1}{2}&\frac{3}{2}\end{array}\right)&
\left(\begin{array}{@{\hspace{-2pt}}c@{\;\;}c@{\;\;}c@{\;\;}c@{\;\;}c@{}}\frac{1}{56}&\frac{1}{4}&\frac{1}{4}&\dzero&\dzero\\[-0.5pt]
\frac{1}{56}&\tmi\frac{1}{4}&\frac{1}{4}&\dzero&\dzero\\[-0.5pt]
\frac{1}{152}&\dzero&\tmi\frac{7}{20}&\dzero&\frac{1}{4}\\[-0.5pt]
1&\dzero&\dzero&\dzero&\dzero\\[-0.5pt]
\frac{1}{133}&\tmi\frac{1}{3}&\frac{9}{5}&\tmi\frac{1}{2}&3\\[-0.5pt]
\frac{1}{133}&\frac{1}{3}&\frac{9}{5}&\frac{1}{2}&3\end{array}\right)&
\left(\begin{array}{@{\hspace{-2pt}}c@{\;\;}c@{\;\;}c@{\;\;}c@{\;\;}c@{}}\frac{1}{248}&\frac{1}{4}&\frac{1}{4}&\dzero&\frac{1}{60}\\[-0.5pt]
\frac{1}{248}&\tmi\frac{1}{4}&\frac{1}{4}&\dzero&\frac{1}{60}\\[-0.5pt]
\frac{1}{496}&\dzero&\tmi\frac{3}{8}&\dzero&\frac{4}{15}\\[-0.5pt]
1&\dzero&\dzero&\dzero&\dzero\\[-0.5pt]
\frac{1}{248}&\tmi1&\frac{25}{4}&\tmi1&15\\[-0.5pt]
\frac{1}{248}&1&\frac{25}{4}&1&15\end{array}\right)&
\left(\begin{array}{@{\hspace{-2pt}}c@{\;\;}c@{\;\;}c@{\;\;}c@{\;\;}c@{}}\frac{1}{26}&\frac{1}{4}&\frac{1}{4}&\dzero&\dzero\\[-0.5pt]
\frac{1}{26}&\tmi\frac{1}{4}&\frac{1}{4}&\dzero&\dzero\\[-0.5pt]
\frac{1}{104}&\dzero&\tmi\frac{2}{7}&\dzero&\frac{1}{4}\\[-0.5pt]
1&\dzero&\dzero&\dzero&\dzero\\[-0.5pt]
\frac{1}{52}&\tmi\frac{1}{3}&\frac{9}{7}&\tmi\frac{1}{2}&\frac{3}{2}\\[-0.5pt]
\frac{1}{52}&\frac{1}{3}&\frac{9}{7}&\frac{1}{2}&\frac{3}{2}\end{array}\right)&
\left(\begin{array}{@{\hspace{-2pt}}c@{\;\;}c@{\;\;}c@{\;\;}c@{\;\;}c@{}}\frac{1}{7}&\frac{1}{4}&\frac{1}{4}&\dzero&\dzero\\[-0.5pt]
\frac{1}{7}&\tmi\frac{1}{4}&\frac{1}{4}&\dzero&\dzero\\[-0.5pt]
\dzero&\dzero&\tmi\frac{7}{32}&\dzero&\frac{1}{4}\\[-0.5pt]
1&\dzero&\dzero&\dzero&\dzero\\[-0.5pt]
\frac{1}{14}&\tmi\frac{1}{4}&\frac{9}{16}&\tmi\frac{1}{2}&\frac{1}{2}\\[-0.5pt]
\frac{1}{14}&\frac{1}{4}&\frac{9}{16}&\frac{1}{2}&\frac{1}{2}\end{array}\right)
\end{array}}
\end{array}\vspace{-20pt}$$\end{table}

\vspace{-10pt}\eq{\fwbox{0pt}{\fwboxL{280pt}{(\hspace{-0pt}\mathfrak{d}_{\r{4}})}}\hspace{-15pt}\mathbf{c}[\mathfrak{d}_{\r{4}}]=\left(\begin{array}{@{\hspace{-2pt}}c@{\;\;}c@{\;\;}c@{\;\;}c@{\;\;}c@{\;\;}c@{\;\;}c@{}}
\frac{1}{8}&\frac{1}{4}&\frac{1}{4}&\dzero&\dzero&\dzero&\dzero\\[-0.5pt]
\frac{1}{8}&\tmi\frac{1}{4}&\frac{1}{4}&\dzero&\dzero&\dzero&\dzero\\[-0.5pt]
\frac{1}{56}&\dzero&\frac{1}{12}&\dzero&\frac{1}{4}&\tmi\frac{1}{8}&\tmi\frac{1}{8}\\[-0.5pt]
1&\dzero&\dzero&\dzero&\dzero&\dzero&\dzero\\[-0.5pt]
\frac{1}{28}&\tmi\frac{1}{6}&\frac{1}{6}&\tmi\frac{1}{2}&\frac{1}{2}&\frac{1}{8}&\frac{1}{8}\\[-0.5pt]
\frac{1}{28}&\frac{1}{6}&\frac{1}{6}&\frac{1}{2}&\frac{1}{2}&\frac{1}{8}&\frac{1}{8}%\\[-0.5pt]
\end{array}\right)\hspace{15pt}}
An independent colour tensor could be defined by either 
\vspace{-6pt}\eq{
\mathrm{tr}_{\mathbf{\b{S}}}(\r{1\,2\,3\,4})=\frac{1}{8}\mathcal{B}^{\b{1}}{+}\frac{1}{4}\mathcal{B}^{\b{2}}{+}\frac{3}{16}\mathcal{B}^{\b{6}}\quad\text{or}\quad\mathrm{tr}_{\b{\bar{\mathbf{{S}}}}}(\r{1\,2\,3\,4})=\frac{1}{8}\mathcal{B}^{\b{1}}{+}\frac{1}{4}\mathcal{B}^{\b{2}}{+}\frac{3}{16}\mathcal{B}^{\b{7}}\vspace{-5pt}}
where the spinor representations $\mathbf{\b{S}}$ and $\b{\bar{\mathbf{S}}}$ would have Dynkin labels $[0001]$ and $[0010]$, respectively. While we know (from our discussion of the Fierz identity for $\mathfrak{d}_{\r{k}}$ algebras) that all $\r{f}$-graphs are expressible in terms of multi-traces over any representation, it is interesting to note that multi-traces involving a single (fundamental) representation do not suffice to capture all (non-perturbative, or anomalous) possible tensors. 

\begin{table}[t]\vspace{-9pt}\caption{Numbers of independent five-particle colour tensors for simple Lie algebras.}\label{five_particle_tensor_rank_table}\vspace{-25pt}$$\begin{array}{c}%
\\
\begin{array}{r@{$\,\,$}|@{}r@{$\,$}r@{$\,$}r@{$\,$}r@{$\,$}|r@{$\,$}|r@{$\,$}|@{$\,$}r@{$\,$}r@{$\,$}r@{$\,$}|@{}r@{}r@{}r@{$\,$}|@{}r@{}r@{$\,$}|}\cline{2-15}
&\fwbox{14pt}{\mathfrak{a}_{\r{1}}\!\!}&\fwbox{18pt}{\mathfrak{a}_{\r{2}}\!\!}&\fwbox{18pt}{\mathfrak{a}_{\r{3}}\!\!}&\fwbox{18pt}{\mathfrak{a}_{\r{k}>3}}&\fwbox{18pt}{\mathfrak{b}_{\r{k}}\!\!}&\fwbox{18pt}{\mathfrak{c}_{\r{k}}\!\!}&\fwbox{18pt}{\mathfrak{d}_{\r{4}}\!\!}&\fwbox{18pt}{\mathfrak{d}_{\r{5}}\!\!}&\fwbox{18pt}{\mathfrak{d}_{\r{k}>5}}&\fwbox{16pt}{\mathfrak{e}_{\r{6}}\!\!}&\fwbox{16pt}{\mathfrak{e}_{\r{7}}\!\!}&\fwbox{16pt}{\mathfrak{e}_{\r{8}}\!\!}&\fwbox{16pt}{\mathfrak{f}_{\r{4}}\!\!}&\fwbox{16pt}{\mathfrak{g}_{\r{2}}\!\!}\\\cline{2-15}
(\mathbf{\b{F}}\,\mathbf{\b{F}}\,\mathbf{\r{ad}}|\mathbf{\b{F}}\,\mathbf{\b{F}})\hspace{-1pt}\hspace{-1pt}&
3&4&4&4&6&6&6&6&6&7&9&16&12&9\\\cline{2-15}
(\mathbf{\b{F}}\,\mathbf{\r{ad}}\,\mathbf{\r{ad}}|\mathbf{\r{ad}}\,\mathbf{\b{F}})\hspace{-1pt}\hspace{-1pt}&
4&10&11&11&10&10&11&10&10&10&10&16&10&10\\\cline{2-15}
\hspace{-20pt}(\mathbf{\r{ad}}\,\mathbf{\r{ad}}\,\mathbf{\r{ad}}|\mathbf{\r{ad}}\,\mathbf{\r{ad}})\hspace{-1pt}\hspace{-1pt}&
6&32&43&44&22&22&28&23&22&17&16&16&16&16\\\cline{2-15}
\end{array}\end{array}\vspace{-18pt}$$\end{table}

\subsection{Novel Colour Tensors for Five Particle Scattering Amplitudes}

For five (or more) particles, it is straightforward to construct various Clebsch colour bases, the duality maps between them, and to expand more familiar colour tensors into these bases. For each of the examples discussed below (through six particles), we have done this exercise explicitly. However, the notational complexity involved in labeling bases for even all simple Lie algebras, defining other possible tensors and comparing them, grows considerably. Therefore, in this section and the following, we would like to simply sketch the construction of possible bases in physically interesting cases and discuss some of their most salient features.

In \mbox{Table~\ref{five_particle_tensor_rank_table}}, we list the ranks of spaces spanned by colour tensors involving five particles coloured in the adjoint or fundamental representations for each of the simple Lie algebras.

\subsubsection{\texorpdfstring{Clebsch Colour Bases for $C(\mathbf{\b{F}}\,\mathbf{\b{F}}\,\mathbf{\r{ad}}|\mathbf{\b{F}}\,\mathbf{\b{F}})$}{Clebsch Colour Bases for C(FFg|FF)}}

For the scattering of four fundamental-charged particles with one adjoint, one choice of Clebsch colour tensors would be given by  
\vspace{-5pt}\eq{\fwboxR{0pt}{\mathcal{B}^{\hspace{1pt}\mathbf{\g{q}}_{\g{i}}\mathbf{\g{q}}_{\g{j}}}(\mathbf{\b{F}}\,\mathbf{\b{F}}\,\mathbf{\r{ad}}|\mathbf{\b{F}}\,\mathbf{\b{F}})\;\bigger{\Leftrightarrow}\;}\tikzBox[-5.25pt]{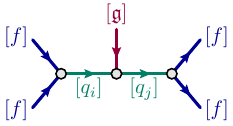}{\arrowTo[hblue]{0,0}{-130};\node[anchor=10,inner sep=2pt] at(in){{\footnotesize$\b{[f]}$}};\arrowTo[hblue]{0,0}{130}\node[anchor=-10,inner sep=2pt] at(in){{\footnotesize$\b{[f]}$}};\arrowFrom[hgreen]{0,0}[1.25]{0}\node[anchor=90,inner sep=2pt] at(arrownode){{\footnotesize$\g{[q_i]}$}};\node[clebsch]at(in){};\arrowTo[hred]{end}{90}\node[anchor=-90,inner sep=2pt] at(in){{\footnotesize$\r{[\adR]}$}};\arrowFrom[hgreen]{end}[1.25]{0}\node[clebsch]at(in){};\node[anchor=90,inner sep=2pt] at(arrownode){{\footnotesize$\g{[q_j]}$}};\arrowFrom[hblue]{end}{50}
\node[anchor=-170,inner sep=2pt] at(end){{\footnotesize$\b{[f]}$}};\arrowFrom[hblue]{in}{-50}\node[anchor=170,inner sep=2pt] at(end){{\footnotesize$\b{[f]}$}};\node[clebsch]at(in){};
}
\vspace{-5pt}}
spanned by all the possible representations $(\g{\mathbf{q}_{i}\,\mathbf{q}_{j}})$ appearing in the decomposition of $\b{\mathbf{F}}\!\otimes\!\mathbf{\b{F}}$ defined in \mbox{Table~\ref{qReps_table}} for which $\mathbf{\g{q}}_{\g{j}}\!\in\!\mathbf{\g{q}}_{\g{i}}\!\otimes\!\mathbf{\r{ad}}$. None of these Clebsch-Gordan coefficients have multiplicity greater than 1, and so no discrete labels are required on any of the vertices. The possible pairs which span the space of colour tensors for each of the simple Lie algebras is given in \mbox{Table~\ref{ffgff_basis_b_label_table}}. 
\begin{table}[b]\vspace{-28pt}$$\hspace{-200pt}\begin{array}{lr@{$\;\;$}|rc@{$\,\,$}c@{$\,\,$}c@{$\,\,$}c@{$\,\,$}c@{$\,\,$}l@{$\,\,$}|l}
%&\multicolumn{1}{r}{\fwboxR{0pt}{\mathfrak{g}\hspace{16pt}}}&\multicolumn{8}{r}{\fwboxL{0pt}{\hspace{3pt}\text{\#}}}\\\cline{3-9} 
&\multicolumn{1}{r}{\fwboxR{0pt}{\mathfrak{g}\hspace{22pt}}}&\multicolumn{8}{r}{\fwboxL{0pt}{\hspace{13pt}\text{\#}}}\\[-10pt]\cline{3-9}
&\fwboxR{0pt}{\mathfrak{a}_{\r{1}}}&&\fwboxR{0pt}{\big\{\!}\rule{0pt}{14pt}(\hspace{-1pt}\mathbf{\g{q}}_{\g{1}}\hspace{1pt}\mathbf{\g{q}}_{\g{2}}\hspace{-1pt})\fwboxL{0pt}{,}&(\hspace{-1pt}\mathbf{\g{q}}_{\g{2}}\hspace{1pt}\mathbf{\g{q}}_{\g{1}}\hspace{-1pt})\fwboxL{0pt}{,}&(\hspace{-1pt}\mathbf{\g{q}}_{\g{2}}\hspace{1pt}\mathbf{\g{q}}_{\g{2}}\hspace{-1pt})\fwboxL{0pt}{\!\big\}}&&&&\fwboxL{0pt}{\hspace{4pt}3}\\[2pt]\cline{3-9}
&\fwboxR{0pt}{\mathfrak{a}_{\r{k}>1}}&\fwboxR{5pt}{\big\{\!}\rule{0pt}{14pt}(\hspace{-1pt}\mathbf{\g{q}}_{\g{1}}\hspace{1pt}\mathbf{\g{q}}_{\g{1}}\hspace{-1pt})\fwboxL{0pt}{,}&(\hspace{-1pt}\mathbf{\g{q}}_{\g{1}}\hspace{1pt}\mathbf{\g{q}}_{\g{2}}\hspace{-1pt})\fwboxL{0pt}{,}&(\hspace{-1pt}\mathbf{\g{q}}_{\g{2}}\hspace{1pt}\mathbf{\g{q}}_{\g{1}}\hspace{-1pt})\fwboxL{0pt}{,}&(\hspace{-1pt}\mathbf{\g{q}}_{\g{2}}\hspace{1pt}\mathbf{\g{q}}_{\g{2}}\hspace{-1pt})\fwboxL{0pt}{\!\big\}}&&&&\fwboxL{0pt}{\hspace{4pt}4}\\[2pt]\cline{3-9}
&\fwboxR{0pt}{\mathfrak{b}_{\r{k}},\mathfrak{c}_{\r{k}},\mathfrak{d}_{\r{k}}}&&\fwboxR{0pt}{\big\{\!}\rule{0pt}{14pt}(\hspace{-1pt}\mathbf{\g{q}}_{\g{1}}\hspace{1pt}\mathbf{\g{q}}_{\g{2}}\hspace{-1pt})\fwboxL{0pt}{,}&(\hspace{-1pt}\mathbf{\g{q}}_{\g{2}}\hspace{1pt}\mathbf{\g{q}}_{\g{1}}\hspace{-1pt})\fwboxL{0pt}{,}&(\hspace{-1pt}\mathbf{\g{q}}_{\g{2}}\hspace{1pt}\mathbf{\g{q}}_{\g{2}}\hspace{-1pt})\fwboxL{0pt}{,}&(\hspace{-1pt}\mathbf{\g{q}}_{\g{2}}\hspace{1pt}\mathbf{\g{q}}_{\g{3}}\hspace{-1pt})\fwboxL{0pt}{,}&(\hspace{-1pt}\mathbf{\g{q}}_{\g{3}}\hspace{1pt}\mathbf{\g{q}}_{\g{2}}\hspace{-1pt})\fwboxL{0pt}{,}&(\hspace{-1pt}\mathbf{\g{q}}_{\g{3}}\hspace{1pt}\mathbf{\g{q}}_{\g{3}}\hspace{-1pt})\fwboxL{5pt}{\!\big\}}&\fwboxL{0pt}{\hspace{4pt}6}\\[2pt]\cline{3-9}
&\fwboxR{0pt}{\mathfrak{e}_{\r{6}}}&\fwboxR{5pt}{\big\{\!}\rule{0pt}{14pt}(\hspace{-1pt}\mathbf{\g{q}}_{\g{1}}\hspace{1pt}\mathbf{\g{q}}_{\g{1}}\hspace{-1pt})\fwboxL{0pt}{,}&(\hspace{-1pt}\mathbf{\g{q}}_{\g{1}}\hspace{1pt}\mathbf{\g{q}}_{\g{2}}\hspace{-1pt})\fwboxL{0pt}{,}&(\hspace{-1pt}\mathbf{\g{q}}_{\g{2}}\hspace{1pt}\mathbf{\g{q}}_{\g{1}}\hspace{-1pt})\fwboxL{0pt}{,}&(\hspace{-1pt}\mathbf{\g{q}}_{\g{2}}\hspace{1pt}\mathbf{\g{q}}_{\g{2}}\hspace{-1pt})\fwboxL{0pt}{,}&(\hspace{-1pt}\mathbf{\g{q}}_{\g{2}}\hspace{1pt}\mathbf{\g{q}}_{\g{3}}\hspace{-1pt})\fwboxL{0pt}{,}&(\hspace{-1pt}\mathbf{\g{q}}_{\g{3}}\hspace{1pt}\mathbf{\g{q}}_{\g{2}}\hspace{-1pt})\fwboxL{0pt}{,}&(\hspace{-1pt}\mathbf{\g{q}}_{\g{3}}\hspace{1pt}\mathbf{\g{q}}_{\g{3}}\hspace{-1pt})\fwboxL{5pt}{\!\big\}}&\fwboxL{0pt}{\hspace{4pt}7}\\[2pt]\cline{3-9}
&\multirow{2}{*}{$\fwboxR{0pt}{\mathfrak{e}_{\r{7}},\mathfrak{g}_{\r{2}}}$}&&\fwboxR{0pt}{\big\{\!}\rule{0pt}{14pt}(\hspace{-1pt}\mathbf{\g{q}}_{\g{1}}\hspace{1pt}\mathbf{\g{q}}_{\g{2}}\hspace{-1pt})\fwboxL{0pt}{,}&(\hspace{-1pt}\mathbf{\g{q}}_{\g{2}}\hspace{1pt}\mathbf{\g{q}}_{\g{1}}\hspace{-1pt})\fwboxL{0pt}{,}&(\hspace{-1pt}\mathbf{\g{q}}_{\g{2}}\hspace{1pt}\mathbf{\g{q}}_{\g{2}}\hspace{-1pt})\fwboxL{0pt}{,}&(\hspace{-1pt}\mathbf{\g{q}}_{\g{2}}\hspace{1pt}\mathbf{\g{q}}_{\g{3}}\hspace{-1pt})\fwboxL{0pt}{,}&(\hspace{-1pt}\mathbf{\g{q}}_{\g{3}}\hspace{1pt}\mathbf{\g{q}}_{\g{2}}\hspace{-1pt})\fwboxL{0pt}{,}&(\hspace{-1pt}\mathbf{\g{q}}_{\g{3}}\hspace{1pt}\mathbf{\g{q}}_{\g{3}}\hspace{-1pt})\fwboxL{5pt}{,}&\multirow{2}{*}{$\fwboxL{0pt}{\hspace{4pt}9}$}\\[2pt]
&&&(\hspace{-1pt}\mathbf{\g{q}}_{\g{3}}\hspace{1pt}\mathbf{\g{q}}_{\g{4}}\hspace{-1pt})\fwboxL{0pt}{,}&(\hspace{-1pt}\mathbf{\g{q}}_{\g{4}}\hspace{1pt}\mathbf{\g{q}}_{\g{3}}\hspace{-1pt})\fwboxL{0pt}{,}&(\hspace{-1pt}\mathbf{\g{q}}_{\g{4}}\hspace{1pt}\mathbf{\g{q}}_{\g{4}}\hspace{-1pt})\fwboxL{0pt}{\!\big\}}&&&\\[2pt]\cline{3-9}
&\multirow{2}{*}{$\fwboxR{0pt}{\mathfrak{f}_{\r{4}}}$}&&\fwboxR{0pt}{\big\{\!}\rule{0pt}{14pt}(\hspace{-1pt}\mathbf{\g{q}}_{\g{1}}\hspace{1pt}\mathbf{\g{q}}_{\g{2}}\hspace{-1pt})\fwboxL{0pt}{,}&(\hspace{-1pt}\mathbf{\g{q}}_{\g{2}}\hspace{1pt}\mathbf{\g{q}}_{\g{1}}\hspace{-1pt})\fwboxL{0pt}{,}&(\hspace{-1pt}\mathbf{\g{q}}_{\g{2}}\hspace{1pt}\mathbf{\g{q}}_{\g{2}}\hspace{-1pt})\fwboxL{0pt}{,}&(\hspace{-1pt}\mathbf{\g{q}}_{\g{2}}\hspace{1pt}\mathbf{\g{q}}_{\g{3}}\hspace{-1pt})\fwboxL{0pt}{,}&(\hspace{-1pt}\mathbf{\g{q}}_{\g{3}}\hspace{1pt}\mathbf{\g{q}}_{\g{2}}\hspace{-1pt})\fwboxL{0pt}{,}&(\hspace{-1pt}\mathbf{\g{q}}_{\g{3}}\hspace{1pt}\mathbf{\g{q}}_{\g{3}}\hspace{-1pt})\fwboxL{0pt}{,}&\multirow{2}{*}{$\fwboxL{0pt}{\hspace{4pt}12}$}\\
&&&(\hspace{-1pt}\mathbf{\g{q}}_{\g{3}}\hspace{1pt}\mathbf{\g{q}}_{\g{4}}\hspace{-1pt})\fwboxL{0pt}{,}&(\hspace{-1pt}\mathbf{\g{q}}_{\g{4}}\hspace{1pt}\mathbf{\g{q}}_{\g{3}}\hspace{-1pt})\fwboxL{0pt}{,}&(\hspace{-1pt}\mathbf{\g{q}}_{\g{4}}\hspace{1pt}\mathbf{\g{q}}_{\g{4}}\hspace{-1pt})\fwboxL{0pt}{,}&(\hspace{-1pt}\mathbf{\g{q}}_{\g{4}}\hspace{1pt}\mathbf{\g{q}}_{\g{5}}\hspace{-1pt})\fwboxL{0pt}{,}&(\hspace{-1pt}\mathbf{\g{q}}_{\g{5}}\hspace{1pt}\mathbf{\g{q}}_{\g{4}}\hspace{-1pt})\fwboxL{0pt}{,}&(\hspace{-1pt}\mathbf{\g{q}}_{\g{5}}\hspace{1pt}\mathbf{\g{q}}_{\g{5}}\hspace{-1pt})\fwboxL{0pt}{\!\big\}}\\[2pt]\cline{3-9}
\end{array}\hspace{-200pt}$$\vspace{-18pt}\caption{Labels $(\hspace{-1pt}\g{\mathbf{q}_i\,\mathbf{q}_j}\hspace{-1pt})$ for basis tensors $\mathcal{B}^{\hspace{1pt}\mathbf{\g{q}}_{\g{i}}\mathbf{\g{q}}_{\g{j}}}(\mathbf{\b{F}}\,\mathbf{\b{F}}\,\mathbf{\r{ad}}|\mathbf{\b{F}}\,\mathbf{\b{F}})$ of simple Lie algebras.}\label{ffgff_basis_b_label_table}\vspace{-18pt}\end{table}

\newpage
An alternate choice of Clebsch colour tensors would be given by  
\vspace{-5pt}\eq{\fwboxR{0pt}{\mathcal{C}^{\hspace{1pt}\mathbf{\t{r}}_{\t{i}}\mathbf{\g{s}}_{\g{j}}}(\mathbf{\b{F}}\,\mathbf{\b{F}}\,\mathbf{\r{ad}}|\mathbf{\b{F}}\,\mathbf{\b{F}})\;\bigger{\Leftrightarrow}\;}\tikzBox{ffgff_alt_basis_diagram}{
\arrowTo[hblue]{0,0}{-150}\node[anchor=10,inner sep=2pt] at(in){{\footnotesize$\b{[f]}$}};\arrowFrom[hblue]{0,0}{-30};\node[anchor=170,inner sep=2pt] at(end){{\footnotesize$\b{[f]}$}};\arrowFrom[hteal]{0,0}[1]{90}\node[clebsch]at(in){};\node[anchor=180,inner sep=2pt] at(arrownode){{\footnotesize$\t{[r_i]}$}};\arrowTo[hblue]{end}{180}\node[anchor=0,inner sep=0pt] at(in){{\footnotesize$\b{[f]}$}};\arrowFrom[hgreen]{end}{90}\node[anchor=180,inner sep=2pt] at(arrownode){{\footnotesize$\g{[s_j]}$}};\node[clebsch]at(in){};\arrowTo[hred]{end}{150};\node[anchor=-10,inner sep=2pt] at(in){{\footnotesize$\r{[\adR]}$}};\arrowFrom[hblue]{end}{30};\node[anchor=190,inner sep=2pt] at(end){{\footnotesize$\b{[f]}$}};\node[clebsch]at(in){};
}\,.\vspace{-10pt}}
Here, $\t{\mathbf{r}_{i}}$ are defined in (\ref{rReps_defined}) and $\g{\mathbf{s}_{j}}$ defined in (\ref{sReps_defined}) and given in \mbox{Table~\ref{sReps_table}}. The possible pairs which label the complete basis of tensors for each simple Lie algebra are given in \mbox{Table~\ref{ffgff_basis_c_label_table}}.

\begin{table}[t]\vspace{-9pt}\caption{Labels $(\hspace{-1pt}\t{\mathbf{r}_i}\,\g{\mathbf{s}_j}\hspace{-1pt})$ for basis tensors $\mathcal{C}^{\hspace{1pt}\mathbf{\t{r}}_{\t{i}}\mathbf{\g{s}}_{\g{j}}}(\mathbf{\b{F}}\,\mathbf{\b{F}}\,\mathbf{\r{ad}}|\mathbf{\b{F}}\,\mathbf{\b{F}})$ of simple Lie algebras.}\label{ffgff_basis_c_label_table}
\vspace{-15pt}$$\hspace{-200pt}\begin{array}{lr@{$\;\;$}|rc@{$\,\,$}c@{$\,\,$}c@{$\,\,$}c@{$\,\,$}c@{$\,\,$}c@{$\,\,$}c@{$\,\,$}l|l}%\multicolumn{11}{c}{~}\\[
&\multicolumn{1}{r}{\fwboxR{0pt}{\mathfrak{g}\hspace{22pt}}}&\multicolumn{10}{r}{\fwboxL{0pt}{\hspace{13pt}\text{\#}}}\\[-10pt]\cline{3-11}
&\mathfrak{a}_{\r{1}}%
&\fwboxR{0pt}{\big\{\!}\rule{0pt}{14pt}(\hspace{-1pt}\t{\mathbf{r}_{1}}\hspace{1pt}\g{\mathbf{s}_{1}}\hspace{-1pt})\fwboxL{0pt}{,}&(\hspace{-1pt}\t{\mathbf{r}_{2}}\hspace{1pt}\g{\mathbf{s}_{1}}\hspace{-1pt})\fwboxL{0pt}{,}&(\hspace{-1pt}\t{\mathbf{r}_{2}}\hspace{1pt}\g{\mathbf{s}_{2}}\hspace{-1pt})\fwboxL{0pt}{\!\big\}}&&&&&&&\multirow{1}{*}{$\fwboxL{0pt}{\hspace{4pt}3}$}\\[2pt]\cline{3-11}
&\mathfrak{a}_{\r{k}>1}%
&\fwboxR{0pt}{\big\{\!}\rule{0pt}{14pt}(\hspace{-1pt}\t{\mathbf{r}_{1}}\hspace{1pt}\g{\mathbf{s}_{1}}\hspace{-1pt})\fwboxL{0pt}{,}&(\hspace{-1pt}\t{\mathbf{r}_{2}}\hspace{1pt}\g{\mathbf{s}_{1}}\hspace{-1pt})\fwboxL{0pt}{,}&(\hspace{-1pt}\t{\mathbf{r}_{2}}\hspace{1pt}\g{\mathbf{s}_{2}}\hspace{-1pt})\fwboxL{0pt}{,}&(\hspace{-1pt}\t{\mathbf{r}_{2}}\hspace{1pt}\g{\mathbf{s}_{3}}\hspace{-1pt})\fwboxL{0pt}{\!\big\}}&&&&&&\multirow{1}{*}{$\fwboxL{0pt}{\hspace{4pt}4}$}\\[2pt]\cline{3-11}
&\fwboxR{0pt}{\mathfrak{b}_{\r{k}},\mathfrak{c}_{\r{k}},\mathfrak{d}_{\r{k}}}
&\fwboxR{0pt}{\big\{\!}\rule{0pt}{14pt}(\hspace{-1pt}\t{\mathbf{r}_{1}}\hspace{1pt}\g{\mathbf{s}_{1}}\hspace{-1pt})\fwboxL{0pt}{,}&(\hspace{-1pt}\t{\mathbf{r}_{2}}\hspace{1pt}\g{\mathbf{s}_{1}}\hspace{-1pt})\fwboxL{0pt}{,}&(\hspace{-1pt}\t{\mathbf{r}_{2}}\hspace{1pt}\g{\mathbf{s}_{2}}\hspace{-1pt})\fwboxL{0pt}{,}&(\hspace{-1pt}\t{\mathbf{r}_{2}}\hspace{1pt}\g{\mathbf{s}_{3}}\hspace{-1pt})\fwboxL{0pt}{,}&(\hspace{-1pt}\t{\mathbf{r}_{3}}\hspace{1pt}\g{\mathbf{s}_{1}}\hspace{-1pt})\fwboxL{0pt}{,}&(\hspace{-1pt}\t{\mathbf{r}_{3}}\hspace{1pt}\g{\mathbf{s}_{3}}\hspace{-1pt})\fwboxL{0pt}{\!\big\}}&&&&\multirow{1}{*}{$\fwboxL{0pt}{\hspace{4pt}6}$}\\[2pt]\cline{3-11}
&\fwboxR{0pt}{\mathfrak{e}_{\r{6}}}&
\fwboxR{0pt}{\big\{\!}\rule{0pt}{14pt}(\hspace{-1pt}\t{\mathbf{r}_{1}}\hspace{1pt}\g{\mathbf{s}_{1}}\hspace{-1pt})\fwboxL{0pt}{,}&(\hspace{-1pt}\t{\mathbf{r}_{2}}\hspace{1pt}\g{\mathbf{s}_{1}}\hspace{-1pt})\fwboxL{0pt}{,}&(\hspace{-1pt}\t{\mathbf{r}_{2}}\hspace{1pt}\g{\mathbf{s}_{2}}\hspace{-1pt})\fwboxL{0pt}{,}&(\hspace{-1pt}\t{\mathbf{r}_{2}}\hspace{1pt}\g{\mathbf{s}_{3}}\hspace{-1pt})\fwboxL{0pt}{,}&(\hspace{-1pt}\t{\mathbf{r}_{3}}\hspace{1pt}\g{\mathbf{s}_{1}}\hspace{-1pt})\fwboxL{0pt}{,}&(\hspace{-1pt}\t{\mathbf{r}_{3}}\hspace{1pt}\g{\mathbf{s}_{2}}\hspace{-1pt})\fwboxL{0pt}{,}&(\hspace{-1pt}\t{\mathbf{r}_{3}}\hspace{1pt}\g{\mathbf{s}_{3}}\hspace{-1pt})\fwboxL{0pt}{\!\big\}}&&&\multirow{1}{*}{$\fwboxL{0pt}{\hspace{4pt}7}$}\\[2pt]\cline{3-11}
&\multirow{1}{*}{$\fwboxR{0pt}{\mathfrak{e}_{\r{7}}}$}&
\fwboxR{5pt}{\big\{\!}\rule{0pt}{14pt}(\hspace{-1pt}\t{\mathbf{r}_{1}}\hspace{1pt}\g{\mathbf{s}_{1}}\hspace{-1pt})\fwboxL{0pt}{,}&(\hspace{-1pt}\t{\mathbf{r}_{2}}\hspace{1pt}\g{\mathbf{s}_{1}}\hspace{-1pt})\fwboxL{0pt}{,}&(\hspace{-1pt}\t{\mathbf{r}_{2}}\hspace{1pt}\g{\mathbf{s}_{2}}\hspace{-1pt})\fwboxL{0pt}{,}&(\hspace{-1pt}\t{\mathbf{r}_{2}}\hspace{1pt}\g{\mathbf{s}_{3}}\hspace{-1pt})\fwboxL{0pt}{,}&(\hspace{-1pt}\t{\mathbf{r}_{3}}\hspace{1pt}\g{\mathbf{s}_{1}}\hspace{-1pt})\fwboxL{0pt}{,}&(\hspace{-1pt}\t{\mathbf{r}_{3}}\hspace{1pt}\g{\mathbf{s}_{2}}\hspace{-1pt})\fwboxL{0pt}{,}&(\hspace{-1pt}\t{\mathbf{r}_{3}}\hspace{1pt}\g{\mathbf{s}_{3}}\hspace{-1pt})\fwboxL{0pt}{,}&(\hspace{-1pt}\t{\mathbf{r}_{4}}\hspace{1pt}\g{\mathbf{s}_{1}}\hspace{-1pt})\fwboxL{0pt}{,}&(\hspace{-1pt}\t{\mathbf{r}_{4}}\hspace{1pt}\g{\mathbf{s}_{3}}\hspace{-1pt})\fwboxL{5pt}{\!\big\}}&\multirow{1}{*}{$\fwboxL{0pt}{\hspace{4pt}9}$}\\[2pt]\cline{3-11}
&\multirow{2}{*}{$\fwboxR{0pt}{\mathfrak{f}_{\r{4}}}$}&
\fwboxR{5pt}{\big\{\!}\rule{0pt}{14pt}(\hspace{-1pt}\t{\mathbf{r}_{1}}\hspace{1pt}\g{\mathbf{s}_{1}}\hspace{-1pt})\fwboxL{0pt}{,}&(\hspace{-1pt}\t{\mathbf{r}_{2}}\hspace{1pt}\g{\mathbf{s}_{1}}\hspace{-1pt})\fwboxL{0pt}{,}&(\hspace{-1pt}\t{\mathbf{r}_{2}}\hspace{1pt}\g{\mathbf{s}_{2}}\hspace{-1pt})\fwboxL{0pt}{,}&(\hspace{-1pt}\t{\mathbf{r}_{2}}\hspace{1pt}\g{\mathbf{s}_{3}}\hspace{-1pt})\fwboxL{0pt}{,}&(\hspace{-1pt}\t{\mathbf{r}_{3}}\hspace{1pt}\g{\mathbf{s}_{1}}\hspace{-1pt})\fwboxL{0pt}{,}&(\hspace{-1pt}\t{\mathbf{r}_{3}}\hspace{1pt}\g{\mathbf{s}_{2}}\hspace{-1pt})\fwboxL{0pt}{,}&(\hspace{-1pt}\t{\mathbf{r}_{3}}\hspace{1pt}\g{\mathbf{s}_{3}}\hspace{-1pt})\fwboxL{0pt}{,}&(\hspace{-1pt}\t{\mathbf{r}_{4}}\hspace{1pt}\g{\mathbf{s}_{1}}\hspace{-1pt})\fwboxL{0pt}{,}&(\hspace{-1pt}\t{\mathbf{r}_{4}}\hspace{1pt}\g{\mathbf{s}_{2}}\hspace{-1pt})&\multirow{2}{*}{$\fwboxL{0pt}{\hspace{4pt}12}$}\\[2pt]
&&(\hspace{-1pt}\t{\mathbf{r}_{4}}\hspace{1pt}\g{\mathbf{s}_{3}}\hspace{-1pt})\fwboxL{0pt}{,}&(\hspace{-1pt}\t{\mathbf{r}_{5}}\hspace{1pt}\g{\mathbf{s}_{1}}\hspace{-1pt})\fwboxL{0pt}{,}&(\hspace{-1pt}\t{\mathbf{r}_{5}}\hspace{1pt}\g{\mathbf{s}_{2}}\hspace{-1pt})\fwboxL{0pt}{\!\big\}}&&&&&&\\[2pt]\cline{3-11}
&\fwboxR{0pt}{\mathfrak{g}_{\r{2}}}&
\fwboxR{5pt}{\big\{\!}\rule{0pt}{14pt}(\hspace{-1pt}\t{\mathbf{r}_{1}}\hspace{1pt}\g{\mathbf{s}_{1}}\hspace{-1pt})\fwboxL{0pt}{,}&(\hspace{-1pt}\t{\mathbf{r}_{2}}\hspace{1pt}\g{\mathbf{s}_{1}}\hspace{-1pt})\fwboxL{0pt}{,}&(\hspace{-1pt}\t{\mathbf{r}_{2}}\hspace{1pt}\g{\mathbf{s}_{2}}\hspace{-1pt})\fwboxL{0pt}{,}&(\hspace{-1pt}\t{\mathbf{r}_{2}}\hspace{1pt}\g{\mathbf{s}_{3}}\hspace{-1pt})\fwboxL{0pt}{,}&(\hspace{-1pt}\t{\mathbf{r}_{3}}\hspace{1pt}\g{\mathbf{s}_{1}}\hspace{-1pt})\fwboxL{0pt}{,}&(\hspace{-1pt}\t{\mathbf{r}_{3}}\hspace{1pt}\g{\mathbf{s}_{2}}\hspace{-1pt})\fwboxL{0pt}{,}&(\hspace{-1pt}\t{\mathbf{r}_{3}}\hspace{1pt}\g{\mathbf{s}_{3}}\hspace{-1pt})\fwboxL{0pt}{,}&(\hspace{-1pt}\t{\mathbf{r}_{4}}\hspace{1pt}\g{\mathbf{s}_{1}}\hspace{-1pt})\fwboxL{0pt}{,}&(\hspace{-1pt}\t{\mathbf{r}_{4}}\hspace{1pt}\g{\mathbf{s}_{2}}\hspace{-1pt})\fwboxL{0pt}{\!\big\}}&\multirow{1}{*}{$\fwboxL{0pt}{\hspace{4pt}9}$}\\[2pt]\cline{3-11}
\end{array}\hspace{-200pt}$$\vspace{-20pt}\end{table}

It is straightforward to compute the duality matrices
\eq{\mathcal{C}^{\t{i}}\equivL\sum_{\g{j}}\mathbf{c}[\mathfrak{g}]\indices{\t{i}}{\g{j}}\mathcal{B}^{\g{j}}\,.}
For $\mathfrak{a}_{\r{1}}$, we find
\eq{\fwbox{0pt}{\fwboxL{567pt}{(\hspace{-1.25pt}%
\mathfrak{a}_{\r{1}}
)}}\hspace{-100pt}\fwboxR{0pt}{\mathbf{c}[\mathfrak{a}_{\r{1}}]=}\left(\begin{array}{c@{$\;\;$}c@{$\;\;$}c}
\frac{1}{2}&\frac{1}{2}&\tmi\frac{1}{2}\\[-1pt]
\frac{3}{4}&\tmi\frac{1}{4}&\frac{1}{4}\\[-1pt]
\dzero&\frac{2}{3}&\frac{1}{3}\end{array}\right).\hspace{-100pt}\label{ffgff_a1_duality_matrix}}

\subsubsection{\texorpdfstring{Clebsch Colour Bases for $C(\mathbf{\b{F}}\,\mathbf{\r{ad}}\,\mathbf{\r{ad}}|\mathbf{\r{ad}}\,\mathbf{\b{F}})$}{Clebsch Colour Bases for C(Fgg|gF)}}

For two fundamental-charged particles and three adjoints, we may consider the basis of colour tensors generated by
\eq{\fwboxR{0pt}{\mathcal{B}_{\hspace{3.5pt}\mu}^{\hspace{1pt}\mathbf{\g{s}}_{\g{i}}\mathbf{\g{s}}_{\g{j}}}(\mathbf{\b{F}}\,\mathbf{\r{ad}}\,\mathbf{\r{ad}}|\mathbf{\r{ad}}\,\mathbf{\b{F}})\;\bigger{\Leftrightarrow}\;}\tikzBox[-5.25pt]{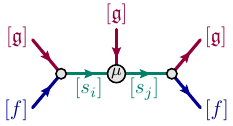}{\arrowTo[hblue]{0,0}{-130};\node[anchor=10,inner sep=2pt] at(in){{\footnotesize$\b{[f]}$}};\arrowTo[hred]{0,0}{130}\node[anchor=-10,inner sep=2pt] at(in){{\footnotesize$\r{[\adR]}$}};\arrowFrom[hgreen]{0,0}[1.25]{0}\node[anchor=90,inner sep=2pt] at(arrownode){{\footnotesize$\g{[s_i]}$}};\node[clebsch]at(in){};\arrowTo[hred]{end}{90}\node[anchor=-90,inner sep=2pt] at(in){{\footnotesize$\r{[\adR]}$}};\arrowFrom[hgreen]{end}[1.25]{0}\node[clebsch]at(in){{\scriptsize{$\phantom{\nu}$}}};\node[]at(in){{\scriptsize{$\mu$}}};\node[anchor=90,inner sep=2pt] at(arrownode){{\footnotesize$\g{[s_j]}$}};\arrowFrom[hred]{end}{50}
\node[anchor=-170,inner sep=2pt] at(end){{\footnotesize$\r{[\adR]}$}};\arrowFrom[hblue]{in}{-50}\node[anchor=170,inner sep=2pt] at(end){{\footnotesize$\b{[f]}$}};\node[clebsch]at(in){};
}.
}
These tensors sometimes involve a multiplicity index $\mu\!\in\![m\indices{\g{\mathbf{s}_{i}}\mathbf{\r{ad}}}{\g{\mathbf{s}_{j}}}]$. The labels which generate the basis for each of the simple Lie algebras are enumerated in \mbox{Table~\ref{fgggf_basis_b_label_table}}.

\begin{table}[t]\vspace{-9pt}\caption{Labels $(\hspace{-1pt}\g{\mathbf{s}_i}\,\g{\mathbf{s}_j}\hspace{-1pt})$ for basis tensors $\mathcal{B}_{\hspace{3.5pt}\mu}^{\hspace{1pt}\mathbf{\g{s}}_{\g{i}}\mathbf{\g{s}}_{\g{j}}}(\mathbf{\b{F}}\,\mathbf{\r{ad}}\,\mathbf{\r{ad}}|\mathbf{\r{ad}}\,\mathbf{\b{F}})$ of simple Lie algebras.}\label{fgggf_basis_b_label_table}
\vspace{-15pt}
$$\hspace{-200pt}\begin{array}{lr@{$\;\;$}|rl@{$\,\,$}l@{$\,\,$}l@{$\,\,$}l|l}
%&\multicolumn{1}{r}{\fwboxR{0pt}{\mathfrak{g}\hspace{16pt}}}&\multicolumn{6}{r}{\fwboxL{0pt}{\hspace{3pt}\text{\#}}}\\\cline{3-7}
&\multicolumn{1}{r}{\fwboxR{0pt}{\mathfrak{g}\hspace{22pt}}}&\multicolumn{6}{r}{\fwboxL{0pt}{\hspace{13pt}\text{\#}}}\\[-10pt]\cline{3-7}
&\multirow{1}{*}{$\fwboxR{0pt}{\mathfrak{a}_{\r{1}}}$}&%
\fwboxR{5pt}{\big\{\!}\rule{0pt}{14pt}(\hspace{-1pt}\g{\mathbf{s}_{1}}\hspace{1pt}\g{\mathbf{s}_{1}}\hspace{-1pt})\fwboxL{0pt}{\hspace{-1pt},}&(\hspace{-1pt}\g{\mathbf{s}_{1}}\hspace{1pt}\g{\mathbf{s}_{2}}\hspace{-1pt})\fwboxL{0pt}{\hspace{-1pt},}&(\hspace{-1pt}\g{\mathbf{s}_{2}}\hspace{1pt}\g{\mathbf{s}_{1}}\hspace{-1pt})\fwboxL{0pt}{\hspace{-1pt},}&(\hspace{-1pt}\g{\mathbf{s}_{2}}\hspace{1pt}\g{\mathbf{s}_{2}}\hspace{-1pt})\fwboxL{0pt}{\!\big\}}&&\multirow{1}{*}{$\fwboxL{0pt}{\hspace{4pt}4}$}\\[2pt]\cline{3-7}
&\multirow{2}{*}{$\fwboxR{0pt}{\mathfrak{a}_{\r{2}}}$}&%
\fwboxR{5pt}{\big\{\!}\rule{0pt}{14pt}(\hspace{-1pt}\g{\mathbf{s}_{1}}\hspace{1pt}\g{\mathbf{s}_{1}}\hspace{-1pt})\fwboxL{0pt}{\hspace{-1pt},}&(\hspace{-1pt}\g{\mathbf{s}_{1}}\hspace{1pt}\g{\mathbf{s}_{2}}\hspace{-1pt})\fwboxL{0pt}{\hspace{-1pt},}&(\hspace{-1pt}\g{\mathbf{s}_{1}}\hspace{1pt}\g{\mathbf{s}_{3}}\hspace{-1pt})\fwboxL{0pt}{\hspace{-1pt},}&(\hspace{-1pt}\g{\mathbf{s}_{2}}\hspace{1pt}\g{\mathbf{s}_{1}}\hspace{-1pt})\fwboxL{0pt}{\hspace{-1pt},}&(\hspace{-1pt}\g{\mathbf{s}_{2}}\hspace{1pt}\g{\mathbf{s}_{2}}\hspace{-1pt})\fwboxL{22pt}{\hspace{0pt}{}_{\mu\in[2]}}\fwboxL{0pt}{\hspace{-2pt},}&\multirow{2}{*}{$\fwboxL{0pt}{\hspace{4pt}10}$}\\[2pt]
&&(\hspace{-1pt}\g{\mathbf{s}_{2}}\hspace{1pt}\g{\mathbf{s}_{3}}\hspace{-1pt})\fwboxL{0pt}{\hspace{-1pt},}&(\hspace{-1pt}\g{\mathbf{s}_{3}}\hspace{1pt}\g{\mathbf{s}_{1}}\hspace{-1pt})\fwboxL{0pt}{\hspace{-1pt},}&(\hspace{-1pt}\g{\mathbf{s}_{3}}\hspace{1pt}\g{\mathbf{s}_{2}}\hspace{-1pt})\fwboxL{0pt}{\hspace{-1pt},}&(\hspace{-1pt}\g{\mathbf{s}_{3}}\hspace{1pt}\g{\mathbf{s}_{3}}\hspace{-1pt})\fwboxL{0pt}{\hspace{0pt}\!\big\}}&\\[2pt]\cline{3-7}
&\multirow{2}{*}{$\fwboxR{0pt}{\begin{array}{@{}r@{}}\mathfrak{a}_{\r{k}>2}\fwboxL{0pt}{,}\\\mathfrak{d}_{\r{4}}\end{array}}$}&
\fwboxR{5pt}{\big\{\!}\rule{0pt}{14pt}(\hspace{-1pt}\g{\mathbf{s}_{1}}\hspace{1pt}\g{\mathbf{s}_{1}}\hspace{-1pt})\fwboxL{0pt}{\hspace{-1pt},}&(\hspace{-1pt}\g{\mathbf{s}_{1}}\hspace{1pt}\g{\mathbf{s}_{2}}\hspace{-1pt})\fwboxL{0pt}{\hspace{-1pt},}&(\hspace{-1pt}\g{\mathbf{s}_{1}}\hspace{1pt}\g{\mathbf{s}_{3}}\hspace{-1pt})\fwboxL{0pt}{\hspace{-1pt},}&(\hspace{-1pt}\g{\mathbf{s}_{2}}\hspace{1pt}\g{\mathbf{s}_{1}}\hspace{-1pt})\fwboxL{0pt}{\hspace{-1pt},}&(\hspace{-1pt}\g{\mathbf{s}_{2}}\hspace{1pt}\g{\mathbf{s}_{2}}\hspace{-1pt})\fwboxL{12pt}{\hspace{0pt}{}_{\mu\in[2]}}\fwboxL{0pt}{\hspace{7pt},}&\multirow{2}{*}{$\fwboxL{0pt}{\hspace{4pt}11}$}\\
&&(\hspace{-1pt}\g{\mathbf{s}_{2}}\hspace{1pt}\g{\mathbf{s}_{3}}\hspace{-1pt})\fwboxL{0pt}{\hspace{-1pt},}&(\hspace{-1pt}\g{\mathbf{s}_{3}}\hspace{1pt}\g{\mathbf{s}_{1}}\hspace{-1pt})\fwboxL{0pt}{\hspace{-1pt},}&(\hspace{-1pt}\g{\mathbf{s}_{3}}\hspace{1pt}\g{\mathbf{s}_{2}}\hspace{-1pt})\fwboxL{0pt}{\hspace{-1pt},}&(\hspace{-1pt}\g{\mathbf{s}_{3}}\hspace{1pt}\g{\mathbf{s}_{3}}\hspace{-1pt})\fwboxL{0pt}{\hspace{0pt}{}_{\mu\in[2]}}\fwboxL{0pt}{\hspace{19pt}\big\}}&\\[2pt]\cline{3-7}
&\multirow{2}{*}{$\fwboxR{0pt}{\begin{array}{@{}r@{}}\mathfrak{b}_{\r{k}},\mathfrak{c}_{\r{k}}\mathfrak{d}_{\r{k}>4}\fwboxL{0pt}{,}\\\mathfrak{e}_{\r{6}},\mathfrak{e}_{\r{7}},\mathfrak{f}_{\r{4}},\mathfrak{g}_{\r{2}}\end{array}}$}&
\fwboxR{5pt}{\big\{\!}\rule{0pt}{14pt}(\hspace{-1pt}\g{\mathbf{s}_{1}}\hspace{1pt}\g{\mathbf{s}_{1}}\hspace{-1pt})\fwboxL{0pt}{\hspace{-1pt},}&(\hspace{-1pt}\g{\mathbf{s}_{1}}\hspace{1pt}\g{\mathbf{s}_{2}}\hspace{-1pt})\fwboxL{0pt}{\hspace{-1pt},}&(\hspace{-1pt}\g{\mathbf{s}_{1}}\hspace{1pt}\g{\mathbf{s}_{3}}\hspace{-1pt})\fwboxL{0pt}{\hspace{-1pt},}&(\hspace{-1pt}\g{\mathbf{s}_{2}}\hspace{1pt}\g{\mathbf{s}_{1}}\hspace{-1pt})\fwboxL{0pt}{\hspace{-1pt},}&(\hspace{-1pt}\g{\mathbf{s}_{2}}\hspace{1pt}\g{\mathbf{s}_{2}}\hspace{-1pt})\fwboxL{0pt}{\hspace{-1pt},}&\multirow{2}{*}{$\fwboxL{0pt}{\hspace{4pt}10}$}\\
&&(\hspace{-1pt}\g{\mathbf{s}_{2}}\hspace{1pt}\g{\mathbf{s}_{3}}\hspace{-1pt})\fwboxL{0pt}{\hspace{-1pt},}&(\hspace{-1pt}\g{\mathbf{s}_{3}}\hspace{1pt}\g{\mathbf{s}_{1}}\hspace{-1pt})\fwboxL{0pt}{\hspace{-1pt},}&(\hspace{-1pt}\g{\mathbf{s}_{3}}\hspace{1pt}\g{\mathbf{s}_{2}}\hspace{-1pt})\fwboxL{0pt}{\hspace{-1pt},}&(\hspace{-1pt}\g{\mathbf{s}_{3}}\hspace{1pt}\g{\mathbf{s}_{3}}\hspace{-1pt})\fwboxL{0pt}{\hspace{0pt}{}_{\mu\in[2]}}\fwboxL{0pt}{\hspace{19pt}\big\}}&\\[2pt]\cline{3-7}
\end{array}\hspace{-200pt}$$\vspace{-20pt}\end{table}

\newpage
Alternatively, one could consider a basis of Clebsch colour tensors generated by the tree graph:
\vspace{-5pt}\eq{\fwboxR{0pt}{\mathcal{C}_{\hspace{2.5pt}\mu\nu}^{\hspace{1pt}\mathbf{\t{r}}_{\t{i}}\mathbf{\b{t}}_{\b{j}}}(\mathbf{\b{F}}\,\mathbf{\r{ad}}\,\mathbf{\r{ad}}|\mathbf{\r{ad}}\,\mathbf{\b{F}})\;\bigger{\Leftrightarrow}\;}\tikzBox{fgggf_alt_basis_diagram}{
\arrowTo[hblue]{0,0}{-150}\node[anchor=10,inner sep=2pt] at(in){{\footnotesize$\b{[f]}$}};\arrowFrom[hblue]{0,0}{-30};\node[anchor=170,inner sep=2pt] at(end){{\footnotesize$\b{[f]}$}};\arrowFrom[hteal]{0,0}[1]{90}\node[clebsch]at(in){};\node[anchor=180,inner sep=2pt] at(arrownode){{\footnotesize$\t{[r_i]}$}};\arrowTo[hred]{end}{180}\node[anchor=0,inner sep=0pt] at(in){{\footnotesize$\r{[\adR]}$}};\arrowFrom[hblue]{end}{90}\node[anchor=180,inner sep=2pt] at(arrownode){{\footnotesize$\b{[t_j]}$}};\node[clebsch]at(in){{\scriptsize$\phantom{\nu}$}};\node[]at(in){{\scriptsize${\mu}$}};\arrowTo[hred]{end}{150};\node[anchor=-10,inner sep=2pt] at(in){{\footnotesize$\r{[\adR]}$}};\arrowFrom[hred]{end}{30};\node[anchor=190,inner sep=2pt] at(end){{\footnotesize$\r{[\adR]}$}};\node[clebsch]at(in){{\scriptsize$\phantom{\nu}$}};\node[]at(in){{\scriptsize${\nu}$}};
}\,.
\vspace{-2.5pt}}
These colour tensors are generally labelled by both the pair of irreducible representations $(\t{\mathbf{r}_i}\,\b{\mathbf{t}_j})$, with $\t{\mathbf{r}_i}$ defined in (\ref{rReps_defined}) and $\b{\mathbf{t}_j}$ enumerated for the simple Lie algebras in \mbox{Table~\ref{tReps_table}}, and also the multiplicity indices $\mu\!\in\![m\indices{\t{\mathbf{r}_i}\,\r{\mathbf{ad}}}{\b{\mathbf{t}_j}}]$ and $\nu\!\in\![m\indices{\b{\mathbf{t}_j}\,\mathbf{\r{ad}}}{\mathbf{\r{ad}}}]$. While it would be somewhat cumbersome to enumerate all the label sets for the $\mathcal{C}$-basis in general, for $\mathfrak{a}_{\r{1}}$, these are given by:
%a1 C-basis:
\eq{\fwbox{0pt}{\fwboxL{435pt}{(\hspace{-1.25pt}%
\mathfrak{a}_{\r{1}}
)}}\fwbox{0pt}{\begin{array}{c}
\rule{0pt}{14pt}%
\big\{(\hspace{-1pt}\mathbf{\t{1}}\hspace{-0pt}\mathbf{\b{3}}\hspace{-1pt}),(\hspace{-1pt}\mathbf{\t{3}}\hspace{-0pt}\mathbf{\b{1}}\hspace{-1pt}),(\hspace{-1pt}\mathbf{\t{3}}\hspace{-0pt}\mathbf{\b{3}}\hspace{-1pt}),(\hspace{-1pt}\mathbf{\t{3}}\hspace{-0pt}\mathbf{\b{5}}\hspace{-1pt})\big\}\fwboxL{0pt}{\,.}
%
%\\~
\end{array}}\label{fgggf_labels_c_basis_a1}\vspace{-5pt}}

It is interesting to consider the duality matrices
\eq{\mathcal{C}^{\t{i}}\equivL\sum_{\g{j}}\mathbf{c}[\mathfrak{g}]\indices{\t{i}}{\g{j}}\mathcal{B}^{\g{j}}\,.}
For $\mathfrak{a}_{\r{1}}$, we find
\eq{\fwbox{0pt}{\fwboxL{550pt}{(\hspace{-1.25pt}%
\mathfrak{a}_{\r{1}}
)}}\hspace{-100pt}\fwboxR{0pt}{\mathbf{c}[\mathfrak{a}_{\r{1}}]=}\displaystyle\left(\begin{array}{c@{$\;\;$}c@{$\;\;$}c@{$\;\;$}c}
\tmi\frac{8}{9}&\frac{2}{3}&\frac{2}{3}&\tmi\frac{2}{3}\\[-1pt]
\frac{2}{3}&1&\dzero&\dzero\\[-1pt]
\frac{16}{9}&\tmi\frac{4}{3}&\frac{2}{3}&\tmi\frac{2}{3}\\[-1pt]
\dzero&\dzero&\frac{5}{6}&\frac{1}{6}\end{array}\right).\hspace{-100pt}\label{fgggf_a1_duality_matrix}}

\newpage

\subsubsection{\texorpdfstring{Clebsch Colour Bases for $C(\mathbf{\r{ad}}\,\mathbf{\r{ad}}\,\mathbf{\r{ad}}|\mathbf{\r{ad}}\,\mathbf{\r{ad}})$}{Clebsch Colour Bases for C(ggg|gg)}}

For the scattering of 5 adjoint-coloured particles, one may consider the basis of Clebsch colour tensors given by
\eq{\fwboxR{0pt}{\mathcal{B}^{\hspace{2pt}\mathbf{\b{t}}_{\b{i}}\mathbf{\b{t}}_{\b{j}}}_{\mu\nu\rho}(\mathbf{\r{ad}}\,\mathbf{\r{ad}}\,\mathbf{\r{ad}}|\mathbf{\r{ad}}\,\mathbf{\r{ad}})\;\bigger{\Leftrightarrow}\;}\tikzBox[-5.25pt]{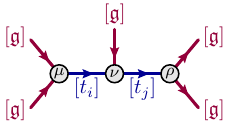}{\arrowTo[hred]{0,0}{-130};\node[anchor=10,inner sep=2pt] at(in){{\footnotesize$\r{[\adR]}$}};\arrowTo[hred]{0,0}{130}\node[anchor=-10,inner sep=2pt] at(in){{\footnotesize$\r{[\adR]}$}};\arrowFrom[hblue]{0,0}[1.25]{0}\node[anchor=90,inner sep=2pt] at(arrownode){{\footnotesize$\b{[t_i]}$}};\node[clebsch]at(in){{\scriptsize$\phantom{\nu}$}};\node[]at(in){{\scriptsize$\mu$}};\arrowTo[hred]{end}{90}\node[anchor=-90,inner sep=2pt] at(in){{\footnotesize$\r{[\adR]}$}};\arrowFrom[hblue]{end}[1.25]{0}\node[clebsch]at(in){{\scriptsize$\phantom{\nu}$}};\node[]at(in){{\scriptsize$\nu$}};\node[anchor=90,inner sep=2pt] at(arrownode){{\footnotesize$\b{[t_j]}$}};\arrowFrom[hred]{end}{50}
\node[anchor=-170,inner sep=2pt] at(end){{\footnotesize$\r{[\adR]}$}};\arrowFrom[hred]{in}{-50}\node[anchor=170,inner sep=2pt] at(end){{\footnotesize$\r{[\adR]}$}};\node[clebsch]at(in){{\scriptsize$\phantom{\nu}$}};\node[]at(in){{\scriptsize$\rho$}};
}\,.
}
These are labelled both by a pair of irreducible representations $(\b{\mathbf{t}_i\,\mathbf{t}_j})$ defined in \mbox{Table~\ref{tReps_table}} and also multiplicity indices $\mu\!\in\![m\indices{\r{\mathbf{ad\,ad}}}{\b{\mathbf{t}_i}}]$, $\nu\!\in\![m\indices{\b{\mathbf{t}_i}\,\r{\mathbf{ad}}}{\b{\mathbf{t}_j}}]$, and $\rho\!\in\![m\indices{\b{\mathbf{t}_j}\,\r{\mathbf{ad}}}{\r{\mathbf{ad}}}]$. Conveniently, for most simple Lie algebras, few multiplicity indices are actually required. The complete list of these (according to the indexing given in \mbox{Table~\ref{tReps_table}}) are listed in \mbox{Table~\ref{ggggg_basis_b_label_table}}.

\begin{table}[t]\vspace{-9pt}\caption{Labels $(\hspace{-1pt}\b{\mathbf{t}_i}\,\b{\mathbf{t}_j}\hspace{-1pt})$ for basis tensors $\mathcal{B}^{\hspace{2pt}\mathbf{\b{t}}_{\b{i}}\mathbf{\b{t}}_{\b{j}}}_{\mu\nu\rho}(\mathbf{\r{ad}}\,\mathbf{\r{ad}}\,\mathbf{\r{ad}}|\mathbf{\r{ad}}\,\mathbf{\r{ad}})$ of simple Lie algebras.}\label{ggggg_basis_b_label_table}
\vspace{-15pt}
$$\hspace{-200pt}\begin{array}{lr@{$\;\;$}|rl@{$\,\,$}l@{$\,\,$}l@{$\,\,$}l@{$\,\,$}l@{$\,\,$}l@{$\,\,$}l@{$\,\,$}l|l}
%&\multicolumn{1}{r}{\fwboxR{0pt}{\mathfrak{g}\hspace{16pt}}}&\multicolumn{10}{r}{\fwboxL{0pt}{\hspace{3pt}\text{\#}}}\\\cline{3-11}
&\multicolumn{1}{r}{\fwboxR{0pt}{\mathfrak{g}\hspace{22pt}}}&\multicolumn{10}{r}{\fwboxL{0pt}{\hspace{13pt}\text{\#}}}\\[-10pt]\cline{3-11}
&\multirow{1}{*}{$\fwboxR{0pt}{\begin{array}{@{}r@{}}\mathfrak{a}_{\r{1}}\end{array}}$}&
\fwboxR{5pt}{\big\{\!}\rule{0pt}{14pt}(\hspace{-1pt}\b{\mathbf{t}_{1}}\hspace{1pt}\b{\mathbf{t}_{2}}\hspace{-1pt})\fwboxL{0pt}{\hspace{-1pt},}&(\hspace{-1pt}\b{\mathbf{t}_{2}}\hspace{1pt}\b{\mathbf{t}_{1}}\hspace{-1pt})\fwboxL{0pt}{\hspace{-1pt},}&(\hspace{-1pt}\b{\mathbf{t}_{2}}\hspace{1pt}\b{\mathbf{t}_{2}}\hspace{-1pt})\fwboxL{0pt}{\hspace{-1pt},}&(\hspace{-1pt}\b{\mathbf{t}_{2}}\hspace{1pt}\b{\mathbf{t}_{3}}\hspace{-1pt})\fwboxL{0pt}{\hspace{-1pt},}&(\hspace{-1pt}\b{\mathbf{t}_{3}}\hspace{1pt}\b{\mathbf{t}_{2}}\hspace{-1pt})\fwboxL{0pt}{\hspace{-1pt},}&(\hspace{-1pt}\b{\mathbf{t}_{3}}\hspace{1pt}\b{\mathbf{t}_{3}}\hspace{-1pt})\fwboxL{0pt}{\!\big\}}&&&&\multirow{1}{*}{$\fwboxL{0pt}{\hspace{4pt}6}$}\\[2pt]\cline{3-11}
&\multirow{3}{*}{$\mathfrak{d}_{\r{4}}$}&%
\fwboxR{5pt}{\big\{\!}\rule{0pt}{14pt}(\hspace{-1pt}\b{\mathbf{t}_{1}}\hspace{1pt}\b{\mathbf{t}_{2}}\hspace{-1pt})\fwboxL{0pt}{\hspace{-1pt},}&(\hspace{-1pt}\b{\mathbf{t}_{2}}\hspace{1pt}\b{\mathbf{t}_{1}}\hspace{-1pt})\fwboxL{0pt}{\hspace{-1pt},}&(\hspace{-1pt}\b{\mathbf{t}_{2}}\hspace{1pt}\b{\mathbf{t}_{2}}\hspace{-1pt})\fwboxL{0pt}{\hspace{-1pt},}&(\hspace{-1pt}\b{\mathbf{t}_{2}}\hspace{1pt}\b{\mathbf{t}_{3}}\hspace{-1pt})\fwboxL{0pt}{\hspace{-1pt},}&(\hspace{-1pt}\b{\mathbf{t}_{2}}\hspace{1pt}\b{\mathbf{t}_{4}}\hspace{-1pt})\fwboxL{0pt}{\hspace{-1pt},}&(\hspace{-1pt}\b{\mathbf{t}_{2}}\hspace{1pt}\b{\mathbf{t}_{5}}\hspace{-1pt})\fwboxL{0pt}{\hspace{-1pt},}&(\hspace{-1pt}\b{\mathbf{t}_{2}}\hspace{1pt}\b{\mathbf{t}_{6}}\hspace{-1pt})\fwboxL{0pt}{\hspace{-1pt},}&(\hspace{-1pt}\b{\mathbf{t}_{2}}\hspace{1pt}\b{\mathbf{t}_{7}}\hspace{-1pt})\fwboxL{0pt}{\hspace{-1pt},}&(\hspace{-1pt}\b{\mathbf{t}_{3}}\hspace{1pt}\b{\mathbf{t}_{2}}\hspace{-1pt})\fwboxL{0pt}{\hspace{-1pt},}&\multirow{3}{*}{$\fwboxL{0pt}{\hspace{4pt}28}$}\\[2pt]
&&(\hspace{-1pt}\b{\mathbf{t}_{3}}\hspace{1pt}\b{\mathbf{t}_{3}}\hspace{-1pt})\fwboxL{0pt}{\hspace{-1pt},}&(\hspace{-1pt}\b{\mathbf{t}_{3}}\hspace{1pt}\b{\mathbf{t}_{4}}\hspace{-1pt})\fwboxL{0pt}{\hspace{-1pt},}&(\hspace{-1pt}\b{\mathbf{t}_{4}}\hspace{1pt}\b{\mathbf{t}_{2}}\hspace{-1pt})\fwboxL{0pt}{\hspace{-1pt},}&(\hspace{-1pt}\b{\mathbf{t}_{4}}\hspace{1pt}\b{\mathbf{t}_{3}}\hspace{-1pt})\fwboxL{0pt}{\hspace{-1pt},}&(\hspace{-1pt}\b{\mathbf{t}_{4}}\hspace{1pt}\b{\mathbf{t}_{5}}\hspace{-1pt})\fwboxL{0pt}{\hspace{-1pt},}&(\hspace{-1pt}\b{\mathbf{t}_{4}}\hspace{1pt}\b{\mathbf{t}_{6}}\hspace{-1pt})\fwboxL{0pt}{\hspace{-1pt},}&(\hspace{-1pt}\b{\mathbf{t}_{4}}\hspace{1pt}\b{\mathbf{t}_{7}}\hspace{-1pt})\fwboxL{0pt}{\hspace{-1pt},}&(\hspace{-1pt}\b{\mathbf{t}_{4}}\hspace{1pt}\b{\mathbf{t}_{4}}\hspace{-1pt})\fwboxL{0pt}{\hspace{0pt}{}_{\nu\in[3]}}\fwboxL{0pt}{\hspace{19pt},}&\\[2pt]
&&(\hspace{-1pt}\b{\mathbf{t}_{5}}\hspace{1pt}\b{\mathbf{t}_{2}}\hspace{-1pt})\fwboxL{0pt}{\hspace{-1pt},}&(\hspace{-1pt}\b{\mathbf{t}_{5}}\hspace{1pt}\b{\mathbf{t}_{4}}\hspace{-1pt})\fwboxL{0pt}{\hspace{-1pt},}&(\hspace{-1pt}\b{\mathbf{t}_{5}}\hspace{1pt}\b{\mathbf{t}_{5}}\hspace{-1pt})\fwboxL{0pt}{\hspace{-1pt},}&(\hspace{-1pt}\b{\mathbf{t}_{6}}\hspace{1pt}\b{\mathbf{t}_{2}}\hspace{-1pt})\fwboxL{0pt}{\hspace{-1pt},}&(\hspace{-1pt}\b{\mathbf{t}_{6}}\hspace{1pt}\b{\mathbf{t}_{4}}\hspace{-1pt})\fwboxL{0pt}{\hspace{-1pt},}&(\hspace{-1pt}\b{\mathbf{t}_{6}}\hspace{1pt}\b{\mathbf{t}_{6}}\hspace{-1pt})\fwboxL{0pt}{\hspace{-1pt},}&(\hspace{-1pt}\b{\mathbf{t}_{7}}\hspace{1pt}\b{\mathbf{t}_{2}}\hspace{-1pt})\fwboxL{0pt}{\hspace{-1pt},}&(\hspace{-1pt}\b{\mathbf{t}_{7}}\hspace{1pt}\b{\mathbf{t}_{4}}\hspace{-1pt})\fwboxL{0pt}{\hspace{-1pt},}&(\hspace{-1pt}\b{\mathbf{t}_{7}}\hspace{1pt}\b{\mathbf{t}_{7}}\hspace{-1pt})\fwboxL{5pt}{\!\big\}}\\[2pt]\cline{3-11}
%\\[2pt]\cline{3-10}
%
&\multirow{3}{*}{$\fwboxR{0pt}{\mathfrak{d}_{\r{5}}}$}&%
\fwboxR{5pt}{\big\{\!}\rule{0pt}{14pt}(\hspace{-1pt}\b{\mathbf{t}_{1}}\hspace{1pt}\b{\mathbf{t}_{2}}\hspace{-1pt})\fwboxL{0pt}{\hspace{-1pt},}&(\hspace{-1pt}\b{\mathbf{t}_{2}}\hspace{1pt}\b{\mathbf{t}_{1}}\hspace{-1pt})\fwboxL{0pt}{\hspace{-1pt},}&(\hspace{-1pt}\b{\mathbf{t}_{2}}\hspace{1pt}\b{\mathbf{t}_{2}}\hspace{-1pt})\fwboxL{0pt}{\hspace{-1pt},}&(\hspace{-1pt}\b{\mathbf{t}_{2}}\hspace{1pt}\b{\mathbf{t}_{3}}\hspace{-1pt})\fwboxL{0pt}{\hspace{-1pt},}&(\hspace{-1pt}\b{\mathbf{t}_{2}}\hspace{1pt}\b{\mathbf{t}_{4}}\hspace{-1pt})\fwboxL{0pt}{\hspace{-1pt},}&(\hspace{-1pt}\b{\mathbf{t}_{2}}\hspace{1pt}\b{\mathbf{t}_{5}}\hspace{-1pt})\fwboxL{0pt}{\hspace{-1pt},}&(\hspace{-1pt}\b{\mathbf{t}_{2}}\hspace{1pt}\b{\mathbf{t}_{6}}\hspace{-1pt})\fwboxL{0pt}{\hspace{-1pt},}&&&\multirow{3}{*}{$\fwboxL{0pt}{\hspace{4pt}23}$}\\[2pt]
&&(\hspace{-1pt}\b{\mathbf{t}_{3}}\hspace{1pt}\b{\mathbf{t}_{2}}\hspace{-1pt})\fwboxL{0pt}{\hspace{-1pt},}&(\hspace{-1pt}\b{\mathbf{t}_{3}}\hspace{1pt}\b{\mathbf{t}_{3}}\hspace{-1pt})\fwboxL{0pt}{\hspace{-1pt},}&(\hspace{-1pt}\b{\mathbf{t}_{3}}\hspace{1pt}\b{\mathbf{t}_{4}}\hspace{-1pt})\fwboxL{0pt}{\hspace{-1pt},}&(\hspace{-1pt}\b{\mathbf{t}_{4}}\hspace{1pt}\b{\mathbf{t}_{2}}\hspace{-1pt})\fwboxL{0pt}{\hspace{-1pt},}&(\hspace{-1pt}\b{\mathbf{t}_{4}}\hspace{1pt}\b{\mathbf{t}_{3}}\hspace{-1pt})\fwboxL{0pt}{\hspace{-1pt},}&(\hspace{-1pt}\b{\mathbf{t}_{4}}\hspace{1pt}\b{\mathbf{t}_{5}}\hspace{-1pt})\fwboxL{0pt}{\hspace{-1pt},}&(\hspace{-1pt}\b{\mathbf{t}_{4}}\hspace{1pt}\b{\mathbf{t}_{4}}\hspace{-1pt})\fwboxL{0pt}{\hspace{0pt}{}_{\nu\in[2]}}\fwboxL{0pt}{\hspace{19pt},}&&\\[2pt]
&&(\hspace{-1pt}\b{\mathbf{t}_{4}}\hspace{1pt}\b{\mathbf{t}_{6}}\hspace{-1pt})\fwboxL{0pt}{\hspace{-1pt},}&(\hspace{-1pt}\b{\mathbf{t}_{5}}\hspace{1pt}\b{\mathbf{t}_{2}}\hspace{-1pt})\fwboxL{0pt}{\hspace{-1pt},}&(\hspace{-1pt}\b{\mathbf{t}_{5}}\hspace{1pt}\b{\mathbf{t}_{4}}\hspace{-1pt})\fwboxL{0pt}{\hspace{-1pt},}&(\hspace{-1pt}\b{\mathbf{t}_{5}}\hspace{1pt}\b{\mathbf{t}_{5}}\hspace{-1pt})\fwboxL{0pt}{\hspace{-1pt},}&(\hspace{-1pt}\b{\mathbf{t}_{6}}\hspace{1pt}\b{\mathbf{t}_{2}}\hspace{-1pt})\fwboxL{0pt}{\hspace{-1pt},}&(\hspace{-1pt}\b{\mathbf{t}_{6}}\hspace{1pt}\b{\mathbf{t}_{4}}\hspace{-1pt})\fwboxL{0pt}{\hspace{-1pt},}&(\hspace{-1pt}\b{\mathbf{t}_{6}}\hspace{1pt}\b{\mathbf{t}_{6}}\hspace{-1pt})\fwboxL{0pt}{\hspace{0pt}{}_{\nu\in[2]}}\fwboxL{0pt}{\hspace{19pt}\big\}}&&\\[2pt]\cline{3-11}
&\multirow{3}{*}{$\fwboxR{0pt}{\begin{array}{@{}r@{}}\mathfrak{b}_{\r{k}},\mathfrak{c}_{\r{k}}\fwboxL{0pt}{,}\\\mathfrak{d}_{\r{k}>5}\end{array}}$}&%
\fwboxR{5pt}{\big\{\!}\rule{0pt}{14pt}(\hspace{-1pt}\b{\mathbf{t}_{1}}\hspace{1pt}\b{\mathbf{t}_{2}}\hspace{-1pt})\fwboxL{0pt}{\hspace{-1pt},}&(\hspace{-1pt}\b{\mathbf{t}_{2}}\hspace{1pt}\b{\mathbf{t}_{1}}\hspace{-1pt})\fwboxL{0pt}{\hspace{-1pt},}&(\hspace{-1pt}\b{\mathbf{t}_{2}}\hspace{1pt}\b{\mathbf{t}_{2}}\hspace{-1pt})\fwboxL{0pt}{\hspace{-1pt},}&(\hspace{-1pt}\b{\mathbf{t}_{2}}\hspace{1pt}\b{\mathbf{t}_{3}}\hspace{-1pt})\fwboxL{0pt}{\hspace{-1pt},}&(\hspace{-1pt}\b{\mathbf{t}_{2}}\hspace{1pt}\b{\mathbf{t}_{4}}\hspace{-1pt})\fwboxL{0pt}{\hspace{-1pt},}&(\hspace{-1pt}\b{\mathbf{t}_{2}}\hspace{1pt}\b{\mathbf{t}_{5}}\hspace{-1pt})\fwboxL{0pt}{\hspace{-1pt},}&(\hspace{-1pt}\b{\mathbf{t}_{2}}\hspace{1pt}\b{\mathbf{t}_{6}}\hspace{-1pt})\fwboxL{0pt}{\hspace{-1pt},}&&&\multirow{3}{*}{$\fwboxL{0pt}{\hspace{4pt}22}$}\\[2pt]
&&(\hspace{-1pt}\b{\mathbf{t}_{3}}\hspace{1pt}\b{\mathbf{t}_{2}}\hspace{-1pt})\fwboxL{0pt}{\hspace{-1pt},}&(\hspace{-1pt}\b{\mathbf{t}_{3}}\hspace{1pt}\b{\mathbf{t}_{3}}\hspace{-1pt})\fwboxL{0pt}{\hspace{-1pt},}&(\hspace{-1pt}\b{\mathbf{t}_{3}}\hspace{1pt}\b{\mathbf{t}_{4}}\hspace{-1pt})\fwboxL{0pt}{\hspace{-1pt},}&(\hspace{-1pt}\b{\mathbf{t}_{4}}\hspace{1pt}\b{\mathbf{t}_{2}}\hspace{-1pt})\fwboxL{0pt}{\hspace{-1pt},}&(\hspace{-1pt}\b{\mathbf{t}_{4}}\hspace{1pt}\b{\mathbf{t}_{3}}\hspace{-1pt})\fwboxL{0pt}{\hspace{-1pt},}&(\hspace{-1pt}\b{\mathbf{t}_{4}}\hspace{1pt}\b{\mathbf{t}_{5}}\hspace{-1pt})\fwboxL{0pt}{\hspace{-1pt},}&(\hspace{-1pt}\b{\mathbf{t}_{4}}\hspace{1pt}\b{\mathbf{t}_{4}}\hspace{-1pt})\fwboxL{0pt}{\hspace{0pt}{}_{\nu\in[2]}}\fwboxL{0pt}{\hspace{19pt},}&&\\[2pt]
&&(\hspace{-1pt}\b{\mathbf{t}_{4}}\hspace{1pt}\b{\mathbf{t}_{6}}\hspace{-1pt})\fwboxL{0pt}{\hspace{-1pt},}&(\hspace{-1pt}\b{\mathbf{t}_{5}}\hspace{1pt}\b{\mathbf{t}_{2}}\hspace{-1pt})\fwboxL{0pt}{\hspace{-1pt},}&(\hspace{-1pt}\b{\mathbf{t}_{5}}\hspace{1pt}\b{\mathbf{t}_{4}}\hspace{-1pt})\fwboxL{0pt}{\hspace{-1pt},}&(\hspace{-1pt}\b{\mathbf{t}_{5}}\hspace{1pt}\b{\mathbf{t}_{5}}\hspace{-1pt})\fwboxL{0pt}{\hspace{-1pt},}&(\hspace{-1pt}\b{\mathbf{t}_{6}}\hspace{1pt}\b{\mathbf{t}_{2}}\hspace{-1pt})\fwboxL{0pt}{\hspace{-1pt},}&(\hspace{-1pt}\b{\mathbf{t}_{6}}\hspace{1pt}\b{\mathbf{t}_{4}}\hspace{-1pt})\fwboxL{0pt}{\hspace{-1pt},}&(\hspace{-1pt}\b{\mathbf{t}_{6}}\hspace{1pt}\b{\mathbf{t}_{6}}\hspace{-1pt})\fwboxL{0pt}{\!\big\}}&&\\[2pt]\cline{3-11}
&\multirow{2}{*}{$\fwboxR{0pt}{\mathfrak{e}_{\r{6}}}$}&
\fwboxR{5pt}{\big\{\!}\rule{0pt}{14pt}(\hspace{-1pt}\b{\mathbf{t}_{1}}\hspace{1pt}\b{\mathbf{t}_{2}}\hspace{-1pt})\fwboxL{0pt}{\hspace{-1pt},}&(\hspace{-1pt}\b{\mathbf{t}_{2}}\hspace{1pt}\b{\mathbf{t}_{1}}\hspace{-1pt})\fwboxL{0pt}{\hspace{-1pt},}&(\hspace{-1pt}\b{\mathbf{t}_{2}}\hspace{1pt}\b{\mathbf{t}_{2}}\hspace{-1pt})\fwboxL{0pt}{\hspace{-1pt},}&(\hspace{-1pt}\b{\mathbf{t}_{2}}\hspace{1pt}\b{\mathbf{t}_{3}}\hspace{-1pt})\fwboxL{0pt}{\hspace{-1pt},}&(\hspace{-1pt}\b{\mathbf{t}_{2}}\hspace{1pt}\b{\mathbf{t}_{4}}\hspace{-1pt})\fwboxL{0pt}{\hspace{-1pt},}&(\hspace{-1pt}\b{\mathbf{t}_{2}}\hspace{1pt}\b{\mathbf{t}_{5}}\hspace{-1pt})\fwboxL{0pt}{\hspace{-1pt},}&(\hspace{-1pt}\b{\mathbf{t}_{3}}\hspace{1pt}\b{\mathbf{t}_{2}}\hspace{-1pt})\fwboxL{0pt}{\hspace{-1pt},}&(\hspace{-1pt}\b{\mathbf{t}_{3}}\hspace{1pt}\b{\mathbf{t}_{3}}\hspace{-1pt})\fwboxL{0pt}{\hspace{0pt}{}_{\nu\in[2]}}\fwboxL{0pt}{\hspace{19pt},}&&\multirow{2}{*}{$\fwboxL{0pt}{\hspace{4pt}17}$}\\[2pt]
&&(\hspace{-1pt}\b{\mathbf{t}_{3}}\hspace{1pt}\b{\mathbf{t}_{4}}\hspace{-1pt})\fwboxL{0pt}{\hspace{-1pt},}&(\hspace{-1pt}\b{\mathbf{t}_{4}}\hspace{1pt}\b{\mathbf{t}_{2}}\hspace{-1pt})\fwboxL{0pt}{\hspace{-1pt},}&(\hspace{-1pt}\b{\mathbf{t}_{4}}\hspace{1pt}\b{\mathbf{t}_{3}}\hspace{-1pt})\fwboxL{0pt}{\hspace{-1pt},}&(\hspace{-1pt}\b{\mathbf{t}_{4}}\hspace{1pt}\b{\mathbf{t}_{4}}\hspace{-1pt})\fwboxL{0pt}{\hspace{-1pt},}&(\hspace{-1pt}\b{\mathbf{t}_{4}}\hspace{1pt}\b{\mathbf{t}_{5}}\hspace{-1pt})\fwboxL{0pt}{\hspace{-1pt},}&(\hspace{-1pt}\b{\mathbf{t}_{5}}\hspace{1pt}\b{\mathbf{t}_{2}}\hspace{-1pt})\fwboxL{0pt}{\hspace{-1pt},}&(\hspace{-1pt}\b{\mathbf{t}_{5}}\hspace{1pt}\b{\mathbf{t}_{4}}\hspace{-1pt})\fwboxL{0pt}{\hspace{-1pt},}&(\hspace{-1pt}\b{\mathbf{t}_{5}}\hspace{1pt}\b{\mathbf{t}_{5}}\hspace{-1pt})\fwboxL{0pt}{\!\big\}}&\\[2pt]\cline{3-11}
&\multirow{2}{*}{$\fwboxR{0pt}{\begin{array}{@{}r@{}}\mathfrak{e}_{\r{7}}\fwboxL{0pt}{,}\\\mathfrak{f}_{\r{4}},\mathfrak{g}_{\r{2}}\end{array}}$}&
\fwboxR{5pt}{\big\{\!}\rule{0pt}{14pt}(\hspace{-1pt}\b{\mathbf{t}_{1}}\hspace{1pt}\b{\mathbf{t}_{2}}\hspace{-1pt})\fwboxL{0pt}{\hspace{-1pt},}&(\hspace{-1pt}\b{\mathbf{t}_{2}}\hspace{1pt}\b{\mathbf{t}_{1}}\hspace{-1pt})\fwboxL{0pt}{\hspace{-1pt},}&(\hspace{-1pt}\b{\mathbf{t}_{2}}\hspace{1pt}\b{\mathbf{t}_{2}}\hspace{-1pt})\fwboxL{0pt}{\hspace{-1pt},}&(\hspace{-1pt}\b{\mathbf{t}_{2}}\hspace{1pt}\b{\mathbf{t}_{3}}\hspace{-1pt})\fwboxL{0pt}{\hspace{-1pt},}&(\hspace{-1pt}\b{\mathbf{t}_{2}}\hspace{1pt}\b{\mathbf{t}_{4}}\hspace{-1pt})\fwboxL{0pt}{\hspace{-1pt},}&(\hspace{-1pt}\b{\mathbf{t}_{2}}\hspace{1pt}\b{\mathbf{t}_{5}}\hspace{-1pt})\fwboxL{0pt}{\hspace{-1pt},}&(\hspace{-1pt}\b{\mathbf{t}_{3}}\hspace{1pt}\b{\mathbf{t}_{2}}\hspace{-1pt})\fwboxL{0pt}{\hspace{-1pt},}&(\hspace{-1pt}\b{\mathbf{t}_{3}}\hspace{1pt}\b{\mathbf{t}_{3}}\hspace{-1pt})\fwboxL{0pt}{\hspace{-1pt},}&&\multirow{2}{*}{$\fwboxL{0pt}{\hspace{4pt}16}$}\\[2pt]
&&(\hspace{-1pt}\b{\mathbf{t}_{3}}\hspace{1pt}\b{\mathbf{t}_{4}}\hspace{-1pt})\fwboxL{0pt}{\hspace{-1pt},}&(\hspace{-1pt}\b{\mathbf{t}_{4}}\hspace{1pt}\b{\mathbf{t}_{2}}\hspace{-1pt})\fwboxL{0pt}{\hspace{-1pt},}&(\hspace{-1pt}\b{\mathbf{t}_{4}}\hspace{1pt}\b{\mathbf{t}_{3}}\hspace{-1pt})\fwboxL{0pt}{\hspace{-1pt},}&(\hspace{-1pt}\b{\mathbf{t}_{4}}\hspace{1pt}\b{\mathbf{t}_{4}}\hspace{-1pt})\fwboxL{0pt}{\hspace{-1pt},}&(\hspace{-1pt}\b{\mathbf{t}_{4}}\hspace{1pt}\b{\mathbf{t}_{5}}\hspace{-1pt})\fwboxL{0pt}{\hspace{-1pt},}&(\hspace{-1pt}\b{\mathbf{t}_{5}}\hspace{1pt}\b{\mathbf{t}_{2}}\hspace{-1pt})\fwboxL{0pt}{\hspace{-1pt},}&(\hspace{-1pt}\b{\mathbf{t}_{5}}\hspace{1pt}\b{\mathbf{t}_{4}}\hspace{-1pt})\fwboxL{0pt}{\hspace{-1pt},}&(\hspace{-1pt}\b{\mathbf{t}_{5}}\hspace{1pt}\b{\mathbf{t}_{5}}\hspace{-1pt})\fwboxL{0pt}{\!\big\}}&\\[2pt]\cline{3-11}
\end{array}\hspace{-200pt}$$\vspace{-20pt}\end{table}

For $\mathfrak{a}_{\r{1}}$ gauge theory, we may compare this basis to one generated by independent multi-trace tensors $\{\mathbf{T}_1,\ldots,\mathbf{T}_6\}$ given by
\eq{\fwbox{0pt}{\fwbox{0pt}{\fwboxL{435pt}{(\hspace{-1.25pt}%
\mathfrak{a}_{\r{1}}
)}}\fwbox{0pt}{\hspace{-10pt}\Big\{
\fwboxL{55pt}{\mathrm{tr}_{\mathbf{\b{F}}}(\hspace{-1pt}\r{12345})},
\fwboxL{55pt}{\mathrm{tr}_{\mathbf{\b{F}}}(\hspace{-1pt}\r{12}|\r{345})},
\fwboxL{55pt}{\mathrm{tr}_{\mathbf{\b{F}}}(\hspace{-1pt}\r{23}|\r{451})},
\fwboxL{55pt}{\mathrm{tr}_{\mathbf{\b{F}}}(\hspace{-1pt}\r{34}|\r{512})},
\fwboxL{55pt}{\mathrm{tr}_{\mathbf{\b{F}}}(\hspace{-1pt}\r{45}|\r{123})},
\fwboxL{55pt}{\mathrm{tr}_{\mathbf{\b{F}}}(\hspace{-1pt}\r{51}|\r{234})}\,\Big\}\fwboxL{0pt}{.}}}}
It is easy to see that these are far from diagonal in colour by computing the overlap
\eq{\fwbox{0pt}{\fwboxL{510pt}{(\hspace{-1.25pt}%
\mathfrak{a}_{\r{1}}
)}}\hspace{-100pt}\fwboxR{0pt}{\langle\mathbf{T}_i|\mathbf{T}_j\rangle=}\frac{1}{2}\!\left(\begin{array}{c@{$\;$}c@{$\;$}c@{$\;$}c@{$\;$}c@{$\;$}c}
15&9&9&9&9&9\\[-4pt]
9&\dzero&\dzero&6&18&6\\[-4pt]
9&\dzero&6&18&6&\dzero\\[-4pt]
9&6&18&6&\dzero&\dzero\\[-4pt]
9&18&6&\dzero&\dzero&6\\[-4pt]
9&6&\dzero&\dzero&6&18\end{array}\right)\fwboxL{0pt}{.}\hspace{-100pt}}
In terms of the basis tensors $\mathcal{B}$, these can be expanded
\eq{\mathbf{T}^i\equivR\sum_{\b{j}}\mathbf{c}[\mathfrak{a}_{\r{1}}]\indices{i}{\b{j}}\,\mathcal{B}^{\b{j}}}
where the matrix is given by:
\eq{\fwbox{0pt}{\fwboxL{530pt}{(\hspace{-1.25pt}%
\mathfrak{a}_{\r{1}}
)}}\hspace{-100pt}\fwboxR{0pt}{\mathbf{c}[\mathfrak{a}_{\r{1}}]\indices{i}{\b{j}}=}\left(\begin{array}{c@{$\;\;$}c@{$\;\;$}c@{$\;\;$}c@{$\;\;$}c@{$\;\;$}c}
\frac{1}{4}&\frac{1}{4}&\frac{1}{8}&\dzero&\dzero&\dzero\\[-1pt]
\frac{1}{2}&\dzero&\dzero&\dzero&\dzero&\dzero\\[-1pt]
\frac{1}{6}&\dzero&\frac{1}{8}&\dzero&\frac{1}{2}&\dzero\\[-1pt]
\dzero&\frac{1}{6}&\frac{1}{8}&\frac{1}{2}&\dzero&\dzero\\[-1pt]
\dzero&\frac{1}{2}&\dzero&\dzero&\dzero&\dzero\\[-1pt]
\frac{1}{6}&\frac{1}{6}&\frac{1}{16}&\tmi\frac{1}{4}&\tmi\frac{1}{4}&\frac{1}{4}\end{array}\right)\fwboxL{0pt}{.}\hspace{-100pt}}

We can also check the sequence (\ref{graphical_5pt_duality_sequence}) by comparing the basis $\mathcal{B}$ to those of
\eq{\begin{array}{@{}c@{$\;\;\;$}c@{$\;\;\;$}c@{$\;\;\;$}c@{}}\tikzBox{ggggg_alt_basis_4_diagram}{
\arrowTo[hred]{0,0}{-130}\node[anchor=10,inner sep=2pt] at(in){{\footnotesize$\r{[\adR]}$}};
\arrowTo[hred]{0,0}{130}\node[anchor=-10,inner sep=2pt] at(in){{\footnotesize$\r{[\adR]}$}};
\arrowFrom[hblue]{0,0}[1]{0}\node[clebsch]at(in){{\scriptsize$\phantom{\nu}$}};\node[]at(in){{\scriptsize${\mu}$}};\node[anchor=90,inner sep=2pt] at(arrownode){{\footnotesize$\b{[t_j]}$}};
\arrowFrom[hred]{end}{-50}\node[anchor=170,inner sep=0pt] at(end){{\footnotesize$\r{[\adR]}$}};
\arrowFrom[hblue]{in}{90}\node[anchor=180,inner sep=2pt] at(arrownode){{\footnotesize$\b{[t_i]}$}};\node[clebsch]at(in){{\scriptsize$\phantom{\nu}$}};\node[]at(in){{\scriptsize${\nu}$}};
\arrowTo[hred]{end}{150}\node[anchor=-10,inner sep=2pt] at(in){{\footnotesize$\r{[\adR]}$}};
\arrowFrom[hred]{end}{30}\node[anchor=-170,inner sep=2pt] at(end){{\footnotesize$\r{[\adR]}$}};
\node[clebsch]at(in){{\scriptsize$\phantom{\nu}$}};\node[]at(in){{\scriptsize${\rho}$}};}&
\tikzBox{ggggg_alt_basis_3_diagram}{
\arrowTo[hred]{0,0}{-150}\node[anchor=10,inner sep=2pt] at(in){{\footnotesize$\r{[\adR]}$}};
\arrowFrom[hred]{0,0}{-30}\node[anchor=170,inner sep=2pt] at(end){{\footnotesize$\r{[\adR]}$}};
\arrowFrom[hblue]{0,0}[1]{90}\node[clebsch]at(in){{\scriptsize$\phantom{\nu}$}};\node[]at(in){{\scriptsize${\mu}$}};
\node[anchor=180,inner sep=2pt] at(arrownode){{\footnotesize$\b{[t_i]}$}};
\arrowTo[hred]{end}{180}\node[anchor=0,inner sep=0pt] at(in){{\footnotesize$\r{[\adR]}$}};
\arrowFrom[hblue]{end}{90}\node[anchor=180,inner sep=2pt] at(arrownode){{\footnotesize$\b{[t_j]}$}};\node[clebsch]at(in){{\scriptsize$\phantom{\nu}$}};\node[]at(in){{\scriptsize${\nu}$}};
\arrowTo[hred]{end}{150}\node[anchor=-10,inner sep=2pt] at(in){{\footnotesize$\r{[\adR]}$}};
\arrowFrom[hred]{end}{30}\node[anchor=-170,inner sep=2pt] at(end){{\footnotesize$\r{[\adR]}$}};
\node[clebsch]at(in){{\scriptsize$\phantom{\nu}$}};\node[]at(in){{\scriptsize${\rho}$}};}&
\tikzBox{ggggg_alt_basis_2_diagram}{
\arrowTo[hred]{0,0}{-150}\node[anchor=10,inner sep=2pt] at(in){{\footnotesize$\r{[\adR]}$}};
\arrowFrom[hred]{0,0}{-30}\node[anchor=170,inner sep=2pt] at(end){{\footnotesize$\r{[\adR]}$}};
\arrowFrom[hblue]{0,0}[1]{90}\node[clebsch]at(in){{\scriptsize$\phantom{\nu}$}};\node[]at(in){{\scriptsize${\rho}$}};
\node[anchor=180,inner sep=2pt] at(arrownode){{\footnotesize$\b{[t_j]}$}};
\arrowFrom[hred]{end}{0}\node[anchor=180,inner sep=0pt] at(end){{\footnotesize$\r{[\adR]}$}};
\arrowFrom[hblue]{in}{90}\node[anchor=180,inner sep=2pt] at(arrownode){{\footnotesize$\b{[t_i]}$}};\node[clebsch]at(in){{\scriptsize$\phantom{\nu}$}};\node[]at(in){{\scriptsize${\nu}$}};
\arrowTo[hred]{end}{150}\node[anchor=-10,inner sep=2pt] at(in){{\footnotesize$\r{[\adR]}$}};
\arrowTo[hred]{end}{30}\node[anchor=-170,inner sep=2pt] at(in){{\footnotesize$\r{[\adR]}$}};
\node[clebsch]at(end){{\scriptsize$\phantom{\nu}$}};\node[]at(end){{\scriptsize${\mu}$}};}&
\tikzBox{ggggg_alt_basis_1_diagram}{
\arrowTo[hred]{0,0}{-150}\node[anchor=10,inner sep=2pt] at(in){{\footnotesize$\r{[\adR]}$}};
\arrowFrom[hblue]{0,0}[1]{0}
\node[anchor=90,inner sep=2pt] at(arrownode){{\footnotesize$\b{[t_i]}$}};
\arrowFrom[hred]{end}{-50}\node[anchor=170,inner sep=2pt] at(end){{\footnotesize$\r{[\adR]}$}};
\arrowFrom[hred]{in}{50}\node[anchor=-170,inner sep=2pt] at(end){{\footnotesize$\r{[\adR]}$}};
\node[clebsch]at(in){{\scriptsize$\phantom{\nu}$}};\node[]at(in){{\scriptsize${\mu}$}};
\arrowTo[hblue]{0,0}[1]{90}
\node[anchor=0,inner sep=2pt] at(arrownode){{\footnotesize$\b{[t_j]}$}};
\arrowTo[hred]{in}{150}\node[anchor=-10,inner sep=0pt] at(in){{\footnotesize$\r{[\adR]}$}};
\arrowTo[hred]{end}{30}\node[anchor=-170,inner sep=0pt] at(in){{\footnotesize$\r{[\adR]}$}};
\node[clebsch]at(0,0){{\scriptsize$\phantom{\nu}$}};\node[]at(0,0){{\scriptsize${\nu}$}};
\node[clebsch]at(end){{\scriptsize$\phantom{\nu}$}};\node[]at(end){{\scriptsize${\rho}$}};}\\
\equivR \mathbf{d}.\mathcal{B}&=(\mathbf{d}^2\hspace{-1pt}).\mathcal{B}&=(\mathbf{d}^3\hspace{-1pt}).\mathcal{B}&= (\mathbf{d}^4\hspace{-1pt}).\mathcal{B}
\end{array}}
where the duality matrix for $\mathfrak{a}_{\r{1}}$ is given by
\eq{\fwbox{0pt}{\fwboxL{530pt}{(\hspace{-1.25pt}%
\mathfrak{a}_{\r{1}}
)}}\hspace{-100pt}\fwboxR{0pt}{\mathbf{d}\equivR}\left(\begin{array}{c@{$\;\;$}c@{$\;\;$}c@{$\;\;$}c@{$\;\;$}c@{$\;\;$}c}
\dzero&\frac{1}{3}&\frac{1}{4}&1&\dzero&\dzero\\[-1pt]
1&\dzero&\dzero&\dzero&\dzero&\dzero\\[-1pt]
\dzero&\frac{4}{3}&\frac{1}{2}&\tmi2&\dzero&\dzero\\[-1pt]
\dzero&\dzero&\dzero&\dzero&\tmi\frac{1}{2}&\frac{1}{2}\\[-1pt]
\dzero&\frac{5}{9}&\tmi\frac{5}{24}&\frac{1}{6}&\dzero&\dzero\\[-1pt]
\dzero&\dzero&\dzero&\dzero&\frac{3}{2}&\frac{1}{2}\end{array}\right)\hspace{-100pt}}
for which $\mathbf{d}^5{=}\mathbbm{1}_{6}$ is a rather non-trivial identity.

\newpage
\subsection{Novel Colour Tensors for Six Particle Scattering Amplitudes}

For six external particles transforming in the adjoint or fundamental representations, the number of basis tensors grows much more considerably. The number of independent tensors which exist for each of the simple Lie algebras is given in \mbox{Table~\ref{six_particle_tensor_rank_table}}.

Below, we describe possible Clebsch colour tensor bases in each case, identifying how each basis element would be labelled by irreducible representations and multiplicities. This rapidly becomes too cumbersome to warrant a thorough treatment, but explicit results for $\mathfrak{a}_{\r{1}}$ are described in some detail.\\

\begin{table}[t]\vspace{-9pt}\caption{Numbers of independent six-particle colour tensors for simple Lie algebras.}\label{six_particle_tensor_rank_table}\vspace{-20pt}$$\begin{array}{c}%
\begin{array}{r@{$\,\,$}|@{}r@{$\,$}r@{$\,$}r@{$\,$}r@{$\,$}r|r@{$\,$}r|r|@{$\,$}r@{$\,$}r@{$\,$}r@{$\,$}r|@{}r@{}r@{}r|@{}r@{}r|}\cline{2-18}
&\fwbox{14pt}{\mathfrak{a}_{\r{1}}\!\!}&\fwbox{18pt}{\mathfrak{a}_{\r{2}}\!\!}&\fwbox{18pt}{\mathfrak{a}_{\r{3}}\!\!}&\fwbox{18pt}{\mathfrak{a}_{\r{4}}\!\!}&\fwbox{18pt}{\mathfrak{a}_{\r{k}>4}}&\fwbox{18pt}{\mathfrak{b}_{\r{2}}\!\!}&\fwbox{18pt}{\mathfrak{b}_{\r{k}>2}}&\fwbox{18pt}{\mathfrak{c}_{\r{k}}\!\!}&\fwbox{18pt}{\mathfrak{d}_{\r{4}}\!\!}&\fwbox{18pt}{\mathfrak{d}_{\r{5}}\!\!}&\fwbox{18pt}{\mathfrak{d}_{\r{6}}\!\!}&\fwbox{18pt}{\mathfrak{d}_{\r{k}>6}}&\fwbox{16pt}{\mathfrak{e}_{\r{6}}\!\!}&\fwbox{16pt}{\mathfrak{e}_{\r{7}}\!\!}&\fwbox{16pt}{\mathfrak{e}_{\r{8}}\!\!}&\fwbox{16pt}{\mathfrak{f}_{\r{4}}\!\!}&\fwbox{16pt}{\mathfrak{g}_{\r{2}}\!\!}\\\cline{2-18}
\hspace{-10pt}(\mathbf{\b{F}}\,\mathbf{\b{F}}\,\mathbf{\b{F}}|\mathbf{\b{F}}\,\mathbf{\b{F}}\,\mathbf{\b{F}})\hspace{-1pt}\hspace{-1pt}&
5&6&6&6&6&15&15&15&15&15&15&15&20&35&79&70&35\\\cline{2-18}\hspace{-10pt}(\mathbf{\b{F}}\,\mathbf{\b{F}}\,\mathbf{\r{ad}}|\mathbf{\r{ad}}\,\mathbf{\b{F}}\,\mathbf{\b{F}})\hspace{-1pt}\hspace{-1pt}&
6&13&14&14&14&21&21&21&22&21&21&21&26&34&79&47&33\\\cline{2-18}
\hspace{-10pt}(\mathbf{\b{F}}\mathbf{\r{ad}}\,\mathbf{\r{ad}}|\mathbf{\r{ad}}\,\mathbf{\r{ad}}\,\mathbf{\b{F}})\hspace{-1pt}\hspace{-1pt}&
9&40&52&53&53&46&46&46&56&47&46&46&46&45&79&46&45\\\cline{2-18}
\hspace{-15pt}(\mathbf{\r{ad}}\,\mathbf{\r{ad}}\,\mathbf{\r{ad}}|\mathbf{\r{ad}}\,\mathbf{\r{ad}}\,\mathbf{\r{ad}})\hspace{-1pt}\hspace{-1pt}&15&145&245&264&265&115&130&130&185&140&131&130&90&80&79&80&80\\\cline{2-18}
\end{array}\end{array}\hspace{-10pt}\vspace{-15pt}$$\end{table}

\subsubsection{\texorpdfstring{Clebsch Colour Bases for $C(\mathbf{\b{F}}\,\mathbf{\b{F}}\,\mathbf{\b{F}}|\mathbf{\b{F}}\,\mathbf{\b{F}}\,\mathbf{\b{F}})$}{Clebsch Colour Bases for C(FFF|FFF)}}

For the case of six external particles transforming under the fundamental representation, a natural choice of Clebsch colour tensors would be given by 
\vspace{-5pt}\eq{\fwboxR{0pt}{\mathcal{B}_{~}^{\,\smash{\mathbf{\g{q}}_{\g{i}}\,\mathbf{{u}}_{{j}}\,\mathbf{\g{q}}_{\g{k}}}}(\mathbf{\b{F}}\,\mathbf{\b{F}}\,\mathbf{\b{F}}|\mathbf{\b{F}}\,\mathbf{\b{F}}\,\mathbf{\b{F}})\;\bigger{\Leftrightarrow}\;}\tikzBox[-5.25pt]{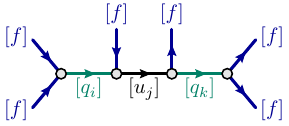}{\arrowTo[hblue]{0,0}{-130};\node[anchor=10,inner sep=2pt] at(in){{\footnotesize$\b{[f]}$}};\arrowTo[hblue]{0,0}{130}\node[anchor=-10,inner sep=2pt] at(in){{\footnotesize$\b{[f]}$}};\arrowFrom[hgreen]{0,0}[1.25]{0}\node[anchor=90,inner sep=2pt] at(arrownode){{\footnotesize$\g{[q_i]}$}};\node[clebsch]at(in){};\arrowTo[hblue]{end}{90}\node[anchor=-90,inner sep=2pt] at(in){{\footnotesize$\b{[f]}$}};\arrowFrom[black]{end}[1.25]{0}\node[clebsch]at(in){};\node[anchor=90,inner sep=2pt] at(arrownode){{\footnotesize${[u_j]}$}};\arrowFrom[hblue]{end}{90}\node[anchor=-90,inner sep=2pt] at(end){{\footnotesize$\b{[f]}$}};\arrowFrom[hgreen]{in}[1.25]{0}\node[clebsch]at(in){};\node[anchor=90,inner sep=2pt] at(arrownode){{\footnotesize$\g{[q_k]}$}};
\arrowFrom[hblue]{end}{50}
\node[anchor=-170,inner sep=2pt] at(end){{\footnotesize$\b{[f]}$}};\arrowFrom[hblue]{in}{-50}\node[anchor=170,inner sep=2pt] at(end){{\footnotesize$\b{[f]}$}};\node[clebsch]at(in){};
}\fwboxL{0pt}{.}\vspace{-5pt}\label{ffffff_b_basis_defined}}
These would be labelled by the three irreducible representations $(\g{\mathbf{q}_i},\mathbf{u}_j\,\g{\mathbf{q}_k})$ where $\g{\mathbf{q}_i}$ were defined in \mbox{Table~\ref{qReps_table}} and the new representations $\mathbf{u}_j$ are defined for the classical Lie algebras in \mbox{Table~\ref{uReps_table}}, and the triples of labels which span the basis (\ref{ffffff_b_basis_defined}) are enumerated in \mbox{Table~\ref{ffffff_basis_b_label_table}}.

\begin{table}[b]\vspace{-10pt}$$\begin{array}{|l@{$\,$}|cccc|}\multicolumn{1}{c}{~}&\fwbox{40pt}{{\mathbf{u}_{1}}}&\fwbox{80pt}{{\mathbf{u}_{2}}}&\multicolumn{1}{c}{\fwbox{80pt}{{\mathbf{u}_{3}}}}&\multicolumn{1}{c}{\fwbox{80pt}{{\mathbf{u}_{4}}}}\\\hline\hline
\raisebox{-4pt}{$\mathfrak{a}_{\r{k}}$}&\dynkLabel{\frac{1}{3}\r{k}(\hspace{-1pt}\r{k}\pl1\hspace{-1pt})(\hspace{-1pt}\r{k}\pl2\hspace{-1pt})}{110\cdots0}&\dynkLabel{\frac{1}{6}\r{k}(\hspace{-1pt}\r{k}\pl1\hspace{-1pt})(\hspace{-1pt}\r{k}\pl2\hspace{-1pt})(\hspace{-1pt}\r{k}\pl3\hspace{-1pt})}{30\cdots0}&\dynkLabel{\frac{1}{6}\r{k}(\hspace{-1pt}\r{k}\pl1\hspace{-1pt})(\hspace{-1pt}\r{k}\mi1\hspace{-1pt})}{0010\cdots0}&\\\hline
\raisebox{-4pt}{$\mathfrak{b}_{\r{k}}$}&\dynkLabel{\mathbf{\b{F}}}{10\cdots0}&\dynkLabel{\frac{1}{3}(\hspace{-1pt}2\r{k}\pl1\hspace{-1pt})(\hspace{-1pt}2\r{k}\mi1\hspace{-1pt})(\hspace{-1pt}2\r{k}\pl3\hspace{-1pt})}{110\cdots0}&\dynkLabel{\frac{1}{3}\r{k}(\hspace{-1pt}2\r{k}\pl1\hspace{-1pt})(\hspace{-1pt}2\r{k}\pl5\hspace{-1pt})}{30\cdots0}&\dynkLabel{\frac{1}{3}\r{k}(\hspace{-1pt}2\r{k}\pl1\hspace{-1pt})(\hspace{-1pt}2\r{k}\mi1\hspace{-1pt})}{0010\cdots0}\\\hline
\raisebox{-4pt}{$\mathfrak{c}_{\r{k}}$}&\dynkLabel{\mathbf{\b{F}}}{10\cdots0}&\dynkLabel{\frac{8}{3}\r{k}(\hspace{-1pt}\r{k}\pl1\hspace{-1pt})(\hspace{-1pt}\r{k}\mi1\hspace{-1pt})}{110\cdots0}&\dynkLabel{\frac{2}{3}\r{k}(\hspace{-1pt}\r{k}\mi2\hspace{-1pt})(\hspace{-1pt}2\r{k}\pl1\hspace{-1pt})}{0010\cdots0}&\dynkLabel{\frac{2}{3}\r{k}(\hspace{-1pt}\r{k}\pl1\hspace{-1pt})(\hspace{-1pt}2\r{k}\pl1\hspace{-1pt})}{30\cdots0}\\\hline
\raisebox{-4pt}{$\mathfrak{d}_{\r{k}}$}&\dynkLabel{\mathbf{\b{F}}}{10\cdots0}&\dynkLabel{\frac{8}{3}\r{k}(\hspace{-1pt}\r{k}\pl1\hspace{-1pt})(\hspace{-1pt}\r{k}\mi1\hspace{-1pt})}{110\cdots0}&\dynkLabel{\frac{2}{3}\r{k}(\hspace{-1pt}\r{k}\pl2\hspace{-1pt})(\hspace{-1pt}2\r{k}\mi1\hspace{-1pt})}{30\cdots0}&\dynkLabel{\frac{2}{3}\r{k}(\hspace{-1pt}\r{k}\mi1\hspace{-1pt})(\hspace{-1pt}2\r{k}\mi1\hspace{-1pt})}{0010\cdots0}\\\hline
\end{array}
$$\vspace{-22pt}\vspace{-0pt}\caption{Irreducible representations appearing in the decomposition of the tensor product $\g{\mathbf{q}_{j}}\!\otimes\!\mathbf{\b{F}}\!\supset\!\bigoplus_{{i}}\mathbf{{u}}_{{i}}$ for simple Lie algebras $\mathfrak{a}_{\r{k}},\mathfrak{b}_{\r{k}},\mathfrak{c}_{\r{k}},\mathfrak{d}_{\r{k}}$.}
\vspace{-00pt}\vspace{-0pt}\label{uReps_table}\end{table}

\begin{table}[t]\vspace{-10pt}\caption{Labels $(\hspace{-1pt}\g{\mathbf{q}_i}\,\mathbf{u}_j\,\g{\mathbf{q}_k}\hspace{-1pt})$ for basis tensors $\mathcal{B}_{~}^{\,\smash{\mathbf{\g{q}}_{\g{i}}\,\mathbf{{u}}_{{j}}\,\mathbf{\g{q}}_{\g{k}}}}(\mathbf{\b{F}}\,\mathbf{\b{F}}\,\mathbf{\b{F}}|\mathbf{\b{F}}\,\mathbf{\b{F}}\,\mathbf{\b{F}})$ of simple Lie algebras.}\label{ffffff_basis_b_label_table}
\vspace{-15pt}
$$\hspace{-200pt}\begin{array}{r@{$\;\;$}|rl@{$\,\,$}l@{$\,\,$}l@{$\,\,$}l@{$\,\,$}l|l}
%\multicolumn{1}{r}{\fwboxR{0pt}{\mathfrak{g}\hspace{16pt}}}&\multicolumn{7}{r}{\fwboxL{0pt}{\hspace{3pt}\text{\#}}}\\\cline{2-7}
\multicolumn{1}{r}{\fwboxR{0pt}{\mathfrak{g}\hspace{22pt}}}&\multicolumn{7}{r}{\fwboxL{0pt}{\hspace{13pt}\text{\#}}}\\[-10pt]\cline{2-7}
\multirow{1}{*}{$\fwboxR{0pt}{\begin{array}{@{}r@{}}\mathfrak{a}_{\r{1}}\end{array}}$}&
\fwboxR{5pt}{\big\{\!}\rule{0pt}{14pt}(\hspace{-1pt}\g{\mathbf{q}_{1}}\hspace{1pt}{\mathbf{u}_{1}}\hspace{1pt}\g{\mathbf{q}_{1}}\hspace{-1pt})\fwboxL{0pt}{\hspace{-1pt},}&(\hspace{-1pt}\g{\mathbf{q}_{1}}\hspace{1pt}{\mathbf{u}_{1}}\hspace{1pt}\g{\mathbf{q}_{2}}\hspace{-1pt})\fwboxL{0pt}{\hspace{-1pt},}&(\hspace{-1pt}\g{\mathbf{q}_{2}}\hspace{1pt}{\mathbf{u}_{1}}\hspace{1pt}\g{\mathbf{q}_{1}}\hspace{-1pt})\fwboxL{0pt}{\hspace{-1pt},}&(\hspace{-1pt}\g{\mathbf{q}_{2}}\hspace{1pt}{\mathbf{u}_{1}}\hspace{1pt}\g{\mathbf{q}_{2}}\hspace{-1pt})\fwboxL{0pt}{\hspace{-1pt},}&(\hspace{-1pt}\g{\mathbf{q}_{2}}\hspace{1pt}{\mathbf{u}_{2}}\hspace{1pt}\g{\mathbf{q}_{2}}\hspace{-1pt})\fwboxL{0pt}{\!\big\}}&&\multirow{1}{*}{$\fwboxL{0pt}{\hspace{4pt}5}$}\\[2pt]\cline{2-7}
\multirow{1}{*}{$\fwboxR{0pt}{\mathfrak{a}_{\r{k}>1}}$}&%
\fwboxR{5pt}{\big\{\!}\rule{0pt}{14pt}(\hspace{-1pt}\g{\mathbf{q}_{1}}\hspace{1pt}{\mathbf{u}_{1}}\hspace{1pt}\g{\mathbf{q}_{1}}\hspace{-1pt})\fwboxL{0pt}{\hspace{-1pt},}&(\hspace{-1pt}\g{\mathbf{q}_{1}}\hspace{1pt}{\mathbf{u}_{1}}\hspace{1pt}\g{\mathbf{q}_{2}}\hspace{-1pt})\fwboxL{0pt}{\hspace{-1pt},}&(\hspace{-1pt}\g{\mathbf{q}_{1}}\hspace{1pt}{\mathbf{u}_{3}}\hspace{1pt}\g{\mathbf{q}_{1}}\hspace{-1pt})\fwboxL{0pt}{\hspace{-1pt},}&(\hspace{-1pt}\g{\mathbf{q}_{2}}\hspace{1pt}{\mathbf{u}_{1}}\hspace{1pt}\g{\mathbf{q}_{1}}\hspace{-1pt})\fwboxL{0pt}{\hspace{-1pt},}&(\hspace{-1pt}\g{\mathbf{q}_{2}}\hspace{1pt}{\mathbf{u}_{1}}\hspace{1pt}\g{\mathbf{q}_{2}}\hspace{-1pt})\fwboxL{0pt}{\hspace{-1pt},}&(\hspace{-1pt}\g{\mathbf{q}_{2}}\hspace{1pt}{\mathbf{u}_{2}}\hspace{1pt}\g{\mathbf{q}_{2}}\hspace{-1pt})\fwboxL{5pt}{\!\big\}}&\multirow{1}{*}{$\fwboxL{0pt}{\hspace{4pt}6}$}\\[2pt]\cline{2-7}
\multirow{3}{*}{$\fwboxR{0pt}{\begin{array}{@{}r@{}}\mathfrak{b}_{\r{k}}\fwboxL{0pt}{,}\\\mathfrak{c}_{\r{k}}\fwboxL{0pt}{,}\\\mathfrak{d}_{\r{k}}\end{array}}$}&%
\fwboxR{5pt}{\big\{\!}\rule{0pt}{14pt}(\hspace{-1pt}\g{\mathbf{q}_{1}}\hspace{1pt}{\mathbf{u}_{1}}\hspace{1pt}\g{\mathbf{q}_{1}}\hspace{-1pt})\fwboxL{0pt}{\hspace{-1pt},}&(\hspace{-1pt}\g{\mathbf{q}_{1}}\hspace{1pt}{\mathbf{u}_{1}}\hspace{1pt}\g{\mathbf{q}_{2}}\hspace{-1pt})\fwboxL{0pt}{\hspace{-1pt},}&(\hspace{-1pt}\g{\mathbf{q}_{1}}\hspace{1pt}{\mathbf{u}_{1}}\hspace{1pt}\g{\mathbf{q}_{3}}\hspace{-1pt})\fwboxL{0pt}{\hspace{-1pt},}&(\hspace{-1pt}\g{\mathbf{q}_{2}}\hspace{1pt}{\mathbf{u}_{1}}\hspace{1pt}\g{\mathbf{q}_{1}}\hspace{-1pt})\fwboxL{0pt}{\hspace{-1pt},}&(\hspace{-1pt}\g{\mathbf{q}_{2}}\hspace{1pt}{\mathbf{u}_{1}}\hspace{1pt}\g{\mathbf{q}_{2}}\hspace{-1pt})\fwboxL{0pt}{\hspace{-1pt},}&(\hspace{-1pt}\g{\mathbf{q}_{2}}\hspace{1pt}{\mathbf{u}_{1}}\hspace{1pt}\g{\mathbf{q}_{3}}\hspace{-1pt})\fwboxL{0pt}{\hspace{-1pt},}&\multirow{3}{*}{$\fwboxL{0pt}{\hspace{4pt}15}$}\\[2pt]
&(\hspace{-1pt}\g{\mathbf{q}_{2}}\hspace{1pt}{\mathbf{u}_{2}}\hspace{1pt}\g{\mathbf{q}_{2}}\hspace{-1pt})\fwboxL{0pt}{\hspace{-1pt},}&(\hspace{-1pt}\g{\mathbf{q}_{2}}\hspace{1pt}{\mathbf{u}_{2}}\hspace{1pt}\g{\mathbf{q}_{3}}\hspace{-1pt})\fwboxL{0pt}{\hspace{-1pt},}&(\hspace{-1pt}\g{\mathbf{q}_{2}}\hspace{1pt}{\mathbf{u}_{4}}\hspace{1pt}\g{\mathbf{q}_{2}}\hspace{-1pt})\fwboxL{0pt}{\hspace{-1pt},}&(\hspace{-1pt}\g{\mathbf{q}_{3}}\hspace{1pt}{\mathbf{u}_{1}}\hspace{1pt}\g{\mathbf{q}_{1}}\hspace{-1pt})\fwboxL{0pt}{\hspace{-1pt},}&(\hspace{-1pt}\g{\mathbf{q}_{3}}\hspace{1pt}{\mathbf{u}_{1}}\hspace{1pt}\g{\mathbf{q}_{2}}\hspace{-1pt})\fwboxL{0pt}{\hspace{-1pt},}&(\hspace{-1pt}\g{\mathbf{q}_{3}}\hspace{1pt}{\mathbf{u}_{1}}\hspace{1pt}\g{\mathbf{q}_{3}}\hspace{-1pt})\fwboxL{0pt}{\hspace{-1pt},}\\[2pt]
&(\hspace{-1pt}\g{\mathbf{q}_{3}}\hspace{1pt}{\mathbf{u}_{2}}\hspace{1pt}\g{\mathbf{q}_{2}}\hspace{-1pt})\fwboxL{0pt}{\hspace{-1pt},}&(\hspace{-1pt}\g{\mathbf{q}_{3}}\hspace{1pt}{\mathbf{u}_{2}}\hspace{1pt}\g{\mathbf{q}_{3}}\hspace{-1pt})\fwboxL{0pt}{\hspace{-1pt},}&(\hspace{-1pt}\g{\mathbf{q}_{3}}\hspace{1pt}{\mathbf{u}_{3}}\hspace{1pt}\g{\mathbf{q}_{3}}\hspace{-1pt})\fwboxL{0pt}{\!\big\}}&&&\\[2pt]\cline{2-7}
\end{array}\hspace{-200pt}$$\vspace{-14pt}\end{table}

\newpage
Alternatively, one could consider a basis of Clebsch colour tensors generated by the tree graph:
\eq{\fwboxR{0pt}{\mathcal{C}_{\hspace{11.5pt}\mu}^{\,\smash{\mathbf{\g{q}}_{\g{i}}\,\mathbf{\t{r}}_{\t{j}}\,\mathbf{\g{q}}_{\g{k}}}}(\mathbf{\b{F}}\,\mathbf{\b{F}}\,\mathbf{\b{F}}|\mathbf{\b{F}}\,\mathbf{\b{F}}\,\mathbf{\b{F}})\;\bigger{\Leftrightarrow}\;}\tikzBox[-10pt]{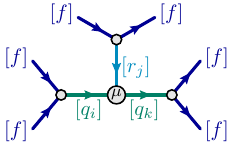}{\arrowTo[hblue]{0,0}{-130};\node[anchor=10,inner sep=2pt] at(in){{\footnotesize$\b{[f]}$}};\arrowTo[hblue]{0,0}{130}\node[anchor=-10,inner sep=2pt] at(in){{\footnotesize$\b{[f]}$}};\arrowFrom[hgreen]{0,0}[1.25]{0}\node[anchor=90,inner sep=2pt] at(arrownode){{\footnotesize$\g{[q_i]}$}};\node[clebsch]at(in){};\arrowTo[hteal]{end}[1.25]{90}\node[anchor=180,inner sep=2pt] at(arrownode){{\footnotesize${\t{[r_j]}}$}};\coordinate(v0)at(end);\arrowTo[hblue]{in}{150};\node[anchor=-10,inner sep=2pt] at(in){{\footnotesize$\b{[f]}$}};\arrowFrom[hblue]{end}{30}\node[anchor=-170,inner sep=2pt] at(end){{\footnotesize$\b{[f]}$}};\node[clebsch]at(in){};
\arrowFrom[hgreen]{v0}[1.25]{0}\node[clebsch]at(in){{\scriptsize$\phantom{\nu}$}};\node[]at(in){{\scriptsize$\mu$}};
\node[anchor=90,inner sep=2pt] at(arrownode){{\footnotesize$\g{[q_k]}$}};\arrowFrom[hblue]{end}{50}
\node[anchor=-170,inner sep=2pt] at(end){{\footnotesize$\b{[f]}$}};\arrowFrom[hblue]{in}{-50}\node[anchor=170,inner sep=2pt] at(end){{\footnotesize$\b{[f]}$}};\node[clebsch]at(in){};
}\fwboxL{0pt}{.}}
(There are of course many inequivalent tree graphs that one could use to define a basis of Clebsch colour tensors.) In the case of $\mathfrak{a}_{\r{1}}$ gauge theory, the five independent colour tensors in the $\mathcal{C}$ basis would be labelled by
%
%a1 C-basis:
\vspace{-4pt}\eq{\fwbox{0pt}{\fwboxL{435pt}{(\hspace{-1.25pt}%
\mathfrak{a}_{\r{1}}
)}}\fwbox{0pt}{\begin{array}{c}
%\multicolumn{1}{c}{\text{Spanning labels %
%$(\g{\mathbf{q}_{i}}\,\t{\mathbf{r}_j}\,\g{\mathbf{q}_{k}})$ of 
%%
%$%
%%
%\mathcal{C}_{\hspace{11.5pt}}^{\,\smash{\mathbf{\g{q}}_{\g{i}}\,\mathbf{\t{r}}_{\t{j}}\,\mathbf{\g{q}}_{\g{k}}}}(\mathbf{\b{F}}\,\mathbf{\b{F}}\,\mathbf{\b{F}}|\mathbf{\b{F}}\,\mathbf{\b{F}}\,\mathbf{\b{F}})
%%
%$
% for %
%$\mathfrak{a}_{\r{1}}$%
%%
%:}}\\\hline\hline
\rule{0pt}{14pt}%
\big\{(\hspace{-1pt}\mathbf{\g{1}}\hspace{-0pt}\mathbf{\t{1}}\hspace{-0pt}\mathbf{\g{1}}\hspace{-1pt}),(\hspace{-1pt}\mathbf{\g{1}}\hspace{-0pt}\mathbf{\t{3}}\hspace{-0pt}\mathbf{\g{3}}\hspace{-1pt}),(\hspace{-1pt}\mathbf{\g{3}}\hspace{-0pt}\mathbf{\t{1}}\hspace{-0pt}\mathbf{\g{3}}\hspace{-1pt}),(\hspace{-1pt}\mathbf{\g{3}}\hspace{-0pt}\mathbf{\t{3}}\hspace{-0pt}\mathbf{\g{1}}\hspace{-1pt}),(\hspace{-1pt}\mathbf{\g{3}}\hspace{-0pt}\mathbf{\t{3}}\hspace{-0pt}\mathbf{\g{3}}\hspace{-1pt})\big\}\fwboxL{0pt}{\,.}
%
%\\~
\end{array}}\label{ffffff_labels_b_basis_a1}\vspace{-4pt}}
The duality between these basis is encoded by the relation
\eq{\mathcal{C}^{\t{i}}\equivR\sum_{\g{j}}\mathbf{c}[\mathfrak{g}]\indices{\t{i}}{\g{j}}\mathcal{B}^{\g{j}}\,,\vspace{-10pt}}
for which the coefficients can be found to be
\vspace{5pt}\eq{\fwbox{0pt}{\fwboxL{550pt}{(\hspace{-1.25pt}%
\mathfrak{a}_{\r{1}}
)}}\hspace{-100pt}\fwboxR{0pt}{\mathbf{c}[\mathfrak{a}_{\r{1}}]\indices{\t{i}}{\g{j}}=}\left(\begin{array}{c@{$\;\;$}c@{$\;\;$}c@{$\;\;$}c@{$\;\;$}c}
\tmi1&\dzero&\dzero&\dzero&\dzero\\[-1pt]
\dzero&\tmi1&\dzero&\dzero&\dzero\\[-1pt]
\dzero&\dzero&\dzero&\tmi\frac{2}{3}&\tmi1\\[-1pt]
\dzero&\dzero&\tmi1&\dzero&\dzero\\[-1pt]
\dzero&\dzero&\dzero&\frac{4}{3}&\tmi1\end{array}\right)\fwboxL{0pt}{.}\hspace{-100pt}\label{ffffff_a1_duality_c_into_b}}

\newpage
\subsubsection{\texorpdfstring{Clebsch Colour Bases for $C(\mathbf{\b{F}}\,\mathbf{\b{F}}\,\mathbf{\r{ad}}|\mathbf{\r{ad}}\,\mathbf{\b{F}}\,\mathbf{\b{F}})$}{Clebsch Colour Bases for C(FFg|gFF)}}

For four fundamentals and two adjoints, one may consider the basis of Clebsch colour tensors given by
\eq{\fwboxR{0pt}{\mathcal{B}^{\,\mathbf{\g{q}}_{\g{i}}\,\mathbf{v}_j\,\mathbf{\g{q}}_{\g{k}}}(\mathbf{\b{F}}\,\mathbf{\b{F}}\,\mathbf{\r{ad}}|\mathbf{\r{ad}}\,\mathbf{\b{F}}\,\mathbf{\b{F}})\;\bigger{\Leftrightarrow}\;}\tikzBox[-5.25pt]{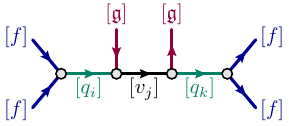}{\arrowTo[hblue]{0,0}{-130};\node[anchor=10,inner sep=2pt] at(in){{\footnotesize$\b{[f]}$}};\arrowTo[hblue]{0,0}{130}\node[anchor=-10,inner sep=2pt] at(in){{\footnotesize$\b{[f]}$}};\arrowFrom[hgreen]{0,0}[1.25]{0}\node[anchor=90,inner sep=2pt] at(arrownode){{\footnotesize$\g{[q_i]}$}};\node[clebsch]at(in){};\arrowTo[hred]{end}{90}\node[anchor=-90,inner sep=2pt] at(in){{\footnotesize$\r{[\adR]}$}};\arrowFrom[black]{end}[1.25]{0}\node[clebsch]at(in){};\node[anchor=90,inner sep=2pt] at(arrownode){{\footnotesize${[v_j]}$}};\arrowFrom[hred]{end}{90}\node[anchor=-90,inner sep=2pt] at(end){{\footnotesize$\r{[\adR]}$}};\arrowFrom[hgreen]{in}[1.25]{0}\node[clebsch]at(in){};\node[anchor=90,inner sep=2pt] at(arrownode){{\footnotesize$\g{[q_k]}$}};
\arrowFrom[hblue]{end}{50}
\node[anchor=-170,inner sep=2pt] at(end){{\footnotesize$\b{[f]}$}};\arrowFrom[hblue]{in}{-50}\node[anchor=170,inner sep=2pt] at(end){{\footnotesize$\b{[f]}$}};\node[clebsch]at(in){};
}\fwboxL{0pt}{.}
\label{ffggff_basis_defined}}
These are labelled by the irreducible representations $(\g{\mathbf{q}_i}\,\mathbf{v}_j\,\g{\mathbf{q}_k})$. (For none of the simple Lie algebras do any Clebsch coefficients require multiplicity indices.) The \emph{new} representations $\mathbf{v}_j$ appearing in the tensor product $\g{\mathbf{q}_i}\!\otimes\!\mathbf{\r{ad}}$ (not already identified) are defined in \mbox{Table~\ref{vReps_table}} for each of the simple Lie algebras, and the sets of labels spanning the basis (\ref{ffggff_basis_defined}) are given in \mbox{Table~\ref{ffggff_basis_b_label_table}}.

\begin{table}[t]\vspace{-10pt}\caption{Irreducible representations appearing in the decomposition of the tensor product $\g{\mathbf{q}_{j}}\!\otimes\!\mathbf{\r{ad}}\!\supset\!\bigoplus_{{i}}\mathbf{{v}}_{{i}}$---beyond those of $\mathbf{\g{q}}_{\g{j}}$---for simple Lie algebras $\mathfrak{a}_{\r{k}},\mathfrak{b}_{\r{k}},\mathfrak{c}_{\r{k}},\mathfrak{d}_{\r{k}}$.}
\vspace{-25pt}$$\begin{array}{|l@{$\,$}|cccc|}\multicolumn{1}{c}{~}&\fwbox{40pt}{{\mathbf{v}_{1}}}&\fwbox{80pt}{{\mathbf{v}_{2}}}&\multicolumn{1}{c}{\fwbox{80pt}{{\mathbf{v}_{3}}}}&\multicolumn{1}{c}{\fwbox{80pt}{{\mathbf{v}_{4}}}}\\\hline\hline
\raisebox{-4pt}{$\mathfrak{a}_{\r{k}}$}&\dynkLabel{\frac{1}{6}\r{k}(\hspace{-1pt}\r{k}\pl1\hspace{-1pt})(\hspace{-1pt}\r{k}\pl2\hspace{-1pt})(\hspace{-1pt}\r{k}\pl4\hspace{-1pt})}{30\cdots01}&\dynkLabel{\frac{1}{3}(\hspace{-1pt}\r{k}\pl1\hspace{-1pt})^2(\hspace{-1pt}\r{k}\mi1\hspace{-1pt})(\hspace{-1pt}\r{k}\pl3\hspace{-1pt})}{110\cdots01}&\dynkLabel{\frac{1}{6}\r{k}(\hspace{-1pt}\r{k}\pl1\hspace{-1pt})(\hspace{-1pt}\r{k}\pl2\hspace{-1pt})(\hspace{-1pt}\r{k}\mi2\hspace{-1pt})}{0010\cdots01}&\\\hline
\raisebox{-4pt}{$\mathfrak{b}_{\r{k}}$}&\dynkLabel{\frac{1}{2}\r{k}(\hspace{-1pt}2\r{k}\pl1\hspace{-1pt})(\hspace{-1pt}2\r{k}\pl3\hspace{-1pt})(\hspace{-1pt}\r{k}\mi1\hspace{-1pt})}{1010\cdots0}&\dynkLabel{\frac{1}{3}(\hspace{-1pt}2\r{k}\pl1\hspace{-1pt})(\hspace{-1pt}2\r{k}\pl3\hspace{-1pt})(\hspace{-1pt}\r{k}^2\mi1\hspace{-1pt})}{020\cdots0}&\dynkLabel{\frac{1}{6}\r{k}(\hspace{-1pt}4\r{k}^2\mi1\hspace{-1pt})(\hspace{-1pt}\r{k}\mi1\hspace{-1pt})}{00010\cdots0}&\dynkLabel{\frac{1}{2}\r{k}(\hspace{-1pt}\r{k}\pl1\hspace{-1pt})(\hspace{-1pt}2\r{k}\mi1\hspace{-1pt})(\hspace{-1pt}2\r{k}\pl5\hspace{-1pt})}{210\cdots0}
\\\hline
\raisebox{-4pt}{$\mathfrak{c}_{\r{k}}$}&\dynkLabel{\frac{1}{2}\r{k}(\hspace{-1pt}2\r{k}\pl1\hspace{-1pt})(\hspace{-1pt}2\r{k}\pl3\hspace{-1pt})(\hspace{-1pt}\r{k}\mi1\hspace{-1pt})}{210\cdots0}&\dynkLabel{\frac{1}{3}\r{k}(\hspace{-1pt}4\r{k}^3\mi7\r{k}\pl3\hspace{-1pt})}{020\cdots0}&\dynkLabel{\frac{1}{6}\r{k}(\hspace{-1pt}\r{k}\pl1\hspace{-1pt})(\hspace{-1pt}2\r{k}\pl1\hspace{-1pt})(\hspace{-1pt}2\r{k}\pl3\hspace{-1pt})}{40\cdots0}&\dynkLabel{\frac{1}{2}(\hspace{-1pt}\r{k}\pl1\hspace{-1pt})(\hspace{-1pt}\r{k}\mi2\hspace{-1pt})(\hspace{-1pt}4\r{k}^2\mi1\hspace{-1pt})}{1010\cdots0}
\\\hline
\raisebox{-4pt}{$\mathfrak{d}_{\r{k}}$}&\dynkLabel{\frac{1}{2}\r{k}(\hspace{-1pt}\r{k}\pl1\hspace{-1pt})(\hspace{-1pt}2\r{k}\mi1\hspace{-1pt})(\hspace{-1pt}2\r{k}\mi3\hspace{-1pt})}{1010\cdots0}
&\dynkLabel{\frac{1}{3}\r{k}(\hspace{-1pt}4\r{k}^3\mi7\r{k}\mi3\hspace{-1pt})}{020\cdots0}
&\dynkLabel{\frac{1}{6}\r{k}(\hspace{-1pt}2\r{k}\mi1\hspace{-1pt})(\hspace{-1pt}2\r{k}\mi3\hspace{-1pt})(\hspace{-1pt}\r{k}\mi1\hspace{-1pt})}{00010\cdots0}
&\dynkLabel{\frac{1}{2}(\hspace{-1pt}\r{k}\pl1\hspace{-1pt})(\hspace{-1pt}\r{k}\mi1\hspace{-1pt})(\hspace{-1pt}4\r{k}^2\mi1\hspace{-1pt})}{2110\cdots0}
\\\hline
\end{array}
$$\vspace{-15pt}\vspace{-0pt}\label{vReps_table}\end{table}

\begin{table}[b]\vspace{-10pt}
\vspace{-15pt}
$$\hspace{-200pt}\begin{array}{r@{$\;\;$}|rl@{$\,\,$}l@{$\,\,$}l@{$\,\,$}l@{$\,\,$}l@{$\,\,$}l@{$\,\,$}l|l}
%\multicolumn{1}{r}{\fwboxR{0pt}{\mathfrak{g}\hspace{16pt}}}&\multicolumn{9}{r}{\fwboxL{0pt}{\hspace{3pt}\text{\#}}}\\\cline{2-9}
\multicolumn{1}{r}{\fwboxR{0pt}{\mathfrak{g}\hspace{22pt}}}&\multicolumn{9}{r}{\fwboxL{0pt}{\hspace{13pt}\text{\#}}}\\[-10pt]\cline{2-9}
\multirow{1}{*}{$\fwboxR{0pt}{\begin{array}{@{}r@{}}\mathfrak{a}_{\r{1}}\end{array}}$}&
\fwboxR{5pt}{\big\{\!}\rule{0pt}{14pt}(\hspace{-1pt}\g{\mathbf{q}_{1}}\hspace{1pt}{\mathbf{q}_{2}}\hspace{1pt}\g{\mathbf{q}_{1}}\hspace{-1pt})\fwboxL{0pt}{\hspace{-1pt},}&(\hspace{-1pt}\g{\mathbf{q}_{1}}\hspace{1pt}{\mathbf{q}_{2}}\hspace{1pt}\g{\mathbf{q}_{2}}\hspace{-1pt})\fwboxL{0pt}{\hspace{-1pt},}&(\hspace{-1pt}\g{\mathbf{q}_{2}}\hspace{1pt}{\mathbf{q}_{1}}\hspace{1pt}\g{\mathbf{q}_{2}}\hspace{-1pt})\fwboxL{0pt}{\hspace{-1pt},}&(\hspace{-1pt}\g{\mathbf{q}_{2}}\hspace{1pt}{\mathbf{q}_{2}}\hspace{1pt}\g{\mathbf{q}_{1}}\hspace{-1pt})\fwboxL{0pt}{\hspace{-1pt},}&(\hspace{-1pt}\g{\mathbf{q}_{2}}\hspace{1pt}{\mathbf{q}_{2}}\hspace{1pt}\g{\mathbf{q}_{2}}\hspace{-1pt})\fwboxL{0pt}{\hspace{-1pt},}&(\hspace{-1pt}\g{\mathbf{q}_{2}}\hspace{1pt}{\mathbf{v}_{1}}\hspace{1pt}\g{\mathbf{q}_{2}}\hspace{-1pt})\fwboxL{0pt}{\!\big\}}&&&\multirow{1}{*}{$\fwboxL{0pt}{\hspace{4pt}6}$}\\[2pt]\cline{2-9}
\multirow{2}{*}{$\fwboxR{0pt}{\mathfrak{a}_{\r{2}}}$}&%
\fwboxR{5pt}{\big\{\!}\rule{0pt}{14pt}(\hspace{-1pt}\g{\mathbf{q}_{1}}\hspace{1pt}{\mathbf{q}_{1}}\hspace{1pt}\g{\mathbf{q}_{1}}\hspace{-1pt})\fwboxL{0pt}{\hspace{-1pt},}&(\hspace{-1pt}\g{\mathbf{q}_{1}}\hspace{1pt}{\mathbf{q}_{1}}\hspace{1pt}\g{\mathbf{q}_{2}}\hspace{-1pt})\fwboxL{0pt}{\hspace{-1pt},}&(\hspace{-1pt}\g{\mathbf{q}_{1}}\hspace{1pt}{\mathbf{q}_{2}}\hspace{1pt}\g{\mathbf{q}_{1}}\hspace{-1pt})\fwboxL{0pt}{\hspace{-1pt},}&(\hspace{-1pt}\g{\mathbf{q}_{1}}\hspace{1pt}{\mathbf{q}_{2}}\hspace{1pt}\g{\mathbf{q}_{2}}\hspace{-1pt})\fwboxL{0pt}{\hspace{-1pt},}&(\hspace{-1pt}\g{\mathbf{q}_{1}}\hspace{1pt}{\mathbf{v}_{2}}\hspace{1pt}\g{\mathbf{q}_{1}}\hspace{-1pt})\fwboxL{0pt}{\hspace{-1pt},}&(\hspace{-1pt}\g{\mathbf{q}_{1}}\hspace{1pt}{\mathbf{v}_{2}}\hspace{1pt}\g{\mathbf{q}_{2}}\hspace{-1pt})\fwboxL{0pt}{\hspace{-1pt},}&(\hspace{-1pt}\g{\mathbf{q}_{2}}\hspace{1pt}{\mathbf{q}_{1}}\hspace{1pt}\g{\mathbf{q}_{1}}\hspace{-1pt})&&\multirow{2}{*}{$\fwboxL{0pt}{\hspace{4pt}13}$}\\[2pt]
&(\hspace{-1pt}\g{\mathbf{q}_{2}}\hspace{1pt}{\mathbf{q}_{1}}\hspace{1pt}\g{\mathbf{q}_{2}}\hspace{-1pt})\fwboxL{0pt}{\hspace{-1pt},}&(\hspace{-1pt}\g{\mathbf{q}_{2}}\hspace{1pt}{\mathbf{q}_{2}}\hspace{1pt}\g{\mathbf{q}_{1}}\hspace{-1pt})\fwboxL{0pt}{\hspace{-1pt},}&(\hspace{-1pt}\g{\mathbf{q}_{2}}\hspace{1pt}{\mathbf{q}_{2}}\hspace{1pt}\g{\mathbf{q}_{2}}\hspace{-1pt})\fwboxL{0pt}{\hspace{-1pt},}&(\hspace{-1pt}\g{\mathbf{q}_{2}}\hspace{1pt}{\mathbf{v}_{1}}\hspace{1pt}\g{\mathbf{q}_{2}}\hspace{-1pt})\fwboxL{0pt}{\hspace{-1pt},}&(\hspace{-1pt}\g{\mathbf{q}_{2}}\hspace{1pt}{\mathbf{v}_{2}}\hspace{1pt}\g{\mathbf{q}_{1}}\hspace{-1pt})\fwboxL{0pt}{\hspace{-1pt},}&(\hspace{-1pt}\g{\mathbf{q}_{2}}\hspace{1pt}{\mathbf{v}_{2}}\hspace{1pt}\g{\mathbf{q}_{2}}\hspace{-1pt})\fwboxL{0pt}{\!\big\}}&&\\[2pt]\cline{2-9}
\multirow{2}{*}{$\fwboxR{0pt}{\mathfrak{a}_{\r{k}>2}}$}&%
\fwboxR{5pt}{\big\{\!}\rule{0pt}{14pt}(\hspace{-1pt}\g{\mathbf{q}_{1}}\hspace{1pt}{\mathbf{q}_{1}}\hspace{1pt}\g{\mathbf{q}_{1}}\hspace{-1pt})\fwboxL{0pt}{\hspace{-1pt},}&(\hspace{-1pt}\g{\mathbf{q}_{1}}\hspace{1pt}{\mathbf{q}_{1}}\hspace{1pt}\g{\mathbf{q}_{2}}\hspace{-1pt})\fwboxL{0pt}{\hspace{-1pt},}&(\hspace{-1pt}\g{\mathbf{q}_{1}}\hspace{1pt}{\mathbf{q}_{2}}\hspace{1pt}\g{\mathbf{q}_{1}}\hspace{-1pt})\fwboxL{0pt}{\hspace{-1pt},}&(\hspace{-1pt}\g{\mathbf{q}_{1}}\hspace{1pt}{\mathbf{q}_{2}}\hspace{1pt}\g{\mathbf{q}_{2}}\hspace{-1pt})\fwboxL{0pt}{\hspace{-1pt},}&(\hspace{-1pt}\g{\mathbf{q}_{1}}\hspace{1pt}{\mathbf{v}_{2}}\hspace{1pt}\g{\mathbf{q}_{1}}\hspace{-1pt})\fwboxL{0pt}{\hspace{-1pt},}&(\hspace{-1pt}\g{\mathbf{q}_{1}}\hspace{1pt}{\mathbf{v}_{2}}\hspace{1pt}\g{\mathbf{q}_{2}}\hspace{-1pt})\fwboxL{0pt}{\hspace{-1pt},}&(\hspace{-1pt}\g{\mathbf{q}_{1}}\hspace{1pt}{\mathbf{v}_{3}}\hspace{1pt}\g{\mathbf{q}_{1}}\hspace{-1pt})\fwboxL{0pt}{\hspace{-1pt},}&&\multirow{2}{*}{$\fwboxL{0pt}{\hspace{4pt}14}$}\\[2pt]
&(\hspace{-1pt}\g{\mathbf{q}_{2}}\hspace{1pt}{\mathbf{q}_{1}}\hspace{1pt}\g{\mathbf{q}_{1}}\hspace{-1pt})\fwboxL{0pt}{\hspace{-1pt},}&(\hspace{-1pt}\g{\mathbf{q}_{2}}\hspace{1pt}{\mathbf{q}_{1}}\hspace{1pt}\g{\mathbf{q}_{2}}\hspace{-1pt})\fwboxL{0pt}{\hspace{-1pt},}&(\hspace{-1pt}\g{\mathbf{q}_{2}}\hspace{1pt}{\mathbf{q}_{2}}\hspace{1pt}\g{\mathbf{q}_{1}}\hspace{-1pt})\fwboxL{0pt}{\hspace{-1pt},}&(\hspace{-1pt}\g{\mathbf{q}_{2}}\hspace{1pt}{\mathbf{q}_{2}}\hspace{1pt}\g{\mathbf{q}_{2}}\hspace{-1pt})\fwboxL{0pt}{\hspace{-1pt},}&(\hspace{-1pt}\g{\mathbf{q}_{2}}\hspace{1pt}{\mathbf{v}_{1}}\hspace{1pt}\g{\mathbf{q}_{2}}\hspace{-1pt})\fwboxL{0pt}{\hspace{-1pt},}&(\hspace{-1pt}\g{\mathbf{q}_{2}}\hspace{1pt}{\mathbf{v}_{2}}\hspace{1pt}\g{\mathbf{q}_{1}}\hspace{-1pt})\fwboxL{0pt}{\hspace{-1pt},}&(\hspace{-1pt}\g{\mathbf{q}_{2}}\hspace{1pt}{\mathbf{v}_{2}}\hspace{1pt}\g{\mathbf{q}_{2}}\hspace{-1pt})\fwboxL{0pt}{\!\big\}}&\\[2pt]\cline{2-9}
\multirow{3}{*}{$\fwboxR{0pt}{\begin{array}{@{}r@{}}\mathfrak{b}_{\r{k}}\fwboxL{0pt}{,}\\\mathfrak{c}_{\r{k}}\fwboxL{0pt}{,}\\\mathfrak{d}_{\r{k}>4}\end{array}}$}&%
\fwboxR{5pt}{\big\{\!}\rule{0pt}{14pt}(\hspace{-1pt}\g{\mathbf{q}_{1}}\hspace{1pt}{\mathbf{q}_{2}}\hspace{1pt}\g{\mathbf{q}_{1}}\hspace{-1pt})\fwboxL{0pt}{\hspace{-1pt},}&(\hspace{-1pt}\g{\mathbf{q}_{1}}\hspace{1pt}{\mathbf{q}_{2}}\hspace{1pt}\g{\mathbf{q}_{2}}\hspace{-1pt})\fwboxL{0pt}{\hspace{-1pt},}&(\hspace{-1pt}\g{\mathbf{q}_{1}}\hspace{1pt}{\mathbf{q}_{2}}\hspace{1pt}\g{\mathbf{q}_{3}}\hspace{-1pt})\fwboxL{0pt}{\hspace{-1pt},}&(\hspace{-1pt}\g{\mathbf{q}_{2}}\hspace{1pt}{\mathbf{q}_{1}}\hspace{1pt}\g{\mathbf{q}_{2}}\hspace{-1pt})\fwboxL{0pt}{\hspace{-1pt},}&(\hspace{-1pt}\g{\mathbf{q}_{2}}\hspace{1pt}{\mathbf{q}_{2}}\hspace{1pt}\g{\mathbf{q}_{1}}\hspace{-1pt})\fwboxL{0pt}{\hspace{-1pt},}&(\hspace{-1pt}\g{\mathbf{q}_{2}}\hspace{1pt}{\mathbf{q}_{2}}\hspace{1pt}\g{\mathbf{q}_{2}}\hspace{-1pt})\fwboxL{0pt}{\hspace{-1pt},}&(\hspace{-1pt}\g{\mathbf{q}_{2}}\hspace{1pt}{\mathbf{q}_{2}}\hspace{1pt}\g{\mathbf{q}_{3}}\hspace{-1pt})\fwboxL{0pt}{\hspace{-1pt},}&&\multirow{3}{*}{$\fwboxL{0pt}{\hspace{4pt}21}$}\\[2pt]
&(\hspace{-1pt}\g{\mathbf{q}_{2}}\hspace{1pt}{\mathbf{q}_{3}}\hspace{1pt}\g{\mathbf{q}_{2}}\hspace{-1pt})\fwboxL{0pt}{\hspace{-1pt},}&(\hspace{-1pt}\g{\mathbf{q}_{2}}\hspace{1pt}{\mathbf{q}_{3}}\hspace{1pt}\g{\mathbf{q}_{3}}\hspace{-1pt})\fwboxL{0pt}{\hspace{-1pt},}&(\hspace{-1pt}\g{\mathbf{q}_{2}}\hspace{1pt}{\mathbf{v}_{1}}\hspace{1pt}\g{\mathbf{q}_{2}}\hspace{-1pt})\fwboxL{0pt}{\hspace{-1pt},}&(\hspace{-1pt}\g{\mathbf{q}_{2}}\hspace{1pt}{\mathbf{v}_{1}}\hspace{1pt}\g{\mathbf{q}_{3}}\hspace{-1pt})\fwboxL{0pt}{\hspace{-1pt},}&(\hspace{-1pt}\g{\mathbf{q}_{2}}\hspace{1pt}{\mathbf{v}_{2}}\hspace{1pt}\g{\mathbf{q}_{2}}\hspace{-1pt})\fwboxL{0pt}{\hspace{-1pt},}&(\hspace{-1pt}\g{\mathbf{q}_{2}}\hspace{1pt}{\mathbf{v}_{3}}\hspace{1pt}\g{\mathbf{q}_{2}}\hspace{-1pt})\fwboxL{0pt}{\hspace{-1pt},}&(\hspace{-1pt}\g{\mathbf{q}_{3}}\hspace{1pt}{\mathbf{q}_{2}}\hspace{1pt}\g{\mathbf{q}_{1}}\hspace{-1pt})\fwboxL{0pt}{\hspace{-1pt},}&&\\[2pt]
&(\hspace{-1pt}\g{\mathbf{q}_{3}}\hspace{1pt}{\mathbf{q}_{2}}\hspace{1pt}\g{\mathbf{q}_{2}}\hspace{-1pt})\fwboxL{0pt}{\hspace{-1pt},}&(\hspace{-1pt}\g{\mathbf{q}_{3}}\hspace{1pt}{\mathbf{q}_{2}}\hspace{1pt}\g{\mathbf{q}_{3}}\hspace{-1pt})\fwboxL{0pt}{\hspace{-1pt},}&(\hspace{-1pt}\g{\mathbf{q}_{3}}\hspace{1pt}{\mathbf{q}_{3}}\hspace{1pt}\g{\mathbf{q}_{2}}\hspace{-1pt})\fwboxL{0pt}{\hspace{-1pt},}&(\hspace{-1pt}\g{\mathbf{q}_{3}}\hspace{1pt}{\mathbf{q}_{3}}\hspace{1pt}\g{\mathbf{q}_{3}}\hspace{-1pt})\fwboxL{0pt}{\hspace{-1pt},}&(\hspace{-1pt}\g{\mathbf{q}_{3}}\hspace{1pt}{\mathbf{v}_{1}}\hspace{1pt}\g{\mathbf{q}_{2}}\hspace{-1pt})\fwboxL{0pt}{\hspace{-1pt},}&(\hspace{-1pt}\g{\mathbf{q}_{3}}\hspace{1pt}{\mathbf{v}_{1}}\hspace{1pt}\g{\mathbf{q}_{3}}\hspace{-1pt})\fwboxL{0pt}{\hspace{-1pt},}&(\hspace{-1pt}\g{\mathbf{q}_{3}}\hspace{1pt}{\mathbf{v}_{4}}\hspace{1pt}\g{\mathbf{q}_{3}}\hspace{-1pt})\fwboxL{0pt}{\!\big\}}&\\[2pt]\cline{2-9}
\multirow{3}{*}{$\fwboxR{0pt}{\begin{array}{@{}r@{}}\mathfrak{d}_{\r{4}}\end{array}}$}&%
\fwboxR{5pt}{\big\{\!}\rule{0pt}{14pt}(\hspace{-1pt}\g{\mathbf{q}_{1}}\hspace{1pt}{\mathbf{q}_{2}}\hspace{1pt}\g{\mathbf{q}_{1}}\hspace{-1pt})\fwboxL{0pt}{\hspace{-1pt},}&(\hspace{-1pt}\g{\mathbf{q}_{1}}\hspace{1pt}{\mathbf{q}_{2}}\hspace{1pt}\g{\mathbf{q}_{2}}\hspace{-1pt})\fwboxL{0pt}{\hspace{-1pt},}&(\hspace{-1pt}\g{\mathbf{q}_{1}}\hspace{1pt}{\mathbf{q}_{2}}\hspace{1pt}\g{\mathbf{q}_{3}}\hspace{-1pt})\fwboxL{0pt}{\hspace{-1pt},}&(\hspace{-1pt}\g{\mathbf{q}_{2}}\hspace{1pt}{\mathbf{q}_{1}}\hspace{1pt}\g{\mathbf{q}_{2}}\hspace{-1pt})\fwboxL{0pt}{\hspace{-1pt},}&(\hspace{-1pt}\g{\mathbf{q}_{2}}\hspace{1pt}{\mathbf{q}_{2}}\hspace{1pt}\g{\mathbf{q}_{1}}\hspace{-1pt})\fwboxL{0pt}{\hspace{-1pt},}&(\hspace{-1pt}\g{\mathbf{q}_{2}}\hspace{1pt}{\mathbf{q}_{2}}\hspace{1pt}\g{\mathbf{q}_{2}}\hspace{-1pt})\fwboxL{0pt}{\hspace{-1pt},}&(\hspace{-1pt}\g{\mathbf{q}_{2}}\hspace{1pt}{\mathbf{q}_{2}}\hspace{1pt}\g{\mathbf{q}_{3}}\hspace{-1pt})\fwboxL{0pt}{\hspace{-1pt},}&(\hspace{-1pt}\g{\mathbf{q}_{2}}\hspace{1pt}{\mathbf{q}_{3}}\hspace{1pt}\g{\mathbf{q}_{2}}\hspace{-1pt})\fwboxL{5pt}{\hspace{-1pt},}&\multirow{3}{*}{$\fwboxL{0pt}{\hspace{4pt}22}$}\\[2pt]
&(\hspace{-1pt}\g{\mathbf{q}_{2}}\hspace{1pt}{\mathbf{q}_{3}}\hspace{1pt}\g{\mathbf{q}_{3}}\hspace{-1pt})\fwboxL{0pt}{\hspace{-1pt},}&(\hspace{-1pt}\g{\mathbf{q}_{2}}\hspace{1pt}{\mathbf{v}_{1}}\hspace{1pt}\g{\mathbf{q}_{2}}\hspace{-1pt})\fwboxL{0pt}{\hspace{-1pt},}&(\hspace{-1pt}\g{\mathbf{q}_{2}}\hspace{1pt}{\mathbf{v}_{1}}\hspace{1pt}\g{\mathbf{q}_{3}}\hspace{-1pt})\fwboxL{0pt}{\hspace{-1pt},}&(\hspace{-1pt}\g{\mathbf{q}_{2}}\hspace{1pt}{\mathbf{v}_{2}}\hspace{1pt}\g{\mathbf{q}_{2}}\hspace{-1pt})\fwboxL{0pt}{\hspace{-1pt},}&(\hspace{-1pt}\g{\mathbf{q}_{2}}\hspace{1pt}{\mathbf{t}_{6}}\hspace{1pt}\g{\mathbf{q}_{2}}\hspace{-1pt})\fwboxL{0pt}{\hspace{-1pt},}&(\hspace{-1pt}\g{\mathbf{q}_{2}}\hspace{1pt}{\mathbf{t}_{7}}\hspace{1pt}\g{\mathbf{q}_{2}}\hspace{-1pt})\fwboxL{0pt}{\hspace{-1pt},}&(\hspace{-1pt}\g{\mathbf{q}_{3}}\hspace{1pt}{\mathbf{q}_{2}}\hspace{1pt}\g{\mathbf{q}_{1}}\hspace{-1pt})\fwboxL{0pt}{\hspace{-1pt},}&(\hspace{-1pt}\g{\mathbf{q}_{3}}\hspace{1pt}{\mathbf{q}_{2}}\hspace{1pt}\g{\mathbf{q}_{2}}\hspace{-1pt})\fwboxL{0pt}{\hspace{-1pt},}&\\[2pt]
&(\hspace{-1pt}\g{\mathbf{q}_{3}}\hspace{1pt}{\mathbf{q}_{2}}\hspace{1pt}\g{\mathbf{q}_{3}}\hspace{-1pt})\fwboxL{0pt}{\hspace{-1pt},}&(\hspace{-1pt}\g{\mathbf{q}_{3}}\hspace{1pt}{\mathbf{q}_{3}}\hspace{1pt}\g{\mathbf{q}_{2}}\hspace{-1pt})\fwboxL{0pt}{\hspace{-1pt},}&(\hspace{-1pt}\g{\mathbf{q}_{3}}\hspace{1pt}{\mathbf{q}_{3}}\hspace{1pt}\g{\mathbf{q}_{3}}\hspace{-1pt})\fwboxL{0pt}{\hspace{-1pt},}&(\hspace{-1pt}\g{\mathbf{q}_{3}}\hspace{1pt}{\mathbf{v}_{1}}\hspace{1pt}\g{\mathbf{q}_{2}}\hspace{-1pt})\fwboxL{0pt}{\hspace{-1pt},}&(\hspace{-1pt}\g{\mathbf{q}_{3}}\hspace{1pt}{\mathbf{v}_{1}}\hspace{1pt}\g{\mathbf{q}_{3}}\hspace{-1pt})\fwboxL{0pt}{\hspace{-1pt},}&(\hspace{-1pt}\g{\mathbf{q}_{3}}\hspace{1pt}{\mathbf{v}_{4}}\hspace{1pt}\g{\mathbf{q}_{3}}\hspace{-1pt})\fwboxL{0pt}{\!\big\}}&&\\[2pt]\cline{2-9}
\end{array}\hspace{-200pt}$$\vspace{-20pt}\caption{Labels $(\hspace{-1pt}\g{\mathbf{q}_i}\,\mathbf{v}_j\,\g{\mathbf{q}_k}\hspace{-1pt})$ for tensors $\mathcal{B}^{\,\mathbf{\g{q}}_{\g{i}}\,\mathbf{v}_j\,\mathbf{\g{q}}_{\g{k}}}(\mathbf{\b{F}}\,\mathbf{\b{F}}\,\mathbf{\r{ad}}|\mathbf{\r{ad}}\,\mathbf{\b{F}}\,\mathbf{\b{F}})$ of simple Lie algebras.}\label{ffggff_basis_b_label_table}\end{table}

\newpage

Alternatively, one could consider the basis of Clebsch colour tensors defined by the tree graph
\eq{\fwboxR{0pt}{\mathcal{C}_{\hspace{10.5pt}\mu}^{\,\smash{\mathbf{\t{r}}_{\t{i}}\,\mathbf{\g{s}}_{\g{j}}\,\mathbf{\g{s}}_{\g{k}}}}(\mathbf{\b{F}}\,\mathbf{\b{F}}\,\mathbf{\r{ad}}|\mathbf{\r{ad}}\,\mathbf{\b{F}}\,\mathbf{\b{F}})\;\bigger{\Leftrightarrow}\;}\tikzBox{ffggff_alt_basis_diagram}{
\arrowTo[hblue]{0,0}{-150}\node[anchor=10,inner sep=2pt] at(in){{\footnotesize$\b{[f]}$}};\arrowFrom[hblue]{0,0}{-30};\node[anchor=170,inner sep=2pt] at(end){{\footnotesize$\b{[f]}$}};\arrowFrom[hteal]{0,0}[1]{90}\node[clebsch]at(in){};\node[anchor=180,inner sep=2pt] at(arrownode){{\footnotesize$\t{[r_i]}$}};\arrowTo[hgreen]{end}{180}\node[anchor=-90,inner sep=2pt] at(arrownode){{\footnotesize$\g{[s_j]}$}};%\node[anchor=0,inner sep=0pt] at(in){{\footnotesize$\r{[\adR]}$}};
\coordinate(up)at(end);
\arrowTo[hblue]{in}{230};\node[anchor=10,inner sep=2pt] at(in){{\footnotesize$\b{[f]}$}};
\arrowTo[hred]{end}{130};\node[anchor=-10,inner sep=2pt] at(in){{\footnotesize$\r{[\adR]}$}};\node[clebsch]at(end){};
\arrowFrom[hgreen]{up}{0}\node[anchor=-90,inner sep=2pt] at(arrownode){{\footnotesize$\g{[s_k]}$}};
\arrowFrom[hred]{end}{50};\node[anchor=-170,inner sep=2pt] at(end){{\footnotesize$\r{[\adR]}$}};
\arrowFrom[hblue]{in}{-50};\node[anchor=170,inner sep=2pt] at(end){{\footnotesize$\b{[f]}$}};\node[clebsch]at(in){};
\node[clebsch]at(up){{\scriptsize$\phantom{\nu}$}};\node[]at(up){{\scriptsize${\mu}$}};
}\fwboxL{0pt}{.}}
Although it would be cumbersome to enumerate the spanning sets of labels for each of the Lie algebras, in the case of $\mathfrak{a}_{\r{1}}$ the 6 independent colour tensors would be labelled in the $\mathcal{C}$ basis by the irreducible representations
%a1 C-basis:
\vspace{-4pt}\eq{\fwbox{0pt}{\fwboxL{435pt}{(\hspace{-1.25pt}%
\mathfrak{a}_{\r{1}}
)}}\fwbox{0pt}{\begin{array}{c}
\rule{0pt}{14pt}%
\big\{(\hspace{-1pt}\mathbf{\t{1}}\hspace{-0pt}\mathbf{\g{2}}\hspace{-0pt}\mathbf{\g{2}}\hspace{-1pt}),(\hspace{-1pt}\mathbf{\t{1}}\hspace{-0pt}\mathbf{\g{4}}\hspace{-0pt}\mathbf{\g{4}}\hspace{-1pt}),(\hspace{-1pt}\mathbf{\t{3}}\hspace{-0pt}\mathbf{\g{2}}\hspace{-0pt}\mathbf{\g{2}}\hspace{-1pt}),(\hspace{-1pt}\mathbf{\t{3}}\hspace{-0pt}\mathbf{\g{2}}\hspace{-0pt}\mathbf{\g{4}}\hspace{-1pt}),(\hspace{-1pt}\mathbf{\t{3}}\hspace{-0pt}\mathbf{\g{4}}\hspace{-0pt}\mathbf{\g{2}}\hspace{-1pt}),(\hspace{-1pt}\mathbf{\t{3}}\hspace{-0pt}\mathbf{\g{4}}\hspace{-0pt}\mathbf{\g{4}}\hspace{-1pt})\big\}\fwboxL{0pt}{\,.}
%
%\\~
\end{array}}\label{ffggff_labels_c_basis_a1}\vspace{-4pt}}

The duality between these basis is encoded by the relation
\eq{\mathcal{C}^{\t{i}}\equivR\sum_{\g{j}}\mathbf{c}[\mathfrak{g}]\indices{\t{i}}{\g{j}}\mathcal{B}^{\g{j}}\,,}
for which the coefficients can be found to be
\vspace{5pt}\eq{\fwbox{0pt}{\fwboxL{520pt}{(\hspace{-1.25pt}%
\mathfrak{a}_{\r{1}}
)}}\hspace{-100pt}\fwboxR{0pt}{\mathbf{c}[\mathfrak{a}_{\r{1}}]\indices{\t{i}}{\g{j}}=}\left(\begin{array}{c@{$\;\;$}c@{$\;\;$}c@{$\;\;$}c@{$\;\;$}c@{$\;\;$}c}
\tmi\frac{1}{4}&\frac{1}{4}&\tmi\frac{1}{2}&\frac{1}{4}&\tmi\frac{1}{4}&\dzero\\[-1pt]
\tmi\frac{1}{3}&\tmi\frac{1}{6}&\dzero&\tmi\frac{1}{6}&\tmi\frac{1}{12}&\tmi1\\[-1pt]
\tmi\frac{1}{8}&\frac{1}{8}&\frac{3}{4}&\frac{1}{8}&\tmi\frac{1}{8}&\dzero\\[-1pt]
\frac{1}{3}&\frac{1}{6}&\dzero&\tmi\frac{1}{3}&\tmi\frac{1}{6}&\dzero\\[-1pt]
\frac{1}{3}&\tmi\frac{1}{3}&\dzero&\frac{1}{6}&\tmi\frac{1}{6}&\dzero\\[-1pt]
\frac{5}{6}&\frac{5}{12}&\dzero&\frac{5}{12}&\frac{5}{24}&\tmi\frac{3}{2}\end{array}\right)\fwboxL{0pt}{.}\hspace{-100pt}\label{ffggff_a1_duality_c_into_b}}

\subsubsection{\texorpdfstring{Clebsch Colour Bases for $C(\mathbf{\b{F}}\,\mathbf{\r{ad}}\,\mathbf{\r{ad}}|\mathbf{\r{ad}}\,\mathbf{\r{ad}}\,\mathbf{\b{F}})$}{Clebsch Colour Bases for C(Fgg|ggF)}}

For the case of six external particles, two of which transform under the fundamental representation and four in the adjoint, a natural choice of Clebsch colour tensors would be given by 
\eq{\fwboxR{0pt}{\mathcal{B}_{\hspace{9.05pt}\mu\nu}^{\hspace{1pt}\mathbf{\g{s}}_{\g{i}}\,\mathbf{w}_j\,\mathbf{\g{s}}_{\g{k}}}(\mathbf{\b{F}}\,\mathbf{\r{ad}}\,\mathbf{\r{ad}}|\mathbf{\r{ad}}\,\mathbf{\r{ad}}\,\mathbf{\b{F}})\;\bigger{\Leftrightarrow}\;}\tikzBox{fggggf_basis_diagram}{\arrowTo[hblue]{0,0}{-130};\node[anchor=50,inner sep=2pt] at(in){{\footnotesize$\b{[f]}$}};\arrowTo[hred]{0,0}{130}\node[anchor=-50,inner sep=2pt] at(in){{\footnotesize$\r{[\adR]}$}};\arrowFrom[hgreen]{0,0}[1.25]{0}\node[anchor=90,inner sep=2pt] at(arrownode){{\footnotesize$\g{[s_i]}$}};\node[clebsch]at(in){};\arrowTo[hred]{end}{90}\node[anchor=-90,inner sep=2pt] at(in){{\footnotesize$\r{[\adR]}$}};\arrowFrom[black]{end}[1.25]{0}\node[clebsch]at(in){{\scriptsize$\phantom{\nu}$}};\node[]at(in){{\scriptsize$\mu$}};\node[anchor=90,inner sep=2pt] at(arrownode){{\footnotesize${[w_j]}$}};\arrowFrom[hred]{end}{90}\node[anchor=-90,inner sep=2pt] at(end){{\footnotesize$\r{[\adR]}$}};\arrowFrom[hgreen]{in}[1.25]{0}\node[clebsch]at(in){{\scriptsize$\phantom{\nu}$}};\node[]at(in){{\scriptsize${\nu}$}};\node[anchor=90,inner sep=2pt] at(arrownode){{\footnotesize$\g{[s_k]}$}};
\arrowFrom[hred]{end}{50}
\node[anchor=-130,inner sep=2pt] at(end){{\footnotesize$\r{[\adR]}$}};\arrowFrom[hblue]{in}{-50}\node[anchor=130,inner sep=2pt] at(end){{\footnotesize$\b{[f]}$}};\node[clebsch]at(in){};
}\fwboxL{0pt}{.}
\label{fggggf_basis_defined}}
These would be labelled by the three irreducible representations $(\g{\mathbf{s}_i},\mathbf{w}_j\,\g{\mathbf{s}_k})$ where $\g{\mathbf{s}_i}$ was defined in (\ref{sReps_defined}) and $\mathbf{w}_j$ appears in the decomposition of the tensor product $\g{\mathbf{s}_{i}}\!\otimes\!\mathbf{\r{ad}}$ into irreducible representations; for these, there are also possible multiplicity labels $\mu\!\in\![m\indices{\g{\mathbf{s}_i}\,\r{\mathbf{ad}}}{\mathbf{w}_j}]$ and $\nu\!\in\![m\indices{\mathbf{w}_j\,\r{\mathbf{ad}}}{\g{\mathbf{s}_k}}]$.

It would require too much notation to identify the many representations $\mathbf{w}_j$ that appear for each of the Lie algebras; but in the case of $\mathfrak{a}_{\r{1}}$, the basis consists of nine tensors labelled by the irreducible representations
\eq{\fwbox{0pt}{\fwboxL{435pt}{(\hspace{-1.25pt}%
\mathfrak{a}_{\r{1}}
)}}\fwbox{0pt}{\begin{array}{c}
%\multicolumn{1}{c}{\text{Spanning labels %
%$(\g{\mathbf{s}_{i}}\,{\mathbf{{w}}_{{j}}}\,\g{\mathbf{s}_{k}})$ of 
%%
%$%
%%
%\mathcal{B}_{\hspace{9.05pt}~}^{\hspace{1pt}\mathbf{\g{s}}_{\g{i}}\,\mathbf{w}_j\,\mathbf{\g{s}}_{\g{k}}}(\mathbf{\b{F}}\,\mathbf{\r{ad}}\,\mathbf{\r{ad}}|\mathbf{\r{ad}}\,\mathbf{\r{ad}}\,\mathbf{\b{F}})
%%
%$
% for %
%$\mathfrak{a}_{\r{1}}$%
%%
%:}}\\\hline\hline
\rule{0pt}{14pt}%
\big\{(\hspace{-1pt}\mathbf{\g{2}}\hspace{-0pt}\mathbf{{2}}\hspace{-0pt}\mathbf{\g{2}}\hspace{-1pt}),(\hspace{-1pt}\mathbf{\g{2}}\hspace{-0pt}\mathbf{{2}}\hspace{-0pt}\mathbf{\g{4}}\hspace{-1pt}),(\hspace{-1pt}\mathbf{\g{2}}\hspace{-0pt}\mathbf{{4}}\hspace{-0pt}\mathbf{\g{2}}\hspace{-1pt}),(\hspace{-1pt}\mathbf{\g{2}}\hspace{-0pt}\mathbf{{4}}\hspace{-0pt}\mathbf{\g{4}}\hspace{-1pt}),(\hspace{-1pt}\mathbf{\g{4}}\hspace{-0pt}\mathbf{{2}}\hspace{-0pt}\mathbf{\g{2}}\hspace{-1pt}),(\hspace{-1pt}\mathbf{\g{4}}\hspace{-0pt}\mathbf{{2}}\hspace{-0pt}\mathbf{\g{4}}\hspace{-1pt}),(\hspace{-1pt}\mathbf{\g{4}}\hspace{-0pt}\mathbf{{4}}\hspace{-0pt}\mathbf{\g{2}}\hspace{-1pt}),(\hspace{-1pt}\mathbf{\g{4}}\hspace{-0pt}\mathbf{{4}}\hspace{-0pt}\mathbf{\g{4}}\hspace{-1pt}),(\hspace{-1pt}\mathbf{\g{4}}\hspace{-0pt}\mathbf{{6}}\hspace{-0pt}\mathbf{\g{4}}\hspace{-1pt})\big\}\fwboxL{0pt}{.}
%
%\\~
\end{array}}\label{fggggf_labels_b_basis_a1}\vspace{-6pt}}
%Melia would be 24-dimensional. 

An alternative choice of colour tensors could be given be generated by the tree graph:
\eq{\fwboxR{0pt}{\mathcal{C}_{\hspace{6.55pt}\mu\nu\rho}^{\,\smash{\mathbf{\t{r}}_{\t{i}}\,\mathbf{\b{t}}_{\b{j}}\,\mathbf{\b{t}}_{\b{k}}}}(\mathbf{\b{F}}\,\mathbf{\r{ad}}\,\mathbf{\r{ad}}|\mathbf{\r{ad}}\,\mathbf{\r{ad}}\,\mathbf{\b{F}})\;\bigger{\Leftrightarrow}\;}\tikzBox{fggggf_alt_basis_diagram}{
\arrowTo[hblue]{0,0}{-150}\node[anchor=10,inner sep=2pt] at(in){{\footnotesize$\b{[f]}$}};\arrowFrom[hblue]{0,0}{-30};\node[anchor=170,inner sep=2pt] at(end){{\footnotesize$\b{[f]}$}};\arrowFrom[hteal]{0,0}[1]{90}\node[clebsch]at(in){};\node[anchor=180,inner sep=2pt] at(arrownode){{\footnotesize$\t{[r_i]}$}};\arrowTo[hblue]{end}{180}\node[anchor=-90,inner sep=2pt] at(arrownode){{\footnotesize$\b{[t_j]}$}};\coordinate(up)at(end);\arrowTo[hred]{in}{230};\node[anchor=10,inner sep=2pt] at(in){{\footnotesize$\r{[\adR]}$}};
\arrowTo[hred]{end}{130};\node[anchor=-10,inner sep=2pt] at(in){{\footnotesize$\r{[\adR]}$}};\node[clebsch]at(end){{\scriptsize$\phantom{\nu}$}};\node[]at(end){{\scriptsize${\mu}$}};
\arrowFrom[hblue]{up}{0}\node[anchor=-90,inner sep=2pt] at(arrownode){{\footnotesize$\b{[t_k]}$}};
\arrowFrom[hred]{end}{50};\node[anchor=-170,inner sep=2pt] at(end){{\footnotesize$\r{[\adR]}$}};
\arrowFrom[hred]{in}{-50};\node[anchor=170,inner sep=2pt] at(end){{\footnotesize$\r{[\adR]}$}};\node[clebsch]at(in){{\scriptsize$\phantom{\nu}$}};\node[]at(in){{\scriptsize${\rho}$}};
\node[clebsch]at(up){{\scriptsize$\phantom{\nu}$}};\node[]at(up){{\scriptsize${\nu}$}};
}\fwboxL{0pt}{.}}
For this basis, the nine elements for $\mathfrak{a}_{\r{1}}$ gauge theory would be labelled by the irreducible representations
%
%
%a1 C-basis:
\eq{\fwbox{0pt}{\fwboxL{435pt}{(\hspace{-1.25pt}%
\mathfrak{a}_{\r{1}}
)}}\fwbox{0pt}{\begin{array}{c}
%\multicolumn{1}{c}{\text{Spanning labels %
%$(\t{\mathbf{r}_{i}}\,\b{\mathbf{{t}}_{{j}}}\,\b{\mathbf{t}_{k}})$ of 
%%
%$%
%%
%\mathcal{C}_{\hspace{6.55pt}~}^{\,\smash{\mathbf{\t{r}}_{\t{i}}\,\mathbf{\b{t}}_{\b{j}}\,\mathbf{\b{t}}_{\b{k}}}}(\mathbf{\b{F}}\,\mathbf{\r{ad}}\,\mathbf{\r{ad}}|\mathbf{\r{ad}}\,\mathbf{\r{ad}}\,\mathbf{\b{F}})
%%
%$
% for %
%$\mathfrak{a}_{\r{1}}$%
%%
%:}}\\\hline\hline
\rule{0pt}{14pt}%
\big\{(\hspace{-1pt}\mathbf{\t{1}}\hspace{-0pt}\mathbf{\b{1}}\hspace{-0pt}\mathbf{\b{1}}\hspace{-1pt}),(\hspace{-1pt}\mathbf{\t{1}}\hspace{-0pt}\mathbf{\b{3}}\hspace{-0pt}\mathbf{\b{3}}\hspace{-1pt}),(\hspace{-1pt}\mathbf{\t{1}}\hspace{-0pt}\mathbf{\b{5}}\hspace{-0pt}\mathbf{\b{5}}\hspace{-1pt}),(\hspace{-1pt}\mathbf{\t{3}}\hspace{-0pt}\mathbf{\b{1}}\hspace{-0pt}\mathbf{\b{3}}\hspace{-1pt}),(\hspace{-1pt}\mathbf{\t{3}}\hspace{-0pt}\mathbf{\b{3}}\hspace{-0pt}\mathbf{\b{1}}\hspace{-1pt}),(\hspace{-1pt}\mathbf{\t{3}}\hspace{-0pt}\mathbf{\b{3}}\hspace{-0pt}\mathbf{\b{3}}\hspace{-1pt}),(\hspace{-1pt}\mathbf{\t{3}}\hspace{-0pt}\mathbf{\b{3}}\hspace{-0pt}\mathbf{\b{5}}\hspace{-1pt}),(\hspace{-1pt}\mathbf{\t{3}}\hspace{-0pt}\mathbf{\b{5}}\hspace{-0pt}\mathbf{\b{3}}\hspace{-1pt}),(\hspace{-1pt}\mathbf{\t{3}}\hspace{-0pt}\mathbf{\b{5}}\hspace{-0pt}\mathbf{\b{5}}\hspace{-1pt})\big\}\fwboxL{0pt}{.}
%
%\\~
\end{array}}\label{fggggf_labels_c_basis_a1}\vspace{-2pt}}

It is interesting to consider the duality matrices
\eq{\mathcal{C}^{\t{i}}\equivL\sum_{\g{j}}\mathbf{c}[\mathfrak{g}]\indices{\t{i}}{\g{j}}\mathcal{B}^{\g{j}}\,.}
For $\mathfrak{a}_{\r{1}}$, we find
\eq{\fwbox{0pt}{\fwboxL{550pt}{(\hspace{-1.25pt}%
\mathfrak{a}_{\r{1}}
)}}\hspace{-200pt}\fwboxR{0pt}{\mathbf{c}[\mathfrak{a}_{\r{1}}]=}\displaystyle\left(\begin{array}{c@{$\;\;$}c@{$\;\;$}c@{$\;\;$}c@{$\;\;$}c@{$\;\;$}c@{$\;\;$}c@{$\;\;$}c@{$\;\;$}c}
\frac{4}{9}&\frac{2}{3}&\dzero&\dzero&\frac{2}{3}&1&\dzero&\dzero&\dzero\\[-1pt]
\frac{32}{27}&\tmi\frac{8}{9}&\frac{4}{9}&\tmi\frac{4}{9}&\tmi\frac{8}{9}&\frac{2}{3}&\tmi\frac{4}{9}&\frac{4}{9}&\dzero\\[-1pt]
\dzero&\dzero&\frac{5}{9}&\frac{1}{9}&\dzero&\dzero&\frac{1}{9}&\frac{1}{45}&1\\[-1pt]
\tmi\frac{8}{9}&\frac{2}{3}&\dzero&\dzero&\tmi\frac{4}{3}&1&\dzero&\dzero&\dzero\\[-1pt]
\tmi\frac{8}{9}&\tmi\frac{4}{3}&\dzero&\dzero&\frac{2}{3}&1&\dzero&\dzero&\dzero\\[-1pt]
\tmi\frac{64}{27}&\frac{16}{9}&\frac{4}{9}&\tmi\frac{4}{9}&\frac{16}{9}&\tmi\frac{4}{3}&\tmi\frac{4}{9}&\frac{4}{9}&\dzero\\[-1pt]
\dzero&\dzero&\frac{5}{9}&\frac{1}{9}&\dzero&\dzero&\tmi\frac{5}{9}&\tmi\frac{1}{9}&\dzero\\[-1pt]
\dzero&\dzero&\frac{5}{9}&\tmi\frac{5}{9}&\dzero&\dzero&\frac{1}{9}&\tmi\frac{1}{9}&\dzero\\[-1pt]
\dzero&\dzero&\tmi\frac{5}{3}&\tmi\frac{1}{3}&\dzero&\dzero&\tmi\frac{1}{3}&\tmi\frac{1}{15}&2\end{array}\right)\hspace{-100pt}.\hspace{-100pt}\label{fggggf_a1_duality_matrix}}

\newpage
\subsubsection{\texorpdfstring{Clebsch Colour Bases for $C(\mathbf{\r{ad}}\,\mathbf{\r{ad}}\,\mathbf{\r{ad}}|\mathbf{\r{ad}}\,\mathbf{\r{ad}}\,\mathbf{\r{ad}})$}{Clebsch Colour Bases for C(ggg|ggg)}}%

Finally, let us consider the case of scattering six adjoint-coloured particles. In this case, a basis of Clebsch colour tensors could be generated by the tree graph:
\eq{\fwboxR{0pt}{\mathcal{B}_{\hspace{4.05pt}\mu\nu\rho\sigma}^{\hspace{1pt}\mathbf{\b{t}}_{\b{i}}\,\mathbf{y}_j\,\mathbf{\b{t}}_{\b{k}}}(\mathbf{\r{ad}}\,\mathbf{\r{ad}}\,\mathbf{\r{ad}}|\mathbf{\r{ad}}\,\mathbf{\r{ad}}\,\mathbf{\r{ad}})\;\bigger{\Leftrightarrow}\hspace{-5pt}}\tikzBox[-5.25pt]{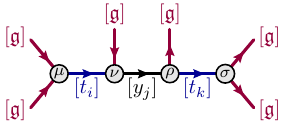}{\arrowTo[hred]{0,0}{-130};\node[anchor=10,inner sep=2pt] at(in){{\footnotesize$\r{[\adR]}$}};\arrowTo[hred]{0,0}{130}\node[anchor=-10,inner sep=2pt] at(in){{\footnotesize$\r{[\adR]}$}};\arrowFrom[hblue]{0,0}[1.25]{0}\node[anchor=90,inner sep=2pt] at(arrownode){{\footnotesize$\b{[t_i]}$}};\node[clebsch]at(in){{\scriptsize$\phantom{\nu}$}};\node[]at(in){{\scriptsize${\mu}$}};\arrowTo[hred]{end}{90}\node[anchor=-90,inner sep=2pt] at(in){{\footnotesize$\r{[\adR]}$}};\arrowFrom[black]{end}[1.25]{0}\node[clebsch]at(in){{\scriptsize$\phantom{\nu}$}};\node[]at(in){{\scriptsize$\nu$}};\node[anchor=90,inner sep=2pt] at(arrownode){{\footnotesize${[y_j]}$}};\arrowFrom[hred]{end}{90}\node[anchor=-90,inner sep=2pt] at(end){{\footnotesize$\r{[\adR]}$}};\arrowFrom[hblue]{in}[1.25]{0}\node[clebsch]at(in){{\scriptsize$\phantom{\nu}$}};\node[]at(in){{\scriptsize$\rho$}};\node[anchor=90,inner sep=2pt] at(arrownode){{\footnotesize$\b{[t_k]}$}};
\arrowFrom[hred]{end}{50}
\node[anchor=-170,inner sep=2pt] at(end){{\footnotesize$\r{[\adR]}$}};\arrowFrom[hred]{in}{-50}\node[anchor=170,inner sep=2pt] at(end){{\footnotesize$\r{[\adR]}$}};\node[clebsch]at(in){{\scriptsize$\phantom{\sigma}$}};\node[]at(in){{\scriptsize$\sigma$}};
}\fwboxL{0pt}{.}
\label{gggggg_basis_defined}}
While it would be cumbersome to identify the irreducible representations $\mathbf{y}_j$ arising for each of the Lie algebras, the fifteen basis elements in the case of $\mathfrak{a}_{\r{1}}$ gauge theory would be labelled by the irreducible representations:
%
%a1 B-basis:
\vspace{-5pt}\eq{\fwbox{0pt}{\fwboxL{435pt}{(\hspace{-1.25pt}%
\mathfrak{a}_{\r{1}}
)}}\fwbox{0pt}{\begin{array}{c}
\rule{0pt}{14pt}%
\fwboxR{0pt}{\big\{}(\hspace{-1pt}\mathbf{\b{1}}\hspace{-0pt}\mathbf{{3}}\hspace{-0pt}\mathbf{\b{1}}\hspace{-1pt}),(\hspace{-1pt}\mathbf{\b{1}}\hspace{-0pt}\mathbf{{3}}\hspace{-0pt}\mathbf{\b{3}}\hspace{-1pt}),(\hspace{-1pt}\mathbf{\b{1}}\hspace{-0pt}\mathbf{{3}}\hspace{-0pt}\mathbf{\b{5}}\hspace{-1pt}),(\hspace{-1pt}\mathbf{\b{3}}\hspace{-0pt}\mathbf{{1}}\hspace{-0pt}\mathbf{\b{3}}\hspace{-1pt}),(\hspace{-1pt}\mathbf{\b{3}}\hspace{-0pt}\mathbf{{3}}\hspace{-0pt}\mathbf{\b{1}}\hspace{-1pt}),\\
(\hspace{-1pt}\mathbf{\b{3}}\hspace{-0pt}\mathbf{{3}}\hspace{-0pt}\mathbf{\b{3}}\hspace{-1pt}),(\hspace{-1pt}\mathbf{\b{3}}\hspace{-0pt}\mathbf{{3}}\hspace{-0pt}\mathbf{\b{5}}\hspace{-1pt}),(\hspace{-1pt}\mathbf{\b{3}}\hspace{-0pt}\mathbf{{5}}\hspace{-0pt}\mathbf{\b{3}}\hspace{-1pt}),(\hspace{-1pt}\mathbf{\b{3}}\hspace{-0pt}\mathbf{{5}}\hspace{-0pt}\mathbf{\b{5}}\hspace{-1pt}),(\hspace{-1pt}\mathbf{\b{5}}\hspace{-0pt}\mathbf{{3}}\hspace{-0pt}\mathbf{\b{1}}\hspace{-1pt}),\\
(\hspace{-1pt}\mathbf{\b{5}}\hspace{-0pt}\mathbf{{3}}\hspace{-0pt}\mathbf{\b{3}}\hspace{-1pt}),(\hspace{-1pt}\mathbf{\b{5}}\hspace{-0pt}\mathbf{{3}}\hspace{-0pt}\mathbf{\b{5}}\hspace{-1pt}),(\hspace{-1pt}\mathbf{\b{5}}\hspace{-0pt}\mathbf{{5}}\hspace{-0pt}\mathbf{\b{3}}\hspace{-1pt}),(\hspace{-1pt}\mathbf{\b{5}}\hspace{-0pt}\mathbf{{5}}\hspace{-0pt}\mathbf{\b{5}}\hspace{-1pt}),(\hspace{-1pt}\mathbf{\b{5}}\hspace{-0pt}\mathbf{{7}}\hspace{-0pt}\mathbf{\b{5}}\hspace{-1pt})\phantom{,}\fwboxL{0pt}{\hspace{-3pt}\big\}}\fwboxL{0pt}{\,\,.}
%
%\\~
\end{array}}\label{gggggg_labels_b_basis_a1}\vspace{-6pt}}
We may compare these colour-diagonal basis tensors to a choice of fifteen independent multi-trace tensors, $\{\mathbf{T}^1,\ldots,\mathbf{T}^{15}\}$ given by
\eq{\begin{split}\Big\{&
\fwboxL{62pt}{\mathrm{tr}_{\mathbf{\b{F}}}(\hspace{-2pt}\r{123456})},
\fwboxL{62pt}{\mathrm{tr}_{\mathbf{\b{F}}}(\hspace{-2pt}\r{123465})},
\fwboxL{62pt}{\mathrm{tr}_{\mathbf{\b{F}}}(\hspace{-2pt}\r{123564})},
\fwboxL{62pt}{\mathrm{tr}_{\mathbf{\b{F}}}(\hspace{-2pt}\r{124563})},
\fwboxL{62pt}{\mathrm{tr}_{\mathbf{\b{F}}}(\hspace{-2pt}\r{123546})},\\[-5pt]
\fwbox{0pt}{\fwboxL{69pt}{(\hspace{-1.25pt}%
\mathfrak{a}_{\r{1}}
)}}&
\fwboxL{62pt}{\mathrm{tr}_{\mathbf{\b{F}}}(\hspace{-2pt}\r{132456})},
\fwboxL{62pt}{\mathrm{tr}_{\mathbf{\b{F}}}(\hspace{-2pt}\r{132546})},
\fwboxL{62pt}{\mathrm{tr}_{\mathbf{\b{F}}}(\hspace{-2pt}\r{12}|\r{3546})},
\fwboxL{62pt}{\mathrm{tr}_{\mathbf{\b{F}}}(\hspace{-2pt}\r{124}|\r{356})},
\fwboxL{62pt}{\mathrm{tr}_{\mathbf{\b{F}}}(\hspace{-2pt}\r{125}|\r{346})},\\[-5pt]
&
\fwboxL{62pt}{\mathrm{tr}_{\mathbf{\b{F}}}(\hspace{-2pt}\r{134}|\r{256})},
\fwboxL{62pt}{\mathrm{tr}_{\mathbf{\b{F}}}(\hspace{-2pt}\r{1324}|\r{56})},
\fwboxL{62pt}{\mathrm{tr}_{\mathbf{\b{F}}}(\hspace{-2pt}\r{1526}|\r{34})},
\fwboxL{62pt}{\mathrm{tr}_{\mathbf{\b{F}}}(\hspace{-2pt}\r{1425}|\r{36})},
\fwboxL{62pt}{\mathrm{tr}_{\mathbf{\b{F}}}(\hspace{-2pt}\r{12}|\r{34}|\r{56})}\,\Big\}\fwboxL{0pt}{.}\end{split}\label{a1_gggggg_independent_multitraces}}
These can be expressed in terms of the basis tensors
\eq{\mathbf{T}^i\equivR\sum_{\b{j}}\mathbf{c}[\mathfrak{a}_{\r{1}}]\indices{i}{\b{j}}\,\mathcal{B}^{\b{j}}}
where the coefficients are given by
\vspace{5pt}\eq{\fwbox{0pt}{\fwboxL{784pt}{(\hspace{-1.25pt}%
\mathfrak{a}_{\r{1}}
)}}\hspace{-300pt}\fwboxR{0pt}{\mathbf{c}[\mathfrak{a}_{\r{1}}]=}\left(\begin{array}{c@{$\;\;$}c@{$\;\;$}c@{$\;\;$}c@{$\;\;$}c@{$\;\;$}c@{$\;\;$}c@{$\;\;$}c@{$\;\;$}c@{$\;\;$}c@{$\;\;$}c@{$\;\;$}c@{$\;\;$}c@{$\;\;$}c@{$\;\;$}c}
\frac{1}{4}&\frac{1}{8}&\dzero&\frac{1}{8}&\frac{1}{8}&\frac{1}{16}&\dzero&\dzero&\dzero&\dzero&\dzero&\dzero&\dzero&\dzero&\dzero\\[-1pt]
\frac{1}{4}&\tmi\frac{1}{8}&\dzero&\tmi\frac{1}{8}&\frac{1}{8}&\tmi\frac{1}{16}&\dzero&\dzero&\dzero&\dzero&\dzero&\dzero&\dzero&\dzero&\dzero\\[-1pt]
\frac{1}{4}&\tmi\frac{1}{8}&\dzero&\frac{1}{8}&\frac{1}{8}&\tmi\frac{1}{16}&\dzero&\dzero&\dzero&\dzero&\dzero&\dzero&\dzero&\dzero&\dzero\\[-1pt]
\frac{1}{4}&\frac{1}{8}&\dzero&\frac{1}{8}&\tmi\frac{1}{8}&\tmi\frac{1}{16}&\dzero&\dzero&\dzero&\dzero&\dzero&\dzero&\dzero&\dzero&\dzero\\[-1pt]
\tmi\frac{1}{12}&\dzero&\frac{1}{2}&\tmi\frac{1}{8}&\tmi\frac{1}{24}&\dzero&\frac{1}{4}&\dzero&\dzero&\dzero&\dzero&\dzero&\dzero&\dzero&\dzero\\[-1pt]
\tmi\frac{1}{12}&\tmi\frac{1}{24}&\dzero&\tmi\frac{1}{8}&\dzero&\dzero&\dzero&\dzero&\dzero&\frac{1}{2}&\frac{1}{4}&\dzero&\dzero&\dzero&\dzero\\[-1pt]
\frac{1}{36}&\dzero&\tmi\frac{1}{6}&\frac{1}{8}&\dzero&\dzero&\dzero&\dzero&\dzero&\tmi\frac{1}{6}&\dzero&1&\dzero&\dzero&\dzero\\[-1pt]
\tmi\frac{1}{6}&\dzero&1&\dzero&\dzero&\dzero&\dzero&\dzero&\dzero&\dzero&\dzero&\dzero&\dzero&\dzero&\dzero\\[-1pt]
\dzero&\dzero&\dzero&\frac{1}{12}&\dzero&\tmi\frac{1}{16}&\dzero&\frac{1}{4}&\dzero&\dzero&\dzero&\dzero&\dzero&\dzero&\dzero\\[-1pt]
\dzero&\dzero&\dzero&\tmi\frac{1}{12}&\frac{1}{12}&\tmi\frac{1}{32}&\tmi\frac{1}{8}&\frac{1}{8}&\frac{1}{8}&\dzero&\dzero&\dzero&\dzero&\dzero&\dzero\\[-1pt]
\dzero&\frac{1}{12}&\dzero&\tmi\frac{1}{12}&\dzero&\tmi\frac{1}{32}&\dzero&\frac{1}{8}&\dzero&\dzero&\tmi\frac{1}{8}&\dzero&\frac{1}{8}&\dzero&\dzero\\[-1pt]
\tmi\frac{1}{6}&\dzero&\dzero&\dzero&\dzero&\dzero&\dzero&\dzero&\dzero&1&\dzero&\dzero&\dzero&\dzero&\dzero\\[-1pt]
\tmi\frac{1}{6}&\dzero&\dzero&\dzero&\dzero&\dzero&\dzero&\dzero&\dzero&\dzero&\dzero&\frac{3}{5}&\dzero&\frac{1}{12}&1\\[-1pt]
\tmi\frac{1}{18}&\tmi\frac{1}{24}&\tmi\frac{1}{6}&\dzero&\dzero&\dzero&\dzero&\dzero&\dzero&\frac{1}{3}&\tmi\frac{1}{8}&\frac{1}{10}&\tmi\frac{1}{8}&\tmi\frac{1}{24}&1\\[-1pt]
1&\dzero&\dzero&\dzero&\dzero&\dzero&\dzero&\dzero&\dzero&\dzero&\dzero&\dzero&\dzero&\dzero&\dzero\end{array}\right)\fwboxL{0pt}{.}\hspace{-300pt}\label{gggggg_a1_tr_to_b}}

Alternatively, one may consider the Clebsch colour-tensor bases generated by the graphs
\eq{\fwboxR{0pt}{\mathcal{C}_{\hspace{4.05pt}\mu\nu\rho\sigma}^{\,\smash{\mathbf{\b{t}}_{\b{i}}\,\mathbf{\g{t}}_{\g{j}}\,\mathbf{\t{t}}_{\t{k}}}}(\mathbf{\r{ad}}\,\mathbf{\r{ad}}\,\mathbf{\r{ad}}|\mathbf{\r{ad}}\,\mathbf{\r{ad}}\,\mathbf{\r{ad}})\;\bigger{\Leftrightarrow}}\tikzBox{gggggg_alt_basis_diagram_1}{\arrowTo[hred]{0,0}{-130};\node[anchor=10,inner sep=2pt] at(in){{\footnotesize$\r{[\adR]}$}};\arrowTo[hred]{0,0}{130}\node[anchor=-10,inner sep=2pt] at(in){{\footnotesize$\r{[\adR]}$}};\arrowFrom[hblue]{0,0}[1.25]{0}\node[anchor=90,inner sep=2pt] at(arrownode){{\footnotesize$\b{[t_i]}$}};\node[clebsch]at(in){{\scriptsize$\phantom{\nu}$}};\node[]at(in){{\scriptsize$\mu$}};\arrowTo[hgreen]{end}[1.25]{90}\node[anchor=180,inner sep=2pt] at(arrownode){{\footnotesize${\g{[t_j]}}$}};\coordinate(v0)at(end);\arrowTo[hred]{in}{150};\node[anchor=-10,inner sep=2pt] at(in){{\footnotesize$\r{[\adR]}$}};\arrowFrom[hred]{end}{30}\node[anchor=-170,inner sep=2pt] at(end){{\footnotesize$\r{[\adR]}$}};\node[clebsch]at(in){{\scriptsize$\phantom{\nu}$}};\node[]at(in){{\scriptsize${\nu}$}};
\arrowFrom[hteal]{v0}[1.25]{0}\node[clebsch]at(in){{\scriptsize$\phantom{\nu}$}};\node[]at(in){{\scriptsize$\sigma$}};
\node[anchor=90,inner sep=2pt] at(arrownode){{\footnotesize$\t{[t_k]}$}};\arrowFrom[hred]{end}{50}
\node[anchor=-170,inner sep=2pt] at(end){{\footnotesize$\r{[\adR]}$}};\arrowFrom[hred]{in}{-50}\node[anchor=170,inner sep=2pt] at(end){{\footnotesize$\r{[\adR]}$}};\node[clebsch]at(in){{\scriptsize$\phantom{\nu}$}};\node[]at(in){{\scriptsize${\rho}$}};
}\fwboxL{0pt}{,}}
or
\eq{\fwboxR{0pt}{\mathcal{D}_{\hspace{4.05pt}\mu\nu\rho\sigma}^{\,\smash{\mathbf{\b{t}}_{\b{i}}\,\mathbf{\g{t}}_{\g{j}}\,\mathbf{\t{t}}_{\t{k}}}}(\mathbf{\r{ad}}\,\mathbf{\r{ad}}\,\mathbf{\r{ad}}|\mathbf{\r{ad}}\,\mathbf{\r{ad}}\,\mathbf{\r{ad}})\;\bigger{\Leftrightarrow}}\tikzBox{gggggg_alt_basis_diagram_2_}{
\arrowTo[hred]{0,0}{-150}\node[anchor=10,inner sep=2pt] at(in){{\footnotesize$\r{[\adR]}$}};\arrowFrom[hred]{0,0}{-30};\node[anchor=170,inner sep=2pt] at(end){{\footnotesize$\r{[\adR]}$}};\arrowFrom[hblue]{0,0}[1]{90}\node[clebsch]at(in){{\scriptsize$\phantom{\nu}$}};\node[]at(in){{\scriptsize${\mu}$}};\node[anchor=180,inner sep=2pt] at(arrownode){{\footnotesize$\b{[t_i]}$}};\arrowTo[hgreen]{end}{180}\node[anchor=-90,inner sep=2pt] at(arrownode){{\footnotesize$\g{[t_j]}$}};%\node[anchor=0,inner sep=0pt] at(in){{\footnotesize$\r{[\adR]}$}};
\coordinate(up)at(end);
\arrowTo[hred]{in}{230};\node[anchor=10,inner sep=2pt] at(in){{\footnotesize$\r{[\adR]}$}};
\arrowTo[hred]{end}{130};\node[anchor=-10,inner sep=2pt] at(in){{\footnotesize$\r{[\adR]}$}};\node[clebsch]at(end){{\scriptsize$\phantom{\nu}$}};\node[]at(end){{\scriptsize${\nu}$}};
\arrowFrom[hteal]{up}{0}\node[anchor=-90,inner sep=2pt] at(arrownode){{\footnotesize$\t{[t_k]}$}};
\arrowFrom[hred]{end}{50};\node[anchor=-170,inner sep=2pt] at(end){{\footnotesize$\r{[\adR]}$}};
\arrowFrom[hred]{in}{-50};\node[anchor=170,inner sep=2pt] at(end){{\footnotesize$\r{[\adR]}$}};\node[clebsch]at(in){{\scriptsize$\phantom{\nu}$}};\node[]at(in){{\scriptsize${\rho}$}};
\node[clebsch]at(up){{\scriptsize$\phantom{\nu}$}};\node[]at(up){{\scriptsize${\sigma}$}};
}\fwboxL{0pt}{.}}
These two cases would be labelled by the same sets of irreducible representations and multiplicities, but would of course generate distinct sets of colour tensors. 

In the case of $\mathfrak{a}_{\r{1}}$ gauge theory, the fifteen independent colour tensors in the $\mathcal{C}$ or $\mathcal{D}$ bases would be labelled by the irreducible representations
%
%a1 CD-basis:
\vspace{-4pt}\eq{\fwbox{0pt}{\fwboxL{435pt}{(\hspace{-1.25pt}%
\mathfrak{a}_{\r{1}}
)}}\fwbox{0pt}{\begin{array}{c}
%\multicolumn{1}{c}{\text{Spanning labels %
%$(\b{\mathbf{t}_{i}}\,\g{\mathbf{{t}}_{{j}}}\,\t{\mathbf{t}_{k}})$ of 
%%
%$%
%%
%\mathcal{C}_{\hspace{4.05pt}}^{\,\smash{\mathbf{\b{t}}_{\b{i}}\,\mathbf{\g{t}}_{\g{j}}\,\mathbf{\t{t}}_{\t{k}}}}(\mathbf{\r{ad}}\,\mathbf{\r{ad}}\,\mathbf{\r{ad}}|\mathbf{\r{ad}}\,\mathbf{\r{ad}}\,\mathbf{\r{ad}})
%%
%$
% for %
%$\mathfrak{a}_{\r{1}}$%
%%
%:}}\\\hline\hline
\rule{0pt}{14pt}%
\fwboxR{0pt}{\big\{}(\hspace{-1pt}\mathbf{\b{1}}\hspace{-0pt}\mathbf{\g{1}}\hspace{-0pt}\mathbf{\t{1}}\hspace{-1pt}),(\hspace{-1pt}\mathbf{\b{1}}\hspace{-0pt}\mathbf{\g{3}}\hspace{-0pt}\mathbf{\t{3}}\hspace{-1pt}),(\hspace{-1pt}\mathbf{\b{1}}\hspace{-0pt}\mathbf{\g{5}}\hspace{-0pt}\mathbf{\t{5}}\hspace{-1pt}),(\hspace{-1pt}\mathbf{\b{3}}\hspace{-0pt}\mathbf{\g{1}}\hspace{-0pt}\mathbf{\t{3}}\hspace{-1pt}),(\hspace{-1pt}\mathbf{\b{3}}\hspace{-0pt}\mathbf{\g{3}}\hspace{-0pt}\mathbf{\t{1}}\hspace{-1pt}),\\
(\hspace{-1pt}\mathbf{\b{3}}\hspace{-0pt}\mathbf{\g{3}}\hspace{-0pt}\mathbf{\t{3}}\hspace{-1pt}),(\hspace{-1pt}\mathbf{\b{3}}\hspace{-0pt}\mathbf{\g{3}}\hspace{-0pt}\mathbf{\t{5}}\hspace{-1pt}),(\hspace{-1pt}\mathbf{\b{3}}\hspace{-0pt}\mathbf{\g{5}}\hspace{-0pt}\mathbf{\t{3}}\hspace{-1pt}),(\hspace{-1pt}\mathbf{\b{3}}\hspace{-0pt}\mathbf{\g{5}}\hspace{-0pt}\mathbf{\t{5}}\hspace{-1pt}),(\hspace{-1pt}\mathbf{\b{5}}\hspace{-0pt}\mathbf{\g{1}}\hspace{-0pt}\mathbf{\t{5}}\hspace{-1pt}),\\
(\hspace{-1pt}\mathbf{\b{5}}\hspace{-0pt}\mathbf{\g{3}}\hspace{-0pt}\mathbf{\t{3}}\hspace{-1pt}),(\hspace{-1pt}\mathbf{\b{5}}\hspace{-0pt}\mathbf{\g{3}}\hspace{-0pt}\mathbf{\t{5}}\hspace{-1pt}),(\hspace{-1pt}\mathbf{\b{5}}\hspace{-0pt}\mathbf{\g{5}}\hspace{-0pt}\mathbf{\t{1}}\hspace{-1pt}),(\hspace{-1pt}\mathbf{\b{5}}\hspace{-0pt}\mathbf{\g{5}}\hspace{-0pt}\mathbf{\t{3}}\hspace{-1pt}),(\hspace{-1pt}\mathbf{\b{5}}\hspace{-0pt}\mathbf{\g{5}}\hspace{-0pt}\mathbf{\t{5}}\hspace{-1pt})
\phantom{,}\fwboxL{0pt}{\hspace{-3pt}\big\}}\fwboxL{0pt}{\,\,.}
%
%\\~
\end{array}}\label{gggggg_labels_cd_basis_a1}\vspace{-6pt}}

The expansion of one basis into another generates the duality matrices
\eq{\mathcal{C}^{\t{i}}\equivR\sum_{\b{j}}\mathbf{c}[\mathfrak{g}]\indices{\t{i}}{\b{j}}\,\mathcal{B}^{\b{j}}}
or
\eq{\mathcal{D}^{\t{i}}\equivR\sum_{\b{j}}\mathbf{d}[\mathfrak{g}]\indices{\t{i}}{\b{j}}\,\mathcal{C}^{\b{j}}\,.}
In the case of $\mathfrak{a}_{\r{1}}$, 
\eq{\fwbox{0pt}{\fwboxL{809pt}{(\hspace{-1.25pt}%
\mathfrak{a}_{\r{1}}
)}}\hspace{-300pt}\fwboxR{0pt}{\mathbf{c}[\mathfrak{a}_{\r{1}}]=}\left(\begin{array}{c@{$\;\;$}c@{$\;\;$}c@{$\;\;$}c@{$\;\;$}c@{$\;\;$}c@{$\;\;$}c@{$\;\;$}c@{$\;\;$}c@{$\;\;$}c@{$\;\;$}c@{$\;\;$}c@{$\;\;$}c@{$\;\;$}c@{$\;\;$}c}
1&\dzero&\dzero&\dzero&\dzero&\dzero&\dzero&\dzero&\dzero&\dzero&\dzero&\dzero&\dzero&\dzero&\dzero\\[-1pt]
\dzero&1&\dzero&\dzero&\dzero&\dzero&\dzero&\dzero&\dzero&\dzero&\dzero&\dzero&\dzero&\dzero&\dzero\\[-1pt]
\dzero&\dzero&1&\dzero&\dzero&\dzero&\dzero&\dzero&\dzero&\dzero&\dzero&\dzero&\dzero&\dzero&\dzero\\[-1pt]
\dzero&\dzero&\dzero&\frac{1}{3}&\dzero&\frac{1}{4}&\dzero&1&\dzero&\dzero&\dzero&\dzero&\dzero&\dzero&\dzero\\[-1pt]
\dzero&\dzero&\dzero&\dzero&1&\dzero&\dzero&\dzero&\dzero&\dzero&\dzero&\dzero&\dzero&\dzero&\dzero\\[-1pt]
\dzero&\dzero&\dzero&\frac{4}{3}&\dzero&\frac{1}{2}&\dzero&\tmi2&\dzero&\dzero&\dzero&\dzero&\dzero&\dzero&\dzero\\[-1pt]
\dzero&\dzero&\dzero&\dzero&\dzero&\dzero&\tmi\frac{1}{2}&\dzero&\frac{1}{2}&\dzero&\dzero&\dzero&\dzero&\dzero&\dzero\\[-1pt]
\dzero&\dzero&\dzero&\frac{5}{9}&\dzero&\tmi\frac{5}{24}&\dzero&\frac{1}{6}&\dzero&\dzero&\dzero&\dzero&\dzero&\dzero&\dzero\\[-1pt]
\dzero&\dzero&\dzero&\dzero&\dzero&\dzero&\frac{3}{2}&\dzero&\frac{1}{2}&\dzero&\dzero&\dzero&\dzero&\dzero&\dzero\\[-1pt]
\dzero&\dzero&\dzero&\dzero&\dzero&\dzero&\dzero&\dzero&\dzero&\dzero&\dzero&\frac{3}{5}&\dzero&\frac{1}{12}&1\\[-1pt]
\dzero&\dzero&\dzero&\dzero&\dzero&\dzero&\dzero&\dzero&\dzero&\dzero&\tmi\frac{1}{2}&\dzero&\frac{1}{2}&\dzero&\dzero\\[-1pt]
\dzero&\dzero&\dzero&\dzero&\dzero&\dzero&\dzero&\dzero&\dzero&\dzero&\dzero&\frac{18}{5}&\dzero&\frac{1}{6}&\tmi4\\[-1pt]
\dzero&\dzero&\dzero&\dzero&\dzero&\dzero&\dzero&\dzero&\dzero&1&\dzero&\dzero&\dzero&\dzero&\dzero\\[-1pt]
\dzero&\dzero&\dzero&\dzero&\dzero&\dzero&\dzero&\dzero&\dzero&\dzero&\frac{3}{2}&\dzero&\frac{1}{2}&\dzero&\dzero\\[-1pt]
\dzero&\dzero&\dzero&\dzero&\dzero&\dzero&\dzero&\dzero&\dzero&\dzero&\dzero&\frac{21}{10}&\dzero&\tmi\frac{7}{24}&1\end{array}\right)\hspace{-300pt}\label{gggggg_a1_c_into_b_duality}}
and
\eq{\fwbox{0pt}{\fwboxL{855pt}{(\hspace{-1.25pt}%
\mathfrak{a}_{\r{1}}
)}}\hspace{-300pt}\hspace{-60pt}\fwboxR{0pt}{\mathbf{d}[\mathfrak{a}_{\r{1}}]=}\left(\begin{array}{c@{$\;\;$}c@{$\;\;$}c@{$\;\;$}c@{$\;\;$}c@{$\;\;$}c@{$\;\;$}c@{$\;\;$}c@{$\;\;$}c@{$\;\;$}c@{$\;\;$}c@{$\;\;$}c@{$\;\;$}c@{$\;\;$}c@{$\;\;$}c}
\frac{1}{9}&\frac{1}{12}&\frac{1}{3}&\frac{1}{12}&\frac{1}{12}&\frac{1}{32}&\tmi\frac{1}{8}&\tmi\frac{1}{8}&\frac{1}{8}&\frac{1}{3}&\tmi\frac{1}{8}&\frac{1}{8}&\frac{1}{3}&\frac{1}{8}&\frac{1}{6}\\[-1pt]
\frac{4}{9}&\frac{1}{6}&\tmi\frac{2}{3}&\frac{1}{6}&\frac{1}{3}&\frac{1}{16}&\frac{1}{4}&\tmi\frac{1}{4}&\tmi\frac{1}{4}&\tmi\frac{2}{3}&\tmi\frac{1}{4}&\tmi\frac{1}{4}&\frac{4}{3}&\frac{1}{4}&\tmi\frac{1}{3}\\[-1pt]
\frac{5}{27}&\tmi\frac{5}{72}&\frac{1}{18}&\tmi\frac{5}{72}&\frac{5}{36}&\tmi\frac{5}{192}&\tmi\frac{1}{48}&\frac{5}{48}&\frac{1}{48}&\frac{1}{18}&\frac{5}{48}&\frac{1}{48}&\frac{5}{9}&\tmi\frac{5}{48}&\frac{1}{36}\\[-1pt]
\frac{4}{9}&\frac{1}{3}&\frac{4}{3}&\frac{1}{6}&\frac{1}{6}&\frac{1}{16}&\tmi\frac{1}{4}&\tmi\frac{1}{4}&\frac{1}{4}&\tmi\frac{2}{3}&\frac{1}{4}&\tmi\frac{1}{4}&\tmi\frac{2}{3}&\tmi\frac{1}{4}&\tmi\frac{1}{3}\\[-1pt]
\frac{4}{9}&\frac{1}{6}&\tmi\frac{2}{3}&\frac{1}{3}&\frac{1}{6}&\frac{1}{16}&\tmi\frac{1}{4}&\frac{1}{4}&\tmi\frac{1}{4}&\frac{4}{3}&\tmi\frac{1}{4}&\frac{1}{4}&\tmi\frac{2}{3}&\tmi\frac{1}{4}&\tmi\frac{1}{3}\\[-1pt]
\frac{8}{9}&\frac{1}{3}&\tmi\frac{4}{3}&\frac{1}{3}&\frac{1}{3}&\frac{1}{2}&1&1&\dzero&\tmi\frac{4}{3}&1&\dzero&\tmi\frac{4}{3}&\dzero&\frac{4}{3}\\[-1pt]
\tmi\frac{10}{27}&\tmi\frac{5}{36}&\frac{5}{9}&\frac{5}{36}&\tmi\frac{5}{36}&\frac{5}{48}&\tmi\frac{1}{6}&\frac{2}{3}&\frac{1}{12}&\tmi\frac{1}{9}&\tmi\frac{1}{6}&\tmi\frac{1}{12}&\frac{5}{9}&\frac{1}{12}&\tmi\frac{2}{9}\\[-1pt]
\tmi\frac{10}{27}&\frac{5}{36}&\tmi\frac{1}{9}&\tmi\frac{5}{36}&\tmi\frac{5}{36}&\frac{5}{48}&\tmi\frac{1}{6}&\tmi\frac{1}{6}&\tmi\frac{1}{12}&\frac{5}{9}&\frac{2}{3}&\frac{1}{12}&\frac{5}{9}&\frac{1}{12}&\tmi\frac{2}{9}\\[-1pt]
\frac{10}{9}&\tmi\frac{5}{12}&\frac{1}{3}&\tmi\frac{5}{12}&\frac{5}{12}&\dzero&\tmi\frac{1}{4}&\frac{1}{4}&\dzero&\frac{1}{3}&\frac{1}{4}&\dzero&\tmi\frac{5}{3}&\frac{1}{2}&\tmi\frac{1}{3}\\[-1pt]
\frac{5}{27}&\frac{5}{36}&\frac{5}{9}&\tmi\frac{5}{72}&\tmi\frac{5}{72}&\tmi\frac{5}{192}&\frac{5}{48}&\frac{5}{48}&\tmi\frac{5}{48}&\frac{1}{18}&\tmi\frac{1}{48}&\frac{1}{48}&\frac{1}{18}&\frac{1}{48}&\frac{1}{36}\\[-1pt]
\tmi\frac{10}{27}&\tmi\frac{5}{36}&\frac{5}{9}&\tmi\frac{5}{36}&\frac{5}{36}&\frac{5}{48}&\frac{2}{3}&\tmi\frac{1}{6}&\frac{1}{12}&\frac{5}{9}&\tmi\frac{1}{6}&\frac{1}{12}&\tmi\frac{1}{9}&\tmi\frac{1}{12}&\tmi\frac{2}{9}\\[-1pt]
\frac{10}{9}&\frac{5}{12}&\tmi\frac{5}{3}&\tmi\frac{5}{12}&\tmi\frac{5}{12}&\dzero&\frac{1}{4}&\frac{1}{4}&\frac{1}{2}&\frac{1}{3}&\tmi\frac{1}{4}&\dzero&\frac{1}{3}&\dzero&\tmi\frac{1}{3}\\[-1pt]
\frac{5}{27}&\tmi\frac{5}{72}&\frac{1}{18}&\frac{5}{36}&\tmi\frac{5}{72}&\tmi\frac{5}{192}&\frac{5}{48}&\tmi\frac{1}{48}&\frac{1}{48}&\frac{5}{9}&\frac{5}{48}&\tmi\frac{5}{48}&\frac{1}{18}&\frac{1}{48}&\frac{1}{36}\\[-1pt]
\frac{10}{9}&\tmi\frac{5}{12}&\frac{1}{3}&\frac{5}{12}&\tmi\frac{5}{12}&\dzero&\frac{1}{4}&\tmi\frac{1}{4}&\dzero&\tmi\frac{5}{3}&\frac{1}{4}&\frac{1}{2}&\frac{1}{3}&\dzero&\tmi\frac{1}{3}\\[-1pt]
\frac{35}{54}&\tmi\frac{35}{144}&\frac{7}{36}&\tmi\frac{35}{144}&\tmi\frac{35}{144}&\frac{35}{192}&\tmi\frac{7}{24}&\tmi\frac{7}{24}&\tmi\frac{7}{48}&\frac{7}{36}&\tmi\frac{7}{24}&\tmi\frac{7}{48}&\frac{7}{36}&\tmi\frac{7}{48}&\frac{2}{9}\end{array}\right)\fwboxL{0pt}{.}\hspace{-60pt}\hspace{-300pt}\label{gggggg_a1_c_to_d_duality}}
The latter of these dualities is more interesting, as a double transformation $\mathbf{d}^2$ would map the basis $\mathcal{D}$ to a rotated image of itself
\eq{r^2(\mathbf{\b{t}}_{\b{i}}\,\mathbf{\g{t}}_{\g{j}}\,\mathbf{\t{t}}_{\t{k}})\mapsto(\mathbf{\g{t}}_{\g{j}}\,\mathbf{\t{t}}_{\t{k}}\,\mathbf{\b{t}}_{\b{i}})\,\label{gggggg_c_basis_rotation}}
Indeed, squaring this relation results in 
\eq{\fwbox{0pt}{\fwboxL{826pt}{(\hspace{-1.25pt}%
\mathfrak{a}_{\r{1}}
)}}\hspace{-300pt}\fwboxR{0pt}{\mathbf{d}.\mathbf{d}=}\left(\begin{array}{c@{$\;\;$}c@{$\;\;$}c@{$\;\;$}c@{$\;\;$}c@{$\;\;$}c@{$\;\;$}c@{$\;\;$}c@{$\;\;$}c@{$\;\;$}c@{$\;\;$}c@{$\;\;$}c@{$\;\;$}c@{$\;\;$}c@{$\;\;$}c}
1&\dzero&\dzero&\dzero&\dzero&\dzero&\dzero&\dzero&\dzero&\dzero&\dzero&\dzero&\dzero&\dzero&\dzero\\[-6pt]
\dzero&\dzero&\dzero&1&\dzero&\dzero&\dzero&\dzero&\dzero&\dzero&\dzero&\dzero&\dzero&\dzero&\dzero\\[-6pt]
\dzero&\dzero&\dzero&\dzero&\dzero&\dzero&\dzero&\dzero&\dzero&1&\dzero&\dzero&\dzero&\dzero&\dzero\\[-6pt]
\dzero&\dzero&\dzero&\dzero&1&\dzero&\dzero&\dzero&\dzero&\dzero&\dzero&\dzero&\dzero&\dzero&\dzero\\[-6pt]
\dzero&1&\dzero&\dzero&\dzero&\dzero&\dzero&\dzero&\dzero&\dzero&\dzero&\dzero&\dzero&\dzero&\dzero\\[-6pt]
\dzero&\dzero&\dzero&\dzero&\dzero&1&\dzero&\dzero&\dzero&\dzero&\dzero&\dzero&\dzero&\dzero&\dzero\\[-6pt]
\dzero&\dzero&\dzero&\dzero&\dzero&\dzero&\dzero&\dzero&\dzero&\dzero&1&\dzero&\dzero&\dzero&\dzero\\[-6pt]
\dzero&\dzero&\dzero&\dzero&\dzero&\dzero&1&\dzero&\dzero&\dzero&\dzero&\dzero&\dzero&\dzero&\dzero\\[-6pt]
\dzero&\dzero&\dzero&\dzero&\dzero&\dzero&\dzero&\dzero&\dzero&\dzero&\dzero&1&\dzero&\dzero&\dzero\\[-6pt]
\dzero&\dzero&\dzero&\dzero&\dzero&\dzero&\dzero&\dzero&\dzero&\dzero&\dzero&\dzero&1&\dzero&\dzero\\[-6pt]
\dzero&\dzero&\dzero&\dzero&\dzero&\dzero&\dzero&1&\dzero&\dzero&\dzero&\dzero&\dzero&\dzero&\dzero\\[-6pt]
\dzero&\dzero&\dzero&\dzero&\dzero&\dzero&\dzero&\dzero&\dzero&\dzero&\dzero&\dzero&\dzero&1&\dzero\\[-6pt]
\dzero&\dzero&1&\dzero&\dzero&\dzero&\dzero&\dzero&\dzero&\dzero&\dzero&\dzero&\dzero&\dzero&\dzero\\[-6pt]
\dzero&\dzero&\dzero&\dzero&\dzero&\dzero&\dzero&\dzero&1&\dzero&\dzero&\dzero&\dzero&\dzero&\dzero\\[-6pt]
\dzero&\dzero&\dzero&\dzero&\dzero&\dzero&\dzero&\dzero&\dzero&\dzero&\dzero&\dzero&\dzero&\dzero&1\end{array}\right)\fwboxL{0pt}{}\hspace{-300pt}\label{gggggg_a1_c_to_d_duality_squared}}
which corresponds to a permutation of basis elements
\eq{\fwboxR{0pt}{\sigma\equivR\!}\left(\begin{array}{@{}ccccccccccccccc@{}}
\fwbox{12pt}{1}&\fwbox{12pt}{2}&\fwbox{12pt}{3}&\fwbox{12pt}{4}&\fwbox{12pt}{5}&\fwbox{12pt}{6}&\fwbox{12pt}{7}&\fwbox{12pt}{8}&\fwbox{12pt}{9}&\fwbox{12pt}{10}&\fwbox{12pt}{11}&\fwbox{12pt}{12}&\fwbox{12pt}{13}&\fwbox{12pt}{14}&\fwbox{12pt}{15}\\[-6pt]
\downarrow&\downarrow&\downarrow&\downarrow&\downarrow&\downarrow&\downarrow&\downarrow&\downarrow&\downarrow&\downarrow&\downarrow&\downarrow&\downarrow&\downarrow\\[-4pt]
\fwbox{12pt}{1}&\fwbox{12pt}{5}&\fwbox{12pt}{13}&\fwbox{12pt}{2}&\fwbox{12pt}{4}&\fwbox{12pt}{6}&\fwbox{12pt}{8}&\fwbox{12pt}{11}&\fwbox{12pt}{14}&\fwbox{12pt}{3}&\fwbox{12pt}{7}&\fwbox{12pt}{9}&\fwbox{12pt}{10}&\fwbox{12pt}{12}&\fwbox{12pt}{15}
\end{array}\right)}
exactly as implied by (\ref{gggggg_c_basis_rotation}).

\newpage
\subsection{\texorpdfstring{Illustrations of High-Multiplicity Scattering Amplitudes in $\mathfrak{a}_{\r{1}}$-Theory}{Illustrations of High-Multiplicity Scattering Amplitudes in a1-Theory}}

The case of scattering amplitudes in $\mathfrak{a}_{\r{1}}$ gauge theory is especially interesting as it is the unique Lie algebra of rank one. Not only is it exceptionally familiar (and relevant to real-world physics), but it is as far as possible from the `large-rank' limit. Scattering amplitudes in $\mathfrak{a}_{\r{1}}$ gauge theory have the fewest independent colour tensors (correspondingly, the most relations among tensors that would be linearly independent for other algebras). 

As is well known to most undergraduate students of quantum mechanics, say, the irreducible representations of $\mathfrak{a}_{\r{1}}$ are uniquely characterized either by its `spin' $\g{j}(\mathbf{\b{r}})\!\in\!\mathbb{Z}_{\geq0}/2$, its one-index Dynkin label $[\t{w}(\mathbf{\b{r}})]$ with $\t{w}(\mathbf{\b{r}})\!\in\!\mathbb{Z}_{\geq0}$, or its dimension; these are related by 
\eq{\mathrm{dim}(\mathbf{\b{r}})=2\,\g{j}(\mathbf{\b{r}}){+}1=\t{w}(\mathbf{\b{r}}){+}1\,.}

The triviality of the case makes it convenient (if slightly abusive) to label irreducible representations directly by their dimension, so that `$\mathrm{dim}(\mathbf{\b{r}})$'=$\mathbf{\b{r}}$. For example, $\r{\mathbf{ad}}$ would be $\r{\mathbf{3}}$, and tensor products involving $\r{\mathbf{3}}$ can be very simply characterized by $\r{\mathbf{3}}\!\otimes\!{\mathbf{1}}\simeq \r{\mathbf{3}}$ and
\eq{\r{\mathbf{3}}\otimes\b{\mathbf{r}}\simeq\b{\mathbf{(r{-}2})}\oplus\b{\mathbf{r}}\oplus\b{\mathbf{(r{+}2)}}, \qquad\b{\mathbf{r}}\!\geq\!2\,.}
Using these it is trivial to compute tensor powers $\r{\mathbf{3}}^{\otimes n}$, and the multiplicity of the $\mathbf{1}$ in this tensor product gives the number of tensors in the Clebsch basis. This number turns out to be the $n^{\text{th}}$ \emph{Riordan number} \cite{oeis-riordan},
\eq{\r{\mathbf{3}}^{\otimes n} \simeq \mathbf{{1}}^{\oplus \text{Riordan($n$)}}\oplus \cdots}
which are given by the recurrence with initial data , $\text{Riordan($0$)}\!=\!1,~\text{Riordan($1$)}\!=\!0$
\eq{\text{Riordan($n$)} =  \frac{(n{-}1)}{(n{+}1)}(2\,\text{Riordan($n{-}1$)} + 3\,\text{Riordan($n{-}2$)})\,.}
For large $n$ we find the asymptotic behavior
\eq{\text{Riordan($n$)} 
  \xrightarrow{n\to\infty} \frac {3^{n{+}2}}{\sqrt{3 n \pi}\, 8 n}\Big(1 - \frac {21}{16 n} + \mathcal{O}(n^{{-}2})\Big).
}
At finite ranks, asymptotic growth characterized by an exponential of the dimension of the adjoint is typical, and so for $\mathfrak{a}_1$ there is the slowest asymptotic growth of all simple algebras \cite{Bourjaily:2024jbt}. Since this is the \textit{largest} number of tensors that can appear at \textit{any} order of perturbation theory, it is interesting to ask how the number of tensors at any \textit{particular} order of perturbation theory compares. In pure-adjoint scattering at tree-level, for instance, one must ask how large is the independent span of the $(2n{-}5)!!$ tensors associated with the trivalent cubic plane graphs constructed from contractions of structure constants. For generic Lie algebras these $(2n{-}5)!!$ tensors are reduced to $(n{-}2)!$ by the Jacobi identity. For $n\!\geq\!5$ one has $\text{Riordan($n$)}\!\leq\!(n{-}2)!$, making the $(n{-}2)!$ assuredly redundant. Even in these cases though, it is possible that the $(n{-}2)!$ do not span the full $\text{Riordan($n$)}$-dimensional space. In a small surprise, by explicit checks up to $n{=}14$, we find that for $n$-even the span of the $(n{-}2)!$ tensors are always $(\text{Riordan($n$)}{-}1)$-dimensional, while for $n$-odd the span is $\text{Riordan($n$)}$-dimensional. From this one can conclude that all of the complexity in colour-space is already there for odd-$n$ at tree-level, and for even-$n$ by explicit checks we find that adding \textit{any} 1-loop $\r{f}$-graph renders the span full-rank. Thus, all $\r{f}$-graphs produced at higher orders in the perturbative expansion generate no new independent tensors.

We can further make an interesting observation about the one redundancy that exists for even-$n$. As is well known, relations amongst colour tensors imply relations amongst the associated partial amplitudes\footnote{To be clear, the relations are amongst the $(n{-}2)!$ tensors, and these imply relations amongst the $\text{Riordan(n)}$ \emph{partial amplitudes}. The Clebsch tensor basis of size $\text{Riordan($n$)}$ is, of course, independent and satisfies no relations.} \cite{Naculich_2012, naculich2024}. In our case, we can rephrase things as: for $n$-even there is exactly one linear relation among the $\text{Riordan($n$)}$ partial amplitudes, which is the vanishing sum of some subset of the partial amplitudes. An interesting observation is \textit{which} partial amplitudes appear in this relation. From the Clebsch basis in a half-ladder topology we can label the tensors by the dimensions of the irreducible representations on the internal edges. We can then observe that the partial amplitudes appearing in the identity are exactly the ones for which the \emph{spin} $\g{j}(\mathbf{\b{r}})$ of the internal representations alternate `$\text{even/odd/even/odd/}\dots$'. We have explicitly verified this curious fact for $n\!\leq\!14$, and leave it as a conjecture to be checked (or better yet, proved) for even $n\!>\!14$. It is remarkable that this relation can not only be detected, but diagnosed with such precision using the Clebsch basis to make the curious even/odd alternating pattern apparent, facilitating an all-$n$ conjecture on the tree-level partial amplitudes of $\mathfrak{a}_{\r{1}}$ gauge theory.

%================================================================================================================
%    Conclusions 
%================================================================================================================
\newpage
\section{Conclusions and Future Directions}\label{sec:conclusions}

We have seen how spanning sets of colour-orthogonal tensors can be used to describe the colour-dependence of scattering processes involving particles coloured by arbitrary representations of arbitrary Lie algebras. These tensors manifestly span all possible colour dependence to arbitrary orders of perturbation (and beyond). 

Interestingly, these tensors bear little resemblance to those that arise directly or naturally from Feynman diagrams in perturbative quantum field theory. For example, although it is clear that only $\r{f}$-graphs appear directly from the Feynman rules of pure gauge theory, the rank spanned by these tensors (as a function of multiplicity or loop-order, say) remains an open question. Similarly, although the symmetric $\r{d}$-tensor (of $\mathfrak{a}_{\r{k}}$ gauge theory) can only arise through anomalous charged matter, these tensors appear and are very relevant to the Clebsch colour tensors described here---even for entirely non-anomalous scattering. Surely, the number of colour tensors required for amplitudes in a non-anomalous gauge theory should be smaller; but it remains an open problem to characterize the subspace of colour tensors required to represent amplitudes in strictly \emph{non-anomalous} gauge theories. 

More generally, the basis of colour tensors described here are decoupled from any notion of perturbation. This is in sharp contrast to the Feynman expansion which naturally generates precise subspaces of colour tensors according to loop order. The particular subspaces required to represent amplitudes at particular loop orders remains an interesting and open question. 

As we have seen in the concrete examples discussed above, processes involving relatively benign particle content (such as fundamentals and adjoints) rarely involve Clebsch tensors with multiplicity greater than 1. Much of this can be understood from the fact that for all simple Lie algebras besides $\mathfrak{e}_{\r{8}}$ and $\mathfrak{f}_{\r{4}}$, any tensor product involving at least one fundamental representation will have multiplicity 1, \cite{Stembridge2003}. 

In most examples where higher multiplicity does arise, orthogonal Clebsch-Gordan coefficients can be constructed via symmetrization/anti-symmetrization of the tensor products. But this conveniently avoids the more pervasive problem that Clebsch-Gordan coefficients can have (in principle) unbounded multiplicity; and their orthogonalization is just as computationally cumbersome as that of general colour tensors (one of the key problems which motivates this work). While it is quite convenient that these issues can be avoided or postponed for most processes of interest, it is worthwhile to seek a more systematic approach to this general problem.

%================================================================================================================
%    Acknowledgements 
%================================================================================================================
\vspace{\fill}\vspace{-4pt}
\section*{Acknowledgements}%
\vspace{-4pt}
The authors gratefully acknowledge fruitful conversations with JJ\ Carrasco, Adrian Ocneanu, Radu Roiban, and Mark Spradlin. This work was supported in part by a grant from the US Department of Energy (No.\ DE-SC00019066).

%================================================================================================================
%    Appendices 
%================================================================================================================
\newpage
\appendix
\section{Building \emph{Concrete} Representations of Lie Algebras}\label{appendix:concrete_representations}
%\addtocontents{toc}{\protect\newpage} 

For all of the examples discussed in this paper, \emph{concrete} tensors were constructed in \textsc{Mathematica} to verify our results. Even for well-known representations of familiar Lie algebras, this turns out to be \emph{dramatically} complicated (and slowed---often to the point of complete obstruction) by seemingly benign assumptions and common conventions often chosen for a representation's generators. 

One of these assumptions---well motivated by physics---is that Lie algebra representations exponentiate to \emph{unitary} representations of groups. This requires that a representation's generators be \emph{anti-Hermitian} (and therefore involve generally complex numbers). Another common, but computationally-problematic, assumption is that the generators of representations be chosen so that 
\eq{\fwbox{0pt}{\fwboxL{435pt}{\text{[sic]}}}\fwbox{0pt}{\mathrm{tr}_{\mathbf{\b{R}}}(\r{a}\,\r{b})\,\equivR\sum_{\b{r_i}\in\b{[r]}}\mathbf{\b{R}}^{\smash{\b{r_1}\,\r{a}}}_{\phantom{\smash{r_1\,a}}\smash{\b{r_2}}}\mathbf{\b{R}}^{\smash{\b{r_2}\,\r{b}}}_{\phantom{\smash{r_2\,b}}\smash{\b{r_1}}}\propto\delta^{\r{a\,b}}\,.}\nonumber}

These common conventions often \emph{require} that matrix elements of generators take values in some algebraic \emph{field extension} over $\mathbb{Q}$ (whether by involving $i\equivR\sqrt{{-}1}$ or other roots over the rationals). Consider for example the simple field extension $\mathbb{F}\equivR\mathbb{Z}{+}\sqrt{{-}1}\,\mathbb{Z}$; the high-school-level identity that 
\eq{\frac{a{-}b\sqrt{{-}1}}{a^2{+}b^2}=\frac{1}{a{+}b\sqrt{{-}1}}\label{trivial_identity_over_c}}
requires \emph{canonicalization}---which is computationally non-trivial. (Indeed, \textsc{Mathematica} does not assign any truth value to the statement (\ref{trivial_identity_over_c}) unless it is wrapped in \textbf{\texttt{Simplify[]}}, say.) This fairly trivial fact means that even simple linear operations on even extremely sparse generator matrices take considerably longer to execute and require considerably more local memory to work with.\\

We can illustrate this point by comparing two possible `concrete' forms of the 3-dimensional, fundamental representation of $\mathfrak{a}_{\r{2}}{=}\mathfrak{su}_{\r{3}}$. Let us first consider the physicists' favorite concrete representation given by the \emph{Gell-Mann} matrices:
\eq{\hspace{-50pt}\begin{array}{@{}c@{$\!$}c@{$\!$}c@{$\!$}c@{$\!$}c@{$\!$}c@{$\!$}c@{$\!$}c@{$\!$}@{}}
~\\[-24pt]\mathbf{\b\lambda}^{\r{1}}&\mathbf{\b\lambda}^{\r{2}}&\mathbf{\b\lambda}^{\r{3}}&\mathbf{\b\lambda}^{\r{4}}&\mathbf{\b\lambda}^{\r{5}}&\mathbf{\b\lambda}^{\r{6}}&\mathbf{\b\lambda}^{\r{7}}&\mathbf{\b\lambda}^{\r{8}}\\
\left(\begin{array}{@{}c@{$\,\,\,$}c@{$\,\,\,$}c@{}}\dzero&1&\dzero\\[-4pt]1&\dzero&\dzero\\[-4pt]\dzero&\dzero&\dzero\end{array}\right)&
\left(\begin{array}{@{}c@{$\,\,\,$}c@{$\,\,\,$}c@{}}\dzero&\tmi i&\dzero\\[-4pt]i&\dzero&\dzero\\[-4pt]\dzero&\dzero&\dzero\end{array}\right)&
\left(\begin{array}{@{}c@{$\,\,\,$}c@{$\,\,\,$}c@{}}1&\dzero&\dzero\\[-4pt]\dzero&\tmi1&\dzero\\[-4pt]\dzero&\dzero&\dzero\end{array}\right)&
\left(\begin{array}{@{}c@{$\,\,\,$}c@{$\,\,\,$}c@{}}\dzero&\dzero&1\\[-4pt]\dzero&\dzero&\dzero\\[-4pt]1&\dzero&\dzero\end{array}\right)&
\left(\begin{array}{@{}c@{$\,\,\,$}c@{$\,\,\,$}c@{}}\dzero&\dzero&\tmi i\\[-4pt]\dzero&\dzero&\dzero\\[-4pt]i&\dzero&\dzero\end{array}\right)&
\left(\begin{array}{@{}c@{$\,\,\,$}c@{$\,\,\,$}c@{}}\dzero&\dzero&\dzero\\[-4pt]\dzero&\dzero&1\\[-4pt]\dzero&1&\dzero\end{array}\right)&
\left(\begin{array}{@{}c@{$\,\,\,$}c@{$\,\,\,$}c@{}}\dzero&\dzero&\dzero\\[-4pt]\dzero&\dzero&\tmi i\\[-4pt]\dzero&i&\dzero\end{array}\right)&
\,\frac{1}{\sqrt{3}}\!\!\left(\begin{array}{@{}c@{$\,\,\,$}c@{$\,\,\,$}c@{}}1&\dzero&\dzero\\[-4pt]\dzero&1&\dzero\\[-4pt]\dzero&\dzero&\tmi 2\end{array}\right)\fwboxL{0pt}{\!\!.}\phantom{\frac{1}{\sqrt{3}}}
\end{array}\hspace{-40pt}}
In terms of these, the Killing metric is diagonal:
\eq{\mathrm{tr}_{\mathbf{\b{\lambda}}}(\r{a\,b})=2\,\delta^{\r{a\,b}}.\label{gell_mann_killing_metric}}
Each of the generators $\mathbf{\b{\lambda}}^{\r{a}}$ are \emph{Hermitian}, and therefore $i\mathbf{\b{\lambda}}^{\r{a}}$ is \emph{anti}-Hermitian; as such, $\mathrm{exp}(\alpha_{\r{a}}\mathbf{\b{\lambda}}^{\r{a}})$ will be unitary for pure-imaginary $\alpha_{\r{a}}\!\in\!i\,\mathbb{R}^{8}$. Moreover, because the Killing metric is proportional to the identity (\ref{gell_mann_killing_metric}), there is very little distinction between raised and lowered (adjoint) indices: $\mathbf{\b{\lambda}}^{\r{a}}\!\propto\!\mathbf{\b{\lambda}}_{\r{a}}$ and $\r{f}^{\r{a\,b\,c}}\!\propto\!\r{f}\indices{\r{a\,b}}{\r{c}}$.

Let us compare this set of generators to those presented in the `Chevalley' basis:
\eq{\begin{array}{@{}c@{$\!$}c@{$\!$}c@{$\!$}c@{$\!$}c@{$\!$}c@{$\!$}c@{$\!$}c@{$\!$}c@{$\!$}c@{}}
~\\[-24pt]\mathbf{\b{F}}^{\r{\tilde{1}}}&\mathbf{\b{F}}^{\r{\tilde{2}}}&\mathbf{\b{F}}^{\r{\tilde{3}}}&\mathbf{\b{F}}^{\r{\tilde{4}}}&\mathbf{\b{F}}^{\r{\tilde{5}}}&\mathbf{\b{F}}^{\r{\tilde{6}}}&\mathbf{\b{F}}^{\r{\tilde{7}}}&\mathbf{\b{F}}^{\r{\tilde{8}}}\\
\left(\begin{array}{@{}c@{$\,\,\,$}c@{$\,\,\,$}c@{}}\dzero&\dzero&\dzero\\[-4pt]\dzero&\dzero&\dzero\\[-4pt]1&\dzero&\dzero\end{array}\right)&
\left(\begin{array}{@{}c@{$\,\,\,$}c@{$\,\,\,$}c@{}}\dzero&\dzero&\dzero\\[-4pt]1&\dzero&\dzero\\[-4pt]\dzero&\dzero&\dzero\end{array}\right)&
\left(\begin{array}{@{}c@{$\,\,\,$}c@{$\,\,\,$}c@{}}\dzero&\dzero&\dzero\\[-4pt]\dzero&\dzero&\dzero\\[-4pt]\dzero&1&\dzero\end{array}\right)&
\,\frac{1}{3}\!\!\left(\begin{array}{@{}c@{$\,\,\,$}c@{$\,\,\,$}c@{}}2&\dzero&\dzero\\[-4pt]\dzero&\tmi1&\dzero\\[-4pt]\dzero&\dzero&\tmi1\end{array}\right)\phantom{\frac{1}{3}}\!\!\!&
\!\frac{1}{3}\!\!\left(\begin{array}{@{}c@{$\,\,\,$}c@{$\,\,\,$}c@{}}1&\dzero&\dzero\\[-4pt]\dzero&1&\dzero\\[-4pt]\dzero&\dzero&\tmi2\end{array}\right)\phantom{\frac{1}{3}}\!\!\!&
\left(\begin{array}{@{}c@{$\,\,\,$}c@{$\,\,\,$}c@{}}\dzero&\dzero&\dzero\\[-4pt]\dzero&\dzero&1\\[-4pt]\dzero&\dzero&\dzero\end{array}\right)&
\left(\begin{array}{@{}c@{$\,\,\,$}c@{$\,\,\,$}c@{}}\dzero&1&\dzero\\[-4pt]\dzero&\dzero&\dzero\\[-4pt]\dzero&\dzero&\dzero\end{array}\right)&
\left(\begin{array}{@{}c@{$\,\,\,$}c@{$\,\,\,$}c@{}}\dzero&\dzero&1\\[-4pt]\dzero&\dzero&\dzero\\[-4pt]\dzero&\dzero&\dzero\end{array}\right)\fwboxL{0pt}{\!\!.}
\end{array}\label{chevalley_basis_generators}}
In terms of these, the Killing metric is not diagonal, but rather
\eq{\fwboxR{0pt}{\mathrm{tr}_{\mathbf{\b{F}}}(\r{\tilde{a}\,\tilde{b}})=\frac{1}{6}}\!\left(\begin{array}{@{}c@{$\,\,\,$}c@{$\,\,\,$}c@{$\,\,\,$}c@{$\,\,\,$}c@{$\,\,\,$}c@{$\,\,\,$}c@{$\,\,\,$}c@{}}\dzero&\dzero&\dzero&\dzero&\dzero&\dzero&\dzero&6\\[-4pt]
\dzero&\dzero&\dzero&\dzero&\dzero&\dzero&6&\dzero\\[-4pt]
\dzero&\dzero&\dzero&\dzero&\dzero&6&\dzero&\dzero\\[-4pt]
\dzero&\dzero&\dzero&4&2&\dzero&\dzero&\dzero\\[-4pt]
\dzero&\dzero&\dzero&2&4&\dzero&\dzero&\dzero\\[-4pt]
\dzero&\dzero&6&\dzero&\dzero&\dzero&\dzero&\dzero\\[-4pt]
\dzero&6&\dzero&\dzero&\dzero&\dzero&\dzero&\dzero\\[-4pt]
6&\dzero&\dzero&\dzero&\dzero&\dzero&\dzero&\dzero\\[-4pt]
\end{array}\right)\fwboxL{0pt}{\!\!.}\label{chevalley_basis_inv_metric}}
Because the generators $\mathbf{\b{F}}^{\r{\tilde{a}}}$ are \emph{neither} Hermitian nor anti-Hermitian, exponentiation $\mathrm{exp}(\beta_{\r{\tilde{a}}}\mathbf{\b{F}}^{\r{\tilde{a}}})$ would only result in a unitary representation of the group if the coefficients $\beta_{\r{\tilde{a}}}\!\in\!\mathbb{C}^{8}$ satisfied particular conditions (for example, that $\beta_{\r{\tilde{1}}}{=}{-}\bar{\beta_{\r{\tilde{8}}}}$). And because $\mathrm{tr}_{\mathbf{\b{F}}}(\r{\tilde{a}\,\tilde{b}})$ is not proportional to the identity (\ref{chevalley_basis_inv_metric}), there is a non-trivial distinction between raised and lowered adjoint indices.

Of course, it is easy to see that the generators $\mathbf{\b{\lambda}}^{\r{a}}$ are simple linear combinations of the generators $\mathbf{\b{F}}^{\r{\tilde{a}}}$: $\mathbf{\b{\lambda}}^{\r{a}}\equivL\mathbf{M}\indices{\r{a}}{\r{\tilde{b}}}\mathbf{\b{F}}^{\r{\tilde{b}}}$, where the combination of generators is defined by 
\eq{\fwboxR{0pt}{\mathbf{M}\indices{\r{{a}}}{\r{\tilde{a}}}=}\,\left(\begin{array}{@{}c@{$\,\,\,\,$}c@{$\,\,\,\,$}c@{$\,\,\,\,$}c@{$\,\,\,\,$}c@{$\,\,\,\,$}c@{$\,\,\,\,$}c@{$\,\,\,\,$}c@{}}\dzero&1&\dzero&\dzero&\dzero&\dzero&1&\dzero\\[-4pt]
\dzero&i&\dzero&\dzero&\dzero&\dzero&\tmi i&\dzero\\[-4pt]
\dzero&\dzero&\dzero&2&\tmi1&\dzero&\dzero&\dzero\\[-4pt]
1&\dzero&\dzero&\dzero&\dzero&\dzero&\dzero&1\\[-4pt]
i&\dzero&\dzero&\dzero&\dzero&\dzero&\dzero&\tmi i\\[-4pt]
\dzero&\dzero&1&\dzero&\dzero&1&\dzero&\dzero\\[-4pt]
\dzero&\dzero&i&\dzero&\dzero&\tmi i&\dzero&\dzero\\[-4pt]
\dzero&\dzero&\dzero&\dzero&\sqrt{3}&\dzero&\dzero&\dzero
\end{array}\right)\,.}

We can begin to appreciate how the two concrete representations compare by considering how the structure constants (\ref{definition_of_antisymmetric_structure_constants}) appear for the two representations. Taking both representations' Dynkin index to be 1---$T(\mathbf{\b{\lambda}}){=}T(\mathbf{\b{R}}){=}1$---we would find 
\eq{\frac{1}{2}\r{f}_{\mathbf{\b{\lambda}}}^{\r{1\,2\,3}}{=}\r{f}_{\mathbf{\b{\lambda}}}^{\r{1\,4\,7}}{=}\r{f}_{\mathbf{\b{\lambda}}}^{\r{1\,6\,5}}{=}\r{f}_{\mathbf{\b{\lambda}}}^{\r{2\,4\,6}}{=}\r{f}_{\mathbf{\b{\lambda}}}^{\r{2\,5\,7}}{=}\r{f}_{\mathbf{\b{\lambda}}}^{\r{3\,4\,5}}{=}\r{f}_{\mathbf{\b{\lambda}}}^{\r{3\,7\,6}}{=}\,\frac{i}{4},\quad\r{f}_{\mathbf{\b{\lambda}}}^{\r{4\,5\,8}}{=}\r{f}_{\mathbf{\b{\lambda}}}^{\r{6\,7\,8}}{=}\frac{i\sqrt{3}}{4}\,,}
for the Gell-Mann matrices, and 
\eq{\r{f}_{\mathbf{\b{F}}}^{\r{\tilde{1}\,\tilde{4}\,\tilde{8}}}{=}\r{f}_{\mathbf{\b{F}}}^{\r{\tilde{1}\,\tilde{5}\,\tilde{8}}}{=}\r{f}_{\mathbf{\b{F}}}^{\r{\tilde{1}\,\tilde{7}\,\tilde{6}}}{=}\r{f}_{\mathbf{\b{F}}}^{\r{\tilde{2}\,\tilde{8}\,\tilde{3}}}{=}\r{f}_{\mathbf{\b{F}}}^{\r{\tilde{2}\,\tilde{4}\,\tilde{7}}}{=}\r{f}_{\mathbf{\b{F}}}^{\r{\tilde{3}\,\tilde{5}\,\tilde{6}}}{=}\,1,}
for the Chevalley generators; in both cases, all other non-vanishing components dictated by antisymmetry. 

This discrepant complexity may seem trivial, but it compounds substantially when constructing larger tensors. For example, using the Gell-Mann matrices as generators, it would take about \textbf{47 seconds}\footnote{These benchmarks were obtained using a standard-issue laptop running \textsc{Mathematica} with only moderately clever encoding.} to explicitly construct the tensor
\eq{\mathrm{tr}_{\mathbf{\b{\lambda}}}(\r{1\,2\,3\,4\,5\,6\,7\,8\,9\,10})\,;}
this tensor would involve $35,\!446,\!548$ non-vanishing components and require \textbf{7.950\,GB} to store in memory. In contrast, using the generators $\mathbf{\b{F}}$,
\eq{\mathrm{tr}_{\mathbf{\b{F}}}(\r{\tilde{1}\,\tilde{2}\,\tilde{3}\,\tilde{4}\,\tilde{5}\,\tilde{6}\,\tilde{7}\,\tilde{8}\,\tilde{9}\,\tilde{10}})\,,}
would require about \textbf{260 \emph{milli}seconds} to construct, would involve only $1,\!046,\!530$ non-vanishing components and require only \textbf{142 MB} to store in memory. It is easy to appreciate that factors of $\sim\!50$ in memory and $\sim\!200$ in significantly affect one's ability to compute tensors explicitly.\\

For all the simple Lie algebras, irreducible representations can be constructed for which all components $\mathbf{\b{R}}\indices{\b{r_1}\,\r{a}}{\b{r_2}}\!\in\!\mathbb{Q}$, allowing for extremely efficient construction of concrete tensors. These can be obtained via various computer algebra packages (\emph{e.g.}~\cite{magma}), and we plan to release an implementation of these, together with basic tensor manipulations for \textsc{Mathematica} in the near future \cite{lie_algebra_representation_tensor_tools}.

\newpage
\section{Conventions for Weight Systems of Simple Lie Algebras}\label{appendix:weight_system_conventions}
\addtocontents{toc}{\vspace{-10pt}}
\addtocontents{toc}{\protect\enlargethispage{20\baselineskip}}  

Insofar as we must make reference to explicit (irreducible) representations, it is best to label them by their `Dynkin labels' (or highest weights).
This requires an explicit choice of conventions be made for how we organize the weight systems for each of the simple Lie algebras. Interested readers can consult \emph{e.g.}~\cite{Slansky:1981yr,Cornwell:1997ke,DiFrancesco:1997nk,fulton2013}---but our conventions differ slightly from those chosen by other authors. We document these conventional choices here.

\eq{\fwbox{00pt}{\hspace{-30pt}\begin{array}{@{}c@{}c@{}c@{}c@{}}\fwboxL{0pt}{\hspace{2pt}\mathfrak{a}_{\r{k}}\!\!:}\fwbox{140pt}{\text{Dynkin diagram}}&\fwbox{125pt}{\text{Cartan matrix}}&\fwbox{135pt}{\text{weight metric}}\\
%&\rule{155pt}{1pt}&\rule{130pt}{1pt}&\rule{135pt}{1pt}\\
\begin{array}{@{}l@{}}\fwbox{140pt}{\tikzBox{a_series_dynkin_diagram}{\draw[edge](0,0)--(2,0);\draw[dashed,edge](2,0)--(3.175,0);
\node[dynkS](v1)at(0,0){};\node[anchor=90,inner sep=10pt]at(v1){$\alpha_\fwboxL{0pt}{1}$};
\node[dynkS](v2)at(1,0){};\node[anchor=90,inner sep=10pt]at(v2){$\alpha_\fwboxL{0pt}{2}$};
\node[dynkS](v3)at(2,0){};\node[anchor=90,inner sep=10pt]at(v3){$\alpha_\fwboxL{0pt}{3}$};
\node[dynkS](vn)at(3.175,0){};\node[anchor=90,inner sep=10pt]at(vn){$\alpha_\fwboxL{0pt}{\r{k}}$};
}
}\\[-5pt]\hspace{10pt}\fwboxL{140pt}{\begin{array}{@{}l@{$\equivR\!$}l@{}}
w(\mathbf{\r{ad}})&[10\cdots01]\!\equivL\theta\\
w(\mathbf{\b{F}})&[10\cdots00]\\[-5pt]
\end{array}}\end{array}\vspace{-10pt}&
\fwbox{130pt}{\left(\begin{array}{@{\hspace{-2pt}}c@{}c@{}c@{}c@{}c@{}c@{}c@{\hspace{-5pt}}}
2&\mone&\dzero&\dzero&{\color{dim}\cdots}&\dzero\\[-5pt]
\mone&2&\mone&\dzero&{\color{dim}\cdots}&\dzero\\[-7.5pt]
\dzero&\ddots&\ddots&\ddots&{\color{dim}\ddots}&{\color{dim}\smash{\vdots}}\\[-7.5pt]
\raisebox{0pt}{$\smash{{\color{dim}\vdots}}$}&{\color{dim}\ddots}&\mone&2&\mone&\dzero\\[-5pt]
\dzero&{\color{dim}\cdots}&\dzero&\mone&2&\fwbox{4pt}{\text{-}}1\fwbox{4pt}{~}\\[-5pt]
\dzero&{\color{dim}\cdots}&\dzero&\dzero&\mone&2
\end{array}\right)}
&\fwbox{135pt}{\fwboxR{0pt}{\frac{1}{\r{k}\text{+}1}\!\!}\left(\begin{array}{@{\hspace{-5pt}}c@{}c@{\hspace{-3pt}}c@{\hspace{-5pt}}c@{\hspace{-0pt}}c@{\hspace{-0pt}}}
\r{k}&1(\r{k}\text{-}1)\phantom{1}&1(\r{k}\text{-}2)\phantom{1}&\cdots&1\\[-5pt]
(\r{k}\text{-}1)&2(\r{k}\text{-}1)\phantom{2}&2(\r{k}\text{-}2)\phantom{2}&\cdots&2\\[-5pt]
(\r{k}\text{-}2)&2(\r{k}\text{-}2)\phantom{2}&3(\r{k}\text{-}2)\phantom{3}&\cdots&3\\[-7pt]
\smash{\vdots}&\smash{\vdots}&\smash{\vdots}&\ddots&\smash{\vdots}\\[-5pt]
%2&4&6&\cdots&\fwboxR{4pt}{2}(\r{k}\text{-}1)\fwboxL{4pt}{~}&(\r{k}\text{-}1)\\[-3pt]
1&2&3&\cdots&\r{k}\end{array}\right)}\end{array}}\vspace{25pt}\label{ak_weight_conventions}}

\eq{\fwbox{00pt}{\hspace{-30pt}\begin{array}{@{}c@{}c@{}c@{}c@{}}\fwboxL{0pt}{\hspace{2pt}\mathfrak{b}_{\r{k}}\!\!:}\fwbox{140pt}{\text{Dynkin diagram}}&\fwbox{125pt}{\text{Cartan matrix}}&\fwbox{135pt}{\text{weight metric}}\\
%&\rule{155pt}{1pt}&\rule{130pt}{1pt}&\rule{135pt}{1pt}\\
\begin{array}{@{}l@{}}\fwbox{140pt}{\tikzBox{b_series_dynkin_diagram}{\draw[edge](0,0)--(1,0);\draw[dashed,edge](1,0)--(2.175,0);\draw[edge](2.175,0.15)--(3.175,0.15);\draw[edge](2.175,-0.15)--(3.175,-0.15);
\node[dynkS](v1)at(0,0){};\node[anchor=90,inner sep=10pt]at(v1){$\alpha_\fwboxL{0pt}{1}$};
\node[dynkS](v2)at(1,0){};\node[anchor=90,inner sep=10pt]at(v2){$\alpha_\fwboxL{0pt}{2}$};
\node[dynkS](vnm1)at(2.175,0){};\node[anchor=90,inner sep=10pt]at(vnm1){$\alpha_\fwboxL{0pt}{\r{k}\text{-}1}$};
\node[dynkL](vn)at(3.175,0){};\node[anchor=90,inner sep=10pt]at(vn){$\alpha_\fwboxL{0pt}{\r{k}}$};
}
}\\[-5pt]\hspace{10pt}\fwboxL{140pt}{\begin{array}{@{}l@{$\equivR\!$}l@{}}
w(\mathbf{\r{ad}})&[010\cdots]\!\equivL\theta\\
w(\mathbf{\b{F}})&[10\cdots0]\\[-5pt]
\end{array}}\end{array}\vspace{-10pt}&
\fwbox{130pt}{\left(\begin{array}{@{}c@{}c@{}c@{}c@{}c@{}c@{}c@{\hspace{-3pt}}}
2&\mone&\dzero&\dzero&{\color{dim}\cdots}&\dzero\\[-5pt]
\mone&2&\mone&\dzero&{\color{dim}\cdots}&\dzero\\[-7.5pt]
\dzero&\ddots&\ddots&\ddots&{\color{dim}\ddots}&{\color{dim}\smash{\vdots}}\\[-7.5pt]
\raisebox{0pt}{$\smash{{\color{dim}\vdots}}$}&{\color{dim}\ddots}&\mone&2&\mone&\dzero\\[-5pt]
\dzero&{\color{dim}\cdots}&\dzero&\mone&2&\fwbox{4pt}{\text{-}}2\fwbox{4pt}{~}\\[-5pt]
\dzero&{\color{dim}\cdots}&\dzero&\dzero&\mone&2
\end{array}\right)}
&\fwbox{135pt}{\fwboxR{0pt}{\frac{1}{2}\!\!}\left(\begin{array}{@{}cccc@{$\,$}c@{$\,$}c@{\hspace{-3pt}}}
2&2&2&\cdots&2&1\\[-5pt]
2&4&4&\cdots&4&2\\[-5pt]
2&4&6&\cdots&6&3\\[-7pt]
\smash{\vdots}&\smash{\vdots}&\smash{\vdots}&\ddots&\smash{\vdots}&\smash{\vdots}\\[-5pt]
2&4&6&\cdots&\fwboxR{4pt}{2}(\r{k}\text{-}1)\fwboxL{4pt}{~}&(\r{k}\text{-}1)\\[-3pt]
1&2&3&\cdots&(\r{k}\text{-}1)&\r{k}/2\end{array}\right)}\end{array}}\vspace{25pt}\label{bk_weight_conventions}}

\eq{\fwbox{00pt}{\hspace{-30pt}\begin{array}{@{}c@{}c@{}c@{}c@{}}\fwboxL{0pt}{\hspace{2pt}\mathfrak{c}_{\r{k}}\!\!:}\fwbox{140pt}{\text{Dynkin diagram}}&\fwbox{125pt}{\text{Cartan matrix}}&\fwbox{135pt}{\text{weight metric}}\\
%&\rule{155pt}{1pt}&\rule{130pt}{1pt}&\rule{135pt}{1pt}\\
\begin{array}{@{}l@{}}\fwbox{140pt}{\tikzBox{c_series_dynkin_diagram}{\draw[edge](0,0)--(1,0);\draw[dashed,edge](1,0)--(2.175,0);\draw[edge](2.175,0.15)--(3.175,0.15);\draw[edge](2.175,-0.15)--(3.175,-0.15);
\node[dynkL](v1)at(0,0){};\node[anchor=90,inner sep=10pt]at(v1){$\alpha_\fwboxL{0pt}{1}$};
\node[dynkL](v2)at(1,0){};\node[anchor=90,inner sep=10pt]at(v2){$\alpha_\fwboxL{0pt}{2}$};
\node[dynkL](vnm1)at(2.175,0){};\node[anchor=90,inner sep=10pt]at(vnm1){$\alpha_\fwboxL{0pt}{\r{k}\text{-}1}$};
\node[dynkS](vn)at(3.175,0){};\node[anchor=90,inner sep=10pt]at(vn){$\alpha_\fwboxL{0pt}{\r{k}}$};
}
}\\[-5pt]\hspace{10pt}\fwboxL{140pt}{\begin{array}{@{}l@{$\equivR\!$}l@{}}
w(\mathbf{\r{ad}})&[20\cdots0]\!\equivL\theta\\
w(\mathbf{\b{F}})&[10\cdots0]\\[-5pt]
\end{array}}\end{array}\vspace{-10pt}&
\fwbox{130pt}{\left(\begin{array}{@{}c@{}c@{}c@{}c@{}c@{}c@{}c@{\hspace{-3pt}}}
2&\mone&\dzero&\dzero&{\color{dim}\cdots}&\dzero\\[-5pt]
\mone&2&\mone&\dzero&{\color{dim}\cdots}&\dzero\\[-7.5pt]
\dzero&\ddots&\ddots&\ddots&{\color{dim}\ddots}&{\color{dim}\smash{\vdots}}\\[-7.5pt]
\raisebox{0pt}{$\smash{{\color{dim}\vdots}}$}&{\color{dim}\ddots}&\mone&2&\mone&\dzero\\[-5pt]
\dzero&{\color{dim}\cdots}&\dzero&\mone&2&\mone\\[-5pt]
\dzero&{\color{dim}\cdots}&\dzero&\dzero&\fwbox{4pt}{\text{-}}2\fwbox{4pt}{~}&2
\end{array}\right)}
&\fwbox{135pt}{\fwboxR{0pt}{\frac{1}{2}\!\!}\left(\begin{array}{@{}cccc@{$\,$}c@{$\,$}c@{\hspace{-3pt}}}
1&1&1&\cdots&1&1\\[-5pt]
1&2&2&\cdots&2&2\\[-5pt]
1&2&3&\cdots&3&3\\[-7pt]
\smash{\vdots}&\smash{\vdots}&\smash{\vdots}&\ddots&\smash{\vdots}&\smash{\vdots}\\[-5pt]
1&2&3&\cdots&(\r{k}\text{-}1)&(\r{k}\text{-}1)\\[-3pt]
1&2&3&\cdots&(\r{k}\text{-}1)&\r{k}\end{array}\right)}\end{array}}\vspace{25pt}\label{ck_weight_conventions}}

\eq{\fwbox{00pt}{\hspace{-30pt}\begin{array}{@{}c@{}c@{}c@{}c@{}}\fwboxL{0pt}{\hspace{2pt}\mathfrak{d}_{\r{k}}\!\!:}\fwbox{140pt}{\text{Dynkin diagram}}&\fwbox{125pt}{\text{Cartan matrix}}&\fwbox{135pt}{\text{weight metric}}\\
%&\rule{155pt}{1pt}&\rule{130pt}{1pt}&\rule{135pt}{1pt}\\
\begin{array}{@{}l@{}}\fwbox{140pt}{\tikzBox{d_series_dynkin_diagram}{\draw[edge](0,0)--(1*\dynkEdge,0);\draw[dashed,edge](1*\dynkEdge,0)--(2.275*\dynkEdge,0);\draw[edge](2.275*\dynkEdge,0)--(3.275*\dynkEdge,0);\draw[edge](3.275*\dynkEdge,0)--($(3.275*\dynkEdge,0)+(90:1*\dynkEdge)$);\draw[edge](3.275*\dynkEdge,0)--($(3.175*\dynkEdge,0)+(-0:1*\dynkEdge)$);
\node[dynkS](v1)at(0*\dynkEdge,0){};\node[anchor=90,inner sep=10pt]at(v1){$\alpha_\fwboxL{0pt}{1}$};
\node[dynkS](v2)at(1*\dynkEdge,0){};\node[anchor=90,inner sep=10pt]at(v2){$\alpha_\fwboxL{0pt}{2}$};
\node[dynkS](vnm3)at(2.275*\dynkEdge,0){};\node[anchor=90,inner sep=10pt]at(vnm3){$\alpha_\fwboxL{0pt}{\r{k}\text{-}3}$};
\node[dynkS](vnm2)at(3.275*\dynkEdge,0){};\node[anchor=90,inner sep=10pt]at(vnm2){$\alpha_\fwboxL{0pt}{\r{k}\text{-}2}$};
\node[dynkS](vnm1)at(4.275*\dynkEdge,0){};\node[anchor=90,inner sep=10pt]at(vnm1){$\alpha_\fwboxL{0pt}{\r{k}\text{-}1}$};
\node[dynkS](vn)at(3.275*\dynkEdge,1*\dynkEdge){};\node[anchor=180,inner sep=7.5pt]at(vn){$\alpha_\fwboxL{0pt}{\r{k}}$};
}
}\\[-5pt]\hspace{10pt}\fwboxL{140pt}{\begin{array}{@{}l@{$\equivR\!$}l@{}}
w(\mathbf{\r{ad}})&[010\cdots]\!\equivL\theta\\
w(\mathbf{\b{F}})&[10\cdots0]\\[-5pt]
\end{array}}\end{array}\vspace{-10pt}&
\fwbox{130pt}{\left(\begin{array}{@{}c@{}c@{}c@{}c@{}c@{}c@{\hspace{-2pt}}}
2&\mone&\dzero&\dzero&{\color{dim}\cdots}&\dzero\\[-7.5pt]
\mone&2&\mone&\dzero&{\color{dim}\ddots}&\smash{{\color{dim}\vdots}}\\[-7.5pt]
\dzero&\ddots&\ddots&\ddots&{\color{dim}\ddots}&\smash{{\color{dim}0}}\\[-7.5pt]
\raisebox{0pt}{$\smash{{\color{dim}\vdots}}$}&{\color{dim}\ddots}&\mone&2&\mone&\mone\\[-5pt]
\dzero&{\color{dim}\cdots}&\dzero&\mone&2&\dzero\\[-5pt]
\dzero&{\color{dim}\cdots}&\dzero&\mone&\dzero&2
\end{array}\right)}
&\fwbox{135pt}{\fwboxR{0pt}{\frac{1}{2}\!\!}\left(\begin{array}{@{}ccc@{$\hspace{-0pt}$}c@{$\hspace{-15pt}$}c@{$\hspace{-10pt}$}c@{\hspace{-5pt}}}
2&2&\cdots&2&1&1\\[-5pt]
2&4&\cdots&4&2&2\\[-7pt]
\smash{\vdots}&\smash{\vdots}&\ddots&\smash{\vdots}&\smash{\vdots}&\smash{\vdots}\\[-5pt]
2&4&\cdots&2(\r{k}\text{-}2)\phantom{2}&(\r{k}\text{-}2)&(\r{k}\text{-}2)\\[-3pt]
1&2&\cdots&(\r{k}\text{-}2)&\r{\r{k}}/2&\phantom{2/}(\r{k}\text{-}2)/2\\[-3pt]
1&2&\cdots&(\r{k}\text{-}2)&\phantom{2/}(\r{k}\text{-}2)/2&\r{k}/2\end{array}\right)}\\[130pt]~\end{array}}\vspace{-80pt}\label{dk_weight_conventions}}

\eq{\fwbox{00pt}{\hspace{-30pt}\begin{array}{@{}c@{}c@{}c@{}c@{}}\fwboxL{0pt}{\hspace{2pt}\mathfrak{e}_{\r{6}}\!\!:}\fwbox{140pt}{\text{Dynkin diagram}}&\fwbox{125pt}{\text{Cartan matrix}}&\fwbox{135pt}{\text{weight metric}}\\
%&\rule{155pt}{1pt}&\rule{130pt}{1pt}&\rule{135pt}{1pt}\\
\begin{array}{@{}l@{}}\fwbox{140pt}{\tikzBox{e6_dynkin_diagram}{
\draw[edge](0,0)--(4*\dynkEdge,0);\draw[edge](2*\dynkEdge,0)--(2*\dynkEdge,1*\dynkEdge);
\node[dynkS](v1)at(0,0){};\node[anchor=90,inner sep=10pt]at(v1){$\alpha_\fwboxL{0pt}{1}$};
\node[dynkS](v2)at(1*\dynkEdge,0){};\node[anchor=90,inner sep=10pt]at(v2){$\alpha_\fwboxL{0pt}{2}$};
\node[dynkS](v3)at(2*\dynkEdge,0){};\node[anchor=90,inner sep=10pt]at(v3){$\alpha_\fwboxL{0pt}{3}$};
\node[dynkS](v4)at(3*\dynkEdge,0){};\node[anchor=90,inner sep=10pt]at(v4){$\alpha_\fwboxL{0pt}{4}$};
\node[dynkS](v5)at(4*\dynkEdge,0){};\node[anchor=90,inner sep=10pt]at(v5){$\alpha_\fwboxL{0pt}{5}$};
\node[dynkS](v6)at(2*\dynkEdge,1*\dynkEdge){};\node[anchor=180,inner sep=7.5pt]at(v6){$\alpha_\fwboxL{0pt}{6}$};
}
}\\[-5pt]\hspace{10pt}\fwboxL{140pt}{\begin{array}{@{}l@{$\equivR\!$}l@{}}
w(\mathbf{\r{ad}})&[000001]\!\equivL\theta\\
w(\mathbf{\b{F}})&[100000]\\[-20pt]
\end{array}}\end{array}&
\fwbox{130pt}{\left(\begin{array}{@{}cccccc@{\hspace{0pt}}}
2&\mone&\dzero&\dzero&\dzero&\dzero\\[-5pt]
\mone&2&\mone&\dzero&\dzero&\dzero\\[-5pt]
\dzero&\mone&2&\mone&\dzero&\mone\\[-5pt]
\dzero&\dzero&\mone&2&\mone&\dzero\\[-5pt]
\dzero&\dzero&\dzero&\mone&2&\dzero\\[-5pt]
\dzero&\dzero&\mone&\dzero&\dzero&2
\end{array}\right)}
&\fwbox{135pt}{\fwboxR{0pt}{\frac{1}{3}\!\!}\left(\begin{array}{@{}cccccc@{\hspace{-0pt}}}
4&5&6&4&2&3\\[-5pt]
5&10&12&8&4&6\\[-5pt]
6&12&18&12&6&9\\[-5pt]
4&8&12&10&5&6\\[-5pt]
2&4&6&5&4&3\\[-5pt]
3&6&9&6&3&6\end{array}\right)}\end{array}}\vspace{35pt}\label{e6_weight_conventions}}

\eq{\fwbox{00pt}{\hspace{-30pt}\begin{array}{@{}c@{}c@{}c@{}c@{}}\fwboxL{0pt}{\hspace{2pt}\mathfrak{e}_{\r{7}}\!\!:}\fwbox{140pt}{\text{Dynkin diagram}}&\fwbox{125pt}{\text{Cartan matrix}}&\fwbox{135pt}{\text{weight metric}}\\
%&\rule{155pt}{1pt}&\rule{130pt}{1pt}&\rule{135pt}{1pt}\\
\begin{array}{@{}l@{}}\fwbox{140pt}{\tikzBox{e7_dynkin_diagram}{\draw[edge](0,0)--(5*\dynkEdge,0);\draw[edge](3*\dynkEdge,0)--(3*\dynkEdge,1*\dynkEdge);
\node[dynkS](v1)at(0,0){};\node[anchor=90,inner sep=10pt]at(v1){$\alpha_\fwboxL{0pt}{1}$};
\node[dynkS](v2)at(\dynkEdge,0){};\node[anchor=90,inner sep=10pt]at(v2){$\alpha_\fwboxL{0pt}{2}$};
\node[dynkS](v3)at(2*\dynkEdge,0){};\node[anchor=90,inner sep=10pt]at(v3){$\alpha_\fwboxL{0pt}{3}$};
\node[dynkS](v4)at(3*\dynkEdge,0){};\node[anchor=90,inner sep=10pt]at(v4){$\alpha_\fwboxL{0pt}{4}$};
\node[dynkS](v5)at(4*\dynkEdge,0){};\node[anchor=90,inner sep=10pt]at(v5){$\alpha_\fwboxL{0pt}{5}$};
\node[dynkS](v6)at(5*\dynkEdge,0){};\node[anchor=90,inner sep=10pt]at(v6){$\alpha_\fwboxL{0pt}{6}$};
\node[dynkS](v7)at(3*\dynkEdge,\dynkEdge){};
\node[anchor=180,inner sep=7.5pt]at(v7){$\alpha_\fwboxL{0pt}{7}$};
}
}\\[-5pt]\hspace{10pt}\fwboxL{140pt}{\begin{array}{@{}l@{$\equivR\!$}l@{}}
w(\mathbf{\r{ad}})&[0000010]\!\equivL\theta\\
w(\mathbf{\b{F}})&[1000000]\\[-5pt]
\end{array}}\end{array}\vspace{-10pt}&
\fwbox{130pt}{\left(\begin{array}{@{}ccccccc@{}}
2&\mone&\dzero&\dzero&\dzero&\dzero&\dzero\\[-5pt]
\mone&2&\mone&\dzero&\dzero&\dzero&\dzero\\[-5pt]
\dzero&\mone&2&\mone&\dzero&\dzero&\dzero\\[-5pt]
\dzero&\dzero&\mone&2&\mone&\dzero&\mone\\[-5pt]
\dzero&\dzero&\dzero&\mone&2&\mone&\dzero\\[-5pt]
\dzero&\dzero&\dzero&\dzero&\mone&2&\dzero\\[-5pt]
\dzero&\dzero&\dzero&\mone&\dzero&\dzero&2
\end{array}\right)}
&\fwbox{135pt}{\fwboxR{0pt}{\frac{1}{2}\!\!}\left(\begin{array}{@{}ccccccc@{}}
3&4&5&6&4&2&3\\[-5pt]
4&8&10&12&8&4&6\\[-5pt]
5&10&15&18&12&6&9\\[-5pt]
6&12&18&24&16&8&12\\[-5pt]
4&8&12&16&12&6&8\\[-5pt]
2&4&6&8&6&4&4\\[-5pt]
3&6&9&12&8&4&7\end{array}
\right)}\end{array}}\vspace{35pt}\label{e7_weigth_conventions}}

\eq{\fwbox{00pt}{\hspace{-30pt}\begin{array}{@{}c@{}c@{}c@{}c@{}}\fwboxL{0pt}{\hspace{2pt}\mathfrak{e}_{\r{8}}\!\!:}\fwbox{140pt}{\text{Dynkin diagram}}&\fwbox{125pt}{\text{Cartan matrix}}&\fwbox{135pt}{\text{weight metric}}\\
%&\rule{155pt}{1pt}&\rule{130pt}{1pt}&\rule{135pt}{1pt}\\
\begin{array}{@{}l@{}}\fwboxL{140pt}{\tikzBox{e8_dynkin_diagram}{\draw[edge](0,0)--(6*\dynkEdge,0);\draw[edge](4*\dynkEdge,0)--(4*\dynkEdge,1*\dynkEdge);
\node[dynkS](v1)at(0,0){};\node[anchor=90,inner sep=10pt]at(v1){$\alpha_\fwboxL{0pt}{1}$};
\node[dynkS](v2)at(\dynkEdge,0){};\node[anchor=90,inner sep=10pt]at(v2){$\alpha_\fwboxL{0pt}{2}$};
\node[dynkS](v3)at(2*\dynkEdge,0){};\node[anchor=90,inner sep=10pt]at(v3){$\alpha_\fwboxL{0pt}{3}$};
\node[dynkS](v4)at(3*\dynkEdge,0){};\node[anchor=90,inner sep=10pt]at(v4){$\alpha_\fwboxL{0pt}{4}$};
\node[dynkS](v5)at(4*\dynkEdge,0){};\node[anchor=90,inner sep=10pt]at(v5){$\alpha_\fwboxL{0pt}{5}$};
\node[dynkS](v6)at(5*\dynkEdge,0){};\node[anchor=90,inner sep=10pt]at(v6){$\alpha_\fwboxL{0pt}{6}$};
\node[dynkS](v7)at(6*\dynkEdge,0){};\node[anchor=90,inner sep=10pt]at(v7){$\alpha_\fwboxL{0pt}{7}$};
\node[dynkS](v8)at(4*\dynkEdge,1*\dynkEdge){};\node[anchor=180,inner sep=7.5pt]at(v8){$\alpha_\fwboxL{0pt}{8}$};
}}\\[-5pt]\hspace{10pt}\fwboxL{140pt}{\begin{array}{@{}l@{$\equivR\!$}l@{}}w(\mathbf{\r{ad}})&[10000000]\!\equivL\theta\\
w(\mathbf{\b{F}})&[10000000]\end{array}}\end{array}&
\fwbox{130pt}{\left(\begin{array}{@{}cccccccc@{}}
2&\mone&\dzero&\dzero&\dzero&\dzero&\dzero&\dzero\\[-5pt]
\mone&2&\mone&\dzero&\dzero&\dzero&\dzero&\dzero\\[-5pt]
\dzero&\mone&2&\mone&\dzero&\dzero&\dzero&\dzero\\[-5pt]
\dzero&\dzero&\mone&2&\mone&\dzero&\dzero&\dzero\\[-5pt]
\dzero&\dzero&\dzero&\mone&2&\mone&\dzero&\mone\\[-5pt]
\dzero&\dzero&\dzero&\dzero&\mone&2&\mone&\dzero\\[-5pt]
\dzero&\dzero&\dzero&\dzero&\dzero&\mone&2&\dzero\\[-5pt]
\dzero&\dzero&\dzero&\dzero&\mone&\dzero&\dzero&2
\end{array}\right)}&
\fwbox{135pt}{\left(\begin{array}{@{}cccccccc@{}}
2&3&4&5&6&4&2&3\\[-5pt]
3&6&8&10&12&8&4&6\\[-5pt]
4&8&12&15&18&12&6&9\\[-5pt]
5&10&15&20&24&16&8&12\\[-5pt]
6&12&18&24&30&20&10&15\\[-5pt]
4&8&12&16&20&14&7&10\\[-5pt]
2&4&6&8&10&7&4&5\\[-5pt]
3&6&9&12&15&10&5&8\end{array}\right)}\end{array}}\vspace{15pt}\label{e8_weight_conventions}}

\eq{\fwbox{00pt}{\hspace{-30pt}\begin{array}{@{}c@{}c@{}c@{}c@{}}\fwboxL{0pt}{\hspace{2pt}\mathfrak{f}_{\r{4}}\!\!:}\fwbox{140pt}{\text{Dynkin diagram}}&\fwbox{125pt}{\text{Cartan matrix}}&\fwbox{135pt}{\text{weight metric}}\\
%&\rule{155pt}{1pt}&\rule{130pt}{1pt}&\rule{135pt}{1pt}\\
\begin{array}{@{}l@{}}\vspace{-20pt}\fwbox{140pt}{\tikzBox{f4_dynkin_diagram}{\draw[edge](0,0)--(1,0);\draw[edge](2,0)--(3,0);
\draw[edge](1,0.15)--(2,0.15);\draw[edge](1,-0.15)--(2,-0.15);
\node[dynkL](v1)at(0,0){};\node[anchor=90,inner sep=10pt]at(v1){$\alpha_\fwboxL{0pt}{1}$};
\node[dynkL](v2)at(1,0){};\node[anchor=90,inner sep=10pt]at(v2){$\alpha_\fwboxL{0pt}{2}$};
\node[dynkS](v3)at(2,0){};\node[anchor=90,inner sep=10pt]at(v3){$\alpha_\fwboxL{0pt}{3}$};
\node[dynkS](v4)at(3,0){};\node[anchor=90,inner sep=10pt]at(v4){$\alpha_\fwboxL{0pt}{4}$};
}
}\\[30pt]\hspace{10pt}\fwboxL{140pt}{\begin{array}{@{}l@{$\equivR\!$}l@{}}
w(\mathbf{\r{ad}})&[0001]\!\equivL\theta\\
w(\mathbf{\b{F}})&[1000]\\[-25pt]
\end{array}}\end{array}\vspace{-00pt}&
\fwbox{130pt}{\left(\begin{array}{@{}cccc@{}}
2&\mone&\dzero&\dzero\\[-5pt]
\mone&2&\mone&\dzero\\[-5pt]
\dzero&\fwbox{4pt}{\text{-}}2\fwbox{4pt}{~}&2&\mone\\[-5pt]
\dzero&\dzero&\mone&2
\end{array}\right)}
&\fwbox{135pt}{\fwboxR{0pt}{\frac{1}{2}\!\!}\left(\begin{array}{@{}cccc@{}}
2&3&4&2\\[-5pt]
3&6&8&4\\[-5pt]
4&8&12&6\\[-5pt]
2&4&6&4\end{array}\right)}\end{array}}\vspace{30pt}\label{f4_weight_conventions}}

\eq{\fwbox{00pt}{\hspace{-30pt}\begin{array}{@{}c@{}c@{}c@{}c@{}}\fwboxL{0pt}{\hspace{2pt}\mathfrak{g}_{\r{2}}\!\!:}\fwbox{140pt}{\text{Dynkin diagram}}&\fwbox{125pt}{\text{Cartan matrix}}&\fwbox{135pt}{\text{weight metric}}\\
%&\rule{155pt}{1pt}&\rule{130pt}{1pt}&\rule{135pt}{1pt}\\
\begin{array}{@{}l@{}}\vspace{-20pt}\fwbox{140pt}{\tikzBox{g2_dynkin_diagram}{
\draw[edge](1,0.15)--(2,0.15);\draw[edge](1,-0.15)--(2,-0.15);\draw[edge](1,0)--(2,0);
\node[dynkL](v1)at(1,0){};\node[anchor=90,inner sep=10pt]at(v1){$\alpha_\fwboxL{0pt}{1}$};
\node[dynkS](v2)at(2,0){};\node[anchor=90,inner sep=10pt]at(v2){$\alpha_\fwboxL{0pt}{2}$};
}
}\\[10pt]\hspace{10pt}\fwboxL{140pt}{\begin{array}{@{}l@{$\equivR\!$}l@{}}
w(\mathbf{\r{ad}})&[01]\!\equivL\theta\\
w(\mathbf{\b{F}})&[10]\\[-35pt]
\end{array}}\end{array}\vspace{-20pt}&
\fwbox{130pt}{\Big(\begin{array}{@{}cc@{}}
2&\mone\\[-5pt]
\fwbox{4pt}{\text{-}}3\fwbox{4pt}{~}&2
\end{array}\Big)}
&\fwbox{135pt}{\fwboxR{0pt}{\frac{1}{3}}\Big(\begin{array}{@{}cc@{}}6&3\\[-5pt]3&2\end{array}\Big)}\\[35pt]\end{array}}\vspace{-15pt}\label{g2_weight_conventions}}

%================================================================================================================
%    References (& /Document)
%================================================================================================================
\newpage
%\bibliographystyle{physics}
%\bibliography{colour_refs}

\begin{thebibliography}{10}

\bibitem{Bern:1987tw}
Z.~Bern and D.~A. Kosower, ``{A New Approach to One Loop Calculations in Gauge
  Theories},''
\href{http://dx.doi.org/10.1103/PhysRevD.38.1888}{{\em Phys. Rev.} {\bf D38}
  (1988)  1888}.
%%CITATION = PHRVA,D38,1888;%%.

\bibitem{Bern:1991aq}
Z.~Bern and D.~A. Kosower, ``{The Computation of Loop Amplitudes in Gauge
  Theories},'' \href{http://dx.doi.org/10.1016/0550-3213(92)90134-W}{{\em Nucl.
  Phys. B} {\bf 379} (1992)  451--561}.

\bibitem{Bern1991ColorDO}
Z.~Bern and D.~A. Kosower, ``{Color Decomposition of One-Loop Amplitudes in
  Gauge Theories},'' {\em Nuclear Physics} {\bf 362} (1991)  389--448.
  \url{https://api.semanticscholar.org/CorpusID:55474752}.

\bibitem{Bern:1993qk}
Z.~Bern, G.~Chalmers, L.~J. Dixon, and D.~A. Kosower, ``{One-Loop $N$ Gluon
  Amplitudes with Maximal Helicity Violation via Collinear Limits},''
  \href{http://dx.doi.org/10.1103/PhysRevLett.72.2134}{{\em Phys. Rev. Lett.}
  {\bf 72} (1994)  2134--2137},
\href{http://arxiv.org/abs/hep-ph/9312333}{{ arXiv:hep-ph/9312333}}.
%%CITATION = HEP-PH/9312333;%%.

\bibitem{Bern:2004ky}
Z.~Bern, V.~Del~Duca, L.~J. Dixon, and D.~A. Kosower, ``{All Non-Maximally-
  Helicity-Violating One-Loop Seven-Gluon Amplitudes in $\mathcal{N}\!=\!4$
  Super-Yang-Mills Theory},''
  \href{http://dx.doi.org/10.1103/PhysRevD.71.045006}{{\em Phys. Rev.} {\bf
  D71} (2005)  045006},
\href{http://arxiv.org/abs/hep-th/0410224}{{ arXiv:hep-th/0410224}}.
%%CITATION = HEP-TH/0410224;%%.

\bibitem{Bern:2004bt}
Z.~Bern, L.~J. Dixon, and D.~A. Kosower, ``{All Next-to-Maximally
  Helicity-Violating One-Loop Gluon Amplitudes in $\mathcal{N}\!=\!4$
  Super-Yang-Mills Theory},''
  \href{http://dx.doi.org/10.1103/PhysRevD.72.045014}{{\em Phys. Rev.} {\bf
  D72} (2005)  045014},
\href{http://arxiv.org/abs/hep-th/0412210}{{ arXiv:hep-th/0412210}}.
%%CITATION = HEP-TH/0412210;%%.

\bibitem{BCF}
R.~Britto, F.~Cachazo, and B.~Feng, ``{New Recursion Relations for Tree
  Amplitudes of Gluons},''
  \href{http://dx.doi.org/10.1016/j.nuclphysb.2005.02.030}{{\em Nucl.Phys.}
  {\bf B715} (2005)  499--522},
\href{http://arxiv.org/abs/hep-th/0412308}{{ arXiv:hep-th/0412308 [hep-th]}}.
%%CITATION = HEP-TH/0412308;%%.

\bibitem{BCFW}
R.~Britto, F.~Cachazo, B.~Feng, and E.~Witten, ``{Direct Proof of Tree-Level
  Recursion Relation in Yang- Mills Theory},''
  \href{http://dx.doi.org/10.1103/PhysRevLett.94.181602}{{\em Phys. Rev. Lett.}
  {\bf 94} (2005)  181602},
\href{http://arxiv.org/abs/hep-th/0501052}{{ arXiv:hep-th/0501052}}.
%%CITATION = HEP-TH/0501052;%%.

\bibitem{Arkani-Hamed:2008owk}
N.~Arkani-Hamed, F.~Cachazo, and J.~Kaplan, ``{What is the Simplest Quantum
  Field Theory?},'' \href{http://dx.doi.org/10.1007/JHEP09(2010)016}{{\em JHEP}
  {\bf 09} (2010)  016}, \href{http://arxiv.org/abs/0808.1446}{{
  arXiv:0808.1446 [hep-th]}}.

\bibitem{ArkaniHamed:2009dn}
N.~Arkani-Hamed, F.~Cachazo, C.~Cheung, and J.~Kaplan, ``{A Duality For The
  $S$-Matrix},'' \href{http://dx.doi.org/10.1007/JHEP03(2010)020}{{\em JHEP}
  {\bf 1003} (2010)  020},
\href{http://arxiv.org/abs/0907.5418}{{ arXiv:0907.5418 [hep-th]}}.
%%CITATION = ARXIV:0907.5418;%%.

\bibitem{ArkaniHamed:2009sx}
N.~Arkani-Hamed, J.~Bourjaily, F.~Cachazo, and J.~Trnka, ``{Local Spacetime
  Physics from the Grassmannian},''
  \href{http://dx.doi.org/10.1007/JHEP01(2011)108}{{\em JHEP} {\bf 1101} (2011)
   108},
\href{http://arxiv.org/abs/0912.3249}{{ arXiv:0912.3249 [hep-th]}}.
%%CITATION = ARXIV:0912.3249;%%.

\bibitem{ArkaniHamed:2009dg}
N.~Arkani-Hamed, J.~Bourjaily, F.~Cachazo, and J.~Trnka, ``{Unification of
  Residues and Grassmannian Dualities},''
  \href{http://dx.doi.org/10.1007/JHEP01(2011)049}{{\em JHEP} {\bf 1101} (2011)
   049},
\href{http://arxiv.org/abs/0912.4912}{{ arXiv:0912.4912 [hep-th]}}.
%%CITATION = ARXIV:0912.4912;%%.

\bibitem{Arkani-Hamed:2013jha}
N.~Arkani-Hamed and J.~Trnka, ``{The Amplituhedron},''
  \href{http://dx.doi.org/10.1007/JHEP10(2014)030}{{\em JHEP} {\bf 1410} (2014)
   30},
\href{http://arxiv.org/abs/1312.2007}{{ arXiv:1312.2007 [hep-th]}}.
%%CITATION = ARXIV:1312.2007;%%.

\bibitem{Arkani-Hamed:2014via}
N.~Arkani-Hamed, J.~L. Bourjaily, F.~Cachazo, and J.~Trnka, ``{Singularity
  Structure of Maximally Supersymmetric Scattering Amplitudes},''
  \href{http://dx.doi.org/10.1103/PhysRevLett.113.261603}{{\em Phys. Rev.
  Lett.} {\bf 113} (2014) no. 26, 261603},
\href{http://arxiv.org/abs/1410.0354}{{ arXiv:1410.0354 [hep-th]}}.
%%CITATION = ARXIV:1410.0354;%%.

\bibitem{Arkani-Hamed:2016byb}
N.~Arkani-Hamed, J.~L. Bourjaily, F.~Cachazo, A.~B. Goncharov, A.~Postnikov,
  and J.~Trnka, \href{http://dx.doi.org/10.1017/CBO9781316091548}{{\em
  {Grassmannian Geometry of Scattering Amplitudes}}}.
\newblock Cambridge University Press, 4, 2016.
\newblock \href{http://arxiv.org/abs/1212.5605}{{ arXiv:1212.5605 [hep-th]}}.

\bibitem{ArkaniHamed:2010gh}
N.~Arkani-Hamed, J.~L. Bourjaily, F.~Cachazo, and J.~Trnka, ``{Local Integrals
  for Planar Scattering Amplitudes},''
  \href{http://dx.doi.org/10.1007/JHEP06(2012)125}{{\em JHEP} {\bf 1206} (2012)
   125},
\href{http://arxiv.org/abs/1012.6032}{{ arXiv:1012.6032 [hep-th]}}.
%%CITATION = ARXIV:1012.6032;%%.

\bibitem{Bourjaily:2013mma}
J.~L. Bourjaily, S.~Caron-Huot, and J.~Trnka, ``{Dual-Conformal Regularization
  of Infrared Loop Divergences and the {\it Chiral} Box Expansion},''
  \href{http://dx.doi.org/10.1007/JHEP01(2015)001}{{\em JHEP} {\bf 1501} (2015)
   001},
\href{http://arxiv.org/abs/1303.4734}{{ arXiv:1303.4734 [hep-th]}}.
%%CITATION = ARXIV:1303.4734;%%.

\bibitem{Bourjaily:2015jna}
J.~L. Bourjaily and J.~Trnka, ``{Local Integrand Representations of All
  Two-Loop Amplitudes in Planar SYM},''
  \href{http://dx.doi.org/10.1007/JHEP08(2015)119}{{\em JHEP} {\bf 08} (2015)
  119},
\href{http://arxiv.org/abs/1505.05886}{{ arXiv:1505.05886 [hep-th]}}.
%%CITATION = ARXIV:1505.05886;%%.

\bibitem{Bourjaily:2017wjl}
J.~L. Bourjaily, E.~Herrmann, and J.~Trnka, ``{Prescriptive Unitarity},''
  \href{http://dx.doi.org/10.1007/JHEP06(2017)059}{{\em JHEP} {\bf 06} (2017)
  059},
\href{http://arxiv.org/abs/1704.05460}{{ arXiv:1704.05460 [hep-th]}}.
%%CITATION = ARXIV:1704.05460;%%.

\bibitem{Bourjaily:2010wh}
J.~L. Bourjaily, ``{Efficient Tree-Amplitudes in $\mathcal{N}\!=\!4$: Automatic
  BCFW Recursion in {\sc Mathematica}},''
\href{http://arxiv.org/abs/1011.2447}{{ arXiv:1011.2447 [hep-ph]}}.
%%CITATION = ARXIV:1011.2447;%%.

\bibitem{Bourjaily:2023uln}
J.~L. Bourjaily, ``{Computational Tools for Trees in Gauge Theory and
  Gravity},'' \href{http://arxiv.org/abs/2312.17745}{{ arXiv:2312.17745
  [hep-th]}}.

\bibitem{Bourjaily:2011hi}
J.~L. Bourjaily, A.~DiRe, A.~Shaikh, M.~Spradlin, and A.~Volovich, ``{The
  Soft-Collinear Bootstrap: $\mathcal{N}\!=\!4$ Yang-Mills Amplitudes at Six
  and Seven Loops},'' \href{http://dx.doi.org/10.1007/JHEP03(2012)032}{{\em
  JHEP} {\bf 1203} (2012)  032},
\href{http://arxiv.org/abs/1112.6432}{{ arXiv:1112.6432 [hep-th]}}.
%%CITATION = ARXIV:1112.6432;%%.

\bibitem{Bourjaily:2015bpz}
J.~L. Bourjaily, P.~Heslop, and V.-V. Tran, ``{Perturbation Theory at Eight
  Loops: Novel Structures and the Breakdown of Manifest Conformality in
  $\mathcal{N}\!=\!4$ Supersymmetric Yang-Mills Theory},''
  \href{http://dx.doi.org/10.1103/PhysRevLett.116.191602}{{\em Phys. Rev.
  Lett.} {\bf 116} (2016) no. 19, 191602},
\href{http://arxiv.org/abs/1512.07912}{{ arXiv:1512.07912 [hep-th]}}.
%%CITATION = ARXIV:1512.07912;%%.

\bibitem{Bourjaily:2016evz}
J.~L. Bourjaily, P.~Heslop, and V.-V. Tran, ``{Amplitudes and Correlators to
  Ten Loops Using Simple, Graphical Bootstraps},''
  \href{http://dx.doi.org/10.1007/JHEP11(2016)125}{{\em JHEP} {\bf 11} (2016)
  125},
\href{http://arxiv.org/abs/1609.00007}{{ arXiv:1609.00007 [hep-th]}}.
%%CITATION = ARXIV:1609.00007;%%.

\bibitem{He:2024cej}
S.~He, C.~Shi, Y.~Tang, and Y.-Q. Zhang, ``{The Cusp Limit of Correlators and a
  New Graphical Bootstrap for Correlators/Amplitudes to Eleven Loops},''
  \href{http://dx.doi.org/10.1007/JHEP03(2025)192}{{\em JHEP} {\bf 03} (2025)
  192}, \href{http://arxiv.org/abs/2410.09859}{{ arXiv:2410.09859 [hep-th]}}.

\bibitem{Bourjaily:2025iad}
J.~L. Bourjaily, S.~He, C.~Shi, and Y.~Tang, ``{The Four-Point Correlator of
  Planar sYM at Twelve Loops},'' \href{http://arxiv.org/abs/2503.15593}{{
  arXiv:2503.15593 [hep-th]}}.

\bibitem{BCJ}
Z.~Bern, J.~Carrasco, and H.~Johansson, ``{New Relations for Gauge-Theory
  Amplitudes},'' \href{http://dx.doi.org/10.1103/PhysRevD.78.085011}{{\em Phys.
  Rev.} {\bf D78} (2008)  085011},
\href{http://arxiv.org/abs/0805.3993}{{ arXiv:0805.3993 [hep-ph]}}.
%%CITATION = ARXIV:0805.3993;%%.

\bibitem{Bern:2019prr}
Z.~Bern, J.~J.~M. Carrasco, M.~Chiodaroli, H.~Johansson, and R.~Roiban, ``{The
  Duality Between Color and Kinematics and its Applications},''
  \href{http://arxiv.org/abs/1909.01358}{{ arXiv:1909.01358 [hep-th]}}.

\bibitem{Carrasco:2011mn}
J.~J. Carrasco and H.~Johansson, ``{Five-Point Amplitudes in
  $\mathcal{N}\!=\!4$ Super-Yang-Mills Theory and $\mathcal{N}\!=\!8$
  Supergravity},'' \href{http://dx.doi.org/10.1103/PhysRevD.85.025006}{{\em
  Phys. Rev.} {\bf D85} (2012)  025006},
\href{http://arxiv.org/abs/1106.4711}{{ arXiv:1106.4711 [hep-th]}}.
%%CITATION = ARXIV:1106.4711;%%.

\bibitem{Carrasco:2012ca}
J.~J.~M. Carrasco, M.~Chiodaroli, M.~G\"unaydin, and R.~Roiban, ``{One-Loop
  Four-Point Amplitudes in Pure and Matter-Coupled $\mathcal{N}\!\leq\!4$
  Supergravity},'' \href{http://dx.doi.org/10.1007/JHEP03(2013)056}{{\em JHEP}
  {\bf 03} (2013)  056}, \href{http://arxiv.org/abs/1212.1146}{{
  arXiv:1212.1146 [hep-th]}}.

\bibitem{Bjerrum-Bohr:2013iza}
N.~E.~J. Bjerrum-Bohr, T.~Dennen, R.~Monteiro, and D.~O'Connell, ``{Integrand
  Oxidation and One-Loop Colour-Dual Numerators in $\mathcal{N}\!\!=\!\!4$
  Gauge Theory},'' \href{http://dx.doi.org/10.1007/JHEP07(2013)092}{{\em JHEP}
  {\bf 1307} (2013)  092},
\href{http://arxiv.org/abs/1303.2913}{{ arXiv:1303.2913 [hep-th]}}.
%%CITATION = ARXIV:1303.2913;%%.

\bibitem{Monteiro:2013rya}
R.~Monteiro and D.~O'Connell, ``{The Kinematic Algebras from the Scattering
  Equations},'' \href{http://dx.doi.org/10.1007/JHEP03(2014)110}{{\em JHEP}
  {\bf 1403} (2014)  110},
\href{http://arxiv.org/abs/1311.1151}{{ arXiv:1311.1151 [hep-th]}}.
%%CITATION = ARXIV:1311.1151;%%.

\bibitem{He:2015wgf}
S.~He, R.~Monteiro, and O.~Schlotterer, ``{String-Inspired BCJ Numerators for
  One-Loop MHV Amplitudes},''
  \href{http://dx.doi.org/10.1007/JHEP01(2016)171}{{\em JHEP} {\bf 01} (2016)
  171},
\href{http://arxiv.org/abs/1507.06288}{{ arXiv:1507.06288 [hep-th]}}.
%%CITATION = ARXIV:1507.06288;%%.

\bibitem{Mogull:2015adi}
G.~Mogull and D.~O'Connell, ``{Overcoming Obstacles to Colour-Kinematics
  Duality at Two Loops},''
  \href{http://dx.doi.org/10.1007/JHEP12(2015)135}{{\em JHEP} {\bf 12} (2015)
  135}, \href{http://arxiv.org/abs/1511.06652}{{ arXiv:1511.06652 [hep-th]}}.

\bibitem{Johansson:2017bfl}
H.~Johansson, G.~K\"alin, and G.~Mogull, ``{Two-Loop Supersymmetric QCD and
  Half-Maximal Supergravity Amplitudes},''
  \href{http://dx.doi.org/10.1007/JHEP09(2017)019}{{\em JHEP} {\bf 09} (2017)
  019}, \href{http://arxiv.org/abs/1706.09381}{{ arXiv:1706.09381 [hep-th]}}.

\bibitem{Bern:2017yxu}
Z.~Bern, J.~J.~M. Carrasco, W.-M. Chen, H.~Johansson, and R.~Roiban, ``{Gravity
  Amplitudes as Generalized Double Copies of Gauge-Theory Amplitudes},''
  \href{http://dx.doi.org/10.1103/PhysRevLett.118.181602}{{\em Phys. Rev.
  Lett.} {\bf 118} (2017) no. 18, 181602},
  \href{http://arxiv.org/abs/1701.02519}{{ arXiv:1701.02519 [hep-th]}}.

\bibitem{Bern:2017ucb}
Z.~Bern, J.~J.~M. Carrasco, W.-M. Chen, H.~Johansson, R.~Roiban, and M.~Zeng,
  ``{The Five-Loop Four-Point Integrand of $\mathcal{N}\!=\!8$ Supergravity as
  a Generalized Double Copy},''
  \href{http://dx.doi.org/10.1103/PhysRevD.96.126012}{{\em Phys. Rev. D} {\bf
  96} (2017) no. 12, 126012}, \href{http://arxiv.org/abs/1708.06807}{{
  arXiv:1708.06807 [hep-th]}}.

\bibitem{Edison:2022jln}
A.~Edison, S.~He, H.~Johansson, O.~Schlotterer, F.~Teng, and Y.~Zhang,
  ``{Perfecting One-Loop BCJ Numerators in SYM and Supergravity},''
  \href{http://dx.doi.org/10.1007/JHEP02(2023)164}{{\em JHEP} {\bf 02} (2023)
  164}, \href{http://arxiv.org/abs/2211.00638}{{ arXiv:2211.00638 [hep-th]}}.

\bibitem{Porkert:2022efy}
F.~Porkert and O.~Schlotterer, ``{One-Loop Amplitudes in Einstein-Yang-Mills
  from Forward Limits},'' \href{http://dx.doi.org/10.1007/JHEP02(2023)122}{{\em
  JHEP} {\bf 02} (2023)  122}, \href{http://arxiv.org/abs/2201.12072}{{
  arXiv:2201.12072 [hep-th]}}.

\bibitem{Arkani-Hamed:2014bca}
N.~Arkani-Hamed, J.~L. Bourjaily, F.~Cachazo, A.~Postnikov, and J.~Trnka,
  ``{On-Shell Structures of MHV Amplitudes Beyond the Planar Limit},''
  \href{http://dx.doi.org/10.1007/JHEP06(2015)179}{{\em JHEP} {\bf 06} (2015)
  179},
\href{http://arxiv.org/abs/1412.8475}{{ arXiv:1412.8475 [hep-th]}}.
%%CITATION = ARXIV:1412.8475;%%.

\bibitem{Bourjaily:2018omh}
J.~L. Bourjaily, E.~Herrmann, and J.~Trnka, ``{Amplitudes at Infinity},''
  \href{http://dx.doi.org/10.1103/PhysRevD.99.066006}{{\em Phys. Rev. D} {\bf
  99} (2019) no. 6, 066006}, \href{http://arxiv.org/abs/1812.11185}{{
  arXiv:1812.11185 [hep-th]}}.

\bibitem{Bourjaily:2019gqu}
J.~L. Bourjaily, E.~Herrmann, C.~Langer, A.~J. McLeod, and J.~Trnka,
  ``{All-Multiplicity Nonplanar Amplitude Integrands in Maximally
  Supersymmetric Yang-Mills Theory at Two Loops},''
  \href{http://dx.doi.org/10.1103/PhysRevLett.124.111603}{{\em Phys. Rev.
  Lett.} {\bf 124} (2020) no. 11, 111603},
  \href{http://arxiv.org/abs/1911.09106}{{ arXiv:1911.09106 [hep-th]}}.

\bibitem{Bourjaily:2021hcp}
J.~L. Bourjaily, C.~Langer, and Y.~Zhang, ``{Illustrations of Integrand-Basis
  Building at Two Loops},'' \href{http://arxiv.org/abs/2112.05157}{{
  arXiv:2112.05157 [hep-th]}}.

\bibitem{Bourjaily:2021iyq}
J.~L. Bourjaily, C.~Langer, and Y.~Zhang, ``{All Two-Loop, Color-Dressed,
  Six-Point Amplitude Integrands in sYM},''
  \href{http://dx.doi.org/10.1103/PhysRevD.105.105015}{{\em Phys. Rev. D} {\bf
  105} (2022) no. 10, 105015}, \href{http://arxiv.org/abs/2112.06934}{{
  arXiv:2112.06934 [hep-th]}}.

\bibitem{Bourjaily:2024jbt}
J.~L. Bourjaily, M.~Plesser, and C.~Vergu, ``{The Many Colours of
  Amplitudes},'' \href{http://arxiv.org/abs/2412.21189}{{ arXiv:2412.21189
  [hep-th]}}.

\bibitem{Benincasa:2007xk}
P.~Benincasa and F.~Cachazo, ``{Consistency Conditions on the $S$-Matrix of
  Massless Particles},''
\href{http://arxiv.org/abs/0705.4305}{{ arXiv:0705.4305 [hep-th]}}.
%%CITATION = ARXIV:0705.4305;%%.

\bibitem{DelDuca:1999rs}
V.~Del~Duca, L.~J. Dixon, and F.~Maltoni, ``{New Color Decompositions for Gauge
  Amplitudes at Tree and Loop Level},''
  \href{http://dx.doi.org/10.1016/S0550-3213(99)00809-3}{{\em Nucl. Phys.} {\bf
  B571} (2000)  51--70},
\href{http://arxiv.org/abs/hep-ph/9910563}{{ arXiv:hep-ph/9910563 [hep-ph]}}.
%%CITATION = HEP-PH/9910563;%%.

\bibitem{Johansson:2015oia}
H.~Johansson and A.~Ochirov, ``{Color-Kinematics Duality for QCD Amplitudes},''
  \href{http://dx.doi.org/10.1007/JHEP01(2016)170}{{\em JHEP} {\bf 01} (2016)
  170}, \href{http://arxiv.org/abs/1507.00332}{{ arXiv:1507.00332 [hep-ph]}}.

\bibitem{Melia:2015ika}
T.~Melia, ``{Proof of a New Colour Decomposition for QCD Amplitudes},''
  \href{http://dx.doi.org/10.1007/JHEP12(2015)107}{{\em JHEP} {\bf 12} (2015)
  107}, \href{http://arxiv.org/abs/1509.03297}{{ arXiv:1509.03297 [hep-ph]}}.

\bibitem{Ochirov:2019mtf}
A.~Ochirov and B.~Page, ``{Multi-Quark Colour Decompositions from Unitarity},''
  \href{http://dx.doi.org/10.1007/JHEP10(2019)058}{{\em JHEP} {\bf 10} (2019)
  058}, \href{http://arxiv.org/abs/1908.02695}{{ arXiv:1908.02695 [hep-ph]}}.

\bibitem{Zeppenfeld:1988bz}
D.~Zeppenfeld, ``{Diagonalization of Color Factors},''
  \href{http://dx.doi.org/10.1142/S0217751X88000916}{{\em Int. J. Mod. Phys. A}
  {\bf 3} (1988)  2175--2179}.

\bibitem{Keppeler2012OrthogonalMB}
S.~Keppeler and M.~Sj{\"o}dahl, ``{Orthogonal Multiplet Bases in $SU(N_c)$
  Color Space},'' {\em Journal of High Energy Physics} {\bf 2012} (2012)
  1--50. \url{https://api.semanticscholar.org/CorpusID:6133148}.

\bibitem{Sjodahl2015}
M.~Sjodahl and J.~Thoren, ``{Decomposing Color Structure into Multiplet
  Bases},'' \href{http://arxiv.org/abs/1507.03814}{{ arXiv:1507.03814
  [hep-ph]}}.% \url{https://arxiv.org/abs/1507.03814}.

\bibitem{Sjodahl2024}
M.~Sjodahl, ``{Orthogonal Color Bases for Exotic Representations},''
  \href{http://arxiv.org/abs/2412.07390}{{ arXiv:2412.07390 [hep-ph]}}.
%  \url{https://arxiv.org/abs/2412.07390}.

\bibitem{Cvitanovi1976GroupTF}
P.~Cvitanovi{\'c}, ``{Group Theory for Feynman Diagrams in non-Abelian Gauge
  Theories},'' {\em Physical Review D} {\bf 14} (1976)  1536--1553.
  \url{https://api.semanticscholar.org/CorpusID:6985555}.

\bibitem{Slansky:1981yr}
R.~Slansky, ``{Group Theory for Unified Model Building},''
  \href{http://dx.doi.org/10.1016/0370-1573(81)90092-2}{{\em Phys. Rept.} {\bf
  79} (1981)  1--128}.

\bibitem{Cornwell:1997ke}
J.~F. Cornwell, {\em {Group Theory in Physics: An Introduction}}.
\newblock 1997.

\bibitem{DiFrancesco:1997nk}
P.~Di~Francesco, P.~Mathieu, and D.~Senechal,
  \href{http://dx.doi.org/10.1007/978-1-4612-2256-9}{{\em {Conformal Field
  Theory}}}.
\newblock Graduate Texts in Contemporary Physics. Springer-Verlag, New York,
  1997.

\bibitem{fulton2013}
W.~Fulton and J.~Harris, {\em {Representation Theory: a First Course}},
  vol.~129.
\newblock Springer Science \& Business Media, 2013.

\bibitem{MR3674995}
V.~Turaev and A.~Virelizier,
  \href{http://dx.doi.org/10.1007/978-3-319-49834-8}{{\em {Monoidal Categories
  and Topological Field Theory}}}, vol.~322 of {\em Progress in Mathematics}.
\newblock Birkh\"{a}user/Springer, Cham, 2017.
%\newblock \url{https://doi.org/10.1007/978-3-319-49834-8}.

\bibitem{beer2018categories}
K.~Beer, D.~Bondarenko, A.~Hahn, M.~Kalabakov, N.~Knust, L.~Niermann, T.~J.
  Osborne, C.~Schridde, S.~Seckmeyer, D.~E. Stiegemann, and R.~Wolf, ``{From
  Categories to Anyons: a Travelogue}.''
  \url{https://arxiv.org/abs/1811.06670}, 2018.

\bibitem{MR182649}
G.~M. Kelly, ``{On {M}ac{L}ane's Conditions for Coherence of Natural
  Associativities, Commutativities, etc.},''
  \href{http://dx.doi.org/10.1016/0021-8693(64)90018-3}{{\em J. Algebra} {\bf
  1} (1964)  397--402}.% \url{https://doi.org/10.1016/0021-8693(64)90018-3}.

\bibitem{Selinger2011}
P.~Selinger, {\em {A Survey of Graphical Languages for Monoidal Categories}},
  \href{http://dx.doi.org/10.1007/978-3-642-12821-9_4}{pp.~289--355}.
\newblock Springer Berlin Heidelberg, Berlin, Heidelberg, 2011.
%\newblock \url{https://doi.org/10.1007/978-3-642-12821-9_4}.

\bibitem{FockGoncharov2007}
V.~V. Fock and A.~B. Goncharov, ``{Cluster Ensembles, Quantization and the
  Dilogarithm II: The Intertwiner},''
  \href{http://arxiv.org/abs/math/0702398}{{ arXiv:math/0702398 [math.QA]}}.
%  \url{https://arxiv.org/abs/math/0702398}.

\bibitem{Golden:2013xva}
J.~Golden, A.~B. Goncharov, M.~Spradlin, C.~Vergu, and A.~Volovich, ``{Motivic
  Amplitudes and Cluster Coordinates},''
  \href{http://dx.doi.org/10.1007/JHEP01(2014)091}{{\em JHEP} {\bf 1401} (2014)
   091},
\href{http://arxiv.org/abs/1305.1617}{{ arXiv:1305.1617 [hep-th]}}.
%%CITATION = ARXIV:1305.1617;%%.

\bibitem{Biedenharn1971}
J.~D. Louck and L.~C. Biedenharn, ``{Identity Satisfied by the Racah
  Coefficients of $\mathfrak{u}_n$},''
  \href{http://dx.doi.org/10.1063/1.1665571}{{\em J. Math. Phys.} {\bf 12}
  (1971)  173--177}.

\bibitem{birdtracks}
P.~Cvitanovi\'{c}, \href{http://dx.doi.org/10.1515/9781400837670}{{\em {Group
  Theory}}}.
\newblock Princeton University Press, Princeton, NJ, 2008.
\newblock \url{https://doi.org/10.1515/9781400837670}.
\newblock Birdtracks, Lie's, and exceptional groups.

\bibitem{Searle1988}
B.~G. Searle, ``{Calculation of $6j$ Symbols},''
\newblock 1988.
\newblock \url{https://api.semanticscholar.org/CorpusID:125000678}.

\bibitem{Bern:2007ct}
Z.~Bern, J.~Carrasco, H.~Johansson, and D.~Kosower, ``{Maximally Supersymmetric
  Planar Yang-Mills Amplitudes at Five Loops},''
  \href{http://dx.doi.org/10.1103/PhysRevD.76.125020}{{\em Phys. Rev. D} {\bf
  D76} (2007)  125020},
\href{http://arxiv.org/abs/0705.1864}{{ arXiv:0705.1864 [hep-th]}}.
%%CITATION = ARXIV:0705.1864;%%.

\bibitem{Bourjaily:2019iqr}
J.~L. Bourjaily, E.~Herrmann, C.~Langer, A.~J. McLeod, and J.~Trnka,
  ``{Prescriptive Unitarity for Non-Planar Six-Particle Amplitudes at Two
  Loops},'' \href{http://dx.doi.org/10.1007/JHEP12(2019)073}{{\em JHEP} {\bf
  12} (2019)  073}, \href{http://arxiv.org/abs/1909.09131}{{ arXiv:1909.09131
  [hep-th]}}.

\bibitem{Carrasco:2021otn}
J.~J.~M. Carrasco, A.~Edison, and H.~Johansson, ``{Maximal Super-Yang-Mills at
  Six Loops via Novel Integrand Bootstrap},''
  \href{http://arxiv.org/abs/2112.05178}{{ arXiv:2112.05178 [hep-th]}}.

\bibitem{KK}
R.~Kleiss and H.~Kuijf, ``{Multi-Gluon Cross-Sections and Five Jet Production
  at Hadron Colliders},''
\href{http://dx.doi.org/10.1016/0550-3213(89)90574-9}{{\em Nucl. Phys.} {\bf
  B312} (1989)  616}.
%%CITATION = NUPHA,B312,616;%%.

\bibitem{Berends:1987me}
F.~A. Berends and W.~T. Giele, ``{Recursive Calculations for Processes with $n$
  Gluons},''
\href{http://dx.doi.org/10.1016/0550-3213(88)90442-7}{{\em Nucl. Phys.} {\bf
  B306} (1988)  759}.
%%CITATION = NUPHA,B306,759;%%.

\bibitem{Mangano:1987xk}
M.~L. Mangano, S.~J. Parke, and Z.~Xu, ``{Duality and Multi-Gluon
  Scattering},''
\href{http://dx.doi.org/10.1016/0550-3213(88)90001-6}{{\em Nucl. Phys.} {\bf
  B298} (1988)  653--672}.
%%CITATION = NUPHA,B298,653;%%.

\bibitem{Melia:2013bta}
T.~Melia, ``{Dyck Words and Multiquark Primitive Amplitudes},''
  \href{http://dx.doi.org/10.1103/PhysRevD.88.014020}{{\em Phys. Rev. D} {\bf
  88} (2013) no. 1, 014020}, \href{http://arxiv.org/abs/1304.7809}{{
  arXiv:1304.7809 [hep-ph]}}.

\bibitem{Melia:2013epa}
T.~Melia, ``{Getting More Flavor out of One-Flavor QCD},''
  \href{http://dx.doi.org/10.1103/PhysRevD.89.074012}{{\em Phys. Rev. D} {\bf
  89} (2014) no. 7, 074012}, \href{http://arxiv.org/abs/1312.0599}{{
  arXiv:1312.0599 [hep-ph]}}.

\bibitem{Melia:2013xok}
T.~Melia, ``{Dyck Words and Multi-Quark Amplitudes},''
  \href{http://dx.doi.org/10.22323/1.197.0031}{{\em PoS} {\bf RADCOR2013}
  (2013)  031}.

\bibitem{tools_for_multiflavour_amps}
J.~L. Bourjaily, M.~Plesser, and P.~Velie, ``{Arbitrary Multi-Quark Amplitudes
  in QCD via sYM}.'' In preparation.

\bibitem{oeis-riordan}
{OEIS Foundation Inc. (2024)}, ``{The Riordan Numbers, Entry A005043 in The
  On-Line Encyclopedia of Integer Sequences}.''
\newblock \url{https://oeis.org/A005043}.

\bibitem{Naculich_2012}
S.~G. Naculich,
  \href{http://dx.doi.org/10.1016/j.physletb.2011.12.010}{``All-Loop
  Group-Theory Constraints for Color-Ordered $\mathfrak{su}_{n}$ Gauge-Theory
  amplitudes,''{\em Physics Letters B} {\bf 707} (Jan., 2012)  191--197}.
%  \url{http://dx.doi.org/10.1016/j.physletb.2011.12.010}.

\bibitem{naculich2024}
S.~G. Naculich and A.~Osathapan, ``All-Loop Group-Theory Constraints for
  Four-Point Amplitudes of $\mathfrak{su}_{n}$, $\mathfrak{so}_{n}$, and $\mathfrak{sp}_n$ Gauge Theories,''
  2024.
\newblock \url{https://arxiv.org/abs/2407.03403}.

\bibitem{Stembridge2003}
J.~R. Stembridge, ``{Multiplicity-Free Products and Restrictions of Weyl
  Characters},'' {\em Representation Theory of The American Mathematical
  Society} {\bf 7} (2003)  404--439.
  \url{https://api.semanticscholar.org/CorpusID:14305161}.

\bibitem{magma}
W.~Bosma, J.~Cannon, and C.~Playoust, ``{The \texttt{{M}agma} Algebra System.
  {I}. {T}he User Language},''
  \href{http://dx.doi.org/10.1006/jsco.1996.0125}{{\em J. Symbolic Comput.}
  {\bf 24} (1997) no. 3-4, 235--265}.
%  \url{http://dx.doi.org/10.1006/jsco.1996.0125}. Computational Algebra and
%  Number Theory (London, 1993).

\bibitem{lie_algebra_representation_tensor_tools}
J.~L. Bourjaily, M.~Plesser, and P.~Velie, ``{Rational Representations of
  Simple Lie Algebras}.'' In preparation.

\end{thebibliography}
\providecommand{\href}[2]{#2}\begingroup\raggedright\endgroup

\end{document}